# The Pierre Auger Observatory: Contributions to the 33rd International Cosmic Ray Conference (ICRC 2013)

## The Pierre Auger Collaboration

A. Aab[40], P. Abreu[62], M. Aglietta[51], M. Ahlers[92], E.J. Ahn[80], I.F.M. Albuquerque[16], I. Allekotte[1], J. Allen[84], P. Allison[86], A. Almela[11,8], J. Alvarez Castillo[55], J. Alvarez-Muñiz[73], R. Alves Batista[39], M. Ambrosio[44], A. Aminaei[56], L. Anchordoqui[93], S. Andringa[62], T. Antičić[23], C. Aramo[44], F. Arqueros[70], H. Asorey[1], P. Assis[62], J. Aublin[29], M. Ave[73], M. Avenier[30], G. Avila[10], A.M. Badescu[66], K.B. Barber[12], R. Bardenet[28], J. Bäuml[34], C. Baus[36], J.J. Beatty[86], K.H. Becker[33], A. Bellétoile[32], J.A. Bellido[12], S. BenZvi[92], C. Berat[30], X. Bertou[1], P.L. Biermann[37], P. Billoir[29], F. Blanco[70], M. Blanco[29], C. Bleve[33], H. Blümer[36,34], M. Boháčová[25], D. Boncioli[50,45], C. Bonifazi[21], R. Bonino[51], N. Borodai[60], J. Brack[78], I. Brancus[63], P. Brogueira[62], W.C. Brown[79], P. Buchholz[40], A. Bueno[72], R.E. Burton[76], M. Buscemi[44], K.S. Caballero-Mora[73,87], B. Caccianiga[43], L. Caccianiga[29], M. Candusso[45], L. Caramete[37], R. Caruso[46], A. Castellina[51], G. Cataldi[48], L. Cazon[62], R. Cester[47], S.H. Cheng[87], A. Chiavassa[51], J.A. Chinellato[17], J. Chudoba[25], M. Cilmo[44], R.W. Clay[12], G. Cocciolo[48], R. Colalillo[44], L. Collica[43], M.R. Coluccia[48], R. Conceição[62], F. Contreras[9], H. Cook[74], M.J. Cooper[12], S. Coutu[87], C.E. Covault[76], A. Criss[87], J. Cronin[88], A. Curutiu[37], R. Dallier[32,31], B. Daniel[17], S. Dasso[5,3], K. Daumiller[34], B.R. Dawson[12], R.M. de Almeida[22], M. De Domenico[46], S.J. de Jong[56,58], G. De La Vega[7], W.J.M. de Mello Junior[17], J.R.T. de Mello Neto[21], I. De Mitri[48], V. de Souza[15], K.D. de Vries[57], L. del Peral[71], O. Deligny[27], H. Dembinski[34], N. Dhital[83], C. Di Giulio[45], J.C. Diaz[83], M.L. Díaz Castro[13], P.N. Diep[94], F. Diogo[62], C. Dobrigkeit [17], W. Docters[57], J.C. D'Olivo[55], P.N. Dong[94,27], A. Dorofeev[78], J.C. dos Anjos[13], M.T. Dova[4], J. Ebr[25], R. Engel[34], M. Erdmann[38], C.O. Escobar[80,17], J. Espadanal[62], A. Etchegoyen[8,11], P. Facal San Luis[88], H. Falcke[56,59,58], K. Fang[88], G. Farrar[84], A.C. Fauth[17], N. Fazzini[80], A.P. Ferguson[76], B. Fick[83], J.M. Figueira[8,34], A. Filevich[8], A. Filipčič[67,68], N. Foerster[40], B.D. Fox[89], C.E. Fracchiolla[78], E.D. Fraenkel[57], O. Fratu[66], U. Fröhlich[40], B. Fuchs[36], R. Gaior[29], R.F. Gamarra[8], S. Gambetta[41], B. García[7], S.T. Garcia Roca[73], D. García-Gámez[28], D. Garcia-Pinto[70], G. Garilli[46], A. Gascon Bravo[72], H. Gemmeke[35], P.L. Ghia[29], M. Giller[61], J. Gitto[7], C. Glaser[38], H. Glass[80], F. Gomez Albarracin[4], M. Gómez Berisso[1], P.F. Gómez Vitale[10], P. Gonçalves[62], J.G. Gonzalez[36], B. Gookin[78], A. Gorgi[51], P. Gorham[89], P. Gouffon[16], S. Grebe[56,58], N. Griffith[86], A.F. Grillo[50], T.D. Grubb[12], Y. Guardincerri[3], F. Guarino[44], G.P. Guedes[18], P. Hansen[4], D. Harari[1], T.A. Harrison[12], J.L. Harton[78], A. Haungs[34], T. Hebbeker[38], D. Heck[34], A.E. Herve[12], G.C. Hill[12], C. Hojvat[80], N. Hollon[88], P. Homola[40,60], J.R. Hörandel[56,58], P. Horvath[26], M. Hrabovský[26,25], D. Huber[36], T. Huege[34], A. Insolia[46], P.G. Isar[64], S. Jansen[56,58], C. Jarne[4], M. Josebachuili[8,34], K. Kadija[23], O. Kambeitz[36], K.H. Kampert[33], P. Karhan[24], P. Kasper[80], I. Katkov[36], B. Kégl[28], B. Keilhauer[34], A. Keivani[82], E. Kemp[17], R.M. Kieckhafer[83], H.O. Klages[34], M. Kleifges[35], J. Kleinfeller[9,34], J. Knapp[74 d], R. Krause[38], N. Krohm[33], O. Krömer[35], D. Kruppke-Hansen[33], D. Kuempel[38], N. Kunka[35], G. La Rosa[49], D. LaHurd[76], L. Latronico[51], R. Lauer[91], M. Lauscher[38], P. Lautridou[32], S. Le Coz[30], M.S.A.B. Leão[14], D. Lebrun[30], P. Lebrun[80], M.A. Leigui de Oliveira[20], A. Letessier-Selvon[29], I. Lhenry-Yvon[27], K. Link[36], R. López[52], A. Lopez Agüera[73], K. Louedec[30], J. Lozano Bahilo[72], L. Lu[33,74], A. Lucero[8], M. Ludwig[36], H. Lyberis[21], M.C. Maccarone[49], C. Macolino[29], M. Malacari[12], S. Maldera[51], J. Maller[32], D. Mandat[25], P. Mantsch[80], A.G. Mariazzi[4], V. Marin[32], I.C. Mariş[29], H.R. Marquez Falcon[54], G. Marsella[48], D. Martello[48], L. Martin[32,31], H. Martinez[53], O. Martínez Bravo[52], D. Martraire[27], J.J. Masías Meza[3], H.J. Mathes[34], J. Matthews[82], J.A.J. Matthews[91], G. Matthiae[45], D. Maurel[34], D. Maurizio[13], E. Mayotte[77], P.O. Mazur[80], C. Medina[77], G. Medina-Tanco[55],

M. Melissas[36], D. Melo[8], E. Menichetti[47], A. Menshikov[35], S. Messina[57], R. Meyhandan[89], S. Mićanović[23], M.I. Micheletti[6], L. Middendorf[38], I.A. Minaya[70], L. Miramonti[43], B. Mitrica[63], L. Molina-Bueno[72], S. Mollerach[1], M. Monasor[88], D. Monnier Ragaigne[28], F. Montanet[30], B. Morales[55], C. Morello[51], J.C. Moreno[4], M. Mostafá[78], C.A. Moura[20], M.A. Muller[17], G. Müller[38], M. Münchmeyer[29], R. Mussa[47], G. Navarra[51 ‡], J.L. Navarro[72], S. Navas[72], P. Necesal[25], L. Nellen[55], A. Nelles[56, 58], J. Neuser[33], P.T. Nhung[94], M. Niechciol[40], L. Niemietz[33], T. Niggemann[38], D. Nitz[83], D. Nosek[24], L. Nožka[25], J. Oehlschläger[34], A. Olinto[88], M. Oliveira[62], M. Ortiz[70], N. Pacheco[71], D. Pakk Selmi-Dei[17], M. Palatka[25], J. Pallotta[2], N. Palmieri[36], G. Parente[73], A. Parra[73], S. Pastor[69], T. Paul[93, 85], M. Pech[25], J. Pękala[60], R. Pelayo[52], I.M. Pepe[19], L. Perrone[48], R. Pesce[41], E. Petermann[90], S. Petrera[42], A. Petrolini[41], Y. Petrov[78], R. Piegaia[3], T. Pierog[34], P. Pieroni[3], M. Pimenta[62], V. Pirronello[46], M. Platino[8], M. Plum[38], M. Pontz[40], A. Porcelli[34], T. Preda[64], P. Privitera[88], M. Prouza[25], E.J. Quel[2], S. Querchfeld[33], S. Quinn[76], J. Rautenberg[33], O. Ravel[32], D. Ravignani[8], B. Revenu[32], J. Ridky[25], S. Riggi[49, 73], M. Risse[40], P. Ristori[2], H. Rivera[43], V. Rizi[42], J. Roberts[84], W. Rodrigues de Carvalho[73], I. Rodriguez Cabo[73], G. Rodriguez Fernandez[45, 73], J. Rodriguez Martino[9], J. Rodriguez Rojo[9], M.D. Rodríguez-Frías[71], G. Ros[71], J. Rosado[70], T. Rossler[26], M. Roth[34], B. Rouillé-d'Orfeuil[88], E. Roulet[1], A.C. Rovero[5], C. Rühle[35], S.J. Saffi[12], A. Saftoiu[63], F. Salamida[27], H. Salazar[52], F. Salesa Greus[78], G. Salina[45], F. Sánchez[8], P. Sanchez-Lucas[72], C.E. Santo[62], E. Santos[62], E.M. Santos[21], F. Sarazin[77], B. Sarkar[33], R. Sato[9], N. Scharf[38], V. Scherini[43], H. Schieler[34], P. Schiffer[39], A. Schmidt[35], O. Scholten[57], H. Schoorlemmer[89, 56, 58], P. Schovánek[25], F.G. Schröder[34, 8], A. Schulz[34], J. Schulz[56], S.J. Sciutto[4], M. Scuderi[46], A. Segreto[49], M. Settimo[40, 48], A. Shadkam[82], R.C. Shellard[13], I. Sidelnik[1], G. Sigl[39], O. Sima[65], A. Śmiałkowski[61], R. Šmída[34], G.R. Snow[90], P. Sommers[87], J. Sorokin[12], H. Spinka[75, 80], R. Squartini[9], Y.N. Srivastava[85], S. Stanič[68], J. Stapleton[86], J. Stasielak[60], M. Stephan[38], M. Straub[38], A. Stutz[30], F. Suarez[8], T. Suomijärvi[27], A.D. Supanitsky[5], T. Šuša[23], M.S. Sutherland[82], J. Swain[85], Z. Szadkowski[61], M. Szuba[34], A. Tapia[8], M. Tartare[30], O. Taşcău[33], R. Tcaciuc[40], N.T. Thao[94], J. Tiffenberg[3], C. Timmermans[58, 56], W. Tkaczyk[61 ‡], C.J. Todero Peixoto[15], G. Toma[63], L. Tomankova[34], B. Tomé[62], A. Tonachini[47], G. Torralba Elipe[73], D. Torres Machado[32], P. Travnicek[25], D.B. Tridapalli[16], E. Trovato[46], M. Tueros[73], R. Ulrich[34], M. Unger[34], J.F. Valdés Galicia[55], I. Valiño[73], L. Valore[44], G. van Aar[56], A.M. van den Berg[57], S. van Velzen[56], A. van Vliet[39], E. Varela[52], B. Vargas Cárdenas[55], G. Varner[89], J.R. Vázquez[70], R.A. Vázquez[73], D. Veberič[68, 67], V. Verzi[45], J. Vicha[25], M. Videla[7], L. Villaseñor[54], H. Wahlberg[4], P. Wahrlich[12], O. Wainberg[8, 11], D. Walz[38], A.A. Watson[74], M. Weber[35], K. Weidenhaupt[38], A. Weindl[34], F. Werner[34], S. Westerhoff[92], B.J. Whelan[87], A. Widom[85], G. Wieczorek[61], L. Wiencke[77], B. Wilczyńska[60 ‡], H. Wilczyński[60], M. Will[34], C. Williams[88], T. Winchen[38], B. Wundheiler[8], S. Wykes[56], T. Yamamoto[88 a], T. Yapici[83], P. Younk[81], G. Yuan[82], A. Yushkov[73], B. Zamorano[72], E. Zas[73], D. Zavrtanik[68, 67], M. Zavrtanik[67, 68], I. Zaw[84 c], A. Zepeda[53 b], J. Zhou[88], Y. Zhu[35], M. Zimbres Silva[17], M. Ziolkowski[40]

[1] Centro Atómico Bariloche and Instituto Balseiro (CNEA-UNCuyo-CONICET), San Carlos de Bariloche, Argentina

[2] Centro de Investigaciones en Láseres y Aplicaciones, CITEDEF and CONICET, Argentina

[3] Departamento de Física, FCEyN, Universidad de Buenos Aires y CONICET, Argentina

[4] IFLP, Universidad Nacional de La Plata and CONICET, La Plata, Argentina

[5] Instituto de Astronomía y Física del Espacio (CONICET-UBA), Buenos Aires, Argentina

[6] Instituto de Física de Rosario (IFIR) - CONICET/U.N.R. and Facultad de Ciencias Bioquímicas y Farmacéuticas U.N.R., Rosario, Argentina

[7] Instituto de Tecnologías en Detección y Astropartículas (CNEA, CONICET, UNSAM), and National Technological University, Faculty Mendoza (CONICET/CNEA), Mendoza, Argentina

[8] Instituto de Tecnologías en Detección y Astropartículas (CNEA, CONICET, UNSAM), Buenos Aires, Argentina

[9] Observatorio Pierre Auger, Malargüe, Argentina

[10] Observatorio Pierre Auger and Comisión Nacional de Energía Atómica, Malargüe, Argentina

[11] Universidad Tecnológica Nacional - Facultad Regional Buenos Aires, Buenos Aires, Argentina

[12] University of Adelaide, Adelaide, S.A., Australia

[13] Centro Brasileiro de Pesquisas Fisicas, Rio de Janeiro, RJ, Brazil

[14] Faculdade Independente do Nordeste, Vitória da Conquista, Brazil
[15] Universidade de São Paulo, Instituto de Física, São Carlos, SP, Brazil
[16] Universidade de São Paulo, Instituto de Física, São Paulo, SP, Brazil
[17] Universidade Estadual de Campinas, IFGW, Campinas, SP, Brazil
[18] Universidade Estadual de Feira de Santana, Brazil
[19] Universidade Federal da Bahia, Salvador, BA, Brazil
[20] Universidade Federal do ABC, Santo André, SP, Brazil
[21] Universidade Federal do Rio de Janeiro, Instituto de Física, Rio de Janeiro, RJ, Brazil
[22] Universidade Federal Fluminense, EEIMVR, Volta Redonda, RJ, Brazil
[23] Rudjer Bošković Institute, 10000 Zagreb, Croatia
[24] Charles University, Faculty of Mathematics and Physics, Institute of Particle and Nuclear Physics, Prague, Czech Republic
[25] Institute of Physics of the Academy of Sciences of the Czech Republic, Prague, Czech Republic
[26] Palacky University, RCPTM, Olomouc, Czech Republic
[27] Institut de Physique Nucléaire d'Orsay (IPNO), Université Paris 11, CNRS-IN2P3, Orsay, France
[28] Laboratoire de l'Accélérateur Linéaire (LAL), Université Paris 11, CNRS-IN2P3, France
[29] Laboratoire de Physique Nucléaire et de Hautes Energies (LPNHE), Universités Paris 6 et Paris 7, CNRS-IN2P3, Paris, France
[30] Laboratoire de Physique Subatomique et de Cosmologie (LPSC), Université Joseph Fourier Grenoble, CNRS-IN2P3, Grenoble INP, France
[31] Station de Radioastronomie de Nançay, Observatoire de Paris, CNRS/INSU, France
[32] SUBATECH, École des Mines de Nantes, CNRS-IN2P3, Université de Nantes, France
[33] Bergische Universität Wuppertal, Wuppertal, Germany
[34] Karlsruhe Institute of Technology - Campus North - Institut für Kernphysik, Karlsruhe, Germany
[35] Karlsruhe Institute of Technology - Campus North - Institut für Prozessdatenverarbeitung und Elektronik, Karlsruhe, Germany
[36] Karlsruhe Institute of Technology - Campus South - Institut für Experimentelle Kernphysik (IEKP), Karlsruhe, Germany
[37] Max-Planck-Institut für Radioastronomie, Bonn, Germany
[38] RWTH Aachen University, III. Physikalisches Institut A, Aachen, Germany
[39] Universität Hamburg, Hamburg, Germany
[40] Universität Siegen, Siegen, Germany
[41] Dipartimento di Fisica dell'Università and INFN, Genova, Italy
[42] Università dell'Aquila and INFN, L'Aquila, Italy
[43] Università di Milano and Sezione INFN, Milan, Italy
[44] Università di Napoli "Federico II" and Sezione INFN, Napoli, Italy
[45] Università di Roma II "Tor Vergata" and Sezione INFN, Roma, Italy
[46] Università di Catania and Sezione INFN, Catania, Italy
[47] Università di Torino and Sezione INFN, Torino, Italy
[48] Dipartimento di Matematica e Fisica "E. De Giorgi" dell'Università del Salento and Sezione INFN, Lecce, Italy
[49] Istituto di Astrofisica Spaziale e Fisica Cosmica di Palermo (INAF), Palermo, Italy
[50] INFN, Laboratori Nazionali del Gran Sasso, Assergi (L'Aquila), Italy
[51] Osservatorio Astrofisico di Torino (INAF), Università di Torino and Sezione INFN, Torino, Italy
[52] Benemérita Universidad Autónoma de Puebla, Puebla, Mexico
[53] Centro de Investigación y de Estudios Avanzados del IPN (CINVESTAV), México, Mexico
[54] Universidad Michoacana de San Nicolas de Hidalgo, Morelia, Michoacan, Mexico
[55] Universidad Nacional Autonoma de Mexico, Mexico, D.F., Mexico
[56] IMAPP, Radboud University Nijmegen, Netherlands
[57] Kernfysisch Versneller Instituut, University of Groningen, Groningen, Netherlands
[58] Nikhef, Science Park, Amsterdam, Netherlands
[59] ASTRON, Dwingeloo, Netherlands
[60] Institute of Nuclear Physics PAN, Krakow, Poland
[61] University of Łódź, Łódź, Poland
[62] LIP and Instituto Superior Técnico, Technical University of Lisbon, Portugal

[63] 'Horia Hulubei' National Institute for Physics and Nuclear Engineering, Bucharest- Magurele, Romania
[64] Institute of Space Sciences, Bucharest, Romania
[65] University of Bucharest, Physics Department, Romania
[66] University Politehnica of Bucharest, Romania
[67] J. Stefan Institute, Ljubljana, Slovenia
[68] Laboratory for Astroparticle Physics, University of Nova Gorica, Slovenia
[69] Institut de Física Corpuscular, CSIC-Universitat de València, Valencia, Spain
[70] Universidad Complutense de Madrid, Madrid, Spain
[71] Universidad de Alcalá, Alcalá de Henares (Madrid), Spain
[72] Universidad de Granada and C.A.F.P.E., Granada, Spain
[73] Universidad de Santiago de Compostela, Spain
[74] School of Physics and Astronomy, University of Leeds, United Kingdom
[75] Argonne National Laboratory, Argonne, IL, USA
[76] Case Western Reserve University, Cleveland, OH, USA
[77] Colorado School of Mines, Golden, CO, USA
[78] Colorado State University, Fort Collins, CO, USA
[79] Colorado State University, Pueblo, CO, USA
[80] Fermilab, Batavia, IL, USA
[81] Los Alamos National Laboratory, Los Alamos, NM, USA
[82] Louisiana State University, Baton Rouge, LA, USA
[83] Michigan Technological University, Houghton, MI, USA
[84] New York University, New York, NY, USA
[85] Northeastern University, Boston, MA, USA
[86] Ohio State University, Columbus, OH, USA
[87] Pennsylvania State University, University Park, PA, USA
[88] University of Chicago, Enrico Fermi Institute, Chicago, IL, USA
[89] University of Hawaii, Honolulu, HI, USA
[90] University of Nebraska, Lincoln, NE, USA
[91] University of New Mexico, Albuquerque, NM, USA
[92] University of Wisconsin, Madison, WI, USA
[93] University of Wisconsin, Milwaukee, WI, USA
[94] Institute for Nuclear Science and Technology (INST), Hanoi, Vietnam

[‡] Deceased
[a] Now at Konan University
[b] Also at the Universidad Autonoma de Chiapas on leave of absence from Cinvestav
[c] Now at NYU Abu Dhabi
[d] now at DESY Zeuthen

# The Energy Scale of the Pierre Auger Observatory

Valerio Verzi[1] for the Pierre Auger Collaboration[2]

[1] INFN, Sezione di Roma "Tor Vergata", via della Ricerca Scientifica 1, 00133 Roma, Italia
[2] Full author list: http://www.auger.org/archive/authors_2013_05.html

auger_spokespersons@fnal.gov

**Abstract:** The energy scale of the Pierre Auger Observatory is derived from fluorescence observations of extensive air showers, an intrinsically calorimetric technique. Taking advantage of more precise measurements of the fluorescence yield, of a deeper understanding of the detector and consequently improved event reconstruction and of a better estimate of the *invisible energy*, we present an update of the method used to determine the energy scale. Differences in energy with respect to earlier measurements and the systematic uncertainties associated with the new energy scale are discussed.

## 1 Introduction

The Pierre Auger Observatory [1] has been designed to study ultra-high energy cosmic rays with unprecedented statistics and with low systematic uncertainties. It comprises an array of 1660 water-Cherenkov detectors deployed over 3000 km$^2$, collectively called the surface detector array (SD) with the atmosphere above it viewed by the Fluorescence Detector (FD) [2]. The FD consists of 27 telescopes located at 5 sites on the periphery of the SD array. Each telescope contains 440 photomultiplier pixels that detect light focused by a large spherical mirror.

The *hybrid* combination of the FD and the SD has an enormous advantage in the determination of the energy scale. The FD provides a nearly calorimetric energy measurement as the fluorescence light is produced in proportion to the energy dissipation by a shower in the atmosphere. These measurements are performed with a duty cycle of about 13%, as the FD can only operate during clear nights with little moonlight. The SD measures the distribution of particles on the ground with a duty cycle of almost 100%. By means of showers viewed by the FD in coincidence with the SD (*hybrid* events), the signal detected by the SD at 1000 m from the shower axis is calibrated against the calorimetric energy measured with the FD [3]. The advantage of the hybrid detector is therefore that the energy assignment is largely independent of air shower simulations.

The reconstruction of the fluorescence events is a complex process that requires the knowledge of several parameters. The FD measures the number of fluorescence photons produced from the de-excitation of atmospheric nitrogen molecules excited by the charged particles of the shower. The emission of these photons is isotropic and mostly in the wavelength range between 300 and 430 nm. The fluorescence yield is the proportionality factor between the number of photons emitted and the energy deposited in the atmosphere. It is therefore a key ingredient for the reconstruction: the light collected by the FD telescopes as a function of the atmospheric depth $X$ can be converted to the longitudinal profile of the energy deposit ($dE/dX$) of the air shower. An accurate reconstruction of $dE/dX$ requires continuous monitoring of the atmospheric conditions. This is particularly important for estimating the attenuation of the light due to molecular and aerosol scattering as it travels from the shower to the telescopes. Another key ingredient is the absolute calibration of the telescopes. Finally, the integral $\int dE/dX dX$ represents essentially the electromagnetic energy of the shower. The total energy is obtained from the calorimetric energy by adding the so-called *invisible energy* which accounts for the energy carried into the ground by high energy muons and neutrinos.

Using new knowledge both at the level of the detector and of the fluorescence process, we have updated the reconstruction of fluorescence events. In sections 2-6 we describe the changes made to the different parts of the reconstruction chain. For each we address the effects on FD energy determination and related systematic uncertainties, distinguishing between correlated and non-correlated errors between different showers. This is crucial to correctly propagate the uncertainties from FD measurements to SD energies in the calibration of SD events, which is updated in section 7 where we discuss the differences with respect to the previous energy scale and we summarise the total uncertainty of the new determination of the energy.

## 2 The fluorescence yield

The parameters characterising the fluorescence yield include an absolute normalisation of the wavelength spectrum, the relative intensities in different spectral bands, and their dependencies on pressure, temperature and humidity.

Previously the Auger collaboration used all of the parameters measured in the Airfly experiment [4, 5] with the exception of the absolute normalisation of the spectrum. This is parameterised by the intensity of the 337 nm spectral band and in the past we used the measurement of Nagano et al. [6] which has an uncertainty of 14%. The uncertainty of the absolute yield made the largest contribution to the overall uncertainty of the energy scale (22%).

We have now adopted a precise measurement of the absolute yield of the 337 nm band made by the Airfly collaboration with an uncertainty of 4% [7]. The Airfly measurement is the most precise available and it is compatible with the analysis presented in [8]. Its impact in the reconstruction of FD events is very important. Shower energies are lowered by about 8% and the precision due to the uncertainty of the measurement of the absolute yield is on average 3.4%.

As the yield is now known with high precision, we have also evaluated the uncertainties arising from the other fluorescence parameters. These uncertainties have been calculated by changing the fluorescence parameters by their uncertainties according to their degree of correlation reported in the Airfly papers [4, 5]. The uncertainty in shower energies arising from the relative intensities of the bands of the fluorescence spectrum is 1%. Those arising from the pressure, temperature and humidity dependencies are respectively 0.1%, 0.3% and 0.1%.

## 3 The atmosphere

An extensive array of instruments were designed and are deployed to monitor the atmosphere at the Pierre Auger Observatory [9]. The aerosol monitoring uses two lasers placed near the center of the array, four elastic scattering lidar stations, two optical telescopes and two systems which monitor the differential angular distribution of the aerosol scattering cross section (the *phase function*). Four infra-red cameras are used for cloud detection.

We have improved the hourly estimates of the aerosol optical depth profile [10] used to calculate the aerosol transmission factor [11] . The uncertainty on these profiles has two components, one correlated and another uncorrelated between different showers, components giving rise to an uncertainty in the shower energies which increases with energy from 3% to 6%. Other correlated uncertainties related to aerosols are those from the measurements of the *phase function* and from the wavelength dependence of the scattering cross-section. They are 1% and 0.5% respectively. Another uncorrelated uncertainty of 1% is associated with the spatial variability of the aerosols across the site [9].

The density profiles of the atmosphere are estimated using the Global Data Assimilation System (GDAS) meteorological model [12]. The day-to-day fluctuations of the pressure, temperature and humidity around the GDAS model have been estimated using meteorological radio-sondes launched locally. From studies of these fluctuations, we have identified an uncorrelated uncertainty in the energies of about 1% and a correlated one of about 1%.

## 4 The absolute calibration of the telescopes

Periodically the FD telescopes are calibrated absolutely with a drum-shaped light source (*drum*) placed in front of the diaphragm [2]. In this way we perform an end-to-end calibration of all elements of the telescope. The absolute calibration is made at 375nm with an uncertainty of 9%. This is fully correlated between different showers. Following the progress reported in [13], the Collaboration is working to reduce this uncertainty to the 5% level.

The short and long term changes of the detector response are tracked by a relative optical calibration system [2]. The response of the photomultipliers (PMTs) to the relative calibration performed during the *drum* operation and before and after each night of data taking are used to track the absolute calibration in the periods between the *drum* measurements. Two uncertainties are associated with this tracking, 3% for the uncorrelated part and 2% for the correlated one.

A new feature of the event reconstruction is the treatment of the calibration constants of the pixels. The optical properties of the telescopes have been studied using an isotropic point-like source put in the field of view of a telescope using a flying platform [14]. For a fixed position of the light source, we have discovered that the reflectivity of the PMT surface causes an optical *halo* extending over the full focal surface of the telescope. The *drum* calibration constants have been corrected for this effect with shower energy increasing by about 3%.

A further improvement concerns the relative FD response at various wavelengths. We have introduced a more precise optical efficiency curve measured using the *drum* device while in the past we used the optical efficiencies of the each component of the telescopes. This revised efficiency increases the shower energies by about 4%. The uncertainty in the measurement propagated to the shower energies introduces a correlated uncertainty of 3.5%.

## 5 Reconstruction of the longitudinal profile of the showers

A further change in the event reconstruction is due to an improved technique for the determination of the energy deposit in atmosphere [2, 15]. In a FD telescope a shower is observed as a sequence of pixels triggered at different times. The pointing direction of the pixels and FD and SD timing information are used to determine the position of the shower axis in the sky. The longitudinal profile of the light is derived from the time traces of the PMTs. The pixel selection is made by maximising the signal-to-noise ratio, excluding the night-sky light that dominates off the image axis. Knowing the shower geometry, the FD absolute calibration, the attenuation of the light flux in the atmosphere and by estimating the number of Cherenkov photons detected by the FD, it is possible to calculate the energy deposit with a fit to the $dE/dX$ data being made using Gaisser-Hillas [16] function. This enables an estimate of the energy deposit even outside the field of view of the telescopes and therefore yields the energy deposited in the atmosphere.

Because of the intrinsic shower width and the optical point spread function of the telescopes, part of the incoming light is spread away from the image axis, in the field of view of the non-selected pixels. The contribution of this light to the $dE/dX$ is calculated by estimating the size of the shower image at the telescope diaphragm. Two models are used for the fluorescence [17] and Cherenkov [18] light. We have now introduced a further correction which takes into account the angular spread close to the shower axis produced by the optical elements of the FD telescope [14]. The folding of this point spread function with the intrinsic shower width spreads the light more than predicted by the two models that only take into account the shower width. This effect has been parameterised by analysing shower data and it increases the shower energy by an amount ranging from 5% to 9% (the correction is larger at lower energies). We assign, conservatively, a correlated systematic uncertainty in the light collection of about 5%.

A further complication arises from the light which reaches the telescopes after multiple scatterings in the atmosphere. To avoid an overestimation of the shower energy, this light must be subtracted from the profile of detected photons. The multiple scattering contribution has been parameterised using [19] and the uncertainty of it affects the shower energies in a fully correlated way by about 1% [9].

In a further update we have developed a maximum likelihood fit taking into account realistic fluctuations of the signal in the PMTs. This increases the shower energies by about 2%.

To improve the fit of $dE/dX$, a Gaussian constraint is imposed on the parameters that define the Gaisser-Hillas function [15]. Changing these constraints by one standard deviation, we have evaluated a further correlated uncertainty in the shower energy which ranges from 3.5% to 1% (it decreases with energy). Other errors on the energies arise from the statistical error of the $dE/dX$ fit which decreases with energy from 5% to 3%, and an average uncertainty of 1.5% that arises from the uncertainty in the shower axis geometry. Both effects are uncorrelated.

The full reconstruction technique has been tested using Monte Carlo simulations. On average, the reconstructed energies differ from the true ones by about 2%. This bias has been considered as another correlated uncertainty.

## 6 The invisible energy

The final update in the reconstruction concerns the estimate of the *invisible energy* [20]. Previously we used an estimate based entirely on simulated showers [21] while now it is derived from data. This significantly reduces the dependence on the hadronic interaction models and mass composition. The *invisible energy* ($E_{inv}$) can be calculated for each shower using the FD measurement of the longitudinal profile and the SD signal at 1000 m from the axis, $S(1000)$. $E_{inv}$ can be reliably estimated only above $3 \times 10^{18}$ eV (the energy above which the SD array is fully efficient) as below this energy $S(1000)$ is biased by upward fluctuations of the shower signals. As the FD detects showers at lower energies and since we want to update the *invisible energy* for all FD events, $E_{inv}$ is parameterised with an analytical function above $3 \times 10^{18}$ eV, with the function being extrapolated to $10^{17}$ eV.

The same set of hybrid showers used to calibrate the SD energies (see below) is used to find the relation between $E_{inv}$ and the calorimetric energy $E_{cal}$: $E_{inv} = a_0(E_{cal}[\mathrm{EeV}])^{a_1}$. The fit is performed by minimising a $\chi^2$ function which takes into account the fluctuations of both FD and SD measurements, yielding the parameters: $a_0 = (0.174 \pm 0.001) \times 10^{18}$ eV and $a_1 = 0.914 \pm 0.008$. The correlation coefficient of the two parameters is -1.

The number of muons measured with the SD [22] is higher than predicted by the simulations formerly used to derive the *invisible energy* [21]. This contribution to the primary energy now ranges between 15% at $10^{18}$ eV and 11% at the highest energies (before we had $11\% \div 8\%$) with total shower energies increasing by about 4%. Analysis of the systematic uncertainties on the *invisible energy* [20] shows a correlated uncertainty in the total energy which decreases with energy from 3% to 1.5%. With the old parameterisation the overall uncertainty was 4%.

Due to the stochastic nature of air showers, the *invisible energy* is also affected by shower-to-shower fluctuations. These are parameterised according to [15] and an uncorrelated uncertainty of about 1.5% is introduced.

## 7 Impact on the energy scale and on its systematic uncertainty

The changes in the event reconstruction described in the previous sections have an impact on the energy determination and associated uncertainty for both FD and SD events. Concerning FD energies, all changes are summarised in table 1, for a reference energy of $10^{18}$ eV. Figure 1 shows the cumulative energy shift as a function of the shower energy when we introduce the effects described in sections 2, 4, 5 and 6. The update of the analysis of the aerosol optical depth profiles described in section 3 does not change the shower energies significantly. The overall change ranges from about +16% at $10^{18}$ eV to +12% at $10^{19}$ eV . We note that the new energy scale is consistent with the old one for which we gave an overall systematic uncertainty of 22% [3]. Moreover the changes are also consistent within each sector of the reconstruction. Indeed in [3] we quoted uncertainties of 14% for the fluorescence yield, 9.5% for the FD calibration, 10% for the longitudinal profile reconstruction and 4% for the *invisible energy*.

| **Changes in FD energies at $10^{18}$ eV** | |
|---|---|
| Absolute fluorescence yield (sec. 2) | **-8.2%** |
| New optical efficiency | 4.3% |
| Calibr. database update | 3.5% |
| Sub total (FD calibration - sec. 4) | **7.8%** |
| Likelihood fit of the profile | 2.2% |
| Folding with the point spread function | 9.4% |
| Sub total (FD profile reconstruc. - sec. 5) | **11.6%** |
| New invisible energy (sec. 6) | **4.4%** |
| Total | **15.6%** |

**Table 1**: Changes to the energy of showers at $10^{18}$ eV.

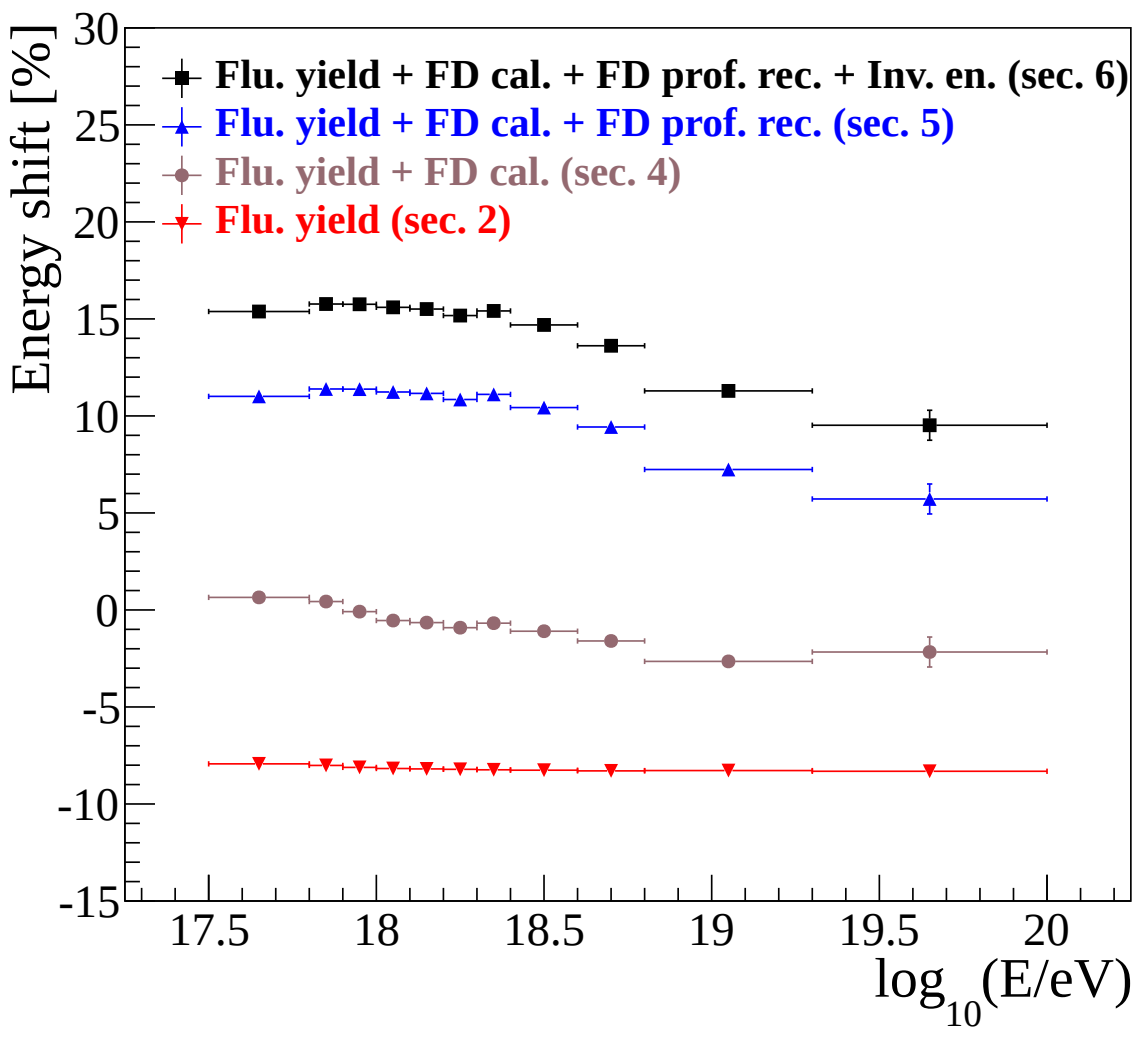

**Figure 1**: Cumulative energy shift as a function of the shower energy when we introduce the various effects.

The SD energies are obtained on the basis of the analysis presented in [3] with the new selection criteria described in [23]. The SD energy estimator, $S_{38}$, may be regarded as the signal $S(1000)$ that the shower would have produced had it arrived with a zenith angle, $\theta = 38°$.The relation between $S_{38}$ and the FD energy $E_{FD}$ is well described by a single power-law function, $E_{FD} = AS_{38}^{B}$. The parameters have been updated with a fit to a subset of high-quality hybrid events with $\theta < 60°$ detected between 1 January 2004 and 31

December 2012. The number of showers above $3 \times 10^{18}$ eV is 1475. The fit takes into account the resolutions of both $E_{FD}$ and $S_{38}$ (see table 2). The resolution of $E_{FD}$ is determined using all uncorrelated uncertainties described above. The fit yields: $A = (0.190 \pm 0.005) \times 10^{18}$ eV and $B = 1.025 \pm 0.007$ and with a correlation coefficient of -0.98. The root-mean-square deviation of the distribution of $AS_{38}^{B}/E_{FD}$ is about 18.5%. It is dominated by low-energy showers and is compatible with the expected resolution obtained from the quadratic sum of all the uncertainties listed in table 2 (18% at $3 \times 10^{18}$ eV).

| **Uncertainties entering into the SD calibration fit** | |
|---|---|
| Aerosol optical depth | 3%÷6% |
| Horizontal uniformity | 1% |
| Atmosphere variability | 1% |
| Nightly relative calibration | 3% |
| Statistical error of the profile fit | 5%÷3% |
| Uncertainty in shower geometry | 1.5% |
| Invis. energy (shower-to-shower fluc.) | 1.5% |
| Sub total FD energy resolution | **7% ÷ 8%** |
| Statistical error of the $S(1000)$ fit [3] | 12% ÷ 3% |
| Uncert. in lateral distrib. function [3] | 5% |
| shower-to-shower fluctuations [3] | 10% |
| Sub total SD energy resolution | **17% ÷ 12%** |

**Table 2**: Uncertainties uncorrelated between different showers and affecting the SD energy estimator.

The large number of hybrid showers detected over 9 years has allowed several consistency checks [24]. The SD energy estimator ($E_{SD} = AS_{38}^{B}$ for a given value of $S_{38}$) has been studied by making calibration fits to data collected during different time periods and/or under different conditions. We have found that $E_{SD}$ is stable within 5%, significantly above the statistical uncertainties. Even though these variations of $E_{SD}$ are consistent with the quoted systematic uncertainties, we use them conservatively to introduce another uncertainty of 5%.

| **Systematic uncertainties on the energy scale** | |
|---|---|
| Absolute fluorescence yield | 3.4% |
| Fluor. spectrum and quenching param. | 1.1% |
| Sub total (Fluorescence yield - sec. 2) | **3.6%** |
| Aerosol optical depth | 3%÷6% |
| Aerosol phase function | 1% |
| Wavelength depend. of aerosol scatt. | 0.5% |
| Atmospheric density profile | 1% |
| Sub total (Atmosphere - sec. 3) | **3.4%÷6.2%** |
| Absolute FD calibration | 9% |
| Nightly relative calibration | 2% |
| Optical efficiency | 3.5% |
| Sub total (FD calibration - sec. 4) | **9.9%** |
| Folding with point spread function | 5% |
| Multiple scattering model | 1% |
| Simulation bias | 2% |
| Constraints in the Gaisser-Hillas fit | 3.5% ÷ 1% |
| Sub total (FD profile rec. - sec. 5) | **6.5% ÷5.6%** |
| Invisible energy (sec. 6) | **3%÷1.5%** |
| Stat. error of the SD calib. fit (sec. 7) | **0.7%÷1.8%** |
| Stability of the energy scale (sec. 7) | **5%** |
| **Total** | **14%** |

**Table 3**: Systematic uncertainties on the energy scale.

The FD uncertainties correlated between different showers should be propagated to the SD energy scale by shifting all FD energies coherently by their uncertainties. This means that the correlated uncertainties propagate entirely to the SD energies. Table 3 lists all uncertainties on the Auger energy scale. Most of them have a mild dependence on energy. When this dependence is non-negligible, we report the variation of the uncertainty in the energy range between $3 \times 10^{18}$ eV and $10^{20}$ eV. The total uncertainty is about 14% and approximately independent of energy. We stress that we have made a significant improvement by comparison with the total 22% uncertainty reported previously [3].

# Estimate of the non-calorimetric energy of showers observed with the fluorescence and surface detectors of the Pierre Auger Observatory

MATIAS J. TUEROS[1] FOR THE PIERRE AUGER COLLABORATION[2]

[1]*Departamento de Física de Partículas, Universidad de Santiago de Compostela, España*
[2] *Full author list: http://www.auger.org/archive/authors_2013_05.html*

*auger_spokespersons@fnal.gov*

**Abstract:** The determination of the primary energy of extensive air showers using the fluorescence technique requires an estimation of the energy carried away by particles that do not deposit all of their energy in the atmosphere. This estimation is typically made using Monte Carlo simulations and thus depends on the assumed primary particle composition and model predictions for neutrino and muon production. In this contribution we introduce a new method to obtain the invisible energy directly from events measured simultaneously with the fluorescence and the surface detectors of the Pierre Auger Observatory. The robustness of the method, which is based on the correlation of the invisible energy with the muon number at ground, is demonstrated by applying it to different sets of Monte Carlo events. An event-by-event estimate of the invisible energy is given for the hybrid data set used for the energy calibration of the surface detector of the Pierre Auger Observatory.

## 1 Introduction

When an ultra-high energy cosmic ray interacts in the atmosphere a cascade of particles is generated. In this cascade, an important fraction of the energy is deposited in the atmosphere as ionisation of the air molecules and atoms. A fraction of the deposited energy is then re-emitted during the de-excitation of the ionized molecules as fluorescence light that can be detected by fluorescence telescopes.

Since the fluorescence intensity is proportional to the deposited energy, the integral of the fluorescence profile yields an accurate measurement of the energy of the primary particle ($E_0$) that was deposited in the atmosphere by the charged particles due to electromagnetic energy losses. This is usually referred to as the calorimetric energy ($E_{\text{Cal}}$).

The remaining energy, carried away mostly by neutrinos and high-energy muons that do not deposit all their energy in the atmosphere, is *a priori* unknown. An estimation of this "invisible" energy is required to derive the primary energy ($E_0$) from the measured $E_{\text{Cal}}$. Historically, this non-calorimetric energy has been called "missing energy" [1]. However, we will use the name "invisible energy" ($E_{\text{Inv}}$) deeming it more appropriate.

Generally, the invisible energy correction is parameterized as a function of $E_{\text{Cal}}$ ($E_{\text{Inv}}(E_{\text{Cal}})$) and it is typically estimated using Monte Carlo simulations averaging over many showers. The average value depends on the high-energy hadronic interaction model and on the primary mass, ranging from 8.5 to 17% of the primary energy at 1 EeV and from 7 to 13.5 % at 10 EeV.

Selecting a particular interaction model when analysing real events could introduce a bias to the reconstruction of the primary energy that is ultimately unknowable. An accurate knowledge of the invisible energy is thus essential in experiments using the fluorescence technique if a reliable measurement of the primary energy of cosmic rays is to be obtained.

In a previous work [2] we have described a method that relies on the properties of shower universality and a simple model of extensive air showers to find a parameterization of $E_{\text{Inv}}$. This method is robust to changes in the hadronic interaction models used in Monte Carlo simulations. In this work, the method has been updated to take into account the fact that the signal attenuation curve and the muon content measured in extensive air showers may not be properly described simultaneously by current Monte Carlo simulations.[3, 4, 5].

## 2 A simple model for the invisible energy

In the Heitler model extended to hadronic cascades by Matthews [6], the primary energy is distributed between the electromagnetic and muonic components of the air shower as

$$E_0 = \xi_c^e N_e^{\max} + \xi_c^\pi N_\mu, \tag{1}$$

where $E_0$ is the primary energy, $N_e^{\max}$ is the number of electrons at the shower maximum, and $\xi_c^e$ is the critical energy for the electromagnetic particles. The second term is the energy transferred to the muonic component of the cascade and is considered to be proportional to the total number of muons ($N_\mu$). The critical energy of the pion, $\xi_c^\pi$, is chosen as the proportionality factor to account for the fact that, in this model, the muons are considered to originate from pion decays with an associated muon neutrino (or muon antineutrino), transferring all of the energy into the non-calorimetric channel independently of how much energy goes to each muon. With these considerations, the second term of Eq.(1) can be identified directly with the invisible energy.

The model presented is clearly an oversimplification, as there are also muons being produced by other processes, the next in importance being kaon decay (roughly 10 times less frequent). Therefore, $\xi_c^\pi$ should be considered as an "effective" critical energy, that averages the different contributions to the muonic component. If we pick cascades at the same stage of shower development at ground level, measured by the slant depth from shower maximum to ground level ($DX$), the number of high-energy muons is correlated with the number of muons at ground level.

The number of muons is not measured directly at the Pierre Auger Observatory. We will use an observable that can be related to the muon content, namely the signal at 1000m from the core $S(1000)$. Based on universality studies [7, 8], the relationship between $S(1000)$ and the muon content of the shower is universal when expressed as a function of the stage of development of the cascade at ground level.

The primary energy $E_0$ can be parameterized as a power law of $S(1000)$

$$E_0 = \gamma_0(DX)\,[S(1000)]^{\gamma} \qquad (2)$$

for a fixed zenith angle ($S_{38^\circ}$) [9], or for a fixed stage of shower development measured by $DX$. The value of $\gamma_0(DX)$ is closely related to the attenuation of $S(1000)$ with $DX$.

In the Heitler-Matthews model it is also inferred that the total number of muons follows a power law with the primary energy,

$$N_\mu = \beta_0 \left(\frac{E_0}{\xi_c^\pi}\right)^{\beta}. \qquad (3)$$

Here $\beta$ depends mildly on the pion multiplicity, on the inelasticity of the first interactions in the cascade, and on the energy partition between charged and neutral pions. $\beta_0$ is a muon scale factor introduced to account for these effects. This parameter can be considered to be independent of $DX$ since the atmospheric depth of the maximum should have little influence on how many muons are generated in the shower.

Combining our correlation of $E_{\rm Inv}$ with the total number of muons and equations (3) and (2), we get a power law dependence of the invisible energy on $S(1000)$

$$E_{\rm Inv} = \xi_c^\pi N_\mu = \xi_c^\pi \beta_0 \left(\frac{\gamma_0(DX)\,S(1000)^{\gamma}}{\xi_c^\pi}\right)^{\beta}. \qquad (4)$$

Based on these results, the invisible energy can be estimated as a function of $S(1000)$ and $DX$

$$\begin{aligned} \log(E_{\rm Inv}) &= A(DX) + B\log(S(1000)) \qquad (5)\\ A(DX) &= (1-\beta)\log(\xi_c^\pi) + \\ &\quad \log(\beta_0) + \beta\log(\gamma_0(DX)) \\ B &= \beta\gamma. \end{aligned}$$

where $A(DX)$ and $B$ can be determined from fits of $\log(E_{\rm Inv})$ vs $\log(S(1000))$ from full Monte Carlo simulations, in order to capture any further dependences with $DX$ that are not described by this simplified model. To ensure a good reconstruction of $S(1000)$, only events in which the detector with the highest signal has all its 6 closest neighbours working at the time of the event (the 6T5 selection cut, see [10]) are used. This new parameterization of the invisible energy will be called $E_{\rm Inv}(S(1000), DX)$ from now on.

With current hadronic models, $\beta$ is usually found to be within 10% of 0.9 [6] and $\gamma$ is in the 1.06 - 1.09 range [11]. The constant $B$ was fixed to 0.98 in this work, considering that the values of $\beta$ and $\gamma$ that best describes our QGSJET-II simulations are 0.925 and 1.0594 respectively. Other models will have slightly different values. However, the product $\beta\gamma$ is close to 0.98 for all the hadronic models considered, varying within 2% of this value. It is important to point out that this model for the invisible energy can be extended to include the effect of a primary nucleus using the superposition principle. The extended model gives a very good description of the change in the invisible energy associated with a change in the primary mass.

Differences in $\gamma_0$, $\beta_0$ and $\xi_c^\pi$ for different hadronic models and masses tend to compensate each other to give a similar parameterizations of the $E_{\rm Inv}(S(1000), DX)$. However, this compensation is not complete and each hadronic model has a slightly different $A(DX)$. Since $\gamma_0$ and $\beta_0$ are the parameters that show greater variability between models, we will implement a correction to $A(DX)$ to compensate for the differences. This correction will be crucial when the method is applied to data, as it is known that the measured attenuation curve (closely related to $\gamma_0$) and the shower muon content (closely related to $\beta_0$) cannot be reproduced simultaneously [3, 4, 5] using the hadronic models currently available.

If we take the expression for $A(DX)$ for a reference model (in our case QGSJet-II [16] 50% proton/ 50% iron mixed composition) and we ignore the small variations in $\gamma$, $\beta$ and $\xi_c^\pi$ among the various models, the difference in $A(DX)$ from a different MC model is

$$\begin{aligned} A^{\rm MC}(DX) &= A^{\rm QII}(DX) + \qquad (6)\\ &\quad \log_{10}\left(\left(\frac{\gamma_0^{\rm MC}(DX)}{\gamma_0^{\rm QII}(DX)}\right)^{\beta_{\rm QII}} \frac{\beta_0^{\rm MC}}{\beta_0^{\rm QII}}\right). \end{aligned}$$

The correction clearly has two separate contributions: one arising from the difference in the attenuation curve and another one from the difference in the muon normalization. The contributions tend to compensate each other, and the final correction is relatively small. The effect of this correction on the estimation of the invisible energy for each model is less than 15% in the most extreme case and is usually within 5%.

To illustrate the performance of the correction we plot in Fig. 1 the reconstruction of the average $E_{\rm Inv}$ done with the QGSJET-II reference parameterization on events generated with other hadronic models, namely EPOS 1.6 and 1.99 [15], using the corresponding correction for each hadronic model and primary mass. Note that the correction is capable of recovering the correct invisible energy even for simulations done with EPOS 1.6, a hadronic model that gives a significantly different number of muons with respect to QGSJET-II.

## 3 $E_{\rm Inv}(S(1000), DX)$ from observations

To make an estimation of the invisible energy using experimental data, we use high-quality hybrid events that trigger the surface detector (SD) and the fluorescence detector (FD) of the Pierre Auger Observatory independently. These are known as "golden hybrid" events.

The correction due to the difference in the attenuation curve between data and Monte Carlo simulations can be measured at the Pierre Auger Observatory simply by making the corresponding fits of $E_0$ vs $S(1000)$ in bins of $DX$. From these fits, $\gamma_0^{\rm data}(DX)$ can be obtained. As there are still insufficient events to make reliable fits of Eq. (2) be-low $DX$=200 g/cm$^2$ and above $DX$=900 g/cm$^2$ for sufficiently small bins of $DX$, the applicability of our method is presently limited to this range.

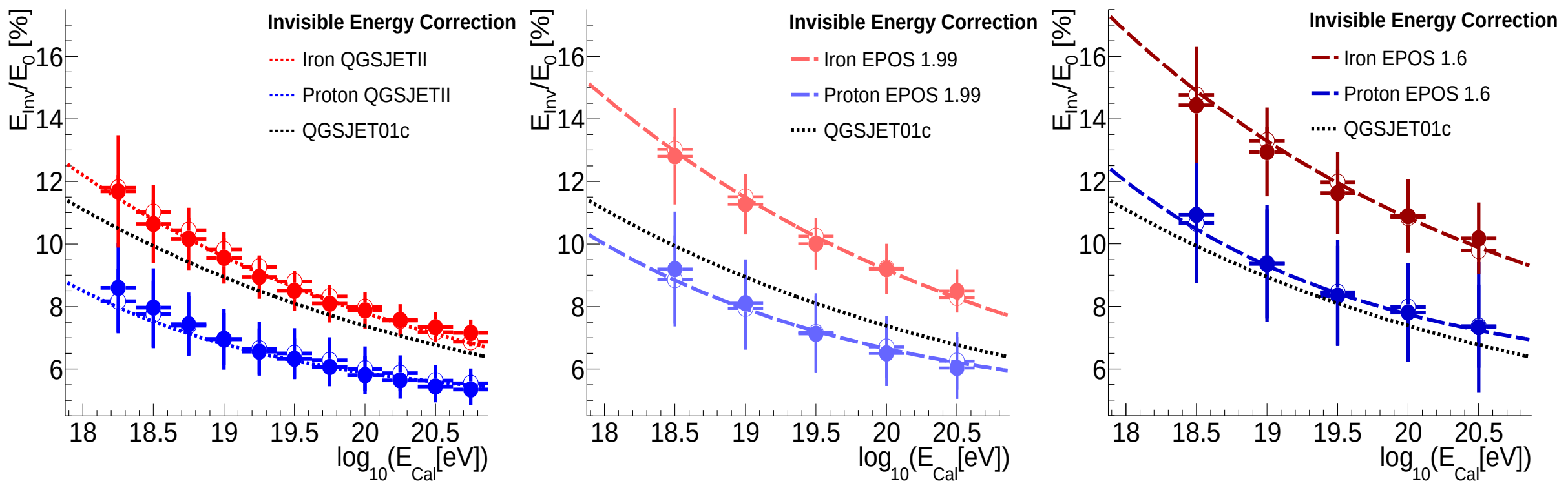

**Figure 1**: Average invisible energy as a function of the calorimetric energy. Open symbols represent the average invisible energy for a given energy bin, as obtained from Monte Carlo simulations performed with CORSIKA [13]. Filled symbols represent the average invisible energy for a given energy bin, obtained using $E_{\rm Inv}(S(1000), DX)$ with $A(DX)$ given by Eq. (5) for fits to QGSjet-II events and corrected by Eq. (7) (see text). The $E_{\rm Inv}(E_{\rm Cal})$ parameterization for QGSJet01c [14] 50% proton 50% iron from [1] is shown for reference in each of the plots.

The factor $\beta_0^{\rm data}/\beta_0^{\rm MC}$ can be estimated using the $N_{19}$ muon scale factor obtained in [12]. In that article, an estimation of the number of muons on inclined events with respect to a reference Monte Carlo is used to make a power law fit equivalent to Eq. (3). To estimate $\beta_0^{\rm data}/\beta_0^{\rm MC}$ we use the multiplicative constant of the power law in Eq. (3) of [12] (1.81, that represents the number of muons at 10 EeV with respect to QGSJET-II simulations for proton primaries) and re-scale it to our reference 50% proton/ 50% iron mixture. The obtained value is 1.56.

Finally, since there is no way to estimate $\xi_c^\pi$ from hybrid events, we will continue to treat this parameter as a constant taken from the reference Monte Carlo. The uncertainty associated with the possibility of having a different $\xi_c^\pi$ in data was estimated to be around 1%, assuming that $\xi_c^\pi$ is consistent with any of the hadronic models used in this work.

In Fig. 2 the invisible energy is shown for a random sample of golden hybrid events that pass the quality cuts. The error bars indicate the statistical uncertainty arising from the uncertainty in the fit parameters, plus a propagation of the systematic uncertainty in the values $S(1000)$ and $DX$ that are used as input for the parameterization. The small dotted lines identify the bands within which the average invisible energy can vary due to its systematic uncertainty.

The total uncertainty has been estimated by considering the uncertainty in the $N_{19}$ measurement and the estimation of $\beta_0$, the uncertainty due to possible difference in $\xi_c^\pi$, the uncertainty due to a deviation from $\beta\gamma = 0.98$ and the uncertainty propagated from the measurement of $S(1000)$ and $DX$. The total systematic uncertainty on the average $E_{\rm Inv}/E_0$ is 5% at $10^{17}$ eV, 3% at $10^{18}$ eV and 2% at $10^{19}$ eV. No further systematic uncertainty should arise from a change in the energy scale of the observatory in the determination of $E_{\rm Inv}$. However, such a change will obviously affect the $E_{\rm Inv}/E_0$ ratio. As the origin of this systematic effect is not intrinsic to the determination of $E_{\rm Inv}$ but to the way of presenting the data, we do not include it here.

In Fig. 2 it is also shown the prediction from $E_{\rm Inv}^{QGSJet01}(E_{\rm Cal})$ 50% proton / 50% iron parameterization, in the event reconstruction procedures used previously. This parameterization underestimated the invisible energy of the primary energy on average by 4 % in units of the primary energy.

It is important to note that below the energy of full trigger efficiency of the Observatory, the trigger is biased towards events with higher number of muons, and thus higher invisible energy. At these low energies, the relative systematic uncertainty in the determination of $S(1000)$ is also higher, resulting in a higher systematic uncertainty in the determination of the invisible energy. To avoid these biases only data at energies above $10^{18.3}$eV were used in this work.

For events in which $S(1000)$ cannot be determined accurately, a parameterization of the average invisible energy as a function of $E_{\rm Cal}$ can be used. In the Heitler-Matthews model, it is also inferred that there exists a power law dependence between $E_{\rm Cal}$ and $E_0$

$$E_{\rm Cal} \approx g\xi_c^e k(E_0)^\alpha \tag{7}$$

where g is the ratio of the total number of electromagnetic particles to the number of electrons and k is a proportionality constant related to the units chosen for the energy.

Combining equations (3),(4) and (7) we get

$$\frac{E_{\rm Inv}}{1{\rm EeV}} = \frac{\beta_0 10^{9(\frac{\beta}{\alpha}-1)}}{(g\xi_c^{em}k)^{\frac{\beta}{\alpha}}\,\xi_c^{\pi\beta-1}}\left(\frac{E_{\rm Cal}}{1{\rm EeV}}\right)^{\frac{\beta}{\alpha}} \tag{8}$$

If we use the values of $\beta = 0.925$, $\beta_0 = 0.4$, $\xi_c^\pi = 12$GeV, that approximately describe QGSJet-II (50% /proton 50% iron) simulations, $\alpha$= 1.011 and $g\xi_c^{em}k$=0.68 GeV$^{1-\alpha}$ to describe data and, expressing the calorimetric energy in EeV, we get 0.117 for the multiplicative constant of the exponential, and 0.915 for the exponent. As the muon normalization for data ($\beta_0$) is a factor 1.56 higher than for QGSJet-II 50% proton /50% iron simulations, we expect the multiplicative constant for data to be close to 0.187.

The fit of a 2-parameter exponential function to the invisible energy on the golden hybrid events above $10^{18.3}$ eV, using a $\chi^2$ function that takes into account the fluctuations of both $E_{\rm Cal}$ and $S(1000)$, is shown in Fig. 3. The result is

$$\frac{E_{\rm Inv}}{1{\rm EeV}} = 0.174\left(\frac{E_{\rm Cal}}{1{\rm EeV}}\right)^{0.914}. \qquad (9)$$

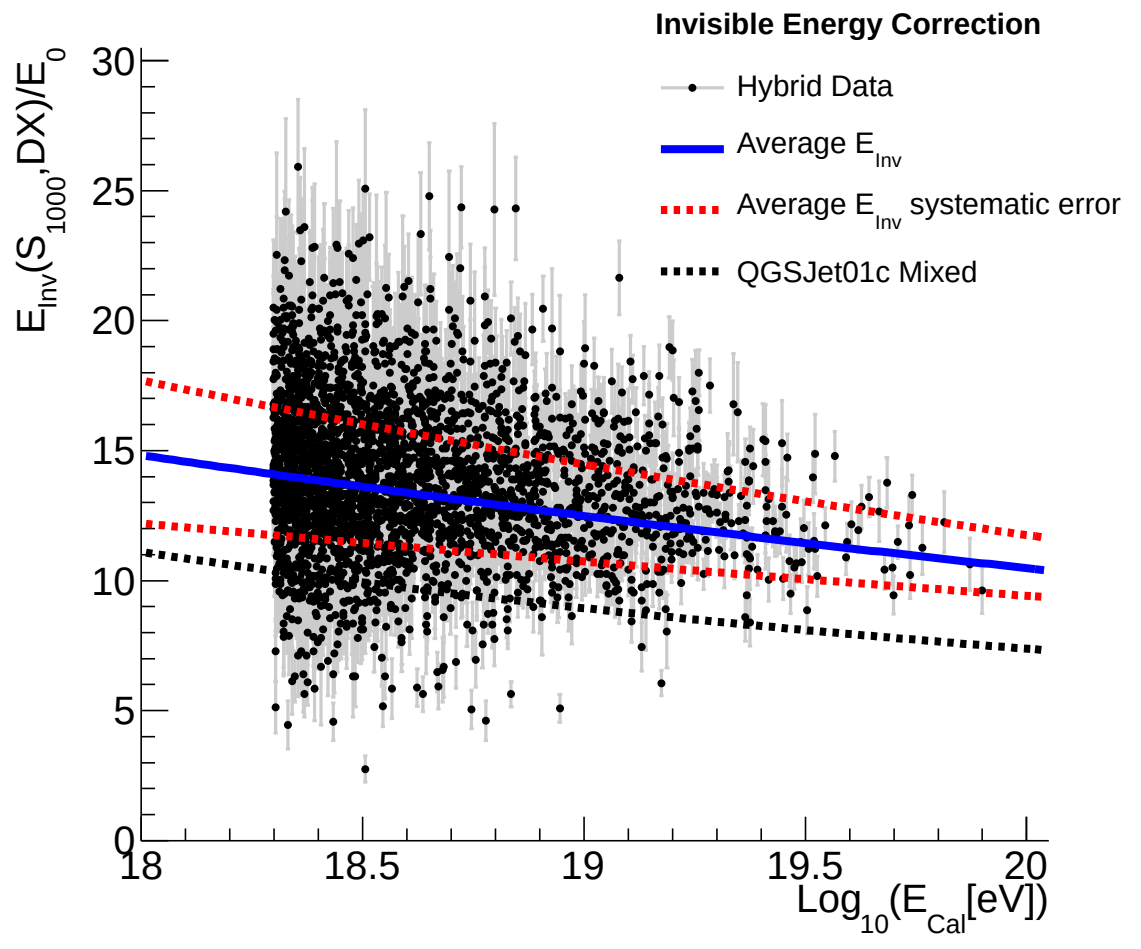

**Figure 2**: Estimation of the invisible energy in golden hybrid events using $E_{\rm Inv}(S(1000),DX)$, as proposed in this paper, superimposed on the $E_{Inv}^{QGSJet01c}(E_{\rm Cal})$ parameterization calculated in [1] for 50% proton 50% iron (dot dashed line). The red dotted line shows the bounds of the systematic uncertainty in the average. The error bars on the points represent uncertainties propagated from the systematic uncertainty in $S_{1000}$ and $DX$ plus the statistical uncertainty from the $A(DX)$ fit.

The good agreement between the model and the fit gives us confidence in using the extrapolation of this fit for events with energies below $10^{18.3}$ eV. Using $E_{\rm Inv}(S(1000),DX)$ introduces a dependence of $E_0$ on $S(1000)$ that complicates the energy calibration of the surface detector of the Pierre Auger Observatory using golden hybrid events, as it correlates the fluctuations in the energy determined with the surface detector with the fluctuations in the energy determined with the fluorescence detector. To avoid these complications, the presented parameterization (9) is used for the determination of the invisible energy instead of $E_{\rm Inv}(S(1000),DX)$ over the whole energy range of the Observatory [17]. The statistical uncertainty of this fit is very small and its impact on the total energy is below 0.5%.

## 4 Conclusions

We have presented a method that allows us to make an unbiased and model-independent determination of the invisible energy. The method is based on a calibration of the invisible energy with $S(1000)$ and $DX$ made with Monte Carlo simulations, that is then corrected with the measured $N_{19}$ from horizontal events and the attenuation curve obtained from golden hybrid events.

The method was successfully applied to measure the average invisible energy of a set of golden hybrid events showing that the correction previously in use [1] underestimated the primary energy by approximately 4% on average, introducing a shift in the energy scale [17].

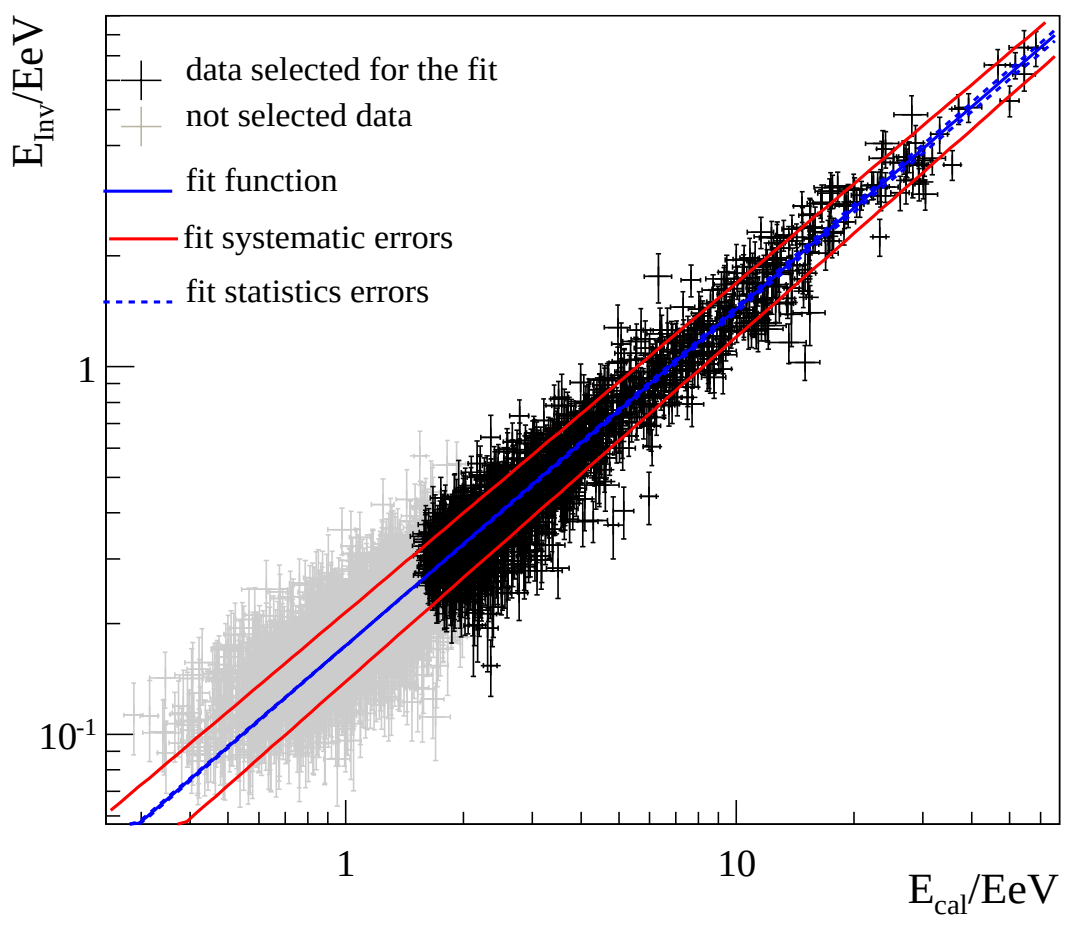

**Figure 3**: Fit of $E_{\rm Inv}(S(1000),DX)$ vs $E_{\rm Cal}$ presented in Eq. (9).

An expression of the invisible energy as a function of $E_{\rm Cal}$ was presented and used to parametrize the data at energies above $10^{18.3}$ eV. Good agreement between the model prediction and the parameters obtained was found. This function is used to calculate the invisible energy correction over the full energy range of the Pierre Auger Observatory.

# Measurement of the Optical Properties of the Auger Fluorescence Telescopes

JULIA BÄUML[1] FOR THE PIERRE AUGER COLLABORATION[2]

[1] *Karlsruhe Institute of Technology, Karlsruhe, Germany*
[2] *Full author list: http://www.auger.org/archive/authors_2013_05.html*

*auger_spokespersons@fnal.gov*

**Abstract:** Fluorescence telescopes are used for the calorimetric measurement of the energy of air showers at the Pierre Auger Observatory. The optical properties of these telescopes have to be known precisely to allow for an absolute energy calibration. We present a method to measure their light distribution function independent of shower data. We have developed an isotropic, point-like light source, that was brought into the field of view of a telescope using an autonomously flying platform, an octocopter. The optical properties of the telescopes are probed in detail by illuminating the telescopes from different angles with respect to the telescope axis and by obscuring or removing certain telescope components. In this contribution, we describe the light source which was developed and the important properties of the octocopter. We further present the first results on the measured light distribution and compare them to detailed telescope simulations.

## 1 Introduction

The Pierre Auger Observatory is a hybrid detector allowing for the measurement of air showers with an array of water-Cherenkov detectors and fluorescence telescopes [1]. The number of fluorescence photons produced by an air shower in the atmosphere is proportional to the energy deposited by the shower. Therefore, the observed fluorescence light profile can be used to obtain an estimate of the calorimetric energy of an air shower and hence the energy of the primary particle. The shower energy derived this way is almost independent of shower simulations and, in particular, hadronic interaction models. A set of showers observed simultaneously with the surface detector array and the fluorescence telescopes on clear dark nights (14% of the overall measurement time) is used to derive an energy calibration for the surface detector, which reaches a duty cycle of $\sim$ 100%.

The fluorescence telescopes [2] are calibrated with an end-to-end method using a drum that covers and uniformly illuminates the entire telescope aperture [3]. An air shower, on the other hand, is equivalent to a localized, almost point-like light source. To analyse the shower profile it is necessary to reconstruct the light distribution at the aperture, accounting not only for the transfer function measured with the drum calibration, but also for the point spread function of the telescope.

In this paper we discuss a novel technique to study the optical properties of fluorescence and atmospheric Cherenkov telescopes and present a first measurement of the point spread function of the Auger telescopes. The same method can be used to measure the telescope pointing or to obtain an absolute calibration of individual pixels of fluorescence telescopes, see [4].

A portable, point-like light source is placed in the field of view of a telescope by an octocopter. The GPS- and pressure-sensor-based positioning system of the octocopter can be programmed to place the light source at different distances and altitudes for a duration of up to 20 min. The light source has an isotropic emission pattern and the intensity of the flashes of UV light is electronically stabilized. The camera of the fluorescence telescope is read out for each flash, including non-triggered pixels. Averaging over many images taken for the same position of the light source gives access to the tail in the point spread function that is otherwise dominated by the noise of the background light.

After describing the custom-built light source and its calibration in laboratory measurements we will briefly summarize the parameters of the commercially available octocopter in Sec. 2. Using this setup an extensive set of measurement campaigns has been carried out and point spread functions for different telescopes have been derived. First results will be presented in Sec. 4 and compared to simulations of the optical properties of the telescopes. Implications for the fraction of light of a shower image within a few degrees of the shower axis as well as the interpretation of the telescope calibration will be discussed in Sec. 5.

## 2 Experimental setup

### 2.1 The light source

Our requirements for the light source were threefold: isotropy, light weight and emittance in the UV range. As light emitters we use LEDs of type H2A1-H375 from Roithner Laser [5]. Their spectrum has its peak at 375 nm and lies within the range of high telescope efficiency (see Fig. 1).

To achieve the best possible homogeneity, a total of 12 LEDs were distributed evenly on a sphere. The centres of the faces of a dodecahedron provide such a distribution. To improve the homogeneity of the light distribution, the body of the light source with mounted LEDs was coated with Tyvek and surrounded by a diffuser bowl. As diffuser we use a clear polystyrene sphere made up of two hemispheres, which we etched with acetone to make them diffuse. To reduce the weight of the light source and to make it as compact as possible, the LEDs were detached from their hexagonal mounting plates. Based on simulations of the resulting light distribution with a diffuser sphere of radius 50 mm, the edge length of the dodecahedron was chosen as 16 mm. The body was printed with a 3D printer from Acrylonitrile Butadiene Styrene (ABS) based on a 3D model of the dodecahedron with embedded sockets for the LEDs and channels for all cables (see Fig. 2).

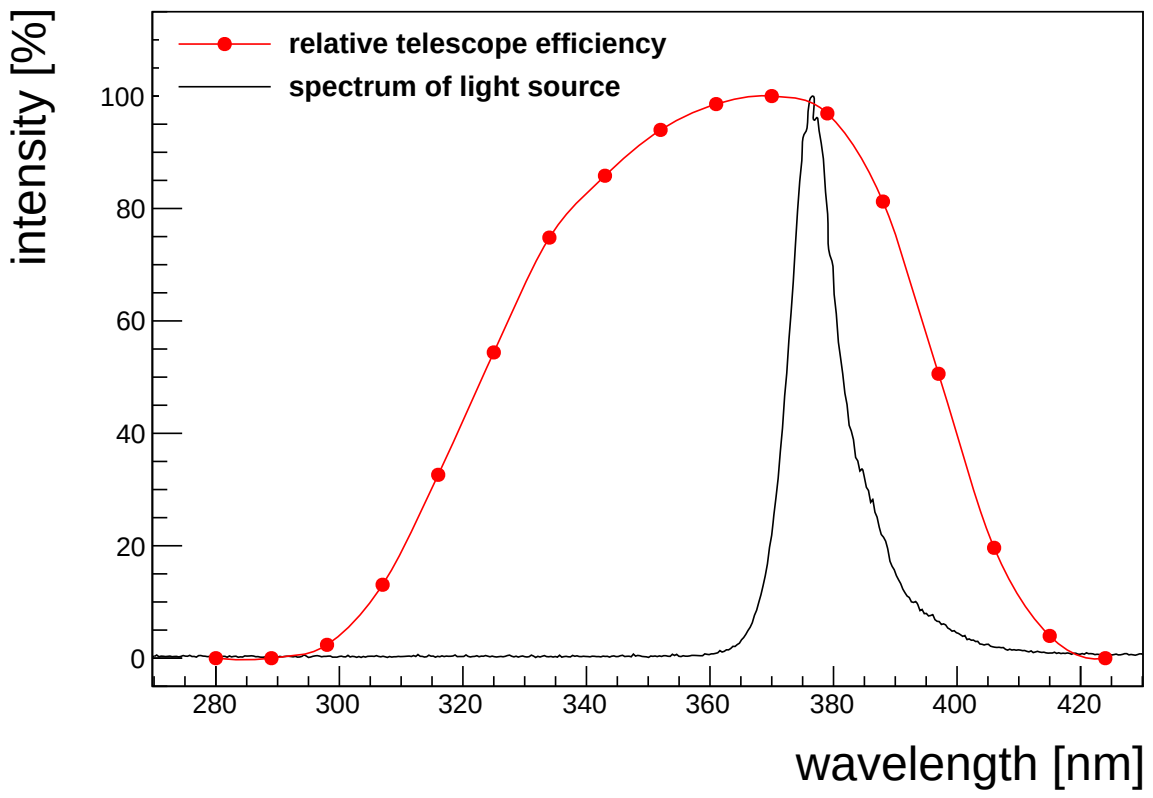

**Figure 1**: Spectrum and telescope efficiency. The spectrum of the LEDs being used lies within the range of high telescope efficiency. The measured spectra of a single LED and the complete light source agree well, there is no spectral effect of the surrounding diffuser bowl.

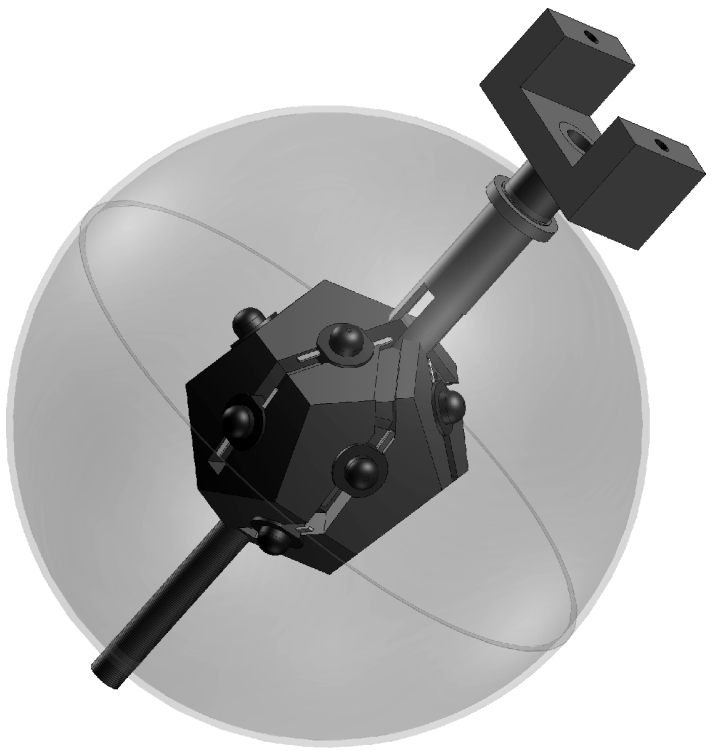

**Figure 2**: Schematic drawing of the light source. The LEDs are mounted on a dodecahedron-shaped body made from ABS. Channels for all cabling are embedded in the body as well as the supporting rod. The diffuser bowl surrounding the dodecahedron is fixed to the supporting rod, which also attaches the light source to the octocopter.

The homogeneity that was achieved with this setup is very good. The maximum difference in intensity between the hottest and the coldest spot over a range of $\pm 18°$ is 3.5%. The actual pointing direction of the light source towards the telescope is monitored during measurement campaigns, which allows for a correction of the intensity differences.

The electronic board driving the LEDs has twelve output channels. Emitted light pulses can have durations between $2\,\mu$s and $64\,\mu$s and one of six adjustable amplitudes. They are triggered by the PPS signal of the on-board GPS receiver with a programmable delay between $50\,\mu$s and $1000\,\mu$s. For other purposes a DC operation is also possible. All properties can be configured via an $I^2C$ interface.

For monitoring, a photodiode and a temperature sensor are placed inside the diffuser bowl, a second temperature sensor is attached to the electronics board.

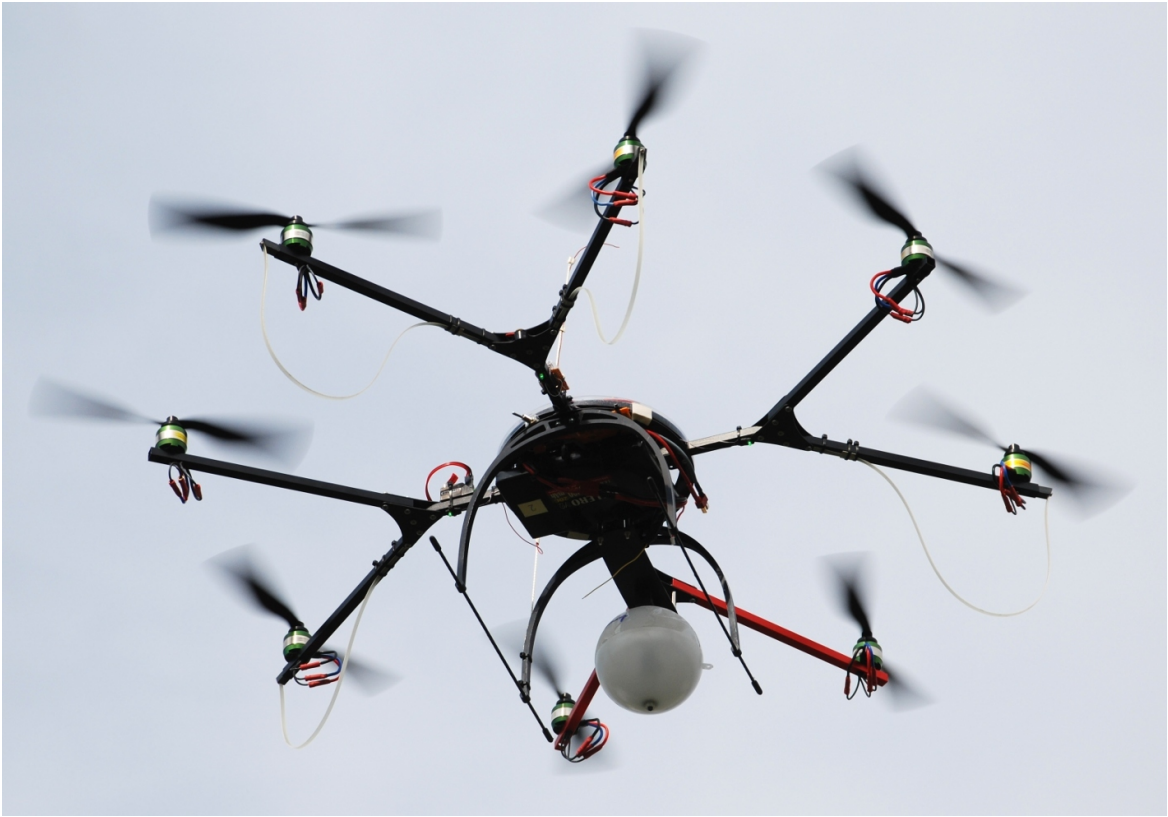

**Figure 3**: Octocopter with light source. The eight engines of the octocopter are arranged on a circle. Their supporting rods are connected to a central platform carrying all circuit boards for operation and communication. The light source is attached to the supporting rods of the two engines that define the forward direction.

The energy emitted by the light source has been measured using a NIST-calibrated photodiode and an electrometer (charge measurement). The accuracy that can be reached in those measurements is as good as $\sim 2\%$. During measurement campaigns at the Telescope Array [6], intensity measurements were made at the University of Utah. Those measurements, with an independent setup using a different NIST-calibrated photodiode and a picoammeter (current measurement), have confirmed the intensity and accuracy values [4].

The value for the absolute intensity of the light source depends slightly on the rate of the emitted light pulses (for feasibility reasons 1 kHz in the lab, 1 Hz in telescope measurements) and the temperature. Furthermore the length of the pulse changes slightly with the amplitude. All of those effects have been studied in great detail and can be corrected for to achieve an absolute calibration.

## 2.2 The octocopter

We employ a flying platform to bring the light source into the field of view of a telescope. The octocopter, a Mikrokopter [7, 8] with eight motors, was chosen for reliability reasons. A failure of any two motors does not affect its self-stabilization capabilities. When given a set of GPS coordinates, timing and pointing directions, the octocopter performs an autonomous way-point flight in 3D, controlled by GPS and a magnetic field sensor. The octocopter is able to keep its position and pointing within $\sim 1$ m (+ GPS uncertainty) and $\sim 5°$, respectively, in moderate wind conditions. Payload weights of up to 1 kg can be lifted and flight times of up to 20 min can be achieved with the light source attached. Due to the open platform, it was possible to integrate the hardware and software of the light source closely into the existing octocopter system.

## 2.3 Measurement campaigns

Measurements with the octocopter and light source device can only take place during the regular data taking periods of the fluorescence detector (moonless nights, no rain) when the wind speed is no higher than 5 m/s. We use an external trigger set to the time of the light pulse, so that the normal data taking is not interrupted, just the dead-time of the telescope is increased. The external trigger mode further allows us to read out all pixels in the triggered telescope. The octocopter is usually flown at distances of 0.5 – 1 km to the telescope. At such distances, the GPS position uncertainty of ±6 m is smaller than the angular size of one telescope pixel (1.5° corresponds to 13 m at 500 m or 26 m at 1 km).

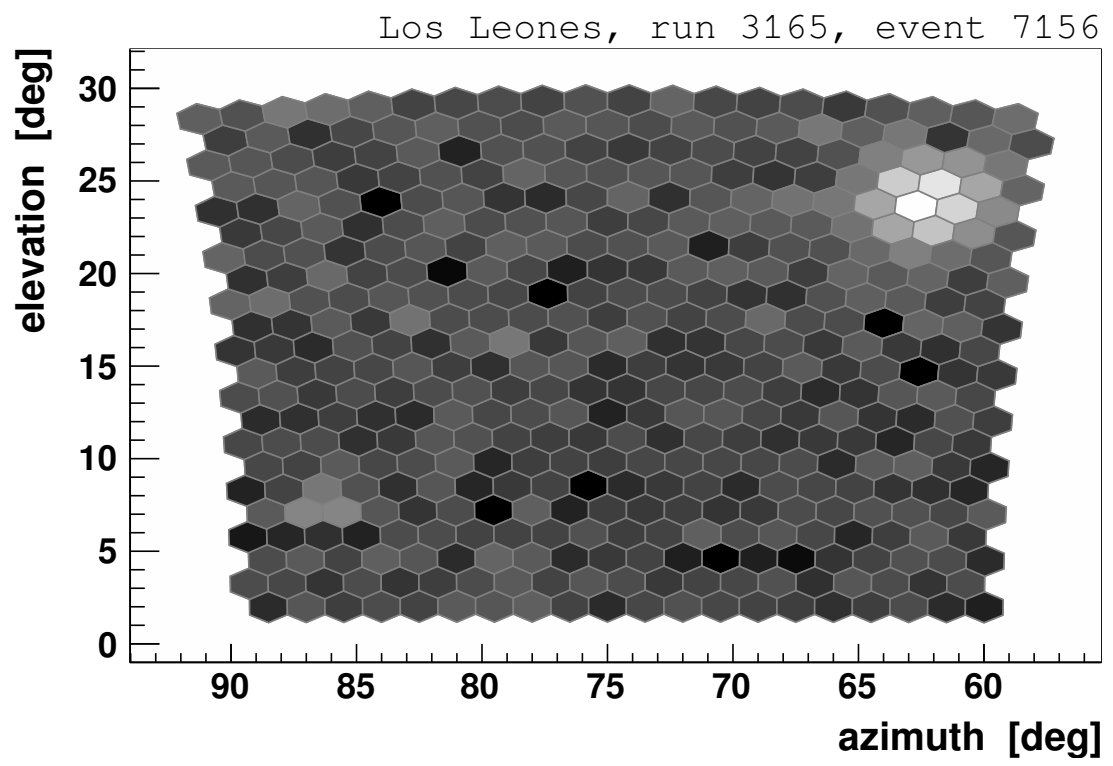

**Figure 4**: Example of a measured flasher event. The hexagons denote the pixels of the FD camera and the main spot is clearly visible in the upper right corner. The logarithmic signal in each pixel is indicated by grey-scale colours.

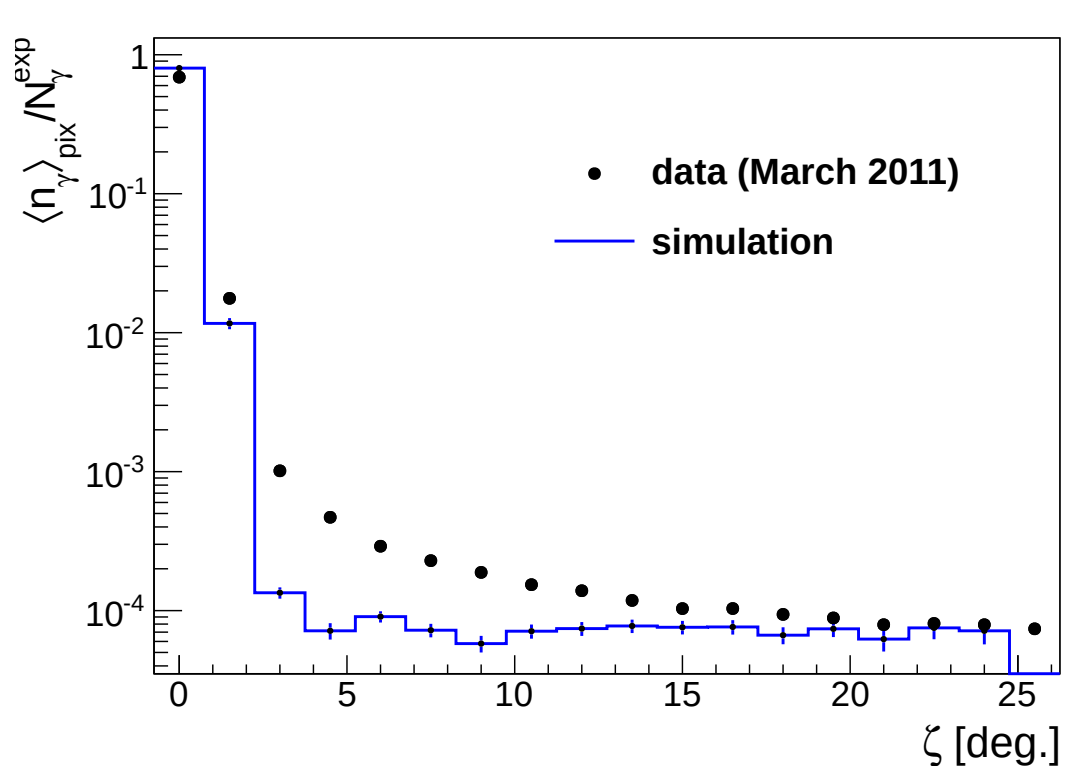

**Figure 5**: Differential light distribution from data (dots) and simulation (histogram). Shown is the average number of photons detected per pixel, $\langle n_\gamma \rangle_{\text{pix}}$, normalised to the expected total number of photons on the aperture, $N_\gamma^{\text{exp}}$.

GPS way-points and the duration per position are programmed into the octocopter to probe the desired pixels in a telescope. The number of light source pulses per way-point is fixed by the pulse frequency of 1 Hz. Over the past few years, campaigns with different setups have been carried out. Several pixels with different positions on the telescope camera have been probed. The position of the light spot on the surface of a pixel has also been varied, changing from a position well centred on a PMT to positions right on top of a light collector between two pixels. For comparison, measurements have been made for several different telescopes and using varying distances between telescope and light source. To study the optics of the telescope in more detail, telescope components like the mirror, the camera, the corrector lens or the filter have been manipulated (e.g. cleaned), covered or removed. An example of a measured flasher event is shown in Fig. 4.

## 3 Simulations

To simulate the response of the telescope to the flasher light source, we use the Auger Off$\underline{\text{line}}$ Framework [9]. It offers two modules for the simulation of the telescope. The standard module is based on simple ray tracing and enables very fast telescope simulations. The ray tracing has been enhanced step by step with more and more knowledge of the telescope optics. Single telescope components can be modified, included or excluded from the photon path, allowing for easy comparison with data taken during octocopter campaigns. The results obtained with the improved ray tracing are confirmed with the second module for telescope simulation that is based on GEANT4 [10]. It allows for a very precise simulation, but due to its precision, its computational requirements are too large to allow usage for standard shower simulations.

## 4 Point spread function

The optical spot diameter of the Auger fluorescence detector is about 0.5° [2]. During measurements it was found that photons from the light source are detected not only in the directly hit pixels but also in most other pixels of the camera. This spread of light, the *halo*, has been subject to research and can partially be explained by detailed simulations of the telescope optics. The measured and simulated angular distributions of light are shown in Fig. 5. They have been obtained by averaging over many events recorded during one octocopter flight in front of telescope 3 at the Los Leones site. For all events the position of the light source, i.e. the centre of the light spot, is assigned to $\zeta = 0°$. The mean value of the signals detected in all pixels with a certain angular distance to the spot centre is plotted versus the corresponding angles. We observe a broadened spot at small angles due to the convolution of the pixel size of 1.5° and the finite spot size of 0.5°. The observed width of the smeared peak depends on the distribution of spot positions on the pixel in the shown data set. With increasing distance to the spot centre the observed signal decreases further and forms a more or less flat tail. The signal however does not decrease as steeply as expected from simulations when going out from the spot centre and about 15% of the light is spread to angles larger than 2°. A second spot, a ghost image point symmetric to the centre of the camera, is observed as well (see Fig. 4). Since its angular distance to the spot centre depends on the position of the spot on the camera, the ghost region has been excluded in the estimation of the point spread function shown.

The point spread function changes only very little for different positions of the spot on the camera and for different distances between the light source and the telescope. Measurements with several telescopes at different fluorescence detector sites show very small changes in the light distribution. A clear difference can be seen between an event sample where the light spot is well within one pixel and another sample where the spot is centred on top of a light collector between two pixels.

The main part of the halo is caused by reflections inside the telescope bay. Most of the large tail of the distribution at angles above $\sim 15^\circ$ is caused by light that has been reflected off the surface of one of the pixel PMTs of the camera (see Fig. 6). The reflectivity of the PMTs has been measured in lab experiments and was found to be about 20%, depending on the position on the PMT, the incidence angle and the wavelength. Parts of the light reflected at the PMT surface, for reasons of symmetry, hit the camera in a broad beam, illuminating a large number of pixels. Due to the camera shadow in the incoming light beam, the actual illumination of the camera depends on the position of the spot. The distributions however resemble each other when viewed with respect to the spot centre. If the spot lies well on top of one of the light collectors surrounding all pixels, the back reflections are different. This is clearly visible in the resulting light distribution.

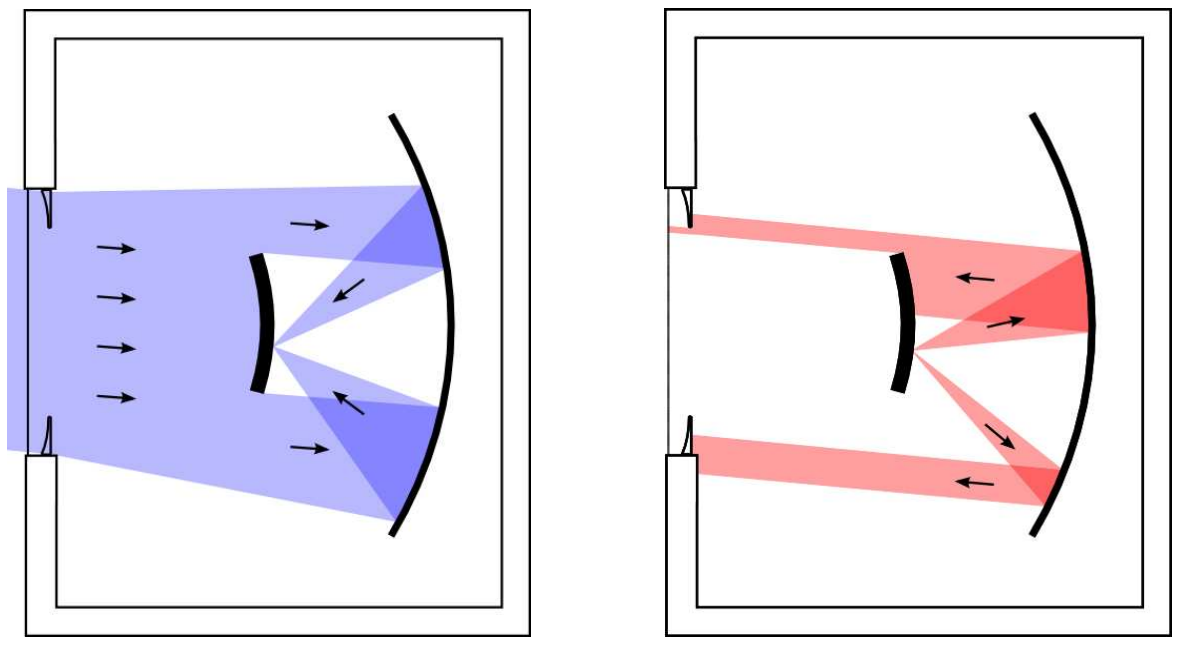

**Figure 6**: Telescope schematics and illustration of PMT reflections. Parallel light entering the telescope from the left hand side traverses the UV-filter at the aperture and the annular corrector lens before it is focused onto the camera (left). Part of the light is reflected on the surface of the camera and is consequently reflected back in the direction of the camera and the aperture by the mirror. This results in a very wide illumination of the camera (right). The part of the light that reaches the aperture of the telescope again can be reflected back to the mirror and is then focused onto a second spot point symmetrically placed about the centre of the camera.

The back-reflected light that does not hit the camera can be reflected again at the aperture of the telescope and is consequently focused in the ghost spot. The intensity of this ghost spot is about 0.4% of the main spot.

Part of the widening of the spot is caused by multiple reflections inside the telescope bay, mainly reflections between the corrector lens and the UV-filter that forms the aperture window. Those reflections disturb the incoming direction of the photon and cause it to be detected in one of the pixels neighbouring the one containing the main spot.

These effects that cause the broadening of the light distribution have been included into the telescope simulations (see Fig. 5). The point spread function derived from simulations describes the measured one well at small angles below $\sim 2^\circ$ and at large angles above $\sim 15^\circ$ to the spot centre. In the intermediate range, some signal is also produced in simulations, although not enough to describe the data.

## 5 Discussion

The measured point spread function is of relevance to shower measurements in two ways: (i) During a drum calibration, all pixels in a camera are simultaneously illuminated. The interpretation of the signals recorded in each pixel therefore directly depends on the point spread function of the optics. (ii) For shower reconstruction, light within an angular region of $\zeta_{\mathrm{opt}}$ is considered. The size of $\zeta_{\mathrm{opt}}$ is determined by the optimal signal to noise ratio. The point spread function of the telescope reduces the fraction of light inside a circle with opening angle $\zeta_{\mathrm{opt}}$.

Despite various cross-checks and measurements of the influence of individual optical components of the telescope on the point spread function, part of the light distribution is still not fully understood, i.e. cannot be reproduced in simulations. This amounts to about $5-6\%$ of the direct light. Multiple scattering in the atmosphere is very unlikely to be a major source: measurements at different distances between the light source and telescopes show no distance dependence. The effect of different aerosol sizes on the light distribution is under investigation. First measurements with a similar light source have been made in 2008. Since then, no significant change in the point spread function has been observed. Current studies focus on measuring individual telescope components in the lab and testing for possible effects of ageing.

Our current understanding of the point spread function requires a change of the drum calibration constants by $3-4\%$. This correction factor has been estimated using simulations and verified by direct measurements inside the telescope bay. Updated calibration constants accounting for this effect are now used for analyses of data from the fluorescence detector. The impact of the measured point spread function on the energy scale of the Pierre Auger Observatory is addressed in detail in [11].

The calibrated point-like light source can also be used to calibrate the fluorescence detector. In contrast to the standard calibration procedure using the drum, the new method results in calibration constants for single pixels within a camera. The calibration of a full camera would be very time consuming and probably not feasible. It provides, however, a systematically independent measurement that can be used to test the current calibration procedure. This method is currently being applied for cross-calibration of the fluorescence telescopes of the Pierre Auger Observatory and the Telescope Array [4].

# The monitoring system of the Pierre Auger Observatory: on-line and long-term data quality controls

CARLA BONIFAZI[1] FOR THE PIERRE AUGER COLLABORATION[2].

[1] *Instituto de Física, Universidade Federal do Rio de Janeiro, Rio de Janeiro, Brazil*
[2] *Full author list: http://www.auger.org/archive/authors_2013_05.html*

*auger_spokespersons@fnal.gov*

**Abstract:** The Pierre Auger Observatory consists of a surface array of 1660 water Cherenkov detectors (SD), overlooked by 27 air fluorescence telescopes (FD) grouped in four sites. A system to monitor the status and the performance of the whole Observatory has been developed to ensure its smooth operation and optimal data quality for physics analysis. In addition to the on-line calculation of the SD exposure and the FD on-time, the available information is used to check the long term stability of key quantities and of data quality, thus defining the performance metrics.

## 1 Introduction

Designed as a hybrid detector, the Pierre Auger Observatory [1], located in Argentina (Pampa Amarilla, 1400 m a.s.l.), uses two techniques to measure the extensive air shower (EAS) properties by observing both their longitudinal development in the atmosphere and their lateral spread at ground level. Charged particles and photons that reach the ground are sampled with the Surface Detector array (SD) which consists of 1660 independent water-Cherenkov detectors (WCDs), filled with 12 tons of pure water each, and equipped with three photomultipliers (PMTs) to detect the Cherenkov light emitted in the water [2]. The WCDs are spread on a triangular grid of 1.5 km spacing over 3000 km$^2$. The fluorescence light generated in the atmosphere by the charged particles of the air shower through excitation of $N_2$ molecules is detected by the Fluorescence Detector (FD) [3] which consists of 27 telescopes, in five different buildings. The field of view of each telescope is 30° in azimuth, and 1.5° to 30° in elevation, except for three of them, for which the elevation is between 30° and 60° (HEAT telescopes [4]). Light is focused with a spherical mirror of 13 m$^2$ on a camera of 440 hexagonal photomultipliers. The FD can only operate during dark nights, which limits its duty cycle to 13% while the SD operates 24 hours per day. Stable data taking with the SD started in January 2004 and the Observatory has been running with its full configuration since 2008.

The operation of the whole Observatory is continuously monitored: information on each detector status, as well as on atmospheric devices, are treated to insure its optimised functioning.

## 2 Monitoring system

The basis of the monitoring system is a database running at the central campus. The front-end is web based using common technologies like PHP, CSS and JavaScript. The replication of the databases and the web site is continuously performed to have a mirror site in Europe. Thanks to this, any Pierre Auger researcher can access the present running conditions as well as the performance of the Observatory from their home institute. On the web site, the main page gives an overview of the general features of the running conditions; any trouble is underlined. Dedicated sections exist for each main part of the Observatory, such as the Central Data Acquisition System (CDAS), the SD, the FD, the communication system, the atmospheric monitoring devices, and the status of the monitoring server itself. In the following the functionalities related to the performance of both the SD and FD are described.

## 3 SD and FD on-line monitoring

Since the SD and the FD are operated differently, the monitoring of their status have different requirements.

### 3.1 SD on-line monitoring

The detectors of the surface array operate constantly in a semi-automated mode. The failures of any WCD component must be detected, and the trigger rates should be controlled. The main page of the SD section displays a summary of the SD array status, where one can have a quick look at trigger rates, at WCDs not sending data ("black tanks") and at malfunctioning ones, noticed via a list of alarms raised by daily checking processes. Several dedicated pages with links between each other allow the user to display more specific information. In particular one can access the slow control data registered every 400 seconds for each of the 1660 WCDs, such as voltages and currents of the solar power system, photomultipliers and CPU board voltages, environmental parameters as well as calibration data and individual trigger rates.

The trigger for the surface detector array is hierarchical [5]. Two levels of trigger (named T1 and T2) are formed at each WCD. The T2 triggers are normally sent to the CDAS; they are combined with those from other detectors and examined for spatial and temporal correlations, leading to an array trigger (T3). The T3 trigger initiates data acquisition and storage. Tables, graphics and maps are available to control T2 triggers, from a page which is regularly refreshed, to get the most up-to-date

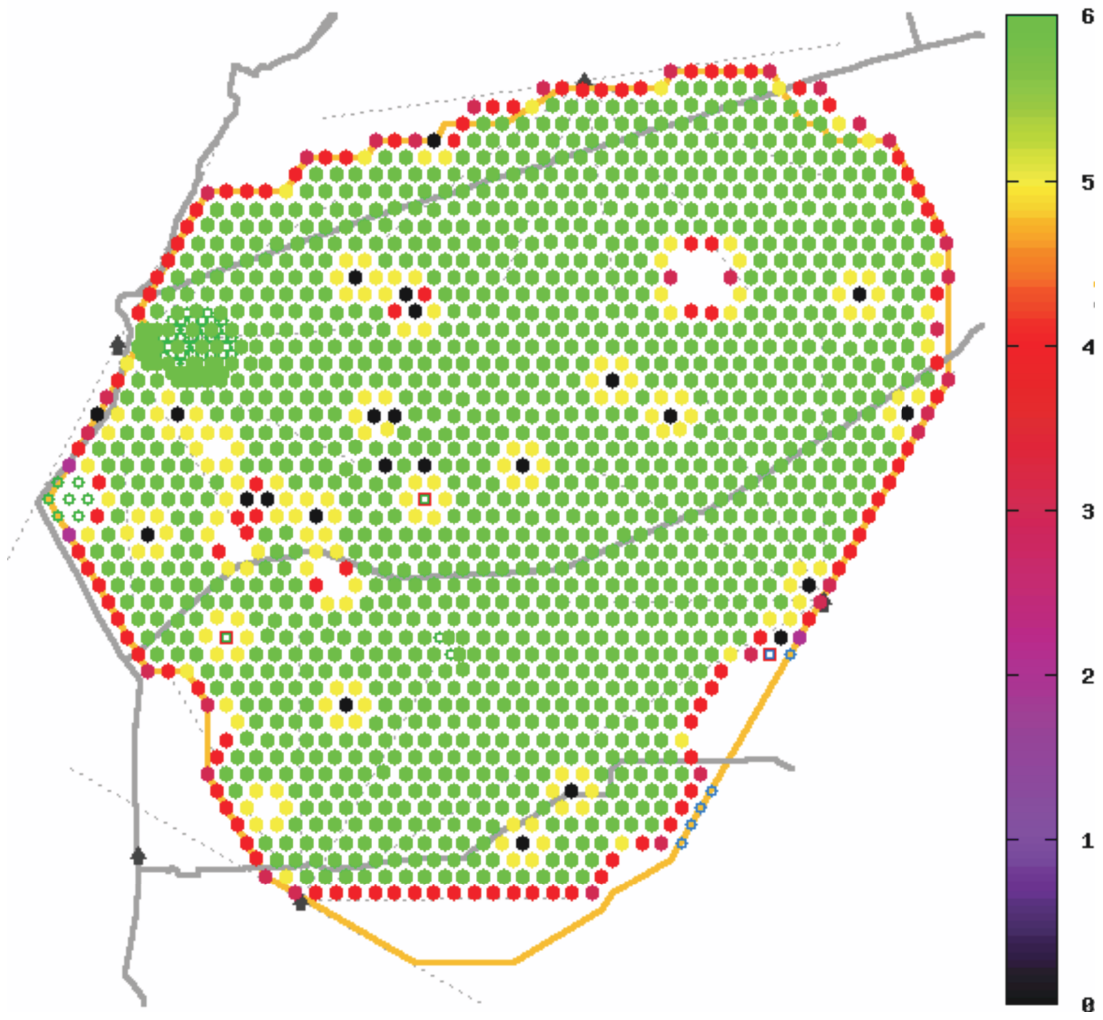

**Figure 1**: Map of the SD array with the current active detectors (colored points). For each working detector, the color indicates the number of surrounding active detectors.

information for displaying the T2 status and its stability over a short period.

The physics trigger, T4, is designed to select real showers from the set of stored T3 data. The high quality SD trigger level is a fiducial trigger to select only events well contained in the array, ensuring the shower core to be properly reconstructed. It requires that six active detectors surround the detector with the highest signal at the time of shower impact, the seven WCDs forming then an active hexagon.

For this trigger the SD is fully efficient for the detection of EAS with energy above $3 \times 10^{18}$ eV and zenith angle below $60^\circ$ [5]. In this range of energies and angles, the SD exposure can be determined on the basis of the geometrical aperture. Due to maintenance operation and "black tanks", the aperture does not reach its nominal value. The number of active hexagons is thus continuously monitored and stored in the database. A dedicated web page of the SD section provides these numbers and a corresponding array map (Fig. 1).

### 3.2 FD on-line monitoring

The data acquisition for the FD telescopes is organised building-wise to ensure against disruption of data collection due to possible communication losses between the CDAS and the remote detectors. For the FD monitoring the data transport is organised via the database internal replication mechanism. This mechanism recognises communication problems and tries to catch up with the submitted database changes when the connection is reestablished, thus guaranteeing completeness of the data-set on the central server.

The data taking of the FD can only take place under specific environmental conditions and is organised in night shifts. The telescopes are not operated when the weather conditions are dangerous (high wind speed, rain, snow, etc.) and when the observed sky brightness (caused mainly by scattered moonlight) is too high. As a consequence, the shifters have to continuously monitor the atmospheric and environmental conditions [6] and judge the operation-mode on the basis of the available information.

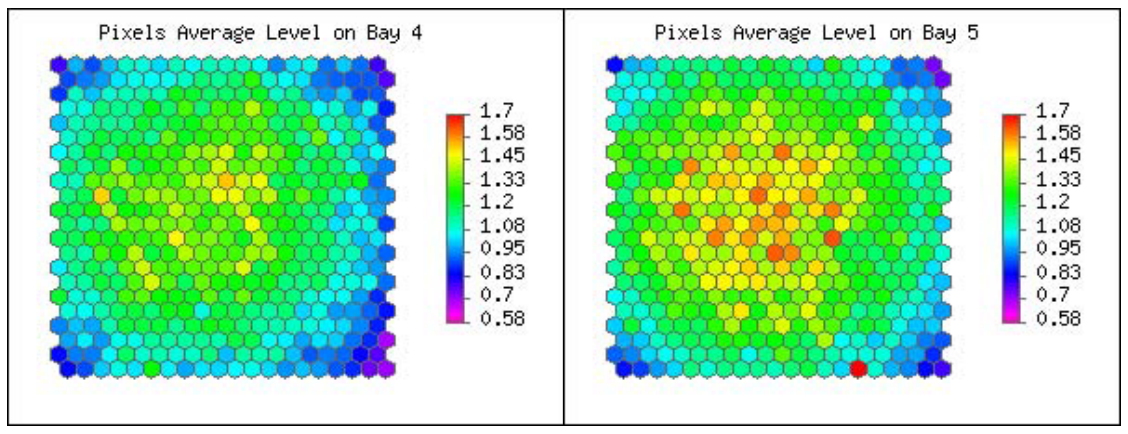

**Figure 2**: Picture from the web-interface showing a selection of FD-calibration data for two of the six cameras in FD-building Coihueco in a dedicated view representing the PMT arrangement.

Alarms, occurrences of states that require immediate action, are first filled into a specified table of the database. The web front-end checks this table for new entries and indicates them on the web page. The shifter is expected to notice and acknowledge the alarm, which then can be declared as resolved once the raised issue is solved.

The information collected for the supervision of the FD operation is split into five sections, dedicated respectively to: i) information from the different levels of calibration (Fig. 2), ii) the background data obtained from each 30 second readout of the full camera, iii) the DAQ and trigger showing the frequency of fired triggers that indicate the status of the telescopes at an advanced stage, iv) the weather conditions and temperatures, and v) the LIDAR monitoring the atmospherical conditions [7] close to the building which is vital for the operation of the telescopes.

## 4 Long-term data quality

### 4.1 Surface Detector

Relevant data useful for long-term studies and for quality checks are stored in the Auger Monitoring database on a one-day basis. Dedicated pages in its SD section web site allow the user to display the evolution with time of the response and of the trigger rates of each Cherenkov detector but also the SD array working status and the quality of the SD data.

Mean values over one day of the number of active SD detectors, of the "black tanks", the number of active hexagons as well as the nominal one (expected value if all the detectors deployed were active), are stored with other metrics in a dedicated table that can be accessed via the web site. From these information, one can check the evolution of the number of active WCDs and of the active hexagons compared to the existing ones. As an example are shown on Fig. 3 the number of active WCDs normalised to the nominal number of WCDs in the array (left) and the number of active hexagons (right) for the last 3 years.

Each WCD has three PMTs, which are balanced such that they produce on average the same output signal. Due to some identified failures, a small percentage of PMTs should not be considered in the analysis. Each PMT has to fulfill several quality criteria to be used in the analysis. Criteria are based on mean values and standard deviations of the PMT baselines, and on parameters used in the WCD calibration [8]. The implementation of the quality cuts is done on a day-by-day basis to provide for each day a list of PMTs showing troubles, stored in a database. The results of the implementation of the quality cut procedure are available via a dedicated SD section. In Fig. 4 we show the percentage of PMTs which do not verify the quality criteria among the functioning ones, since the completion of the array, and this allows us to check the time evolution of the number of rejected PMTs.

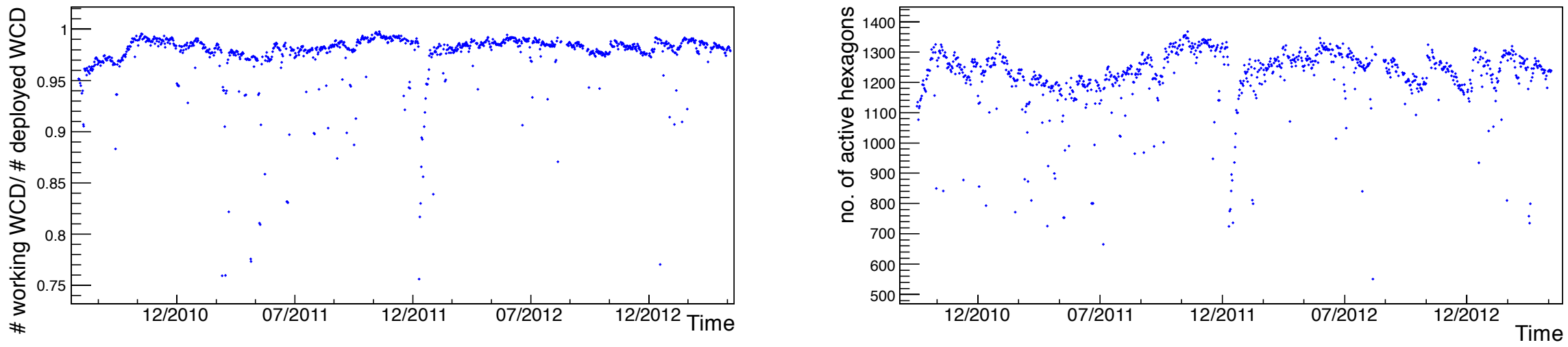

**Figure 3**: Left: number of active WCDs normalised to the nominal number of WCDs in the array, as a function of time. Right: number of active hexagons as a function of time.

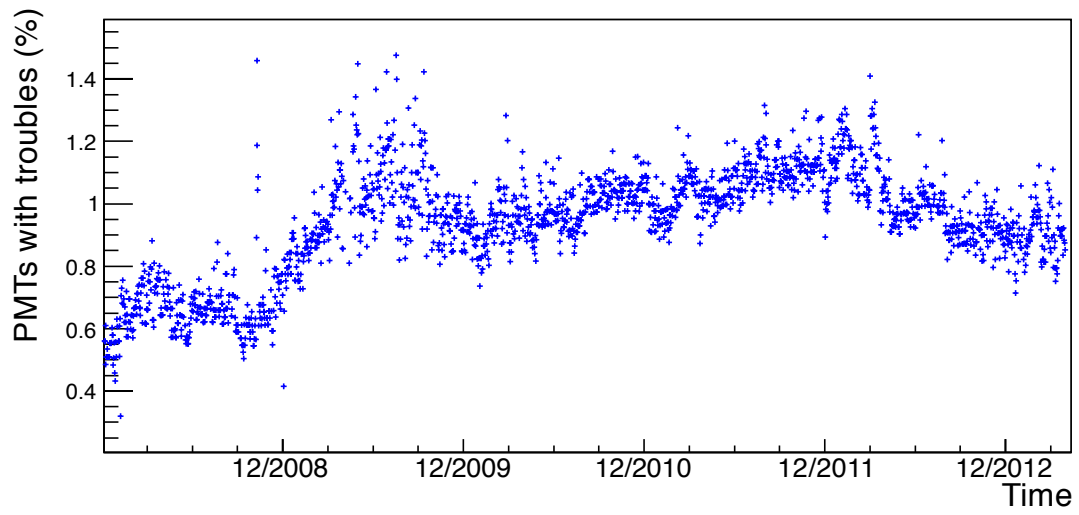

**Figure 4**: Percentage of PMTs which do not verify the quality criteria among the functioning ones, as a function of time.

The most important parameters of the SD calibration [8] are the peak current measured for a vertical muon, $I_{VEM}$[1] (so-called *peak*) and the corresponding charge $Q_{VEM}$ (so-called *area*). The calibration procedure allows the conversion of one VEM in electronics units. $I_{VEM}$ and $Q_{VEM}$ are available from the local station software using the signal produced by the atmospheric muons. To control the uniformity of the detector response, as well as its evolution with time, the distributions of both the *peak* and the *area* can be displayed for all the PMTs of the SD array. Examples of such distributions are shown in Fig. 5, corresponding to one month of data for two different years. The uniformity and the stability of the calibration parameters ensure a stable and uniform response to shower signals. The decrease of the *area* mean value is due to a convolution of water transparency, Tyvek® reflection and electronic response of the WCDs. This does not affect the quality of the data [9].

Beside individual trigger rates and PMT parameters of each WCD, which can be checked over long periods, the T3 trigger rates are also monitored since they reflect the evolution of the SD response. As an example, the T3 trigger rate over past year is shown on Fig. 6.

### 4.2 Fluorescence Detector

The calculation of the on-time for each FD telescope is derived by taking into account the status of the data acquisition, of the telescopes, the camera pixels, the communication system, among others. Details of the

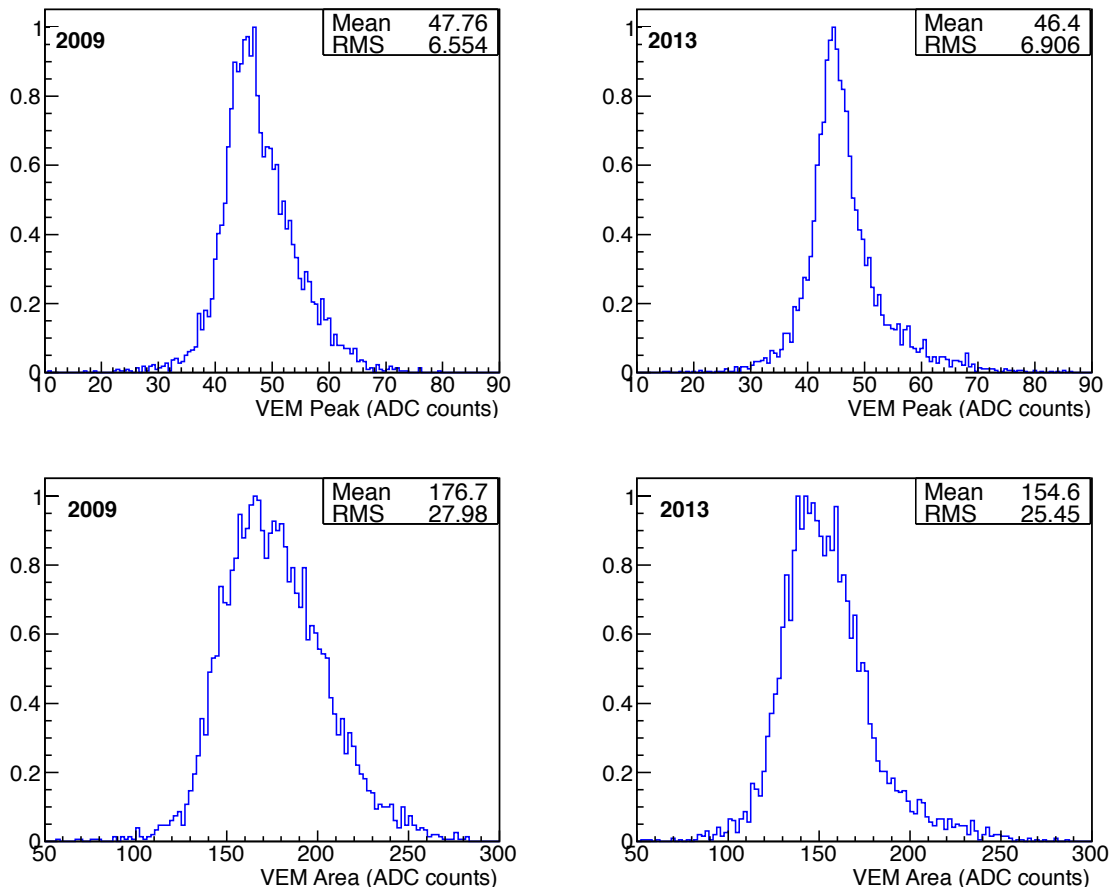

**Figure 5**: Distribution of the *peak* (top) and *area* (bottom) over all working PMTs (one month of data) for 2 different years.

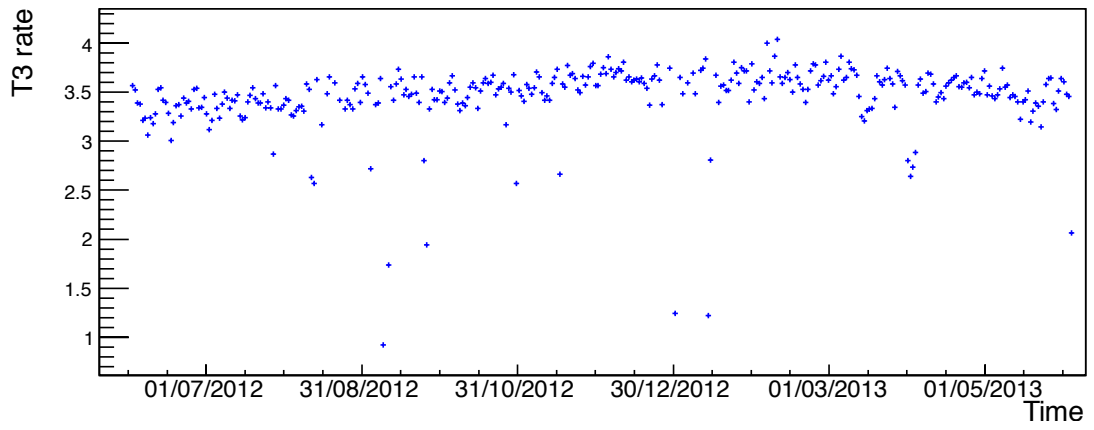

**Figure 6**: T3 trigger rate over past year.

on-time and exposure calculations, necessary ingredients for the measurement of the energy spectrum, are given in [10]. Since July 2007 a tool based on the monitoring system [11] is available for the on-time calculation, accounting also for vetoed time intervals induced by the operation of the LIDAR system or in the case of an excessive rate of FD triggers. The average variances and the on-time-fraction of individual telescopes are calculated for time-intervals of ten minutes, balancing the statistical precision of the calculated on-time with the information frequency. After the initial phase due to the start-up of the running operations the mean on-time is about 13% for all the FD-sites. A program performing the calculation is running on the database server and the appropriate tables are continuously filled

1. VEM: Vertical Equivalent Muon.

in. The web-interface displays the stored quantities. The FD and hybrid[2] on-time of each telescope as well as the accumulated on-time since 1 Jul 2007 for the six telescopes of Coihueco and for the three telescopes of HEAT are plotted on Fig. 7. Similar plots are available for the FD on the monitoring web pages, showing the on-time in quasi real-time for the shifter as a diagnostic and figure of merit.

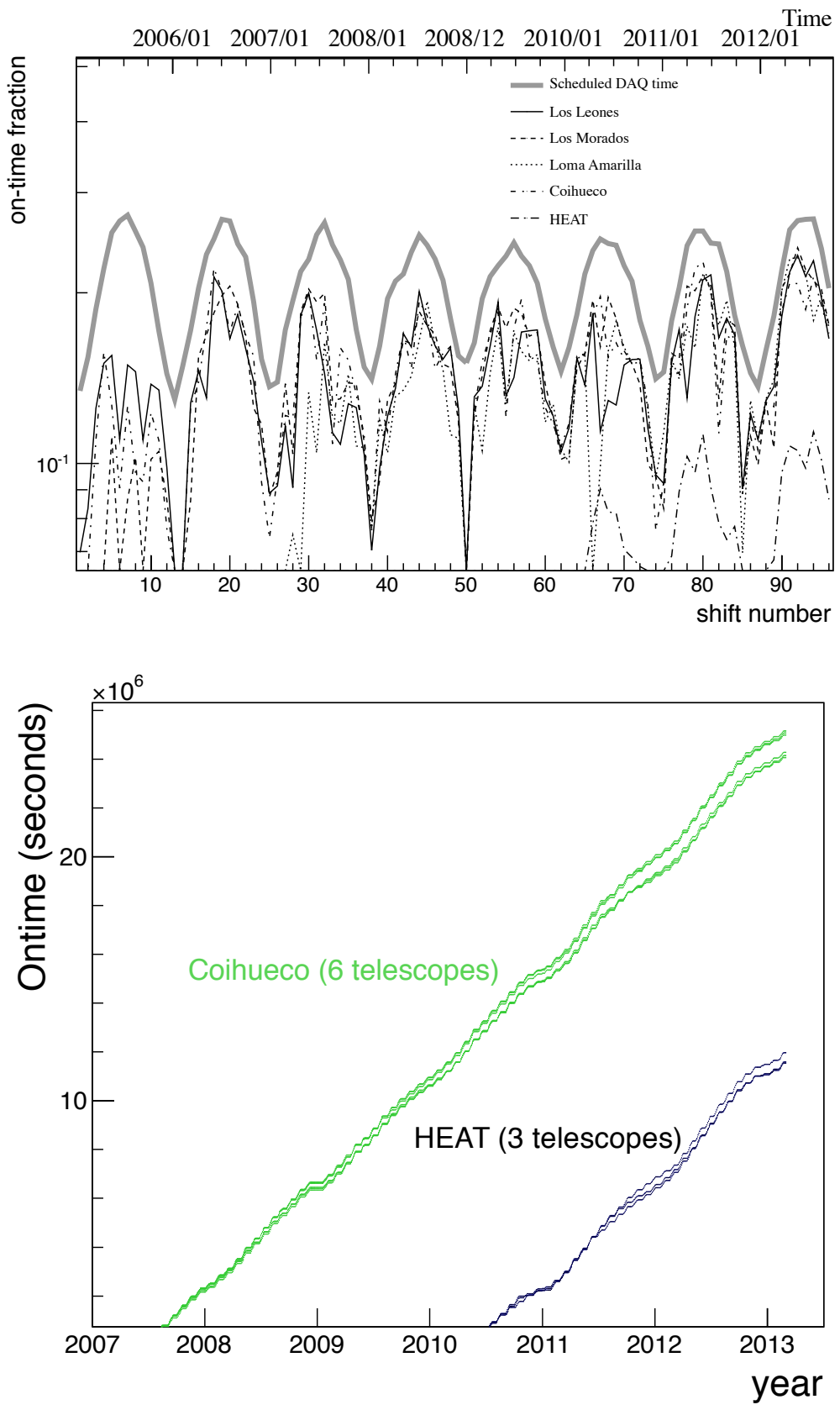

**Figure 7**: Top: time evolution of the average hybrid on-time fraction for the four FD sites and HEAT. The thick gray line defines the scheduled data-taking time fraction limited to the nights with moon-fraction lower than 60%. Bottom: the accumulated on-time since 1 Jul 2007 for the six telescopes of Coihueco and for the three telescopes of HEAT.

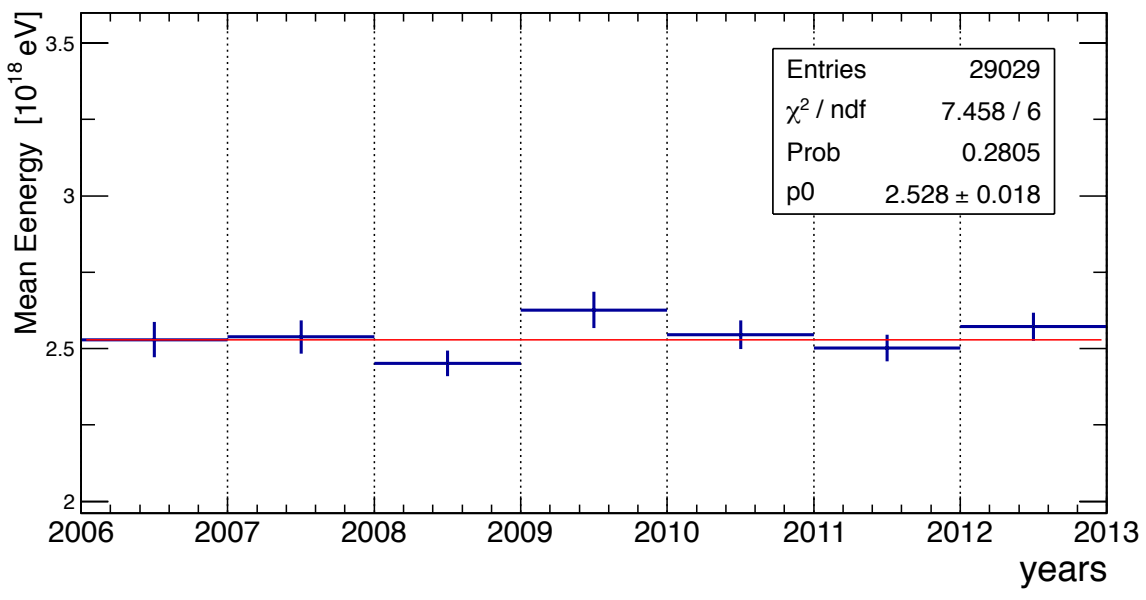

**Figure 8**: Mean energy for reconstructed hybrid events.

### 4.3 Hybrid data quality

Thanks to the smooth running of the Observatory, the performance of the hybrid detector is demonstrated as a function of time using a sample of events fulfilling basic reconstruction requirements, such as a reliable geometrical reconstruction and accurate longitudinal profile and energy measurement.

In Fig. 8, the mean energy of the hybrid events above $10^{18}$ eV with distance to the shower maximum between 7 and 25 km (corresponding to the 90% of the entire hybrid data sample) is shown as a function of time. This plot demonstrates the hybrid data long term stability.

## 5 Conclusions

The Auger Monitoring system is used to control on-line the running of the Observatory and to solve the troubles raised by the alarms. Moreover, it provides also a large number of valuable displays to check the quality of data taking and the long term performances of both the SD and FD.

2. Hybrid events are events measured by both FD and SD.

# Estimation of Signal in Saturated Stations of Pierre Auger Surface Detector

DARKO VEBERIČ[1] FOR THE PIERRE AUGER COLLABORATION[2]

[1] *Laboratory for Astroparticle Physics, University of Nova Gorica, Slovenia*
[2] *Observatorio Pierre Auger, Av. San Martin Norte 304, 5613 Malargüe, Argentina*
*auger_spokespersons@fnal.gov, full author list: http://www.auger.org/archive/authors_2013_05.html*

**Abstract:** The Surface Detector Array of the Pierre Auger Observatory consists of more than 1600 water-Cherenkov stations deployed in a triangular grid with a spacing of 1.5 km and covering an area of 3000 km$^2$. From the recorded signals and their timing we reconstruct the impact point, the axis of extensive air-showers and the lateral distribution of the particles on the ground. When the impact point of the shower at ground is close to a detector, the dynamic range of the recording electronics is smaller than required to record the Cherenkov signal produced by the particles. We present an off-line recovery procedure developed to estimate the signal in case of saturation. We will discuss the performance of this method and the implications for event reconstruction.

## 1 Introduction

In each of the water-Cherenkov stations 12 tonnes of water in a light-tight container is viewed by three 9 inch photomultipliers[1] (PMTs) [1]. To increase the dynamic range, a signal from each of the PMTs is sampled by two 10 bit, 40 MHz FADC readouts, where one is directly connected to the PMT anode, $A$, and the other to the last dynode, $D$, through an amplifier chain [2] with effective signal ratio $D/A \approx 30$ (see schematics in Fig. 1). The effective dynamic range achieved is ~15 bit and thus signals overflowing the digital range only in the dynode readout are trivially recovered using the unsaturated anode readout (for a fraction of such events see gray points in Fig. 2). Nevertheless, showers with energy of 1 EeV also saturate the anode readout of stations that are less than around 200 m from the impact point, while for showers with energy 100 EeV this distance increases to ~500 m (see black points in Fig. 2 for related probability). It is very difficult to cover the full dynamic range since the shower signal rises rapidly when approaching the shower core. However, part of the signal can be recovered even if the PMTs are saturated.

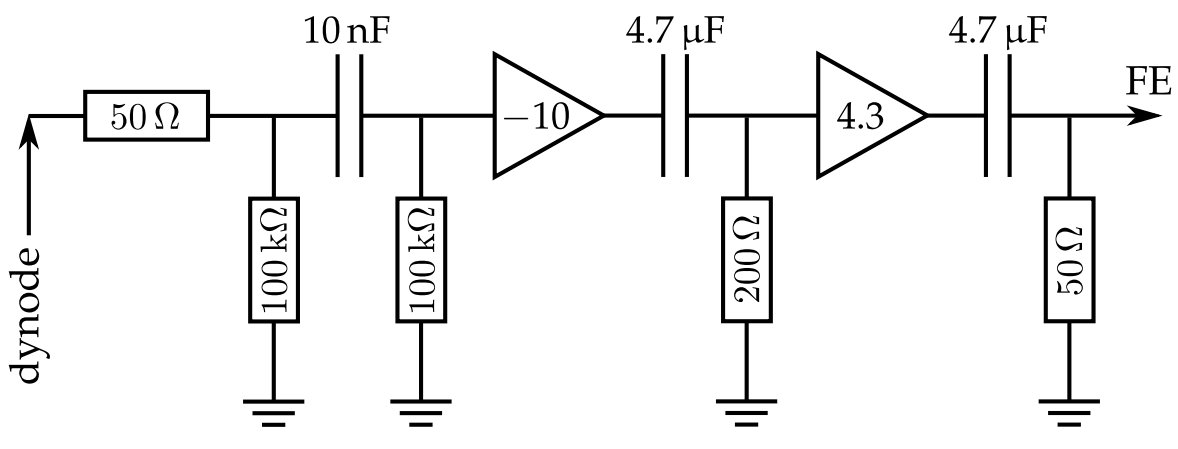

**Figure 1**: Schematics of the readout components between the PMT dynode and the front-end electronics (FE).

A self-calibration procedure is operated at the SD stations based on monitoring of the signals of background muons so that all references to signals are here given in relative units of a single *vertical equivalent muon* (VEM). If not otherwise stated, all plots are made using all 5T5 events [3] with reconstructed zenith angles $\theta < 60^\circ$ and at least one saturated station. The fiducial trigger, 5T5, ensures adequate containment of the event inside the array. Events recorded from 1 January 2004 to 31 December 2012 are used for this analysis.

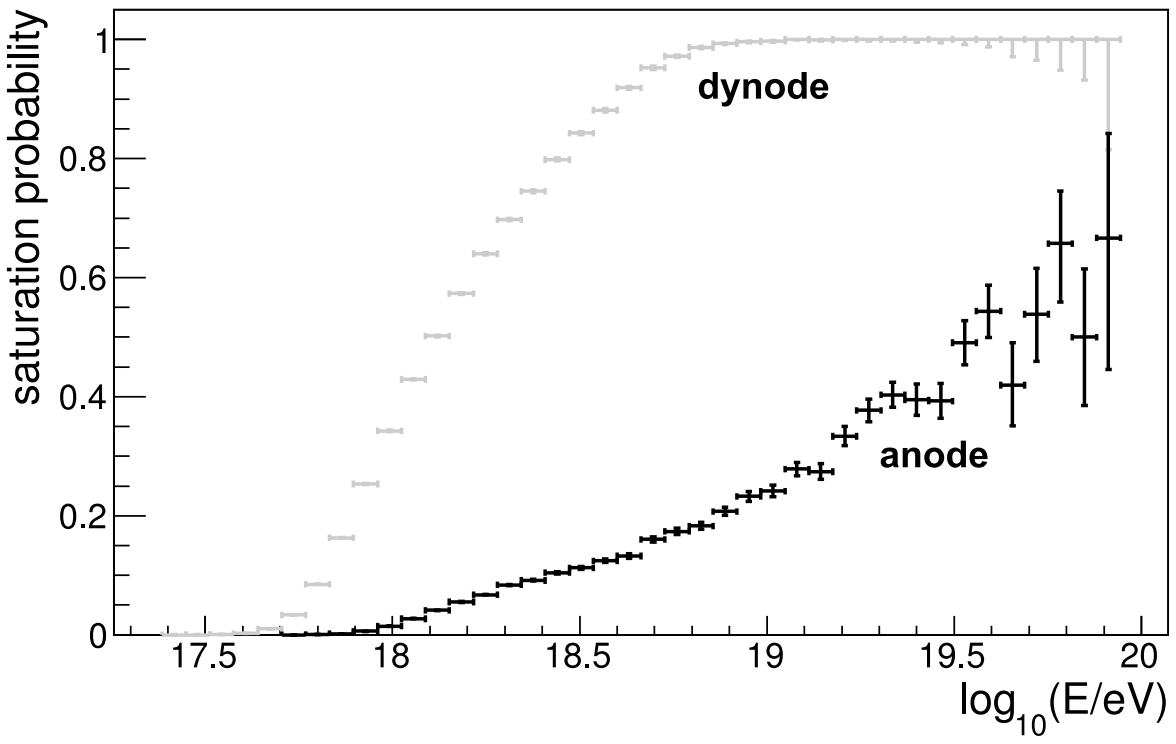

**Figure 2**: Energy dependence of the probability that the station closest to the impact point has a saturated dynode (upper points) or anode signal (lower points).

From the black data-points in Fig. 2 it can be seen that in more than half of the high-energy events the station closest to the shower core is saturated and distorted by one, or a combination of, the following non-trivial possibilities:

**Overflow of the anode FADC** with a dynamic range of 10 bit: with the ~50 ADC channels dedicated for a baseline offset, there are about 950 ADC channels left for the signal range. The typical settings of the SD stations are such that a signal of 1 VEM corresponds to ~50 channels in the dynode readout and ~1.6 in the anode readout, resulting in the range overflow at ~20 VEM in the dynode and 600 VEM in the anode readout.

**PMT non-linearity and saturation:** due to the space-charge effects the response of the PMT for anode currents greater than 100 mA is no longer linear. We have modeled (and measured) the PMT response with a function $g(S)$ which is linear for small input signals $S$ and saturates towards a constant maximum for large signals. With 1 VEM

1. Photonis XP 1805

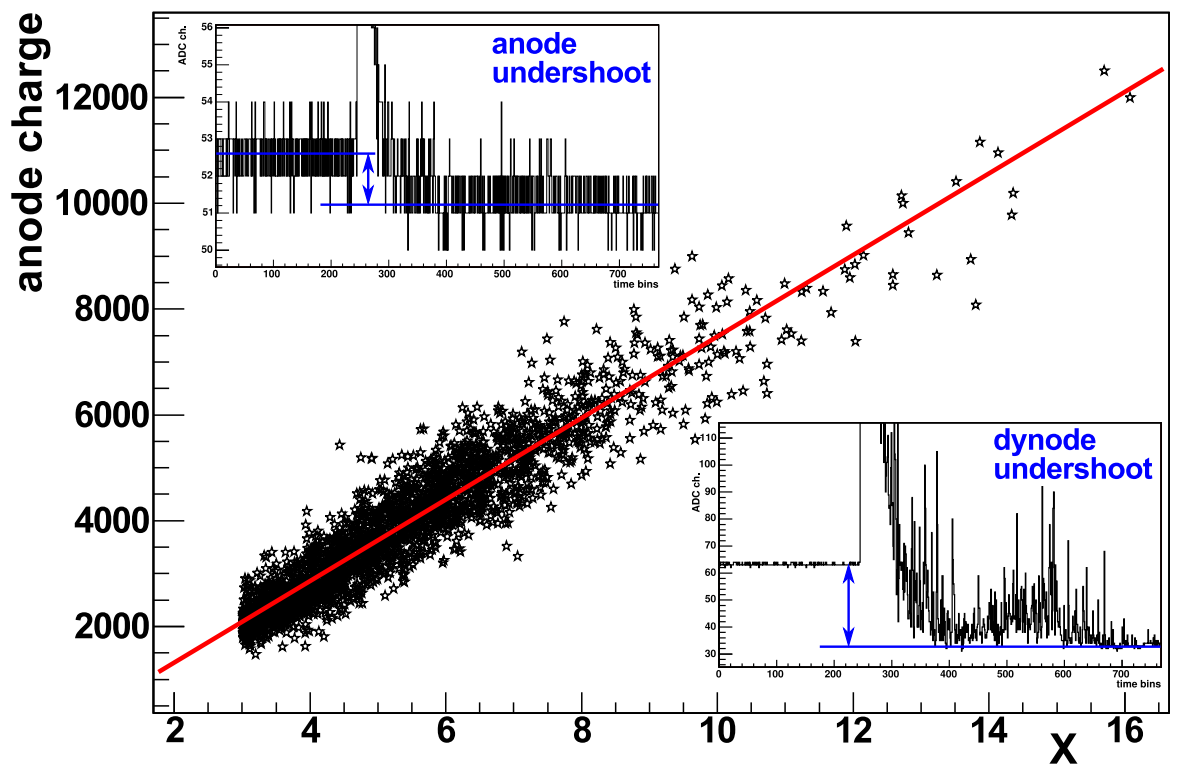

**Figure 3**: To illustrate of the relationship between anode charge $Q_a$ and dynode undershoot $U_d$, values on abscissa were chosen as $X = U_d - 1.35\times10^{-4}Q_d - 0.055K_d$. The two insets show typical saturated anode and dynode traces with lowering of the baseline after the main signal pulse clearly seen.

corresponding roughly to $\sim$100 photoelectrons and with a typical gain of $2\times10^5$, the anode FADC overflow mentioned above corresponds to currents of 50 mA, which is at the onset of the PMT non-linearity specified by the manufacturer (production requirement was less than $\pm5\%$ deviation from linearity below 50 mA).

## 2 Recovery method

The anode and dynode readout chains can be effectively described in terms of one and three RC circuits, respectively. We expect that the amplitudes of the undershoots $U_a$ and $U_d$ appearing in the anode and dynode traces are due to coupling capacitors (see Fig. 1) in a simple relation to the corresponding integral of the anode signal, also known as charge $Q_a$. Note that for large signals $U_d$ itself can have an "underflow" when the undershoot swing of the signal goes below zero. The inclusion of $U_a$ is therefore necessary to extend the useful range of undershoot values. We have employed two approaches: in the first, the time constants of the RC circuits are calculated and the undershoot is obtained analytically; in the second, a heuristic approach was used to identify relevant observables from the data itself. These observables correlate to the anode charge below saturation and the expression obtained is then extrapolated into the saturated regime. Since below saturation the two methods had complementary systematic uncertainties, they were combined into one expression with the difference serving as an estimate of the systematic error. The final prediction of the anode charge from the anode and dynode undershoots can be written as

$$Q_a = \alpha\, U_d^{\beta} + \gamma U_a + \delta\, Q_d + \varepsilon\, K_d + \zeta, \qquad (1)$$

where the values of the coefficients $\alpha$ to $\zeta$ depend on the magnitudes of the undershoots $U_a \gtrless 1$ and $U_d \gtrless 30$, and $K_d$ is the number of overflow bins in the dynode trace.

While individual characteristics of each PMT were thoroughly measured and checked for compliance with the specifications [4] before their deployment in the field, the non-linearity and saturation of some PMTs deployed in the SD array have been, for the purposes of this study, re-measured deep into the saturated region with the two-LED technique [5]. In this method the PMT response function $g$ is obtained from consecutive measurements with increasing strength of light flashes. The deviation from an absolute linear behavior is deduced from the non-linearity estimator

$$\mathrm{nL} = \frac{V_C - (V'_A + V'_B)}{V'_A + V'_B}, \qquad (2)$$

where the $V'_A$ and $V'_B$ signals are already corrected to the expected values, $V'_{A,B} = g^{-1}(V_{A,B})$, using the inverse of the PMT response function obtained at lower intensities. Fig. 4 shows some of the actual curves for nL as a function of the anode voltage. From all the measurements we deduced the mean PMT response $\langle g\rangle$ and established $\pm\sigma$ bounds which are used as estimates of the systematic uncertainties. Due to the large difference of individual PMT responses in the highly non-linear regime, the uncertainties are dominated by these systematic effects. The determination of responses of the individual PMTs is currently under study to reduce this contribution to the uncertainties.

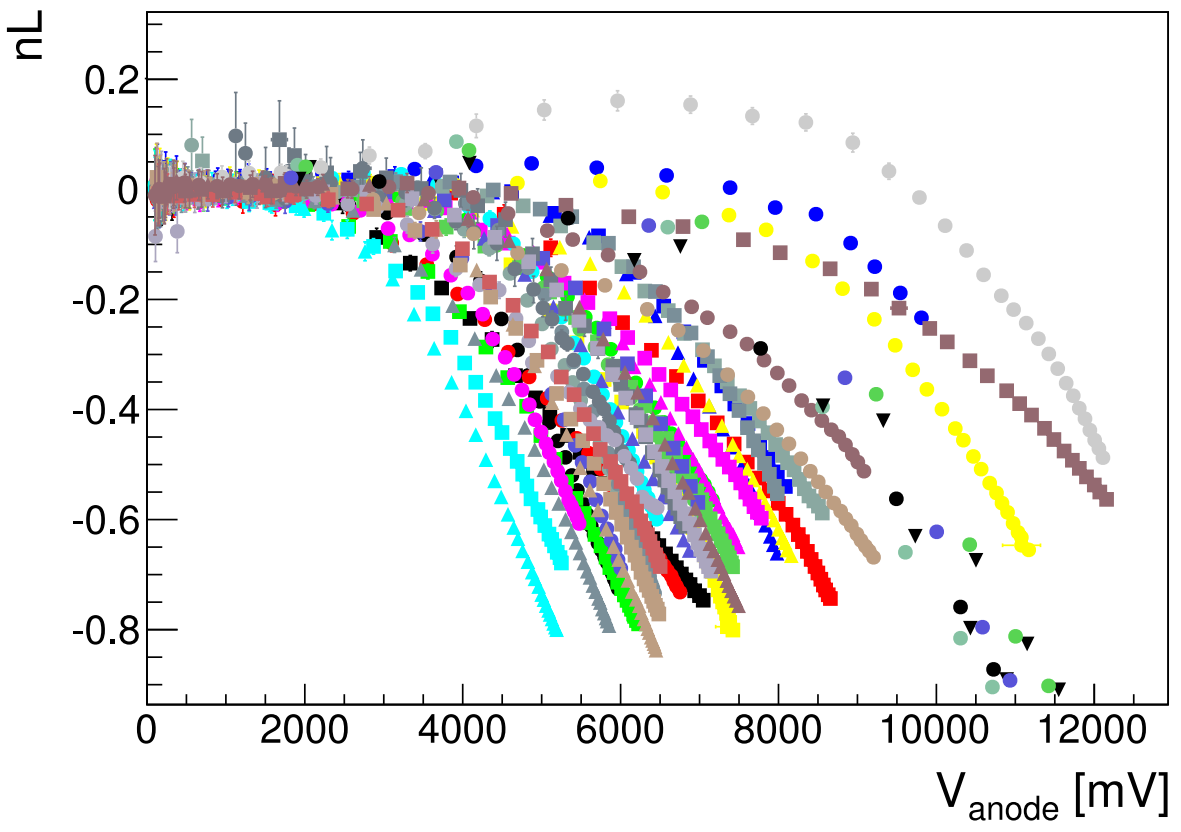

**Figure 4**: Measurements of non-linearity and saturation for several of the PMTs (various colors) deployed in the SD. The values of nL close to zero correspond to the linear regime of the PMT response while a value of $-1$ indicates total saturation (a plateau) of the output voltage. Resistive load is 50 Ω, i.e. 12 V corresponds to 240 mA.

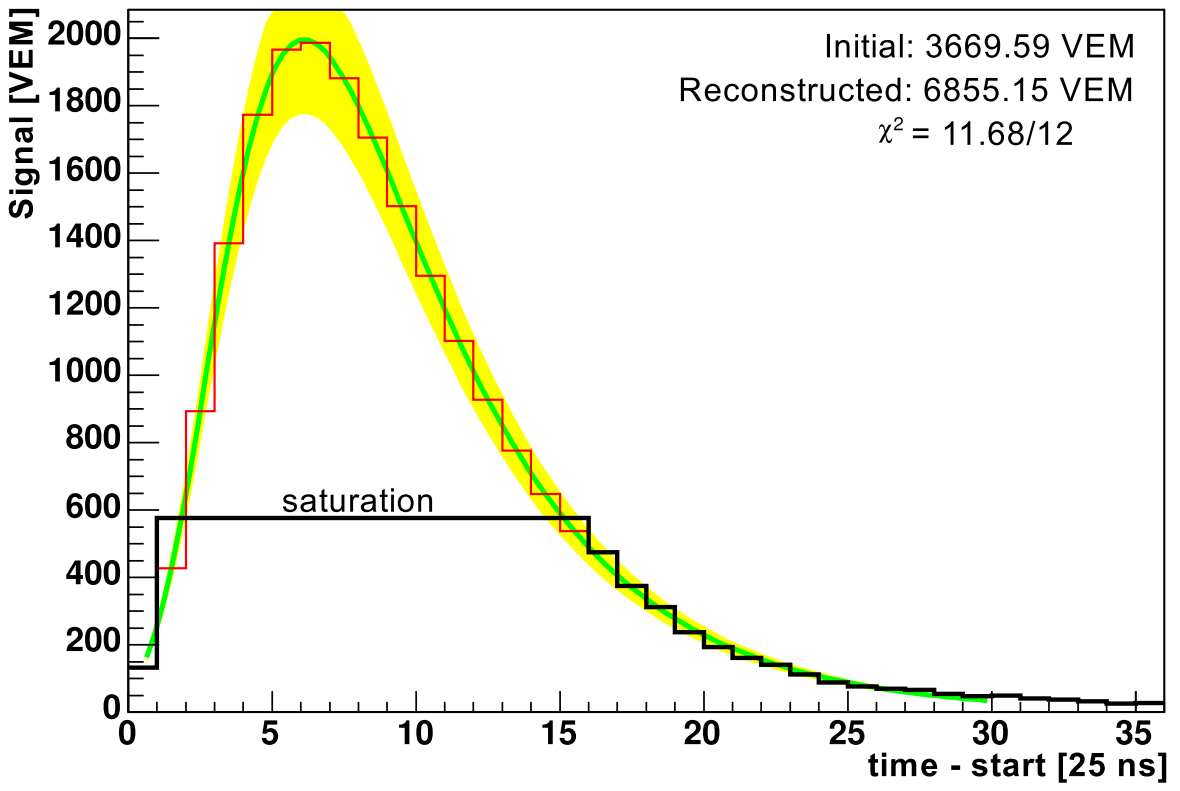

**Figure 5**: Time-dependent signal trace is fitted with the Moyal functional form given in Eq. (4). In this example the overflow of the dynamic range occurs at $\sim$600 VEM and the observed charge of $Q_a^{\rm sat} = 3670$ VEM is increased by $\sim$200% through the recovery method.

Tests with various functional forms showed that the best description of large (but unsaturated) signal shapes is by the functional form of the Moyal distribution [6],

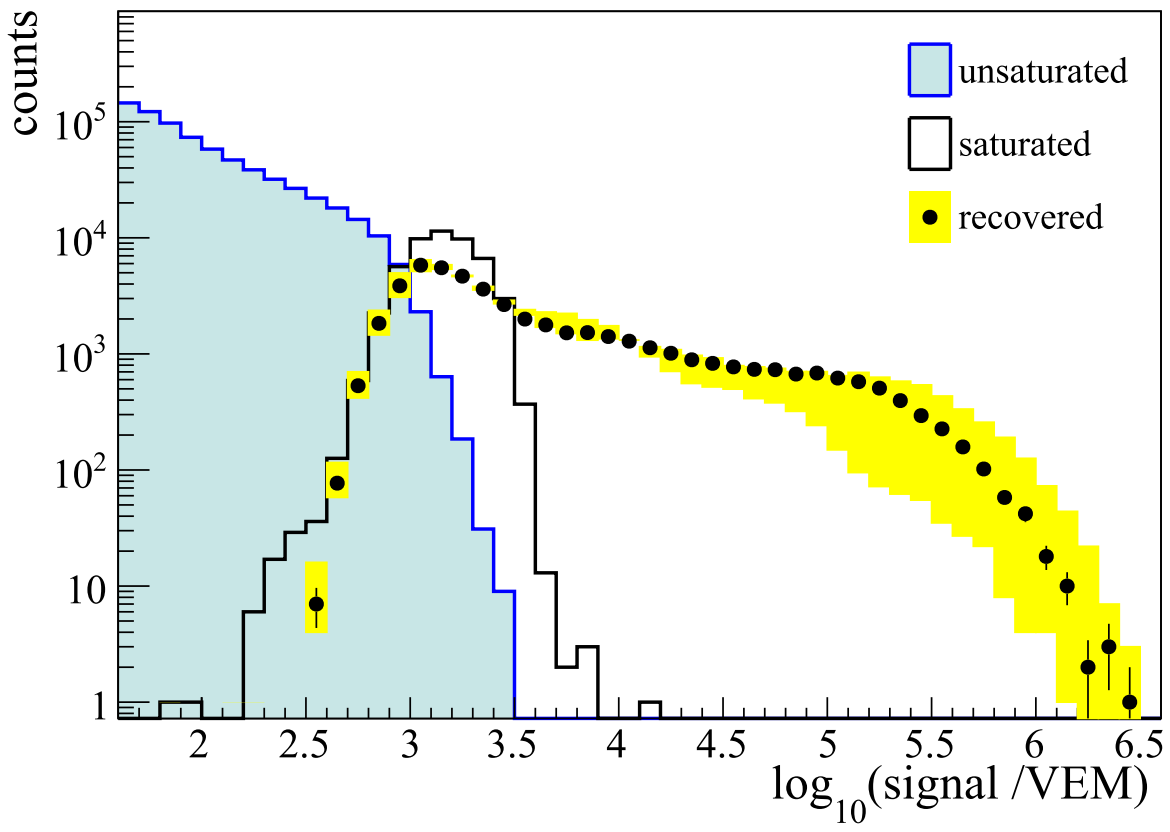

**Figure 6**: Spectrum of different signal categories over all stations in all events. The distribution of the unsaturated signals is shown with the blue-shaded region; the uncorrected saturated signals are denoted with the black line, and the recovered saturated signals are shown with black markers with statistical uncertainties (black bars) and systematic uncertainty (yellow band).

$$\mathrm{M}(x) = \frac{1}{\sqrt{2\pi}} \exp[-\tfrac{1}{2}(x + e^{-x})]. \quad (3)$$

An example of a fit is shown in Fig. 5. In traces with the range overflow we proceed by fitting the remaining parts of the trace with a shifted ($t_0$), broadened ($\sigma_t$) and rescaled function ($a$), wrapped together with the mean PMT response function ($g$),

$$f(t) = g(a\mathrm{M}((t - t_0)/\sigma_t)/\sigma_t), \quad (4)$$

where the charge estimate, $Q_\mathrm{a}$, from Eq. (1) is used as a constraint on the integral of the resulting function,

$$\int f(t)\,\mathrm{d}t \equiv c\,Q_\mathrm{a}. \quad (5)$$

The constant $c$ contains integration and VEM calibration factors. In this way we obtain from the signal overflow and saturation of the PMT response, the final estimate of the true charge

$$Q_\mathrm{a}^\mathrm{rec} = c\,a. \quad (6)$$

The total uncertainty of the recovered signal can be separated into three sources: the uncertainty related to the fitting of the Moyal shape, the uncertainty of the non-linear PMT responses, and the uncertainty originating from the extrapolation of the undershoot relations to large charges. There is also the intrinsic accuracy of the detection of Cherenkov light in the SD stations which is nearly Poissonian (when considered in VEM units). All of the three former uncertainties are increasing functions of the size of recovered signal relative to the observed signal. While the uncertainty from the fitting part is always less than 30%, the PMT non-linearity and the undershoot parts can both reach up to 70% for extremely large fractions of recovered to observed signals but remain below 10% for fractions below 4.

The recovered signal is used in the reconstruction of the lateral distribution of the air-shower only at distances greater than ~60 m. The corresponding term in the maximum-likelihood fit is of log-normal form. At these distances the LDF shape is not well-known (or even unknown) and the assumed power-law descriptions are probably no longer good approximations since most of them are divergent at such small scales and may be a poor description of the flattening of the true LDFs at distances smaller than several Molière radii.

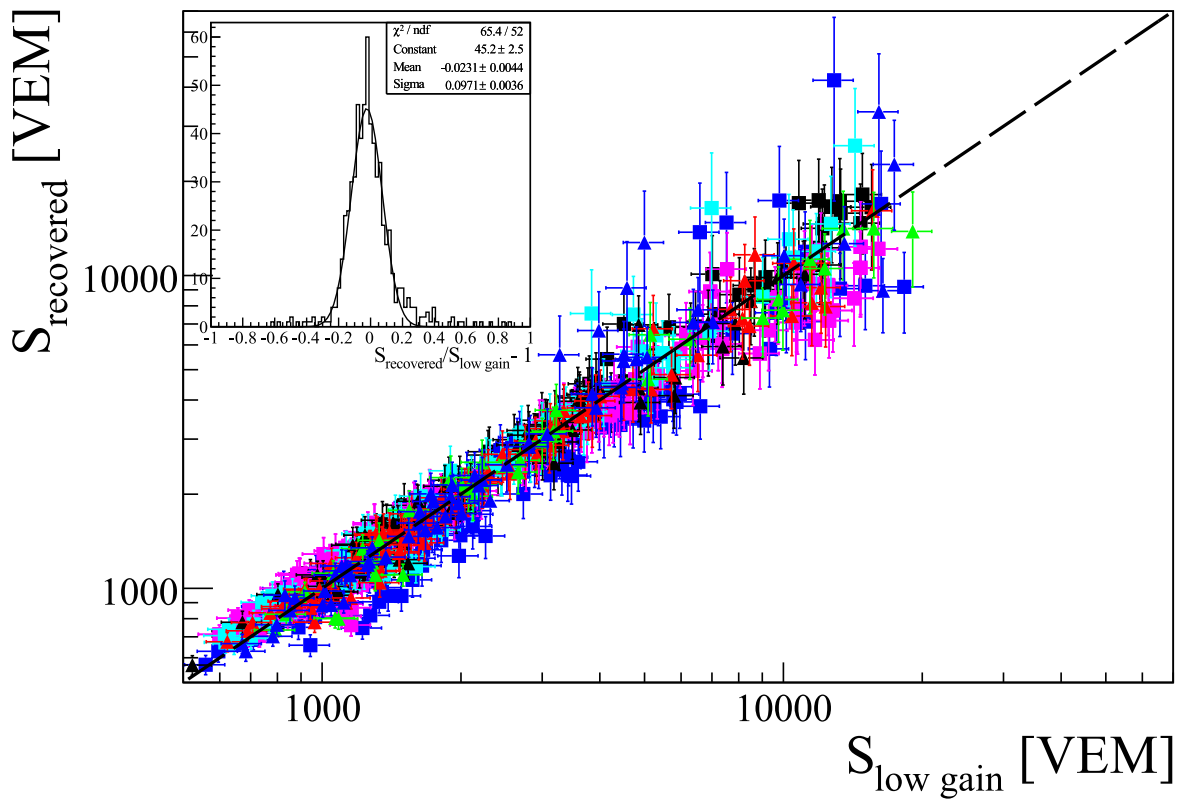

**Figure 7**: Test of the signal recovery method. The scatter plot shows agreement between the direct measurement of the signal with the lowered-gain PMT and the signal estimate from the recovery procedure on the remaining two signals of the nominal PMTs (various colors indicate different stations). The resolution of the method (see inset) is ~10%.

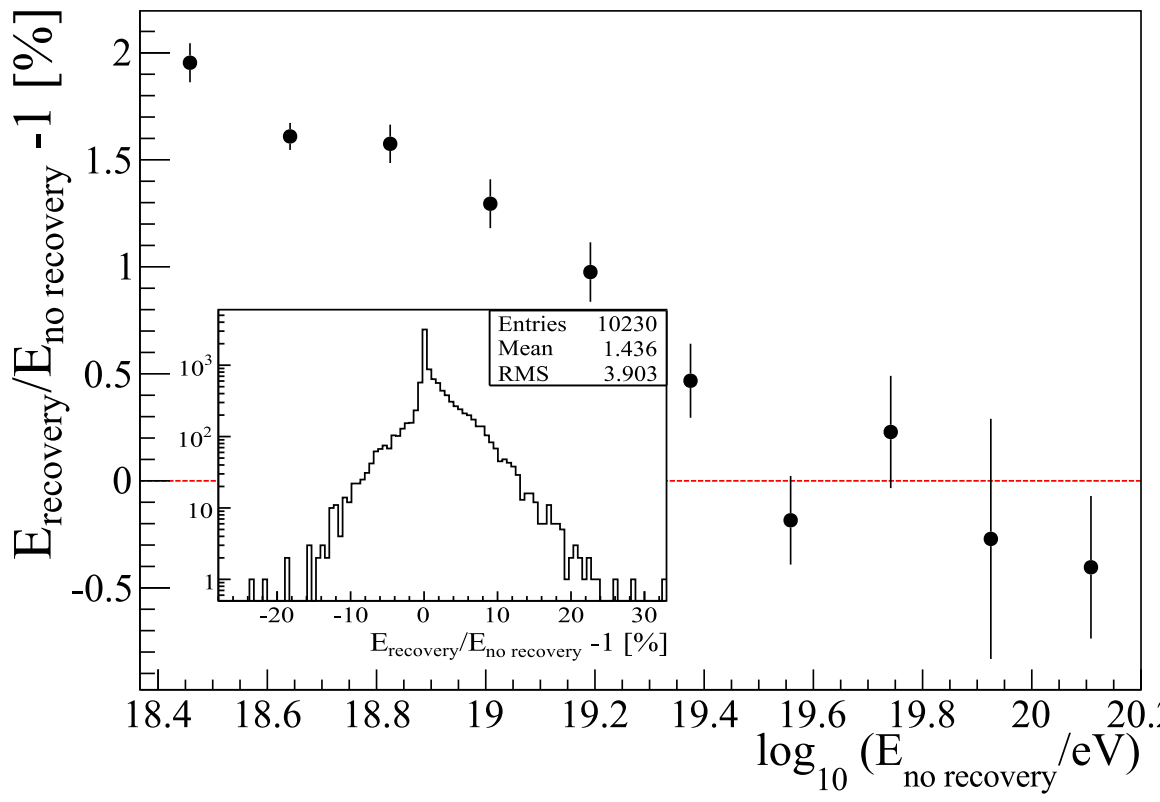

**Figure 8**: Effect of the signal recovery procedure on the energy $E_\mathrm{SD}$ of the events where the change induced by the recovery procedure is evaluated in terms of relative difference. The overall spread of the energies is ~4% (inset) and the energy dependence of the energy difference stays below 2%.

In Fig. 6 a spectrum of signals from all triggered stations in all events is shown for three cases: unsaturated signals, saturated, and recovered signals. As can be clearly seen, the saturated signals that could otherwise have been suppressed are, after the recovery procedure, spread out to larger magnitudes and the resulting spectrum makes a matching continuation into the spectrum of the unsaturated signals below $10^3$ VEM.

## 3 Tests of the method

To validate the recovery procedure, the gain of one of the three PMTs was lowered by a factor of 25 in 12 stations by reduction of the high-voltage setting. In this way, for one of the PMTs, the onset of saturation is extended from ~500 to 12 000 VEM. The comparison of unsaturated signals from this low-gain measurement with the saturated and recovered signals from the other two PMTs enables us to validate the recovery procedure and estimate its resolution and systematic uncertainty.

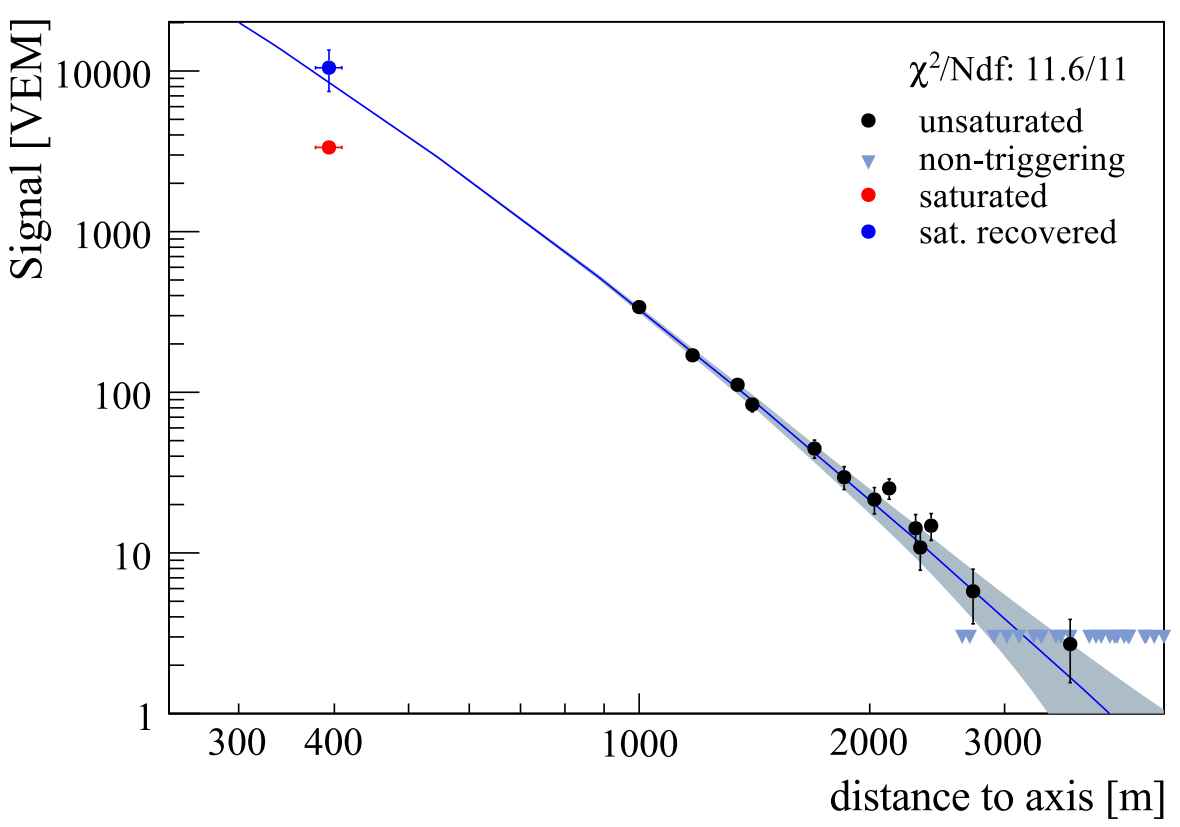

**Figure 9**: An example of a lateral distribution reconstruction. The signal close to the shower axis was recovered from 3500 to 11 000 VEM. The energy of the event is 64 EeV.

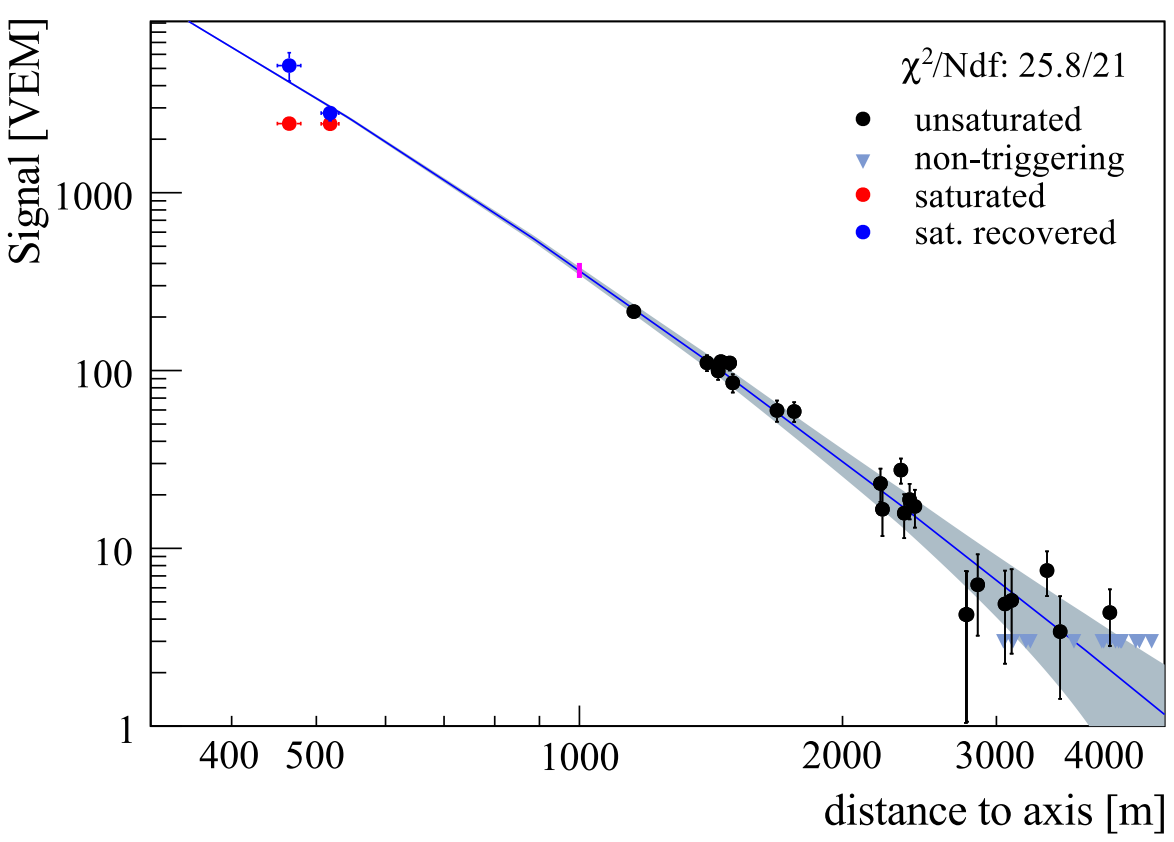

**Figure 10**: Another example of an event with two saturated stations. The signals close to the shower axis were recovered from 2400 to 2800 VEM and from 2500 to 5100 VEM. The energy of the event is 103 EeV.

In Fig. 7 such a comparison is presented for eight selected stations which under lowered gain conditions had less than 5% deviation from the PMT linearity. Using individual nL curves from the measurements (see Fig. 4) and undershoot relations that have been using individually-tuned Eq. (1), the signal recovery performs well and without bias with a resolution of $\sim$10% all the way up to the recovered signals of more than $10^4$ VEM. Nevertheless, until a database of individual nL measurements is established we can only use an average nL parametrization, a mean undershoot relation in our current implementation and the spread of the nL curves must be included in the estimates of the systematic uncertainties.

Fig. 8 shows the effect of the recovery procedure on the reconstructed energy of the events where the relative difference is defined as $\Delta E/E$ and the two energies $\Delta E = E_{\rm rec} - E$ are those obtained with and without the recovery procedure. The mean estimation of the energy of the cosmic rays is changed by less than 2% with an RMS of 4%, but differences of up to more than 15% can occur. The highest energy events are less affected due to the high multiplicity (more than 10 stations above 60 EeV) of stations that were triggered by the shower.

Fig. 9 and 10 show two examples of an event illustrating the effect of the saturation recovery on signal measurements of the stations close to the shower axis. The signals from the unsaturated stations are shown with black points while the saturated observed signals are in red. The recovered signals are shown in blue. While the first event suffered saturation only in the closest station, the second event had the two closest stations saturated. The resulting fitted LDF and its uncertainty are shown with blue line and shaded band. Untriggered stations without a signal are denoted by the blue triangles.

## 4 Summary

We have presented a recovery method used in the off-line reconstruction of the surface detector events of the Pierre Auger Observatory [9]. The method enables inclusion of saturated stations in the lateral distribution-function fits with consequent improvements in energy estimation.

The resolution of the recovered signal is 20% if the PMT responses for very high currents are known. The recovery procedure extends the dynamic range of the PMTs from about $3\times10^3$ VEM up to $10^6$ VEM, currently with uncertainties larger than 60% for recovered signals above $10^5$ VEM. Taking into account the individual PMT responses will reduce this uncertainty in the future.

# The measurement of the energy spectrum of cosmic rays above $3\times10^{17}$ eV with the Pierre Auger Observatory

ALEXANDER SCHULZ [1], FOR THE PIERRE AUGER COLLABORATION [2]

[1] *Institute for nuclear physics, KIT, Karlsruhe*
[2] *Full author list: http://www.auger.org/archive/authors_2013_05.html*

*auger_spokespersons@fnal.gov*

**Abstract:** The flux of cosmic rays above $3\times10^{17}$ eV has been measured with unprecedented precision at the Pierre Auger Observatory based on data in the period between 1 January 2004 and 31 December 2012. The unique combination of different nested detector arrangements has been used to record cosmic ray data spanning over an energy range of almost three decades. The hybrid nature of the instrument has been exploited to determine the energy in a data-driven mode with minimal Monte Carlo input. The spectral features are presented in detail and the impact of systematic uncertainties on these features is addressed.

## 1 Introduction

The measurement of the energy spectrum of ultra-high energy cosmic rays addresses fundamental questions about the origin and propagation of these particles, as well as about physical properties of accelerators and particle cross-sections at the highest energies. The most distinct features of the flux above $10^{18}$ eV are a flattening of the spectrum at $4\times10^{18}$ eV (the *ankle*) and a strong flux suppression above $5\times10^{19}$ eV which is often attributed to the GZK cut-off but might also be due to the maximum source energy [1, 2, 3]. The exact physical explanation of the observed spectral features remains uncertain. Also, the transition from galactic to extra-galactic cosmic rays may occur between $10^{17}$ eV and the ankle. A precise measurement of the flux at energies above $10^{17}$ eV is important for discriminating between different theoretical models [4, 5, 6, 7].

The Pierre Auger Observatory is a hybrid detector employing two complementary detection techniques for the ground-based measurement of air showers induced by UHECRs, a *surface detector array* (SD) and a *fluorescence detector* (FD). The SD is an array of 10 m$^2$ water Cherenkov detectors. 1600 detectors are arranged in a hexagonal grid with spacing of 1500 m, covering a total area of 3000 km$^2$. This array is fully efficient at energies above $3\times10^{18}$ eV [8]. 49 additional detectors with 750 m spacing have been nested within the 1500 m array to cover an area of 25 km$^2$ with full efficiency above $3\times10^{17}$ eV [9]. The SD array is sensitive to electromagnetic and muonic secondary particles of air showers and has a duty cycle of almost 100% [10, 11]. The SD is overlooked by 27 optical telescopes grouped in 5 buildings on the periphery of the array. The FD is sensitive to the fluorescence light emitted by nitrogen molecules that are excited by secondary particles of the shower and to the Cherenkov light induced by these particles. This allows for the observation of the longitudinal development of air showers during clear and moonless nights, resulting in a duty cycle of about 13% [12, 13].

We present the measurement of the flux of cosmic rays above $3\times10^{17}$ eV, obtained by combining data from these detectors. The dataset extends from 1 January 2004 to 31 December 2012, thus updating earlier measurements.

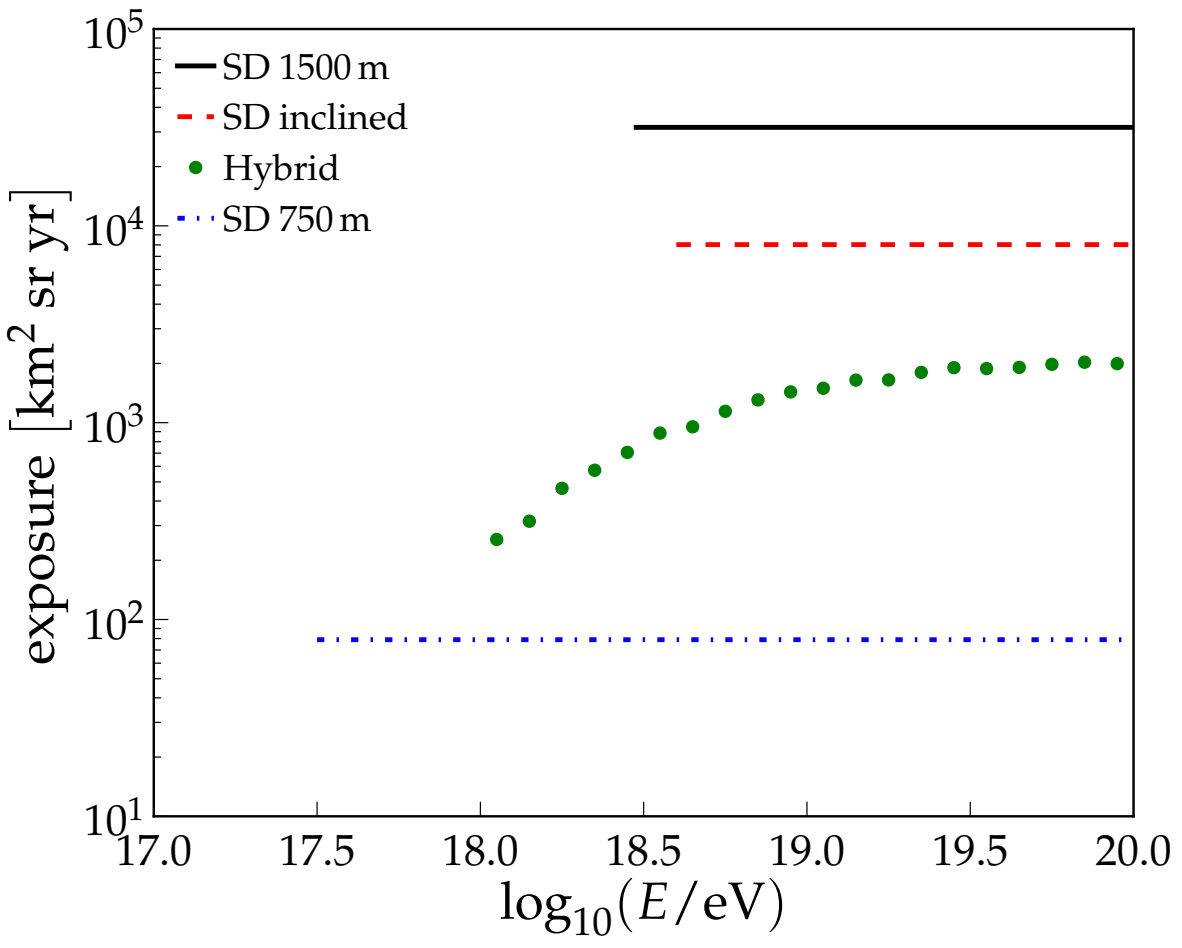

**Figure 1**: The integrated exposure of the different detectors at the Pierre Auger Observatory as a function of energy. The SD exposure in the three cases is flat above the energy corresponding to full trigger efficiency for the surface arrays. Values and zenith angle ranges are given in Table 1.

## 2 Flux measurements with the SD array

The reconstruction of arrival direction and core position of air showers measured with the SD array is performed using the trigger times and signals recorded by individual detector stations. Signals are calibrated in units of VEM, corresponding to the signal produced by a Vertical Equivalent Muon [11, 14]. Different attenuation characteristics of the electromagnetic and muonic shower components lead to different reconstruction methods for different zenith angle ranges. In the following we distinguish between *vertical events* ($\theta < 60°$) and *inclined events* ($62° \leq \theta < 80°$).

The energy reconstruction of vertical events is based on the estimation of the lateral distribution of secondary particles of an air shower reaching ground at an optimal distance to the shower core. The optimal distances are those at which, for a wide range of reasonable lateral distribution functions, the spread in this signal size predicted at that distance is a minimum. For the 1500 m and 750 m arrays the optimal distances, determined empirically, are 1000 m and 450 m respectively. See [15, 16] for details. The signals $S(1000)$ and $S(450)$ are corrected for their zenith angle dependence due to air shower attenuation in the atmosphere with a Constant Intensity Cut (CIC) method [17]. The equivalent signal at median zenith angle of 38° (35°) is used to infer the energy for the 1500 m (750 m) array [9, 18, 19]. Note that for the 750 m array, only events with zenith angle below 55° are accepted. Variations of the shape of the attenuation function due to the change of the average maximum depth of shower development with energy are below 5% for the considered zenith angles.

| | Auger SD | | | Auger hybrid |
|---|---|---|---|---|
| | **1500 m vertical** | **1500 m inclined** | **750 m vertical** | |
| Data taking period | 01/2004 - 12/2012 | 01/2004 - 12/2012 | 08/2008 - 12/2012 | 11/2005 - 12/2012 |
| Exposure $[\text{km}^2 \text{ sr yr}]$ | $31645 \pm 950$ | $8027 \pm 240$ | $79 \pm 4$ | see Fig. 1 |
| Zenith angles $[°]$ | $0-60$ | $62-80$ | $0-55$ | $0-60$ |
| Threshold energy $E_{\text{eff}}$ [eV] | $3 \times 10^{18}$ | $4 \times 10^{18}$ | $3 \times 10^{17}$ | $10^{18}$ |
| No. of events ($E > E_{\text{eff}}$) | 82318 | 11074 | 29585 | 11155 |
| No. of events (golden hybrids) | 1475 | 175 | 414 | - |
| Energy calibration (A) [EeV] | $0.190 \pm 0.005$ | $5.61 \pm 0.1$ | $(1.21 \pm 0.07) \cdot 10^{-2}$ | - |
| Energy calibration (B) | $1.025 \pm 0.007$ | $0.985 \pm 0.02$ | $1.03 \pm 0.02$ | - |

**Table 1**: Summary of the experimental parameters describing data of the different measurements at the Pierre Auger Observatory. Numbers of events are given above the energies corresponding to full trigger efficiency. Missing parameters will be added in the final version of the paper.

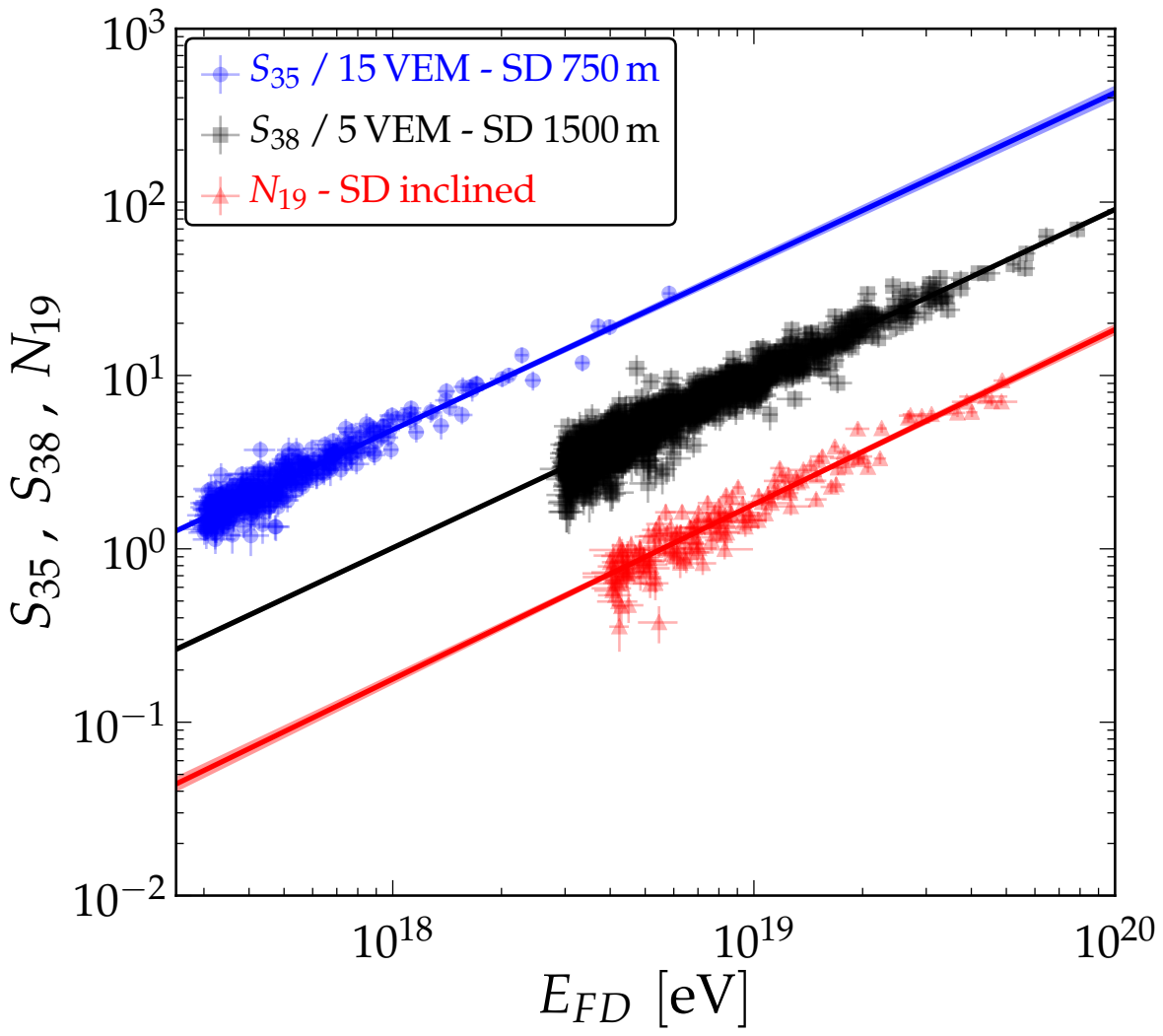

**Figure 2**: The correlation between the different energy estimators $S_{38}$, $S_{35}$ and $N_{19}$ (see text) and the energy determined by FD.

Inclined air-showers are characterized by the dominance of secondary muons at ground, as the electromagnetic component is largely absorbed in the large atmospheric depth traversed by the shower [20]. The reconstruction is based on the estimation of the relative muon content $N_{19}$ with respect to a simulated proton shower with energy $10^{19}$ eV [21]. $N_{19}$ is used to infer the primary energy for inclined events. Due to the limited exposure of the 750 m array only inclined events from the 1500 m array are included in the present analysis.

Events, both vertical and inclined, are selected if the detector with the highest signal is enclosed in a hexagon of six active stations. The exposure is obtained by integrating the effective area (i.e. the sum of the areas of all active hexagons) over observation time [8]. Exposures of the SD array for the different datasets are shown in Fig. 1. Values up to 31 Dec 2012 are given in Table 1 together with their uncertainties and the relevant zenith angle ranges. In case of vertical events measured with the 1500 m array the integrated exposure amounts to an increase of 50% with respect to the previous publication [1, 22]. The number of events above $3 \times 10^{18}$ eV does not fully reflect this increase due to changes in the energy scale and calibration [23].

Events that have independently triggered the SD array and FD telescopes (called golden hybrid events) are used for the energy calibration of SD data. Only a sub-sample of events that pass strict quality and field of view selections are used [9, 18]. The relations between the different energy estimators $\hat{E}$, i.e. $S_{38}$, $S_{35}$, $N_{19}$, and the energies reconstructed from the FD measurements $E_{\text{FD}}$ are well described by power-laws $E_{\text{FD}} = A \cdot \hat{E}^B$. The calibration parameters are given in Table 1 together with the number of golden hybrid events. The correlation between the different estimators and $E_{\text{FD}}$ is shown in Fig. 2 superimposed with the calibration functions resulting from maximum-likelihood fits. For the vertical events of the 1500 m array, the SD energy resolution due to limited sampling statistics decreases from 15% below $6 \times 10^{18}$ eV to less than 12% above $10^{19}$ eV [24]. Physical fluctuations in shower development are the major contribution at highest energies with $\approx 12\%$. In case of inclined events, physical fluctuations are larger, $\approx 16\%$ [21].

To check the energy reconstruction and intrinsic resolutions, the reconstruction was also performed using simulated events. For vertical events of the 1500 m array, the distribution of the ratio of the inferred SD energy $E_{\text{SD}}$ and the reconstructed FD energy $E_{\text{FD}}$ is compared to Monte-Carlo simulations in Fig. 3. Due to the lack of muons in simulations compared to data (e.g. [25]), the SD energy scale of simulations was rescaled by 24% (averaging primaries and energies) to match that of data. Based on this rescaling, the observed distributions are well reproduced by Monte-Carlo simulations.

Due to the steepness of the energy spectrum and the finite resolution of the SD measurements, the measured spectra represent a smearing of the true spectrum due to bin-to-bin migrations. Corrections have been applied to obtain the true energy spectrum [1]. These are below 15% in the energy range of interest.

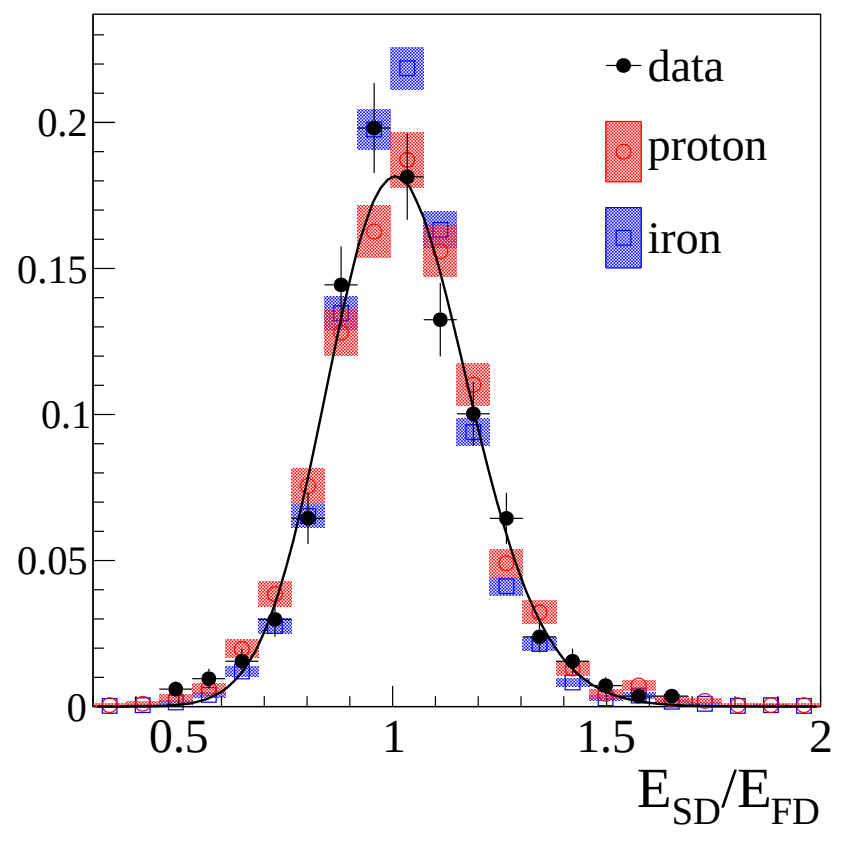

**Figure 3**: Distribution of the ratio between the reconstructed SD and FD energy, $E_{\mathrm{SD}}$ and $E_{\mathrm{FD}}$. Ratios are obtained from data and QGSJet-II.03 simulations [26] (see text).

The energy spectra obtained from the three SD datasets are shown in Fig. 4. Due to the calibration with events observed by the FD, the SD energies share the uncertainty of the FD energy scale of 14%, which will be further explained in the next section.

## 3 Flux measurements with the hybrid detector

The hybrid approach is based on the detection of showers observed by the FD in coincidence with at least one station of the SD array. Although a signal in a single station does not allow an independent trigger and reconstruction in SD, it is a sufficient condition for a very accurate determination of the shower geometry using the hybrid reconstruction.

To ensure good energy reconstruction, only events that satisfy strict quality criteria are accepted [13]. In particular, to avoid a possible bias in event selection due to the differences between shower profiles initiated by primaries of different mass, a shower is retained only if its geometry would allow a reliable measurement of any shower profile that occurs in the full data set. A detailed simulation of the detector response has shown that for zenith angles less than 60°, every FD event above $10^{18}$ eV passing all the selection criteria is triggered by at least one SD station, independent of the mass or direction of the incoming primary particle [13].

The measurement of the flux of cosmic rays using hybrid events relies on the precise determination of the detector exposure that is influenced by several factors. The response of the hybrid detector strongly depends on energy and distance from the relevant fluorescence telescope, as well as atmospheric and data taking conditions. To properly take into account all of these configurations and their time variability, the exposure has been calculated using a sample of simulated events that reproduce the exact conditions of the experiment [13]. The total systematic uncertainty on the calculation of the exposure ranges from 14% at $10^{18}$ eV to below 6% above $10^{19}$ eV [13]. The current hybrid exposure as a function of energy is shown in Fig. 1 compared with the exposures of the surface detectors.

The energy spectrum reconstructed from hybrid events will be presented at the conference and in the updated version of this paper. Data taken in the time period given in Table 1 are included. The main systematic uncertainty is due to the energy assignment which relies on the knowledge of the fluorescence yield (3.6%), atmospheric conditions (3%-6%), absolute detector calibration (9%) and shower reconstruction (6%) [23]. The invisible energy is calculated with a new, simulation-driven but model-independent method with an uncertainty of 1.5%-3% [27].

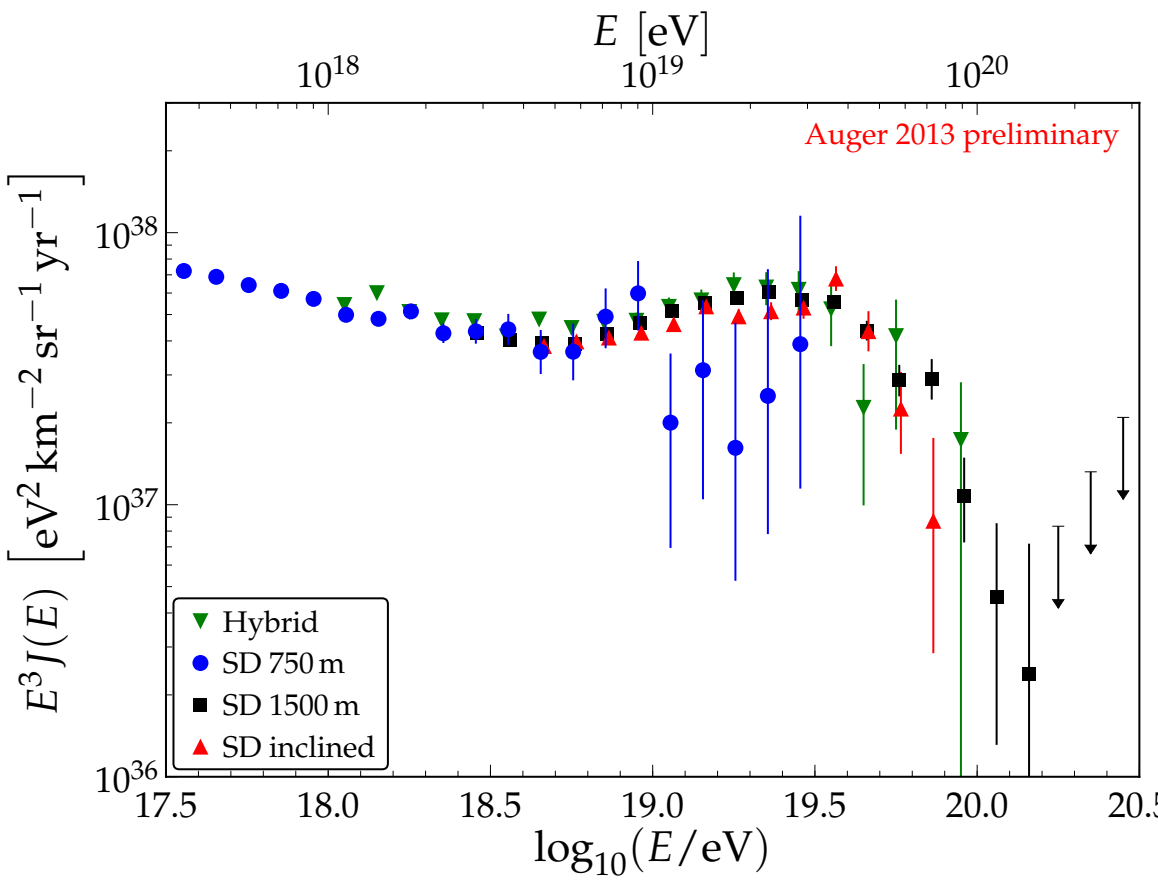

**Figure 4**: Energy spectra, corrected for energy resolution, derived from SD and from hybrid data.

## 4 Combined energy spectrum

The hybrid spectrum extends the SD 1500 m spectrum below the energy of full trigger efficiency of $3\times10^{18}$ eV and overlaps with the spectrum of the 750 m array above $10^{18}$ eV. The latter is fitted up to $3\times10^{18}$ eV and extends the measurement of the energy spectrum below $10^{18}$ eV. The spectrum of inclined events contributes above its full efficiency threshold of $4\times10^{18}$ eV and provides an independent measurement in this energy range. We combine these measurements into a single energy spectrum.

The SD measurements are affected by uncertainties due to the energy calibrations (see Table 1). These uncertainties are taken into account by minimizing the energy calibration likelihoods together with the smearing corrections due to bin-to-bin migrations. In this combined maximum-likelihood fit, the normalizations of the different spectra are allowed to vary within the exposure uncertainties as stated in Table 1.

The combined energy spectrum is shown in Fig. 5 together with the number of observed events within each bin. To characterize the spectral features we describe the data with a power law below the ankle $J(E) \propto E^{-\gamma_1}$ and a power law with smooth suppression above:

$$J(E; E > E_a) \propto E^{-\gamma_2} \left[1 + \exp\left(\frac{\log_{10} E - \log_{10} E_{1/2}}{\log_{10} W_c}\right)\right]^{-1}.$$

$\gamma_1$, $\gamma_2$ are the spectral indices below/above the ankle at $E_a$. $E_{1/2}$ is the energy at which the flux has dropped to half of its peak value before the suppression, the steepness of which is described with $\log_{10} W_c$.

The resulting spectral parameters are given in Table 2. To match the energy spectra, the SD 750 m spectrum has to be scaled up by 2%, the inclined spectrum up by 5% and the hy-brid spectrum down by 6%. Compared to the previous publication, the precision in determining the spectral index below the ankle has increased significantly, mainly due to the addition of the 750 m array. We report a slightly flatter spectrum below the ankle (now: $3.23 \pm 0.01$ (stat) $\pm 0.07$ (sys), previous publication: $3.27 \pm 0.02$) and an increase of $E_a$ (now: $18.72 \pm 0.01$ (stat) $\pm 0.02$ (sys), previous publication: $18.61 \pm 0.01$) [22]. The large systematic uncertainties in $\gamma_1$ are dominated by the uncertainty of the resolution model used for correcting the measured flux. At the same time, the uncertainty in the energy scale of 14% is propagated into the final result.

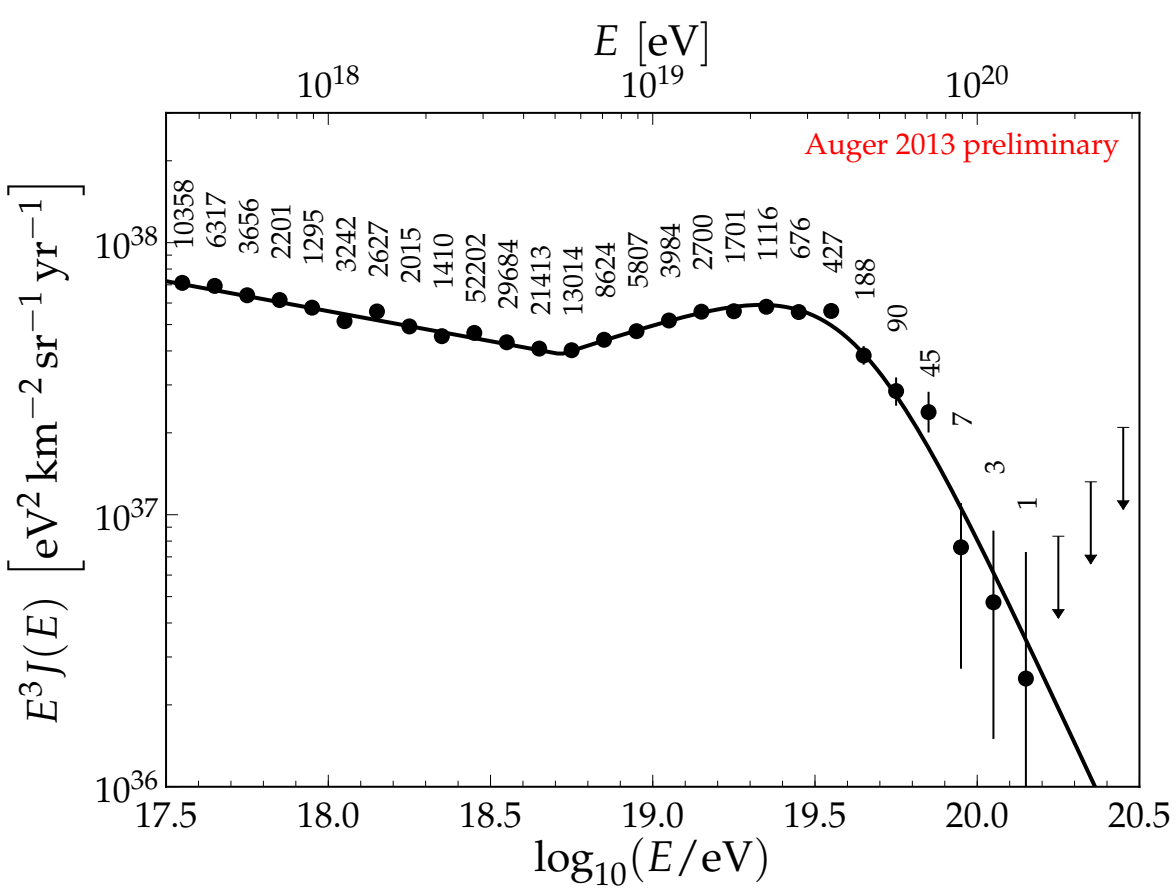

**Figure 5**: The combined energy spectrum of UHECRs as measured at the Pierre Auger Observatory. The numbers give the total number of events inside each bin. The last three arrows represent upper limits at 84% C.L.

| Parameter | Result ($\pm\sigma_{\text{stat}} \pm \sigma_{\text{sys}}$) |
|---|---|
| $\log_{10}(E_a/\text{eV})$ | $18.72 \pm 0.01 \pm 0.02$ |
| $\gamma_1$ | $3.23 \pm 0.01 \pm 0.07$ |
| $\gamma_2$ | $2.63 \pm 0.02 \pm 0.04$ |
| $\log_{10}(E_{1/2}/\text{eV})$ | $19.63 \pm 0.01 \pm 0.01$ |
| $\log_{10} W_c$ | $0.15 \pm 0.01 \pm 0.02$ |

**Table 2**: Parameters, with statistical and systematic uncertainties, of the model describing the combined energy spectrum measured at the Pierre Auger Observatory.

The combined energy spectrum is compared to fluxes from three astrophysical scenarios in Fig. 6. Shown are models assuming pure proton or iron composition. The fluxes result from different assumptions of the spectral index $\beta$ of the source injection spectrum and the source evolution parameter $m$. The model lines have been calculated using CRPropa [30] and validated with SimProp [31].

## 5 Summary

The flux of cosmic rays above $3 \times 10^{17}$ eV has been measured at the Pierre Auger Observatory combining data from surface and fluorescence detectors. The spectral features are determined with unprecedented statistical precision. The fitted parameters are compatible with previous results given the change in the energy scale. There is an overall uncertainty of the revised energy scale of 14% [23]. Current results from $X_{\max}$ measurements and an interpretation of the measurements concerning mass composition are presented in [28, 29]. The spectrum as measured with the SD 750 m array is presented in more detail at this conference in [9].

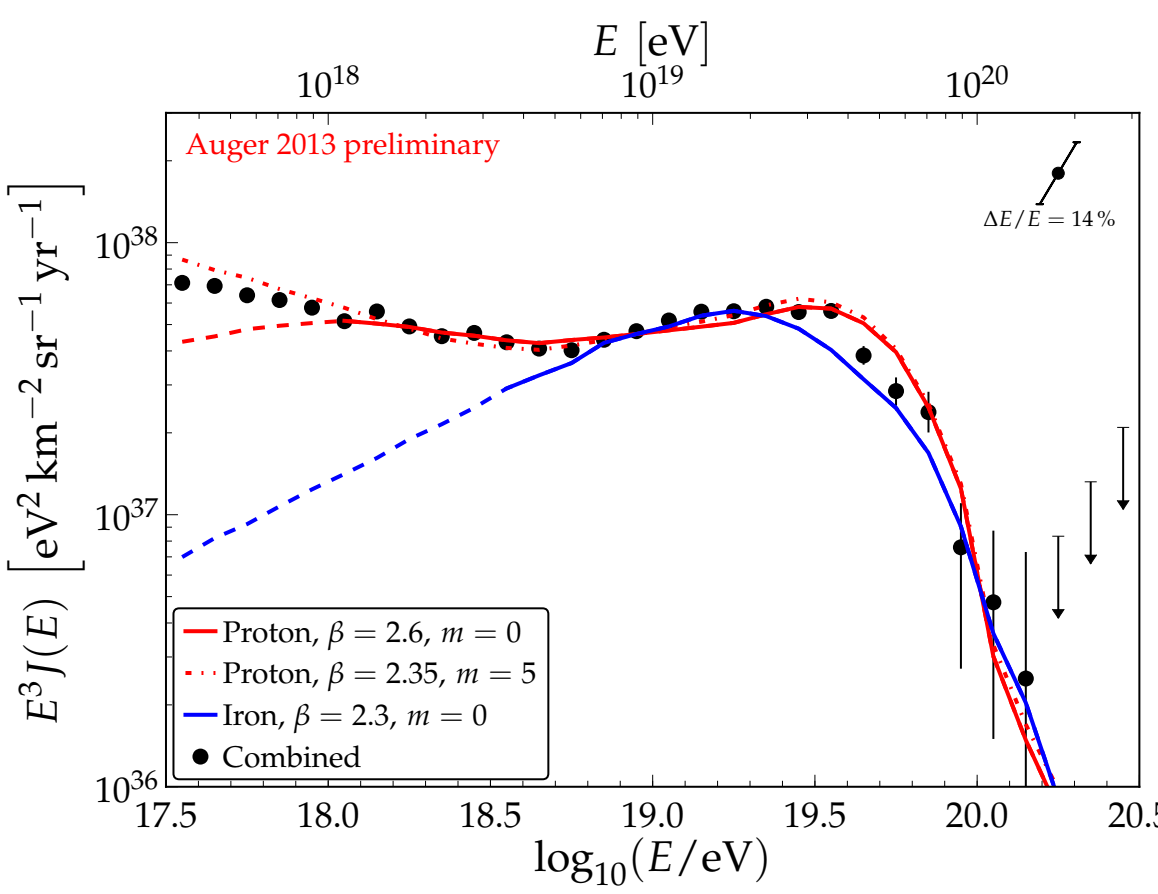

**Figure 6**: The combined energy spectrum compared to energy spectra from different astrophysical scenarios (see text).

# Measurement of the energy spectrum of cosmic rays above $3\times10^{17}$ eV using the AMIGA 750 m surface detector array of the Pierre Auger Observatory

DIEGO RAVIGNANI[1] FOR THE PIERRE AUGER COLLABORATION[2]

[1]*ITeDA (CNEA, CONICET, UNSAM), Buenos Aires, Argentina*
[2]*Full author list: http://www.auger.org/archive/authors_2013_05.html*

*auger_spokespersons@fnal.gov*

**Abstract:** We present a measurement of the cosmic ray energy spectrum above $3\times10^{17}$ eV based on data obtained with the 750 m surface detector array of the Pierre Auger Observatory. We address the steps required to measure the energy spectrum, from the reconstruction of events, through the precise determination of the exposure of the array, up to the determination of the cosmic ray energy. The derived energy spectrum is discussed, and it is compared to those measured by other instruments in the overlapping energy regions.

## 1 Introduction

The Pierre Auger Observatory [1], in Argentina, combines a 3000 km$^2$ surface detector array (SD) of more than 1600 water-Cherenkov detectors, spaced at 1500 m from each other, with 27 fluorescence telescopes to measure extensive air showers initiated by ultra-high energy cosmic rays. This *hybrid* observatory has already been used to measure the energy spectrum of cosmic rays with energies above $10^{18}$ eV, that is, from just below the *ankle* of the spectrum (at about $4\times10^{18}$ eV) up to the highest energies [2].

The energy region below the ankle is also of extreme interest. The study of the evolution of the energy spectrum from below the anticipated second knee up to the ankle, together with that of the primary mass composition and of the large-scale distribution of arrival directions are crucial to explaining the possibility of a transition from a galactic cosmic ray origin to an extragalactic one. This is expected to take place either at the ankle [3] or at the so called second knee [4]. With the aim of extending the energy of observations down to $10^{17}$ eV, the Auger collaboration has started to deploy new instruments. The extensions include a surface array of 750 m spacing with muon detection capabilities (the AMIGA project) [5, 6], and three fluorescence telescopes (the HEAT project) [7]. The 750 m surface detector array and the HEAT telescopes are now fully operational.

We present in this paper the measurement of the energy spectrum with the 750 m array. In section 2 we describe the 750 m surface array. In section 3 the reconstruction of the observed events is explained. We continue in section 4 with the description of the energy calibration of the detector followed by the presentation of the obtained spectrum in section 5. Finally, we outline the conclusions of this work in section 6.

## 2 The 750 m surface detector array

The 750 m SD array includes 71 water-Cherenkov detectors and covers an area of 27 km$^2$. Thirty five of them started taking data in August 2008. Additional detectors were added afterwards until the final setup was completed in September 2012. A layout of the array is shown in figure 1. The 750 m array is nested within the 1500 m array and is overlooked by three telescopes located at the Coihueco fluorescence detector (FD) site and by another three at the HEAT site.

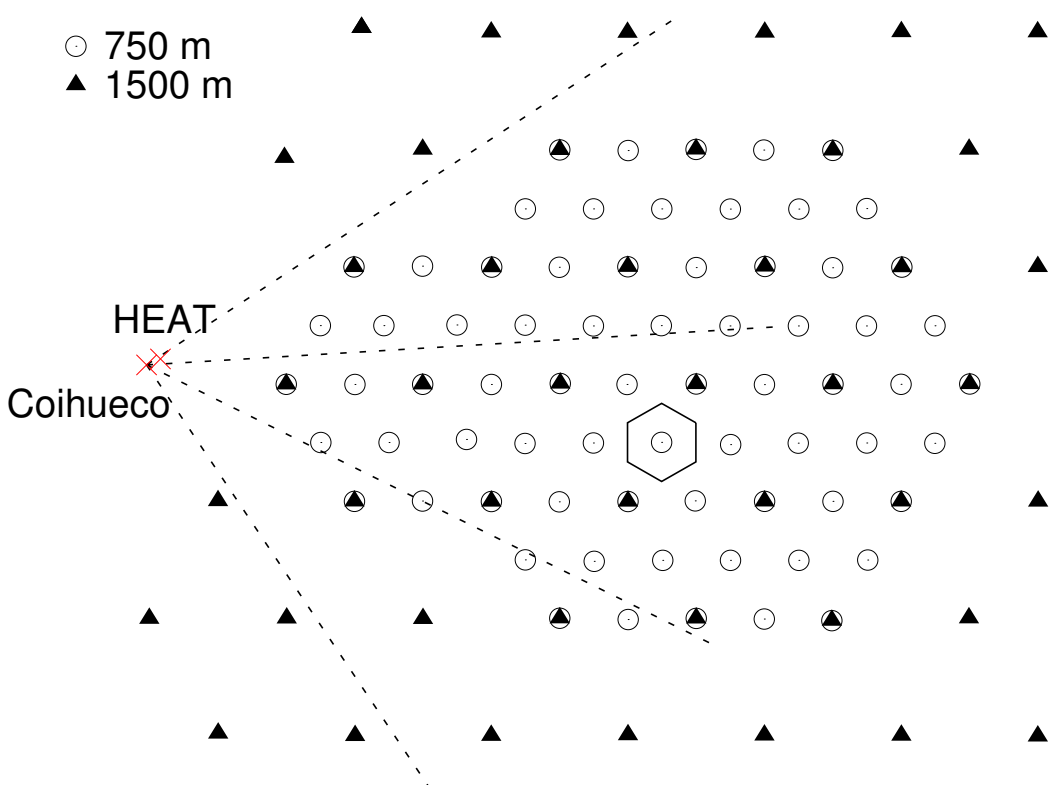

**Figure 1**: Layout of the 750 m nested surface detector array. Detectors of the 1500 m array are shown as filled triangles. Note that some detectors belong to both arrays. An unitary hexagon of the 750 m array is displayed. The lines show the azimuthal acceptance of the three telescopes at Coihueco overlooking the 750 m array.

For the 750 m array we adopt the trigger system, the method and the algorithms to select events and detectors, the calculation of the exposure, the algorithms to reconstruct events and the energy calibration procedure from the 1500 m array. The efficiency of the 750 m array is estimated by means of air-shower simulations performed with CORSIKA [8], using QGSJet-II [9, 10] and FLUKA [11] as hadronic interaction generators. The response of the 750 m array is simulated with the same tools used for the 1500 m array [12]. Simulation results are further validated with data. In figure 2 the simulated efficiencies for selecting events in the most sensitive trigger channel, referred to in the figure as 3ToT [13], are shown for the 750 m and the 1500 m arrays. The smaller spacing between stations of the 750 m array leads to an increase of the trigger efficiency at low energy. Its trigger efficiency is 100% at energies above $3\times10^{17}$ eV for air-showers with a zenith angle smaller than 55°, an energy ten times lower than for the 1500 m array [13].

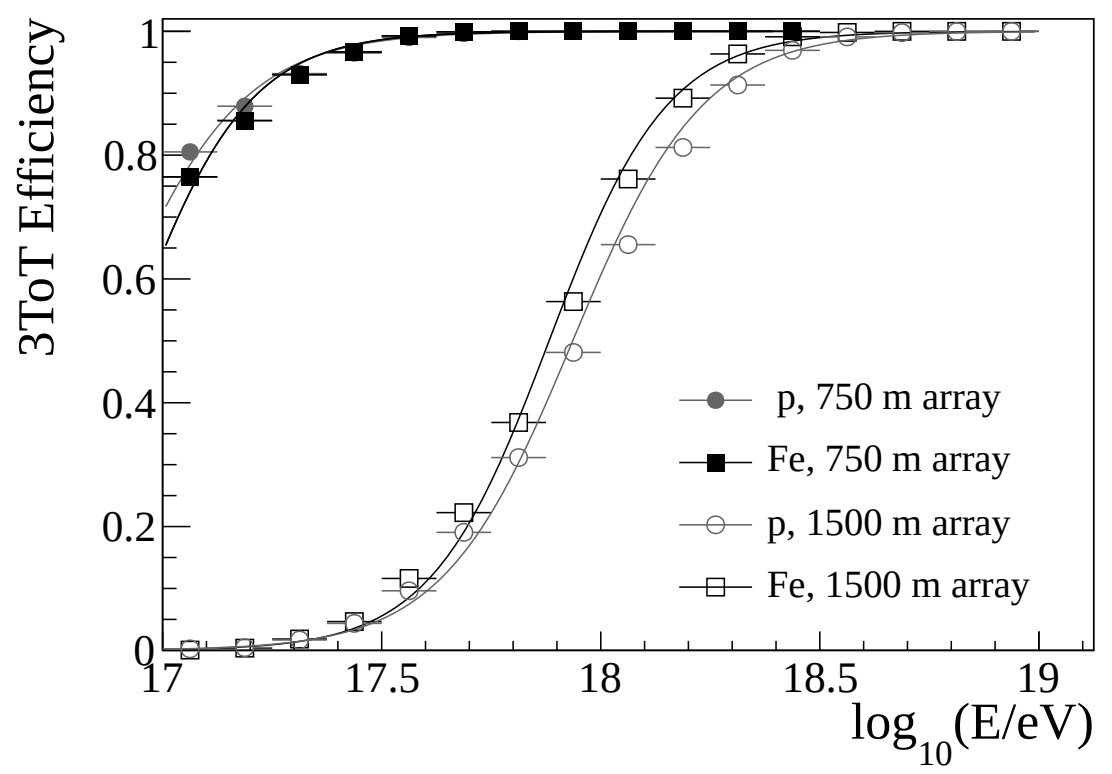

**Figure 2**: 3ToT trigger efficiency of the 750 m array (full symbols) and of the 1500 m array (open symbols) obtained from simulations of iron and proton primaries. Events with zenith angle less than 55° in the case of the 750 m array and less than 60° for the 1500 m array are considered.

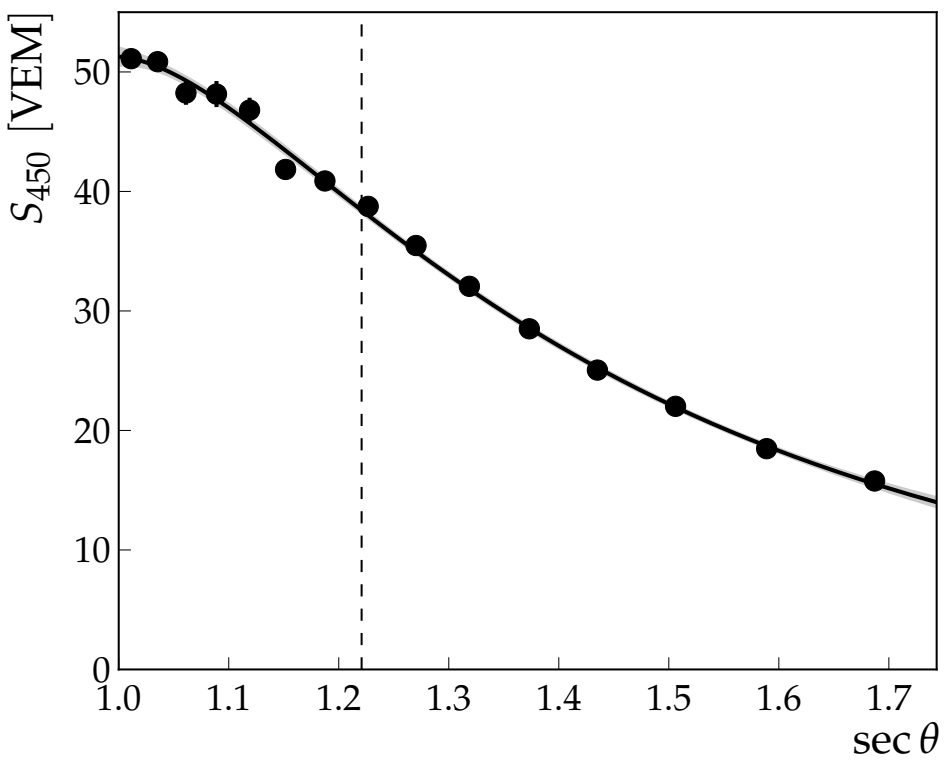

**Figure 3**: $S(450)$ value of the constant intensity cut (see text) in each zenith angle bin. The data are fit with a third degree polynomial in $x = \cos^2\theta - \cos^2 35°$. The vertical dotted line corresponds to the reference zenith angle of 35° and the shadowed region to the fit uncertainty.

Only events that can be reconstructed with high quality are used in the spectrum. Thus it is required for each selected event that the detector with the highest signal is surrounded by 6 working detectors, i.e. it is within an active *hexagon*. This condition also defines an effective detection area for the array given by the set of unitary hexagons centred on the detectors internal to the array (see figure 1). The exposure for energies above full trigger efficiency is obtained by integrating this fiducial area over the observation time [13]. The exposure between August 2008 and December 2012, the period used in this work, is $(79 \pm 4)\,\mathrm{km}^2$ sr yr. The spectrum is almost free of systematic uncertainties coming from the exposure given the simple way it is computed above the full trigger efficiency energy. More details about the calculation of the exposure, the trigger hierarchy and the selection of events of the SD can be found in [13].

## 3 Event reconstruction

The reconstruction of the events of the 750 m array is based on the well-tuned algorithms of the 1500 m array. The direction of a primary particle is obtained from the time of arrival of the air-shower particles at the ground. The angular resolution is better than 1.3° for events that trigger between 3 and 6 stations ($\langle E\rangle \gtrapprox 1.6\times10^{17}$ eV) and better than 1° for those with more than 6 stations ($\langle E\rangle \gtrapprox 4\times10^{17}$ eV). The signals recorded by the surface detectors are converted into units of vertical-equivalent muons (VEM). One VEM is defined as the average of the signals produced in the 3 PMTs of a water-Cherenkov detector by a vertical muon that passes through the detector's centre [14]. The signals in VEM units are fit with a lateral distribution function [15] as in the case of the 1500 m array. The energy reconstruction of the events is based on the estimation of the lateral distribution of secondary air-shower particles reaching ground at an optimal distance. The optimal distance is that at which, for a wide range of reasonable lateral distribution functions, the spread in the signal size predicted at that distance is a minimum [16, 17]. For the 750 m array the optimal distance, determined empirically, is 450 m. The statistical uncertainty in $S(450)$, given the sparse sampling of the shower front and the finite resolution of the surface detectors, decreases from 20% at $\langle E\rangle \approx 1.3\times10^{17}$ eV to 5% at $\langle E\rangle \approx 1.4\times10^{18}$ eV. An additional uncertainty due to fluctuations in shower development is estimated to be of the order of 10% [18].

$S(450)$ is corrected for the zenith angle dependency caused by the attenuation of showers in the atmosphere by means of the constant intensity cut method [19]. The method is based on the hypothesis that showers with energies above a given energy threshold arrive with the same rate from all directions. Following this procedure the data are divided in intervals of $\cos^2\theta$ with $\theta$ being the zenith angle. Events in each bin are ordered by decreasing $S(450)$ values. We choose to study the attenuation of the signal for an intensity of 500 events (corresponding to an energy of $\simeq 5\times10^{17}$ eV). The corresponding values of $S(450)$ versus zenith angle are fit with a third degree polynomial in $x = \cos^2\theta - \cos^2 35°$ as shown in figure 3. In this way an equivalent $S(450)$ at a reference zenith angle of 35° ($S_{35}$), approximately the median angle of an isotropic distribution of cosmic rays between 0° and 55°, is obtained for each event from:

$$S_{35} = \frac{S(450)}{1+ax+bx^2+cx^3}$$

with $a = 1.69\pm0.05$, $b = -1.3\pm0.1$ and $c = -2.3\pm0.7$.

## 4 Energy calibration

The energy of each event is derived from a subset of showers observed by the 750 m array in coincidence with the fluorescence telescopes at Coihueco and HEAT. Only events with energies above the full trigger efficiency threshold are considered. Additional criteria in both the SD and the FD reconstructions are adopted to select only the highest quality events. In the case of the SD the fiducial cuts presented in section 2 are used. For the FD it is required that the SD station used in the hybrid geometry fit is closer than 750 metres to the shower axis; that there exists a contemporary measurement of aerosols [20], and that the vertical aerosol optical depth is less than 0.1; that lidar cloud measurements show less than 25% cloud cover; that the Gaisser-Hillas function fit to the shower profile has a normal-transformed chi-squared ($(\chi^2 - \mathrm{ndf})/\sqrt{2\,\mathrm{ndf}}$) within 2.5 of the mean; that the maximum size of a gap in the measured shower profile is less than 20 g cm$^{-2}$ ; that shower maximum $X_{\rm max}$ is viewed, and measured with an uncertainty of less than 40 g cm$^{-2}$; and that the relative uncertainty in energy is less than 0.18. In addition a fiducial volume cut is applied to ensure that the energy calibration is unbiased with respect to the mass of the cosmic rays [21]. This cut selects shower geometries which would allow an unbiased measurement of any $X_{\rm max}$ across the range of values present at a particular energy, without a significant Cherenkov light component of the signal.

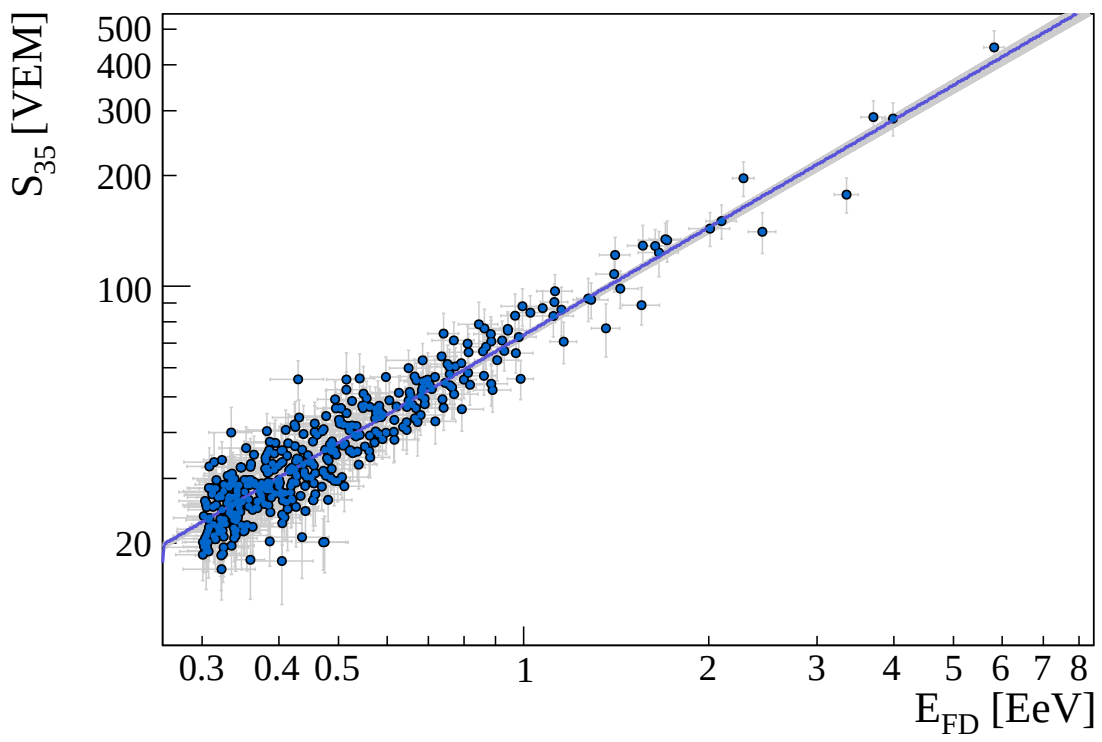

**Figure 4**: Correlation between $S_{35}$ and $E_{\rm FD}$ for the 414 events used in the calibration of the 750 m array. These events are observed by the 750 m array in coincidence with telescopes of the fluorescence detector located at the Coihueco and HEAT sites.

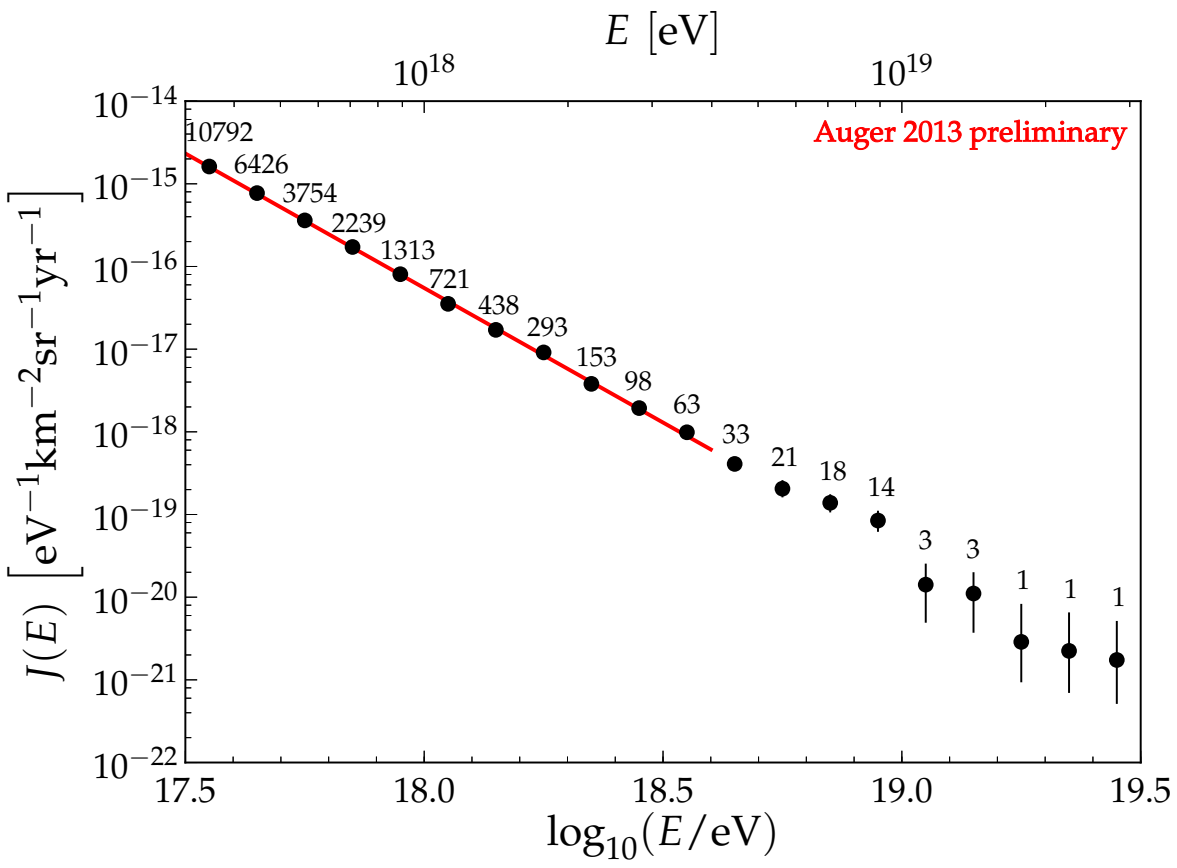

**Figure 5**: Energy spectrum of cosmic rays derived from 26385 events observed by the 750 m array above $10^{17.5}$ eV.

For each event selected for the calibration, $S_{35}$ is correlated with the energy measured by the fluorescence detector ($E_{\rm FD}$) to calibrate the energy of the 750 m array as shown in figure 4. The FD provides a calorimetric measurement of the cosmic ray energy obtained from the integration of the longitudinal profile of air-showers [22]. The energy that is not observed by the FD and which is carried away by neutrinos and muons produced in air-showers, referred to as invisible energy, is added to the calorimetric energy to obtain $E_{\rm FD}$. This invisible energy is estimated for each individual event in the calibration with a new method based on the $S(450)$ measured by the SD [23]. Its average value decreases from 18% at $3\times10^{17}$ eV to 13% at $10^{19}$ eV. The calibration events in figure 4 are fit with the function,

$$E_{SD} = A\,S_{35}^{B}$$

with $A = (12.1\pm0.7)\times10^{15}$ eV and $B = 1.03\pm0.02$. A likelihood function based on a modelled distribution of $S_{35}$ and $E_{\rm FD}$ is maximised to obtain the fit parameters [24]. The resulting conversion from $S_{35}$ to energy is almost linear. The statistical uncertainty in the energy is propagated from that of $S(450)$. Its value is dependent on the energy and improves from $\sim$ 18% at $3\times10^{17}$ eV to $\sim$ 10% above $5\times10^{18}$ eV. The systematic uncertainty in the energy due to the calibration is 1% at $3\times10^{17}$ eV and increases to 3% at $4\times10^{18}$ eV. Other contributions to the systematic uncertainty come from the FD. The 14% uncertainty includes contributions from the absolute fluorescence yield, the FD calibration, the propagation of fluorescence and Cherenkov light in the atmosphere, the FD reconstruction and the invisible energy correction [22].

## 5 Energy spectrum

The energy spectrum of cosmic rays is derived from events observed between August 2008 and December 2012. The exposure presented in section 2 is used to obtain the flux. Given the statistical uncertainty in the energy assigned by the 750 m array, bin-to-bin migrations influence the derived flux and thus the value of the spectral index. To correct for these effects, a forward-folding approach is applied. Monte Carlo simulations of proton and iron showers are used to obtain the energy resolution of the detector in order to derive a bin-to-bin migration matrix. These matrices are then used to find a flux parametrisation that matches the measured data after forward-folding. The ratio of this parametrisation to the folded flux gives a correction factor that is applied to data. The correction to the flux is energy dependent and is less than 10% over the energy range of interest. The energy spectrum corrected for the energy resolution of the 750 m array is shown in figure 5. The exposure, the weather effects and the forward folding procedure contribute to a total systematic uncertainty in the derived flux of 7%.

A power law function is fitted between $3\times10^{17}$ and $4\times10^{18}$ eV, the energy of the ankle according to the 1500 m array [2], using a binned likelihood method. The adjusted spectral index is $3.27\pm0.02(\mathrm{stat})\pm0.07(\mathrm{syst})$. The quoted systematic uncertainty comes from the energy calibration. Spectra built with events falling in the zenith angle bins $[0°,28°]$, $[28°,42°]$ and $[42°,55°]$, all of them having the same acceptance, are compared. The corresponding fluxes corrected for energy resolution are shown in figure 6. The consistency of these spectra highlights the robustness of the reconstruction of events with zenith angles up to 55°.

The spectrum measured by the 750 m array is shown together with the Auger spectra measured by the 1500 m surface array and the fluorescence detector [25] and those from KASCADE-Grande [26], the Telescope Array [27] and Tunka [28] in figure 7. The spectrum of the 750 m array is consistent with the two other Auger spectra and with the result of KASCADE-Grande. The Telescope Array and Tunka, however, measure a higher flux than the 750 m array.

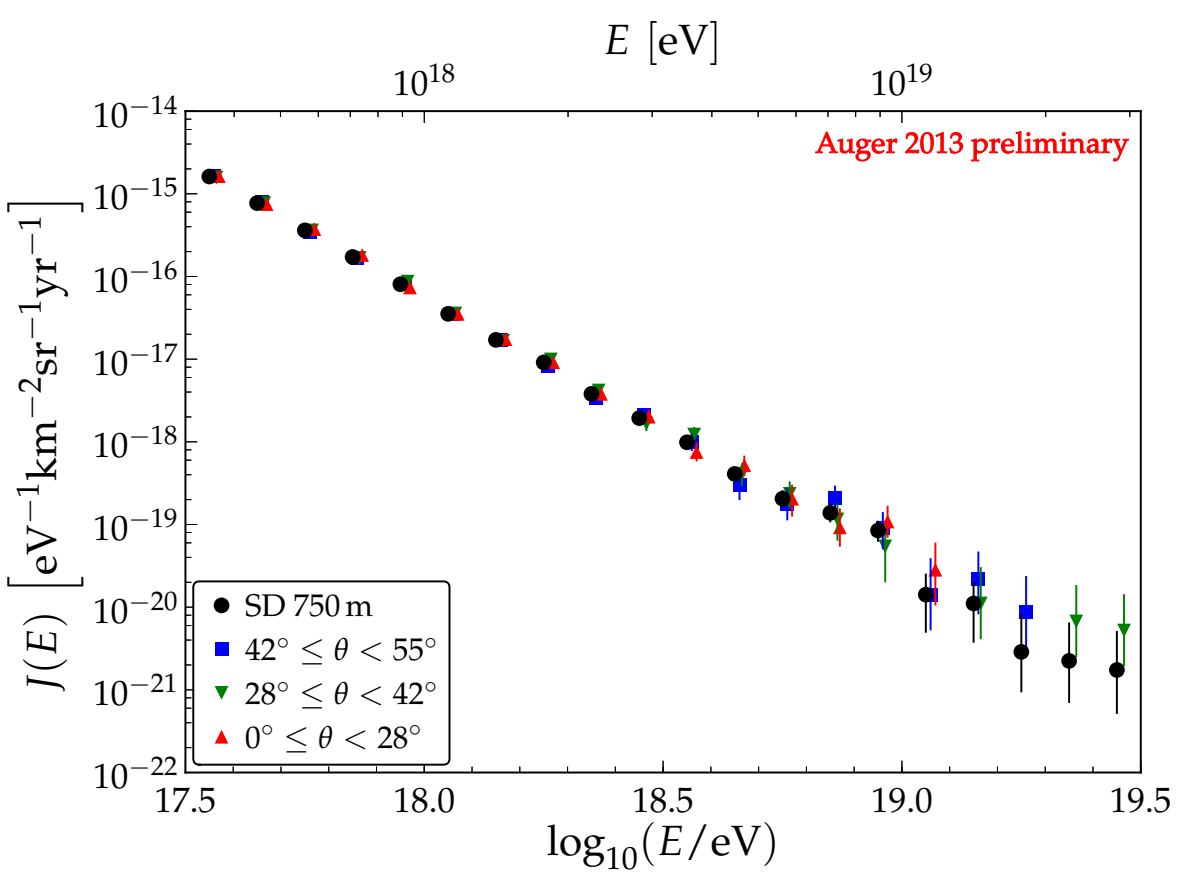

**Figure 6**: Energy spectra corresponding to the zenith angle intervals $[0°, 28°]$, $[28°, 42°]$ and $[42°, 55°]$ containing 8827, 8714 and 8844 events respectively. The overall spectrum up to $55°$ is also shown.

## 6 Summary

We have measured the cosmic ray flux above $3 \times 10^{17}$ eV with the Pierre Auger Observatory using data recorded by the 750 m surface detector array. The energy assigned to events of the 750 m array is unbiased with respect to the zenith angle emphasising the strength of the applied constant intensity cut method. The spectrum follows a power law with a spectral index of $3.27 \pm 0.02(\text{stat}) \pm 0.07(\text{syst})$ up to $4 \times 10^{18}$ eV, the energy of the ankle. This spectral index is in agreement with our previously reported value obtained from observations with the fluorescence detector [29]. As the exposure increases in the future the spectral fit of the ankle feature utilising data of the 750 m array will be in range. The flux observed by the 750 m array is compared with spectra measured by the other Auger detectors, KASCADE-Grande, the Telescope Array and Tunka. The obtained spectrum is combined with those measured with the 1500 m array and the fluorescence detector of the Pierre Auger Observatory in another contribution to this conference [25].

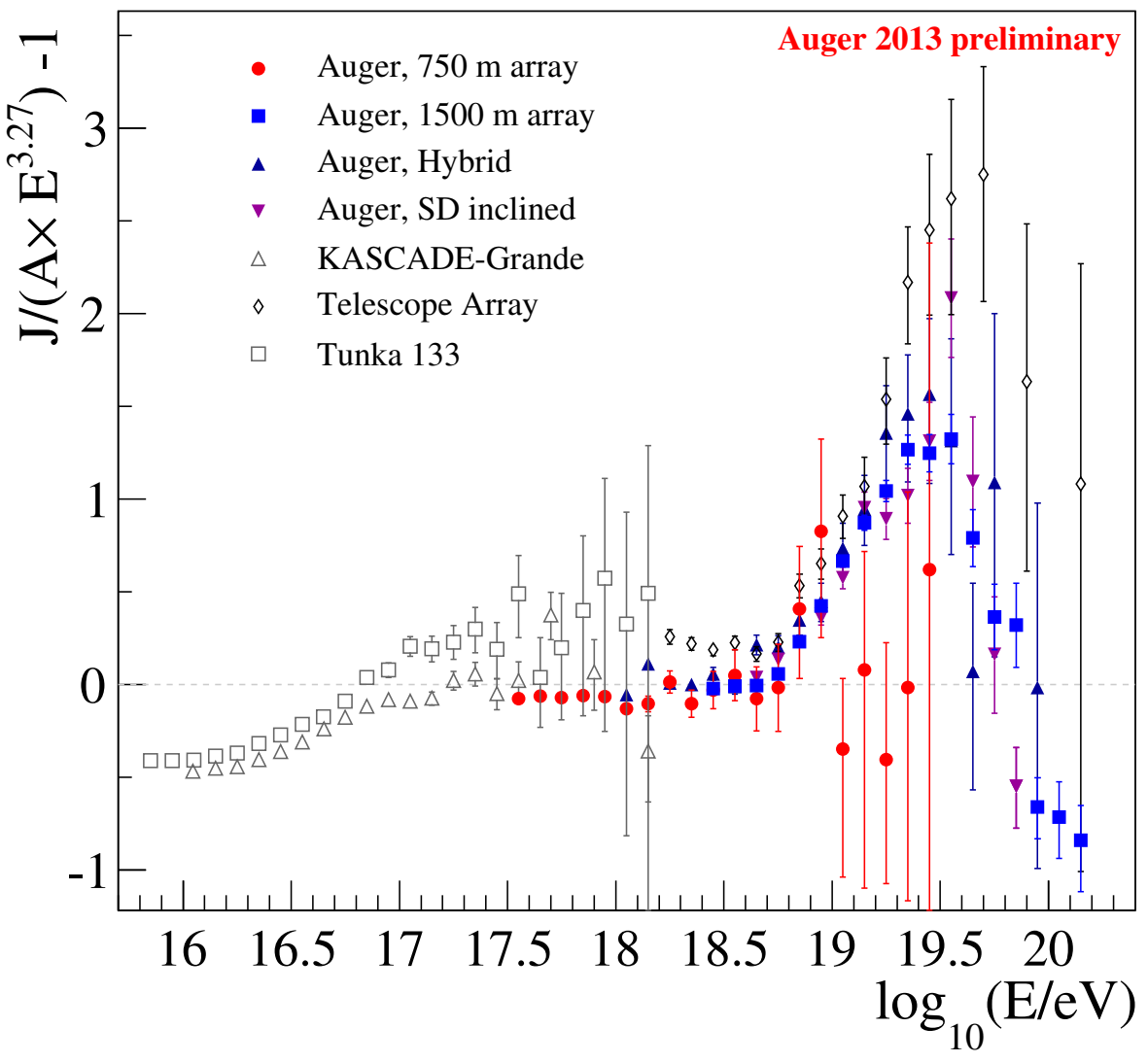

**Figure 7**: Energy spectra of the 750 m and 1500 m surface arrays, the Auger hybrid and inclined spectra [25] and those from KASCADE-Grande [26], the Telescope Array [27] and Tunka [28].

# An update on the measurements of the depth of shower maximum made at the Pierre Auger Observatory

VITOR DE SOUZA[1] FOR THE PIERRE AUGER COLLABORATION[2].

[1] *Instituto de Física de São Carlos, Universidade de São Paulo.*
[2] *Full author list: http://www.auger.org/archive/authors_2013_05.html)*

*auger_spokespersons@fnal.gov*

**Abstract:** Data continue to be collected at The Pierre Auger Observatory. The Fluorescence telescopes of the Pierre Auger Observatory can be used to measure the depth of shower maximum ($X_{\rm max}$) with high precision (20 g/cm$^2$ on average). Thus the measurement of the $X_{\rm max}$ distribution is the best way to infer properties of the primary cosmic ray, such as its composition, and even the interaction cross-section for the proton component. During the Conference we will present our latest measurements of average $X_{\rm max}$ and fluctuations of $X_{\rm max}$ as a function of energy. In this paper we give a general outline of the $X_{\rm max}$ analysis and the improvements made with respect to previously published results.

## 1 Introduction

Data have been collected continuously at the Pierre Auger Observatory since 1 January 2004. In the poster to be displayed at this Conference we will show the results of measurements of the depth of shower maximum from events recorded up to 31 December 2012. The data are described in terms of the first and second moments of the distributions as a function of energy as is common practice in this work. A major aim is to obtain information on the mass composition as a function of energy but interpretation is strongly limited by the uncertainties that remain over features of the hadronic interactions at centre-of-mass energies well-beyond what can be reached at any man-made accelerators. Important parameters are the cross-section for hadronic interactions and the multiplicity and the inelasticity associated with them.

The Pierre Auger Collaboration has faced the composition challenge in several ways: a) $X_{\rm max}$ measurement [1], b) muon production depth [2] and c) rise-time asymmetry measurements [3]. Despite recent advances in composition estimates at ultra-high energies with surface detectors, the traditional $X_{\rm max}$ measurement using fluorescence telescopes remains the one with smallest systematic uncertainties [4] and therefore has been used as the most reliable source of information in these studies. Besides that, $X_{\rm max}$ is directly related to the properties of the interaction of the primary particle with air. This feature increases the value of the $X_{\rm max}$ measurement as it is possible to estimate the proton-air cross-section by studying the tail of the $X_{\rm max}$ distribution.

In this paper we discuss the recent improvements in the event reconstruction and in the procedure followed for creating an unbiased sample of $X_{\rm max}$ distributions. The updated results will be presented at the conference.

## 2 Data analysis

All of the events used for this analysis were obtained from the hybrid mode: that is, events measured by the fluorescence and by surface detectors. The fundamental steps in the analysis procedure are unchanged from those used for our previous publications [1, 5] with information from the surface detector array and the fluorescence detector being used to reconstruct the shower geometry and the longitudinal development [6]. From the latter we obtain both the energy of the shower and the depth of maximum of the shower [7, 8].

We have paid particular attention to: i) choosing selection cuts which guarantee small reconstruction uncertainties, ii) obtaining a data set free from selection biases and iii) making realistic estimates of the uncertainties. The first and second steps are accommodated through selection cuts which reduce the number of events available for the $X_{\rm max}$ analyses while we resort to Monte Carlo calculations to deal with the third point.

The quality cuts have been determined using Monte Carlo simulations of the showers and of the telescopes with the goal of selecting events in which the uncertainty in the measurement of $X_{\rm max}$ is $< 40$ g/cm$^2$. The first quality cut relates to the requirement of having clear nights in which an accurate measurement of the aerosol profile was possible. Dusty periods with the Vertical Atmosphere Optical Atmospheric Depth (VAOD) at 3 km above ground less that 0.1 were excluded. Measurements of the cloud cover and of the aerosol content are made routinely [9, 10].

The second set of quality cuts relates to the reconstruction of the longitudinal profile of the shower. Events which have their axis within a cone of 20° around the direction of orientation of the telescope are rejected as these showers cross the camera at high angular-speed making the systematic offset in time between the fluorescence detector and the relevant water-Cherenkov detector an important source of uncertainty in the reconstruction procedures.

The depth of shower maximum is a key parameter derived from the reconstruction. $X_{\rm max}$ is required to lie within the field of view of the telescope as an extrapolation beyond the measured data would degrade the accuracy of measurement. The quality of the fit to the data is assured by rejecting events with $\chi^2/\mathrm{NDOF} > 2.5$. The statistical precision of the measurement is derived from the fitting procedure: if the uncertainty in $X_{\max}$, after taking into account both geometrical reconstruction and atmospheric conditions, is $> 40$ g/cm$^2$ then the event is rejected.

The detection efficiency of a fluorescence telescope depends upon the geometry of the shower with respect to it and on the nature of the primary particle that initiated the event. To make an unbiased estimate of primary composition one must be certain that the detector has an acceptance that is independent of the mass of the incoming particle. This is achieved by selecting only those events that have, for the given shower geometry and energy, a large enough acceptance range to bracket the $X_{\max}$ distribution. Full details are given in [13, 14].

### 2.1 Improvements in the data analysis

This paper is an update of earlier publications [1, 5]: previous interpretations remain valid [15]. The improvements that will be presented at the Conference are three-fold: i) the increased number of events from 27 additional months of data-taking leads to a reduction in the statistical uncertainties; ii) an improved understanding of our detector which has led to a reduction in the systematic uncertainties in $X_{\max}$ and iii) the development of more detailed methods of analysis with the aim of correcting the acceptance of the detector. The update given at the Conference will use the post-LHC interaction models, QGSJETII-04 and EPOS-LHC, as guidelines for interpretation. The inclusion of the LHC data in the models has had the effect of reducing the differences between them [16] and of predicting a large value of $X_{\max}$ than given previously for a primary of a particular type and energy.

An important result to be presented by the Auger Collaboration at this Conference relates to the energy of the measured events. A better understanding of our detectors has led to change of about 15% at $10^{18}$ eV to the energy scale and this will be included in the update of the $X_{\max}$ results. Details of the new energy scale are given in [17].

Knowledge of the atmospheric conditions at the Observatory has also improved with a model based on GDAS [18] now being used. The new model allows an event-by-event description of the atmosphere. This replaces the monthly average profiles used in the earlier analyses [19]. The GDAS model has been validated using local balloon launches and has therefore been chosen as the standard description of the atmosphere at the site. The use of GDAS has led to a fall in the systematic uncertainties in $\langle X_{\max}\rangle$ of $\sim 0.5$ g/cm$^2$ and in RMS($X_{\max}$) of $\sim 2$ g/cm$^2$ at $10^{18}$ eV and $\sim 3.5$ g/cm$^2$ at higher energies. Recent measurements of the aerosol attenuation have been upgraded and two analyses techniques have been implemented to give a better determination of the uncertainties [10, 11, 12].

Further we now have a better understanding of the lateral spread of the light signal across the photomultipliers in the cameras [20, 17] which leads to changes of the reconstructed $X_{\max}$. In the new analysis procedure, the lateral spread of the light from the shower is convolved with the size of the optical spot of the telescope. The combination of these factors results in light being spread over a large area away from the shower axis. This light loss has been parametrized as a function of shower distance and shower age. The correction will be introduced for the data that will be reported.

Quality and fiducial cuts have been tailored to provide an unbiased measurement of the $X_{\max}$ distribution. Therefore the data set available after the selection cuts should have constant acceptance for most of the $X_{\max}$ values. However, events with very deep $X_{\max}$ values may have a smaller acceptance. To study the effect of the smaller acceptance in the tails of the $X_{\max}$ distribution we have used Monte Carlo events. The effect of the smaller acceptance in the tails of the $X_{\max}$ distribution is less than 5 g/cm$^2$ in the estimated $\langle X_{\max}\rangle$ and in the estimated RMS($X_{\max}$). Despite the effect of this correction being smaller than 5 g/cm$^2$ for both $\langle X_{\max}\rangle$ and RMS($X_{\max}$), we are estimating the appropriate acceptance correction for each energy bin: the results will be shown at the Conference.

## 3 Conclusion

We have presented in this report a short review of the data analysis developed by the Pierre Auger Collaboration. The final analysis will be presented at the Conference and will include:

- new energy scale;
- detailed corrections due to the lateral width of the shower image;
- new aerosol data analysis;
- GDAS atmospheric models;
- acceptance correction.

Besides the improvements in the data analysis briefly described above, the increase in the number of events (27 more months of data) and the new energy scale allow us to introduce additional bins below $10^{18}$ eV and above $10^{19}$ eV.

# Inferences about the mass composition of cosmic rays from data on the depth of maximum at the Auger Observatory

EUN-JOO AHN[1] FOR THE THE PIERRE AUGER COLLABORATION[2]

[1] *Fermilab, Batavia, IL, USA*
[2] *Full author list: http://www.auger.org/archive/authors_2013_05.html*

*auger_spokespersons@fnal.gov*

**Abstract:** In this paper we outline the details of a method used to estimate the mass composition of ultra-high energy cosmic rays, which is based on interpretation of the data on shower maximum, $X_{\max}$, measured using the fluorescence telescopes of the Pierre Auger Observatory. The analysis is performed using a variety of hadronic interaction models. The results of our analyses will be presented at the Conference.

## 1 Introduction

The composition of ultra-high energy cosmic rays (UHECRs) is yet to be fully understood. The atmospheric depth where the longitudinal development of the air shower reaches the maximum number of particles, $X_{\max}$, is a standard observable used to extract composition information as different nuclei produce different mean values of $X_{\max}$ and different dispersions in $X_{\max}$. Conversion of $X_{\max}$ to mass is inferred through air-shower simulations. In particular, the mean and dispersion of $X_{\max}$ are commonly used to infer the mass composition.

We report a method that parameterises the mean value of the depth of shower maximum, $\langle X_{\max}\rangle$, and its dispersion, $\sigma(X_{\max})$, and these observables are converted to the first two moments of the log-mass distribution, $\langle \ln A\rangle$ and $\sigma^2_{\ln A}$. Refinements to the method have been made over those originally proposed [1].

The reliance on hadronic interaction models mean that the mass composition measurement is subject to some level of uncertainty, as different physics assumptions are used to extrapolate interaction properties beyond man-made accelerator energies. Hence results must always be interpreted within the context of the models used.

The Pierre Auger Collaboration will present the results of the updated analysis on the mass composition derived from the $X_{\max}$ data using the updated hadronic interaction models at the Conference. This paper discusses only the method and not the results.

## 2 Utilising the moments of $X_{\max}$

The superposition model allows simple scaling between masses and a log-linear dependence of energy. This model is generalized to include additional energy and mass dependent terms and is better able to accommodate all the hadronic interaction models used. The $\langle X_{\max}\rangle$ is now expressed as

$$\langle X_{\max}\rangle = X_0 + D\log\left(\frac{E}{E_0 A}\right) + \xi \ln A + \delta \ln A \log\left(\frac{E}{E_0}\right), \quad (1)$$

where $X_0$ is the mean depth of proton showers at energy $E_0$ and $D$ is the elongation rate, and the parameters $\xi$ and $\delta$ are derived for each hadronic interaction model.

From Eq. (1), the mean and dispersion of $X_{\max}$ for a mixed composition are derived as

$$\langle X_{\max}\rangle = \langle X_{\max}\rangle_p + f_E \langle \ln A\rangle \quad (2)$$

$$\sigma^2(X_{\max}) = \langle\sigma^2_{\rm sh}\rangle + f_E^2\, \sigma^2_{\ln A}\,. \quad (3)$$

The linearity of the mean with respect to $\langle \ln A\rangle$ is demonstrated in Eq. (2), where the first term is the mean of $X_{\max}$ for a proton. The dispersion of Eq. (3) has two terms, where the first term denotes the shower-to-shower fluctuation is contained in $\langle\sigma^2_{\rm sh}\rangle$ and the second term reflects the dispersion in $\ln A$ arising from the mass distribution of the composition. The energy-dependent parameter $f_E$ is expressed as

$$f_E = \xi - \frac{D}{\ln 10} + \delta \log\left(\frac{E}{E_0}\right). \quad (4)$$

The first two moments of $\ln A$ can be obtained by inverting Eqs. (2, 3);

$$\langle \ln A\rangle = \frac{\langle X_{\max}\rangle - \langle X_{\max}\rangle_p}{f_E} \quad (5)$$

$$\sigma^2_{\ln A} = \frac{\sigma^2(X_{\max}) - \sigma^2_{\rm sh}(\langle \ln A\rangle)}{b\,\sigma_p^2 + f_E^2}\,. \quad (6)$$

To obtain an explicit expression for $\langle\sigma^2_{\rm sh}\rangle$ we need a parameterization for $\sigma^2_{\rm sh}(\ln A)$. We assume a quadratic law in $\ln A$:

$$\sigma^2_{\rm sh}(\ln A) = \sigma_p^2[1 + a\ln A + b(\ln A)^2]\,, \quad (7)$$

where $\sigma_p^2$ is the $X_{\max}$ variance for proton showers. The evolution of $\sigma^2_{\rm sh}(\ln A)$ with energy is included in $\sigma_p^2$ and the parameter $a$:

$$\sigma_p^2 = p_0 + p_1 \log_{10}\left(\frac{E}{E_0}\right) + p_2\left[\log_{10}\left(\frac{E}{E_0}\right)\right]^2 \quad (8)$$

$$a = a_0 + a_1 \log_{10}\left(\frac{E}{E_0}\right). \quad (9)$$

The parameters necessary to calculate Eqs. (5) and (6) for each of the hadronic interaction models have been derived using the air shower generator CONEX [2] and are available in Ref. [3] for EPOS 1.99 [4], QGSJet 01 [5], QGSJet II-03 [6], and Sibyll 2.1 [7]. In this paper we extend the parameterizations to the LHC-tuned hadronic models EPOS-LHC and QGSJet II-04, available in CONEX v4r37. The corresponding paramaters are given in Tables 1 and 2.

| parameter | EPOS-LHC | QGSJet II-04 |
|---|---|---|
| $X_0$ | 806.1 $\pm$ 0.3 | 790.4 $\pm$ 0.3 |
| $D$ | 55.6 $\pm$ 0.5 | 54.4 $\pm$ 0.5 |
| $\xi$ | 0.15 $\pm$ 0.24 | -0.31 $\pm$ 0.24 |
| $\delta$ | 0.83 $\pm$ 0.21 | 0.24 $\pm$ 0.21 |

**Table 1**: Parameters of Eq. (1) for EPOS-LHC and QGSJet II-04, setting $E_0 = 10^{19}$ eV. All values are expressed in g cm$^{-2}$.

| parameter | EPOS-LHC | QGSJet II-04 |
|---|---|---|
| $p_0 \times (\mathrm{g}^{-2}\mathrm{cm}^4)$ | 3284 $\pm$ 51 | 3738 $\pm$ 54 |
| $p_1 \times (\mathrm{g}^{-2}\mathrm{cm}^4)$ | -260 $\pm$ 64 | -375 $\pm$ 66 |
| $p_2 \times (\mathrm{g}^{-2}\mathrm{cm}^4)$ | 132 $\pm$ 108 | -21 $\pm$ 109 |
| $a_0$ | -0.462 $\pm$ 0.006 | -0.397 $\pm$ 0.007 |
| $a_1$ | -0.0008 $\pm$ 0.0016 | 0.0008 $\pm$ 0.0019 |
| $b$ | 0.059 $\pm$ 0.002 | 0.046 $\pm$ 0.002 |

**Table 2**: Parameters of Eqs. (7-9) for EPOS-LHC and QGSJet II-04, setting $E_0 = 10^{19}$ eV.

Equations (5) and (6) can be used to study the evolution of the moments of the log mass distribution, $\langle \ln A \rangle$ and $\sigma^2_{\ln A}$, as shown in Ref. [3] and infer the mass composition in the energy range of the $X_{\max}$ measurements of the Auger Observatory.

## 3 Conclusion

The method using the first two moments of $X_{\max}$ and ln$A$ to infer the mass composition from the $X_{\max}$ data have been discussed. This method can also be used to investigate the validity of hadronic interaction models. The final data analysis using the latest dataset from Auger [8] that also compares the differences between the pre- and post-LHC hadronic interaction models will be presented at the Conference.

# Observations of the longitudinal development of extensive air showers with the surface detectors of the Pierre Auger Observatory

DIEGO GARCÍA-GÁMEZ[1] FOR THE PIERRE AUGER COLLABORATION[2]

[1] *Laboratoire de l'Accélérateur linéaire (LAL), Université Paris 11, CNRS-IN2P3, Orsay, France*
[2] *Full author list: http://www.auger.org/archive/authors_2013_05.html*

*auger_spokespersons@fnal.gov*

**Abstract:** Using the timing information from the FADC traces of surface detectors far from the shower core it is possible to reconstruct a Muon Production Depth distribution (MPD) to provide information about the longitudinal development of the muon component of Extensive Air Showers (EAS). We assess the quality of the MPD reconstruction for zenith angles around 60° and different energies of the primary particle. From these distributions we define $X^{\mu}_{\max}$, the depth, along the shower axis where the number of muons reaches maximum, and explore its potential as a useful observable to infer the mass composition of cosmic rays.

## 1 Introduction

Finding a solution to the question of the origin of the Ultra-High Energy Cosmic Ray (UHECR) requires three experimental feats: finding the mass of the primary particles, measuring the energy spectrum and detecting anisotropies in the distribution of their arrival directions. The energy spectrum is the best known of the three and its main features are well established [1, 2]. More controversial is whether the arrival directions of the highest-energy events are anisotropic [3, 4].

The situation regarding the mass composition of UHECRs is controversial. One way to determine the mass is to study the longitudinal development of the electromagnetic component of a shower. The depth of the shower maximum, $X_{\max}$, is sensitive to the nature of the primary [5]. However $X_{\max}$ measurements suffer from low statistics due to the small duty cycle of the fluorescence detectors and the stringent cuts imposed to avoid a biased data sample [6].

The Auger Collaboration has proposed different methods [7] to infer masses that take advantage of the large statistical sample provided by the high-duty cycle of the surface detector (SD) array. Here we describe one of them. It relies on the study of the longitudinal development of the muonic component of EAS. The surface detectors of the Observatory provide this information through the timing records associated with the muons that reach ground. The muon arrival-times allow the reconstruction of their production points along the shower axis. It is thus possible to reconstruct a distribution of Muon Production Depths [8]. Since muons come from the decay of pions and kaons, the shape of the MPD contains information about the evolution of the hadronic cascade. This information renders interesting the study of MPDs for the following reason: we know that different primaries have distinct hadronic properties (i.e. cross-section and multiplicity) that translate into variations of their respective longitudinal profiles. Therefore it is natural to think that the shape of the MPD must be sensitive to the mass of primary particle.

## 2 The Model for the Muon Arrival Time Distributions

Muons reaching ground have a time structure caused by the importance of different mechanisms during muon propagation. Through a set of simple assumptions, those arrival times can be used to obtain the distribution of muon production distances along the shower axis. The basis of our measurement is a theoretical framework originally developed in [9, 10] and updated in [11]. As a first approximation we assume that muons travel in straight lines at the speed of light $c$ and neglect the delay accumulated by the parent mesons. Muons produced at the position $z$ (along the shower axis) that reach ground at the point defined by $(r,\zeta)$ have travelled a distance $l$ given by the expression:

$$l = ct_{\mu} = \sqrt{r^2 + (z-\Delta)^2} \quad (1)$$

where $t_{\mu}$ is the muon time of flight. $r$ and $\zeta$ are measured in the shower reference frame and represent the distance and the azimuthal position of the point at ground respectively. $\Delta = r\tan\theta\cos\zeta$ is the distance from the point at ground to the shower plane. If we reference the muon time of flight to the arrival time of the shower-front plane for each position $(r,\zeta)$, we obtain the *geometrical delay* $t_g$:

$$ct_g = \sqrt{r^2 + (z-\Delta)^2} - (z-\Delta). \quad (2)$$

Therefore there is a one-to-one correspondence between measured arrival times and muon production distances. The actual muon delay includes effects related to the fact that muon velocities are subluminal (causing a *kinematical delay* $t_{\varepsilon}$). Delays produced by the geomagnetic field and multiple scattering are of lesser importance [11]. The muon production point along the shower axis $z$ is given by the expression:

$$z \simeq \frac{1}{2}\left(\frac{r^2}{ct - c\langle t_{\varepsilon}\rangle} - (ct - c\langle t_{\varepsilon}\rangle)\right) + \Delta \quad (3)$$

where the geometrical delay $t_g$ has been approximated by $t_g \simeq t - \langle t_{\varepsilon}\rangle$.

## 3 Reconstruction of Muon Production Depth

The MPD is reconstructed from the FADC signals from the water-Cherenkov detectors. The production depth $X^{\mu}$ comes from an integration of the atmospheric density, $\rho$, over the range of production distances:

$$X^{\mu} = \int_{z}^{\infty} \rho(z')dz'. \tag{4}$$

The shape of the distribution depends strongly on the observation point at ground as the surface detectors sample the particle cascade at different stages of its development. For discrete detector arrays just a handful of $r$ values are available. This implies that in general the measured MPD will show a severe distortion when compared to the true MPD shape. The MPD observed at ground also depends on the zenith angle. This is mainly a consequence of the muon decay probability. It influences not only the location of the maximum but also shapes the observed MPD.

To gain insight to the physics from the reconstructed shape of the MPD, a fit is made using the Gaisser-Hillas function. Of its four parameters, the one referred to as $X^{\mu}_{\max}$ accounts for the point along the shower axis where the production of muons reaches a maximum. As shown later, this parameter will be our main physics observable for composition studies. The MPD fit is performed in an interval of depths ranging from 0 to 1200 g cm$^{-2}$ that contains the entire range of possible values of $X^{\mu}_{\max}$ (our deepest simulated proton shower has an $X^{\mu}_{\max} \sim 1000$ g cm$^{-2}$). Simulations show that the $X^{\mu}_{\max}$ distribution varies as a function of the mass of the primary particle. For heavier primaries, the average value of $X^{\mu}_{\max}$ is smaller and the distribution narrower compared with that for lighter particles (Figure 1). This behaviour is independent of the energy of the primary cosmic ray. The signals registered by the surface detectors are from a mix of muons and electromagnetic (EM) particles. To build the MPD of an event we are only concerned with the behaviour of the muonic component. The EM signal is a background that must be eliminated. One way to do this is to use inclined events ($\theta >\sim 60°$). For these data, the EM component is heavily absorbed by the atmosphere. Inclined events are also of special interest for this analysis since the dependence of the MPD shape with the distance to the shower axis $r$ drastically decreases as $\theta$ increases. This helps to reduce the impact that the spacing of the Auger surface detectors have in the reconstruction of $X^{\mu}_{\max}$. Therefore, for the sake of simplicity, the present work focuses only on data for which the zenith angles lie in the interval [55°, 65°].

The EM contamination can be further reduced by exploiting the different behaviors of the EM and muonic components. In general, the EM signals are smaller and broader. As a consequence, a cut on signal threshold that rejects all time bins with signals below a certain value ($S_{threshold}$) will help diminish the EM contamination.

To build the production depth distribution, every time bin of the FADC traces is converted to an entry in it by means of equation 3. Since the typical time stamp of a muon does not fall into a single time bin but it rather spans several bins with a known, gamma-like, distribution, we have an uncertainty in the arrival time of muons that must be accounted for. To compensate for this detector effect, we subtract an offset $T_{shift}$. It depends on $S_{threshold}$ and hence simulations must be used to find an appropriate value. Requiring that for $S_{threshold}$= 0.4 VEM the reconstruction of our lower energy events is unbiased, we find $T_{shift}$= 70 ns. We use this value throughout the range of interest. To guarantee that the reconstruction bias in $X^{\mu}_{\max}$ lies within ±10 g cm$^{-2}$ for all energies, $S_{threshold}$ is increased slowly with energy, reaching a maximum value of 0.6 when dealing with the most energetic events.

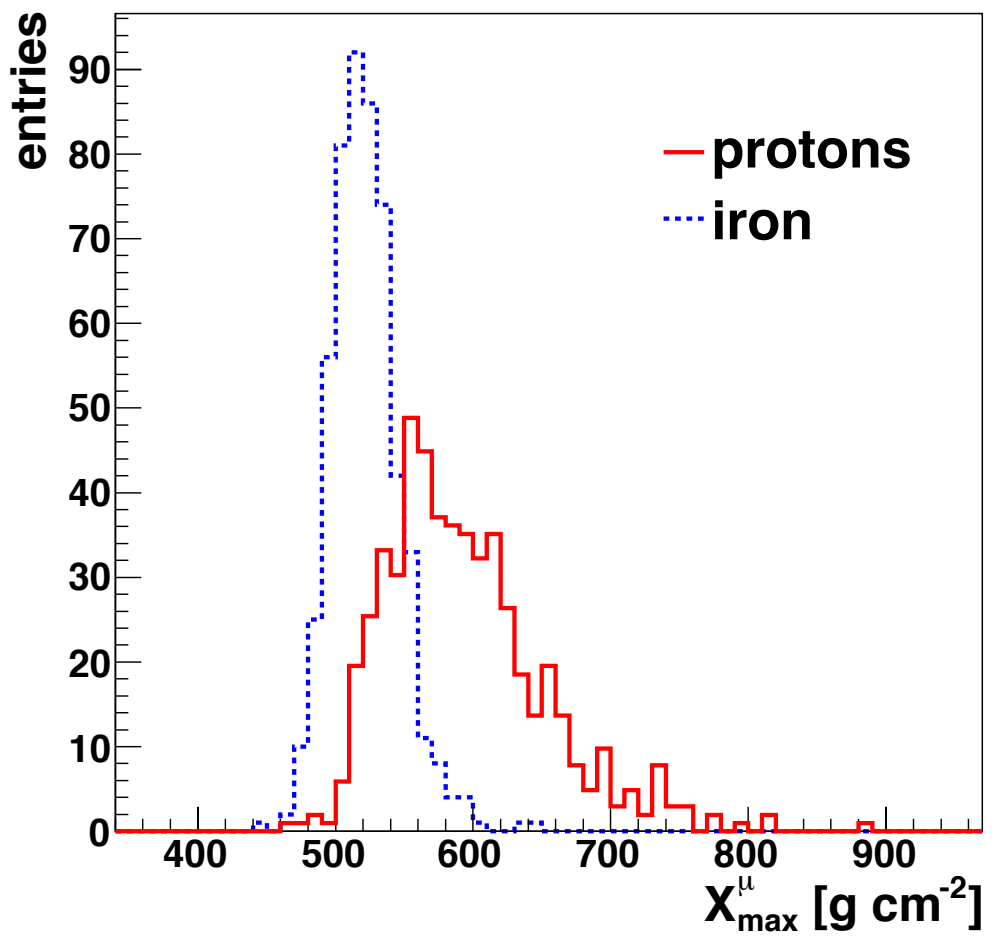

**Figure 1**: $X^{\mu}_{\max}$ distributions for proton and iron showers simulated at 30 EeV with Epos-LHC [12] at zenith angles between 55° and 65°. The mean value and the RMS of the distributions show a clear dependence upon the mass of the primary cosmic ray. For the construction of these MPDs, only muons reaching ground at distances greater than 1700 m have been considered.

For each time bin, the uncertainty introduced in $X^{\mu}$ ($\delta X^{\mu}$ ) is a function of the FADC sampling rate ($\delta t$) and the accuracy of reconstruction of the shower angle and core location. The sampling frequency is 40 MHz, and gives rise to an uncertainty in the $z$ reconstruction [9, 10] , that decreases quadratically with $r$, and increases linearly with $z$ as:

$$\frac{\delta z}{z} \simeq 2\frac{z}{r^2}c\delta t. \tag{5}$$

It is evident that the closer we get to the impact point at ground, the larger the uncertainty in $z$ (and in $X^{\mu}$ through equation 4). The contribution of the geometrical reconstruction to $\delta X^{\mu}$ also increases as we get closer to the core. Thus, to keep the distortions of the reconstructed MPD small, only surface detectors far from the core are useful. A cut in core distance, $r_{cut}$, is therefore mandatory. This cut diminishes the efficiency of the reconstruction and also affects the resolution as it reduces the number of muons in the reconstruction: note that the total uncertainty of the MPD maximum, $\delta X^{\mu}_{\max}$, depends on the number of muons $N_{\mu}$, and therefore decreases as the square root of $N_{\mu}$. The reconstruction efficiency however improves with energy, as the number of muons becomes larger as energy increases. As the number of muons at ground is a function of the mass of the primary, we risk introducing a bias in our selection towards heavier nuclei if the value for $r_{cut}$ is not carefully chosen. Therefore the selection of the distance cut must be a trade off between the resolution of the reconstructed MPD and the number of muons being accepted into such reconstruction [8]. We use Monte Carlo simulations to choose the optimal value for $r_{cut}$. To build the MPD, we consider only those detectors whose distance to the shower core is larger than 1700 m, regardless of the shower energy. Choosing an $r_{cut}$ which is independent of energy implies that any difference in resolution that we find for different energies will be a consequence mainly of the different number of muons detected at ground. To estimate the impact that the distance cut and the undersampling in $r$ have on the determination of $X^{\mu}_{\max}$, we have studied the variation of $X^{\mu}_{\max}$ as a function of $r_{max}$ (upper limit of the distance interval [$r_{cut}$, $r_{max}$] used to integrate the MPD). Our simulations show that the variation of the $X^{\mu}_{\max}$ value amounts to about 10 g cm$^{-2}$ per km shift in $r_{max}$.

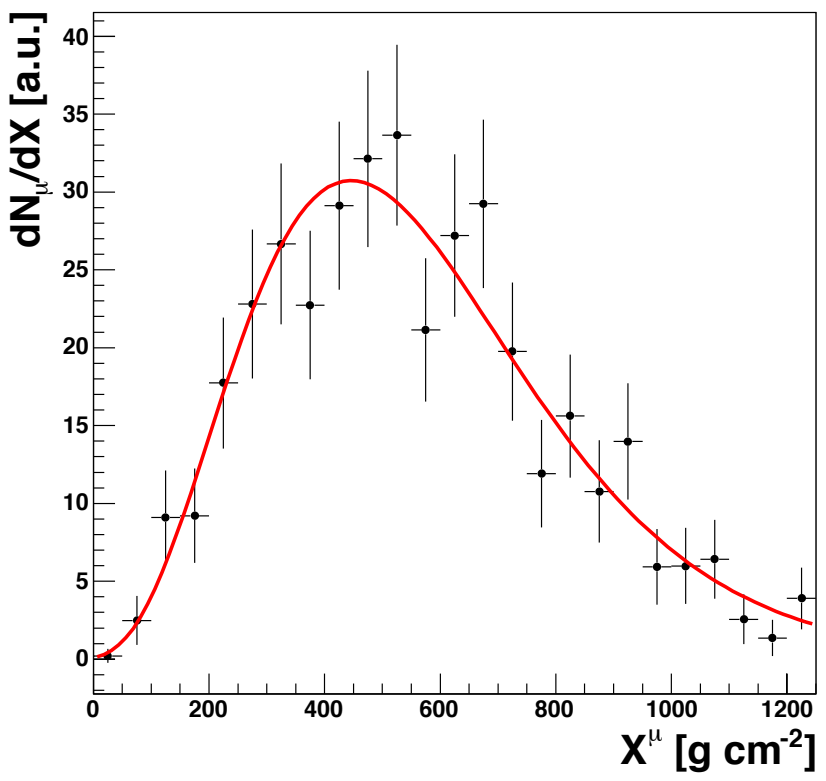

**Figure 2**: Real reconstructed MPD, $\theta$ = (59.06 ± 0.08)° and E = (92 ± 3) EeV, with the fit to a Gaisser-Hillas function.

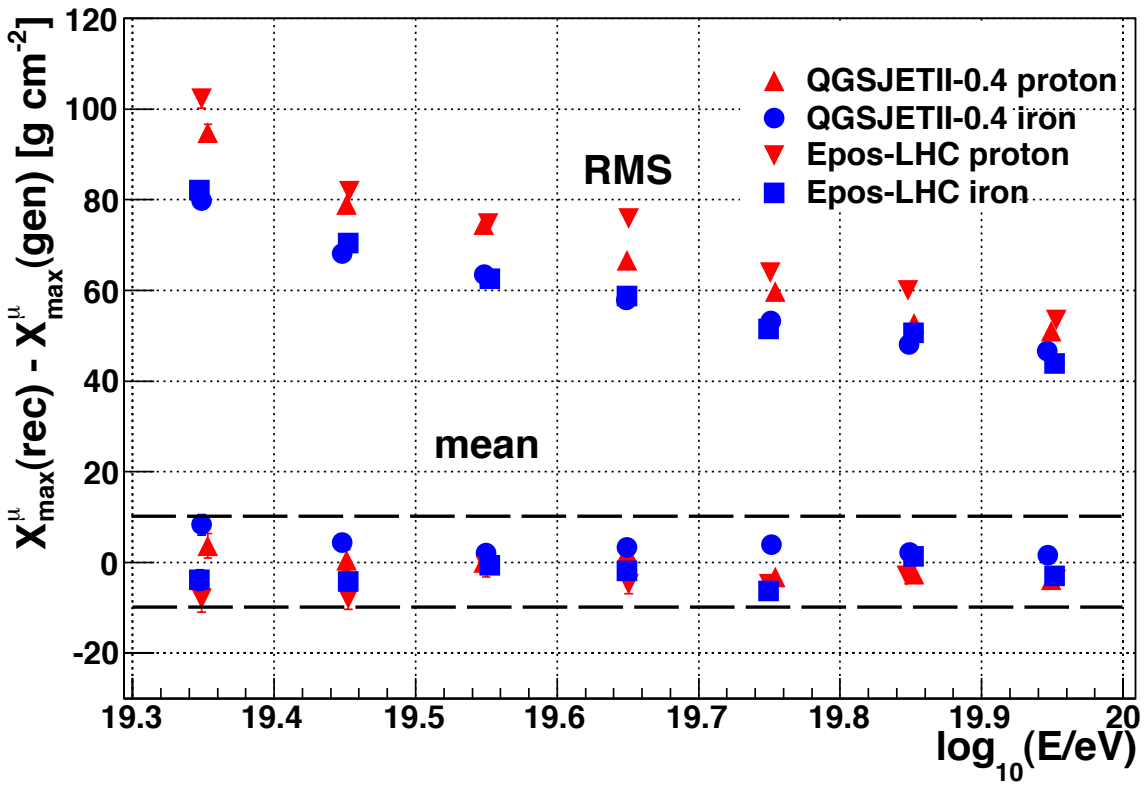

**Figure 3**: Evolution with energy of the RMS of the distribution [$X^{\mu}_{\max}$ (reconstructed) - $X^{\mu}_{\max}$ (true)]. The simulations were made using the QGSJETII-0.4 [13] and EPOS-LHC hadronic models for proton and iron nuclei for 55° ≤ $\theta \leq$ 65°.

The fact that in the selected data we do not use triggered stations further than ∼4000 m implies that we build MPDs by counting muons at ground in the distance range 1700 m ≤ r ≤ 4000 m. The MPD for a single detector is obtained as the average of the three MPDs that each PMT yields. For each event, the final MPD is obtained by adding the individual MPDs observed by each of the selected surface detectors. Figure 2 shows the reconstructed MPD for one of our most energetic events.

We select longitudinal profiles measured using a simple set of criteria: a) **Trigger cut.** We select EAS that fulfill a T5 trigger condition which requires that the detector with the highest signal has all 6 closest neighbours operating; b) **Energy cut.** We restrict our analysis to events with energy larger than 20 EeV as for the less energetic events the population of the MPD is very small, giving a very poor determination of $X^{\mu}_{\max}$; c) $\mathbf{X^{\mu}_{max}}$ **error.** We reject events whose relative error in $X^{\mu}_{\max}$ is bigger than a certain value $\varepsilon_{\max}$, an energy-dependent quantity (see Table 1) since the accuracy in the estimation of $X^{\mu}_{\max}$ improves with energy. This is a natural consequence of the increase in the number of muons that enter the MPD as the energy grows.

The event selection efficiencies after the cut in $X^{\mu}_{\max}$ uncertainty (cut c) are greater than 80%. Monte Carlo studies have shown that the cuts chosen introduce a composition bias smaller than 2 g cm$^{-2}$ (included as a systematic uncertainty). Also, as shown in Figure 3, the absolute value of the mean bias in reconstructions is < 10 g cm$^{-2}$, regardless of the hadronic model, energy and atomic mass of the simulated primary. The resolution, understood as the RMS of the distribution [$X^{\mu}_{\max}$ (reconstructed) - $X^{\mu}_{\max}$ (true)], ranges from 100 (80) g cm$^{-2}$ for proton (iron) at the lower energies to about 50 g cm$^{-2}$ at the highest energy (see Figure 3). The improvement of the resolution with energy is a direct consequence of the increase in the number of muons.

| log$_{10}$E/eV | $\varepsilon_{\max}$(%) |
|---|---|
| [19.3, 19.4] | 15 |
| [19.4, 19.6] | 11 |
| [19.6, 19.7] | 10 |
| [19.7, 19.8] | 8 |
| > 19.8 | 7 |

**Table 1**: Maximum relative errors allowed in the estimation of $X^{\mu}_{\max}$. The value chosen for $\varepsilon_{\max}$ ensures no selection bias between the different primary species.

## 4 Application to data

The data set used in this analysis comprises the events recorded from 1-January 2004 to 31-December 2012. We compute MPDs on an event-by-event basis. We have shown that for events with zenith angles in the interval 55° ≤ $\theta$ ≤ 65°, the total MPD is simply the direct sum of the individual MPDs given by the set of selected water-Cherenkov detector traces. For this angular range, our initial sample is therefore made of 663 events.

To guarantee an accurate reconstruction of the longitudinal profile we impose the selection criteria described in Section 3. Table 2 summarises how the different cuts reduce the number of events.

The evolution of the measured $\langle X^{\mu}_{\max}\rangle$ as a function of energy is shown in Figure 4. The data have been grouped in five energy bins of width 0.1 in log$_{10}$(E/eV), except for the last which contains all events with energy above $\log_{10}$(E/eV)=19.7 (50 EeV). The uncertainties represent the standard error on the mean.

| Cut | Events | Efficiency (%) |
|---|---|---|
| $55° \leq \theta \leq 65°$, E>20 EeV | 663 | 100 |
| T5 trigger* | 500 | 75 |
| $\varepsilon(X^{\mu}_{max})<\varepsilon_{max}$ | 481 | 73 |

**Table 2**: Selection procedure applied to the SD data. * Described in Section 3.

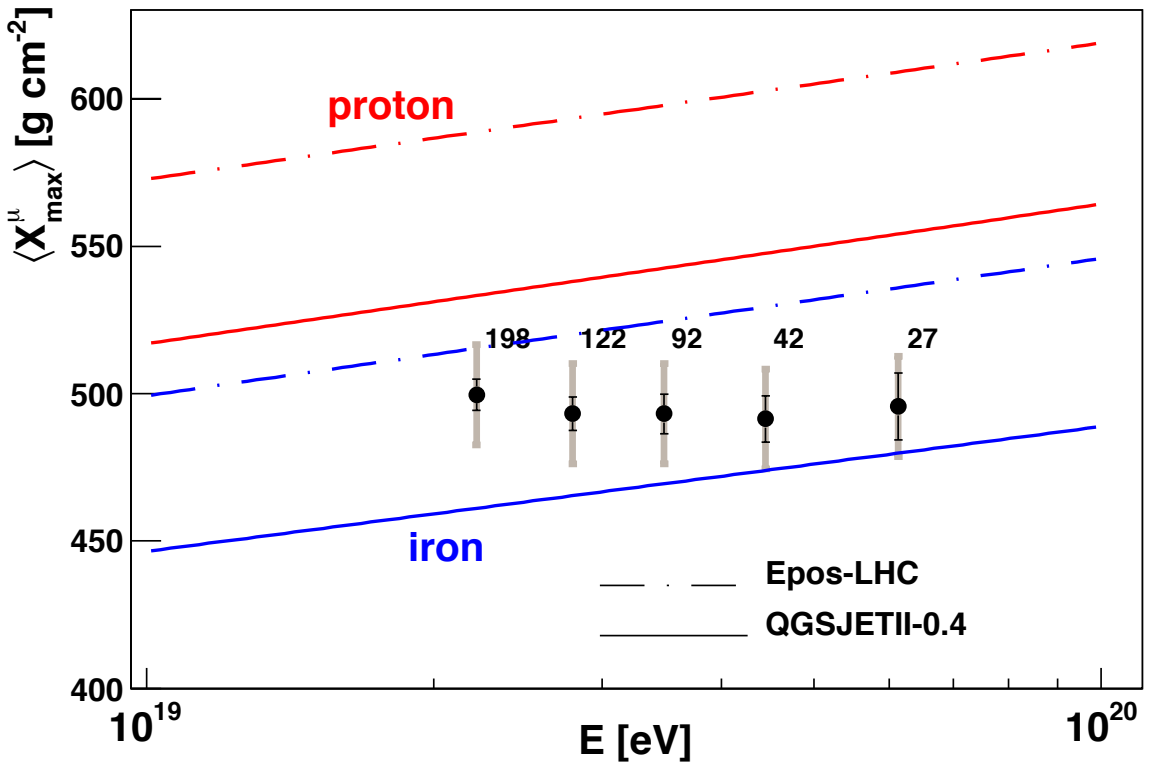

**Figure 4**: $\langle X^{\mu}_{max}\rangle$ as a function of energy. The prediction of different hadronic models for proton and iron are shown. Numbers indicate the amount of selected data in each energy bin and the gray rectangles represent the systematic uncertainty.

Table 3 lists the most relevant sources contributing to the systematic uncertainty. The uncertainties on the MPD reconstruction and event selection translate into an overall systematic uncertainty on $\langle X^{\mu}_{max}\rangle$ of 17 g cm$^{-2}$.

The interpretation of data in terms of mass composition requires a comparison with air shower simulations. For models like those of Figure 4 that assume standard hadronic interactions, the evolution of the mean $X^{\mu}_{max}$ values may suggest a change in composition as the energy increases (flatter trend than pure proton or pure iron predictions). However as the MPD technique currently suffers from small statistics and large resolution measurements, a constant composition would also be acceptable. At this stage, we cannot make conclusive inferences about mass composition. The proposed method can also be used as a tool to investigate the validity of hadronic interaction models. In Figure 4 we can see how QGSJETII-0.4 and EPOS-LHC predict, for both proton and iron, the same *muonic elongation rate* (rate of evolution of $X^{\mu}_{max}$ with energy) but with considerable differences in the absolute value of $X^{\mu}_{max}$. The measurement of muon profiles provides complementary data that set additional constraints to model descriptions and improves understanding of hadronic interactions.

## 5 Conclusions

The FADC traces from the water-Cherenkov detectors at the Auger Observatory located far from the core have been used to make reconstructions of the muon production depth distributions on an event-by-event basis. The maximum of the distribution $X^{\mu}_{max}$ contains information about the mass composition of UHECR. It can be used also to assess the validity of hadronic interaction models at ultra-high energies. With this analysis we have established a novel approach to study the longitudinal development of the hadronic component of EAS.

| Source | Sys. Uncertainty (g cm$^{-2}$) |
|---|---|
| Reconstruction + hadronic model + primary | 10 |
| Core time | 5 |
| Atmospheric profile | 8 |
| Fitting procedure | 3 |
| Selection efficiency | 2 |
| Energy uncertainties | 3 |
| Seasonal | 8 |
| Total | 17 |

**Table 3**: Evaluation of the main sources of systematic uncertainties.

**Acknowledgment:** D. Garcia Gamez was supported by the ANR-2010-COSI-002 grant of the French National Research Agency.

# A measurement of the muon number in showers using inclined events recorded at the Pierre Auger Observatory

INÉS VALIÑO[1] FOR THE PIERRE AUGER COLLABORATION[2].

[1] *Departamento de Física de Partículas e IGFAE, Universidade de Santiago de Compostela, Spain.*
[2] *Full author list: http://www.auger.org/archive/authors_2013_05.html*

*auger_spokespersons@fnal.gov*

**Abstract:** The average muon content of air showers with zenith angles exceeding 62° is obtained as a function of calorimetric energy from events measured simultaneously with the Surface Detector Array and fluorescence telescopes of the Pierre Auger Observatory using a reconstruction method specifically designed for inclined showers. The results are presented in different energy bins above $4 \times 10^{18}$ eV and compared to predictions from current hadronic interaction models for different primary particles.

## 1 Introduction

Understanding the energy dependence of the mass composition of the highest energy cosmic rays is fundamental to unveil their production and propagation mechanisms. Interpretations about observed anisotropies [1, 2] and features in the spectrum such as the break in the power law spectrum around $4 \times 10^{18}$ eV, the ankle, and the flux suppression above $4 \times 10^{19}$ eV [3], lead to very different conclusions depending on the assumed mass composition at Earth.

Because of the low flux of cosmic rays at these energies, their composition cannot be measured directly, but has to be inferred from observations of extensive air showers. As a consequence the most sensitive parameters to mass composition are also dependent on the hadronic interaction properties, which are unknown at very high energies and in phase space regions inaccessible to accelerator experiments. In this context, the estimate of the primary mass can only be made using sets of simulated reference showers, which have been generated with hadronic interaction models based on extrapolation from accelerator data over more than two orders of magnitude in energy in the center-of-mass frame. For this reason, it is advisable to study different observables sensitive to both mass composition and hadronic interaction models to minimise the problem (see e.g. [4] for a recent review). One of the most mass-sensitive observables is the number of muons at the ground. An air shower induced by a nucleus with $A$ nucleons contains approximately $A^{1-\alpha}$ ($\alpha \approx 0.9$) more muons than a proton shower of the same energy. In addition, the number of muons in air showers also depends on several properties of hadronic interactions, including the multiplicity, the charge ratio and the baryon anti-baryon pair production [5, 6].

Cosmic rays arriving with zenith angles exceeding 62° induce extensive air showers characterised by the dominance of secondary energetic muons at ground, because the electromagnetic component has been largely absorbed in the enhanced atmospheric depth crossed by the shower before reaching ground. The study of showers with zenith angles $\theta > 62°$, the so-called inclined showers, provides a direct measurement of the muon content at ground level.

In this work we explain how the muon content is measured in inclined showers detected with the Surface Detector (SD) array of the Pierre Auger Observatory [7]. This paper is organised as follows. In Sec. 2, we describe the reconstruction method of the shower size parameter $N_{19}$ and the associated uncertainties. In Sec. 3, we study $N_{19}$ as a function of calorimetric energy in an unbiased sample of high-quality events measured simultaneously with the SD array and the Fluorescence Detector (FD) of the Auger Observatory. Finally, in Sec. 4, we compare the behaviour of the shower size parameter as a function of the energy above $4 \times 10^{18}$ eV as observed with data to predictions of current hadronic interaction models for different primary masses.

## 2 Reconstruction of $N_{19}$

Inclined showers generate asymmetric and elongated signal patterns in the SD array with narrow pulses in time in detectors, typical of a muonic shower front. Events are selected demanding space-time coincidences of the signal patterns of the triggered surface detectors which must be consistent with the arrival of a shower front [8]. After event selection, the arrival direction $(\theta, \phi)$ of the cosmic-ray is determined from the relative arrival times of the shower front at the triggered stations by fitting a model of the shower front. The angular resolution achieved is better than 0.6° for events of interest [9].

Once the shower direction is established, we define the shower size parameter, $N_{19}$, through the following relation:

$$\rho_\mu = N_{19}\ \rho_{\mu,19}(x, y, \theta, \phi) \tag{1}$$

where $\rho_\mu$[1] is the model prediction for the muon density at ground used to fit the signals recorded at the detectors. $\rho_{\mu,19}$ is a reference profile corresponding to the inferred arrival direction, obtained as a parameterisation [10] of the muon density at ground of proton showers of $10^{19}$ eV simulated using CORSIKA [11] with the QGSJetII-03 [12] and FLUKA [13] interaction models. An example of the reference profile $\rho_{\mu,19}$ for $\theta = 80°$ and $\phi = 0°$ is shown in Fig. 1. It has been found [14] that at a given depth the shape

1. It is defined in the plane perpendicular to the shower axis

and attenuation with the zenith angle of the muon density profile are independent of the cosmic-ray energy $E$ and mass $A$, so the factorisation (1) holds in good approximation.

Introducing $N_\mu$ ($N_{\mu,19}$) as the total number of muons reaching ground as predicted by the integral of Eq. 1 (respectively of $\rho_{\mu,19}$), $N_{19}$ is simply the ratio $N_\mu/N_{\mu,19}$. Hence $N_{19}$ carries the dependence on the energy and mass.

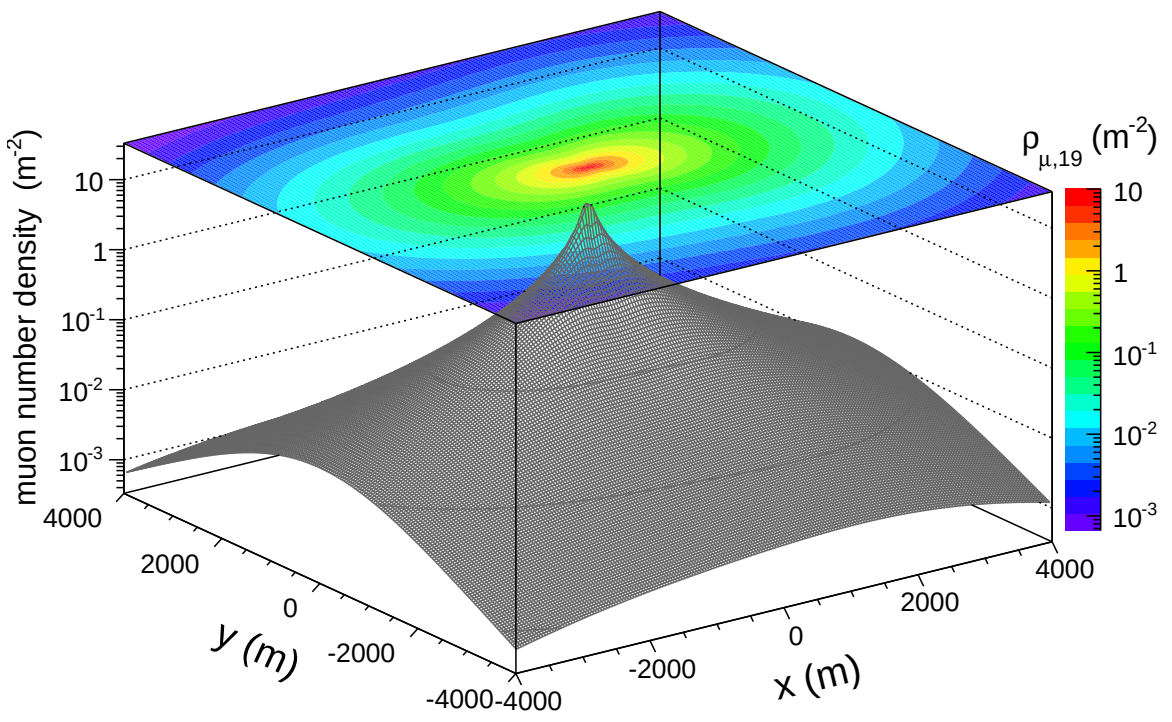

**Figure 1**: Muon number density in the coordinate plane perpendicular to the shower axis at ground level (transverse plane) for proton-induced showers of $10^{19}$ eV at $\theta = 80°$ and $\phi = 0°$ (parallel to $x$-axis) and core at $(x,y) = (0,0)$.

The estimation of $N_{19}$ is done via a maximum-likelihood fit of the predicted $\rho_\mu$ to the measured signals and is based on a detailed model of the detector response to the passage of muons, obtained from GEANT4 [15] simulations within the software framework Offline [16] of the Auger Observatory. To perform the fit, the muonic signal is obtained from the measured signal by subtracting the average contribution of a residual electromagnetic component (typically 20% of the muonic signal) parameterised from simulations [17]. The achieved resolution improves from 20% to 8% in the range $\log_{10}(E/\text{eV})$ = [18.6,19.8] and the systematic uncertainty is smaller than 5%. Further details of the reconstruction procedure and validation tests can be found in [9].

### 2.1 Testing $N_{19}$ as an estimator of the muon number, $R_\mu$

In this section we want, using MC simulations, to test the effectiveness of $N_{19}$ as estimator of the total number of muons reaching ground relative to that contained in the reference distribution. We do so by comparing in simulated showers the $N_{19}$ parameter to the true ratio $R_\mu^{MC} = N_\mu^{true}/N_{\mu,19}$, where $N_\mu^{true}$ is the true number of muons at ground.

Three different sets of simulated events were used in this study. The first set consists of 100000 proton and 100000 iron showers generated using AIRES [18] with QGSJet01 [19] at a relative thinning of $10^{-6}$, following an energy spectrum $E^{-2.6}$ over the energy range $\log_{10}(E/\text{eV})$ = [18.5,20.] and an isotropic angular distribution. The second (third) set consists of 12000 proton and 12000 iron showers generated using CORSIKA with QGSJetII-04 [20] (EPOS LHC [21]) with the same thinning and angular distribution, and an energy spectrum $E^{-1}$ over the energy range $\log_{10}(E/\text{eV})$ = [18.,20.]. Showers subsequently underwent a full simulation of the detector with random core positions on the ground, and were then reconstructed using the procedure adopted for data.

The mean value of the difference between $N_{19}^{MC}$ and $R_\mu^{MC}$ is shown in Fig. 2. We note that the bias is less than 5% for showers with $R_\mu^{MC} > 0.6$, value above which the SD array is over 95% efficient. From this result, we can conclude that $N_{19}$ provides a direct measurement of the relative number of muons with respect to the reference distribution with little bias. To parameterise the average bias as shown in Fig. 2, we have chosen the average of the two extreme cases shown, corresponding to iron showers generated with EPOS LHC and QGSJet01. In the following, we will call $R_\mu$ the measured $N_{19}$ after correction for this average bias. Its uncertainty is estimated to be 5% from the dispersion of the different hadronic interaction models and compositions explored.

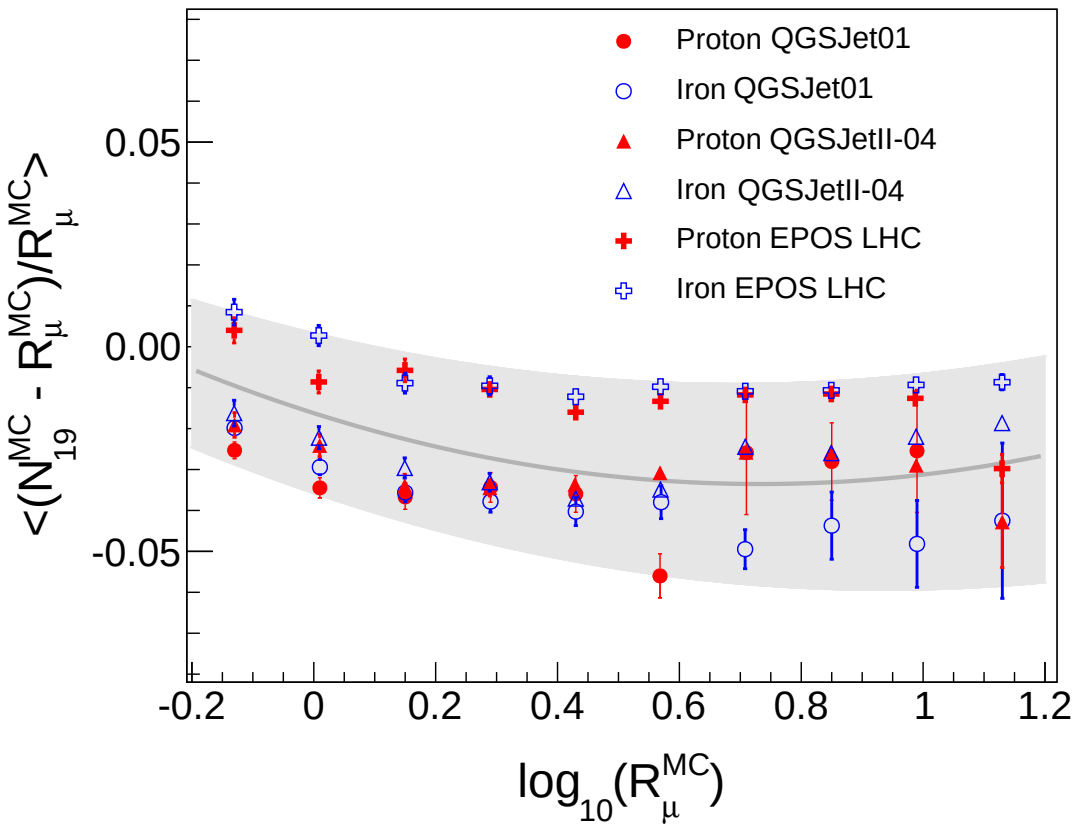

**Figure 2**: Average of the relative difference between $N_{19}^{MC}$ and $R_\mu^{MC}$ for proton and iron showers simulated with QGSJet01, QGSJetII-04 and EPOS LHC. The band indicates the bias region and the solid line indicates the parameterised average bias.

## 3 Energy dependence of $R_\mu$

For inclined showers, we use $N_{19}$ for the energy calibration. This is done using the calorimetric energy $E_{\text{FD}}$ from high-quality events measured simultaneously with SD and FD (golden hybrid events) [22]. In the same way, here, we obtain the correlation of $R_\mu$ with $E_{\text{FD}}$ to study the relative muon content of measured showers as a function of the energy. As noted in the introduction, $R_\mu$ is sensitive to primary composition and to the properties of the hadronic interactions in the shower.

The data set contains hybrid events with zenith angles $62° < \theta < 80°$ with at least four triggered stations and for which the closest station to the fitted core and its six adjacent stations are all active. In addition, these events have to satisfy a set of quality cuts for the FD specifically designed to ensure an accurate reconstruction of the arrival direction and of the longitudinal profile. The cuts are adapted versions of those used in calibration of events with $\theta < 60°$ [23, 24, 25]. The station closest to the shower core that is used for the geometrical reconstruction must be at a distance below 750 m. For a precise estimation of the $E_{\text{FD}}$ we require adequate monitoring of the atmospheric conditions (cloud coverage below 25% in the FD field of view and vertical aerosol optical depth positive and less than 0.1). To obtain accurate values of the shower maximum $X_{\max}$ and of the primary energy we require: a Gaisser-Hillas fit with a $\chi^2$-residual, $(\chi^2 - n_{\rm dof})/\sqrt{2n_{\rm dof}}$, less than 3; a maximum "hole" in the longitudinal profile of 20%. In addition to the quality selection criteria, a fiducial cut on the FD field of view (FOV) is performed ensuring this is large enough to observe all plausible values of $X_{\max}$. This "fiducial FOV cut" includes a cut on the maximum uncertainty of $X_{\max}$ accepted (150 g/cm$^2$) and on the minimum viewing angle of the light in the FD telescope (25°), which avoids cutting on the fraction of Cherenkov light. The uncertainty on $E_{\rm FD}$ is required to be less than 30%. Finally, we only accept FD energies larger than $4\times10^{18}$ eV to assure a trigger probability of 100% for the FD and SD detectors. This selection was applied to inclined events recorded from 1 January 2004 to 31 December 2012; 174 events were kept.

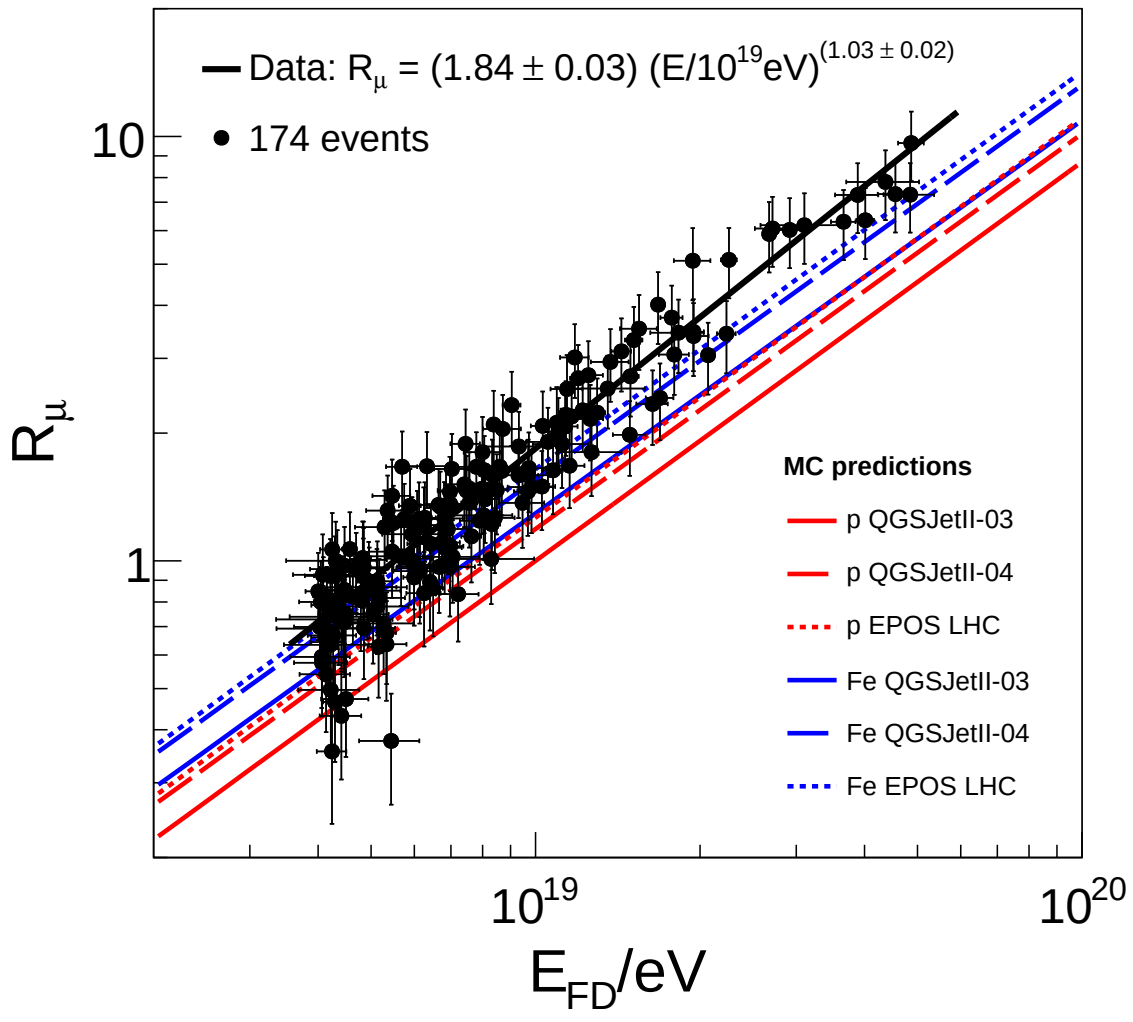

**Figure 3**: Fit of the correlation of $R_\mu = A\ [E_{\rm FD}/10^{19}\,{\rm eV}]^B$ with $E_{\rm FD}$ in high-quality hybrid data. Theoretical curves for proton (blue lines) and iron (red lines) showers simulated with QGSJetII-03 (solid lines), QGSJetII-04 (dashed lines) and EPOS LHC (dotted lines) are shown for comparison.

A power-law fit of $R_\mu$ as a function of the calorimetric energy, $E_{\rm FD}$, gives the $R_\mu/E$ relation as:

$$R_\mu = A\ [E_{\rm FD}/10^{19}\,{\rm eV}]^B \qquad (2)$$

The fit method is based on a maximum-likelihood approach accounting for the estimated uncertainties in the respective $R_\mu$ and $E_{\rm FD}$ reconstructions, and fitting the shower-to-shower fluctuations in the total number of muons to a constant value. The corresponding result of the fit is shown in Fig. 3 and the best fit values are $A = 1.84\pm0.03\pm0.09\,({\rm sys})$ and $B = 1.03\pm0.02\pm0.05\,({\rm sys})$. The systematic uncertainties are estimated from the dispersion of the different models and compositions explored with simulated events (see Sec. 2) and variations of the quality cuts on the FD and of fitting methods applied.

In this work we provide an update of the method presented previously in [9]. The current measurement of the muon number has changed with respect to that made in the past mainly due to upgrades of the reconstruction algorithms of the shower size parameter $N_{19}$ and of the FD energy scale as presented in this conference [26]. The interpretation of data is also affected by the use of the recent released hadronic interaction models QGSJetII-04 and EPOS LHC as guidelines for the analysis.

## 4 Results and discussion

Using the formula for the correlation fit (see Eq. 2), it is possible to derive the number of muons in data compared to the predictions for different models and cosmic ray masses. For example, the number of muons at $10^{19}$ eV deduced from data exceeds that of proton (iron) showers simulated with QGSJetII-03 by a factor 1.8 (1.4). However, the post-LHC versions of the QGSJet and EPOS models, namely QGSJetII-04 and EPOS LHC, predict about 20% more muons than QGSJetII-03 and become more compatible with data.

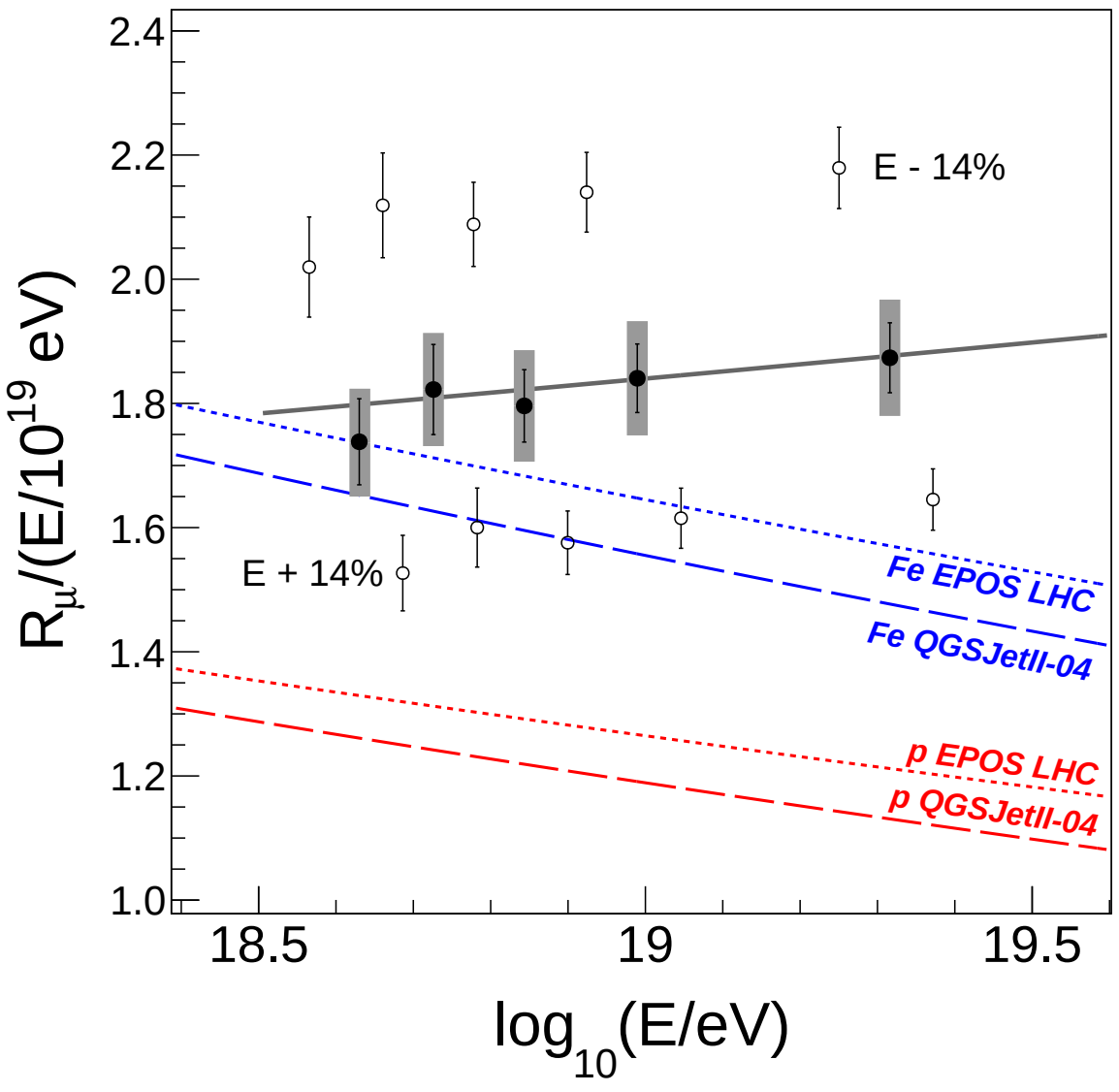

**Figure 4**: Average value of $R_\mu/(E_{FD}/10^{19}$ eV) as a function of shower energy. Theoretical curves for proton and iron showers simulated with QGSJetTII-04 and EPOS LHC are shown for comparison. Open circles indicate the result if the FD energy scale is varied by its systematic uncertainty. The black line represents the calibration fit from Fig. 3. The gray thick error bars indicate the systematic uncertainty of $R_\mu$.

An alternative and interesting comparison of the measured number of muons with predictions is shown in Fig. 4, where the averaged scaled quantity $R_\mu/(E_{FD}/10^{19}$ eV) is shown in five energy bins containing roughly equal number of events. The measurement of $R_\mu$ is rather accurate and can be estimated to have an uncertainty of 5% combining statistical and systematic uncertainties (shown as grey boxes around the points) . The measurement of $R_\mu/(E_{FD}/10^{19}$ eV) is dominated by systematic uncertainties in the energy scale (shown as open circles in the figure). The measured number of muons between $4\times10^{18}$ eV and $2\times10^{19}$ eV is marginally comparable to predictions for iron showers simulated with both QGSJetII-04 and EPOS LHC if we allow the FD energy scale to increase by its systematic uncertainty of about 14% [26].

The slope $\mathrm{d}\ln n_\mu/\mathrm{d}\ln E$ of the muon number, given by the parameter $B = 1.03 \pm 0.02 \pm 0.05\,(\mathrm{sys})$, carries information about possible changes in the average logarithmic mass $\langle \ln A \rangle$. Predictions for proton and iron showers simulated with QGSJetII-04 and EPOS LHC give a slope between 0.93 and 0.94, because the number of muons increases less than linearly with energy if $A$ is constant. In this context, a steeper slope would be indicative of an increase of $\langle \ln A \rangle$ in this energy range. The difference obtained in this work between data and predictions for a constant composition is however not significant (less than 1.7 $\sigma$).

Given that the observed distribution of the depth of shower maximum between $4 \times 10^{18}$ eV and $2 \times 10^{19}$ eV is not compatible with an iron dominated composition [27, 28] we conclude that the observed number of muons is not well reproduced by the shower simulations. This result is compatible with those of independent studies for showers with $\theta < 60^\circ$ [29][2], in which two different methods have been used to derive the fraction of the signal due to muons at 1000 m from the shower core using the temporal distribution of the signals measured with the SD array.

2. We note that in [29] the relative number of muons is given with respect to QGSJetII-04 whereas here the result is given with respect to QGSJetII-03. As the relative muon content of showers generated with QGSJetII-04 is about 20% higher than those of QGSJetII-03, this factor should be taken into account for the comparison.

# Measurement of the muon signal using the temporal and spectral structure of the signals in surface detectors of the Pierre Auger Observatory

BALÁZS KÉGL[1], FOR THE PIERRE AUGER COLLABORATION[2]

[1] *Laboratoire de l'Accélérateur Linéaire, Université Paris Sud, CNRS/IN2P3, Orsay, France*
[2] *Full author list: http://www.auger.org/archive/authors_2013_05.html*
*auger_spokespersons@fnal.gov*

**Abstract:** Applying different filtering techniques to the temporal distribution of the signals measured with the Auger surface detector array (SD), we separate the electromagnetic and muonic signal components of air-showers. The filters are based on the different characteristics of the muonic and electromagnetic components in individual detectors, the former being composed of peaks above a smooth background due to the lower energy deposition of the latter photons and electrons. The muon signal is derived for showers of 10 EeV primary energy at a core distance of 1 km, with the aim of testing the predictions of hadronic interaction models. We compare the fraction of the muonic signal and the total signal to model predictions for proton and iron primaries in a range of zenith angles from 0 to 60°.

## 1 Introduction

Understanding the development of extensive air-showers in the atmosphere is of central importance for deriving information on cosmic rays such as their energy distribution and mass composition. At the same time, detailed measurements of the characteristics of air-showers allow us to probe particle interactions up to the highest energies. While electromagnetic interactions in air-showers are well understood, there are considerable uncertainties in simulating the production of hadronic particles at energies and for phase space regions not accessible in accelerator experiments [1]. This makes the study of the hadronic component of showers particularly important.

Both the electromagnetic and the muonic shower components are fed mainly by pions and kaons produced in the hadronic core of the showers. Photons produced in the decay of neutral pions give rise to electromagnetic subshowers in which further particle multiplication and interactions make the distributions of the initial energy and the production depth inaccessible to ground-based detectors. In contrast, muons suffer only a small energy loss and angular deflection before reaching ground, and so they provide a window to study the hadronic shower core.

In this work we we will use the water Cherenkov detectors of the SD of the Pierre Auger Observatory [2] to measure the number of muons arriving at ground in showers of $E = 10^{19}$ eV at a distance of 1000 m from the shower core. We will compare the data with predictions derived from showers simulated with the interaction models QGSJETII.04 [3] and EPOS LHC [4].

The measurement is based on the time profile and on spectral characteristics of the Cherenkov light signal generated by shower particles in the water of the detectors. We provide updates of methods presented previously [5, 6]. Improved understanding of the essential aspects of these methods and the homogeneity of the time signals in the small energy and distance intervals considered here made it possible to simplify the methods and to better control the systematics due to model uncertainties and the unknown chemical composition of the cosmic rays.

In Section 2 we describe the methods of deriving the signal fraction due to muons and, after discussing the data selection and corresponding Monte Carlo simulations in Section 3, we present the muon measurements in Section 4. The derived muon fraction is then used to obtain the signal of the muonic shower component. A summary is given in Section 5.

## 2 Methods of measuring the muon fraction

The Cherenkov photons produced by the shower particles in the detectors are sampled by three photomultipliers (PMTs) [2]. The analog signal is then digitized with FADCs in 25 ns bins with a 10 bit dynamic range. The raw digital signal of each PMT is calibrated such that the integrated signal of a typical vertical atmospheric muon is 1. The signal in each time bin is thus measured in units of "vertical equivalent muon" or VEM [7]. Finally, the three calibrated traces of the PMTs of each detector are averaged. We will denote the resulting FADC signal by $\mathbf{x} = (x_1, \ldots, x_N)$, where $N$ is the number of time bins, and the total signal by $S = \sum_{j=1}^{N} x_j$. The total muonic and electromagnetic signals will be denoted by $S_\mu$ and $S_{\mathrm{EM}}$, respectively. Note that $S_\mu$ is the pure muonic signal, so the electromagnetic halo produced by muon interactions and muon decay in the atmosphere goes into $S_{\mathrm{EM}}$.

In the following we will derive the fraction $f_\mu$ of the signal that can be attributed to muons relative to the total signal $S$

$$f_\mu = S_\mu / S \qquad (1)$$

by exploiting the information on the temporal structure of the FADC signal at 1000 m from the shower core. Due to the similar energy scaling of the overall and muonic shower signals in the detectors at about 1000 m, this quantity is insensitive to the systematic uncertainty of the energy assigned to air-showers that is of the order of ∼14% [8].

The time response profile of individual particles (a short risetime followed by an exponential decay with decay parameter of about $\tau = 60$ ns) cannot be used to separate the muonic and electromagnetic (EM) signal components since this profile is the same for all particles. There are two population features that can, on average, enable us to separate the two components: the amplitude distribution of the particle responses and the time-of-arrival distributions. The amplitude distribution of muons depends on the zenith angle, but in general it is close to a Gaussian of mean 1 VEM with a lower tail due to short-tracklength corner-clipping muons and a higher tail due to delta rays generated by high-energy muons. The mean amplitude of a single EM particle is much smaller (but with a power-like heavy tail), and the number of EM particles are, on average, an order of magnitude larger than the number of muons. With respect to the time-of-arrival distribution, typically, muons arrive earlier than EM particles. These two features make the muon signal peaky and short and the EM signal smooth and elongated. Both methods presented in Sections 2.1 and 2.2 use these features of the data to measure the muon fraction. There are several limitations of both methods that generate both variance and systematic bias between models and primaries (muon pile-up, small muon peaks due to corner-clipping muons, signal fluctuation), but the main source of uncertainty is due to high-energy photons that can produce a signal similar to that of a muon. Their contribution is estimated to be less than 10% to 15% for proton and iron primaries, respectively, in the considered energy and angular range.

### 2.1 Measuring the muon fraction with a multivariate method

The basic idea of this method is to combine muon-content-sensitive characteristics of the FADC signal to measure the muon fraction. Concretely, we estimate the muon fraction $f_\mu$ by

$$\hat{f}_\mu = a + b\,\hat{\theta} + c\,f_{0.5}^2 + d\,\hat{\theta}\,P_0 + e\,\hat{r}, \tag{2}$$

where $\hat{\theta}$ is the reconstructed zenith angle of the shower and $\hat{r}$ is the distance of the detector from the reconstructed shower axis. $f_{0.5}$ is the portion of the signal in FADC bins larger than 0.5 VEM, that is,

$$f_{0.5} = \frac{1}{S}\sum_{j=1}^{N} x_j\,\mathbb{I}\{x_j > 0.5\}, \tag{3}$$

where the indicator function $\mathbb{I}\{A\}$ is 1 if $A$ is true and 0 otherwise. $P_0$ is the normalized zero-frequency component of the power spectrum, that is,

$$P_0 = \frac{S^2}{N\sum_{j=1}^{N} x_j^2} = \frac{\langle\mathbf{x}\rangle^2}{\langle\mathbf{x}^2\rangle} = \left[1 + \frac{\sigma^2(\mathbf{x})}{\langle\mathbf{x}\rangle^2}\right]^{-1}, \tag{4}$$

where $\langle\mathbf{x}\rangle = S/N$ is the mean of the signal vector $\mathbf{x} = (x_1,\ldots,x_N)$, $\sigma^2(\mathbf{x})$ is the variance of the signal vector, and $\langle\mathbf{x}^2\rangle$ is its second moment. Both $f_{0.5}$ and $P_0$ are sensitive to large relative fluctuation and short signals, which are the signatures of high muon content. Besides the two parametrized families of thresholded signal and binned normalized power spectrum, we also tried other muon-content-sensitive families of variables, namely the time quantiles of the signal $t_q = \min\{t : \sum_{j=1}^{t/25\,\text{ns}} x_j/S \geqslant q\}$ and thresholded discrete derivatives ("jumps" [5]) $J_\delta = \sum_{j=1}^{N-1}(x_{j+1} - x_j)\mathbb{I}\{x_{j+1} - x_j > \delta\}/S$. The formula Eq. (2) was selected by an exhaustive search among all quadratic functions over members of these parametrized families with the objective of minimizing the variance and the sensitivity of the estimator $\hat{f}_\mu$ to models and composition.

We estimate the fit parameters $(a,b,c,d,e)$ using simulations (described in Section 3) in two steps. First we regress the muon fraction $f_\mu$ against $a + b\,\hat{\theta} + c\,f_{0.5}^2 + d\,\hat{\theta}\,P_0$ using a dense ring of 12 artificial detectors, placed at 1000 m from the shower axis. Then we fix $b$, $c$, and $d$, and regress $f_\mu$ against the full estimator $\hat{f}_\mu$ with free $a$ and $e$ using the detectors triggered by the shower. The reason for this two-step procedure is that we have much more dense detectors in the simulations, allowing us to control the statistical error of the fit in the first step, but these detectors are all placed at 1000 m from the core, so we cannot use them to estimate the distance dependence of the muon fraction. The overall bias of $\hat{f}_\mu - f_\mu$ on the different models and primaries is about $\pm 0.02$ and the average resolution is about 0.08.

### 2.2 Measuring the muon fraction with a smoothing method

The basic idea of this method is to run a low-pass filter a few times on the signal to gradually separate the low-frequency smooth EM component from the high-frequency component which is assigned to muons. Formally, we first smooth the signal $\mathbf{x}$ by a moving average

$$\hat{x}_j = \sum_{i=1}^{N} x_i\,p_{ij} \qquad j = 1,\ldots,N, \tag{5}$$

where $p_{ij} = \mathbb{I}\{|i-j| \leqslant L\}/C_j$. $L$ is a tuned window size that depends on the zenith angle $\theta$, and $C_j$ is the size of the set $\{i : |i-j| \leqslant L\}$. The choice of $L$ is driven by the physics of the air-showers: the amount of signal per time bin in the time distribution of the EM component decreases as the zenith angle grows, while the opposite happens to the muonic component. In the smoothing filter, lower frequency cuts correspond to larger convolute ranges. As a consequence, a wider window allows us to easily follow the low frequencies composing the EM signal at large angles, while narrower windows are needed to extract it in vertical showers, where the EM component is more similar to the muonic signal. As a consequence, we let the window size $L$ grow with the zenith angle $\theta$. The exact function $L = 7.83 + 0.09\,\theta/^\circ$ was tuned using simulations.

We assign any positive difference to the muonic signal, that is,

$$S_\mu = \sum_{j=1}^{N} \mathbb{I}\{x_j > \hat{x}_j\}\,(x_j - \hat{x}_j). \tag{6}$$

We repeat the procedure four times, re-smoothing each time the smooth signal $\hat{x}_j$ from Eq. (5), output by the previous iteration. The final muonic signal is the sum of the non-smooth positive differences from Eq. (6). The muon fraction is then estimated by

$$\hat{f}_\mu = S_\mu/S. \tag{7}$$

The overall bias of $\hat{f}_\mu - f_\mu$ on the different models and primaries is about $\pm 0.05$ and the average resolution is about 0.08.

## 3 Simulations and data selection

The methods described in Section 2 were applied to detectors that are part of SD events passing fiducial cuts. The cut requires that six active detectors surround the detector with the highest signal [9], which ensures a reliable core and energy reconstruction. The zenith angle, the energy, and the core position of the shower were reconstructed following the standard Auger SD reconstruction [10, 11]. We first selected events from the time period between Jan 2004 and Dec 2012 with zenith angle $\hat{\theta} < 60°$ and reconstructed energy $\hat{E} \in [10^{18.98}, 10^{19.02}]$ eV, then we selected detectors with a distance from the reconstructed shower axis $\hat{r} \in [950, 1050]$ m, giving us 521 SD signals. At $\hat{E} = 10^{19}$ eV, the resolutions for the core position and the energy are about 50 m and 12%, respectively. The absolute energy scale has a systematic uncertainty of 14% [8].

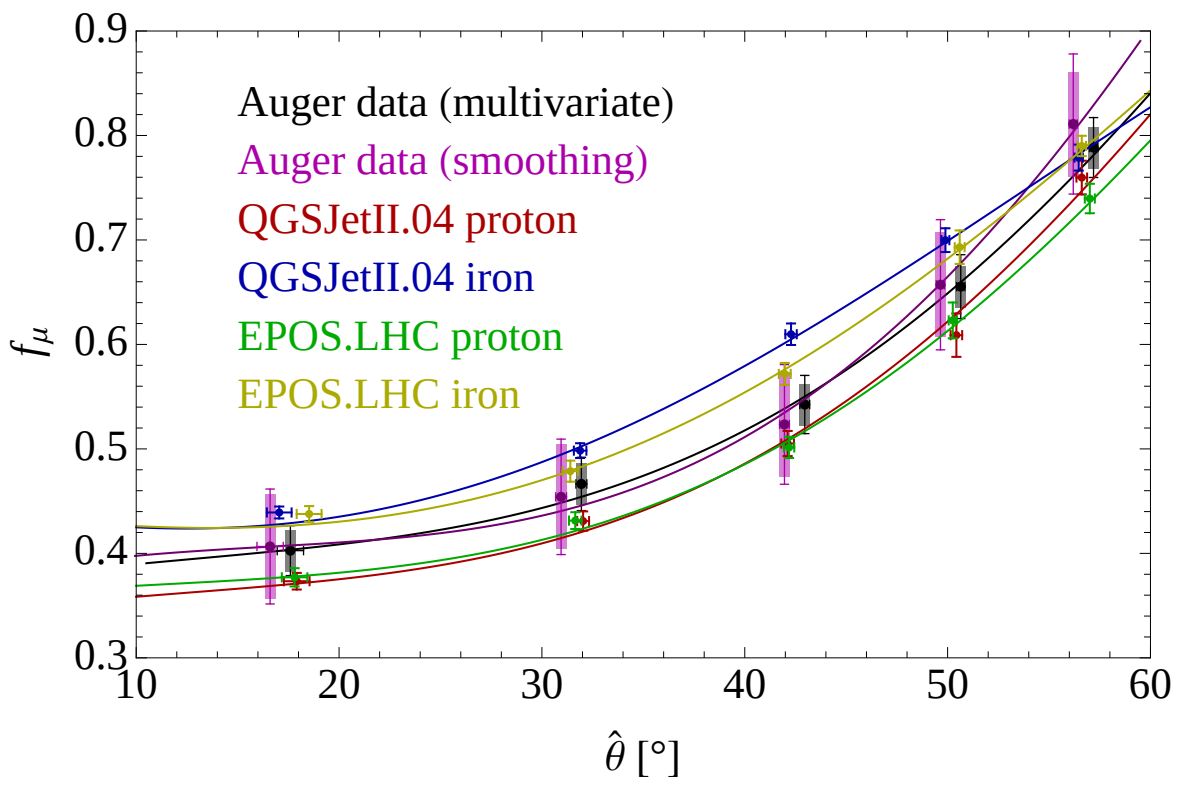

**Figure 1**: The muon fraction for primary energy $E = 10^{19}$ eV in a SD station at 1000 m from the shower axis, as a function of the reconstructed zenith angle $\hat{\theta}$. For Auger data, the rectangles represent the systematic uncertainties, and the error bars represent the statistical uncertainties added to the systematic uncertainties. The points for Auger data are artificially shifted by $\pm 0.5°$ for visibility. See Sections 2.1 and 2.2 for a detailed description.

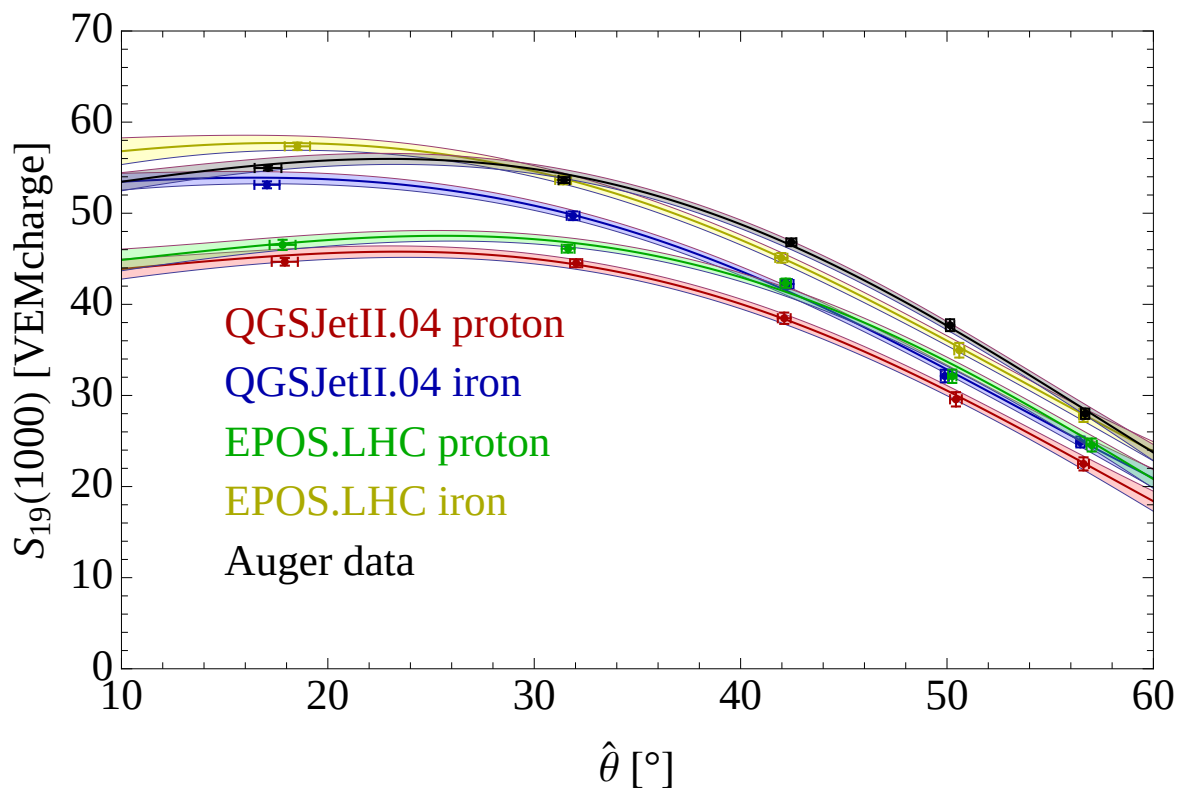

**Figure 2**: The mean total signal for primary energy $E = 10^{19}$ eV in a detector at 1000 m from the shower axis, as a function of the reconstructed zenith angle $\hat{\theta}$. See Section 4.2 for a detailed description.

To tune and test the methods described in Section 2, we used four shower libraries generated with CORSIKA [12]: proton and iron showers using the hadronic models QGSJETII.04 [3] and EPOS LHC [4] with FLUKA [13] as the low-energy interaction model. The detector response of the showers was simulated [14] using GEANT4 [15] simulations within the Offline software framework [16] of the Auger Observatory. We used the same energy, angle, and distance cuts as in the data.

# 4 Results

In the following we will first present the results for the muon fraction $f_\mu$ and compare the Auger data with simulation predictions. In a next step we will derive the overall detector signal and multiply it by the measured muon fraction to derive the muon signal.

## 4.1 Measuring the muon fraction

We estimate the muon fraction $\hat{f}_\mu$ from Eqs. (2) or (7) for every detector in the distance range $\hat{r} \in [950, 1050]$ m. The muon fraction is a very slowly varying function of the lateral distance and energy, which allows us to calculate the mean muon fraction at a given zenith angle by averaging over the selected detectors and showers in a given angular interval.

The results for the muon fraction ($f_\mu$ for simulations and $\hat{f}_\mu$ for Auger data) are shown in Fig. 1 for $E = 10^{19}$ eV and $\hat{r} = 1000$ m.[1] The muon fraction varies between 0.3 and 0.9 as a function of the zenith angle. Good agreement is found for the muon fractions derived with the two analysis methods. The model predictions for proton- and iron-induced showers bracket the measured muon fractions within the systematic uncertainties.

## 4.2 Measuring the total signal

To obtain the total signal $S(1000)$ at 1000 m from the shower axis, we apply

$$S(1000) = S \times \mathrm{LDF}(1000\,\mathrm{m})/\mathrm{LDF}(\hat{r}), \qquad (8)$$

where $\hat{r}$ is the distance from the reconstructed shower axis, and $\mathrm{LDF}(r) = r^\beta$ is the lateral distribution function of the total signal, where $\beta = -3.45$ is obtained by fitting a power law on detector signals of the simulation libraries described in Section 3. We then rescale the total signal to $10^{19}$ eV by further multiplying $S(1000)$ by $C(E) = (E/10^{19}\,\mathrm{eV})^{-0.966}$ to obtain the *projected* total signal

$$S_{19}(1000) = S(1000) \times C(E), \qquad (9)$$

where the exponent $-0.966$ comes from the slope of the energy dependence of $S(1000)$ [10]. To correct the migration effect due to the steep slope of $E^{-2.6}$ of the energy spectrum [17] and the 12% energy resolution [10], we multiply the reconstructed energy by a factor of 0.984 before applying Eq. (9).

Fig. 2 depicts the projected total signal $S_{19}(1000)$ (for data and simulations) as a function of the reconstructed zenith angle. Note that none of the transformations in Eqs. (8) and (9) bias the mean total signal but they do decrease its variance.

For showers with primary energy $E = 10^{19}$ eV, the total signal $S$ in a detector at 1000 m varies between 20 and 60 VEM, depending primarily on the zenith angle but also on the simulation model and on the mass composition of the primary particle (Fig 2). The mean signal in data is significantly higher than that of QGSJETII.04 proton

1. In all figures, showers are binned by their reconstructed zenith angle into bins determined by the borders [0°, 26°, 37°, 47°, 53°, 60°]. The $x$-coordinate of every point is the mean zenith angle in the bin.

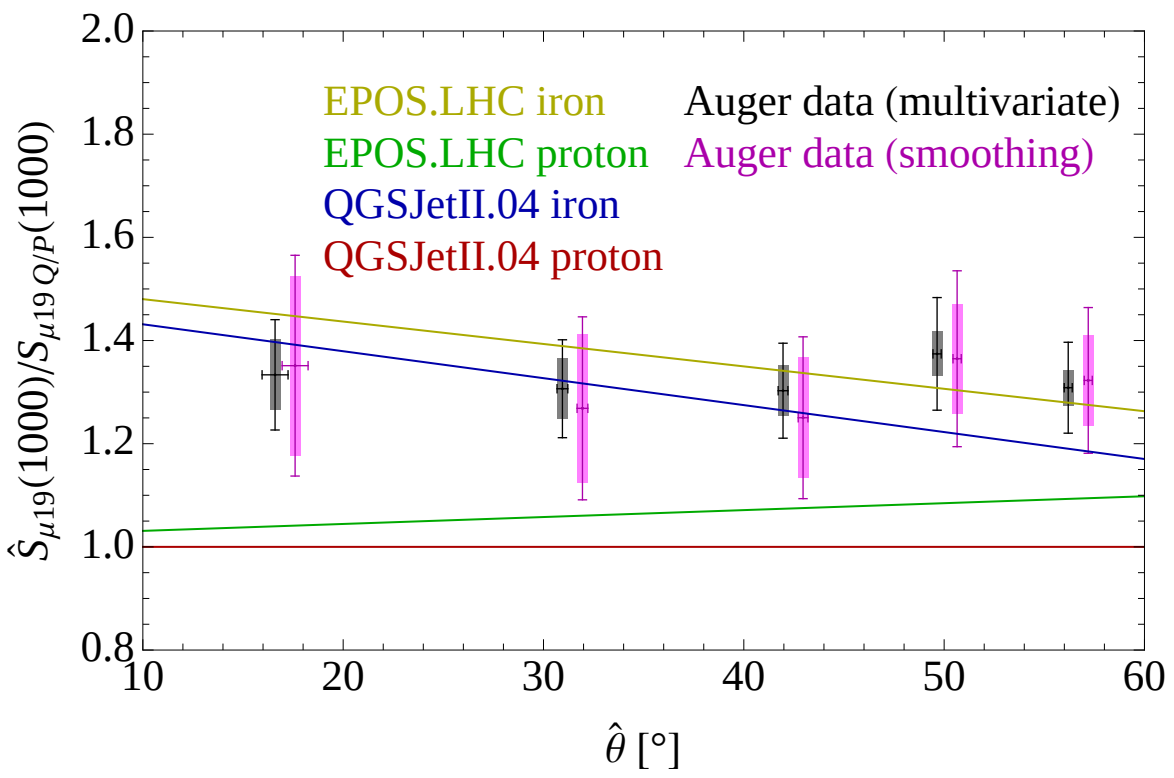

**Figure 3**: The measured muon signal rescaling at $E = 10^{19}$ eV and at 1000 m from the shower axis vs. zenith angle, with respect to QGSJETII.04 proton as baseline. The rectangles represent the systematic uncertainties, and the error bars represent the statistical uncertainties added to the systematic uncertainties. The points for Auger data are artificially shifted by $\pm 0.5°$ for visibility.

simulations and still exceeds somewhat that of iron-induced showers simulated with QGSJETII.04. The discrepancy is possible since the function that relates the ground signal to the primary energy is not determined by Monte Carlo, rather it is calibrated to the calorimetric energy measured by the fluorescence detector [17].

### 4.3 Computing the muon signal rescaling

In the QGSJETII.04 proton simulation, taken as a reference, we compute the muon fraction $f_\mu$ from Eq. (1) for every detector, and multiply it by the projected signal $S_{19}(1000)$ from Eq. (9) to obtain the projected muon signal

$$S_{\mu 19}(1000) = f_\mu \times S_{19}(1000). \qquad (10)$$

In the Auger data set, we compute the muon fraction estimate $\hat{f}_\mu$ from Eqs. (2) or (7) for every detector and multiply the fraction by the projected signal $S_{19}(1000)$ to obtain the estimated projected muon signal

$$\hat{S}_{\mu 19}(1000) = \hat{f}_\mu \times S_{19}(1000). \qquad (11)$$

Then we separate the detectors by the reconstructed zenith angles $\hat{\theta}$ of the corresponding showers into zenith angle bins, and divide the mean estimated muon signal $\langle \hat{S}_{\mu 19}(1000)\rangle$ in data by the mean muon signal $\langle S_{\mu 19}(1000)\rangle$ in the baseline simulation, properly accounting for the small effects of the unequal mean angles and the nonzero variance of the denominator.

The result of the analysis is shown in Fig. 3. The rectangles represent the systematic uncertainties, and the error bars represent the statistical uncertainties added to the systematic uncertainties. We determine that the measured factor of the muon signal in data divided by the muon signal in QGSJETII.04 proton showers at $10^{19}$ eV and at 1000 m in the full angular range of $[0°, 60°]$ is

$$1.33 \pm 0.02\,(\text{stat.}) \pm 0.05\,(\text{sys.}) \quad (\text{multivariate})$$
$$1.31 \pm 0.02\,(\text{stat.}) \pm 0.09\,(\text{sys.}) \quad (\text{smoothing})$$

## 5 Summary

The fraction of the muonic signal measured in the detectors of the Pierre Auger Observatory has been estimated from the time structure of the recorded signal for showers of $10^{19}$ eV in different zenith angle bins between 0° and 60°. Two methods, a multivariate technique and a smoothing technique, have been used to derive the fraction of the signal due to muons. The results of the two methods are in very good agreement. The measured fraction of the muonic to total signal is bracketed by model predictions for proton and iron primaries obtained with CORSIKA and QGSJETII.04 and EPOS LHC.

Combining the estimated muon signal fraction with the measured total signal at 1000 m from the shower core allowed us to derive the part of the detector signal that can be attributed to the muonic shower component. While the measured angular dependence of the muonic signal is found to be similar to the prediction obtained for proton showers and QGSJETII.04, the magnitude of the muonic signal is comparable to the predictions for iron showers.

Given that the observed distribution of the depth of shower maximum at $10^{19}$ eV is not compatible with an iron dominated composition [18] we conclude that the overall detector signal and the muonic signal are not well reproduced by the shower simulations. These results are compatible with that of the independent study for inclined showers whose signal at ground is dominated by muons [19, 20]. Comparing simultaneously the measured longitudinal shower profile and the surface detector signal to simulations provides further constraints on hadronic interaction models [6, 21].

**Acknowledgment:** B. Kégl was supported by the ANR-2010-COSI-002 grant of the French National Research Agency.

# The muon content of hybrid events recorded at the Pierre Auger Observatory

Glennys R. Farrar[1] for the Pierre Auger Collaboration[2]

[1] *Center for Cosmology and Particle Physics, Department of Physics, New York University, NY, NY 10003, USA*
[2] *Full author list: http://www.auger.org/archive/authors_2013_05.html*
*auger_spokespersons@fnal.gov*

**Abstract:** The hybrid events of the Pierre Auger Observatory are used to test the leading, LHC-tuned, hadronic interaction models. For each of 411 well-reconstructed hybrid events collected at the Auger Observatory with energy $10^{18.8}$ – $10^{19.2}$ eV, simulated events with a matching longitudinal profile have been produced using QGSJET-II-04 and EPOS-LHC, for proton, He, N, and Fe primaries. The ground signals of simulated events have a factor 1.3-1.6 deficit of hadronically-produced muons relative to observed showers, depending on which high energy event generator is used, and whether the composition mix is chosen to reproduce the observed $X_{\max}$ distribution or a pure proton composition is assumed. The analysis allows for a possible overall rescaling of the energy, which is found to lie within the systematic uncertainties.

## 1 Introduction

The ground-level muonic component of ultra-high energy (UHE) air showers is sensitive to hadronic particle interactions at all stages in the air shower cascade, and to many properties of hadronic interactions such as the multiplicity, elasticity, fraction of secondary pions which are neutral, and the baryon-to-pion ratio [1]. Air shower simulations rely upon hadronic event generators (HEGs), such as QGSJET-II [2], EPOS [3], and SIBYLL [4]. The HEGs are tuned on accelerator experiments, but when applied to air showers they must be extrapolated to energies inaccessible to accelerators and to phase-space regions not well-covered by existing accelerator experiments. These extrapolations result in a large spread in the predictions of the various HEGs for the muon production in air showers [5].

The hybrid nature of the Pierre Auger Observatory, combining both fluorescence telescopes (FD) [6] and surface detector array (SD) [7], provides an ideal experimental setup for testing and constraining models of high-energy hadronic interactions. Thousands of air showers have been collected which have a reconstructed energy estimator in both the SD and FD. The measurement of the longitudinal profile (LP) constrains the shower development and thus the signal predicted for the SD, at the individual event level.

## 2 Production of Simulated Events

In the present study, we compare the observed ground signal of individual hybrid events to the ground signal of simulated showers with matching LPs.

The data we use for this study are the 411 hybrid events with $10^{18.8} < E < 10^{19.2}$ eV recorded between 1 January 2004 and 31 December 2012 and satisfying the event quality selection cuts in [8, 9]. This energy range is sufficient to have adequate statistics while being small enough that the primary cosmic ray mass composition does not evolve significantly. For each event in this data set we generate Monte Carlo (MC) simulated events with a matching LP, as follows:

• Generate a set of showers with the same geometry and energy, until 12 of them have an $X_{\max}$ value within one sigma of the real event.

• Among those 12 generated showers select, based on the $\chi^2$-fit, the 3 which best reproduce the observed longitudinal profile (LP).

• For each of those 3 showers do a full detector simulation

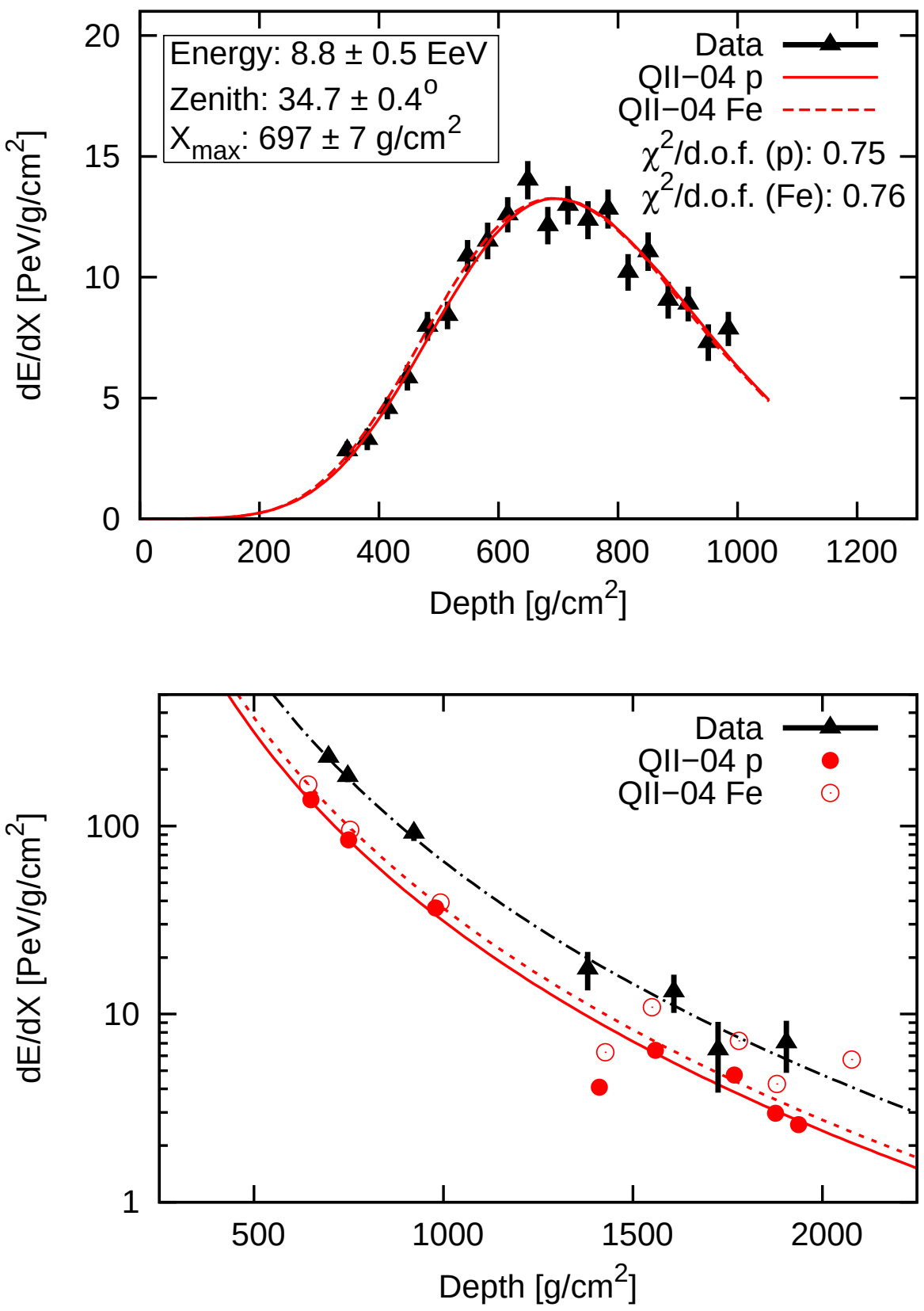

**Figure 1**: Top: The measured longitudinal profile of a typical air shower with two of its matching simulated air showers, for a proton and an iron primary, simulated using QGSJET-II-04. Bottom: The observed and simulated ground signals for the same event.

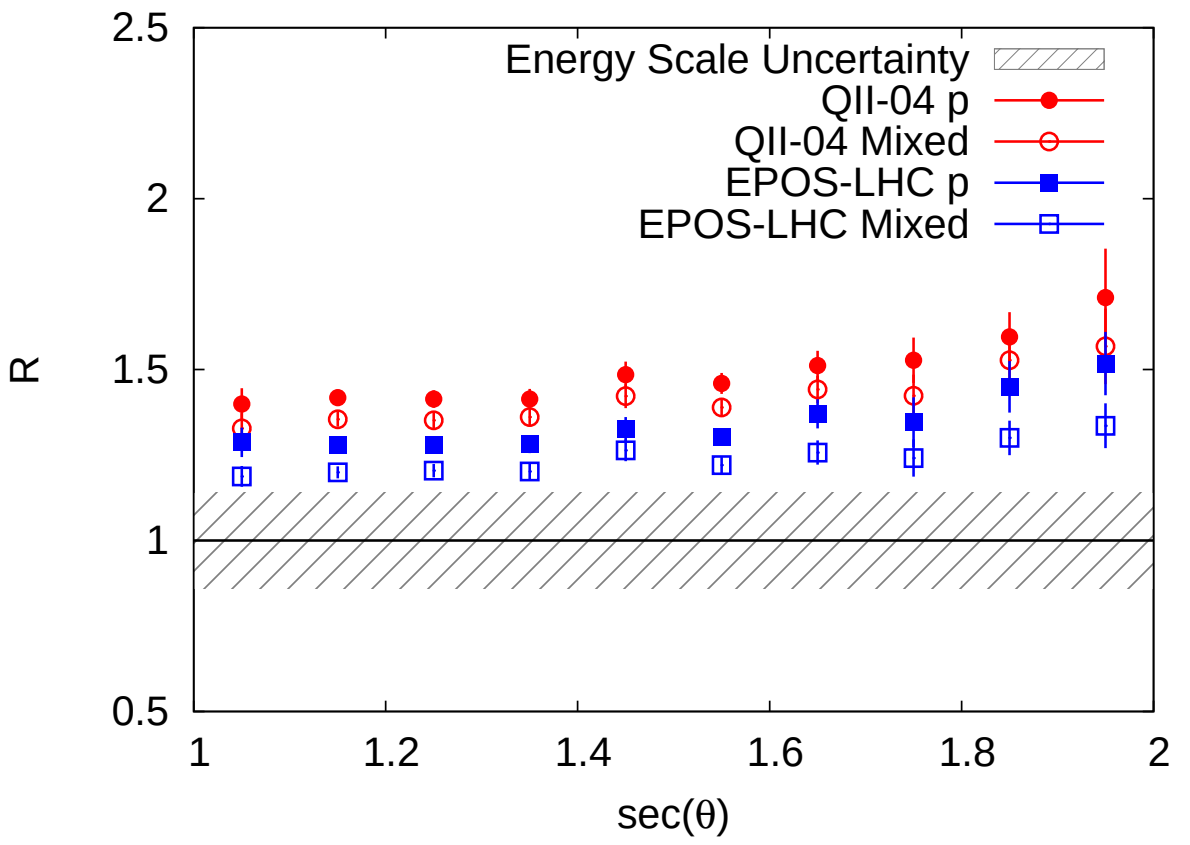

**Figure 2**: The average ratio of the $S(1000)$ of observed events to that in simulated events as a function of zenith angle for mixed or pure proton composition. The gray band represents the impact of the 14% systematic uncertainty in the FD energy scale.

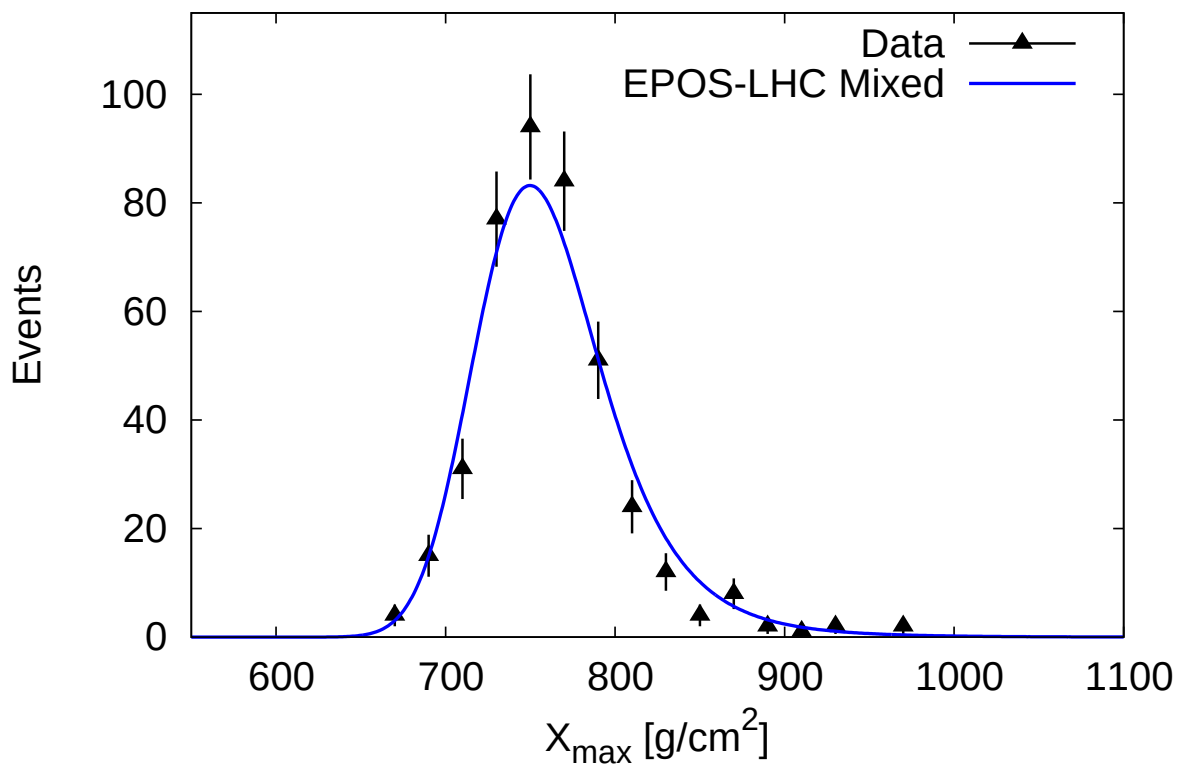

**Figure 3**: The $X_{\rm max}$ distribution of the events used for this study, with the predicted shape from the best-fit $p_j(X_{\rm max})$ functions, for EPOS-LHC.

and generate SD signals for comparison with the data.

We do this for two different HEGS (QGSJET-II-04[10] and EPOS-LHC[11]) and for four different primary cosmic ray types (proton, helium, nitrogen, and iron) for all of the events in the dataset. Note, however, that in some events the $X_{\rm max}$ value is so deep or shallow that the event cannot be reproduced with all four primaries in both HEGs.

Simulation of the detector response is performed with GEANT4 [12] within the software framework $\overline{\rm Off}\underline{\rm line}$ [13] of the Auger Observatory. The MC air shower simulations are performed using the SENECA simulation code [14], with FLUKA [15, 16] as the low-energy HEG. Having three simulated showers which match the LP is sufficient to estimate the mean ground signal for the given LP.

The LP and lateral distribution of the ground signal of a typical event are shown in Fig. 1, along with a matching proton and iron simulated event. A high quality fit to the LP is found for all events for at least one primary type, and the $\chi^2$ distribution of the selected LPs compared to the data is comparable to that found in a Gaisser-Hillas fit to the data.

Fig. 1 illustrates a general feature of the comparison between observed and simulated events: the ground signal of the simulated events is systematically smaller than the ground signal in the data events. Contributing factors to such a discrepancy in the ground signal could be a systematic energy offset, arising due to the 14% systematic uncertainty in the FD energy scale [8], or deficiencies in the HEGs. Elucidating the nature of the discrepancy is the motivation for the present study.

The estimated signal size at 1000 m, $S(1000)$, is the SD energy estimator. Fig. 2 shows the ratio of the $S(1000)$ of observed and simulated events for several HEGs, using a mixed composition that reproduces the $X_{\rm max}$ distribution (Fig. 3), and also using pure protons for comparison. The discrepancy between measured and simulated $S(1000)$ grows with zenith angle for each HEG and is larger than the uncertainty in the FD energy scale at all angles. The growth of the discrepancy with zenith angle suggests that the simulations are predicting too few muons.

## 3 Quantifying the Discrepancy

To explore the potential sources of the discrepancy, the ground signal is modified in the simulated events to fit the ground signal in the data. Two rescaling factors are introduced: $R_E$ and $R_\mu$. $R_E$ acts as a rescaling of the energy of the primary cosmic ray, which rescales the total ground signal of the event uniformly. $R_\mu$ acts as a "muonic" rescaling factor; it rescales only the contribution to the ground signal of inherently hadronic origin. For each event in the dataset, a rescaled simulated $S(1000)$ is calculated as a function of $R_E$, $R_\mu$, and primary particle type. $R_E$ and $R_\mu$ are then fit to minimize the discrepancy between the ensemble of simulated and observed $S(1000)$, for each HEG considered. The likelihood function to be maximized is $\prod_i P_i$, where the contribution of each event is

$$P_i = \sum_j p_j(X_{{\rm max},i})\,\mathcal{N}(S_{\rm resc}(R_E,R_\mu)_{i,j} - S(1000)_i, \sigma_{i,j}).$$

The index $i$ runs over each event in the data set and $j$ labels the primary type; the factor $p_j(X_{{\rm max},i})$ is the probability that the $i$th event comes from primary type $j$, given the $X_{\rm max}$ of the event. We calculate $p_j(X_{\rm max})$ using the mix of $p$, He, N and Fe which best-fits the observed $X_{\rm max}$ distribution, for each HEG. Determination of $\sigma_{i,j}$ and $S_{\rm resc}(R_E,R_\mu)_{i,j}$ are discussed below.

The first step in determining $S_{\rm resc}(R_E,R_\mu)_{i,j}$ is to attribute the ground signal of each simulated particle in the detector to either an electromagnetic (EM) or hadronic origin. To do this, the history of all muons and EM particles ($e^\pm$ and $\gamma$s) reaching ground are tracked during simulation following the description in [17]. EM particles that are produced by muons, through decay or radiative processes, and by low-energy $\pi^0$s are attributed to the muonic signal; muons that are produced through photoproduction are attributed to the electromagnetic signal. Fig. 4 shows the signal produced by each component of a 10 EeV air shower.

Because $S(1000)$ is a reconstructed property of each event, the impact of altering the muonic component or overall energy must be determined using reconstructed showers. To do this, we recalculate the detector response for each simulated shower, increasing the weight of the muonic component by the scale factors $w_\mu = 1.0$, 1.75, and 2.5, and a linear fit is performed to extract the EM and muonic components $S_{\rm EM}$ and $S_\mu$ via $S(1000)(w_\mu) \equiv S_{\rm EM} + w_\mu S_\mu$. The rescaled simulated $S(1000)$ is then

$$S_{\rm resc}(R_E,R_\mu)_{i,j} \equiv R_E\,S_{{\rm EM},i,j} + R_E^\alpha R_\mu\,S_{\mu,i,j}\,, \quad (1)$$

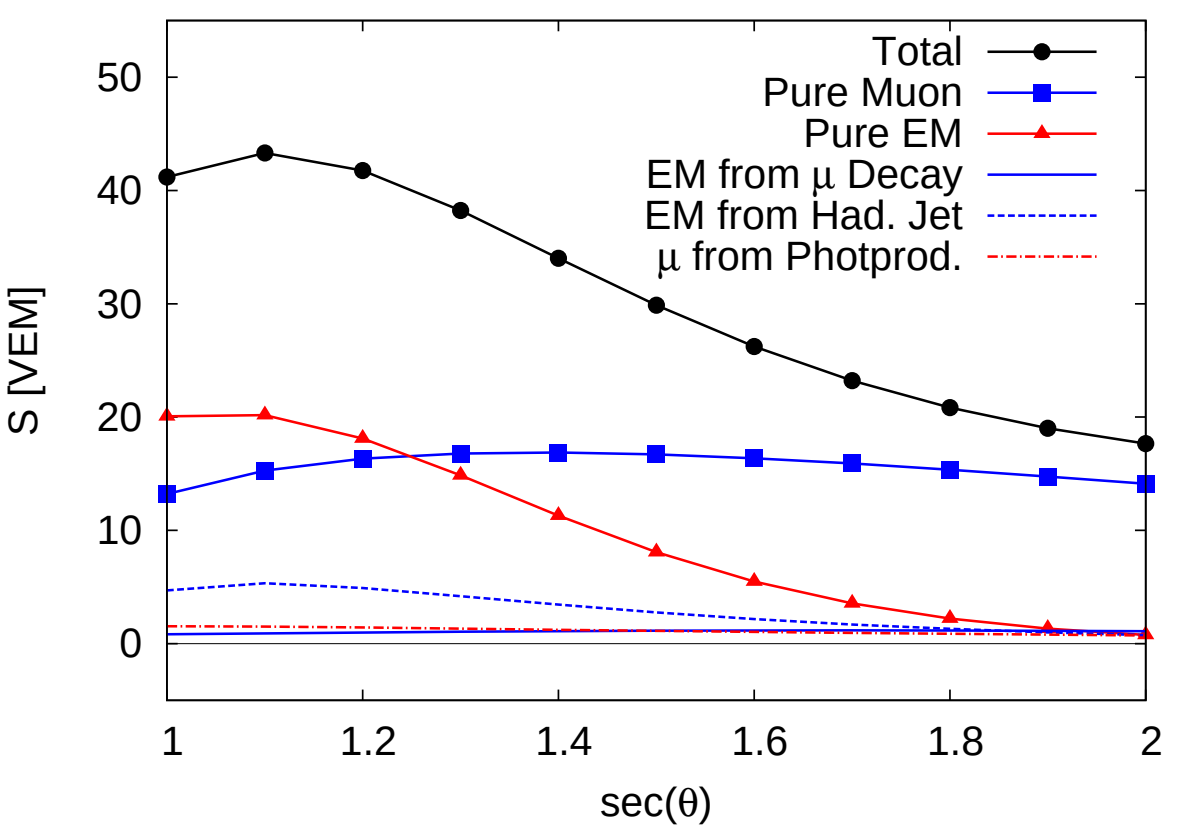

**Figure 4**: The contributions of different components to the average signal as a function of zenith angle, for stations at 1 km from the shower core, in simulated 10 EeV proton air showers illustrated for QGSJET-II-04. The signal size is measured in units of vertical equivalent muons (VEM), the calibrated unit of SD signal size [18].

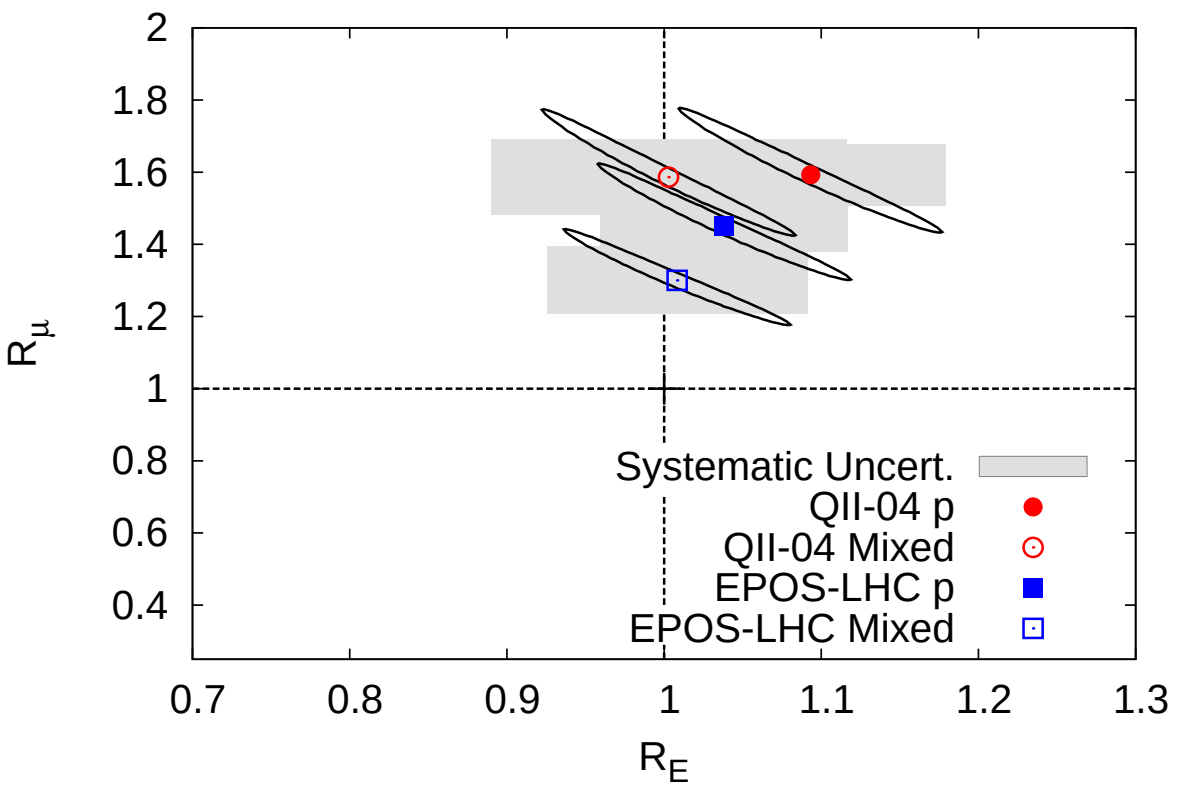

**Figure 5**: The best-fit values of $R_E$ and $R_\mu$ for QGSJET-II-04 and EPOS-LHC, for mixed and pure proton compositions. The ellipses show the one-sigma statistical uncertainties. The grey boxes show the estimated systematic uncertainties as described in the text; these will be refined in a forthcoming journal paper.

where $\alpha$ is the energy scaling of the muonic signal; it has the value 0.89 in both the EPOS and QGSJET-II simulations, independent of composition [19].

Finally, the variance of $S(1000)$ with respect to $S_{\rm resc}$ must be estimated for each event. Contributions to the variance are of two types: the intrinsic shower-to-shower variance in the ground signal for a given LP, $\sigma_{\rm shwr}$, and the variance due to limitations in reconstructing and simulating the shower, $\sigma_{\rm rec}$ and $\sigma_{\rm sim}$. The total variance for event $i$ and primary type $j$, is $\sigma^2_{i,j} = \sigma^2_{\rm rec,i} + \sigma^2_{{\rm sim},i,j} + \sigma^2_{{\rm shwr},i,j}$.

$\sigma_{\rm shwr}$ is the variance in the ground signals of showers with matching LPs. This arises due to shower-to-shower fluctuations in the shower development which result in varying amounts of energy being transferred to the EM and hadronic shower components, even for showers with fixed $X_{\rm max}$ and energy. $\sigma_{\rm shwr}$ is irreducible, as it is independent from the detector resolution and statistics of the simulated showers. It is determined by calculating the variance in the ground signals of the simulated events from their respective means, for each primary type and HEG; it is typically $\approx 16\%$ of $S_{\rm resc}$ for proton initiated showers and 5% for iron initiated showers.

$\sigma_{\rm rec}$ contains i) the uncertainty in the reconstruction of $S(1000)$, ii) the uncertainty in $S_{\rm resc}$ due to the uncertainty in the calorimetric energy measurement, and iii) the uncertainty in $S_{\rm resc}$ due to the uncertainty in $X_{\rm max}$; $\sigma_{\rm rec}$ is typically 12% of $S_{\rm resc}$. $\sigma_{\rm sim}$ contains the uncertainty in $S_{\rm resc}$ due to the uncertainty in $S_\mu$ and $S_{\rm EM}$ from the $S(1000) - w_\mu$ fit and to the limited statistics from having only three simulated events; $\sigma_{\rm sim}$ is typically 10% of $S_{\rm resc}$ for proton initiated showers and 4% for iron initated showers.

The resultant model of $\sigma_{i,j}$ is checked using the 59 events, of the 411, which are observed with two FD eyes whose individual reconstructions pass all required selection cuts for this analysis. The variance in the $S_{resc}$ of each eye is compared to the model for the ensemble of events. All the contributions to $\sigma_{i,j}$ are present in this comparison except for $\sigma_{\rm shwr}$ and the uncertainty in the reconstructed $S(1000)$. The variance of $S_{\rm resc}$ in multi-eye events is well represented by the estimated uncertainties using the model. In addition, the maximum-likelihood fit is also performed where $\sigma_{\rm shwr}$ is a free parameter rather than taken from the models; no significant difference is found between the value of $\sigma_{\rm shwr}$ from the models, and that recovered when it is a fit parameter.

The results of the fit for $R_E$ and $R_\mu$ are shown in Fig. 5 and Table 1 for each HEG. The ellipses show the one-sigma statistical uncertainty region in the $R_E - R_\mu$ plane. The systematic uncertainties in the event reconstruction of $X_{\rm max}$, $E_{\rm FD}$ and $S(1000)$ are propagated through the analysis by shifting the reconstructed central values by their one-sigma systematic uncertainties; this is shown by the grey rectangles.[1] As a benchmark, the results for a purely protonic composition are given as well[2].

The signal deficit is smallest (the best-fit $R_\mu$ is the closest to unity) in the mixed composition case with EPOS. As shown in Fig. 6, the primary difference between the ground signals predicted by the two models is the size of the muonic signal, which is $\approx$15 (20)% larger for EPOS-LHC than QGSJET-II-04, in the pure proton (mixed composition) cases respectively. EPOS benefits more than QGSJET-II when using a mixed composition because the mean primary mass determined from the $X_{\rm max}$ data is larger in EPOS than in QGSJET-II [20].

## 4 Discussion and Summary

In this work, we have used hybrid showers of the Pierre Auger Observatory to quantify the disparity between state-of-the-art hadronic interaction modeling and observed atmospheric air showers of UHECRs. The most important advance with respect to earlier versions of this analysis[21], in addition to now having a much larger hybrid dataset and improved shower reconstruction, is the extension of the anal-

1. The values of $\sigma_{\rm sim}$, $\sigma_{\rm rec}$ and $\sigma_{\rm shwr}$ and the treatment of systematic errors used here will be refined with higher statistics Monte Carlo simulations and using the updated Auger energy and $X_{\rm max}$ uncertainties, for the journal version of this analysis.

2. Respecting the observed $X_{\rm max}$ distribution is essential for evaluating shower modeling discrepancies, since atmospheric attenuation depends on the distance-to-ground. This is automatic in the present analysis, but the simulated LPs – which are selected to match hybrid events – is a biased subset of all simulated events for a pure proton composition since with these HEGs pure proton does not give the observed $X_{\rm max}$ distribution.

**Table 1**: $R_E$ and $R_\mu$ with statistical and systematic uncertainties, for QGSJET-II-04 and EPOS-LHC.

| Model | $R_E$ | $R_\mu$ |
|---|---|---|
| QII-04 *p* | 1.09±0.08±0.09 | 1.59±0.17±0.09 |
| QII-04 Mixed | 1.00±0.08±0.11 | 1.59±0.18±0.11 |
| EPOS *p* | 1.04±0.08±0.08 | 1.45±0.16±0.08 |
| EPOS Mixed | 1.01±0.07±0.08 | 1.30±0.13±0.09 |

ysis method to treat a mixed composition that reproduces the $X_{\max}$ distribution of the data. The previous analysis was restricted to a pure composition, which is inconsistent with the $X_{\max}$ distribution predicted by these same hadronic interaction models. The pure-proton ansatz exaggerates the problem and the pure-Fe ansatz underestimates it.

To give the most basic characterization of the model discrepancies, our analysis introduces only a simple, overall rescaling of the hadronic shower relative to the EM shower, plus a possible overall energy recalibration (which proves not to be needed). In this context, the contributions to the muonic signal due to the hadronic and EM components of the showers can be distinguished, and our $R_\mu$ is the rescaling of the hadronic shower relative to the EM shower. As such, it is not directly comparable to direct muon number determinations provided by Pierre Auger Observatory, obtained from the FADC traces of the surface detector stations and from inclined showers (for which the ground signal is entirely muonic) [23, 24, 25, 26]. The direct methods report a purely experimental observable – the ground signal in muons, for showers in some zenith angle range – whereas $R_\mu$ characterizes the hadronic component of the showers. Nonetheless, all methods indicate that present shower models do not correctly describe the muonic ground signal; the general consistency of the methods is not surprising, since hadronic production is the prime source of muons.

Within the statistics currently available, there is no evidence of a larger event-to-event variance in the ground signal for fixed $X_{\max}$ than predicted by the current models. This means that the muon shortfall cannot be attributed to some exotic phenomenon which produces a very large muon signal in only a fraction of events, such as micro-black hole production.

In summary, the observed hadronic signal in 10 EeV air showers ($E_{\rm CM} = 137$ TeV) is a factor 1.3 to 1.6 larger than predicted using the leading hadronic interaction models tuned to fit LHC and lower energy accelerator data. Relative to the preliminary version of this analysis presented at ICRC2011[21], the central value of $R_\mu$ is closer to one and, with mixed composition, neither HEG calls for an energy rescaling. However the discrepancy between models and observation remains serious because i) the HEGs are now tuned to the LHC, ii) the analysis now allows for a mixed primary composition and has a more sophisticated treatment of fluctuations, and iii) the Auger event reconstruction and energy calibration have been refined[8]. With more than two times as many events, the discrepancy is twice the estimated systematic and statistical uncertainties combined in quadrature, even for the best case of EPOS-LHC with mixed composition.

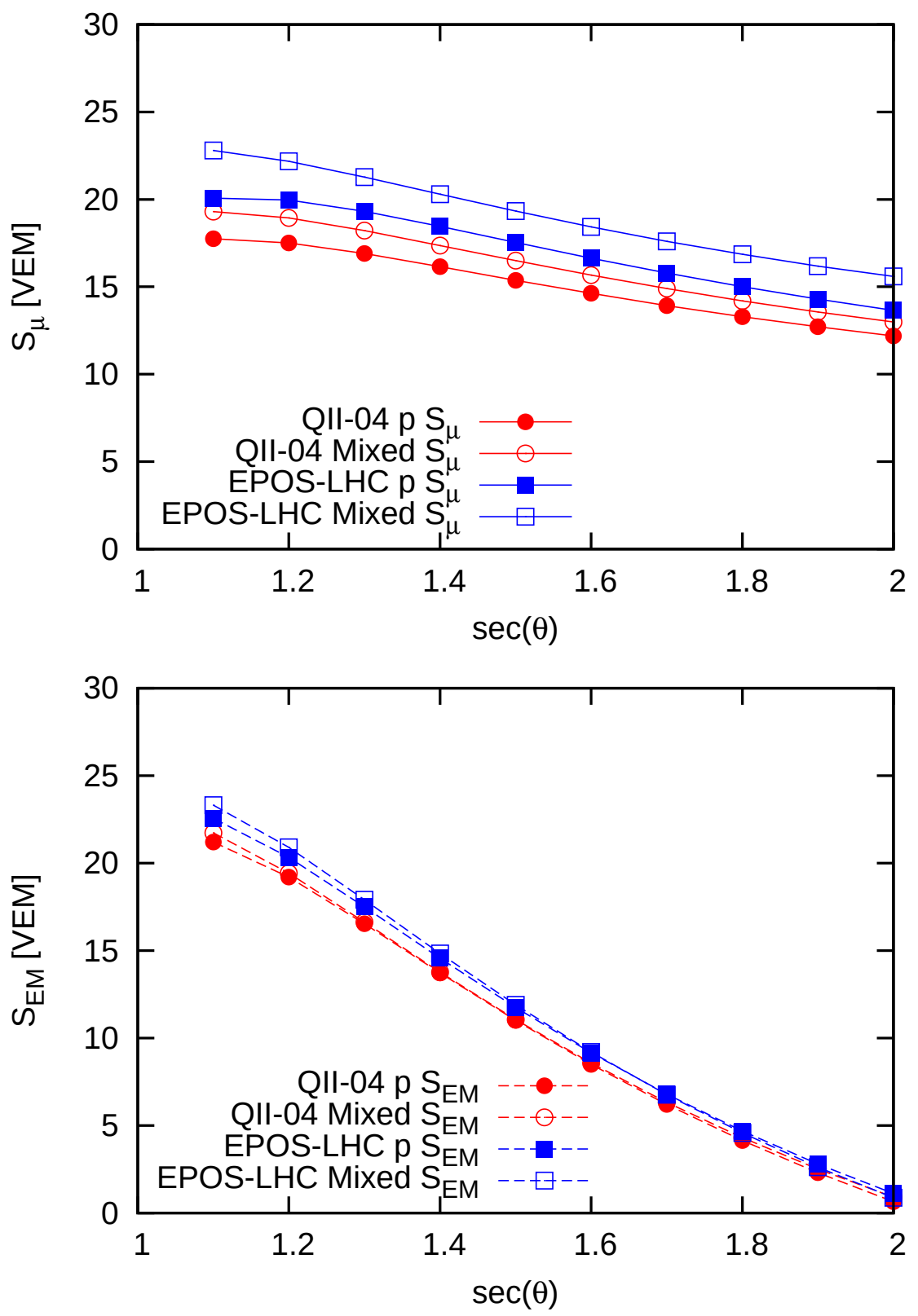

**Figure 6**: Muonic (top) and EM signals (below) at 1000 meters as a function of zenith angle, in the models.

# Measurement of the first harmonic modulation in the right ascension distribution of cosmic rays detected at the Pierre Auger Observatory: towards the detection of dipolar anisotropies over a wide energy range

IVÁN SIDELNIK[1] FOR THE PIERRE AUGER COLLABORATION[2]

[1] *Centro Atómico Bariloche and Instituto Balseiro (CNEA-UNCuyo-CONICET) and U.N. Río Negro, San Carlos de Bariloche, Argentina*
[2] *Full author list: http://www.auger.org/archive/authors_2013_05.html*

*auger_spokespersons@fnal.gov*

**Abstract:** First harmonic analyses of the right ascension distribution of cosmic rays detected at the Pierre Auger Observatory are reported. We here update the upper limits on the dipole component in the equatorial plane and extend the previous results to lower energies by using data recorded by the infill surface detector array. On the other hand, a possible consistency in ordered energy intervals of the phase was observed and reported, consistency that may be indicative of anisotropies whose amplitudes are too small to stand out above the background noise induced by the finite statistics accumulated so far. Based on this posterior observation, a prescribed single shot test was designed on June 25 2011 to establish at 99% CL whether this consistency is real or not. Since the effect has been observed over a wide energy range, the test makes use of data of both the infill and the regular surface detector arrays. At about mid-term, the status of this prescription is reported.

## 1 Introduction

Large scale anisotropy studies are of major importance for cosmic ray physics, because together with the analysis of the spectrum and mass composition can help to understand the nature and origin of these particles. The measurement of the anisotropies at different energies, or the bounds on them, are relevant to constrain different models for the distribution of the sources and for the propagation of cosmic rays. For instance, the transition from a galactic to an extragalactic origin of the cosmic rays should induce a significant change in their large scale angular distribution and there are different theoretical models that locate this transition at different energies.

The Pierre Auger Observatory [1], located in Malargüe, Argentina, was originally designed to study the cosmic rays above $10^{18}$ eV. The Observatory combines two different techniques to detect the extensive air showers that are produced by the interactions of cosmic rays with the atmosphere. The surface detector array (SD) measures the lateral distribution of secondary particles at ground level and the fluorescence detector (FD) measures the longitudinal development of the air shower. The SD consists of an arrangement of 1600 water Cherenkov detectors distributed over an area of 3000 km$^2$ forming a triangular grid, with a detector separation of 1500 m, and has an operation duty cycle of nearly 100%. In order to enhance the capabilities of the Observatory by lowering its energy threshold a 23.5 km$^2$ area of the regular array has been deployed with detectors spaced by 750 m. This *infill* array [2] allows us to detect cosmic rays with energies down to $10^{16}$ eV and has full efficiency above $3\times10^{17}$ eV.

We present in this work a study of the large scale distribution of arrival directions of cosmic rays based on the first harmonic analysis in right ascension with data from the surface detector of the Pierre Auger Observatory. An update of the search above $10^{18}$ eV as a function of both right ascension and declination is presented in [3]. Here we use for the first time the full energy range above $10^{16}$ eV. These energies are accessible thanks to the joint data acquired by both the infill array with 750 m spacing and the regular array with 1.5 km spacing. We use data up to the end of 2012, using events with zenith angles $\theta < 60^\circ$ for the 1.5 km spacing array, and $\theta < 55^\circ$ for the infill array.

## 2 First harmonic analysis in ordered energy intervals

### 2.1 Analysis method

Since the first harmonic modulations are quite small, it is important to account for possible spurious modulations of experimental or atmospheric origin or, alternatively, use methods which are not sensitive to these effects.

In particular, spurious variations due to the evolution of the array size with time or dead periods of each surface detector can be accounted by using the number of unitary cells $n_{cell}(t)$ recorded every second by the trigger system of the Observatory. The fiducial cut applied to the selected events [4] requires that the detector with the highest signal be surrounded by six active detectors, and hence the unitary cells that define the instantaneous exposure of the array to this type of events are defined as an active detector surrounded by six neighbouring active detectors. The same quality cut is used to select events recorded with the infill array. For any periodicity $T$, the total number of unitary cells $N_{cell}(t)$ summed over all periods, and its associated relative variations $\Delta N_{cell}$, are obtained using:

$$N_{cell}(t) = \sum_j n_{cell}(t + jT), \quad \Delta N_{cell}(t) = \frac{N_{cell}}{\langle N_{cell} \rangle} \qquad (1)$$

with $\langle N_{cell} \rangle = 1/T \int_0^T dt N_{cell}$. To perform a first harmonic analysis that accounts for the non-uniform exposure in different parts of the sky we introduce weights in the classical Rayleigh analysis. Each event is weighted with the inverse of the integrated number of unitary cells at the local sidereal time of the event. The Fourier coefficients $a$ and $b$ of the modified Rayleigh analysis are:

$$a = \frac{2}{\mathcal{N}} \sum_{i=1}^{N} w_i \cos(\alpha_i), \quad b = \frac{2}{\mathcal{N}} \sum_{i=1}^{N} w_i \sin(\alpha_i) \qquad (2)$$

where $w_i \equiv [\Delta N_{cell}(\alpha_i^0)]^{-1}$ with $\alpha_i^0$ the local sidereal time of the event with right ascension $\alpha_i$. We express $\alpha_i^0$ in radians and chose it so that it is always equal to the right ascension of the zenith at the center of the array. The sum runs over the number $N$ of events in the energy range considered, and $\mathcal{N} = \sum_{i=1}^{N} w_i$. The amplitude $r$ and phase $\varphi$ are calculated via $r = \sqrt{a^2+b^2}$ and $\varphi = \arctan(b/a)$. They follow a Rayleigh and a uniform distributions, respectively, in the case of underlying isotropy.

Another source of systematic effects is induced by weather variations, leading both to daily and seasonal modulations. To eliminate these variations the conversion of the shower size into energy is performed by relating the observed shower size to the one that would have been measured at reference atmospheric conditions. Above 1 EeV, this procedure is sufficient to control the size of the side-band amplitude to the level of $\simeq 10^{-3}$ [5]. Below 1 EeV weather effects have a significant impact also on the detection efficiency for the regular array with 1.5 km spacing, and hence spurious variations of the counting rates are amplified. Therefore, we adopt in this case the differential $East - West$ method [6]. This takes into account the difference between the event counting rate measured from the East sector, $I_E(\alpha^0)$, and the West sector $I_W(\alpha^0)$. Since the instantaneous exposure for Eastward and Westward events is the same, this difference allows us to remove, at first order in the direction, effects of experimental or atmospheric origin without applying any correction, although at the price of reducing the sensitivity to the first harmonic modulation. For the case of the infill, we will use only the East-West method since we are in this case particularly interested in the very low energies below full efficiency (while above $3 \times 10^{17}$ eV the most sensitive results are obtained from the larger statistics accumulated by with the regular array with 1.5 km spacing). The amplitude $r$ and phase $\varphi$ can be calculated from the arrival times of $N$ events using the standard first harmonic analysis slightly modified to account for the subtraction of the Western sector to the Eastern one. The Fourier coefficients $a_{EW}$ and $b_{EW}$ are defined by:

$$a_{EW} = \frac{2}{N} \sum_{i=1}^{N} \cos(\alpha_i^0 + \xi_i), \quad b_{EW} = \frac{2}{N} \sum_{i=1}^{N} \sin(\alpha_i^0 + \xi_i) \quad (3)$$

where $\xi_i$=0 if the event comes from the East and $\xi_i = \pi$ if it comes from the West (in this way the events from the West are effectively subtracted). The amplitude $r$ of the right ascension modulation determined with the Rayleigh formalism is related to $r_{EW} = \sqrt{a_{EW}^2 + b_{EW}^2}$ through the relation [5] $r = \frac{\pi\langle\cos(\delta)\rangle}{2\langle\sin(\theta)\rangle} r_{EW}$. Note that the phase determined with the East-West method as $\varphi_{EW} = \arctan(b_{EW}/a_{EW})$ is related to the phase determined with the Rayleigh formalism by $\varphi = \varphi_{EW} + \pi/2$.

## 2.2 Analysis at the sidereal frequency

To perform first harmonic analyses as a function of energy, the choice of the size of the energy bins is important to avoid the dilution of a genuine signal with the background noise. The size of the energy bins for the analysis with the array with 1.5 km spacing was chosen to be $\Delta\log_{10}(E) = 0.3$ below 8 EeV (and one single bin for all energies above 8 EeV was used). This is larger than the energy resolution. For the analysis with the infill array a bin size of $\Delta\log_{10}(E) = 0.6$ was used. Data from the larger array was used for energies above 0.25 EeV, and the infill array was used to complement this measurements down to 0.01 EeV.

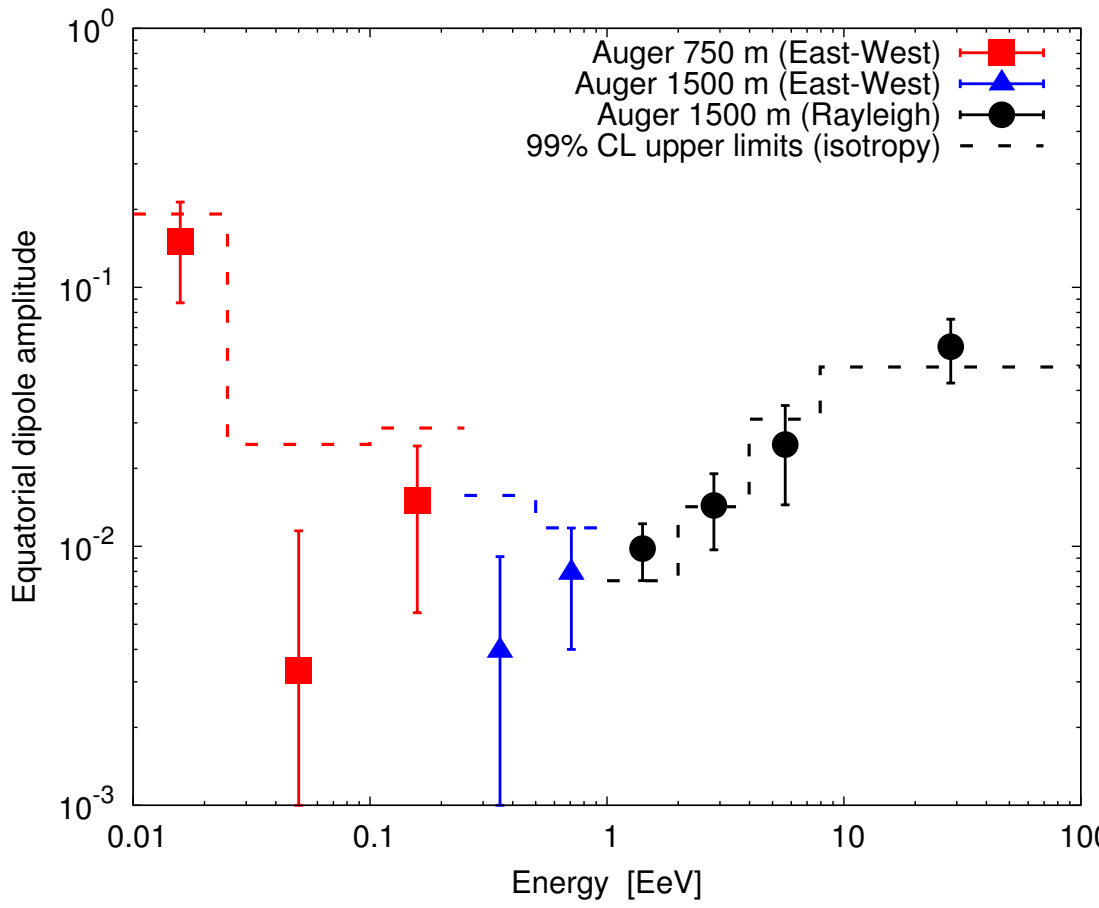

**Figure 1**: Equatorial dipole amplitude as a function of energy. The results of the modified Rayleigh analysis are shown with black circles and blue triangles corresponds to the analysis with East-West method, in both cases using data from the array with 1.5 km spacing. Red squares correspond to data from the infill array using the East-West method. The dashed lines are the 99% CL upper values of the amplitude that could result from fluctuations of an isotropic distribution.

The Rayleigh amplitude $r$ measured by any observatory can be used to reveal (or infer) anisotropies projected on the Earth equatorial plane. In the case of an underlying pure dipole, the relationship between $r$ and the projection of the dipole on the Earth equatorial plane, $d_\perp$, depends on the latitude of the observatory and on the range of zenith angles considered : $d_\perp \simeq r/\langle\cos\delta\rangle$ [5]. $d_\perp$ is the physical quantity of interest to compare the results of different experiments and the pure dipole predictions. For the regular array one has that $\langle\cos\delta\rangle \simeq 0.78$ while for the infill this number results $\langle\cos\delta\rangle \simeq 0.79$. The obtained amplitude $d_\perp$ is shown in Fig. 1 and in Table 1, the dashed line in the plot represents the upper values of the amplitude which may arise from fluctuations in an isotropic distribution at 99% CL, denoted by $d_{\perp 99\%}^{iso}$. Table 1 shows also the number of events, $N$, the phase with its associated uncertainty, the probability $P$ that an amplitude larger or equal than that observed in the data arises by chance from an isotropic distribution ($P(>r) = \exp(-r^2\mathcal{N}/4)$).

Note that in the energy ranges 1-2 and 2-4 EeV the measured amplitudes of $d_\perp$ of $(1.0 \pm 0.2)$% and $(1.4 \pm 0.5)$% have a probability to arise by chance from an isotropic distribution of about 0.03% and 0.9%, while above 8 EeV the measured amplitude of $(5.9 \pm 1.6)$% has chance probabil-ity of only 0.1%. Since several energy bins were searched, these numbers do not represent absolute probabilities. They constitute interesting hints for large scale anisotropies that will be important to further scrutinise with enlarged statistics.

| | ΔE[EeV] | N | $d_\perp \pm \sigma_{d_\perp}$ [%] | $\varphi \pm \Delta_\varphi$ [°] | P($> d_\perp$) [%] | $d^{iso}_{\perp 99\%}$ [%] | $d^{ul}_\perp$ [%] |
|---|---|---|---|---|---|---|---|
| Infill | 0.01 - 0.025 | 11819 | 15 ±6.3 | 334± 25 | 5.9 | 19 | 28.6 |
| East-West | 0.025 - 0.1 | 428028 | 0.3±0.8 | 122±180 | 92 | 2.4 | 2.2 |
| Method | 0.1 - 0.25 | 223342 | 1.4±0.9 | 277± 39 | 28 | 2.9 | 3.5 |
| East-West | 0.25 - 0.5 | 720224 | 0.4±0.5 | 280±180 | 75 | 1.6 | 1.5 |
| Method | 0.5 - 1 | 1081810 | 0.8±0.4 | 258± 30 | 13 | 1.2 | 1.6 |
| | 1 - 2 | 557829 | 1.0±0.2 | 335±14 | 0.03 | 0.7 | 1.5 |
| Modified | 2 - 4 | 148790 | 1.4±0.5 | 8 ±19 | 0.9 | 1.4 | 2.5 |
| Rayleigh | 4 - 8 | 31270 | 2.5±1.0 | 63 ±25 | 5.5 | 3.1 | 4.8 |
| | > 8 | 12292 | 5.9±1.6 | 86 ±16 | 0.1 | 4.9 | 9.4 |

**Table 1**: Results of first harmonic analyses in different energy intervals. Data from the regular SD were used above 0.25 EeV, with the East-West method up to 1 EeV and the modified Rayleigh method above 1 EeV. Data from the infill array was used for energies between 0.01 and 0.25 EeV with the East-West method.

### 2.3 Upper limits on the dipole

The upper limits on $d_\perp$ at 99%CL are given in Table 1 and shown in Fig. 2, together with previous results from EAS-TOP [7], ICE-CUBE [8] KASCADE [9], KASCADE-Grande [10] and AGASA [11], and with some predictions for the anisotropies arising from models of both galactic and extragalactic cosmic ray origin.

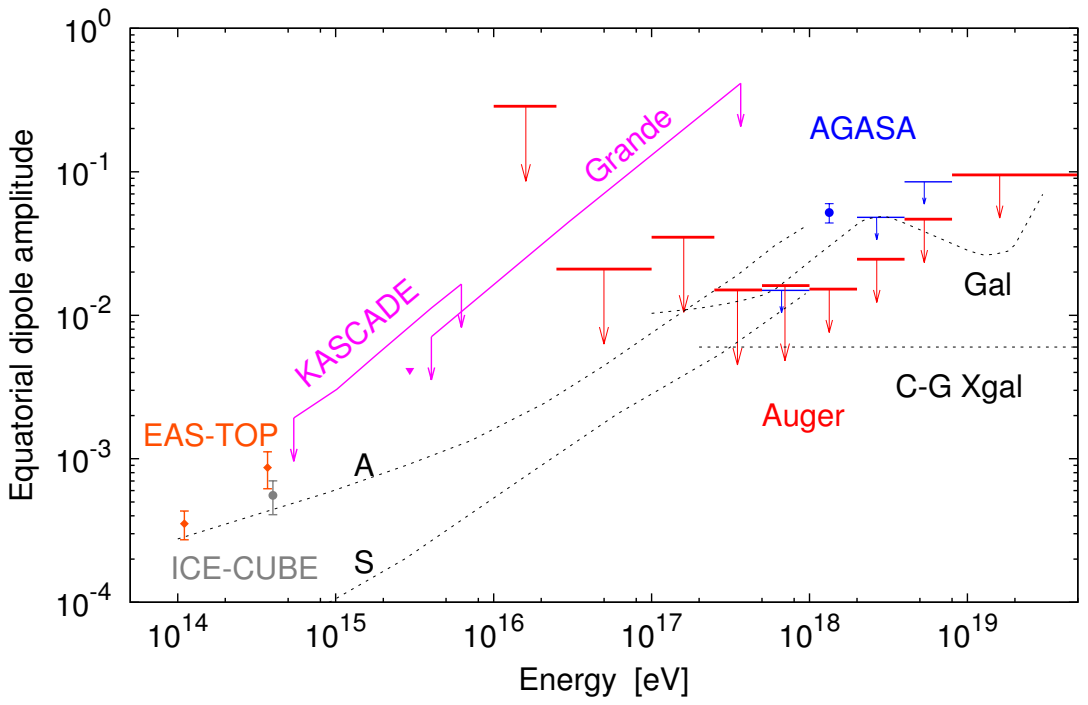

**Figure 2**: Upper limit at 99%CL for the equatorial dipole amplitude as a function of energy. In red are the limits obtained in this work over the full energy range of the Auger Observatory. Results from AGASA are shown in blue, from KASCADE and KASCADE-Grande in magenta, EAS-TOP in orange and ICE-CUBE in grey. Predictions from different models are displayed, labeled as A, S, Gal and C-G Xgal (see text).

The prediction labeled A and S correspond to a model in wich cosmic rays at 1 EeV are predominantly of galactic origin, and their escape from the galaxy by diffusion and drift motion causes the anisotropies. A and S stand for two different galactic magnetic field symmetries (antisymmetric and symmetric)[12]. In the model labeled Gal [13] a purely galactic origin is assumed for cosmic rays up to the highest energies, and the anisotropy is caused by purely diffusive motion due to the turbulent component of the magnetic field. Some of these amplitudes are challenged by our current bounds. The prediction labeled C-G Xgal [14] is the expectation from the Compton-Getting effect for extragalactic cosmic rays due to the motion of our galaxy with respect to the frame of extragalactic isotropy, assumed to be determined by the cosmic microwave background.

The bounds reported here already exclude the particular model with an antisymmetric halo magnetic field (A) above energies of 0.25 EeV and the *Gal* model at few EeV energies, and are starting to become sensitive to the predictions of the model with a symmetric field.

## 3 Phase of the first harmonic and prescription

In previous publications of first harmonic analyses in right ascension [5, 15], the Pierre Auger Collaboration reported the intriguing possibility of a smooth transition from a common phase of $\alpha \simeq 270°$ in the first two bins below 1 EeV to a phase $\alpha \simeq 100°$ above 5 EeV. The phase at lower energies is compatible with the right ascension of the Galactic Center $\alpha_{GC} \simeq 268.4°$. It was pointed out that this consistency of phases in adjacent energy intervals is expected with a smaller number of events than the detection of amplitudes standing out significantly above the background noise in the case of a real underlying anisotropy.

This behaviour motivated us to design a prescription with the intention of establishing at 99% CL whether this consistency in phases in adjacent energy intervals is real. Taking advantage of the wide energy range that the Pierre Auger Observatory is capable to scan thanks to the infill array, the test makes use of all data above $10^{16}$ eV. Thus, once an exposure of 21,000 km$^2$ sr yr is accumulated by the regular SD array from June 25 2011 on, and applying the same first harmonic analyses described in [5] and performed here [1], a positive anisotropy signal will be claimed within a global threshold of 1% if any, or both, of the following tests succeed:

- Using the infill data, an alignment of phases around the value $\varphi = 263°$ is detected by a likelihood ratio test with a chance probability less than 0.5%, assuming an amplitude signal of 0.5% over the whole energy range analysed.
- Using the regular SD data, an alignment of phases around the curve defined by eq. 4 is detected by the likelihood ratio test with a chance probability less than 0.5%, assuming an amplitude signal comparable to the current mean noise in each energy interval (see Tab. 2).

1. Though a change in the binning for the infill has been made to $\Delta\log_{10}(E) = 0.3$ and a single bin between 17.6< $\log_{10}$(E/EeV)<18.3 because of the low statistics.

$$\varphi(E) = \varphi_0 + \varphi_E \arctan\left(\frac{\log_{10}(E/EeV) - \mu}{\sigma}\right) \quad (4)$$

To report the midterm status of the prescription, the phase of the first harmonic is shown in Fig. 3. The top panel shows the phase derived with data from January 1 2004 to December 31 2010 for the larger array, that corresponds to the analysis in [5] and from September 12 2007 to April 11 2011 for the infill. The bottom panel is derived with data since June 25 2011 up to December 31, 2012. At this stage, the values as derived from the analysis applied to the infill array are still affected by large uncertainties. On the other hand, the overall behavior of the points as derived from the analysis applied to the regular array shows good agreement with equation 4, using the same parameters as the ones derived with data prior to 2011. The final result of the prescription is expected for 2015, once the required exposure is reached.

| ΔE[EeV] | mean noise |
|---|---|
| 0.25 - 0.5 | $5 \times 10^{-3}$ |
| 0.5 - 1 | $5 \times 10^{-3}$ |
| 1 - 2 | $3.5 \times 10^{-3}$ |
| 2 - 4 | $6.8 \times 10^{-3}$ |
| 4 - 8 | $1.4 \times 10^{-2}$ |
| $> 8$ | $2.0 \times 10^{-2}$ |

**Table 2**: Mean noise in each energy interval considered in the analysis of the regular array. The analysis performed in the two first energy bins uses the E-W method, which explains why the mean noise is about two times larger than $\sqrt{\pi/N}$.

## 4 Discussion and conclusions

We have searched for large scale patterns in the arrival directions of events recorded at the Pierre Auger Observatory. No statistically significant deviation from isotropy is revealed within the systematic uncertainties. The probabilities for the dipole amplitudes that are measured to arise by chance from an isotropic flux are of about 0.03% in the energy range from 1-2 EeV, 0.9% for 2-4 EeV and 0.1% above 8 EeV.

These are interesting hints for large scale anisotropies that will be important to further scrutinise with independent data. In addition, the intriguing possibility of a smooth transition from a common phase compatible with the right ascension of the Galactic Center at energies below 1 EeV to a phase around 100° above 5 EeV will be specifically tested through a prescribed test.

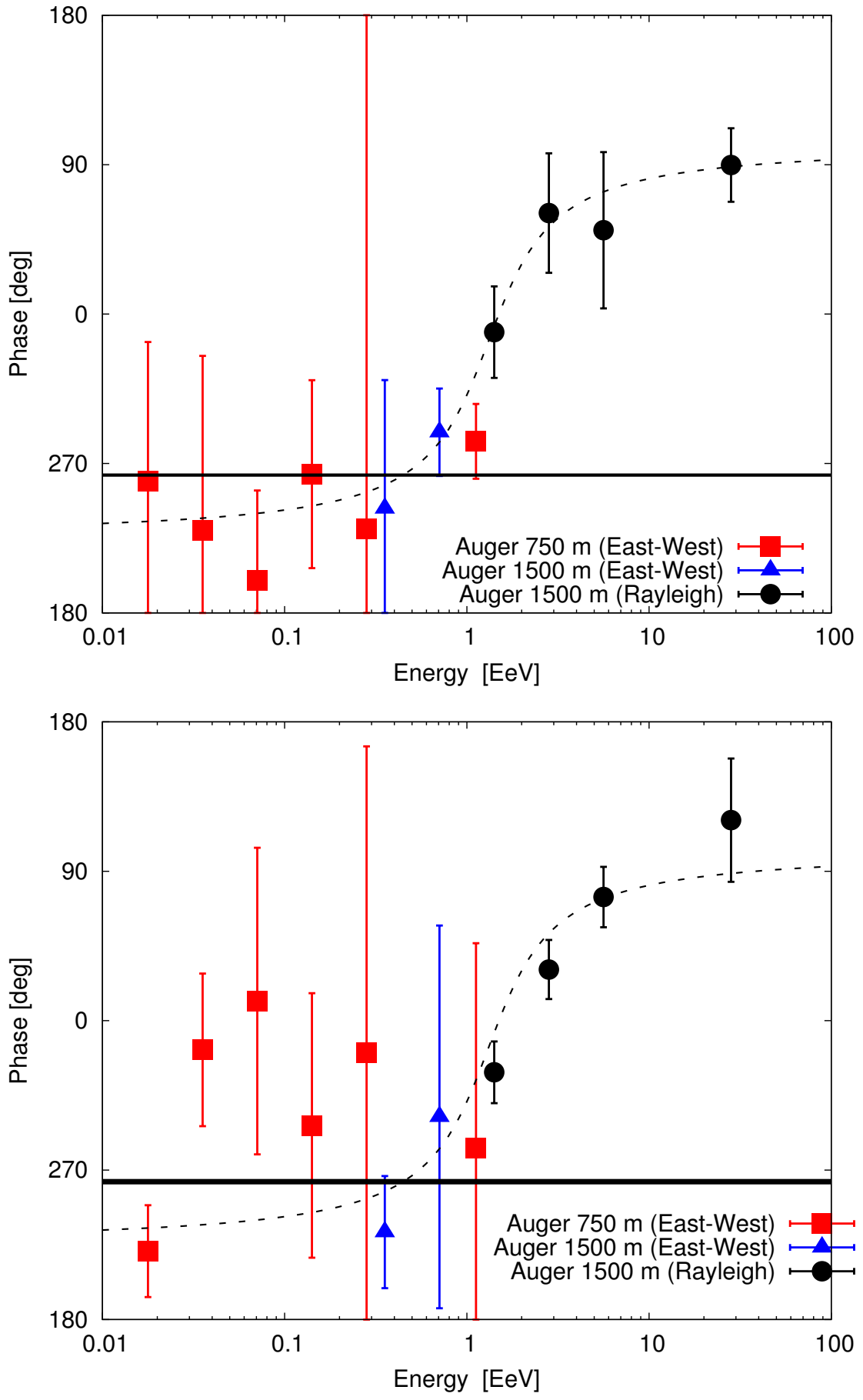

**Figure 3**: Phase of the first harmonic as a function of energy. The top panel shows the phase calculated using data from January 1 2004 to December 31 2010 for the larger array, and from September 12 2007 to April 11 2011 for the infill. The bottom panel shows the phase derived with data since June 25 2011 up to December 31 2012. The continuous line shown in both plots corresponds to the value $\varphi$ =263°, that comes from a fit to the phase measured by the infill of the first period, and the dashed line is the fit performed in [5], defined by eq. 4.

# Constraints on the origin of cosmic rays from large scale anisotropy searches in data of the Pierre Auger Observatory

ROGERIO M. DE ALMEIDA[1] FOR THE PIERRE AUGER COLLABORATION[2].

[1] *Universidade Federal Fluminense, EEIMVR, Volta Redonda, RJ, Brazil*
[2] *Full author list: http://www.auger.org/archive/authors_2013_05.html*

*auger_spokespersons@fnal.gov*

**Abstract:** We update a search for large scale anisotropies in the distribution of arrival directions of cosmic rays detected above $10^{18}$ eV at the Pierre Auger Observatory as a function of both the right ascension and the declination. Within the systematic uncertainties, no significant deviation from isotropy is revealed. Upper limits on dipole and quadrupole amplitudes are updated under the hypothesis that any cosmic ray anisotropy is dominated by such moments in this energy range. These upper limits provide constraints on the production of cosmic rays above $10^{18}$ eV, since they allow us to challenge an origin from stationary galactic sources densely distributed in the galactic disk and emitting predominantly light particles in all directions.

## 1 Introduction

The large scale distribution of arrival directions of Ultra-High Energy Cosmic Rays (UHECRs) as a function of the energy is a key observable to provide further understanding of their origin. As a natural signature of the escape of cosmic rays from the Galaxy [1, 2, 3], large scale anisotropies could be detected at energies below the ankle, a hardening of the energy spectrum located at $\simeq 4$ EeV. On the other hand, if UHECRs above 1 EeV have already a predominant extragalactic origin [4, 5, 6, 7], their angular distribution is expected to be isotropic to a high level. Thus, the study of large scale anisotropies at EeV energies would help in establishing whether the origin of UHECRs is galactic or extragalactic in this energy range.

A thorough search for large scale anisotropies in the distribution of arrival directions of cosmic rays detected above $10^{18}$ eV at the Pierre Auger Observatory was performed for several energy ranges in terms of dipoles and quadrupoles as a function of both the declination and the right ascension with no significant deviation from isotropy [8, 9]. Assuming that the eventual anisotropic component of the angular distribution of cosmic rays is dominated by dipole and quadrupole moments in this energy range, upper limits on their amplitudes were derived, challenging an origin of cosmic rays above $10^{18}$ eV from stationary galactic sources densely distributed in the galactic disk and emitting predominantly light particles in all directions. In this paper, we update this analysis. In section 2, we describe the data set and the procedure performed to control the exposure of the experiment below a 1% level while the results and the method used to derived them are presented in section 3.

## 2 Data set and control of the counting rate

The data set analyzed consists of 679,873 events recorded by the Surface Detector (SD) array of the Pierre Auger Observatory from 1 January 2004 to 31 December 2012, with zenith angles less than 55° and energies above 1 EeV. To ensure good reconstruction, an event is accepted only if all six nearest neighbours of the water-Cherenkov detector with the highest signal were operational at the time of the event [10]. Based on this fiducial cut, any active water-Cherenkov detector with six active neighbours defines an active *elemental cell*. In these conditions, and above the energy at which the detection efficiency saturates, 3 EeV [10], the total exposure of the SD array is 28,130 km$^2$ yr sr.

### 2.1 Influence of atmospheric conditions and geomagnetic field on shower size

Due to the steepness of the energy spectrum, any mild bias in the estimate of the shower energy with time or zenith angle can lead to significant distortions of the event counting rate above a given energy. It is thus critical to control the energy estimate in searching for anisotropies. The energy of each event is determined using the shower size at a reference distance of 1000 m, $S(1000)$. The geomagnetic field deflects the trajectories of charged particles of the shower and breaks the circular symmetry of the lateral spread of the particles inducing a dependence of the $S(1000)$ at a fixed energy in terms of the azimuthal angle. This dependence translates into azimuthal modulations of the estimated event counting rate at a given $S(1000)$ due to the steepness of the energy spectrum.The procedure followed to obtain an unbiased estimate of the shower energy consists in correcting measurements of shower signals for the influence of the geomagnetic field [11]

$$S_{geom}(1000) = \left[1 - g_1 \cos^{-g_2}(\theta) \sin^2(\vec{u} \cdot \vec{b})\right] S(1000) \quad (1)$$

where $g_1 = (4.2 \pm 1.0) \times 10^{-3}$, $g_2 = 2.8 \pm 0.3$, and $\vec{u}$ and $\vec{b} = \vec{B}/\|B\|$ denote the unit vectors in the shower direction and the geomagnetic field direction, respectively.

Besides, the atmospheric conditions also modify the shower sizes: (i) a greater (lower) pressure corresponds to a larger (smaller) matter overburden and implies that the shower is an advanced (old) stage when it reaches the ground level; (ii) the air density (related to temperature) changes the Molière radius and hence the lateral profiles of the showers. Similarly, the procedure to eliminate these variations consists in relating the $S(1000)$, measured at the actual density $\rho$ and pressure $P$, to the one $S_{atm}(1000)$ that would have been measured at reference values $\rho_0$ and $P_0$, chosen as the average values at Malargue [12]

$$S_{atm}(1000) = [1 - \alpha_P(\theta)(P - P_0) - \alpha_\rho(\theta)(\rho_d - \rho_0) - \beta_\rho(\theta)(\rho - \rho_d)]S(1000), \quad (2)$$

where $\rho_d$ is the average daily density at the time when the event was recorded and the coefficients, $\alpha_\rho$ and $\beta_\rho$, reflect respectively the impact of the variations of air density at long and short time scales and the variation of pressure on the shower size, $\alpha_P$.

Once the influence on $S(1000)$ of weather and geomagnetic effects are accounted for, the shower signal is then converted to $S_{38^o}$, the value that would have been expected had the shower arrived at a zenith angle 38°, using the constant intensity cut method (CIC) [13]. This reference shower signal is finally converted into energy using a calibration curve based on hybrid events measured simultaneously by the SD array and the Fluorescence Detector (FD) telescopes through $E_{FD} = AS_{38^o}^B$, since the latter can provide a calorimetric measurement of the energy [14]. The parameters $A$ and $B$ are obtained from a fit to the data [15].

### 2.2 Exposure determination

In searching for anisotropies, it is also critical to know accurately the effective time-integrated collecting area for a flux from each direction of the sky, or in other words, the *directional exposure* $\omega$ of the Observatory. For each elemental cell, this is obtained through the integration over Local Sidereal Time (LST) $\alpha^0$ of $x^{(i)}(\alpha^0) \times a_{\rm cell}(\theta) \times \varepsilon(\theta,\varphi,E)$, with $x^{(i)}(\alpha^0)$ the total operational time of the cell $(i)$ at LST $\alpha^0$, $a_{\rm cell}(\theta) = 1.95 \cos\theta$ km$^2$ the geometric aperture of each elemental cell under incidence zenith angle $\theta$ [10], and $\varepsilon(\theta,\varphi,E)$ the detection efficiency under incidence zenith angle $\theta$ and azimuth angle $\varphi$ at energy $E$.

The zenithal dependence of the detection efficiency $\varepsilon(\theta,\varphi,E)$ can be obtained directly from the data [8] based on the quasi-invariance of the zenithal distribution to large scale anisotropies for zenith angles less than $\sim 60^o$ and for any Observatory whose latitude is far from the poles of the Earth. Since $dN/d\sin^2(\theta)$ is uniform for full efficiency ($E > 3$ EeV), any significant deviation from a uniform behavior in this distribution provides an empirical measurement of the zenithal dependence of the detection efficiency given by

$$\langle\varepsilon(\theta,\varphi,E)\rangle_\varphi = \frac{1}{\mathcal{N}} \frac{dN(\sin^2\theta, E)}{d\sin^2\theta} \quad (3)$$

where the notation $\langle\cdot\rangle_\varphi$ stands for the average over $\varphi$ and the constant $\mathcal{N}$ is the number of events that would have been observed at energy E and for any $\sin^2\theta$ value in case of full efficiency for an energy spectrum $dN/dE = 40(E/EeV)^{-3.27}$ km$^2$yr$^{-1}$sr$^{-1}$ EeV$^{-1}$ as measured between 1 and 4 EeV [16].

Additional effects have an impact on $\omega$, such as the azimuthal dependence of the efficiency due to geomagnetic effects, the corrections to both the geometric aperture of each elemental cell and the detection efficiency due to the tilt of the array, and the corrections due to the spatial extension of the array. A shower under any incident angles $(\theta,\varphi)$ and energy $E$ triggers the SD array with a probability associated with its size which is a function of the azimuth because of the geomagnetic effects. Considering that the energy that would have been obtained without correcting for geomagnetic effects is $E \times (1 + \Delta(\theta,\varphi))^B$,[1] to first order in $\Delta(\theta,\varphi)$, $\varepsilon(\theta,\varphi,E)$ can be estimated as:

$$\varepsilon(\theta,\varphi,E) = \frac{1}{\mathcal{N}} \frac{dN(\sin^2\theta, E(1+\Delta(\theta,\varphi)^B))}{d\sin^2\theta} \simeq \langle\varepsilon(\theta,\varphi,E)\rangle_\varphi + \frac{BE\Delta(\theta,\varphi)}{\mathcal{N}} \frac{\partial\langle\varepsilon(\theta,\varphi,E)\rangle_\varphi}{\partial E}. \quad (4)$$

Thus, it is straightforward to implement the correction to the detection efficiency induced by geomagnetic effects from the knowledge of $\langle\varepsilon(\theta,\varphi,E)\rangle_\varphi$.

The slight tilt of the SD array gives rise to a small azimuthal asymmetry, and consequently, slightly modifies the directional exposure in a twofold way: changing the geometric factor $(\cos\theta)$ of the projected surface under incident angles $(\theta,\varphi)$ for all energy ranges and slightly varying the detection efficiency with azimuth angle $\varphi$ for energies below 3 EeV. The correction of the projected surface is performed replacing the $\cos\theta$ factor in $a_{\rm cell}$ by the geometric directional aperture per cell $\mathrm{a}_{\rm cell}^{(i)}$

$$a_{cell}^{(i)}(\theta,\varphi) = 1.95\hat{n}\cdot\hat{n}_\perp^{(i)} \simeq 1.95[1 + \zeta^{(i)}\tan\theta\cos(\varphi - \varphi_0^{(i)})]\cos\theta \quad (5)$$

where $\zeta^{(i)}$ and $\varphi_0^{(i)}$ are the zenith and azimuth angles of $\hat{n}_\perp^{(i)}$, the normal vector to each elemental cell. The variation of the detection efficiency with azimuth induced by the tilt of the array is because the effective separation between detectors for a given zenith angle depends on the azimuth, since the SD array seen by showers coming from the uphill direction is denser than those coming from the downhill direction. We showed in [8] that this change in the detection efficiency can be estimated by

$$\Delta\varepsilon_{tilt}(\theta,\varphi,E) = \frac{E^3(E_{0.5}^3 - E_{0.5}^{tilt^3}(\theta,\varphi))}{(E^3 + E_{0.5}^3)(E^3 + E_{0.5}^{tilt^3}(\theta,\varphi))} \quad (6)$$

where $E_{0.5}^{tilt}(\theta,\varphi)$ is related to $E_{0.5}$, the zenithal-dependent energy at which $\varepsilon_{notilt}(E,\theta) = 0.5$, through

$$E_{0.5}^{tilt}(\cos\theta,\varphi) \simeq E_{0.5} \times [1 + \zeta^{\rm eff}\tan\theta\cos(\varphi - \varphi_0^{\rm eff})]^{3/2}. \quad (7)$$

Regarding the spatial extension of the array, the range of latitudes covered by all cells reaches $\simeq 0.5^o$ and induces a slightly different directional exposure between the cells located at the northern part of the array and the ones located at the southern part. This can be accounted for using the latitude of each cell $\ell_{cell}^{(i)}$ to perform the conversion from local angles $(\theta,\varphi)$ to equatorial coordinates $(\delta,\alpha)$ in $a_{\rm cell}(\theta(\alpha',\delta))$ before evaluating the integration to determine the exposure.

As in [18] the small modulation of the exposure in local sidereal time $\alpha^0$ due to the variations of the operational time of each cell $x^{(i)}$ can be accounted for by re-weighting

1. The shorthand notation $\Delta(\theta,\varphi)$ stands for $g_1\cos^{-g_2}(\theta)[\sin^2(\vec{u}\cdot\vec{b}) - \langle\sin^2(\vec{u}\cdot\vec{b})\rangle_\varphi]$

the events with the number of elemental cells at the LST of each event $k$, $\Delta N_{\rm cell}(\alpha_k^0)$. Accounting for all these effects, the resulting dependence of $\omega$ on declination is given by

$$\omega(\delta,E) = \sum_{i=1}^{n_{\rm cell}} x^{(i)} \int_0^{24h} \mathrm{d}\alpha' \, a_{\rm cell}^{(i)}(\theta(\alpha',\delta)) \times \\ [\varepsilon(\theta,\varphi,E) + \Delta\varepsilon_{tilt}(\theta,\varphi,E)], \quad (8)$$

where both $\theta$ and $\varphi$ depend on the hour angle $\alpha' = \alpha - \alpha^0$, $\delta$ and $\ell_{cell}^{(i)}$. For a wide range of declinations between $\simeq -89^\circ$ and $\simeq -20^\circ$, the directional exposure is $\simeq 2,990$ km$^2$ yr at 1 EeV, and $\simeq 4,186$ km$^2$ yr for any energy above full efficiency. Then, at higher declinations, it smoothly falls to zero, with no exposure above 20° declination for zenith angles smaller than $55^o$.

## 3 Searches for large scale patterns

Any angular distribution over the sphere $\Phi(\mathbf{n})$ can be expanded in terms of spherical harmonics :

$$\Phi(\mathbf{n}) = \sum_{\ell \geq 0} \sum_{m=-\ell}^{\ell} a_{\ell m} Y_{\ell m}(\mathbf{n}), \quad (9)$$

where $\mathbf{n}$ denotes a unit vector taken in equatorial coordinates. Due to the non-uniform and incomplete coverage of the sky at the Pierre Auger Observatory, the estimated coefficients $\overline{a}_{\ell m}$ are determined in a two-step procedure. First, from any event set with arrival directions $\{\mathbf{n_1},,\mathbf{n_N}\}$ recorded at LST $\{\alpha_1^0,,\alpha_N^0\}$, the multipolar coefficients of the angular distribution coupled to the exposure function are estimated through :

$$\overline{b}_{\ell m} = \sum_{k=1}^{N} \frac{Y_{\ell m}(\mathbf{n}_k)}{\Delta N_{\rm cell}(\alpha_k^0)}. \quad (10)$$

$\Delta N_{\rm cell}(\alpha_k^0)$ corrects for the slightly non-uniform directional exposure in right ascension. Then, assuming that the multipolar expansion of the angular distribution $\Phi(\mathbf{n})$ is *bounded* to $\ell_{\rm max}$, the first $b_{\ell m}$ coefficients with $\ell \leq \ell_{\rm max}$ are related to the non-vanishing $a_{\ell m}$ through :

$$\overline{b}_{\ell m} = \sum_{\ell'=0}^{\ell_{\rm max}} \sum_{m'=-\ell'}^{\ell'} [K]_{\ell m}^{\ell' m'} \overline{a}_{\ell' m'}, \quad (11)$$

where the matrix $K$ is entirely determined by the directional exposure :

$$[K]_{\ell m}^{\ell' m'} = \int_{\Delta\Omega} \mathrm{d}\Omega \; \omega(\mathbf{n}) \; Y_{\ell m}(\mathbf{n}) \; Y_{\ell' m'}(\mathbf{n}). \quad (12)$$

Inverting Eqn. 11 allows us to recover the underlying $\overline{a}_{\ell m}$, with a resolution proportional to $([K^{-1}]_{\ell m}^{\ell m} \, \overline{a}_{00})^{0.5}$ [17]. As a consequence of the incomplete coverage of the sky, this resolution deteriorates by a factor larger than 2 each time $\ell_{\rm max}$ is incremented by 1. With our present statistics, this prevents the recovery of each coefficient with good accuracy as soon as $\ell_{\rm max} \geq 3$, which is why we restrict ourselves to dipole and quadrupole searches.

Assuming that the angular distribution of cosmic rays is modulated by a dipole and a quadrupole, we parameterize the intensity $\Phi(\mathbf{n})$ in any direction as :

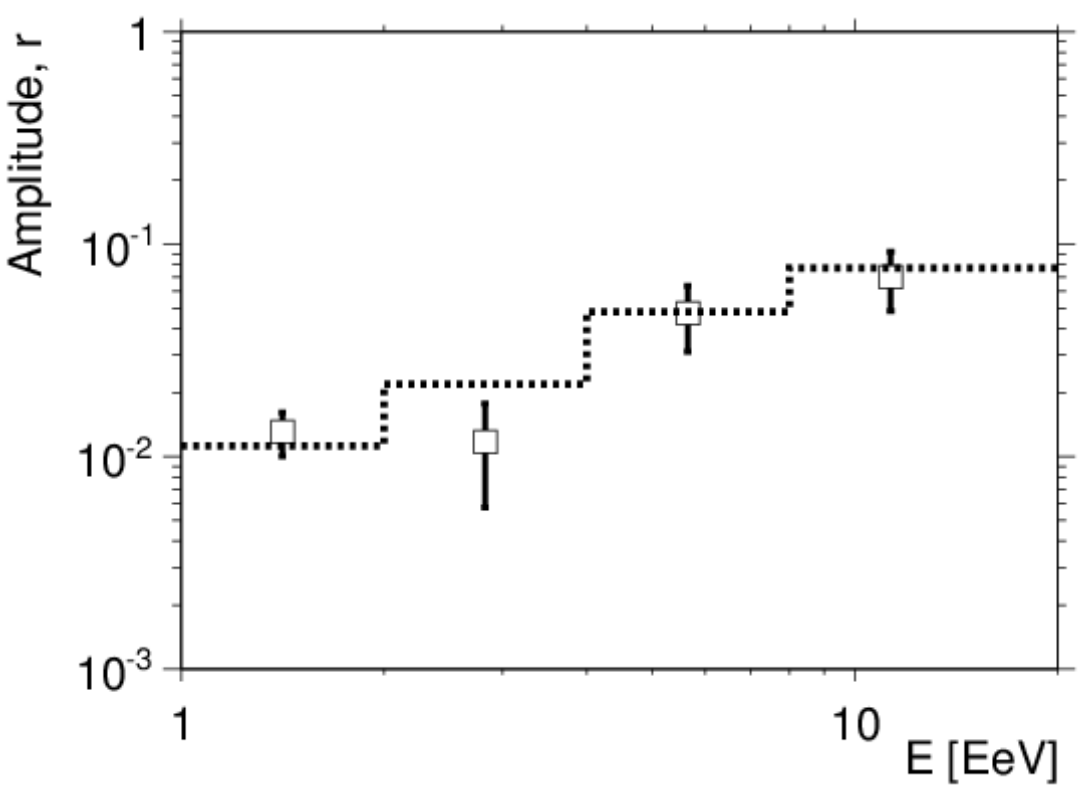

**Fig. 1**: Reconstructed amplitude of the dipole as a function of the energy. The dotted line stands for the 99% *C.L.* upper bounds on the amplitudes that would result from fluctuations of an isotropic distribution.

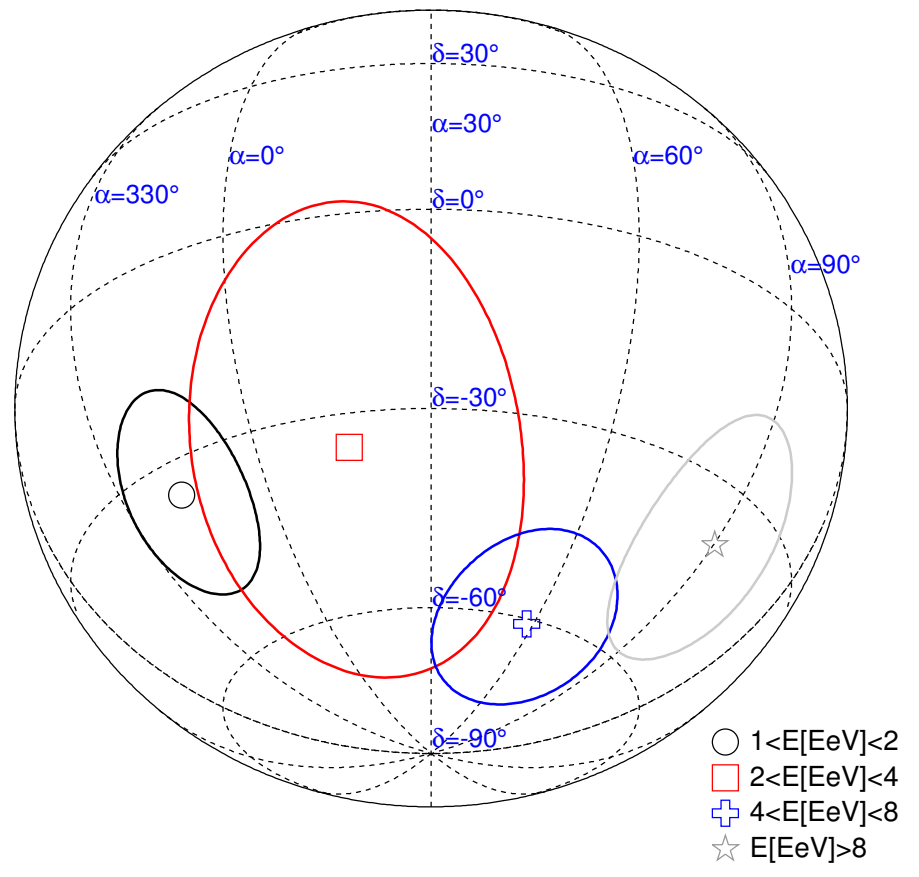

**Fig. 2**: Reconstructed declination and right-ascension of the dipole with corresponding uncertainties, as a function of the energy, in orthographic projection.

$$\Phi(\mathbf{n}) = \frac{\Phi_0}{4\pi} \Big( 1 + r\,\mathbf{d}\cdot\mathbf{n} + \lambda_+ (\mathbf{q}_+ \cdot \mathbf{n})^2 + \\ \lambda_0 (\mathbf{q_0} \cdot \mathbf{n})^2 + \lambda_- (\mathbf{q}_- \cdot \mathbf{n})^2 \Big). \quad (13)$$

The dipole pattern is fully characterized by the dipole unit vector $\mathbf{d}$ corresponding to declination $\delta_d$, right ascension $\alpha_d$ and amplitude $r = (\Phi_{\rm max} - \Phi_{\rm min})/(\Phi_{\rm max} + \Phi_{\rm min})$. Defining the amplitude $\beta \equiv (\lambda_+ - \lambda_-)/(2 + \lambda_+ + \lambda_-)$, which provides a measure of the maximal quadrupolar contrast in the absence of a dipole, any quadrupolar pattern can be fully described by two amplitudes $(\beta, \lambda_+)$ and three angles : $(\delta_+, \alpha_+)$ which define the orientation of $\mathbf{q}_+$ and $(\alpha_-)$ which defines the direction of $\mathbf{q}_-$ in the orthogonal plane to $\mathbf{q}_+$. The third eigenvector $\mathbf{q_0}$ is orthogonal to $\mathbf{q}_+$ and $\mathbf{q}_-$, and its corresponding eigenvalue is such as $\lambda_+ + \lambda_- + \lambda_0 = 0$. All these parameters are determined in a straightforward way from the spherical harmonic coefficients $\overline{a}_{1m}$ and $\overline{a}_{2m}$.

First we consider a case of a pure dipole ($\lambda_{\pm,0} = 0$). The reconstructed amplitudes $\overline{r}$ are shown in Fig. 1 as a function of the energy. The 99% *C.L.* upper bounds on the amplitudes that would result from fluctuations of an isotropic distribution are indicated by the dotted line. One can see, similarly to the results from the analysis in [19], interesting hints for large scale anisotropies that will be important to further scrutinize with independent data. Figure 2 shows the corresponding reconstructed directions in orthographic projection with the associated uncertainties, as a function of the energy. Both angles are expected to be randomly distributed in the case of independent samples whose parent distribution is isotropic. It is thus interesting to note that all reconstructed declinations are in the equatorial southern hemisphere, and to note also the intriguing smooth alignment of the phases in right ascension as a function of the energy. In our previous report on first harmonic analysis in right ascension [18], we already pointed out this alignment, and stressed that such a consistency of phases in adjacent energy intervals is expected with smaller number of events than the detection of amplitudes standing-out significantly above the background noise in the case of a real underlying anisotropy. This motivated us to design a *prescription* aimed at establishing at 99% *C.L.* whether this consistency in phases is real, using the exact same analysis as the one reported in [18]. See [19] for an update of this analysis.

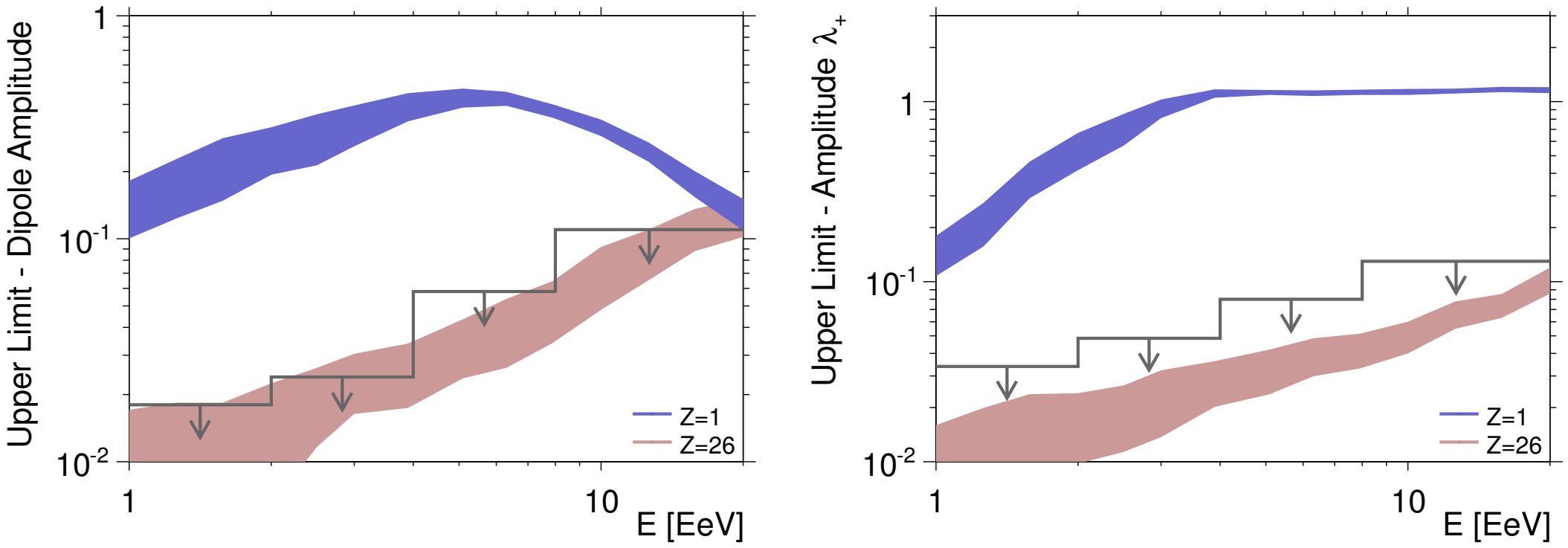

**Fig. 3**: 99% *C.L.* upper limits on dipole and quadrupole amplitudes as a function of the energy. Some generic anisotropy expectations from stationary galactic sources distributed in the disk are also shown, for various assumptions on the cosmic ray composition. The fluctuations (RMS) of the amplitudes due to the stochastic nature of the turbulent component of the magnetic field are sampled from different simulation data sets and are shown by the bands.

Upper bounds on the dipole and quadrupole amplitudes have been obtained at the 99% C.L. The bounds on the dipole amplitudes as a function of energy are shown in the left panel of Figure 3 along with generic estimates of the dipole amplitudes expected from stationary galactic sources distributed in the disk considering two extreme cases of single primaries: protons and iron nuclei. As an illustrative case we consider the Bisymmetric Spiral Structure (BSS) model with anti-symmetric halo with respect to the galactic plane [20] and a turbulent field generated according to a Kolmogorov power spectrum. Furthermore, assuming that the angular distribution of cosmic rays is modulated by a dipole and a quadrupole, the 99% *C.L.* upper bounds on the quadrupole amplitude $\overline{\lambda}_+$ that could result from fluctuations of an isotropic distribution are shown in the right part of Figure 3 together with expectations considering the same astrophysical scenario described before. We will continue monitoring the contribution from higher moments in the flux.

While other magnetic field models, source distributions and emission assumptions must be considered, the example considered here illustrates the potential power of these observational limits on the dipole anisotropy to exclude the hypothesis that the light component of cosmic rays comes from stationary sources densely distributed in the Galactic disk and emitting in all directions.

# Blind searches for localized cosmic ray excesses in the field of view of the Pierre Auger Observatory

BENOÎT REVENU[1] FOR THE PIERRE AUGER COLLABORATION[2]

[1]*SUBATECH, 4 rue Alfred Kastler, BP20722, 44307 Nantes, CEDEX 03, Université de Nantes, École des Mines de Nantes, CNRS/IN2P3, France*
[2]*Full author list: http://www.auger.org/archive/authors_2013_05.html*

*auger_spokespersons@fnal.gov*

**Abstract:** We present the results of a blind search for overdensities with respect to isotropic expectations in the cosmic ray flux detected by the Pierre Auger Observatory. We analyze maps of significances in different energy ranges and for various angular scales. We have also searched for correlations of cosmic ray arrival directions with some promising candidate sources: the directions close to the Galactic Plane and the Galactic Center itself, in the perspective of a galactic origin of cosmic rays, and the Super-Galactic Plane and Centaurus A, in the perspective of an extra-galactic origin.

## 1 Introduction

One of the main goals of ultra-high energy cosmic ray experiments is to search for sources, both at small and large angular scales. We can expect to find localized excesses at small angular scales at energies above 10 EeV (where 1 EeV is $10^{18}$ eV) because the bending of the trajectories of charged particles in the magnetic fields becomes smaller. It is also interesting to search for excesses at large angular scales, that could result either from the spreading of point-like sources by magnetic fields, or by the contribution of clustered sources. The transition from a galactic to an extra-galactic origin could correspond to the observed ankle feature in the cosmic ray energy spectrum around 4 EeV and the escape of the galactic cosmic rays could be manifest through their sky distribution below this energy.

The Pierre Auger Collaboration reported several results on searches for anisotropy, at small and large angular scales, at energies in the EeV range and above, up to the highest energies. The first blind search has been reported in [1] with a result compatible with an isotropic distribution of the cosmic rays. The results of the search for point-like EeV neutron sources at the level of the angular resolution (smaller than 1.4°) have been presented in [2], where upper limits on the neutron flux have been set. A similar search but with stacked targets is presented in this conference in [3]. In the same energy domain, the Galactic Center has been specifically studied in [4] and we reported no deviation from isotropic expectations. Still at small angular scales but at ultra-high energies, we did not find any significant signal using self-clustering studies [5]. At very large angular scales and all energies, we set limits on the dipolar and quadrupolar amplitudes [6, 7] which are updated in this conference [8, 9]. Hints of correlation at the highest energies with nearby extragalactic matter and in particular with the direction towards Centaurus A were reported in [10, 11, 12].

The current paper aims at covering energies above 1 EeV and intermediate angular scales. We report in particular the distributions of significances using top-hat windows of radius 5° and 15° over the full field of view in the same 4 energy ranges used in [6] (1-2 EeV, 2-4 EeV, 4-8 EeV and $\geqslant 8$ EeV) to search for anisotropies at large angular scales. We also search, in these energy ranges, in the direction of the following specific targets: the Galactic Plane (GP), the Galactic center, the Super-Galactic Plane (SGP) and Centaurus A (CenA).

The data set used in this study covers the period 1 January 2004 to 31 December 2012 and is 24 times larger than that of [1].

## 2 The data set

The Pierre Auger Observatory is located in Malargüe, Argentina (35.2°S, 69.5°W) at an altitude of 1400 m asl. We are using two complementary techniques to observe extensive air showers initiated by cosmic rays: a surface detector (SD) [13] and a fluorescence detector (FD) [14]. The SD is composed of 1660 water Cherenkov detectors arranged as an array on a triangular grid with 1.5 km spacing. The SD is fully efficient at $E > 3$ EeV. The FD observes the atmosphere above the SD during dark cloudless nights with 27 telescopes spread over 5 buildings. The fluorescence light emitted by the excited atmospheric nitrogen after the passage of the charged particles of the shower is detected by these telescopes, and permits a calorimetric measurement of the energy of the primary cosmic ray through the observation of the longitudinal profile of the shower. The computation of the exposure of the SD takes into account the growth of the array during construction, from 154 to 1660 water Cherenkov detectors, as well as stations dead times during operation (90% duty cycle). We include events such that the six nearest neighbours of the water Cherenkov detector with the highest signal are fully functional. This defines an active hexagon. These fiducial cuts guarantee good event reconstruction [15]. The total exposure for events that satisfy these cuts and have a zenith angle less than 60° is 31395 km$^2$ yr sr. There are 750181 showers above 1 EeV within that zenith angle range. The distribution of these events with the energy is given in Table 1. The energy of a given shower is determined using first the constant intensity cut method, which provides the shower size at an axis distance of 1000 m that would have been expected if the zenith angle had been 38°, and second, using the calibration curve of the hybrid events independently detected and reconstructed by both SD and FD [16, 17]. The final energy resolution has a statistical uncertainty of 15% and the absolute energy scale has a systematic uncertainty of 14% [18]. Several issues can affect the shower size estimate and have to be corrected for. The influence of the atmospheric conditions has been fully characterized in [19], where we describe the correction to apply to the shower size according to its zenith angle given the measured values of air temperature and pressure at the time of the detection. The geomagnetic field bends the charged secondary particles and broadens their spatial distribution in the direction of the Lorentz force. This effect modifies the lateral distribution function of the particles at the ground level and consequently, the value of the estimated shower size. The correction to apply to the shower size is described in [20]. The data set used in this paper contains all these corrections.

| Energy [EeV] | Number of events |
|---|---|
| 1-2 | 557829 |
| 2-4 | 148790 |
| 4-8 | 31270 |
| $\geqslant 8$ | 12292 |

**Table 1**: Energy distribution of the events used in the considered dataset.

## 3 Directional exposure estimate and events map

The directional exposure provides an estimate of the expected fraction of events within a target solid angle as a function of its direction in the sky, under the assumption of an underlying isotropic distribution of the cosmic rays. We compute the directional exposure from the global acceptance $a_T$ that depends on the shower arrival direction and the time of the detection, using the general formula:

$$W(\alpha,\delta) = \int_{t_{\min}}^{t_{\max}} a_T(t,\theta(\alpha,\delta,t),\phi(\alpha,\delta,t))\,\mathrm{d}t, \qquad (1)$$

as done in [21], where $\alpha, \delta$ are the usual equatorial coordinates, $t$ is the UTC time and $t_{\min}$, $t_{\max}$ define the time period considered. In general, $a_T$ explicitly depends on time through, for instance, the time varying size of the array. In the case of the Pierre Auger Observatory, this dependence is accounted for through the number of active hexagons as a function of time, denoted by $a_{\mathrm{Hex}}(t)$. Concerning the dependence on arrival direction, the acceptance per unit solid angle simply depends on the geometrical factor $\cos\theta$ when the detector is fully efficient, but at low energies ($E < 3$ EeV) and especially at high zenith angles, for which the particles are particularly attenuated by the large atmospheric depth they have crossed, the efficiency becomes less than unity and is zenith dependent.

The total acceptance $a_T$ should therefore take into account this zenith angle modulation $a_\theta$ through a fit of the zenith angle distribution as done in the semi-analytic (SA) method described in [21], where we argued that the zenith angle distribution is poorly sensitive to an actual cosmic-ray sky anisotropy. We have extended the SA method to account for an azimuthal modulation observed in the event counting rate at relatively low energy and large zenith angles. This modulation is due to the hexagonal shape of the SD grid, which introduces an azimuthal dependence on the trigger rate for inclined showers at energies below full efficiency. This can induce relative differences up to 7% in the directional exposure. We take this effect into account in the total acceptance, which is written as: $a_T(t,\theta,\phi) = a_{\mathrm{Hex}}(t)\,a_\theta(\theta)\,a_\phi(\theta,\phi)$, that can be seen as the product of the probability density functions. The specific acceptance $a_\phi(\theta,\phi) = 1+\beta(\theta)\cos(6\phi)$ is shown in Fig. 1 for the energy range 1-2 EeV and for zenith angles $54° \leqslant \theta \leqslant 60°$. It has an amplitude $\beta = -0.068 \pm 0.004$ in this case. The amplitude of the modulation is relevant for energies below 3 EeV and for relatively inclined showers only. Its dependence with zenith angle can be parametrized as $\beta(\theta) = \gamma \times (\theta/\theta_0) \times \exp((\theta-\theta_0)/w)$ with $\theta_0 = 60°$. The fit for the parameters $\gamma$ and $w$ in the energy range 1-2 EeV is $\gamma = -(9 \pm 3)\%$ and $w = 4.2° \pm 1.4°$.

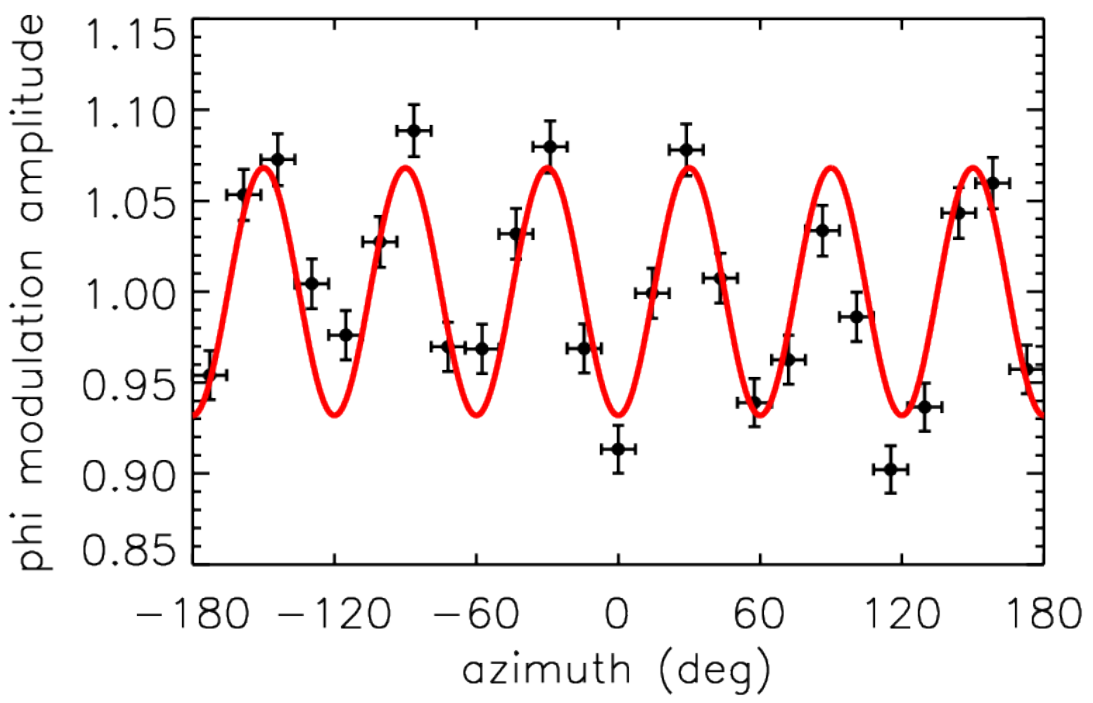

**Figure 1**: Normalized azimuthal distribution of the events with energy in the range 1-2 EeV and $54° \leqslant \theta \leqslant 60°$. The modulation (of the order of 7%) is due to the hexagonal shape of the SD grid. The line corresponds to a fit using the function $\alpha\,(1+\beta\,\cos(6\phi))$ with $\beta = -0.068 \pm 0.004$.

At this level, we have all needed quantities to compute the directional exposure by the numerical integration of Eq. 1. The directional exposure must be computed for each considered energy range and is normalized to the corresponding total number of events. In this analysis, we correct for the azimuthal modulation $a_\phi(\theta,\phi)$ below 3 EeV. For $a_\theta$ we use a fit of the zenith angle distribution in the data, even above 3 EeV where full acceptance implies $a_\theta \propto \sin\theta\cos\theta$, to account for potential residual departures from the geometric distribution due to energy assignment systematics. The explicit time dependence reflects the actual varying size of the SD array $a_{\mathrm{Hex}}(t)$, averaged every five minutes over the 9 years covered by the data set. Note that the directional exposure obtained with this method is very similar to the one provided by the shuffling method [22]. As an example, Fig. 2 presents the integrated directional exposure in galactic coordinates for the energy range 1-2 EeV over circular windows with angular radius of 5°.

For every direction $\alpha, \delta$ at a given angular scale $\omega$, the significance of the difference between the number of observed events $n^{\omega}_{\mathrm{obs}}(\alpha,\delta)$ and the number of expected events from the directional exposure $n^{\omega}_{\mathrm{iso}}(\alpha,\delta)$ is computed using the unbiased Li & Ma estimator [23], where the Li & Ma $\alpha_{\rm LM}$ parameter is taken to be $\alpha_{\rm LM} = n^{\omega}_{\rm iso}(\alpha,\delta)/(N_{\rm tot} - n^{\omega}_{\rm iso}(\alpha,\delta))$, $N_{\rm tot}$ being the total number of events in the considered data set. We compute $n^{\omega}_{\rm obs}(\alpha,\delta)$ (resp. $n^{\omega}_{\rm iso}(\alpha,\delta)$) by integration of the events map (resp. directional exposure) over all directions located at an angular distance smaller than $\omega$ degrees from the direction centered on $(\alpha,\delta)$. We use the HEALPix [24] pixellisation with a pixel width much smaller than 1° ($n_{\rm side} = 512$). From the full-sky significance map we compute the distribution of significances, and compare it with isotropic expectations. For that purpose we simulate 1000 isotropic Monte Carlo datasets, and determine the 68%, 95% and 99.7% dispersion in their distribution of significances.

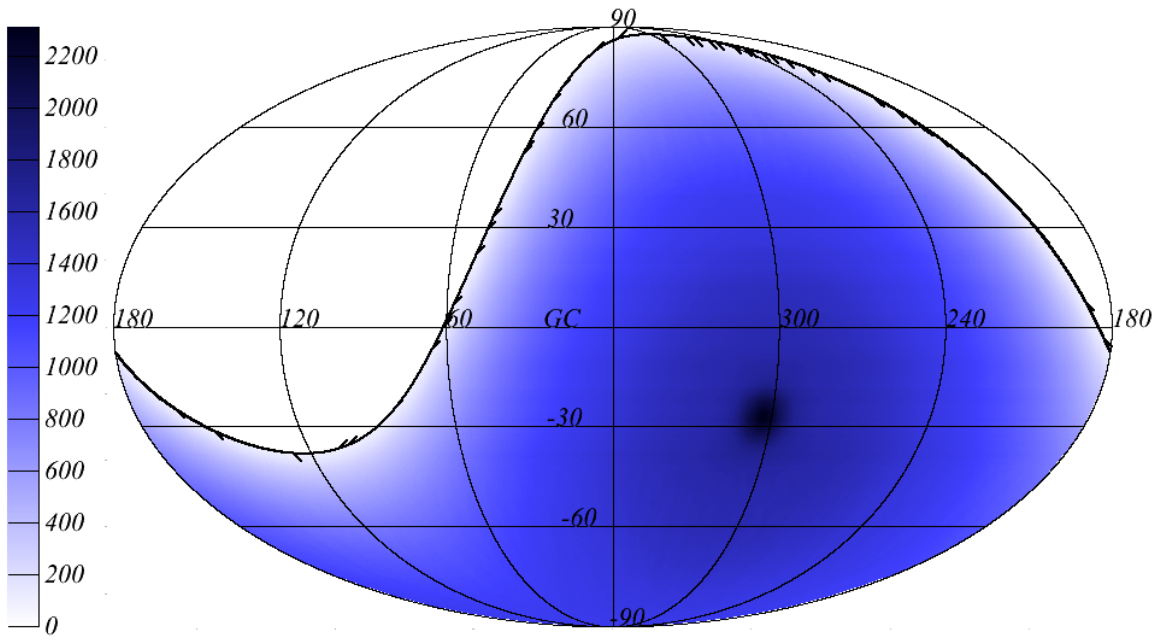

**Figure 2**: Integrated number of events with energy in the range 1-2 EeV expected in windows of radius 5° for an isotropic distribution of arrival directions. The map is in galactic coordinates. The solid line is the border of the field of view for arrival directions with zenith angle $\theta \leqslant 60°$.

## 4 Results

We perform the search for localized excesses in the energy ranges 1-2 EeV, 2-4 EeV, 4-8 EeV, $\geqslant$ 8 EeV at two angular scales: 5° and 15°. The distributions of the Li & Ma significances at an angular scale of 5° are shown in Fig 3. The largest observed significances are compatible with isotropic expectations. The same result holds at an angular scale of 15° and will be reported in a forthcoming paper. We searched for excesses in circular regions centered at the directions toward the Galactic center and CenA by comparing the relative differences $(n_{\rm obs} - n_{\rm iso})/n_{\rm iso}$ between the data and the isotropic expectations using 20000 simulated isotropic data sets. As illustration, the results as a function of the angular radius of the target window centered at the location of CenA for cosmics rays with energies between 4 and 8 EeV (resp. $E \geqslant 8$ EeV) are shown in Fig. 4 (resp. Fig. 5). The results for the location of the Galactic center and cosmic rays with energy in the range 1-2 EeV (resp. 2-4 EeV) are presented in Fig. 6 (resp. Fig. 7).

We have also searched for an excess of arrival directions inside the band within 10° below and above the Galactic and Super-Galactic planes. Table 2 lists the number of events observed with galactic latitudes $-10° < b < 10°$ for cosmic rays in the energy ranges 1-2 EeV and 2-4 EeV, and its ratio with the isotropic expectation, and similarly for arrival directions with Super-Galactic latitudes $-10° < b_{\rm SG} < 10°$ and energies above 8 EeV.

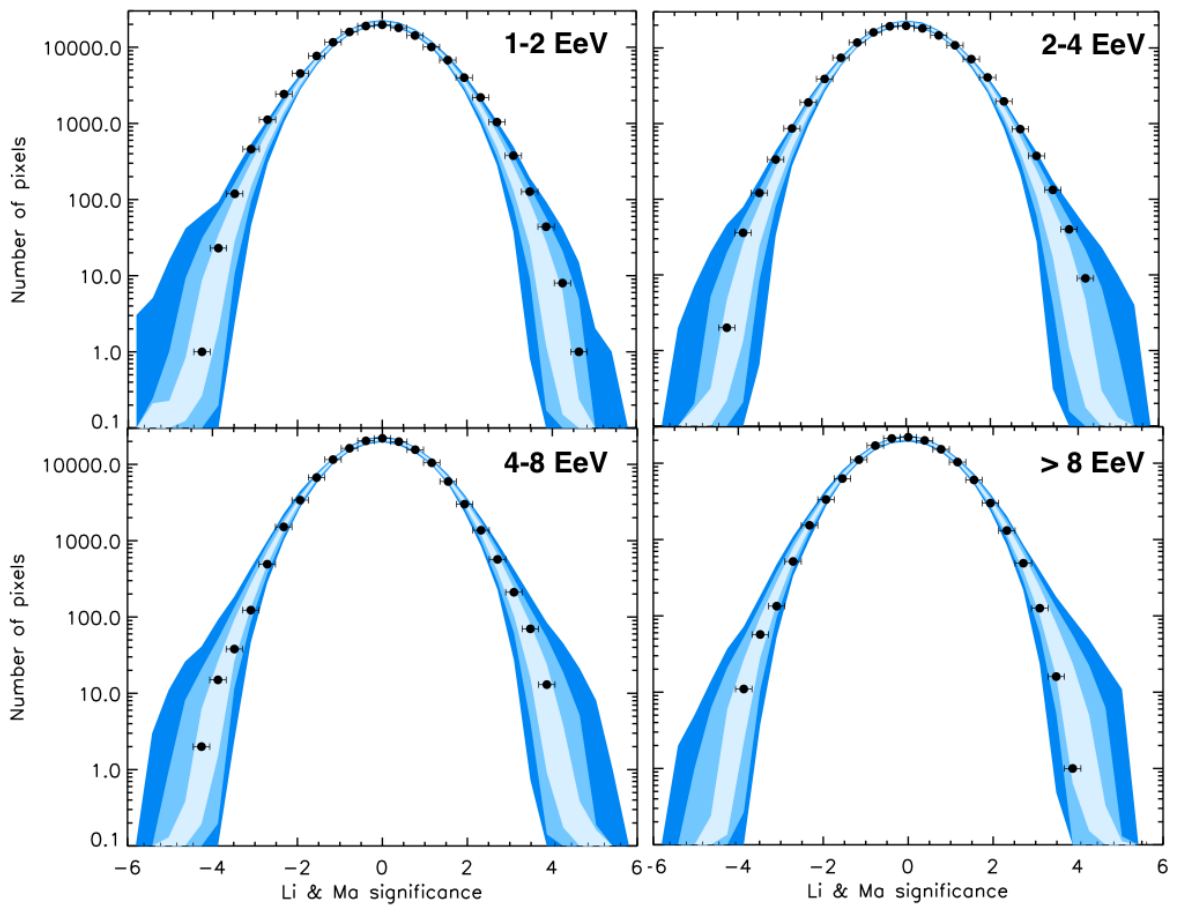

**Figure 3**: Distribution of Li & Ma significances of the difference between the number of arrival directions observed and the isotropic expectation over windows of radius 5° across the exposed sky. The respective energy range is quoted in each plot. The bands correspond to the 68%, 95% and 99.7% dispersion expected for an isotropic flux. The largest observed significances are compatible with isotropic expectations.

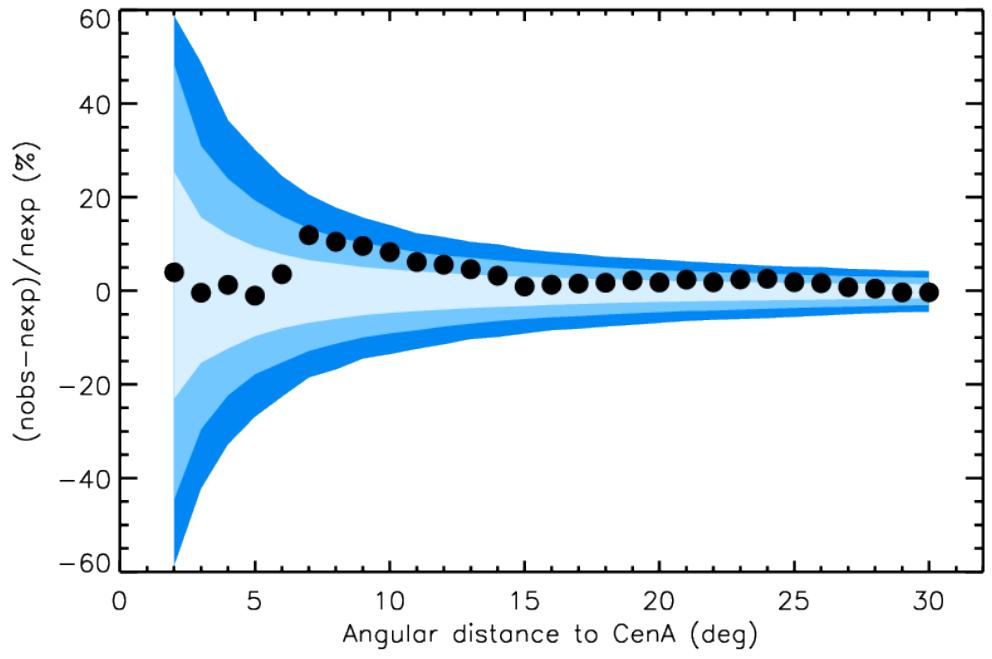

**Figure 4**: Relative difference between the cumulative number of events observed and the isotropic expectation as a function of the angular distance to CenA. The energy range is 4-8 EeV.

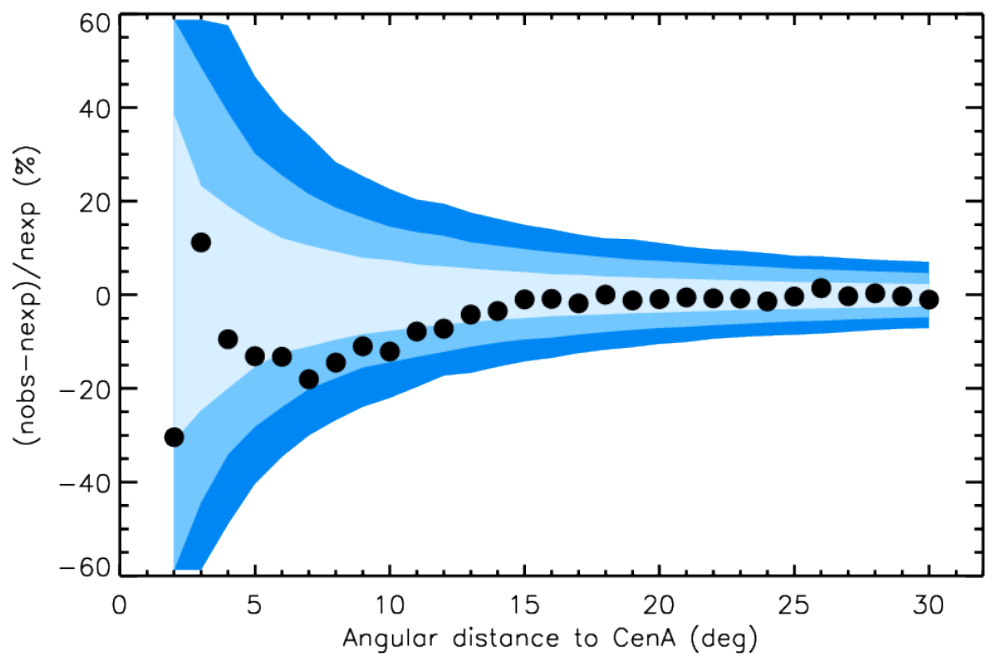

**Figure 5**: Same as Fig. 4 for $E \geqslant 8$ EeV.

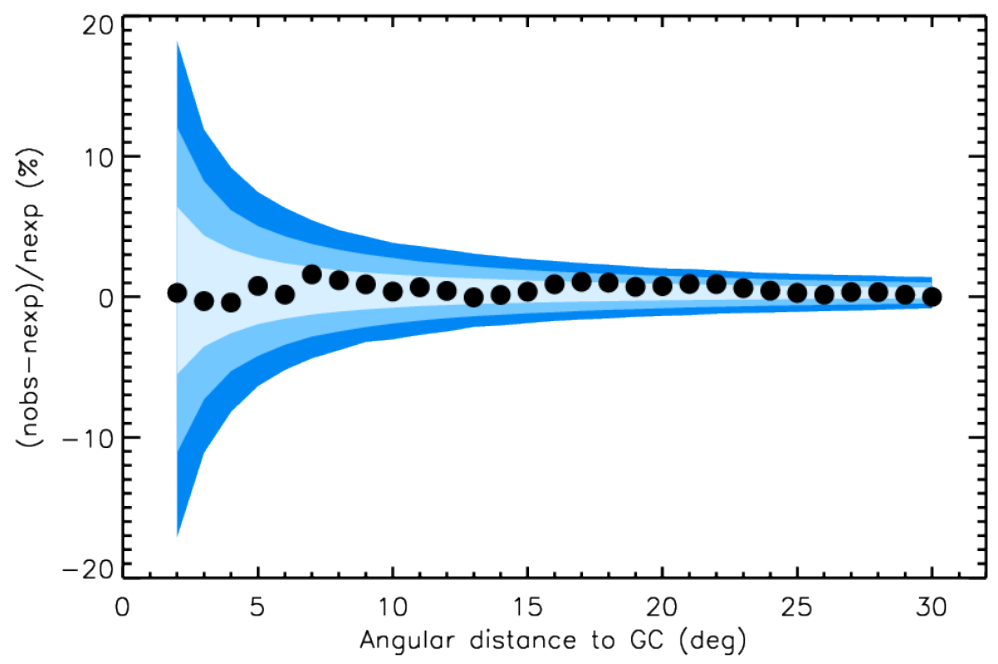

**Figure 6**: Relative difference between the cumulative number of events observed and the isotropic expectation as a function of the angular distance to the Galactic center. The energy range is 1-2 EeV.

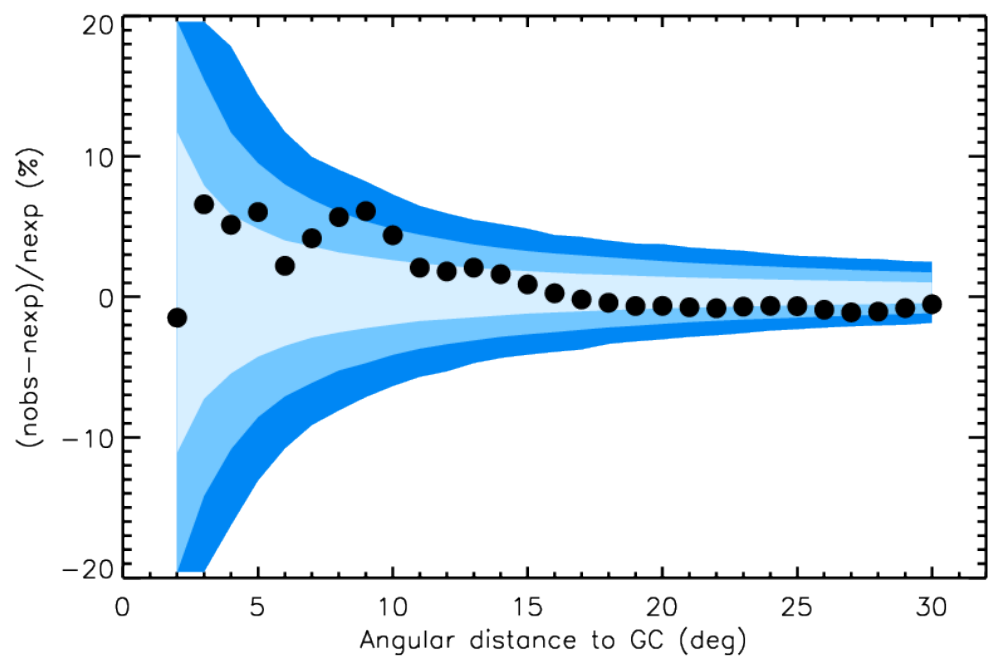

**Figure 7**: Same as Fig. 6 for the energy range 2-4 EeV.

| Target | E [EeV] | $n_{obs}$ | $n_{iso}$ | $n_{obs}/n_{iso}$ |
|---|---|---|---|---|
| GP | 1-2 | 95834 | 95890.4 | $0.999 \pm 0.003$ |
| GP | 2-4 | 25153 | 25491 | $0.986 \pm 0.006$ |
| SGP | $\geqslant 8$ | 2156 | 2135.5 | $1.01 \pm 0.03$ |

**Table 2**: Results of the search for excesses around the Galactic and Super-Galactic planes. A band within latitudes 10° above and below the planes is considered. The number of events observed inside the band is compared to the isotropic expectation. The ratio $n_{obs}/n_{iso}$ is indicated with the 68% dispersion expected for an isotropic flux.

## 5 Conclusion

In this study, we searched for regions of excess flux with respect to isotropic expectations at energies above 1 EeV over the exposed sky at the Pierre Auger Observatory. Data between 2004 and 2012 have been analyzed, considering in particular the same four energy ranges already used for the large scale anisotropy analysis in [6, 7]: 1-2 EeV, 2-4 EeV, 4-8 EeV and $E \geqslant 8$ EeV. Here we searched for anisotropies at intermediate angular scales. We analyzed the distribution of significances for the difference between the number of arrival directions in circular windows of angular radius of 5° and 15° and the isotropic expectation over the full exposed sky. The largest observed significances are compatible with isotropic expectations.

We studied, for the same energy ranges, the distribution of arrival directions with latitudes within 10° of the Galactic and Super-Galactic planes. We analyzed the distribution of arrival directions as a function of the angular distance to the Galactic Center and to the location of the radio-galaxy Centaurus A. We did not find any significant departure from isotropic expectations for these targets in the energy ranges explored.

# Searches for Galactic neutron sources with the Pierre Auger Observatory

FRANCISCO SALESA GREUS[1], FOR THE PIERRE AUGER COLLABORATION[2].

[1] *Colorado State University, Fort Collins, CO, USA*
[2] *Full author list:* `http://www.auger.org/archive/authors_2013_05.html`

*auger_spokespersons@fnal.gov*

**Abstract:** A flux of neutrons from an astrophysical source within our Galaxy would be detected by the Pierre Auger surface detector as an excess of air showers arriving from the direction of the source. In order to reduce the statistical penalty incurred by multiple trials, classes of candidate sources are analyzed collectively as target sets or "stacks". Individual candidate sources are weighted in proportion to their electromagnetic flux and to their exposure to the Auger Observatory. The results are summarized as a combined p-value for each of the stacks, along with information about the candidate source with the minimum individual p-value in each stack. No significant excess flux is found from the targets considered.

## 1 Introduction

The origin of the most energetic (E $>10^{18}$ eV) cosmic rays remains unknown since their discovery 100 years ago. Neutrons produce air showers that are indistinguishable from those produced by protons. However, unlike proton cosmic rays, neutron trajectories are not bent by the magnetic fields. Therefore, a statistically significant clustering of cosmic ray arrival directions would be indicative of a neutron cosmic ray flux. The main drawback of using neutrons as cosmic messengers is that they are unstable outside of the nucleus. Nevertheless, since the neutron mean decay length is $\sim$9.2 kpc (E/EeV) (where EeV=$10^{18}$ eV), at energies above 1 EeV neutrons from Galactic sources can be detected.

The Pierre Auger collaboration has already presented a blind search analysis for neutron sources in the whole exposed sky [1]. However, accounting for the large amount of trials considered, all the excesses in the data sample were shown to be compatible with statistical fluctuations of the background. In order to avoid the statistical penalty for making numerous trials, this paper presents stacked searches using specific classes of potential sources.

Located at latitude 35.2° S and longitude 69.5° W, the Pierre Auger Observatory [2] covers an area of 3000 km$^2$ and is instrumented with 1660 water Cherenkov detectors (WCD). This array of WCDs is known as the Surface Detector Array (SD). Each WCD in the SD is a tank filled with 12,000 liters of water, housing three 9-inch photomultiplier tubes. The WCDs sample at ground level the cascade of secondary particles produced after the interaction of a primary cosmic ray with the atmosphere. Apart from the SD there is also the fluorescence detector (FD) composed of 27 telescopes distributed in five different sites. These installations are located in naturally elevated positions overlooking the space above the SD array. The FD operates during moonless nighttime only. It measures the fluorescence light produced by the interaction of the cascading particles with the atmosphere as the air shower develops.

## 2 Data sample

The data set used for this analysis consists of events collected by the SD from 1 January 2004 to 31 December 2012. The analysis excludes the very inclined events by requiring the reconstructed zenith angles to be smaller than 60°. Therefore, the field of view is limited to declinations from +25° to -90°. Moreover, an event is accepted only if all six nearest neighbors of the station with the highest signal were operational at the time the event was recorded. This cut ensures a good event reconstruction [3]. In addition to these cuts, periods of instability of the array were excluded from the data set. The total exposure is 31,395 km$^2$ yr sr, and the total number of events with E $\geq$ 1 EeV is 750,181.

## 3 Candidate list of sources

Detection of Galactic TeV gamma-rays with energy fluxes near and above 1 eV cm$^{-2}$ s$^{-1}$ have been reported. If these gamma-rays are produced from the pion-photoproducing and nuclear interactions of primary protons near the cosmic source, neutrons should also be produced in the same scenario. The known sources of high energy gamma-rays are therefore likely sources of TeV neutrons, and the flux of neutrons at EeV energies would exceed that same energy flux if the accelerated proton spectrum has a 1/E$^2$ dependence.

This makes the known sources of high energy gamma-rays the most likely candidates for neutron sources. The search presented in this analysis is performed on eight target sets or stacks of astrophysically interesting objects. These directions correspond to HESS sources [4], gamma-ray pulsars [5], low- [6] and high-mass x-ray binaries [7], millisecond and standard radio pulsars [8], microquasars [9], and magnetars [10]. In addition to these target sets, the Galactic Plane and Galactic Center are considered as two additional source stacks for a total of ten target sets. In order to ensure the independence among target sets, a source which appears in two or more is retained only in the most exclusive set, while removed from the others. The search is favored for those candidate sources that have greater electromagnetic flux and that are better exposed. This is achieved by giving each target a weight which is proportional to the product of its directional exposure and the electromagnetic flux recorded for it in the source catalog.

## 4 Method

Four energy ranges were selected to perform the analysis: 1 EeV $\leq$ E <2 EeV (557,829 events), 2 EeV $\leq$ E<3 EeV (113,333 events), E $\geq$ 3 EeV (79,019 events), as well as E $\geq$ 1 EeV. The solid angle size for each target is based on the average angular resolution (AR) for its declination and the energy range as explained in [1]. The solid angle is a "target circle" of radius 1.05 times the AR, centered at the source position. The AR is defined as the angle within which 68% of neutron arrival directions from a candidate source should be included after the event reconstruction. In the particular case of the Galactic Plane, the target is considered to be a band with a thickness of 2 $\times$ 1.05 $\times$ AR along the Galactic Plane.

To recognize the existence of an excess of events in any target circle, it is necessary to know the number that is expected in that circle without the extra source flux. Simulation data sets are used for this. The expected number of events in a given target circle is taken to be the average number found in 10,000 simulated data sets. The simulated data sets are obtained from the actual arrival directions, for each energy range, by a scrambling procedure that thoroughly smooths out any small-scale anisotropy, as explained in [1].

For each target set, a p-value $p_i$ ($i = 1, \ldots, N$, where $N$ is the number of targets in the set) is used to summarize any target $i$ in the set. This p-value $p_i$ is defined as the Poisson probability, given the known expected number, of obtaining a number of events greater than or equal to the one that was actually observed.

The unweighted stacked measure (or unweighted combined p-value) $P$ for a set of $N$ targets is then determined as the fraction of simulations in which the product $\prod_{i=1}^{N} p_i$ is less than or equal to the same product obtained using the actual data. This is illustrated in the left panel of Figure 1 for one of the target sets analysed.

For a weighted set of $N$ targets with weights $w_i$, the p-value $P_w$ is the fraction of simulations in which the weighted product $\prod_{i=1}^{N} p_i^{w_i}$ is equal to, or less than, the same weighted product using the real data (see Figure 1 right). This corresponds to raising each p-value $p_i$ to the power $w_i$ in the product of p-values, so the weight $w_i$ can be regarded as the "number of times" the result for target $i$ is counted relative to other targets of the set.

## 5 Results

The stacking procedure explained in the previous section was applied to the 10 target sets, and repeated for the 4 energy ranges. The results are summarized in Table 1, where the weighted and unweighted combined p-values are shown for comparison. These stacked results reveal no significant excess from any of the target sets.

A neutron flux upper limit has been computed for each target. The method to compute the limits is the same as the one explained in [1]. The definition of the upper limit in the number of neutrons is that of Zech [11] using a 95% confidence level. The upper limit on the flux from a source is the upper limit on the number of neutrons in the top-hat region divided by the directional exposure and by the fraction of the total signal (71.8%) encompassed in the target circle. The directional exposure is defined as the number of events expected from the background in the target circle, divided by the solid angle of this target and the cosmic-ray intensity in (km$^2$ sr yr)$^{-1}$. This cosmic-ray intensity is obtained by integrating the known energy spectrum over the relevant energy range. The energy flux upper limit has been computed assuming an E$^{-2}$ neutron spectra above 1 EeV.

The target with the smallest p-value in each of the target sets is listed in Table 2, which also provides the coordinates of the target, the observed and expected number of events, the particle and energy flux upper limit, and the p-value without and with penalization for the multiple trial targets in each target set. The values reported in this table are only for the energy range E $\geq$ 1 EeV, which corresponds to the inclusive range containing all the other ones.

The p-value of the target penalized for multiple trials is given by:

$$p^* = 1 - (1 - p)^N \tag{1}$$

where $p$ is the original p-value and $N$ the number of targets in the set.

The p-values presented in the tables, for the stacks and for the (penalized) individual targets, are all larger than 2%, which constitute no evidence for neutron fluxes originating from the probed candidates.

## 6 Conclusions

A search for astrophysical neutron sources using data from the SD of the Auger Observatory has been performed using stacked analyses based on catalogs of potential high-energy particle producers in the Galaxy. No significant excess of air showers attributable to neutron fluxes has been detected for any of the catalogs, and flux upper limits were derived. Null results were also derived for the Galactic Plane and the Galactic Center.

| | | Weighted p-value $P_w$ | | | | Unweighted p-value $P$ | | | |
|---|---|---|---|---|---|---|---|---|---|
| Stack | No. | 1-2 EeV | 2-3 EeV | $\geq$ 3 EeV | $\geq$ 1 EeV | 1-2 EeV | 2-3 EeV | $\geq$ 3 EeV | $\geq$ 1 EeV |
| Reg. PSRs | 1326 | 0.95 | 0.06 | 0.49 | 0.67 | 0.80 | 0.43 | 0.48 | 0.89 |
| msec PSRs | 83 | 0.68 | 0.61 | 0.74 | 0.85 | 0.23 | 0.80 | 0.88 | 0.42 |
| $\gamma$-ray PSRs | 75 | 0.02 | 0.90 | 0.14 | 0.089 | 0.59 | 0.49 | 0.71 | 0.60 |
| LMXB | 142 | 0.17 | 0.38 | 0.31 | 0.13 | 0.76 | 0.37 | 0.33 | 0.64 |
| HMXB | 77 | 0.82 | 0.76 | 0.49 | 0.84 | 0.68 | 0.82 | 0.43 | 0.61 |
| HESS | 60 | 0.48 | 0.28 | 0.41 | 0.62 | 0.86 | 0.30 | 0.59 | 0.83 |
| Microquasars | 13 | 0.95 | 0.52 | 0.65 | 0.94 | 0.70 | 0.13 | 0.51 | 0.23 |
| Magnetars | 13 | 0.79 | 0.94 | 0.40 | 0.96 | 0.98 | 0.90 | 0.59 | 0.98 |
| G. Center | 1 | - | - | - | - | 0.77 | 0.41 | 0.45 | 0.73 |
| G. Plane | 1 | - | - | - | - | 0.68 | 0.85 | 0.31 | 0.81 |

**Table 1**: Weighted and unweighted stacked analysis for each target set and each energy range.

| Stack | RA [°] | DEC [°] | Obs | Exp | Flux U.L. [km$^{-2}$yr$^{-1}$] | E-Flux U.L. [eV cm$^{-2}$ s$^{-1}$] | p-value | p* |
|---|---|---|---|---|---|---|---|---|
| Reg. PSRs | 267.44 | -56.09 | 249 | 204 | 0.0161 | 0.117 | 0.0012 | 0.78 |
| msec PSRs | 270.46 | -14.29 | 174 | 146 | 0.0156 | 0.114 | 0.014 | 0.70 |
| $\gamma$-ray PSRs | 195.60 | -32.95 | 222 | 191 | 0.0146 | 0.107 | 0.017 | 0.72 |
| LMXB | 129.35 | -42.90 | 238 | 208 | 0.0135 | 0.0983 | 0.023 | 0.96 |
| HMXB | 249.77 | -46.70 | 237 | 208 | 0.0129 | 0.0945 | 0.028 | 0.88 |
| HESS | 284.58 | 2.09 | 101 | 80.6 | 0.0155 | 0.113 | 0.016 | 0.61 |
| Microquasars | 288.75 | 10.08 | 68 | 53.4 | 0.0161 | 0.118 | 0.030 | 0.33 |
| Magnetars | 248.97 | -47.59 | 224 | 209 | 0.00992 | 0.0724 | 0.15 | 0.88 |
| G. Center | 266.40 | -28.94 | 178 | 186 | 0.0062 | 0.045 | 0.73 | - |
| G. Plane | Galactic lat. = 0° | | 15488 | 15600 | - | - | 0.81 | - |

**Table 2**: List of targets with smallest p-value in each target set for the energy range E $\geq$ 1 EeV. The upper limits are derived at 95% C.L.

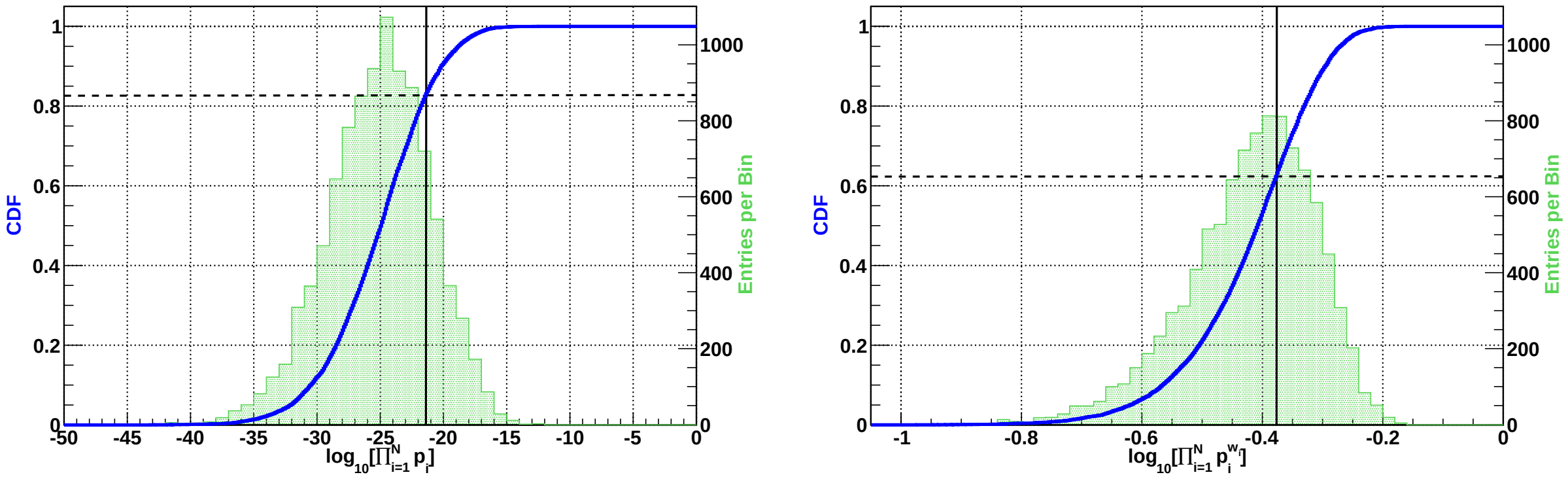

**Figure 1**: Illustration of combined p-values P, unweighted (left plot) and weighted (right plot). This example is for the HESS catalog of 60 exposed candidate sources, and the energy range is ≥1 EeV. The histogram shows the distribution of log(Π) from 10,000 simulation data sets, where Π is the (unweighted or weighted) product of target p-values. The blue curve is the cumulative distribution fraction (*CDF*) of simulations as labeled on the left edge of each plot, and the vertical line is the value of log(Π) obtained using the actual data set. The combined p-value is the CDF value where that line crosses the blue curve.

# Directional search for ultra-high energy photons with the Pierre Auger Observatory

DANIEL KUEMPEL[1] FOR THE PIERRE AUGER COLLABORATION[2]

[1]*RWTH Aachen University, Physikalisches Institut IIIa Otto-Blumenthal-Str., 52056 Aachen, Germany*
[2]*Full author list: http://www.auger.org/archive/authors_2013_05.html*

*auger_spokespersons@fnal.gov*

**Abstract:** The Pierre Auger Observatory, located in Argentina, provides an unprecedented exposure for detecting photons with energies above $10^{17}$ eV over most of the sky. In this work, the information from the surface array of water Cherenkov detectors and from the fluorescence telescopes of the Observatory, are combined in a multivariate analysis to search for photons in the EeV energy range. The arrival directions of candidate photons in the Auger Observatory are here analyzed for the first time. No photon point source is detected. Upper limits on regularly emitting non-beamed photon sources in the Galaxy do not exceed 0.25 eV cm$^{-2}$ s$^{-1}$ and constrain models for the acceleration in the Galaxy of the EeV protons.

## 1 Introduction

The composition of ultra-high energy (UHE) cosmic rays at EeV energies (1 EeV = $10^{18}$ eV) is still unknown. A small fraction might be photons produced in galactic or nearby extragalactic sources. The most prominent production mechanism is the decay of neutral pions produced previously by a "primary process" such as resonant photo-pion production. No UHE photon identification has been reported so far. However, by placing upper limits on the photon flux above EeV energies, severe constraints on "top-down" models were imposed by previous *diffuse* photon searches [1, 2].

In this contribution the search for photons is extended taking into account event arrival directions to search for photon emitting point sources at EeV energies. At these energies, fluxes of photons are attenuated over intergalactic distances by $e^{\pm}$ pair production in collisions of UHE photons with cosmic background photons. The detectable volume of EeV photon sources is small compared to the GZK sphere [3, 4], but large enough to encompass the Local Group of galaxies and possibly Centaurus A given an attenuation length of about 4.5 Mpc at EeV energies [5].

The Pierre Auger Observatory [6] provides an unprecedented sensitivity to search for EeV photon point sources. It encompasses over 1660 individual surface detectors (SD) arranged as an array on a triangular grid with 1500 m spacing. This 3000 km$^2$ array is overlooked by a fluorescence detector (FD) consisting of 27 fluorescence telescopes located at five sites. The SD samples the density of the secondary particles of the air shower at the ground while the FD observes the longitudinal development of the shower. The analysis presented in this work uses hybrid data (detected by at least one FD telescope and one SD station). The hybrid measurement technique provides a precise geometry and energy determination with a low energy threshold for detection. Taking advantage of the two detector systems, several observables are defined and combined in a multivariate analysis (MVA) using advanced boosting techniques to search for photon point sources and to place directional upper limits on the photon flux.

## 2 Directional photon search

The strategy for the directional photon search is based on the selection of a subset of photon-like events, using MVA, to increase the detection probability of photon point sources by reducing the isotropic hadronic background. The selection is optimized direction-wise accounting for the expected background contribution from a given target direction. The *p*-value for the observation of the selected subset is calculated and illustrated in a celestial sky map. Furthermore, directional upper limits on the photon flux are derived.

### 2.1 Multivariate Analysis

The following observables are taken into account in a MVA:
**Depth of shower maximum $X_{\max}$:** It is defined as the atmospheric depth at which the longitudinal development of a shower reaches its maximum in terms of energy deposit. On average, photon induced air showers develop later in the atmosphere compared to hadron induced air showers resulting in larger $X_{\max}$ values.
**Fit of Greisen function to the longitudinal profile**: The Greisen function [11] describes the longitudinal profile of pure electromagnetic showers: a better fit to the longitudinal profile is expected for photon initiated showers if compared to hadronic ones of the same energy. The $\chi^2$/ndof is used to quantify the goodness of the fit.
**Greisen energy**: The only parameter of the Greisen function is the primary energy $E_{\mathrm{gr}}$ which is also influenced by the primary particle. The observable is $E_{\mathrm{gr}}/E_{\mathrm{FD}}$, where $E_{\mathrm{FD}}$ is the energy obtained from the fit of a Gaisser-Hillas function [12] to the longitudinal profile.
$S_3$ **parameter**: This parameter $S_3$ is sensitive to different lateral distribution functions, due to the presence/absence of the flatter muon component [13]. It is defined as

$$S_3 = \sum_{i=1}^{N} \left[ S_i \cdot \left( \frac{r_i}{1000\ \mathrm{m}} \right)^3 \right] , \qquad (1)$$

where the sum extends over all $N$ triggered stations, $S_i$ expresses the signal strength of the $i$–th SD station, and $r_i$ the distance of this station to the shower axis.

**Shape parameter**: The spread of the arrival times of shower particles at a fixed distance from the axis increases for smaller production heights. Consequently, a larger spread is expected in case of deep developing photon primaries compared to hadronic primaries[1]. The shape parameter is the ratio of the early arriving to the late arriving integrated time trace measured in the water Cherenkov tank with the strongest signal

$$\text{ShapeP}(r,\theta) = \frac{S_{\text{early}}(r,\theta)}{S_{\text{late}}(r,\theta)} . \quad (2)$$

The early signal $S_{\text{early}}$ is defined to be signal integrated over time bins smaller than a scaled time $t_i^{\text{scaled}} \leq 0.6\ \mu$s beginning from the signal start slot. The scaled time accounts for different inclination angles $\theta$ and distances to the shower axis $r$ as

$$t_i^{\text{scaled}}(r,\theta) = t_i \cdot \frac{r_0}{r} \cdot \frac{1}{c_1 + c_2 \cdot \cos(\theta)} , \quad (3)$$

where $t_i$ is the real time of bin $i$ and $r_0 = 1000$ m a reference distance. $c_1 = -0.6$ and $c_2 = 1.9$ are scaling parameters to average traces for different inclination angles. Correspondingly, the late signal $S_{\text{late}}$ is the integrated signal over time bins larger than $t_i^{\text{scaled}} > 0.6\ \mu$s until signal end. In a MVA the introduced input observables are combined using boosted decision trees (BDT) as classifier [8, 9]. Since the observables correlate with energy and zenith angle of the primary particle, BDT need to take this correlation into account during the training process. Therefore energy and zenith angle are added to the classification algorithm as additional input observables.

For the classification process boosted decision trees are trained and tested using CORSIKA v. 6.900 [10] simulations. A total number of $\sim 30000$ photon and $\sim 60000$ proton primaries are generated according to a power law spectrum of -2.7 between $10^{17.2}$ eV and $10^{18.5}$ eV using QGSJET-01c [14] and GHEISHA as high and low energy interaction model, respectively. During the classification phase photon and proton showers are reweighted according to a spectral index of -2.0 and -3.0, respectively. The MVA output response value is named $\beta$ and shown in Fig. 1.

### 2.2 Dataset

Hybrid events collected between January 2005 and September 2011 with reconstructed energy between $10^{17.3}$ eV and $10^{18.5}$ eV are selected[2]. The energy refers to the calorimetric energy of the shower including 1% missing energy correction for photons. Air showers with zenith angle smaller than 60° and with a good geometry reconstruction are selected for the analysis. To ensure a reliable profile reconstruction we require: a reduced $\chi^2$ of the longitudinal profile fit to the Gaisser-Hillas function smaller than 2.5, the Cherenkov light contamination smaller than 50% and the uncertainty of the reconstructed energy less than 40%. To reject misreconstructed profiles, only periods with a detected cloud coverage $\leq$ 80% and with a reliable measurement of the vertical optical depth of aerosols [7], are selected. On the SD side we require at least 4 active stations within 2 km from the hybrid reconstructed axis and reject stations with saturated low gain signal. Additionally periods of unstable data taking from the fluorescence detector and surface array have been omitted from the analysis. The final dataset consists of $N_{\text{data}} = 241466$ events.

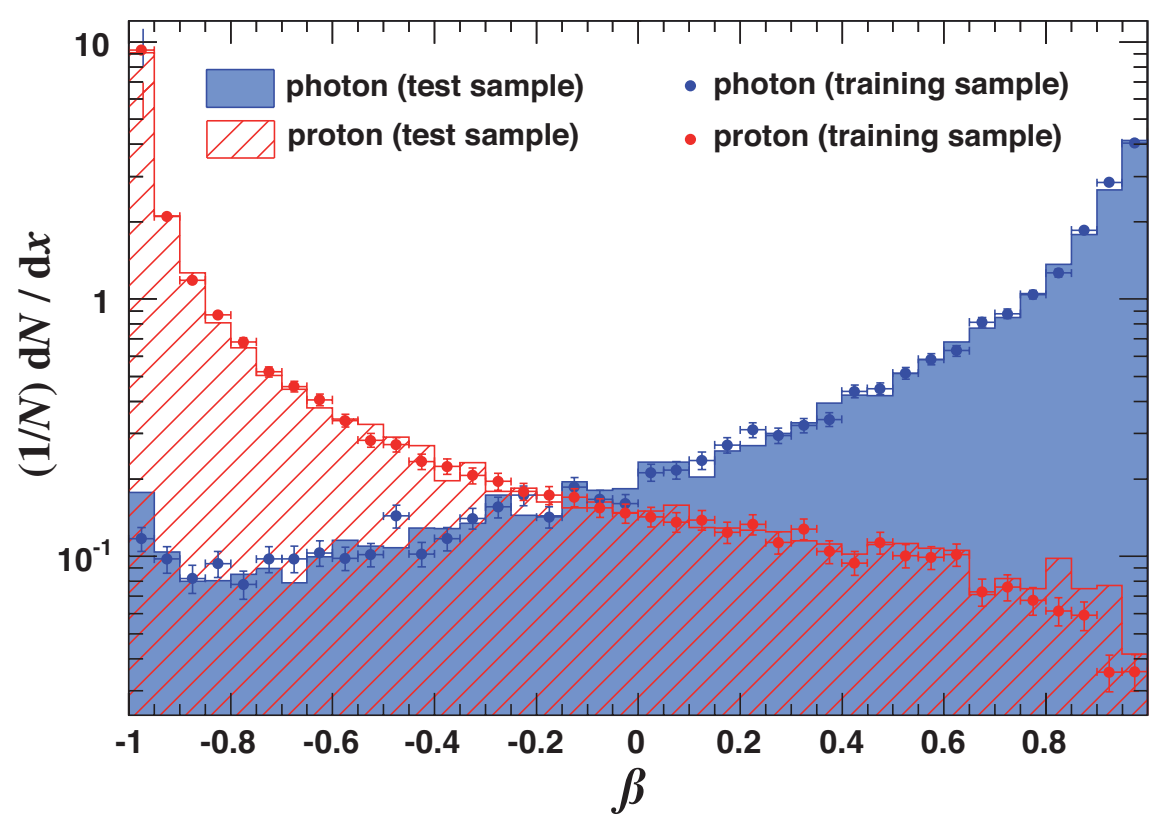

**Figure 1**: MVA Response value $\beta$ for photon and proton primaries using boosted decision trees. During evaluation the MC sample is split into a training (dots) and testing sample (solid line).

The treatment of arrival directions is based on an unbinned analysis. Assuming that the arrival directions of a point source are smeared out by a two-dimensional symmetric point spread function, $n_{\text{inc}} = 90\%$ of the expected signal from a point source is contained in a top-hat counting region of radius $r_{\text{TH}} = 1°$, given an angular resolution of the covered energy range of $\psi = 0.7°$. Sky maps are pixelized using the HEALPix software [15]. Target centers are taken as the central points of a HEALPix grid using $N_{\text{side}} = 256$ (target separation $\sim 0.3°$) resulting in 526200 target centers below a declination of 20°. We limit the analysis to declinations larger than $-85°$ for reasons explained in Sec. 4.

## 3 Search method

To select photon-like air showers an optimized cut on the $\beta$-distribution is performed. The fraction of photon and measured events, $\varepsilon_\gamma^\beta$ and $\varepsilon_{\text{data}}^\beta$, passing a specific cut on the $\beta$-distribution is illustrated in Fig. 2. To estimate $\varepsilon_{\text{data}}^\beta$ more accurately, a declination dependence is taken into account $\varepsilon_{\text{data}}^\beta = \varepsilon_{\text{data}}^\beta(\delta)$.

To improve the detection potential of photons from point sources, the cut on the $\beta$ distribution is optimized, dependent on the direction of a target center. That is, for a given direction the upper limit of photons from a point source is minimized for the case $n_{\text{data}} \stackrel{!}{=} n_b^\beta$, i.e. when the observed number of events is equal to the expected number (cf. Sec. 4). There are alternative ways to define an upper limit on the number of photons $n_s$ at a given confidence level CL in the presence of Poisson distributed background. Here the procedure of Zech [19] is utilized where $n_s$ is given by

$$P(\leq n_{\text{data}}|n_b^\beta + n_s) = \alpha_{\text{CL}} \cdot P(\leq n_{\text{data}}|n_b^\beta) , \quad (4)$$

1. Note that the effect is superimposed also by geometrical effects in the relation between spread and primary composition. Also the competition between the signals from electromagnetic and muonic shower components contributes to this effect.
2. The selected energy range accounts for high statistics and negligible impact of high energy phenomena such as LPM or preshower effects.

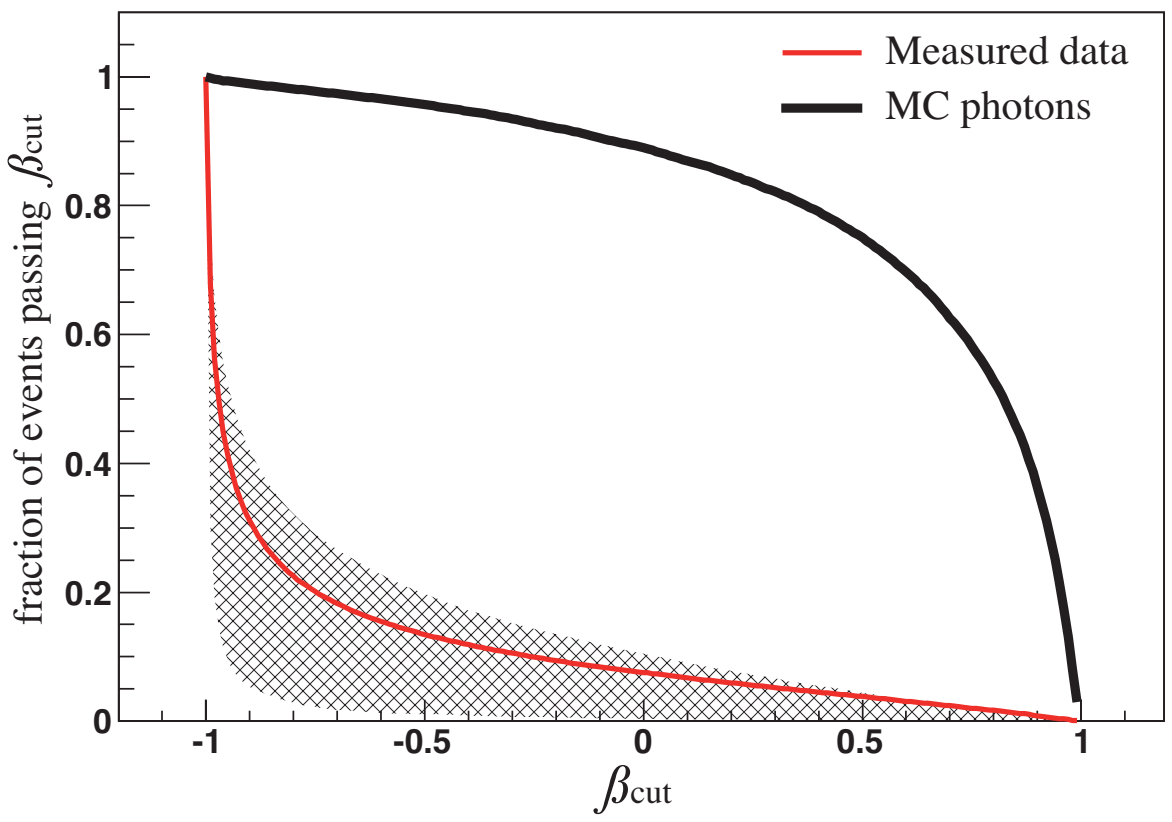

**Figure 2**: Fraction of events passing $\beta_{\text{cut}}$ for primary photons (black) and measured averaged hybrid data (red). The shaded area represents the expectation of a purely hadronic composition derived from MC simulations.

with $\alpha_{\text{CL}} \equiv 1 - \text{CL}$ and the expected background contribution $n_b^\beta$ (cf. [20]). Since $n_b^\beta$ (and hence $n_{\text{data}}$) is not an integer in general, a continuous function of the Poisson expectation is used. The lowest upper limit $n_s^{\text{Zech}}$ is determined by minimizing

$$n_s^{\text{Zech}} = \min\left(\frac{n_s(\beta)}{\varepsilon_\gamma^\beta}\right) . \tag{5}$$

The **directional photon flux upper limit** from a point source is the limit on the number of photons from a given direction divided by the directional acceptance (cf. Sec. 4) from the same target at a confidence level of CL = 95%. The upper limit on the number of photons is calculated using Eqn. (4) and given by

$$f^{\text{UL}} = \frac{n_s^{\text{Zech}}}{n_{\text{inc}} \cdot \mathscr{E}} , \tag{6}$$

where $n_{\text{inc}} = 0.9$ is the expected signal fraction in a top-hat search region and $\mathscr{E}$ the photon exposure (cf. Sec. 4).

When performing a **blind search for photon point sources** the probability $p$ of obtaining a test statistic at least as extreme as the one that was actually observed is calculated, assuming that the hypothesis of an isotropic distribution is true. The test statistic is obtained from the ensemble of scrambled datasets (cf. Sec. 4) assuming a Poisson distributed background. This $p$-value is calculated for a specific target direction as

$$p = 1 - \left(\sum_{i=0}^{n_{\text{data}}-1} \text{Poisson}(i, n_b)\right) , \tag{7}$$

where $\text{Poisson}(i, n_b)$ is the Poisson probability to observe $i$ events expecting a background count of $n_b$. The chance probability $p_{\text{chance}}$ to observe that $p_{\text{min}}$ anywhere in the sky is given by

$$p_{\text{chance}}(p_{\text{min}}^{\text{scr}} \leq p_{\text{min}}) , \tag{8}$$

where $p_{\text{min}}^{\text{scr}}$ is the minimum $p$-value of a simulated scrambled dataset.

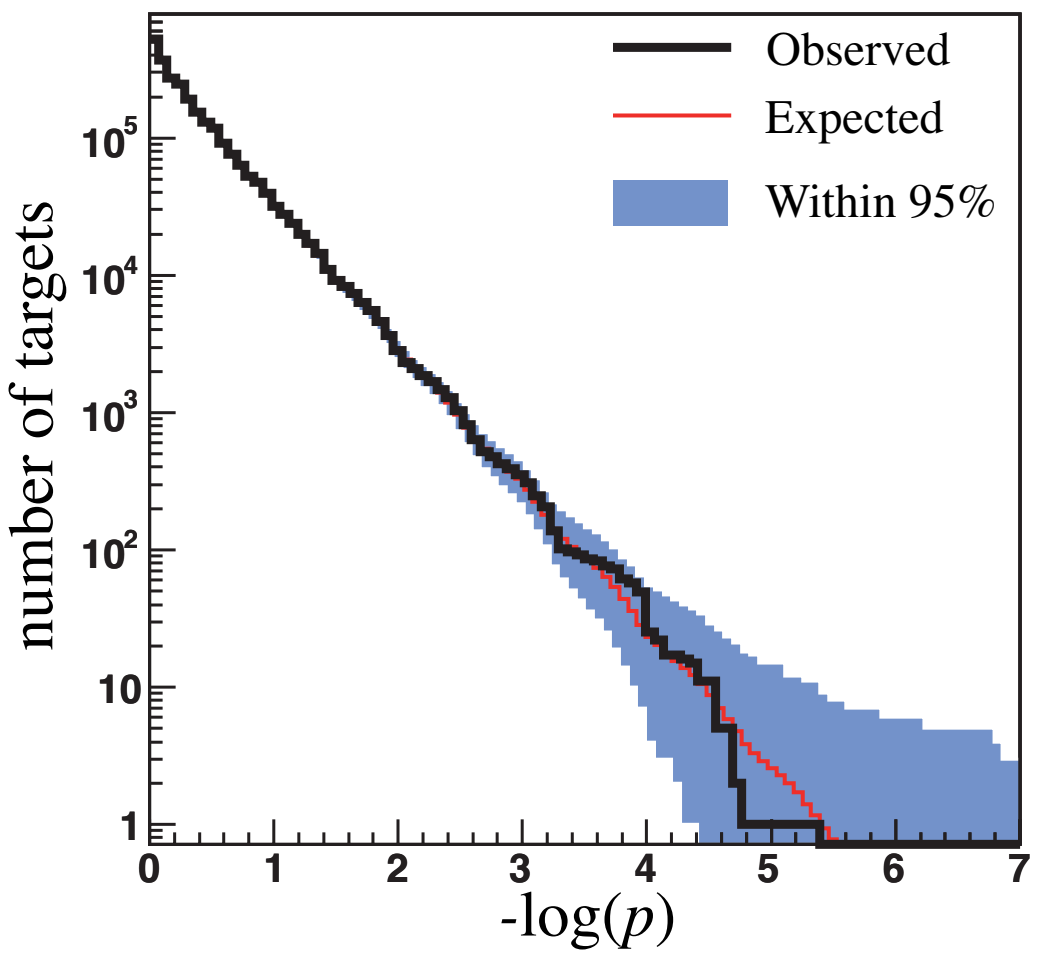

**Figure 3**: Integral distribution of $p$-values. For better visibility $-\log(p)$ is shown. The observed distribution is shown as black line, the mean expected as red line. The blue shaded region corresponds to 95% containment of simulated data sets.

## 4 Background expectation and photon acceptance

The contribution of an isotropic background is obtained using the scrambling technique [18]. In a first step the arrival directions (in local coordinates) of the events are smeared out randomly according to their individual reconstruction uncertainty. In a second step $N_{\text{data}}$ events are formed by choosing randomly a local coordinate and, independently, a Coordinated Universal Time (UTC) from the pool of measured directions and times. This procedure is repeated 5000 times. The mean number of arrival directions within a target is then the expected number for that particular sky location. The expected number of events after cutting in the $\beta$ distribution can be calculated as $n_b^\beta = n_b \cdot \varepsilon_{\text{data}}^\beta(\delta)$. As each telescope bay has different azimuthal trigger probabilities events are binned telescope-wise before scrambling the data. Since the described scrambling technique is less effective to the southern galactic pole region[3], declinations $< -85°$ are omitted from the analysis.

To derive an upper limit on the photon flux an estimate of the exposure of the detector to photon primaries is needed. The exposure for the hybrid detector is not constant with energy and is not uniform in right ascension. Thus, detailed simulations have been performed to take into account the status of the detector and the dependence of its performances with energy and direction (both zenith and azimuth). For the exposure calculation applied here, time dependent simulations have been performed, following the approach described in [17]. The total exposure can be derived as :

$$\mathscr{E} = \mathscr{E}_{\text{prof}} \cdot \varepsilon_\gamma^\beta , \tag{9}$$

where $\mathscr{E}_{\text{prof}}$ indicates the exposure at profile reconstruction level, i.e. before applying a multivariate cut, and $\varepsilon_\gamma^\beta$ the photon efficiency applying a $\beta_{\text{cut}}$.

3. At the pole, the estimated background would always be similar to the observed signal. Therefore a possible excess or deficit of cosmic rays from the pole would always be masked.

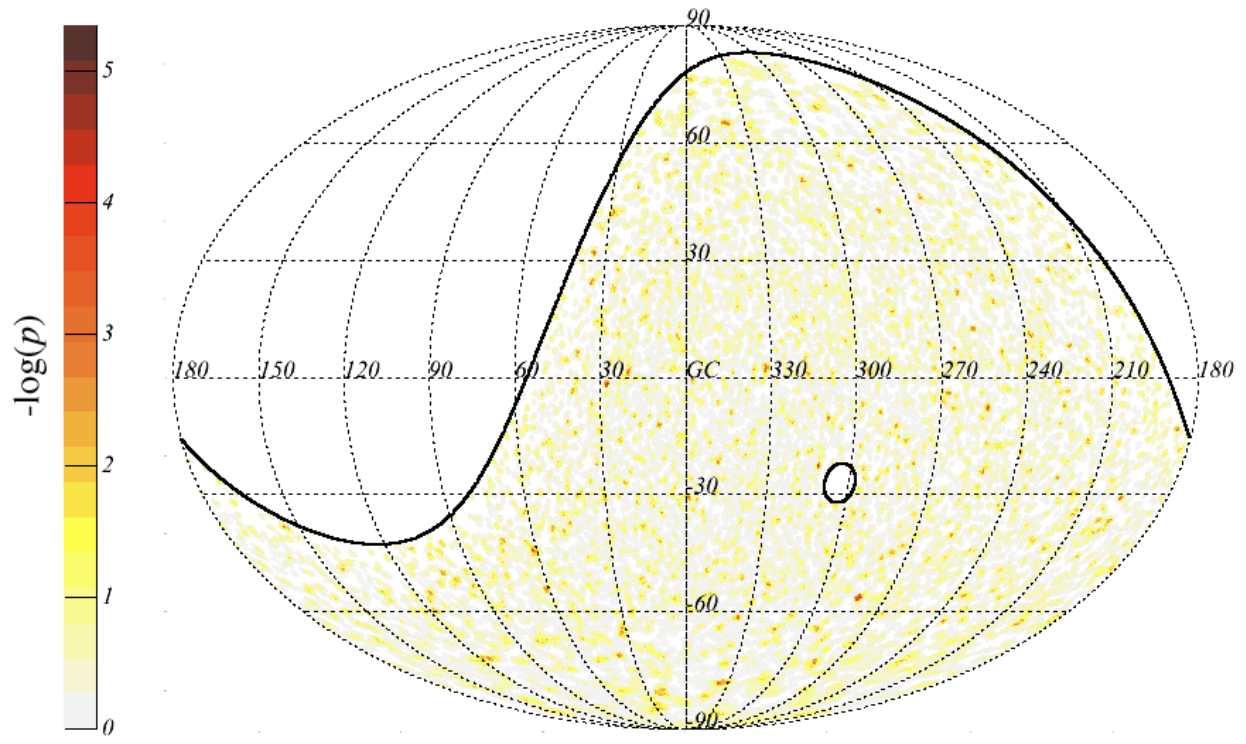

**Figure 4**: Celestial map of $-\log(p)$ values in galactic coordinates.

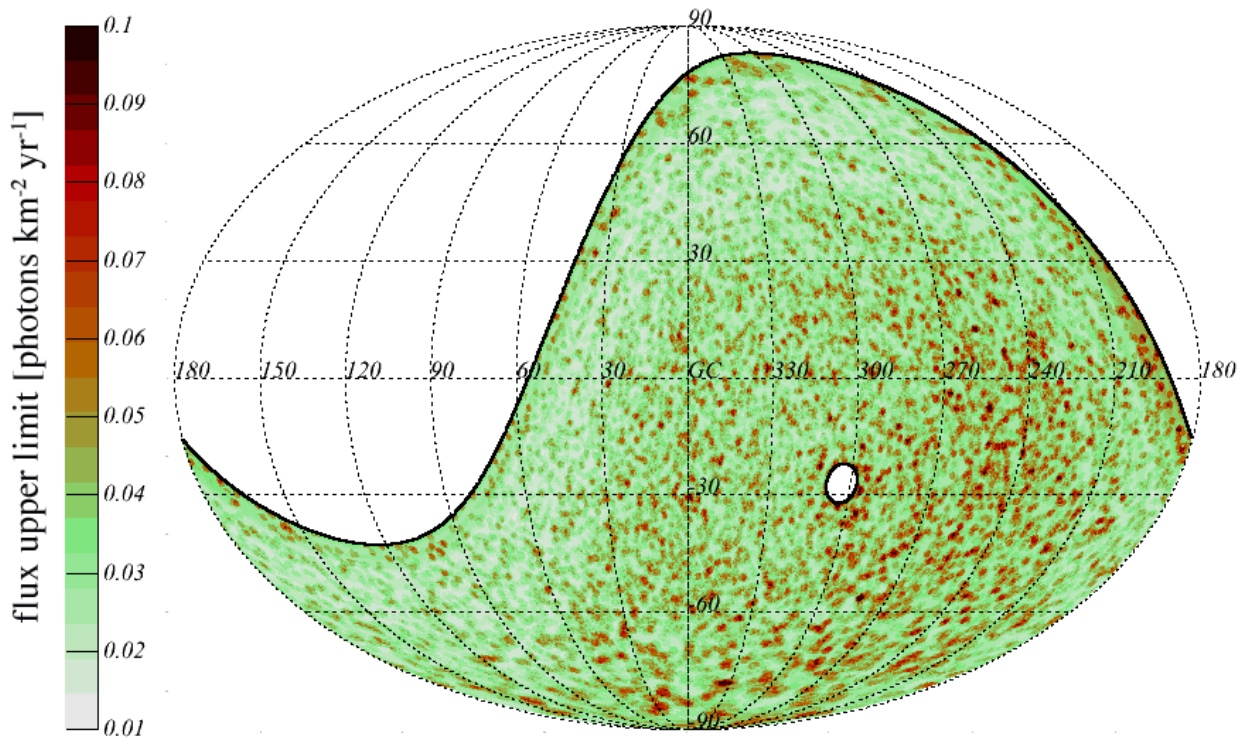

**Figure 5**: Celestial map of photon flux upper limits in $\left[\frac{\text{photons}}{\text{km}^2\cdot\text{yr}}\right]$ illustrated in galactic coordinates.

## 5 Results and discussion

The integral distribution of $-\log(p)$-values is shown in Fig. 3. The corresponding sky map of $-\log(p)$-values is illustrated in Fig. 4. The minimum $p$-value observed is $p_{\rm min} = 4.5 \cdot 10^{-6}$ corresponding to a chance probability that $p_{\rm min}$ is observed anywhere in the sky of $p_{\rm chance} = 36\%$.

Directional photon flux upper limits of point sources (95% confidence level) are derived using Eqn. (6) and shown as a celestial map in Fig. 5. The mean value is 0.035 photons km$^{-2}$ yr$^{-1}$ with a maximum of 0.14 photons km$^{-2}$ yr$^{-1}$ corresponding to an energy flux of 0.06 eV cm$^{-2}$ s$^{-1}$ and 0.25 eV cm$^{-2}$ s$^{-1}$ respectively, assuming an $E^{-2}$ energy spectrum. The energy flux in TeV gamma rays exceeds 1 eV cm$^{-2}$ s$^{-1}$ for some galactic sources with a differential spectral index of $E^{-2}$ [21, 22]. Suppose those gamma rays arise from the decay of $\pi^0$ mesons produced by interactions of protons accelerated at the source. A source with a differential spectral index of $E^{-2}$ puts equal energy in each decade, resulting in an expected energy flux of 1 eV cm$^{-2}$ s$^{-1}$ in the EeV decade. No energy flux that strong in EeV gamma rays is observed from any TeV source in the field of view or from any other target direction. These limits on regularly emitting non-beamed photon sources in the Galaxy constrain models for the acceleration in the Galaxy of the EeV protons that are measured [23].

Various sources of systematic uncertainties have been investigated and the impact on the mean flux upper limit is estimated. A change of the photon flux spectral index by $+0.5$ and $-0.5$ changes the limit by about $-34\%$ and $+51\%$, respectively. Simulation uncertainties of the fraction of photon ($\varepsilon_\gamma^\beta$) and measured events ($\varepsilon_{\rm data}^\beta$) passing a $\beta_{\rm cut}$ including a possible directional dependance contribute less than 6%. A systematic uncertainty of the Auger energy scale of $+20\%$ and $-20\%$ changes the upper limit by about $+11\%$ and $-14\%$, respectively. The size of the top-hat counting region and, correspondingly, the impact of a changing angular resolution results in a $+9\%$ change for a top-hat radius of $0.74°$ (67% containment) and a $+11\%$ change for a top-hat radius of $1.5°$ (99.5% containment). Studying the impact of different hadronic interaction models in the MVA introduces a change of 9% for the mean upper limit of the photon flux.

# Ultra-high energy neutrinos at the Pierre Auger Observatory

PABLO PIERONI[1] FOR THE PIERRE AUGER COLLABORATION[2]

[1]*Facultad de Ciencias Exactas y Naturales, Universidad de Buenos Aires, Argentina*
[2]*Full author list: http://www.auger.org/archive/authors_2013_05.html*

*auger_spokespersons@fnal.gov*

**Abstract:** Neutrinos in the sub-EeV energy range and above can be detected with the Surface Detector array (SD) of the Pierre Auger Observatory. They can be identified through the broad time-structure of the signals expected to be induced in the SD stations. The identification can be efficiently done for neutrinos of all flavours interacting in the atmosphere at large zenith angles, typically above 60° (downward-going), as well as for Earth-skimming neutrino interactions in the case of tau neutrinos (upward-going). The wide angular range calls naturally for three sets of identification criteria designed to search for downward-going neutrinos in the zenith angle bins $60^\circ - 75^\circ$ and $75^\circ - 90^\circ$ as well as for upward-going neutrinos. In this contribution the three searches are combined to give a single limit, providing, in the absence of candidates in data from 1 January 04 until 31 December 12, an updated and stringent limit to the diffuse flux of ultra-high energy neutrinos.

## 1 Introduction

Ultra-high energy (UHE) neutrinos in the EeV range have so far escaped the scrutinity of existing experiments. At these energies, neutrinos may be the only probe of the still-enigmatic sources of UHE cosmic rays at distances further than $\sim 100$ Mpc. UHE neutrinos are produced in the decay of pions created in the interactions of cosmic rays with matter and/or radiation at their potential sources, such as gamma-ray bursts or Active Galactic Nuclei among others [1]. In addition, above $\sim 4\ 10^{19}$ eV cosmic-ray protons interact with cosmic microwave background photons and produce *cosmogenic* neutrinos of energies typically $1/20$ of the proton energy. However, their fluxes are uncertain, and if the primary cosmic-ray flux is dominated by heavy nuclei the UHE neutrino yield would be strongly suppressed [2].

Neutrinos in the sub-EeV energy range and above can be detected with the Surface Detector array (SD) of the Pierre Auger Observatory [3]. A search in Auger data from 1 January 04 up to 31 December 12 has yielded no candidates and updated limits to the UHE neutrino flux are presented.

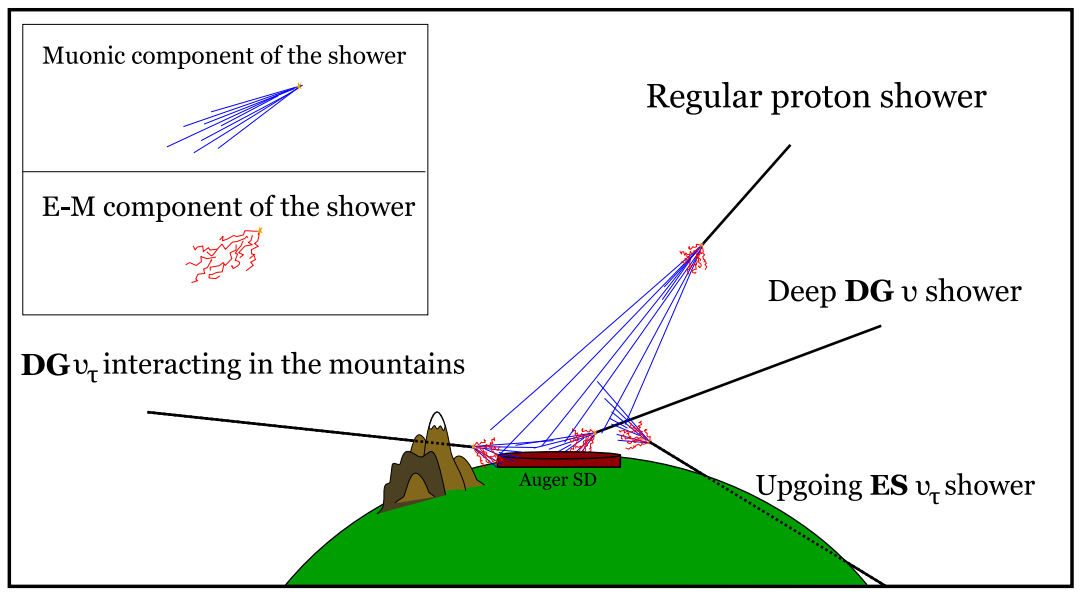

**Figure 1**: Pictorial representation of the different types of inclined showers that can be detected at the SD of the Pierre Auger Observatory. A regular inclined shower induced by a proton interacting high in the atmosphere; a deep downward-going (DG) $\nu$-induced shower (downward-going high angle - DGH, and downward-going low angle - DGL channels, see Table 1); an Earth-skimming (ES) $\nu_\tau$ interacting in the Earth's crust and producing an upward-going $\tau$ lepton decaying in flight and inducing a shower in the atmosphere; and a $\nu_\tau$ interacting in the mountains, producing a downward-going $\tau$ lepton decaying and initiating a shower close to the SD (contributing to the DGH channel).

## 2 Searching for UHE neutrinos in Auger

Although the SD array of the Pierre Auger Observatory is primarily used for the collection of UHE cosmic rays, UHE neutrinos can also be observed. Unlike cosmic-rays, UHE neutrinos can initiate downward-going showers starting at very large depths in the atmosphere as well as upward-going showers close to the ground in interactions in the Earth's crust (see Fig. 1). Due to this, a strong background reduction is possible so that the search for neutrinos with Auger is currently limited not by background but by exposure.

Since the depth at which the shower is initiated cannot be measured directly, surrogate observables are used. In the stations of the SD of the Auger Observatory, the signals produced by the passage of shower particles are digitised with 25 ns resolution. This allows us to distinguish narrow signals in time induced by inclined showers initiated high in the atmosphere, from the broad signals expected in inclined showers initiated close to the ground [4]. From the observational point of view, a Time-over-Threshold (ToT) trigger is usually present in SD stations with signals extended in time, while narrow signals induce other local triggers. Also the Area-over-Peak (AoP) of the signals, defined as the signal divided by its peak value, provides an estimate of the spread-in-time of the traces, and serves as an observable to discriminate broad from narrow shower fronts.

Using Monte Carlo simulations, it has been established that neutrino identification with the SD of the Auger Observatory can be performed efficiently as long as the search is restricted to showers with zenith angles $\theta > 60°$. The search was performed in three angular ranges. For this purpose three different sets of identification criteria were established to maximise the discrimination power. The selections exploit the different characteristics of the showers in each angular bin as determined from Monte Carlo simulations (see Table 1). For instance, at high zenith angles, above $\theta \sim 75°$, Monte Carlo simulations of showers initiated close to ground indicate that their fronts are broader in time than those of conventional cosmic-ray showers, but only in the stations that are triggered first (*early* stations) [6]. This is to be expected as the electromagnetic component is attenuated by the additional air that is traversed by the shower to reach the later water-Cherenkov detectors. The same study performed with simulations of deep, inclined, showers with $\theta \in (60°, 75°)$ indicate that this happens in the early stations closer to the shower core [8].

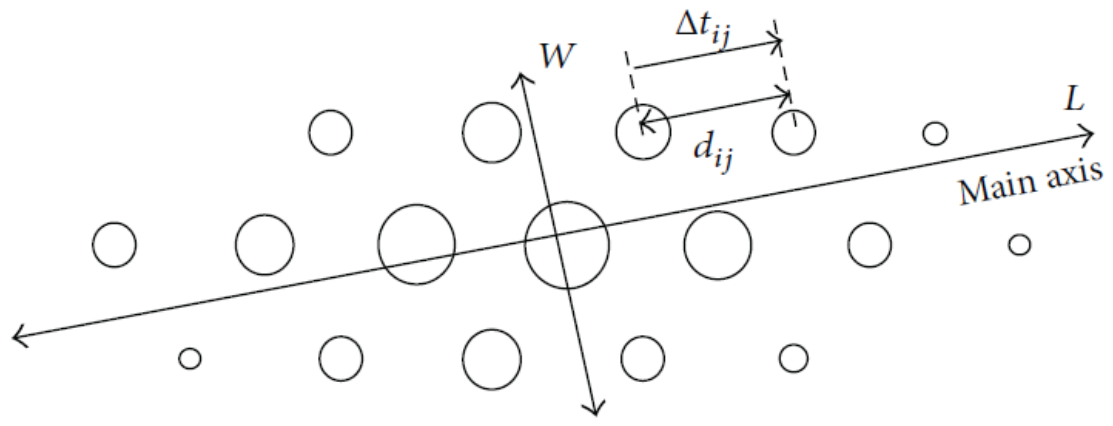

**Figure 2**: Schematic representation of the footprint of a shower triggering the SD array from the left to the right of the figure, along the "main axis". The circles represent the position of the stations, with their sizes being proportional to the signals collected in the PMTs. The length and width of the footprint (see text for details) are also indicated.

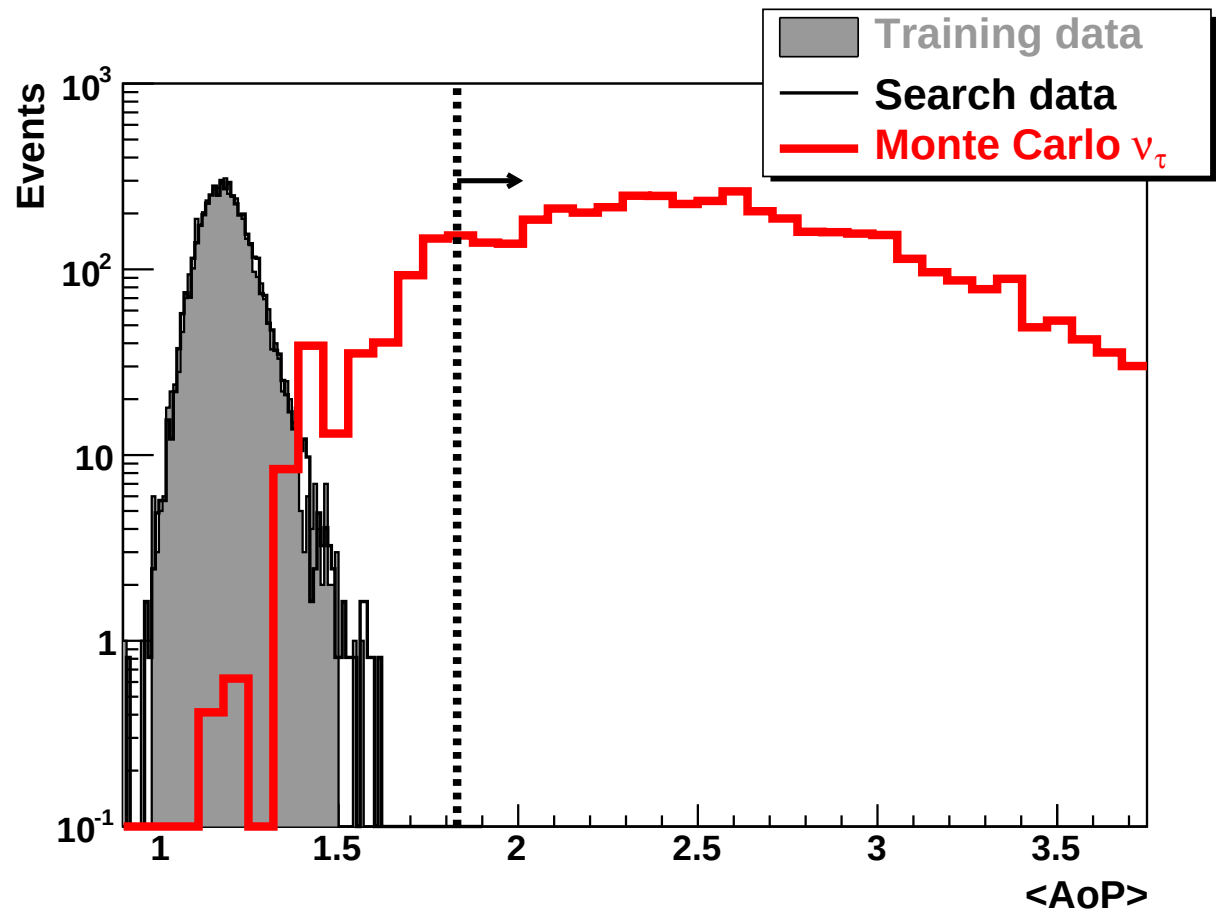

**Figure 3**: Distributions of $\langle$AoP$\rangle$ (the variable used to identifiy neutrinos in the ES selection for data after 1 June 2010 - see Table 1). Gray-filled histogram: distribution of $\langle$AoP$\rangle$ for the data in the training period. Black thin line: distribution of $\langle$AoP$\rangle$ for the data in the search period. The distributions are normalised to the same amount of events for comparison purposes. Red thick line: distribution of $\langle$AoP$\rangle$ for the simulated ES $\nu_\tau$ events. The dashed vertical line represents the cut on $\langle$AoP$\rangle > 1.83$ above which a data event is regarded as a neutrino candidate.

The three selections (denoted as ES, DGH, DGL - see Table 1 and Fig. 1) start with a "trace cleaning" procedure that removes most of the accidental signals (mainly due to atmospheric muons). Also PMTs not passing quality cuts are removed. After that, inclined showers are identified. In these events the triggered stations typically form elongated patterns on the ground along the azimuthal direction of arrival of the event. A length $L$ and a width $W$ can be assigned to the pattern [5], as shown in Fig. 2, and a cut on their ratio $L/W$ is applied in the ES and DGH selections. Also, the apparent speed $V$ of propagation of the signal on ground as the shower front arrives contains information on the inclination of the event [5]. $V$ is used in the ES and DGH selections, and can be calculated from the difference in trigger times of the signals $\Delta t_{ij}$ and the distances $d_{ij}$ between stations projected onto $L$ (Fig. 2). Finally, in the DGH and DGL selections, we reconstruct the event zenith angle $\theta_{rec}$ and place a cut on it as given in Table 1.

The next step is to identify in the data sample containing inclined showers, those with a broad time structure (*young* showers). The strategy to do this is the same for the three selections. First, different fractions of data are used to *train* each selection, assuming the *training data samples* are overwhelmingly, if not totally, made up by background showers. In these samples, the distributions of the observables applied to discriminate young showers, are used to obtain the values of the cuts as explained below. In the ES selections the discriminating variables include the fraction of stations with ToT for data prior to 31 May 10 [5], and the average value of AoP ($\langle$AoP$\rangle$) for data beyond 1 June 10. In the DGH [6] and DGL [8] selections, the AoP of various stations along with other observables constructed from AoP, are combined in a linear Fisher-discriminant polynomial (see Table 1). As an example of the power of the discriminators used, we show in Fig. 3 the distribution of $\langle$AoP$\rangle$ in data after the inclined ES selection is applied, as well as the corresponding distribution in Monte Carlo simulations of Earth-Skimming $\nu_\tau$ interacting in the Earth's crust and producing a $\tau$ that flys through the Earth and decays close to the SD.

We have observed that the tails of the Fisher and $\langle$AoP$\rangle$ distributions are consistent with an exponential shape in all cases, and we fitted and extrapolated them to find the value of the cuts corresponding to less than 1 expected event per 50 yr on the full SD array for each selection [6, 8]. Roughly $\sim 95\%$, $\sim 85\%$, and $\sim 60\%$ of the inclined neutrino events interacting through the charged-current channel, are kept after the ES, DGH and DGL selection of young showers respectively. The smaller efficiencies for the identification of neutrinos in the DGL selection are due to the more stringent criteria in the angular bin $\theta \in (60°, 75°)$ needed to reject the larger contamination from cosmic-ray induced showers.

The three selection criteria are applied to data between 1 January 04 up to 31 December 12 in a search for neutrino candidates. This period includes a new *search sample* corresponding to data from 1 June 10 until 31 December 12 not previously *unblinded* under any of the three selections. For each selection the corresponding training periods, which are different for each channel, as well as periods with problems in the data acquisition [5], are excluded from the search. No neutrino candidates were found with any of the selections in any of the blind search periods and three distinct upper limits on the diffuse flux of UHE neutrinos can be placed.

After the unblinding, we tested the compatibility of the distributions of the young shower discriminating observables in the *search* and *training* samples. These two distributions are shown for the example of the $\langle$AoP$\rangle$ variable in Fig. 3. No statistically significant differences of the shapes are observed. In particular the parameters of exponential fits to the tails of the distributions are compatible within statistical uncertainties ($\sim\ 1\ \sigma$). All the events in the search sample are still rather far from the cut value $\langle$AoP$\rangle = 1.83$. The same is true for the Fisher-discriminant distributions of the training and search samples used in the downward-going selections [6, 8].

| Selection | Earth-skimming (ES) | Downward-going *high* angle (DGH) | Downward-going *low* angle (DGL) |
|---|---|---|---|
| Flavours & Interactions | $\nu_\tau$ CC | $\nu_e,\ \nu_\mu\ ,\nu_\tau$ CC & NC | $\nu_e,\ \nu_\mu\ ,\nu_\tau$ CC & NC |
| Angular range | $\theta > 90°$ | $\theta \in (75°, 90°)$ | $\theta \in (60°, 75°)$ |
| N° of Stations (Nst) | Nst $\geq 3$ | Nst $\geq 4$ | Nst $\geq 4$ |
| Inclined Showers | – <br> $L/W > 5$ <br> $\langle V\rangle \in (0.29,\ 0.31)$ m ns$^{-1}$ <br> RMS$(V) <\ 0.08$ m ns$^{-1}$ | $\theta_{\rm rec} > 75°$ <br> $L/W > 3$ <br> $\langle V\rangle\ <\ 0.313$ m ns$^{-1}$ <br> RMS$(V)/\langle V\rangle < 0.08$ | $\theta_{\rm rec} \in (58.5°,\ 76.5°)$ <br> – <br> – <br> – |
| Young Showers | Data: 1 January 04 - 31 May 10 <br> $\geq 60\%$ of stations with <br> ToT trigger & AoP $> 1.4$ <br> Data: 1 June 10 - 31 December 12 <br> $\langle$AoP$\rangle > 1.83$ <br> AoP$_{\rm min} > 1.4$ if Nst=3 | Fisher discriminant based on AoP of *early* stations | $\geq 75\%$ of stations close to shower core with ToT trigger & Fisher discriminant based on AoP of *early* stations close to shower core |

**Table 1**: Observables and numerical values of cuts applied to select *inclined* and *young* showers for Earth-skimming and downward-going neutrinos. See text for explanation.

## 3 Combined exposure

To make the best use of the Auger data and put the most stringent limits to the diffuse flux of UHE neutrinos, the three individual exposures are combined into a single total exposure by means of a simple procedure. The three sets of selection criteria are applied to each simulated neutrino event regardless of the angular bin in which it was simulated. If the event passes any of the three selections it is identified as a neutrino and contributes to the exposure at the point in the parameter space, where it was simulated (for instance neutrino energy $E_\nu$, zenith angle $\theta$, and interaction depth in the case of the downward-going channels). An integration over the whole parameter space except for energy, yields the exposure $\mathscr{E}_i(E_\nu)$ for each of the selections ($i$ can be ES, DGH or DGL). Then, the total combined exposure $\mathscr{E}_{\rm tot}(E_\nu)$ is obtained by adding the three individual exposures obtained in this way, $\mathscr{E}_{\rm tot} = \mathscr{E}_{\rm ES} + \mathscr{E}_{\rm DGH} + \mathscr{E}_{\rm DGL}$, as shown in Fig. 4. With this procedure the exposure is enhanced as, for instance, an atmospheric shower induced by a tau neutrino interacting in the Earth's crust, might not fulfill the requirements of the ES selection, but might pass the cuts of the DGH and contribute to $\mathscr{E}_{\rm ES}(E_\nu)$. Moreover, with this procedure we avoid double-counting of events: once a simulated neutrino fulfills the criteria of at least one of the selections it contributes only once to the total exposure. Finally, it is worth mentioning that in the calculation of $\mathscr{E}_i(E_\nu)$ we take into account changes in the array configuration due to installation of new stations (up to 2008) and array instabilities (see [5, 6, 8] for details).

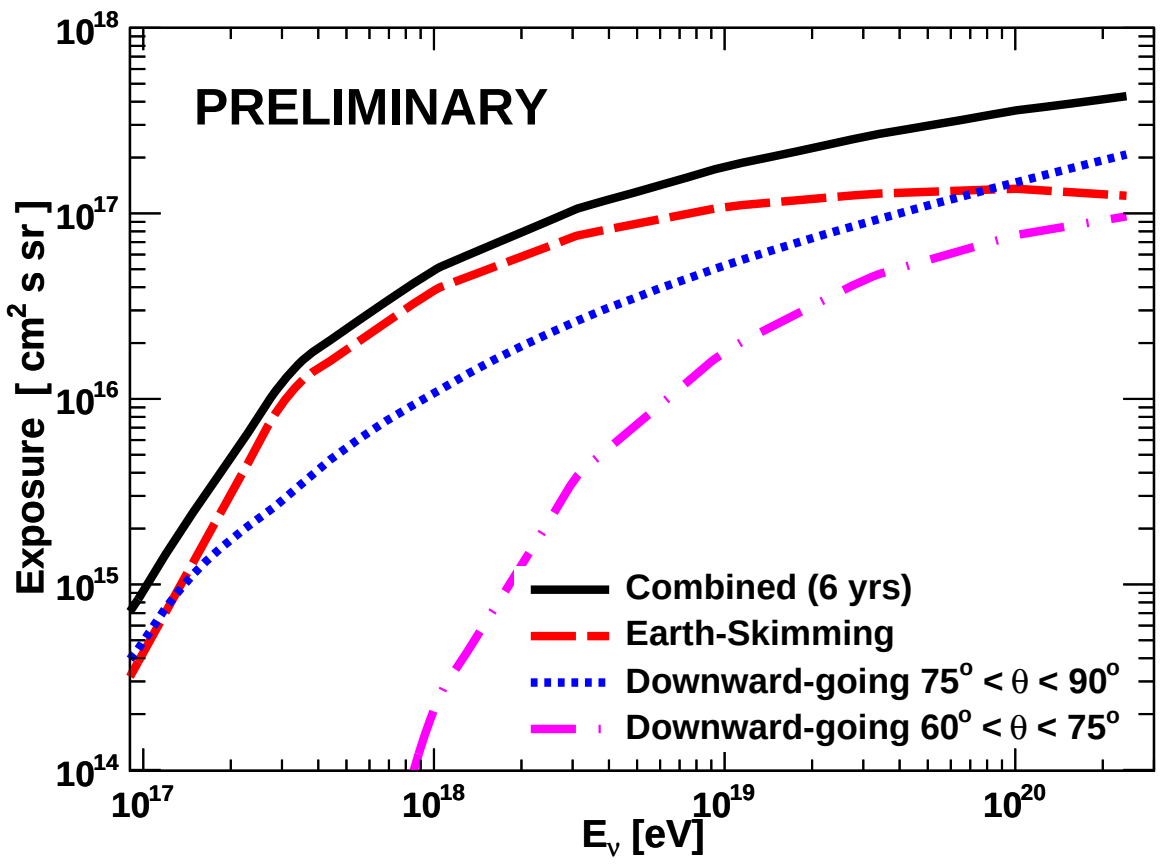

**Figure 4**: Combined exposure of the SD of the Pierre Auger Observatory as a function of neutrino energy after applying the three sets of selection criteria in Table 1 to Monte Carlo simulations of UHE neutrinos. Also shown are the individual exposures corresponding to each of the three selections. For the downward-going channels the exposure represents the sum over the three neutrino flavours as well as CC and NC interactions. For the Earth-Skimming channel, only $\nu_\tau$ CC interactions are relevant. For this channel the exposure falls at the highest energies as there is an increasing probability that the $\tau$ decays high in the atmosphere producing a no triggering shower, or even that the $\tau$ escapes the atmosphere before decaying.

Several sources of systematic uncertainties in the exposure have been investigated [5, 6]. For the downward-going analysis, the major contributions in terms of deviation from a reference exposure come from the knowledge of neutrino-induced shower simulations (+9%, -33%) and of the neutrino cross-section (± 7%) [11]. For the ES analysis, the systematic uncertainties are dominated by the energy losses of the tau (+25%, -10%), the shower simulations (+20%, -5%), and the topography (+18%, 0%).

## 4 Results and conclusions

Using the combined exposure and assuming a $\Phi(E_\nu) = k \cdot E_\nu^{-2}$ differential neutrino flux and a 1:1:1 flavour ratio, an upper limit on the value of $k$ can be obtained as:

$$k = \frac{N_{\rm up}}{\int_{E_{\rm min}}^{E_{\rm max}} E_\nu^{-2}\, \mathscr{E}_{\rm tot}(E_\nu)\, dE_\nu} \quad (1)$$

The actual value of the upper limit on the signal events ($N_{\rm up}$) depends on the number of observed and expected background events as well as on the confidence level required. Using a semi-Bayesian extension [9] of the Feldman-Cousins approach [10] to include the uncertainties in the exposure, $N_{\rm up}$ is different from the nominal value for zero candidates and no expected background ($N_{\rm up} = 2.44$ at 90% C.L.), and is different for each channel depending on the type of systematic uncertainties, and the reference exposure chosen [6, 7].

The updated single-flavour 90% C.L. limit is:

$$k_{90} \ < \ 1.3 \times 10^{-8}\ \text{GeV cm}^{-2}\ \text{s}^{-1}\ \text{sr}^{-1} \quad (2)$$

and applies in the energy interval $\sim 1.0 \times 10^{17}$ eV $- 1.0 \times 10^{20}$ eV where $\sim 90\%$ of the event rate is expected. The result is shown in Fig. 5 along with the limit in different bins of width 0.5 in $\log_{10} E_\nu$ (differential limit) to show at which energies the sensitivity of the SD of the Pierre Auger Observatory peaks. The search period corresponds to an equivalent of almost 6 years of a complete Auger SD array working continuously. The inclusion of the latest data from 1 June 10 until 31 December 12 in the search represents an increase of a factor $\sim 1.7$ in event number with respect to previous searches [6, 7]. The relative contributions of the ES:DGH:DGL channels to the total expected event rate assuming a flux behaving with neutrino energy as $E_\nu^{-2}$, are 0.73:0.23:0.04 respectively.

The current Auger limit is below the Waxman-Bahcall bound on neutrino production in optically thin sources [14]. With data unblinded up to 31 December 12, we are starting to constrain models of cosmogenic $\nu$ fluxes that assume a pure primary proton composition injected at the sources. As an example we expect $\sim 1.4$ cosmogenic neutrino events from a model normalised to Fermi-LAT observations (solid line, bottom right panel in Fig. 4 of [15], also shown in Fig. 5 in this work). The gray shaded area in Fig. 5 brackets the cosmogenic neutrinos fluxes predicted under a wide range of assumptions for the cosmological evolution of the sources, for the transition between the galactic and extragalactic component of cosmic rays, and for the UHECR composition [17]. The corresponding expected number of cosmogenic neutrino events ranges between $\sim 0.2$ and $\sim 0.6$.

The two events in the PeV energy range recently reported by the IceCube collaboration are compatible with a power-law flux which follows $E_\nu^{-2}$ with normalisation $E_\nu^2\, F_\nu = 1.2\ 10^{-8}$ GeV cm$^{-2}$ s$^{-1}$ sr$^{-1}$ for each flavour (see Fig. 5 in [18]). Extending this upper limit to the flux with the same power-law up to $10^{20}$ eV we would expect $\sim$ 2.2 events in Auger while none is observed. The possibility that such a neutrino flux also represents the flux at UHE energies is excluded at close to 90% C.L.

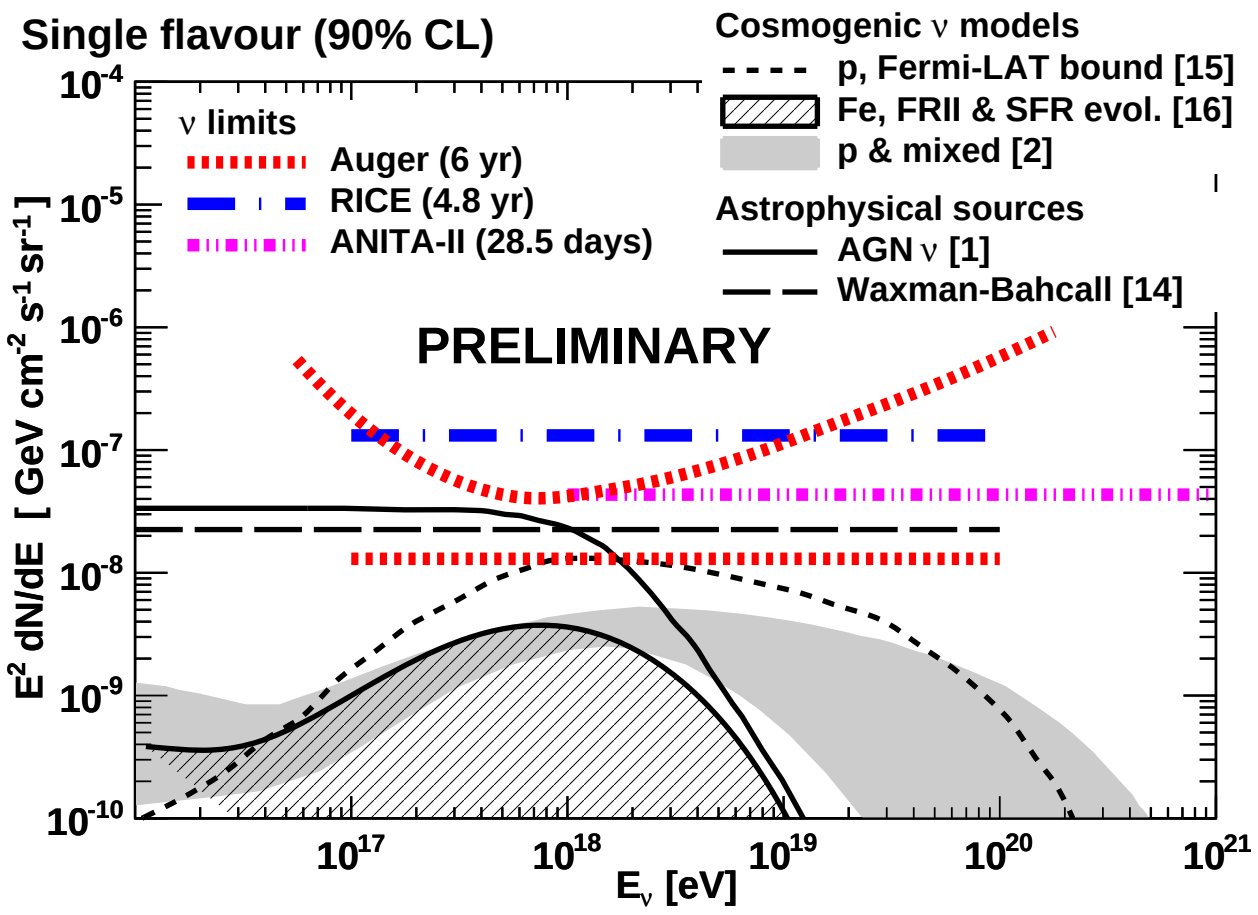

**Figure 5**: Differential and integrated upper limits (at 90% C.L.) from the Pierre Auger Observatory for a diffuse flux of UHE neutrinos. The search period corresponds to $\sim 6$ yr of a complete SD. We also show the integrated limits from ANITAII [12] and RICE [13] experiments, along with expected fluxes for several cosmogenic neutrino models [15, 16, 17] as well as for astrophysical sources [1, 14].

# The AMIGA muon detectors of the Pierre Auger Observatory: overview and status

FEDERICO SUAREZ[1], FOR THE PIERRE AUGER COLLABORATION[2].

[1] *Instituto de Tecnologías en Detección y Astropartículas (CNEA-CONICET-UNSAM), Av. Gral Paz 1499, (1650) Buenos Aires, Argentina.*
[2] *Full author list: http://www.auger.org/archive/authors_2013_05.html*

*auger_spokespersons@fnal.gov*

**Abstract:** The muon detector of the AMIGA (Auger Muons and Infill for the Ground Array) extension of the Pierre Auger Observatory is currently finishing the construction of its engineering array phase. The engineering array consists of seven detectors in a 750 m regular hexagon with buried scintillator counters in each of its vertices and center. The muon counters are buried alongside each Auger surface detector station in the infill area. Two additional twin detectors are being built to study the muon counting accuracy and the design validation. An overview of the construction and deployment of the muon scintillation detector array is presented with an emphasis on the current data analyses.

## 1 Introduction

The AMIGA project [1] is an extension of the Pierre Auger Observatory [2] to provide full efficiency detection of cosmic rays down to $\sim 10^{17}$ eV through an infill of Water Cherenkov Detectors (WCD) of the Auger Surface Detector (SD). This energy region is of great importance because it is the range where the transition from galactic to extragalactic sources of cosmic rays is expected to occur. AMIGA will also improve the cosmic ray mass identification with 30 m$^2$ muon counters buried alongside the surface detectors in the infill to directly measure the muon content of the particle showers produced by the primary particle.

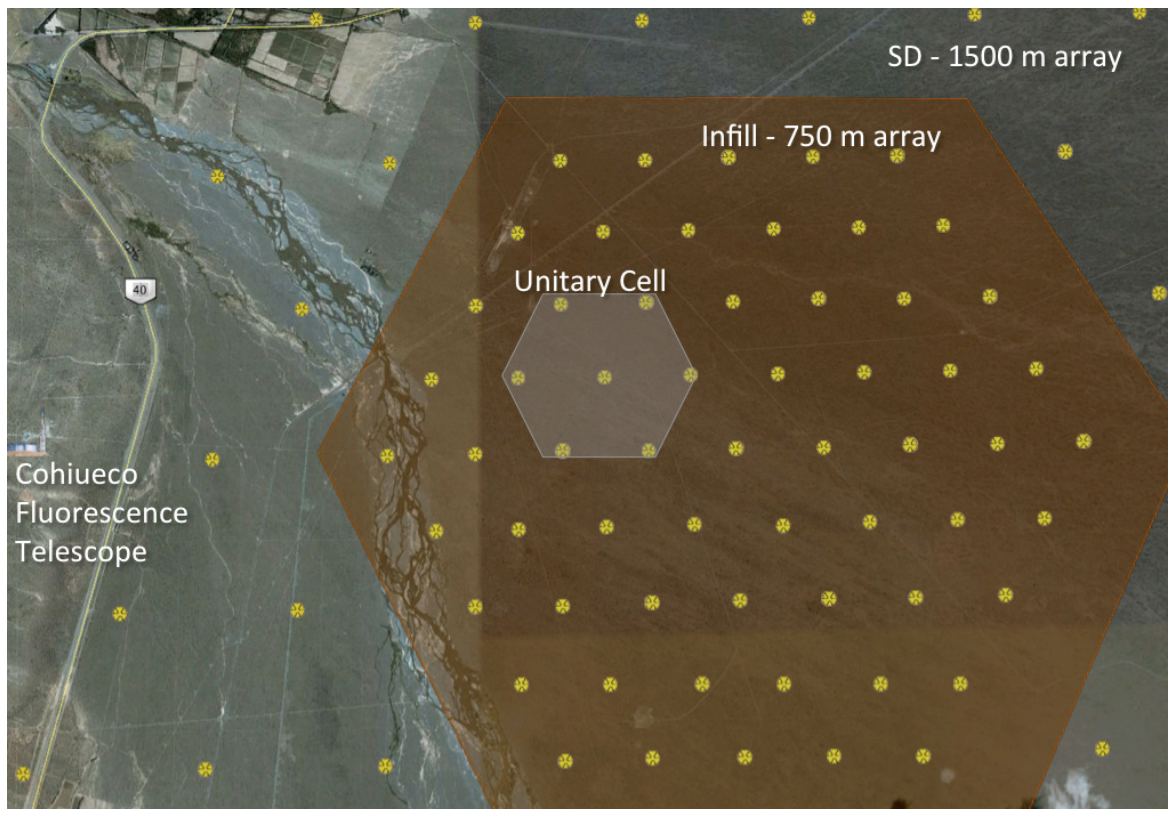

**Fig. 1**: Map of the AMIGA array with brown background. The engineering array positions, where muon counters are already deployed, can be seen highlighted in gray (also called the Unitary Cell).

The muon detectors of AMIGA are being deployed over the infilled area of $\sim$23.5 km$^2$ (fig. 1) which includes 61 stations. The first seven muon detectors are being deployed in an engineering array called Unitary Cell (UC), consisting of 30 m$^2$ counters to validate the detection technique and the detector design. In two positions, two identical 30 m$^2$ detectors are being deployed (the twins) to study the fluctuation in the counting rate for the detector design validation [3]. Finally, the rest of the 54 muon detectors will be deployed in the production phase.

Basically, the AMIGA muon counters have a modular design mainly because of the need to solve some engineering challenges. Therefore, the modules are designed to be water proof, easy, fast, and simple to manufacture, robust enough to resist long and hard transportation conditions, and small enough to fit into regular transportation trucks to reduce costs. Thus, the UC design (see fig. 2) consists of four modules covering the 30 m$^2$ divided into two modules with 10 m$^2$, and two with 5 m$^2$ detection area in each position.

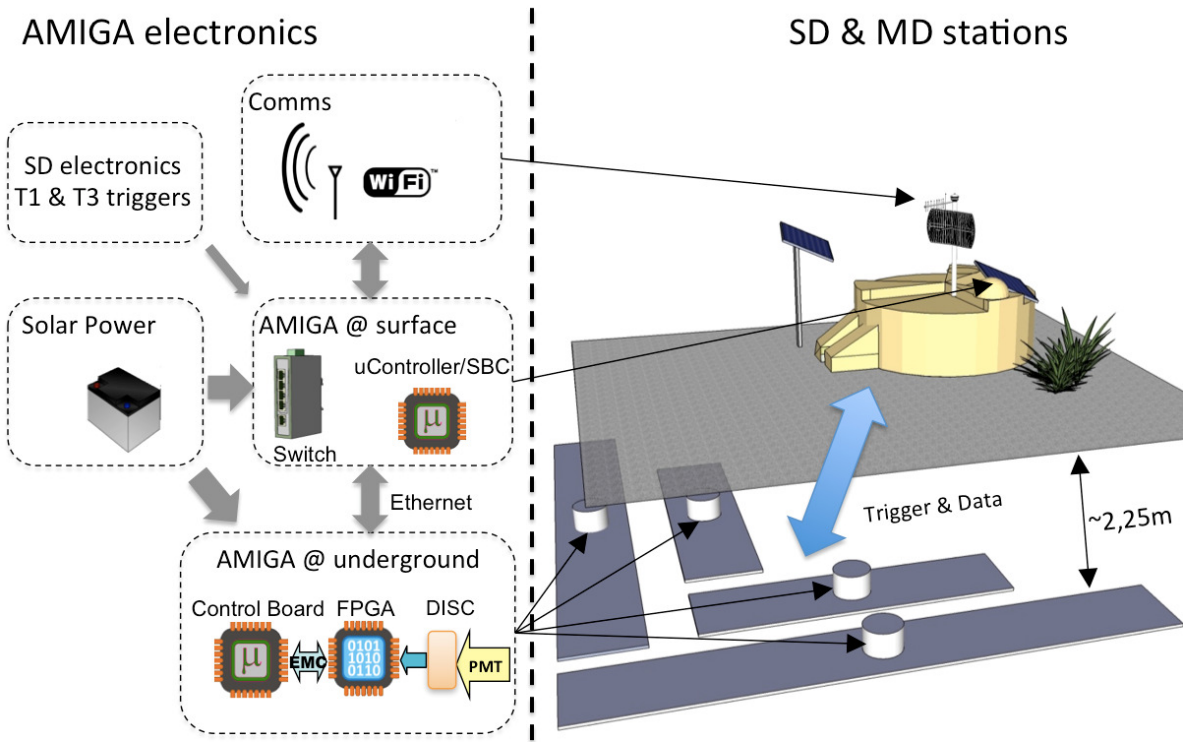

**Fig. 2**: Simple scheme of the AMIGA Unitary Cell muon detector and its electronics [6]. The discrimination levels of the signals from the PMT are adjustable by the calibration algorithm, expected to be set at $\sim$1/3 SPE (Single Photoelectron level).

Each of the UC modules [4] are segmented in 64 scintillation bars produced at Fermilab. The generated light pulses are collected by a WLS (wavelength shifter) optical fiber and then propagated to a multi-anode PMT (Photomultiplier

Tube from Hamamatsu, H8804-200MOD) [5]. As sketched in fig. 2, the readout electronics consists of an analog front-end to amplify and discriminate the pulses coming from the PMT, and a digital board with a FPGA that samples the discriminated signals at 320 MHz to conform a 1-bit digitalization of the signals. Currently, the events consist of a block of 1024 words of 64 bits where each bit corresponds to a module channel and each word to the time bin [6]. The events are stored in a local memory when a first level trigger [7] is received from the surface detector. Finally, the data of the muon counter events are transmitted to the surface through a control and interface board when a third level trigger is broadcasted to the array and re-transmitted through a Wi-Fi system to the Central Data Acquisition System of the Observatory.

## 2 Deployment of the muon detector modules

The detector modules are mostly fabricated at Buenos Aires and then transported ∼1100 km to the Pierre Auger Observatory. Then, they are taken to the field and deployed ∼2.25 m underground (∼540 g cm$^{-2}$ considering local soil) in an ”L” layout (fig. 3 and 4) in an effort to reduce the counting uncertainty produced by inclined muons that could cross two scintillators instead of one (clipping corners). The layer of soil above the detectors is used as a shielding against the electromagnetic component of the particle showers and it was simulated to be enough to avoid punch-through electrons. The modules are placed ∼5 m away from the surface detector to reduce any angular dependence due to a possible ”shading” and to avoid the removal of the surface detector to excavate the pits for the muon detector modules.

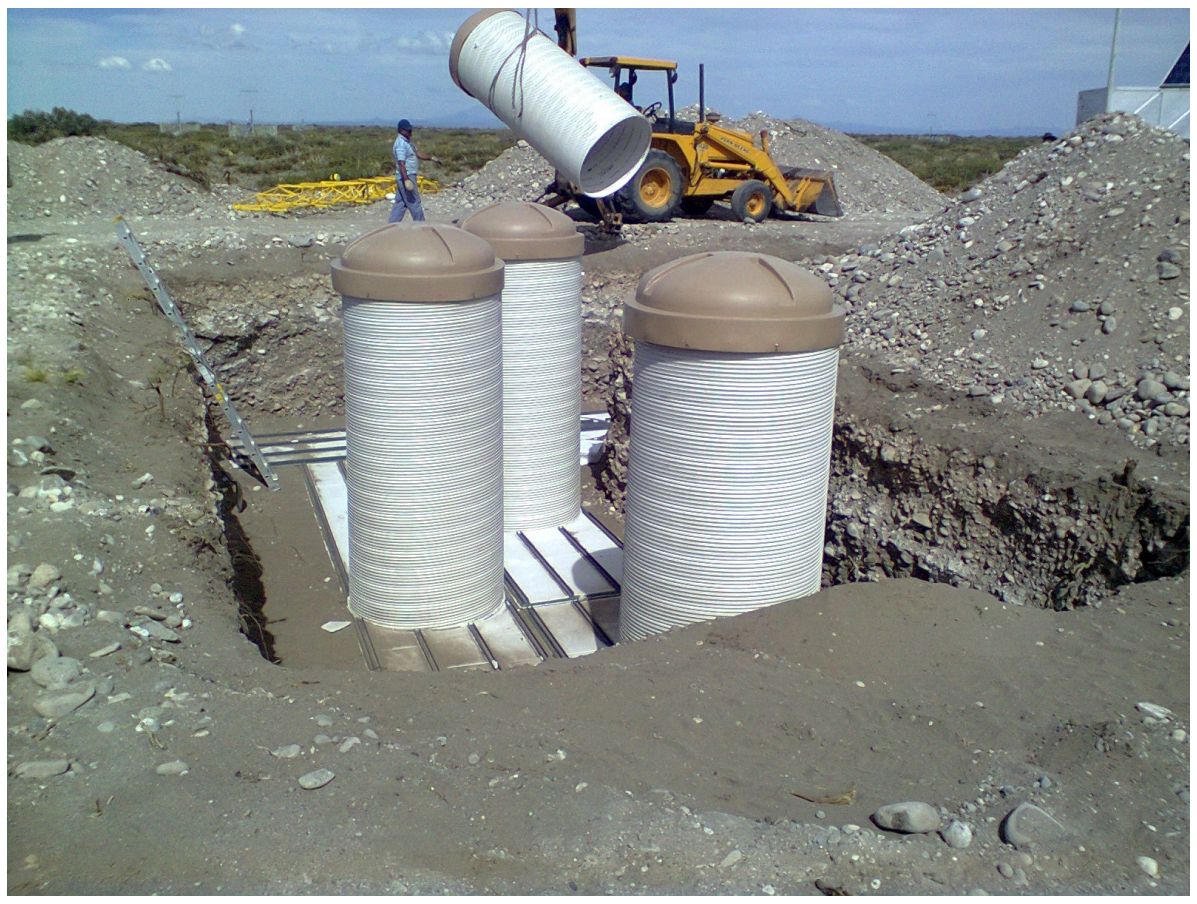

**Fig. 3**: Deployment of the first twin at Kathy-Turner when installing the service-tubes right before burying the modules.

As already mentioned in the previous section, in the case of the UC, the 30 m$^2$ muon detector is divided into four modules. The 5 m$^2$ ones provide better segmentation of the detector to make it suitable to measure particle showers closer to the core where the number of muons is higher, thus reducing the so called pile-up of muons.

From the engineering point of view, the modules and the deployment procedure are designed to provide an easy but safe installation both for the technicians and the module itself. As shown in fig. 3, the modules lay on a sand bed free of rocks in the pit to avoid damages. The pits are excavated with inclined walls at the top to reduce the possibility of collapse. The modules are also fully surrounded by polyfoam as a first protection layer against rocks but also against sun-light exposure that damages the PVC enclosure of the modules while deploying. Then, a ∼10 cm layer of fine sand is placed on top of the modules before proceeding to the final refilling of the pit.

Each of the modules has a service tube used to provide maintenance access to the electronics. The service tubes have a diameter of 1.3 m (comfortable enough for a technician). A special glue is used to connect them to the modules and provides water-tightness. The service tubes are covered with a cap to resist damage by animals, vandalism, and UV exposure. Finally, they are refilled with removable big sand bags (filled with local soil) to make a uniform shielding for the detector.

## 3 Current status of the Unitary Cell construction

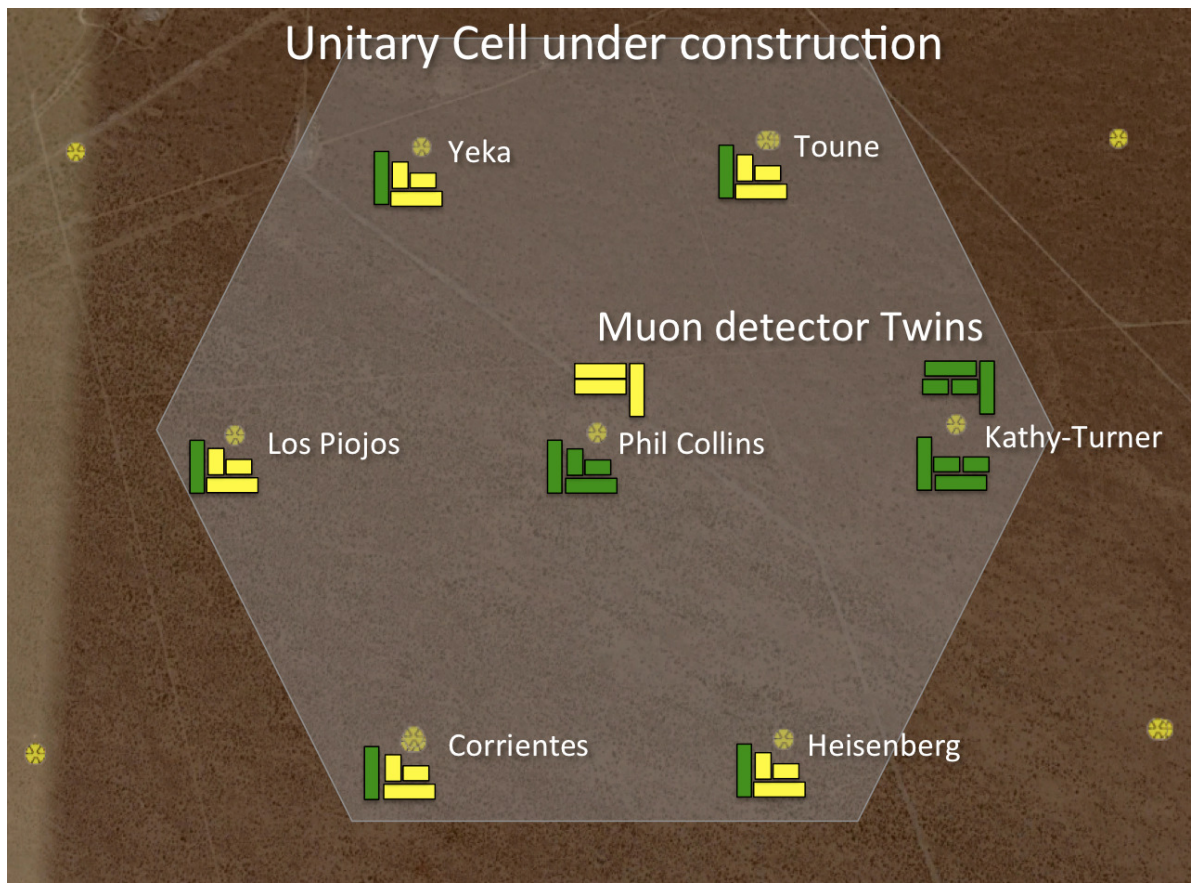

**Fig. 4**: Layout of the AMIGA Unitary Cell. The muon detectors already installed are represented in green, and those to be deployed in near future are shown in yellow. The twin muon detectors at Kathy-Turner are operating and those next to Phil Collins are still under construction. The twin detector at Phil Collins consists of three modules (10 m$^2$ each) instead of four to validate a three modules design for production.

The engineering array is currently growing to form the UC (fig. 4), and it is important to solve the engineering challenges of logistics, to develop the facilities and the module construction procedures for a reasonable construction rate, to finish the mechanical design, to develop the corresponding calibration methods, and to register a reasonable amount of events. Exploiting the UC data, we will be able to get an experimental output to confirm the simulated parameters used as the base-line design of the modules such as the number of muons per shower and their distribution in time and space. A detailed analysis of these events is thus mandatory before getting into the production phase.

The UC also includes two twin positions, i.e. there are two infill surface detectors each one associated with two 30 m$^2$ muon detectors running independently. Currently, one of the two twins is already deployed and taking data; more details about the data analysis of the events registered by the twins can be found in [3].

## 4 First muon counter events

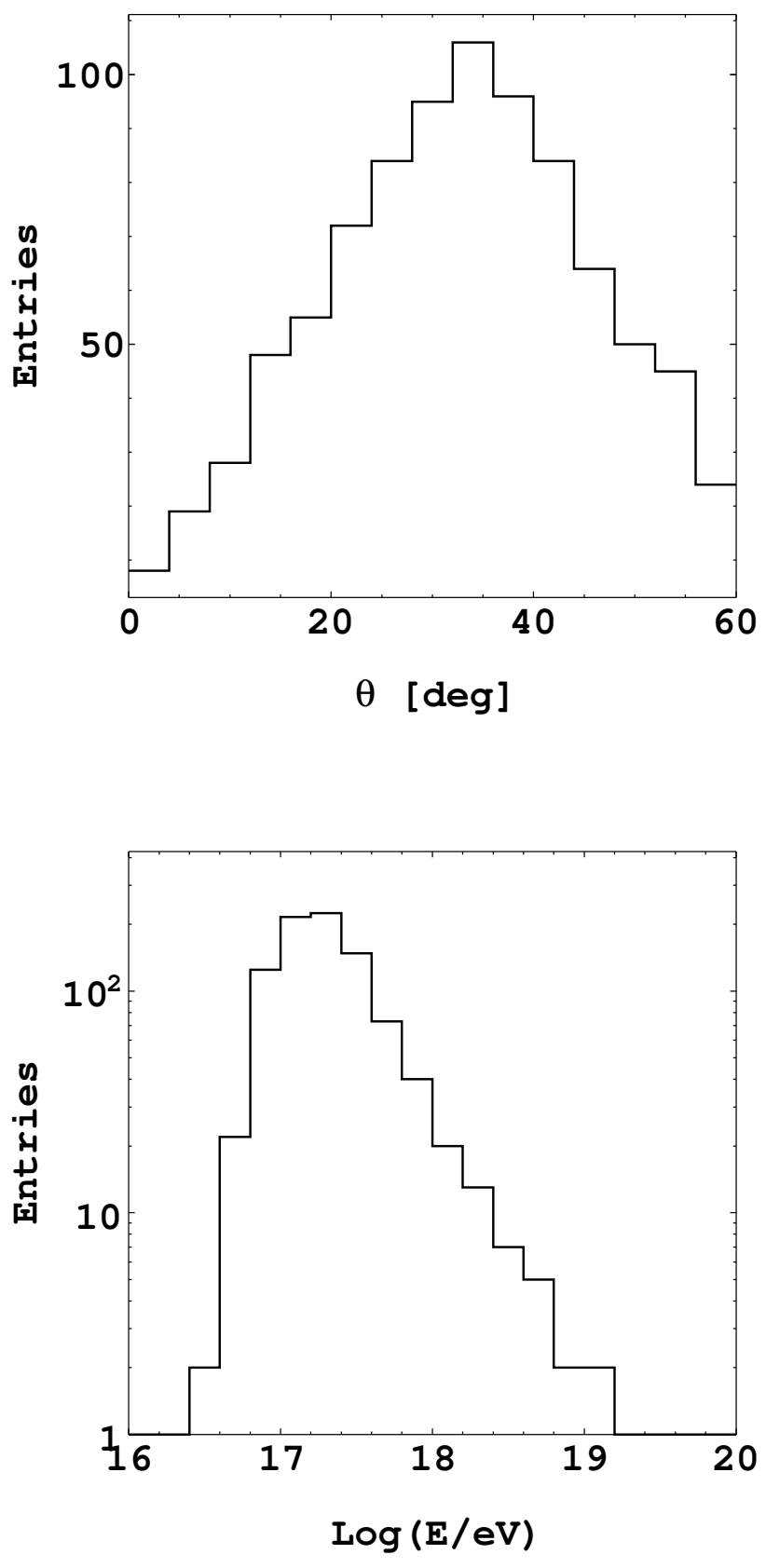

**Fig. 5**: Angular and energy distribution of recorded AMIGA events since Marh 18th 2013

Data taken by the AMIGA muon counters and by the WCD are transmitted to the Central Data Acquisition System where they are merged. The recorded events are analysed with the Offline software to reconstruct the shower parameters using the SD data alone and to extract the number of muons from the buried counters. This is done by applying a certain counting strategy that can be adjusted in order to reduce the miscounting produced by the systematic uncertainties of the detector (e.g. after pulses, channels cross-talk, dark-pulse rate) [9]. On Marh 18th 2013, the UC started the first stable AMIGA acquisition period with parameters of the baseline design, and 901 forth level trigger events were registered with zenith angle below 60° up to May 31st. As expected, most of them (671) are low energy events and only 230 are above the infill full efficiency energy $3 \times 10^{17}$ eV (see fig.5). The footprint on the MD hexagon of a $2.7 \times 10^{18}$ eV shower impinging with $\theta = 39.9°$ zenith angle can be seen in fig. 6.

Given the geometry and the energy of the shower by the SD, a KASCADE-Grande like muon lateral distribution function (MLDF) [10] is fitted to the measured muon densities.

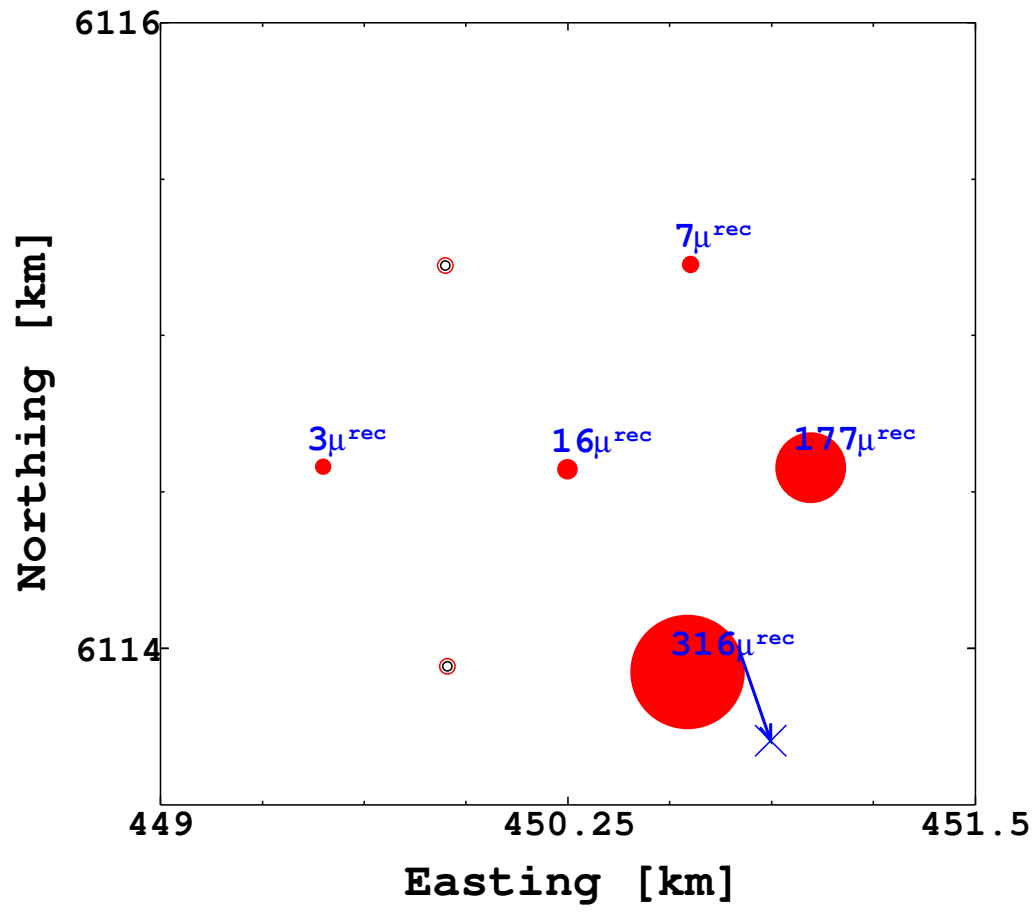

**Fig. 6**: Example of a $2.7 \times 10^{18}$ eV shower impinging with an zenith angle of $\theta = 39.9°$ that hit the border of the muon counters engineering array under construction. The reconstructed core position is marked with the cross sign and the impinging direction is indicated with the arrow. The sizes of the red dots are proportional to the reconstructed number of muons (in blue) per station. Open red circles indicate a *silent* counter, i.e, counters that received the SD triggering signal but counted less than three muons. Open black circles are untriggered counters. The *twin* MD counter is the rightmost vertex of the hexagon. As can be noticed, the 60 m$^2$ area of this counter are responsible for the 177 reconstructed muons, roughly ten times more of its closest 10 m$^2$ companion at the centre of the hexagon. The hottest counter (bottom right vertex), with 10 m$^2$ is saturated (see fig. 7), i.e. more than 21 muons where simultaneously measured in a time window of 25 ns.

$$\rho_\mu(r) = \rho_\mu(450)\left(\frac{r}{r_0}\right)^{-\alpha}\left(1+\frac{r}{r_0}\right)^{-\beta}\left(1+\left(\frac{r}{10r_0}\right)^2\right)^{\gamma} \quad (1)$$

The fitting parameters are $\rho_\mu(450)$ and $\beta$ while the others are fixed at $\alpha = 1$, $\gamma = 1.85$, and $r_0 = 150$ m.

## 5 Data analysis

Although the muon component in the showers is attenuated much less than the electromagnetic component, the shielding of ∼2.25 m of soil adds 540 g cm$^{-2}$ of vertical mass (roughly 60% more than the whole atmosphere at the level of the Auger Observatory, namely, 870 g cm$^{-2}$). However, a detailed study of attenuation is not possible yet due to the low statistics so far achieved. Nevertheless, averaging over all MLDFs normalized to their fitted parameter $\rho_\mu(450)$ allows a qualitatively inspection of the dependence of the $\beta$ parameter with the zenith angle $\theta$ of the impinging shower. In fig. 8 the mean $\rho_\mu(r)/\rho_\mu(450)$ are shown for three angular bins evenly separated in $\cos(\theta)\sin(\theta)$.

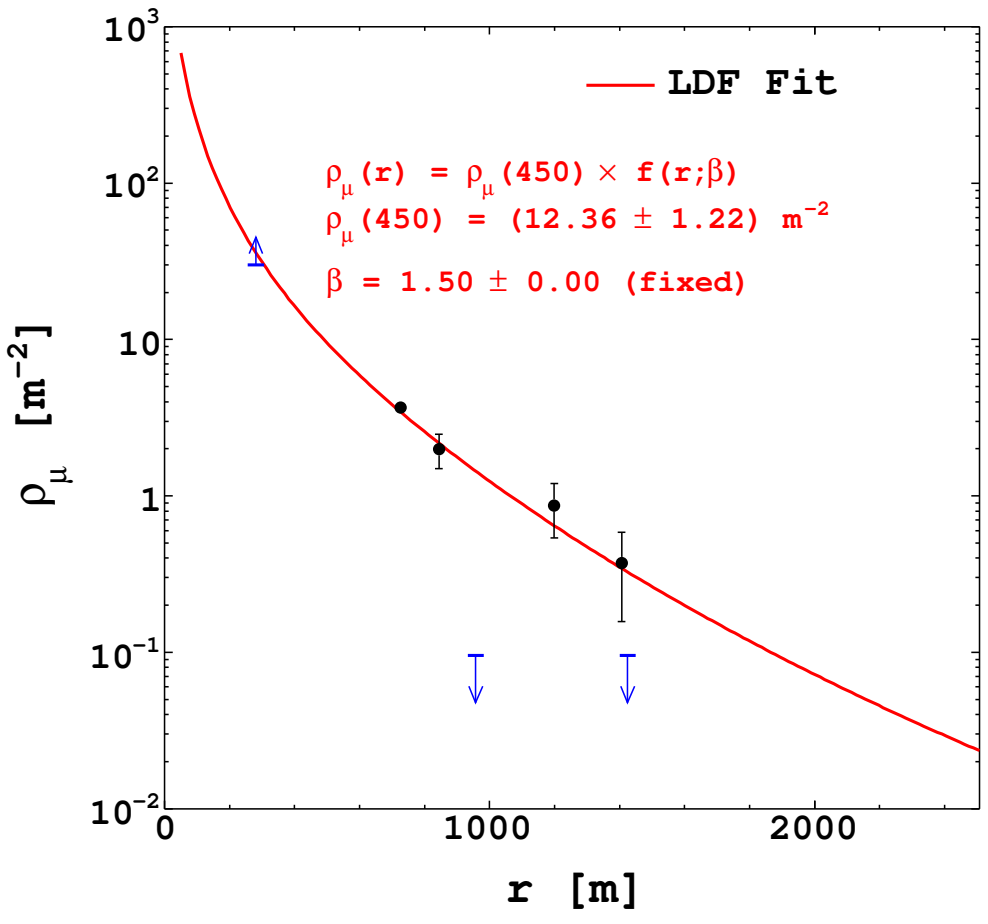

**Fig. 7**: Muon LDF fit (equation 1) corresponding to event of fig. 6. Blue upward and downward arrows are saturated and silent counters respectively.

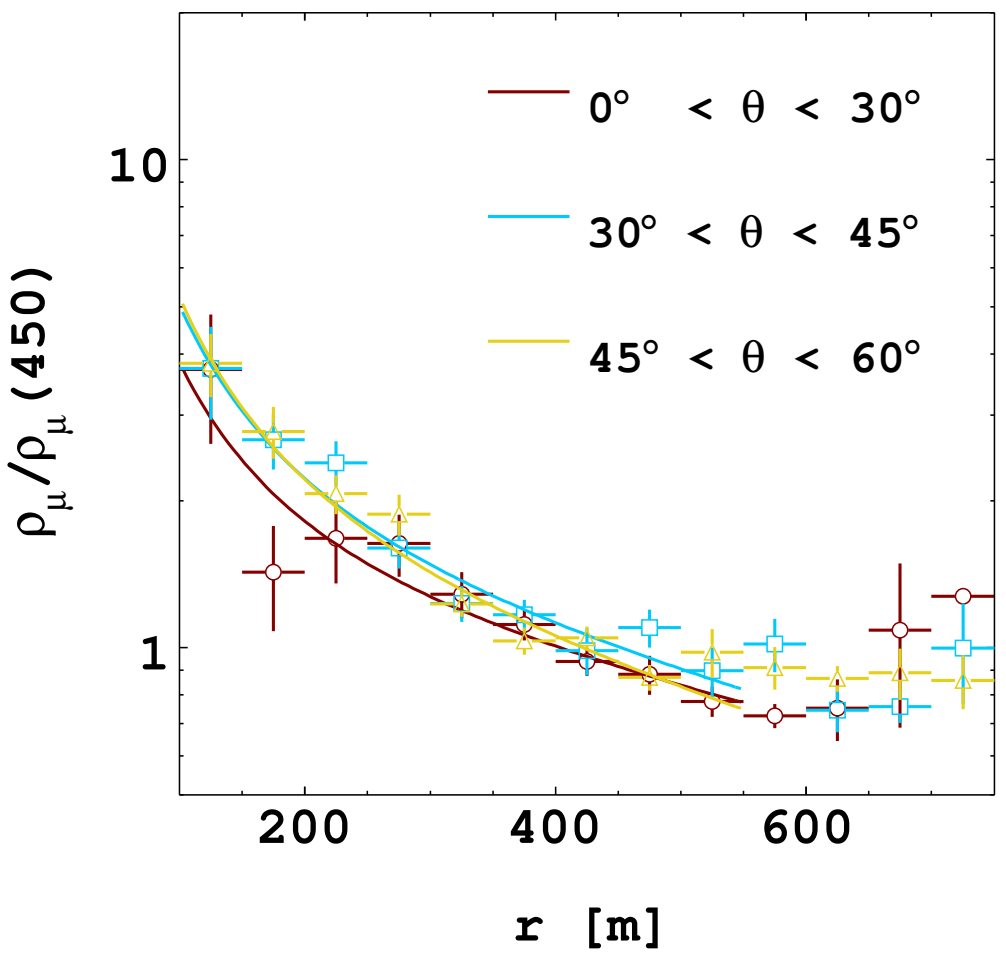

**Fig. 8**: Averaged $\rho_\mu(r)/\rho_\mu(450)$ of 230 measured MD events where at least two counters detected more than two muons above the full efficiency detection. Although still preliminary, the radial dependence of the muon densities with the decreasing zenith angle can be appreciated for the three angular bins evenly separated in $\cos(\theta)\sin(\theta)$.

## 6 Conclusions

The construction of the AMIGA muon detector array is proceeding through the Unitary Cell deployment, and it is growing smoothly. Advances in the engineering of the muon counters fabrication, logistics, procedures, and deployment technique have been achieved with remarkable results concerning the mechanical design and the stability of the detector modules. No mechanical damages nor loss of modules were suffered so far during the construction of the project.

The muon content of the particle showers is being measured with the Unitary Cell under construction and a first collection of events has been analysed. Preliminary muon LDFs have been obtained. A stable and good performance of the AMIGA scintillation modules can be inferred from the analysis of the first events registered.

# Measuring the accuracy of the AMIGA muon counters at the Pierre Auger Observatory

SIMONE MALDERA[1] FOR THE PIERRE AUGER COLLABORATION[2]

[1] *Osservatorio Astrofisico di Torino (INAF), Università di Torino and Sezione INFN, Torino, Italy*
[2] *Full author list: http://www.auger.org/archive/authors_2013_05.html*

*auger_spokespersons@fnal.gov*

**Abstract:** The AMIGA enhancement of the Auger Surface Detector consists of a 23.5 km$^2$ infilled area where the shower particles are sampled by water Cherenkov detectors accompanied by 30 m$^2$ of scintillator counters, buried 2.3 m underground. The accuracy of the muon counting obtained by the buried detectors is a basic element in the reconstruction procedure, and must be determined using experimental air shower data. To perform this measurement, twin muon counters (30+30 m$^2$) have been deployed in two infill locations; their mutual distance being about 10 m, they sample nearly the same region of the air shower. In this paper we discuss the basic properties of the modules as measured during the construction phase and the expected counting performances of the twin counters installed at the experimental site.

## 1 Introduction

AMIGA (Auger Muons and Infill for the Ground Array) [1] is an enhancement of the Pierre Auger Observatory designed to lower the energy threshold of the Auger surface detector array by one order of magnitude, down to $\approx 10^{17}$ eV, i.e. the energy region where the transition from the galactic to the extragalactic component of the cosmic radiation is expected to take place. A detailed study of the features of the energy spectrum and of the mass composition of cosmic rays in that energy range is mandatory to discriminate among the different models proposed to describe that transition ([2] [3] [4]) and advance in the understanding of the origin of cosmic rays.

AMIGA consists of an "infill" of a portion of the Auger surface detector (SD) array, where the spacing between the detectors is reduced to 750 m (half of the spacing in the regular Auger array). Each infill SD station is accompanied by nearby buried muon counters to measure the muonic component of air showers, and to obtain information about the mass of the primary particle. While the infill surface detectors are already deployed and taking data [5], an Engineering Array of muon counters is being developed, consisting of a hexagon of six 30 m$^2$ modules plus one at the center (the "Unitary Cell", UC) [6].

The main goal of the muon Unitary Cell is the validation of the detector design and the complete understanding and optimization of the AMIGA muon counting performances. To evaluate the counting accuracy two complete modules are installed in a *twin* configuration in the Unitary Cell (see Fig. 1). The measurement will be performed by comparing the counts of two "doublets" of two 30 m$^2$ modules located at a short distance ($\approx$10 m) negligible with respect to the dimension of the shower at ground (of the order of 1 km at the energies of interest).

The first of the two twin counters is fully operational since March 18, 2013. In the following sections the basic properties of the muon detectors as measured during the construction of six modules in the mechanical workshop of INFN-Torino (Italy) and the expected counting performances of the twin counters installed at the experimental site will be discussed.

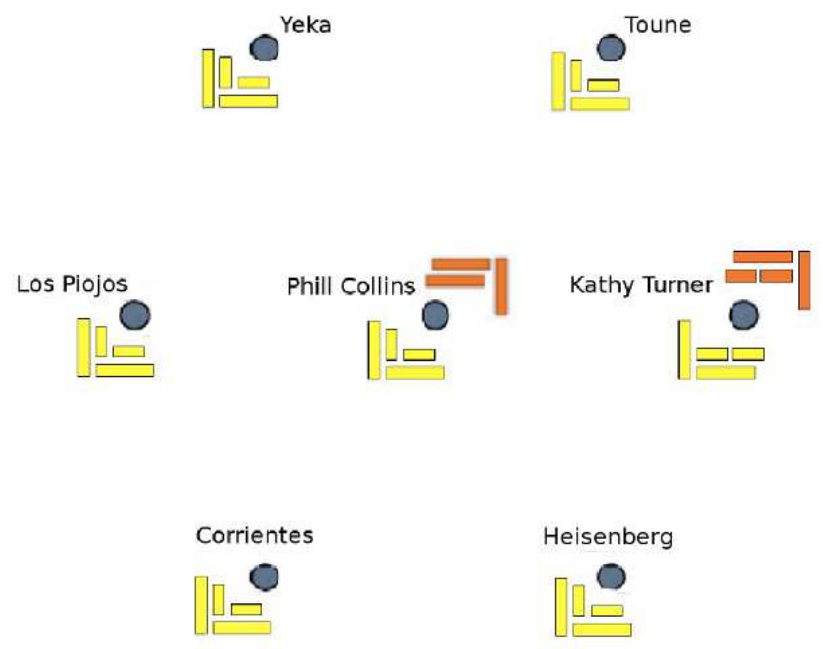

**Figure 1**: Planned layout of a Unitary Cell of muon detectors (in yellow) near the infill surface detectors (green circles). The additional muon counters making up the twins with the Phil Collins and Kathy Turner surface detectors are shown in orange. For information about the status of the deployment see [6].

## 2 Amiga muon detectors

Every muon counter of the Unitary Cell consists of four modules with a total active area of 30 m$^2$, split into two 10 m$^2$ and two 5 m$^2$ units. Each module is composed of 64 plastic scintillator strips 400 cm long (200 cm for the 5 m$^2$ ones), 4.1 cm wide and 1.0 cm high, lodged in a waterproof PVC casing.

The scintillator strips are made of extruded polystyrene doped with fluors (PPO and PPOP), co-extruded with a diffusive titanium dioxide coating. Due to the short light attenuation length of the scintillator a wavelength shifter optical fiber (1.2 mm diameter), hosted in a groove in the mid-dle of the strip, collects the scintillation photons. The scintillator strips are organized into two groups of 32 at each side of a central dome, where a multi-pixel photomultiplier (Hamamatsu H8004-200MOD) and the module electronics are placed. The 64 fibers from the two sides of the module are optically connected to the PMT. The readout electronics produces a digital output counting the pulses above a given threshold. The signal from each pixel is filtered and amplified at a nominal gain of -3.8, then discriminated, digitized with a sampling frequency of 320 MHz and stored in a circular memory. The discrimination level can be adjusted for each one of the 64 channels, and its default level is set to one third of the mean single photoelectron amplitude per pixel. When a trigger signal from the adjacent surface station is received, the digitized traces are transmitted to the central acquisition system. This method, besides strongly reducing the information to be sent to the Auger central data acquisition, is essentially independent of the PMT gain and its fluctuations and of the muon hitting position along the scintillator strip. The optimal depth for the muon detectors has been studied by means of numerical simulations. With a shielding layer of 540 g/cm$^2$ of soil ($\approx$2.3 m) the fraction of counts generated by the electromagnetic component of the shower, is lower than 5%, with an energy threshold for incoming muons of $\approx$1 GeV.

## 3 Characterization of the muon modules

During the construction phase the response of each module has been tested using both atmospheric muons and a radioactive source, using a setup similar to the one described in [7]. In fact a complete calibration of each scintillator strip with a cosmic-ray hodoscope needs about 12 hours of exposure, requiring the implementation of a reliable and fast calibration system for the module production.

After the assembly, the detectors have been exposed to a 0.84 mCi $^{90}$Sr $\beta$ radioactive source, placed at a distance of about 10 cm above the module. The X-Y position of the source was controlled by a robotic arm. A readout board multiplexes the signals from each pixel of the PMT to a charge amplifier and a dedicated data acquisition system. The 64 channels are read out within 100 ms, allowing continuous monitoring of all the scintillator strips.

The source is moved along the direction perpendicular to the strip length at a fixed distance from module median. The signal of each pixel increases as the source is approaching the strip, reaches a maximum value when it is in the center and then decreases. The resulting time profile is fitted with a Gaussian function to get the height of the maximum. Performing such "transversal scans" at different distances to the PMT the light attenuation profile can be derived (see Fig. 2).

The response of the AMIGA modules to through-going muons has been studied using two small detectors consisting of a piece of scintillator (4x10 cm$^2$) and a photomultiplier, placed above and below a given strip. The coincidence of the two small scintillators generates a trigger for the FADC (1 GHz sampling rate, 10 bits) reading the PMT signals from the module. An acquisition time of about one day allows a good measurement of the charge spectrum of the acquired signal. To increase the statistics larger trigger scintillators (10x80 cm$^2$) were also used, allowing more strips to be measured at the same time. In this case a huge background peak appears in the charge spectrum. About 100 measurements on different modules, strips and at different distances from the PMT (computed along the optical fiber) have been performed, allowing one to normalize the results obtained with the radioactive source to the mean collected charge when a muon crosses the detector vertically. Fig. 3 shows the ratio between these two quantities, taken on the same scintillator strip and at the same distance from the PMT.

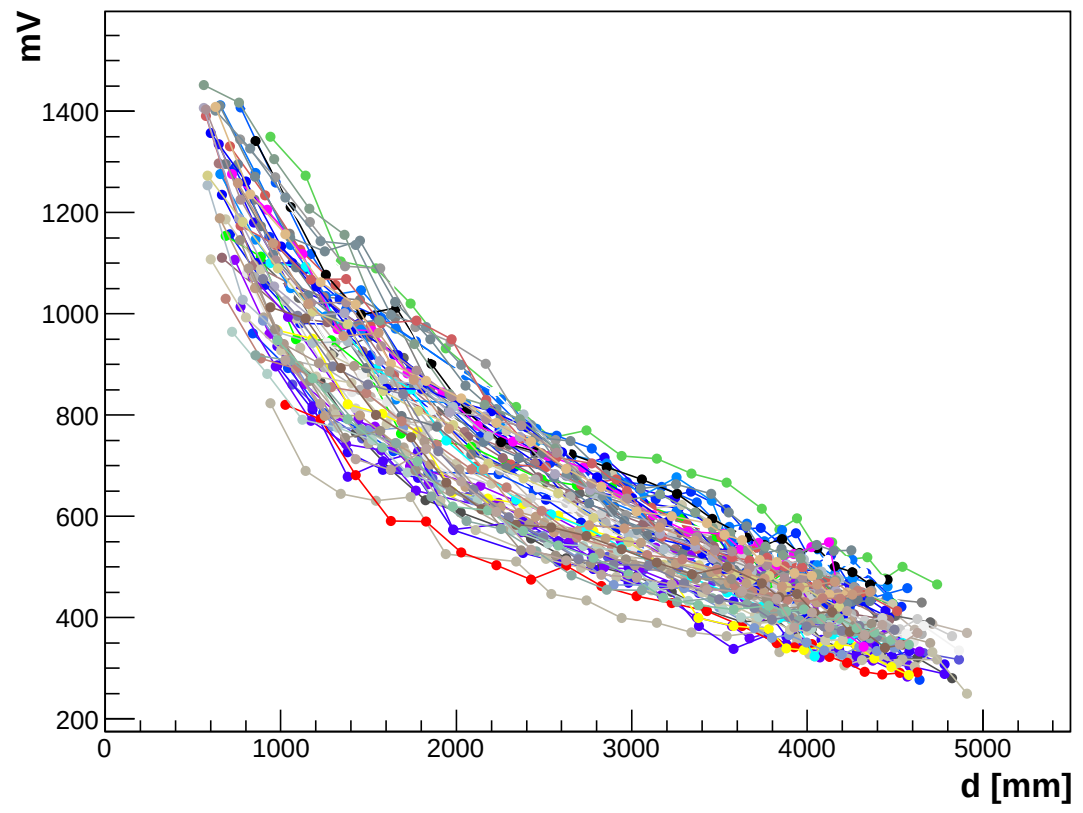

**Figure 2**: Results of the scan with a radioactive $^{90}$Sr source for one 10 m$^2$ module. The maximum of the signal from each scintillator strip is shown as a function of the distance from the PMT (along the fiber).

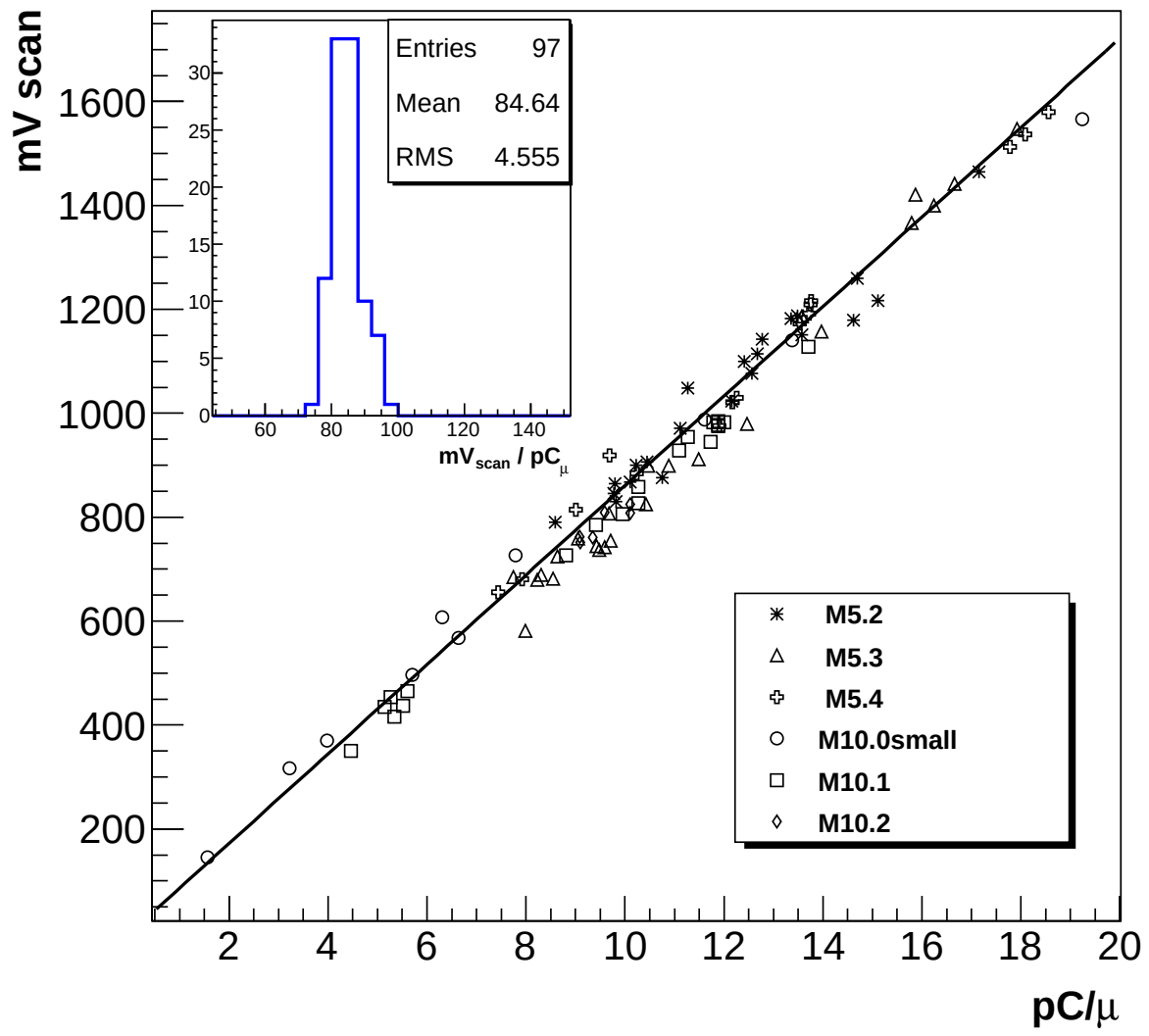

**Figure 3**: Normalization of the signal measured with the radioactive source to the average charge collected per muon crossing the detector vertically, measured on the modules built in Torino (four 5 m$^2$ modules and two 10 m$^2$ ones labeled as M5 and M10 respectively). The histogram of the ratios between all those measurements is shown in the top left corner. The width of this histogram gives the uncertainty in the normalization value.

Finally the number of photoelectrons ($n.p.e.$) produced

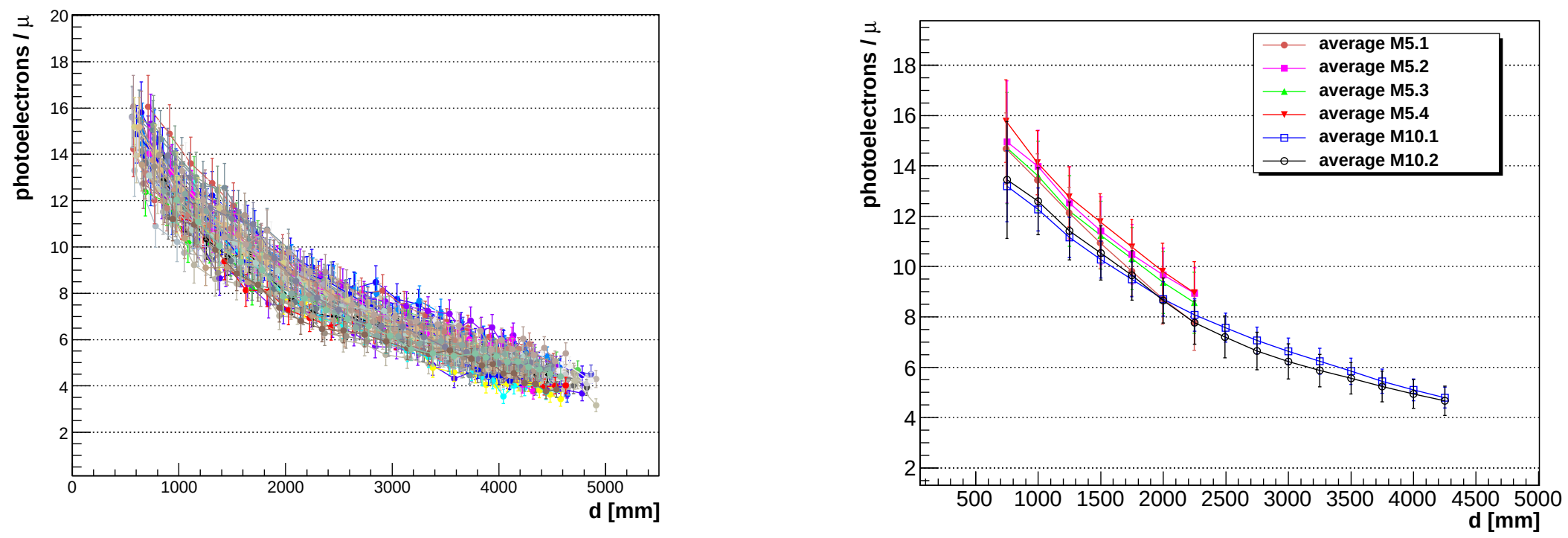

**Figure 4**: **Left:** Number of photoelectrons per muon crossing the detector vertically for the 64 strips of one module, obtained from the measurements with the radioactive source. **Right:** Number of photoelectrons produced by vertically crossing muons, at different distances from the PMT, Each point represents the average of the 64 strips of the module, error-bars correspond to the RMS of their distribution.

at the PMT photocathode by a through-going muon can be derived as:

$$n.p.e. = \frac{Q_\mu}{G_{pixel} \cdot e} = \frac{V_{scan}}{R} \cdot \frac{1}{G_{pixel} \cdot e}$$

where $e$ is the elementary charge, $G_{pixel}$ is the gain of the specific pixel of the PMT, $V_{scan}$ is the signal generated by the radioactive source read out through the charge amplifier, and $R$ is the normalization factor (given by the ratio shown in Fig. 3). The gain of each PMT pixel has been measured using the single photoelectron technique, with an uncertainty of about 7%. In Fig. 4 we show the result of the conversion of the measurements with the radioactive source to the number of photoelectrons, according to the formula above. The uncertainty in the measured $n.p.e.$ is about 12%, and has been derived from the combination of the uncertainties in the peak voltage obtained with the source (about 2%), in the normalization factor ( 8%), and the quoted uncertainty in the pixel gain.

## 4 Simulation of the detector response

The laboratory measurements and results described above have been used to simulate the detector counting performances (similarly to [8]). The energy deposited by a muon in the buried scintillator strip is simulated by means of Geant4 ([9]), and then converted to a number of photoelectrons generated in the PMT given by:

$$n.p.e_{sim} = \frac{E_{dep}}{< E^{\mu}_{dep} >} \times N_{p.e/\mu}(d)$$

being $E_{dep}$ the deposited energy, $< E^{\mu}_{dep} >$ the average energy deposit of a vertically crossing muon (obtained by simulation), and $N_{p.e/\mu}(d)$ the measured average number of photoelectrons per vertically crossing muon (Fig. 4). To reproduce the measured distribution a Poissonian fluctuation is applied to the photoelectron number obtained with the quoted formula. Given the total number of photoelectrons, a corresponding signal shape is extracted from a sample of traces (organized in bins of $n.p.e.$) obtained from the measurements with atmospheric muons described above. To match the conditions of the readout electronics of the muon module, such traces are convolved with a digital low-pass filter with a cutoff frequency of 140 MHz (while the bandwidth of the digitizer used in the laboratory is 500 MHz) and down-sampled to 320 MHz with a simple decimation algorithm. Each time bin of the resulting trace is discriminated at a threshold corresponding to 33% of the photoelectron amplitude, producing a digital data stream similar to the one expected from the real detectors.

Using this simple simulation, the detection efficiency for different counting strategies and discrimination thresholds can be estimated. Fig. 5 shows the ratio between the number of counts obtained with two different counting algorithms and the total number of injected muons, as a function of the distance from the particle position to the module center. Two different counting strategies have been used, one requiring two adjacent positive samples (labeled as *11*), the other one requiring two positive samples spaced by one positive or negative bin (labeled as *1x1*).The results are shown first taking into account only the muons that hit the detector and then considering also the hits of secondary electrons generated in the propagation of muons in the ground. The overall ratio is about 93% for the *11* counting strategy and 82% for the *1x1* strategy considering only muon hits, rising to 98% and 88% respectively when the hits of secondary particles are included.

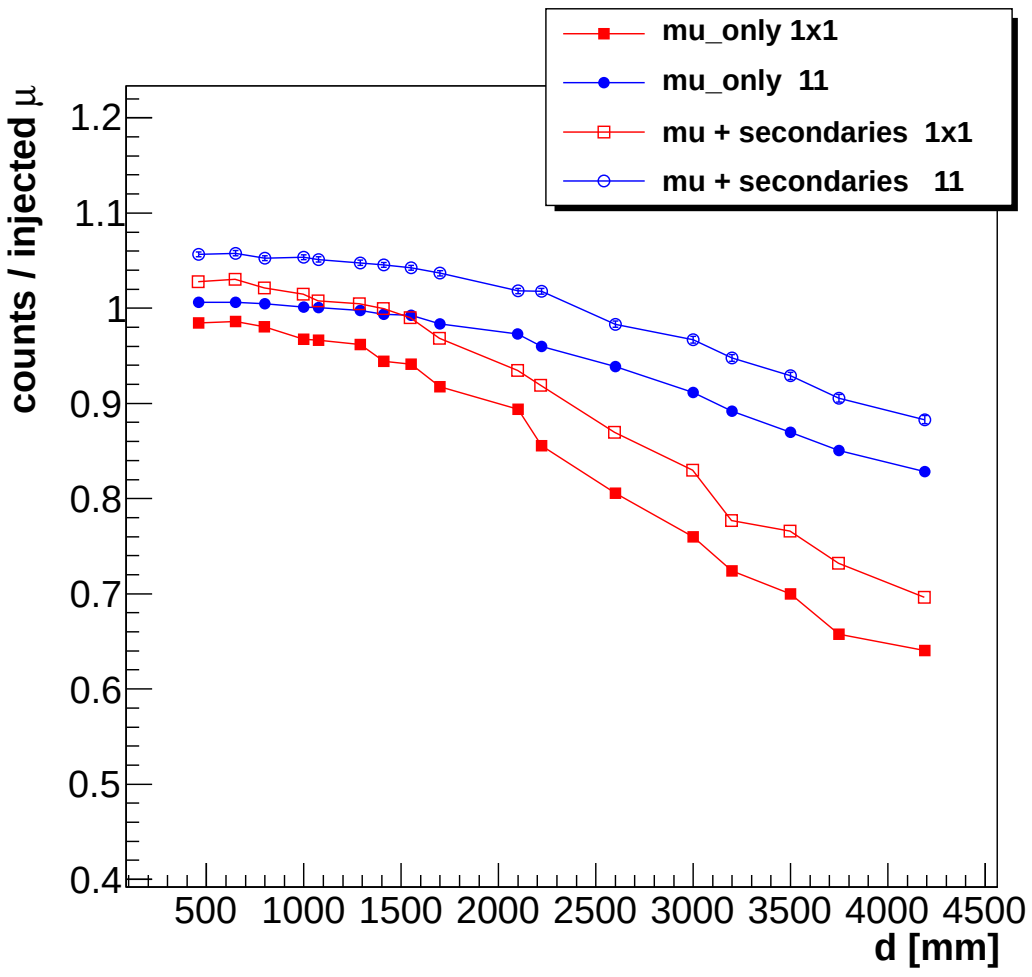

**Figure 5**: Ratio between the number of counts obtained from the module simulation and the total number of injected muons as function of the distance from the particle position to the PMT. Vertical muons of 5 GeV have been considered in the simulation (see text).

## 5 The muon counting accuracy

The muon modules have been doubled at two positions of the UC (as shown in Fig. 1) to directly measure the accuracy in the muon counting. Such accuracy must be determined experimentally using real events measured by the detector. In fact, shower fluctuations are extremely difficult to simulate, due to the large number of particles in the cascades ($> 10^{11}$) and the uncertainties in the numbers of muons, electrons and gamma-rays, which depend on the hadron interactions and on the primary particle type. Moreover, the measurements in the field include environmental and instrumental effects (e.g. background from soil radioactivity and residual punch-through particles, PMT gain fluctuations and other noise effects) which are difficult to be correctly estimated in the simulation.

Since the shower footprint is of the order of several square kilometers, these modules (separated by 10 m, as described above) are virtually measuring the same region of the shower. The muon counting accuracy will be derived from the comparison of the counts in two adjacent counters. In particular the relative fluctuation in the muon number can be defined as:

$$\Delta = \sqrt{2} \cdot \frac{M1 - M2}{M1 + M2}$$

where $M_i$ corresponds to the number of muons measured by the $i$-th counter of the pair. The relative accuracy of a single module is then given by the width of the $\Delta$ distribution, $\delta\Delta = \sigma/M$, where $\sigma$ is the accuracy of a single module. To obtain this expression, it has to be assumed that $M_1 \approx M_2$ and that $\sigma_1 \approx \sigma_2$; quality cuts will be used to ensure that the modules are measuring real EAS events. The detectors are in principle identical, therefore their accuracies should be similar.

The expected results on the counting accuracy of the AMIGA muon modules have been studied by means of simulations. Given the energy threshold of the infill array (full trigger efficiency at $\approx 3 \times 10^{17}$ eV) one year of data taking will allow deviations from a Poissonian behavior (expected for an ideal detector) of the order of 10% to be detected at a level of $2\sigma$. In addition the comparison between the counts in the 5 m$^2$ and the 10 m$^2$ modules will allow us to study the counting efficiency and the effect of pile-up (due to the finite segmentation of the modules).

The first data from the twin counters at the *Kathy Turner* position are shown in Fig. 6. Requiring that the associated SD station is part of an event with reconstructed energy above $10^{17}$ eV, about 280 events have been collected in two month of data taking. The preliminary comparison of the counts of the two counters, already gives a first indication that the detectors are working as expected, allowing the muon counting accuracy to be accessed in the near future.

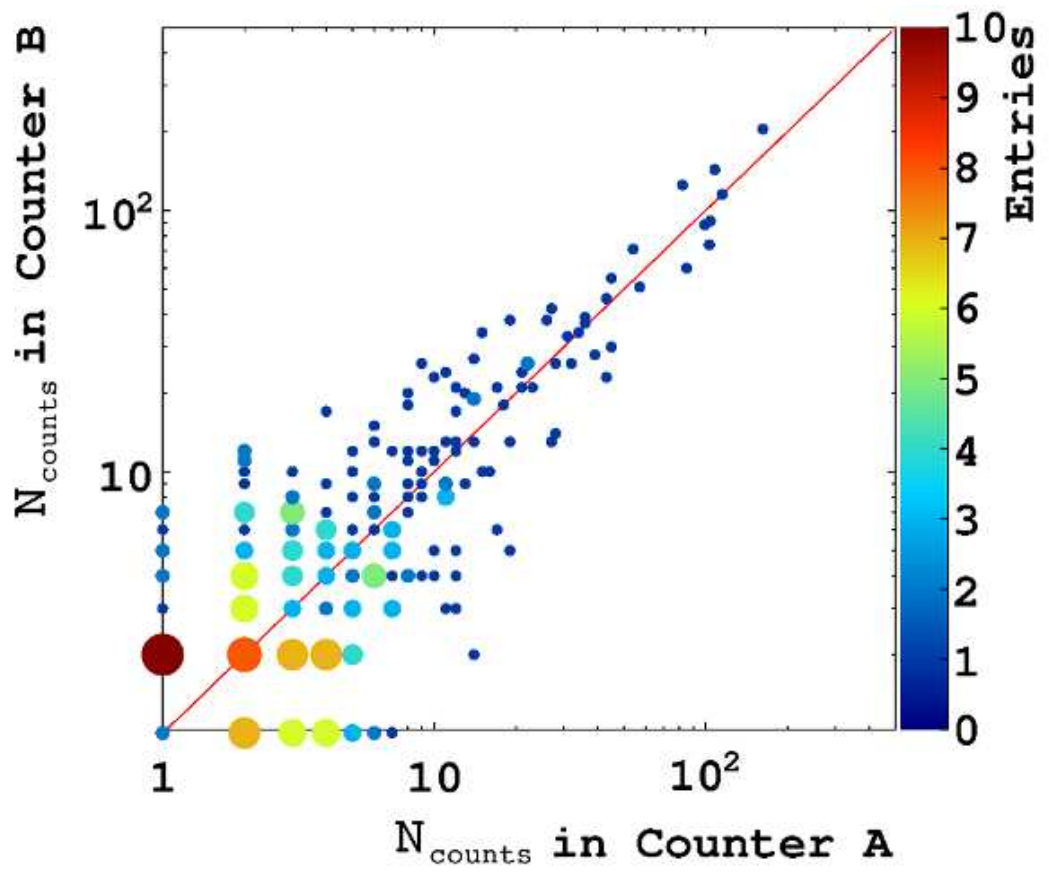

**Figure 6**: Comparison of the counts registered by the two muon counters at the *Kathy Turner* position after the application of the *1x1* counting strategy, for the first two months of data taking (18 March - 18 May 2013). The color code and the dot size are proportional to the number of events in each bin.

## 6 Conclusions

The AMIGA muon Unitary Cell, being deployed at the experimental site, will allow us to validate of the detector design and performances. The counting accuracy will be studied by a couple of twin muon counters buried near the same SD station. The first twin has been taking data since March 2013.

The muon module response has been carefully studied during the construction phase, and the results for the modules built at INFN-Torino have been reported. In particular the average number of photoelectrons in the PMT for a vertical muon crossing the detector has been measured to be between ≈15 and ≈5 according to the position at which the particle crosses the detector.

The laboratory measurements and their results have been used to implement a simple simulation of the detector response, allowing the expected counting performances of the modules to be estimated. Such simulations will be used to further study and optimize the reconstruction algorithm.

# Radio detection of air showers with the Auger Engineering Radio Array

FRANK G. SCHRÖDER[1,2] FOR THE PIERRE AUGER COLLABORATION[3]

[1] *Institut für Kernphysik, Karlsruhe Institute of Technology (KIT), Germany*
[2] *Instituto de Tecnologías en Detección y Astropartículas (CNEA, CONICET, UNSAM), Buenos Aires, Argentina*
[3] *Full author list: http://www.auger.org/archive/authors_2013_05.html*

*auger_spokespersons@fnal.gov*

**Abstract:** The Auger Engineering Radio Array (AERA) is one of the low-energy extensions of the detector systems of the Pierre Auger Observatory. AERA is being used to study the emission of radio waves from extensive air showers. It operates in the frequency range from 30 to 80 MHz. Recently, AERA has been expanded to 124 radio stations over an area of approximately 6 km$^2$. In 2014 we will deploy 36 additional stations, and by this extend the area to at least 10 km$^2$. With this AERA160 setup we will be able to determine the measurement resolution for the arrival direction, the energy, and the mass-composition of primary cosmic rays with energies larger than $10^{17.5}$ eV. In this paper, we describe the setup of AERA. We also present and discuss the first physics results and techniques that have been developed for AERA24, the first phase of AERA consisting of 24 stations distributed over an area of 0.5 km$^2$. In particular, we show a comparison of a measured event with simulations.

## 1 Introduction

Radio detectors for extensive air-shower measurements offer several advantages compared to other techniques. On the one hand, the duty-cycle is close to 100 % like it is for particle detectors, only excluding times of close-by thunderstorms since strong atmospheric electric fields significantly affect the radio emission of the air showers [1]. On the other hand, the radio detectors provide a quasi-calorimetric measurement of the shower energy, as fluorescence and air-Cherenkov detectors do. Moreover, they are sensitive to the shower development [2] and thus to the mass composition of the primary cosmic rays. However, the radio technique is not yet as advanced as the other established techniques, and several key questions are being explored at present:

- Can radio measurements compete in precision for energy and composition with the air-fluorescence technique, which offers approximately only a seventh of the radio duty-cycle?
- Is it possible to operate radio arrays for air-shower detection stand-alone, or are the open physics questions for ultra-high energy cosmic rays better attacked with multi-hybrid observatories?
- Can the radio technique be extended to very large scales. i.e. can it be used for future observatories studying the end of the energy spectrum with high statistics?

The Auger Engineering Radio Array (AERA) is dedicated to answering those questions. At the same time it aims to improve our understanding of the physics processes behind the radio emission. For this purpose, AERA is built in the enhancement area of the Pierre Auger Observatory [3] where each air shower can be measured by many different techniques at the same time. The co-located AMIGA enhancement [4] consists of a surface array of particle detectors with 750 m spacing (with six additional detectors in the center of AERA), and an array of associated muon counters. Moreover, the area is overseen by several fluorescence telescopes. This situation enables a cross calibration between the different techniques, multi-hybrid analyses to improve the reconstruction precision of individual events, as well as the triggering of the different detectors by each other. Thus, the situation offers ideal conditions to ‘engineer’the radio technique, e.g., by optimizing the data-acquisition system, the electronics or the antenna design.

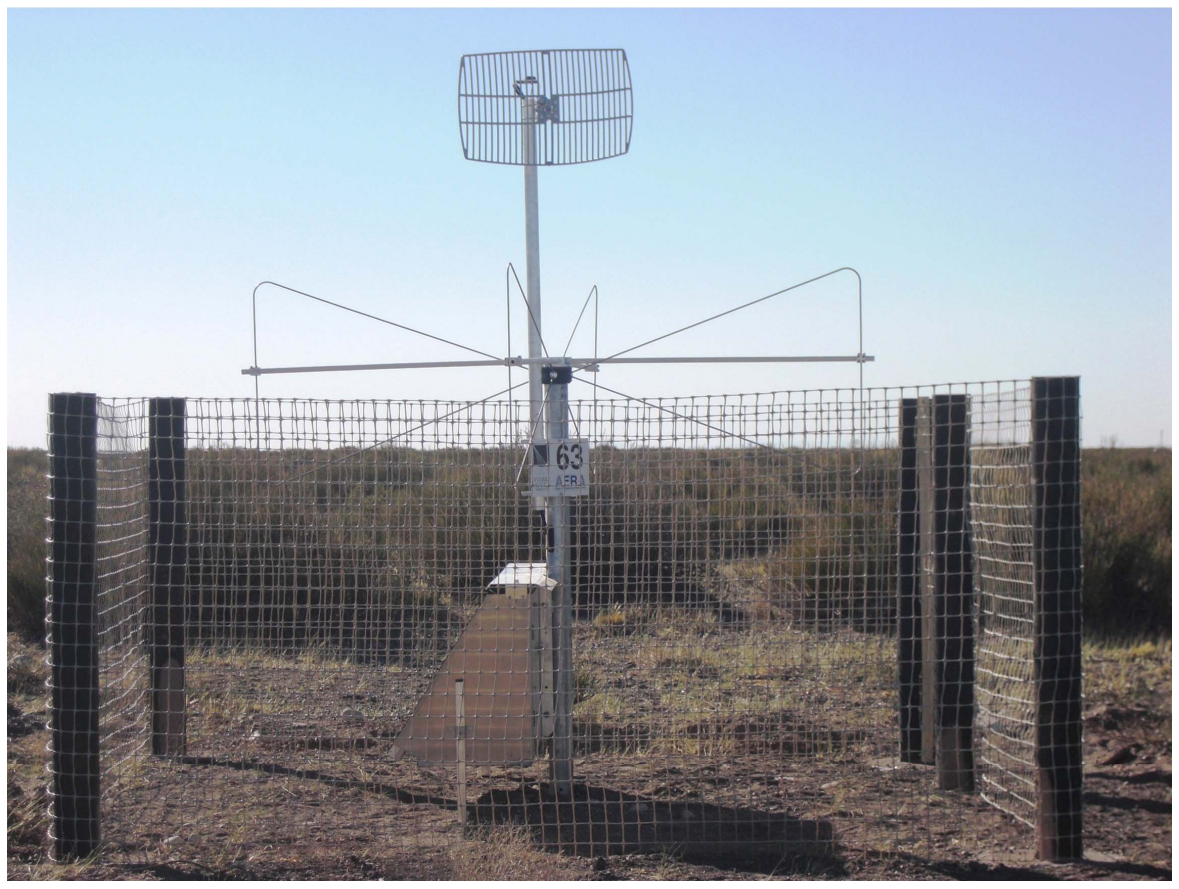

**Figure 1**: Radio station of AERA phase 2: a communication antenna for wireless data transfer, two butterfly antennas for the radio measurements (one aligned east-west and one north-south), a metal box for electronics, a solar panel and a battery for power supply, and a fence protecting against cattle.

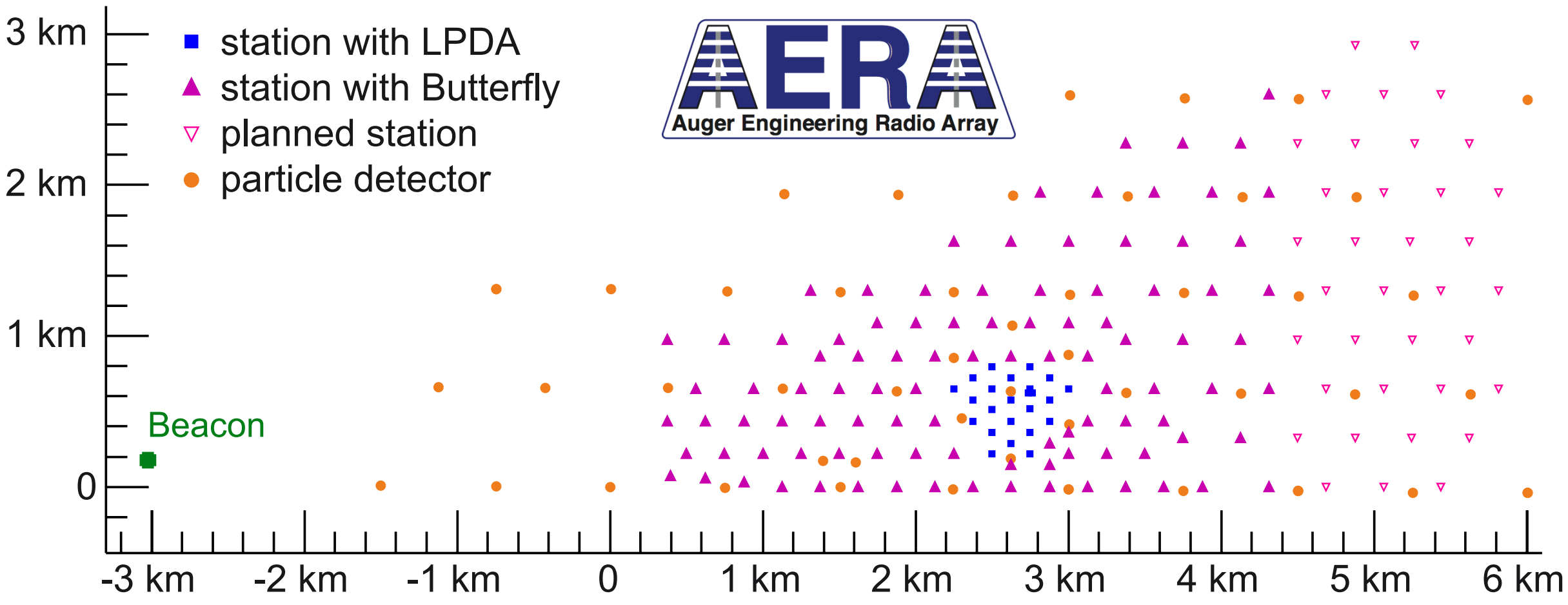

**Figure 2**: Layout of the Auger Engineering Radio Array (AERA). The results presented in this paper are based on the 24 radio stations with Logarithmic Periodic Dipole Antennas (LPDAs). The new stations deployed this year use butterfly antennas. The reference beacon in the west is installed at one of the fluorescence-telescope sites of Auger. Also the particle detectors of the surface array in this area are shown.

## 2 AERA setup

Because AERA aims at a number of technical and physics goals, it is a heterogeneous setup combining different types of hardware for the measurements of the same air showers. Its first phase, AERA24, started operating in April 2011. It consists of 24 stations with logarithmic periodic dipole antennas (LPDAs) distributed on an area of 0.5 km$^2$ with a spacing of 125 m. A more detailed description of AERA24 can be found, e.g., in reference [5]. Here, and also in Ref. [6], we will present an update on recent measurements and physics results obtained with AERA24.

In 2013, we started to extend AERA using a modified station design and a different type of antenna (figure 1). Up-to now we have deployed 100 additional stations covering an area of approximately 6 km$^2$, and plan to deploy an additional 36 stations to extent the total area to at least 10 km$^2$ (figure 2). The new stations use two butterfly antennas for the measurements, because the butterfly antenna is more economical than the LPDA and superior in several technical aspects [7]. Moreover, the new stations only rely on wireless communication for data transfer while, in the much smaller AERA24, data are transmitted via optical fibers. All AERA stations operate autonomously and communicate with a central data-acquisition system to exchange data and trigger information.

In AERA, as an engineering array, we test different technical solutions to optimize the radio technique for future large scale observatories. In particular, we test different systems for the communication and different electronics for the local data acquisition. All stations have the ability to self-trigger on the radio signal. Although we have demonstrated that this is possible, we have not yet achieved a 100 % trigger efficiency. Therefore, we simultaneously use an external trigger. A part of AERA is triggered by the surface detector array, which also allows the read-out of sub-threshold stations. In another part of AERA, we test triggering by the use of a small scintillator integrated directly in the radio stations.

All stations feature two antennas, one aligned in the geomagnetic north-south direction and one in the east-west direction. The signals of both antennas are suppressed in the frequency range below 30 MHz and above 80 MHz by an analog bandpass filter. After amplification, the signals of both antennas are sampled with ADCs, digitally stored and transfered to the central data-acquisition system. The ADC sampling frequency varies between 180 MHz and 200 MHz in different stations. In any case, it is larger than the Nyquist frequency so that a full reconstruction of the time-dependent field strength in our measurement band is possible. Each station features its own GPS clock which is used to tag the recorded data with a time stamp. In this way the measurements from the individual stations can be combined for one single event independent of the trigger sources.

AERA has been calibrated using several methods. In particular we have measured the phase and amplitude behavior of each individual component and correct for it during data analysis. To monitor the relative timing of AERA, we study the phasing of sine waves continuously emitted by a reference beacon [8]. In principle, the beacon can also be used to improve the timing precision of AERA. Our challenging goal is to achieve a relative timing accuracy of 1 ns, i.e., significantly better than the timing precision of the GPS clocks in use. By this we will be able to use AERA with digital radio interferometry, a technique which has already been used by LOPES to lower the detection threshold [9].

For data analysis we use the proprietary software package of the Pierre Auger Collaboration, named Offline [10]. With the radio extension of Offline [11], it features the correction of the measured signals for all hardware properties, the reconstruction of the time-dependent electric-field-strength vector at each station from the measurements of the two individual antennas, and several software modules for high-level physics analysis.

## 3 Results

Since man-made radio background of different types (pulses, constant waves) is present in practically all AERA measurements, the number of triggered events is a poor indicator for the performance of AERA. Although we can effectively filter the background during our data analysis, at the moment we still rely on the coincident surface detector measurements to distinguish real air-shower events from background. Thus, the number of detected events depends strongly on the quality criteria in use, e.g., if one requires a strong radio signal in just one antenna, the event rate is an order of magnitude higher than for high-quality events with a significant signal in at least three antennas and a direction reconstruction which coincides within 20° of the direction measured by the surface detector array.

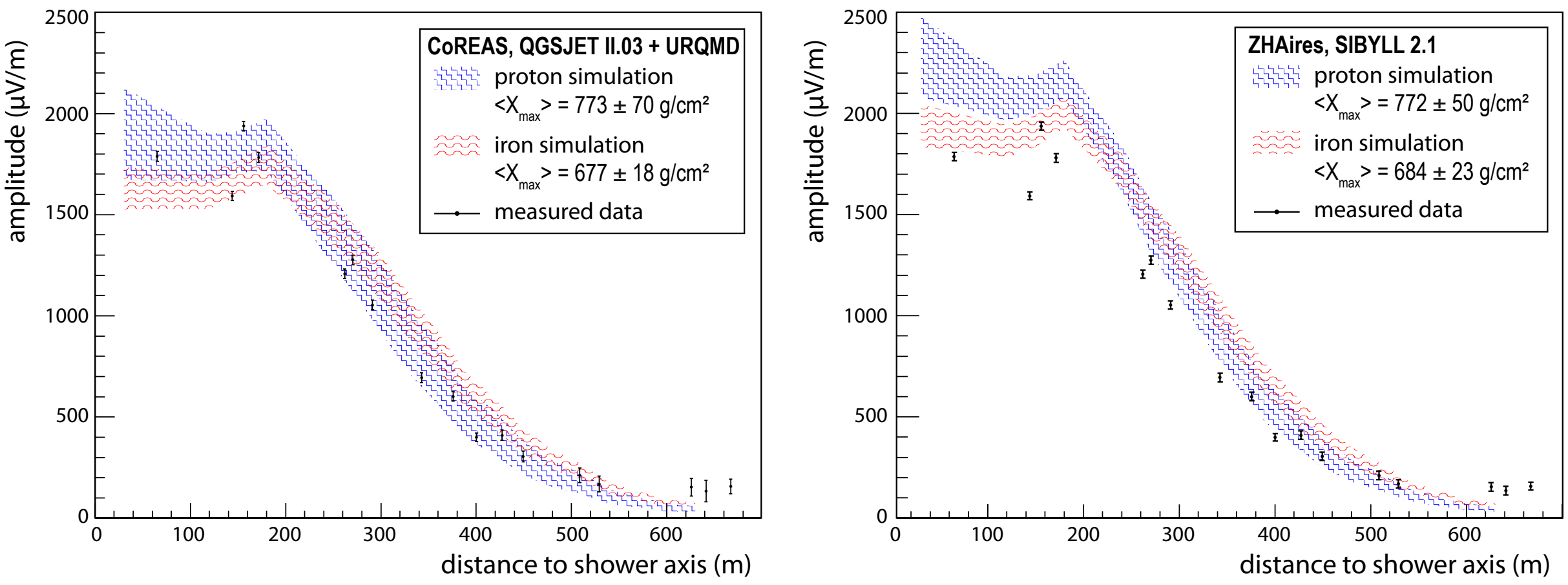

**Figure 4**: Comparison of the radio lateral distribution of the event shown in figure 3 with simulations of two different codes. CoREAS [12] (left) and ZHAires [13] (right) simulations for a proton and an iron nucleus as primary particle, where the bands indicate systematic uncertainties due to the uncertainties of the input parameters for the simulations and due to shower-to-shower fluctuations (strength of geomagnetic field used in simulations: $B = 23\,\mu$T).

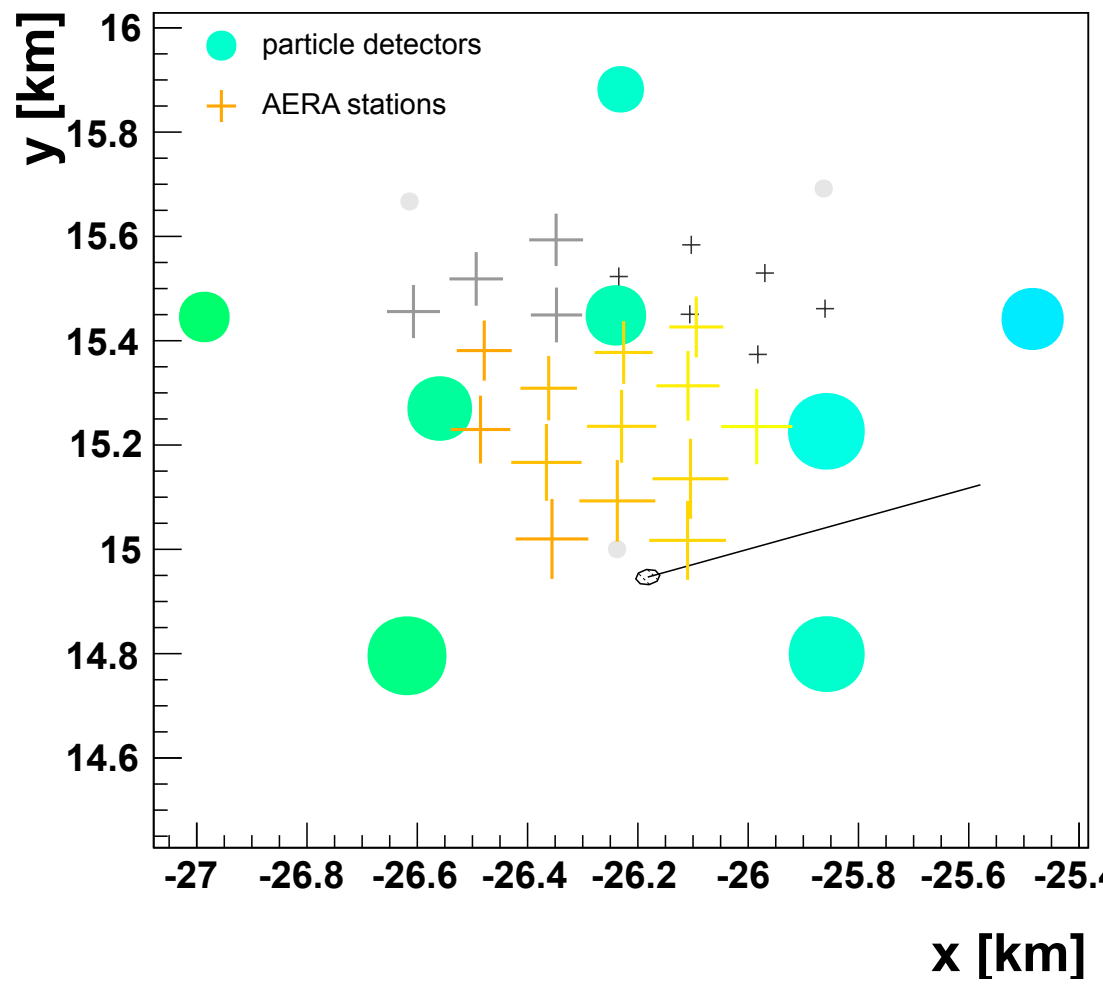

**Figure 3**: Footprint of measured AERA event triggered by the Auger surface detector; each colored cross represents an AERA station with data. The size of the bars represents the amplitude of the radio signal in the north-south and east-west polarization and the color code the arrival time; the circles are the Auger surface detectors in the area of AERA24 where the line indicates the arrival direction and the core of the shower determined with the surface array.

By 27 February 2013, AERA had measured 356 of these high quality events and, for a few of these events, we also have measurements from the fluorescence detector or the muon counters. 229 events have been triggered by the surface detector. 98 events have been self-triggered, and later assigned to the associated surface detector measurement, and 29 events have at least three self-triggered and three externally triggered stations. However, it is difficult to compare the event rates and efficiencies for both triggers from these data, because the exact configuration of AERA24 changed several times. Thus, there are different periods in the data sets for the self-trigger and the external trigger, and also the number of stations equipped with either trigger changed during time. Nevertheless, the number of events indicates the statistics currently available for physics analyses, and gives a lower limit to the event rate which can be expected for the future.

The mean angular deviation between the direction reconstructed with AERA and the surface detector array is approximately 4°. We expect that this number will decrease in future by improving the reconstruction algorithms and the time calibration. The mean energy as reconstructed by the surface detector is in the order of 1 EeV, where some events have an energy below 0.1 EeV. Figure 3 shows one example event with an energy of 4.3 EeV, and a zenith angle of 58.4°.

We analyzed AERA measurements in different ways, and compared them to recent simulation codes for the radio emission from air-showers (figure 4). For this purpose, we chose AERA events containing a large number of antennas with significant signals. For these events we performed air-shower simulations based on the reconstruction parameters of the surface detector, and calculated the radio emission with different codes, e.g., CoREAS [12], ZHAires [13], EVA [15], and SELFAS [14].

So far we have found no contradiction to the following general picture of the origin of the radio emission. The dominant emission process is the geomagnetic deflection of the electrons and positrons in the air shower [18, 19]. The radio emission by the Askaryan effect [20, 6], i.e. the variation of the net charge excess, is for air showers an order of magnitude weaker than the geomagnetic effect, but not negligible. Both processes are affected by the refractive index of the air which changes the coherence conditions of the radio emission. Although the normal Cherenkov radiation due to the excitation of air molecules seems to be negligible for the overall radio signal, the refractive index leads to a Cherenkov-like beaming of the emission generated by the geomagnetic and the Askaryan effect. This explains the lateral distribution of the radio amplitude which first rises until an axis distance of about 100 m is reached and then decreases, as first predicted [21] and indicated by measurements [22] more than 40 years ago.

Moreover, we confirmed that the amplitude of the AERA measurements depends on the energy of the primary particle [16], and we are currently optimizing the reconstruction techniques to maximize the energy precision. To reconstruct the cosmic-ray composition we study three parameters of the radio measurements which are sensitive to the shower development: the slope of the lateral distribution, the shape of the radio wavefront, and the slope of the frequency spectrum. The latter method has the advantage that it needs only a single station as long as the shower geometry is known (e.g., from the surface detector). We have already confirmed that a measurement of the spectral slope is possible in practice [17]. However, the precision for the composition of all three methods is still under study. For this we still need more statistics of AERA measurements in coincidence with the fluorescence detector.

## 4 Conclusion and Outlook

With its extended area AERA will collect the necessary statistics to study radio emission from air showers in the energy range between $10^{17.5}$ to $10^{19}$ eV. By a cross-calibration and comparison with the other detectors in the Auger enhancement area we will be able to determine the precision of AERA for the arrival direction, energy and composition. This is a crucial input for the decision in which way the radio technique can contribute to future cosmic-ray observatories for energies beyond $10^{17}$ eV. If we can confirm that the radio precision is comparable to the precision of the fluorescence techniques, radio detectors could provide an order of magnitude larger statistics for composition studies with the same observatory area. With the high quality measurements of AERA, we can also test the results of other radio observatories for air-showers, e.g., LOPES [9], CODALEMA [23], LOFAR [24], and Tunka-Rex [25].

Moreover, multi-hybrid measurements with different detectors in the enhancement area allow two kind of interesting analyses. First, the combination of complementary measurements can increase the reconstruction accuracy for the primary mass, e.g., the muon number measured by AMIGA and the radio signal measured by AERA depend in different ways on the shower development. Second, with such multi-hybrid measurements we can test the paradigm of shower universality: twin showers with an almost identical measurement in two complementary detectors ought to show also an almost identical signal in a third complementary detector.

Finally, the improved description of the radio measurements by recent simulation codes is promising, and reflects an improved understanding of the underlying physics. In particular, it enables better predictions for the performance of future radio observatories. Moreover, reconstruction techniques of the air shower parameters can be studied using realistic simulations as input. Possibly the AERA measurements can also be used to test hadronic interaction models. The deviations between simulations and measurements of the radio signal are now in the same order of magnitude as the deviations between the simulations and measurements for air-shower particles, especially with respect to the muon content. Thus, any slight mismatch between the simulated and measured radio signal does not necessarily indicate a lack of understanding of the radio emission, but it might be as well caused by a lack of understanding of the particle interactions in the air shower.

# Probing the radio emission from cosmic-ray-induced air showers by polarization measurements

Tim Huege[1] for the Pierre Auger Collaboration[2]

[1]*IKP, Karlsruhe Institute of Technology (KIT), Postfach 3640, 76021 Karlsruhe, Germany*
[2]*Full author list: http://www.auger.org/archive/authors_2013_05.html*

*auger_spokespersons@fnal.gov*

**Abstract:** The emission of radio waves from air showers induced by cosmic rays has been attributed to the so-called geomagnetic emission process. At frequencies around 50 MHz this process leads to coherent radiation which can be observed with rather simple setups. The direction of the electric field vector induced by this emission process depends only on the local magnetic field vector and on the arrival direction of the cosmic ray. We report on measurements of the electric field vector where, in addition to this geomagnetic component, another component has been observed which cannot be described by the geomagnetic emission process. This other electric field component has a radial dependence with respect to the shower axis in agreement with predictions made by Askaryan using a charge-excess model. Our results are compared to calculations based on models that include the radiation mechanism induced by the charge-excess process.

## 1 Introduction

In the last decade, radio detection of cosmic ray air showers has been revived through the use of powerful digital signal processing techniques. The LOPES [1] and CODALEMA [2] experiments in particular have driven these modern developments. However, these "first-generation" modern experiments only covered areas of $< 0.1$ km$^2$ and thus only had a reach in energy of up to $\approx 10^{18}$ eV. The Auger Engineering Radio Array (AERA) [3] situated within the Pierre Auger Observatory has recently been enlarged to an area of $\approx 6$ km$^2$ covered by a total of 124 radio detector stations (RDSs) [4]. Thereby, AERA strives to pave the way for the application of radio detection at ultra-high energies.

It has been known since the experiments in the 1960s [5] and confirmed by modern experiments [1, 2] that radio emission from air showers is strongly correlated with the local geomagnetic field. The emission must thus be dominated by a geomagnetic effect, describable by time-varying transverse currents as originally derived by Kahn & Lerche [6]. However, even before Kahn & Lerche, Askaryan [7, 8] predicted that there should be an emission component related to the time-variation of the negative net charge excess in air showers, and early experiments found evidence that there is indeed a sub-dominant non-geomagnetic contribution to the radio signal [9].

These two predicted emission contributions have distinct polarization characteristics, which are imprinted on the radio signal measured at ground. The geomagnetic emission leads to linearly polarized signals, the electric field vector always being aligned with the Lorentz force. More precisely, the electric field vector points in the direction defined by $(-\vec{v}\times\vec{B})$, where $\vec{v}$ denotes the shower axis and $\vec{B}$ refers to the local geomagnetic field. In contrast, the Askaryan emission is linearly polarized with an electric field vector aligned radially with respect to the shower axis — in other words, the electric field vector orientation varies with the position of the observer. As we will show in the following, the well-calibrated dual-polarized AERA RDSs allow us to exploit polarization characteristics to identify such a radially polarized signal contribution imprinted on the dominating geomagnetic radiation.

## 2 Polarization measurements

In this section, we describe our detector setup and detail two independent methods to quantify the deviation from pure geomagnetic radiation measured with AERA.

### 2.1 AERA data

The analysis presented here rests on data which have been acquired in radio-self-triggered mode with the first 24 AERA RDSs [3] in the period from April to July 2011. At that time, the local geomagnetic field had a strength of 24$\mu$T, an inclination of -36.6° and a declination of 2.7°. The antennas used were dual-polarized logarithmic periodic dipole antennas with an effective bandwidth from 30–78 MHz. After applying a cut for zenith angles $\leq 55°$ and applying appropriate quality cuts for the Auger surface detector (SD) array, a total of 17 events coincident between the RDSs and the SD array have been found. We rely on the SD reconstruction of energy, arrival direction and core position, which are used as input for the analysis. The quoted 17 events have a detected signal in a varying number of RDSs; each of the RDSs with a detected signal contributes a data point to the following analyses.

### 2.2 Polarization angle analysis

In the first analysis (see also [10]), we determine the polarization angle $\phi_p$ for the radio pulse detected in a given RDS. Within our offline analysis software [11] we reconstruct the three-dimensional electric field vector represented in a cartesian coordinate system defined by $(x,y,z)$ = (geographic east, geographic north, vertical up). The relative strength of the electric field components $E_x$ and $E_y$, representing the projection of the electric field vector to the horizontal plane, are quantified in this analysis via the use of Stokes parameters $Q$ and $U$. The measured polarization angle is then given by:

$$\phi_p = \tan^{-1}\left(\frac{E_y}{E_x}\right) = \frac{1}{2}\tan^{-1}\left(\frac{U}{Q}\right). \tag{1}$$

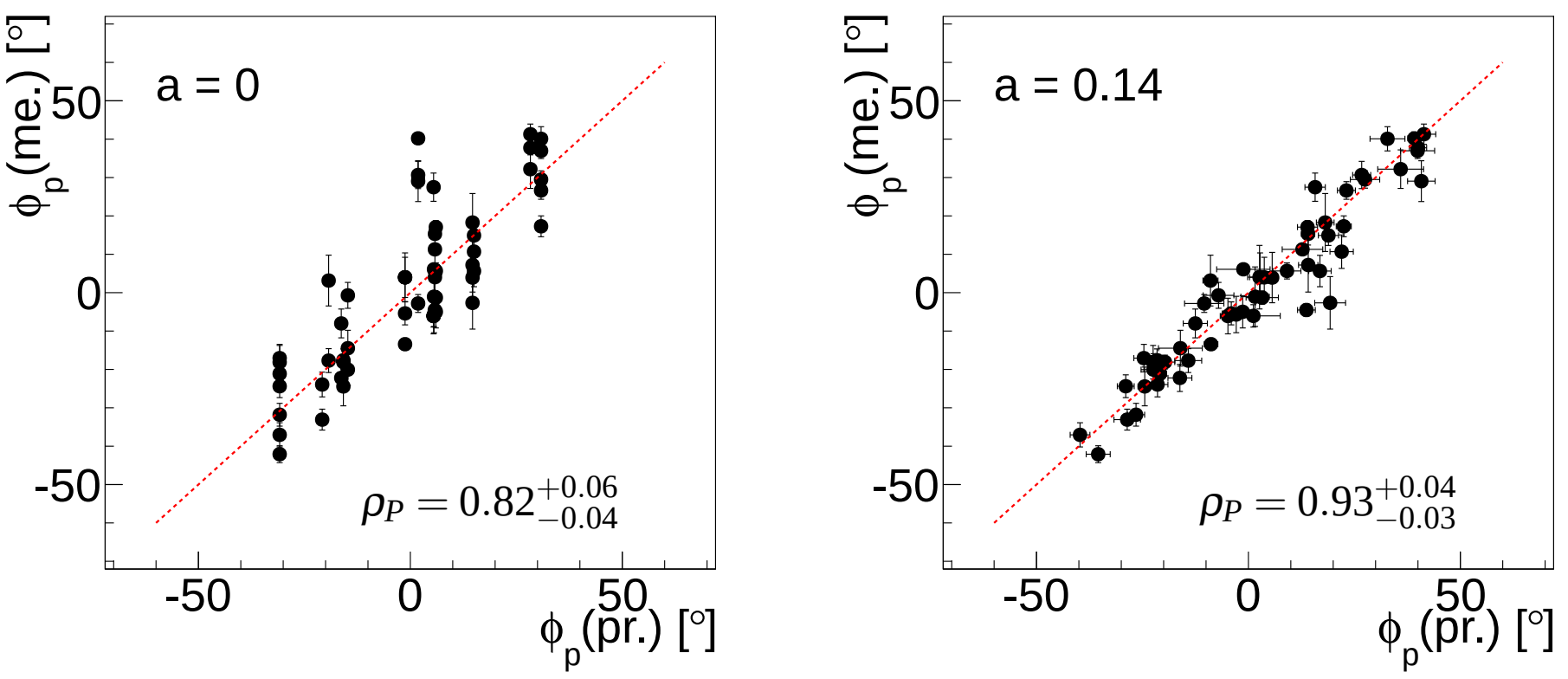

**Fig. 1**: Comparison of the polarization angle measured by AERA with the angle expected for pure geomagnetic radiation (left) and for a superposition of geomagnetic radiation and a contribution with radially aligned electric field vectors of a relative magnitude of 14% (right). The level of agreement is quantified with Pearson correlation coefficients and their 95% confidence levels.

In the left panel of Figure 1, we compare the measured polarization angle $\phi_p(\mathrm{me.})$ with the polarization angle $\phi_p(\mathrm{pr.})$ predicted for pure geomagnetic emission with linear polarization and electric field vectors aligned according to $(-\vec{v}\times\vec{B})$. A clear correlation is visible, as expected for emission dominated by geomagnetic radiation. However, there is significant spread resulting in a $\chi^2/\mathrm{ndf.} = 27$.

In contrast, we can adopt a model for the polarization angles which corresponds to a superposition of the geomagnetic $(-\vec{v}\times\vec{B})$ contribution and a secondary linearly polarized contribution with electric field vectors oriented radially with respect to the shower axis. In this model, the parameter $a$ denotes the relative strength of the radial contribution ($E_\mathrm{r}$) with respect to the geomagnetic emission ($E_\mathrm{g}$), where the latter is normalized by the sine of the angle $\alpha$ between the geomagnetic field and the shower axis:

$$a = \frac{|E_\mathrm{r}|}{|E_\mathrm{g}|/\sin\alpha} \tag{2}$$

A scan has been performed to find the value of $a$ which gives the best agreement with the totality of measured events, the result of which is $a = 0.14 \pm 0.02$. The resulting improvement is illustrated in the right panel of Figure 1; the $\chi^2/\mathrm{ndf.}$ drops to a value of 2. It is noteworthy that this value of $a$ is in agreement with the results of [9], reporting a non-geomagnetic emission component with a strength of $14 \pm 6$%, although it should be kept in mind that these measurements were made at a frequency of 22 MHz and at a different altitude. A detailed look at the individual events (not presented here) reveals that the value of $a$ determined for individual events exhibits scatter at a level not compatible with a constant value of $a$. This additional spread has been incorporated in the scan for the mean $a = 0.14$. Future studies with a larger data set will test a potential dependence of $a$ on air shower parameters such as zenith angle or depth of shower maximum.

### 2.3 R-parameter analysis

In the second analysis (see also [12]), we calculate the so-called $R$-parameter. To determine $R$, the reconstructed electric field vector is projected to the horizontal plane, and is then represented as the components along the axes $\xi$ and $\eta$, where $\xi$ is aligned in the direction of $(\vec{v}\times\vec{B})$ projected on the horizontal plane, and $\eta$ is 90° ahead of that. In this choice of coordinate system, any electric field due to geomagnetic radiation is oriented along $\xi$, so any contribution along $\eta$ is not of geomagnetic origin. To quantify the contributions of the electric field along $\xi$ and $\eta$, a Hilbert-envelope is performed on the bandwidth-limited time-trace, and then a sliding window is used to find the maximum power of an integral over 25 consecutive samples (10.0 $\mu$s). The $R$-parameter is then defined as

$$R(\psi) \equiv \frac{2\,\mathrm{Re}(\mathscr{E}_\xi \mathscr{E}_\eta^*)}{(|\mathscr{E}_\xi|^2 + |\mathscr{E}_\eta|^2)}, \tag{3}$$

where the $\mathscr{E}_i$ denote the $\eta$ and $\xi$ components of the complex-valued integrated Hilbert-envelope of the electric field vector and $\psi$ denotes the observer-angle, i.e., the angle in the horizontal plane by which a given antenna is offset from the axis $\xi$. (For $\psi = 0$, the antenna is located in the direction defined by $\vec{v}\times\vec{B}$ with respect to the shower axis.) $R(\psi)$ thus quantifies deviations from pure geomagnetic emission polarization as a function of observer angle. The resulting distribution of $R$-values for the AERA data is shown in Figure 2. This $R$-distribution is clearly not compatible with $R \equiv 0$, which would correspond to pure geomagnetic radiation. A sinusoidal pattern appears to be visible, indicating that for particular ranges of observer angles, a significant contribution along the $\eta$ axis is present.

## 3 Model comparison

Having shown that the cosmic ray radio emission measured with AERA cannot be explained by pure geomagnetic radiation, we compare the measured $R$-parameters with those derived from simulations performed with various available models. The input parameters for the event simulations are derived from the Auger SD reconstruction and comprise the particle energy, the core position, and the ar-rival direction. To take into account the SD reconstruction uncertainties, each of the 17 air showers was simulated 25 times, varying all input parameters within their respective uncertainties and properly taking parameter correlations into account. The simulated electric field vectors were propagated through a detailed detector simulation taking into account the effects of the logarithmic-periodic dipole antennas, the analog electronics and the digitization. Those simulations were then reconstructed in the same way as the measured data.

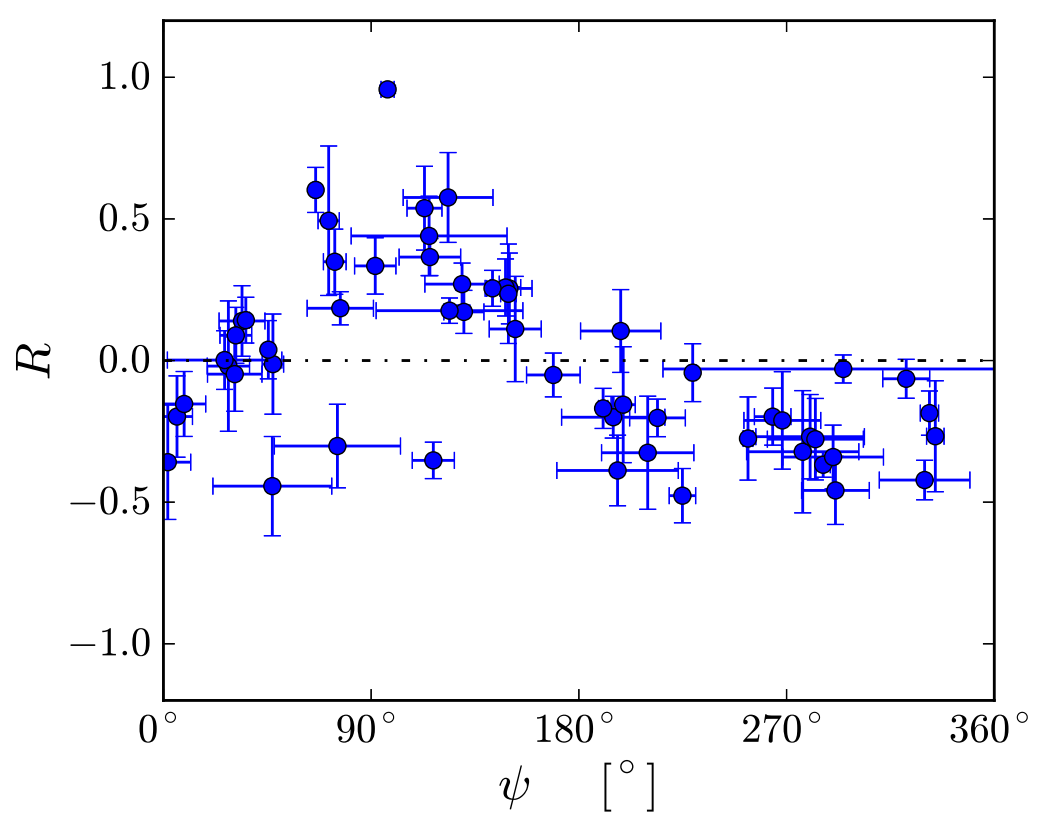

**Fig. 2**: *R*-parameter measured by AERA as a function of antenna observer angle $\psi$ (see text). A horizontal line with $R \equiv 0$ is expected for pure geomagnetic radiation.

In Figure 3, a direct comparison of the $R$-values predicted by the simulations and the measured $R$-values is shown. The measured and simulated $R$-values are clearly correlated, as indicated by the respective Pearson correlation coefficients (a value of 1 signifies full correlation, 0 means uncorrelated data, and -1 signifies full anti-correlation). All of these calculations include the Askaryan charge-excess contribution either explicitly (macroscopic approaches) or implicitly (microscopic approaches). A realistic refractive index of the atmosphere is incorporated for CoREAS [13], REAS [14], EVA [15] and ZHAireS [16], whereas SELFAS [17] and MGMR [18] used a refractive index of unity. In some calculations, the charge-excess contribution can be switched off. This should result in a value of $R \equiv 0$ irrespective of observer angle $\psi$ and is confirmed by the simulation results shown in Figure 4. Without the charge-excess contribution, there is a clear disagreement between the measurements and simulations as indicated by Pearson correlation coefficients compatible with 0. This indicates that the charge-excess contribution is necessary for a proper description of the AERA data.

None of the calculations, however, can describe the measurements completely consistently, and the differences between calculations with respect to the agreement of measured and simulated $R$-parameters are relatively small.

## 4 Conclusion

We have shown with two different analyses that the AERA data, while dominated by linearly polarized geomagnetic emission with electric field vectors oriented along $(-\vec{v} \times B)$, exhibit a systematic deviation in the polarization of the measured signal. This deviation is consistent with a linearly polarized emission contribution with a radially aligned electric field vector. Previously, a systematic shift of the core position reconstructed on the basis of CODALEMA radio data had also indicated the presence of such a contribution with radially oriented electric field vectors [19]. The Askaryan charge-excess emission exhibits this particular polarization pattern, and a comparison of AERA data with simulations demonstrates that calculations including the Askaryan effect can reasonably describe the AERA measurements, while calculations without the Askaryan effect can clearly not. Remaining discrepancies between the modeled and measured polarization characteristics are not yet fully understood and need to be studied in further detail. Such polarization measurements can be used as a tool to test models, ideally in conjunction with other methods such as the comparison of absolute predicted amplitudes and lateral distribution functions.

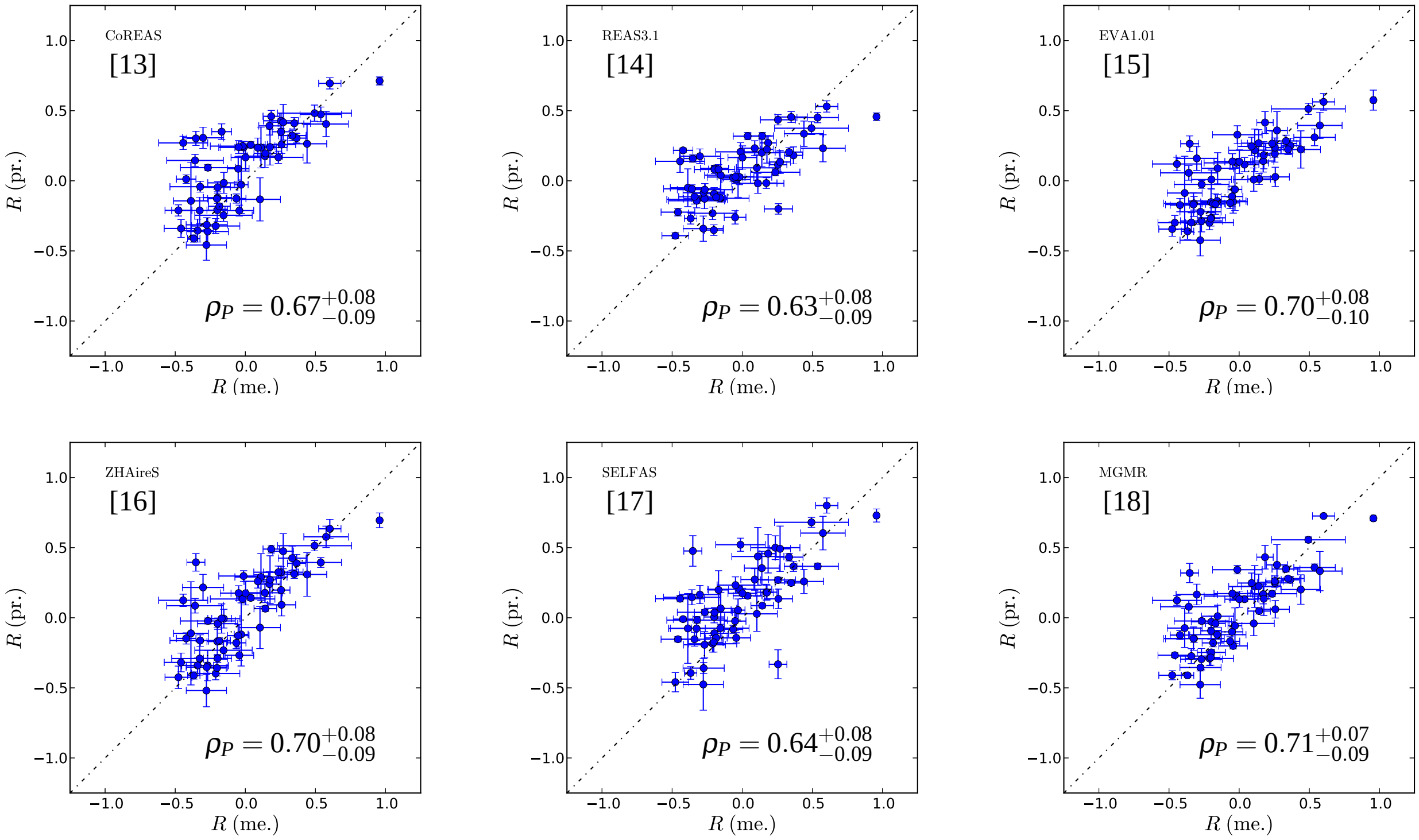

**Fig. 3**: Comparison of the *R*-parameters measured by AERA with six models for radio emission from extensive air showers. These calculations include the Askaryan charge-excess emission mechanism. The Pearson correlation coefficients and their 95% confidence levels quantify the level of agreement between simulation and data.

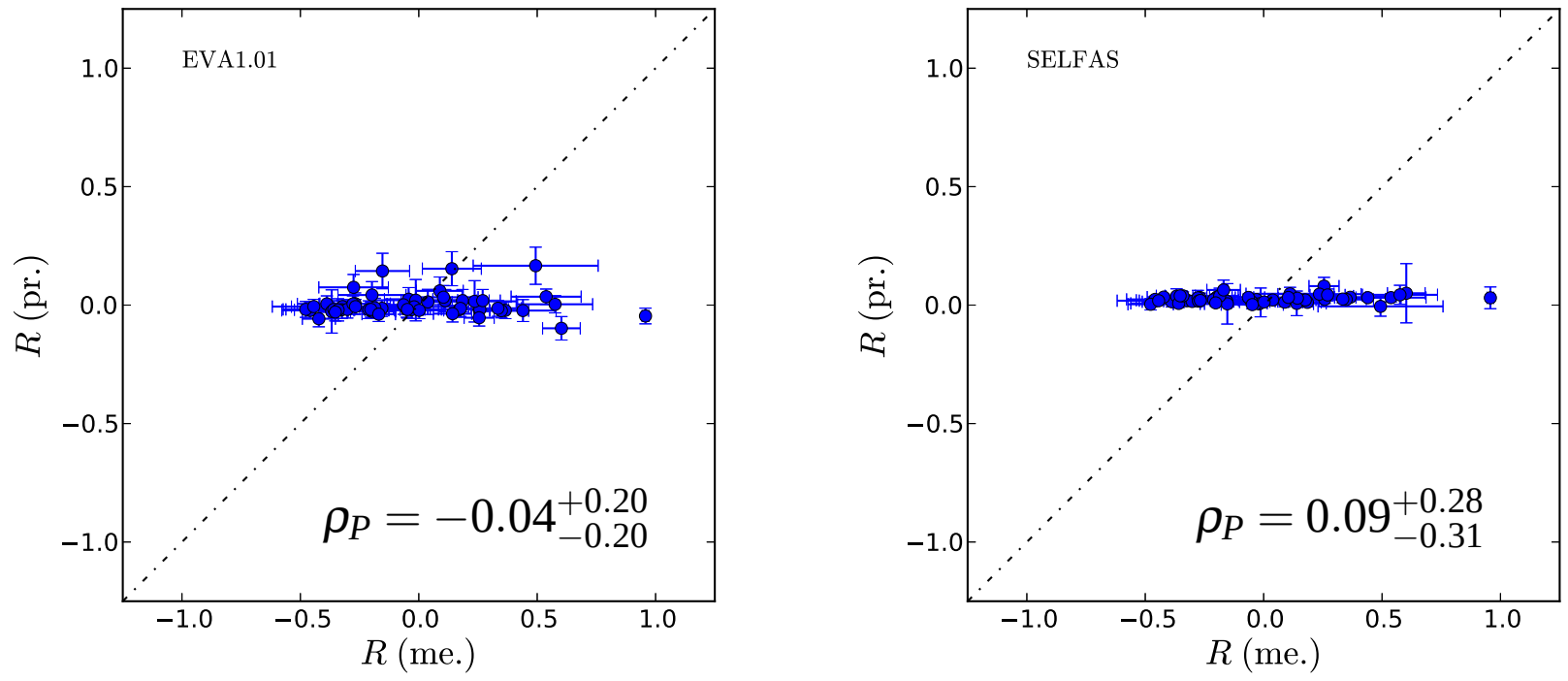

**Fig. 4**: Comparison of the *R*-parameters measured by AERA with two models in which the Askaryan charge-excess emission mechanism has been deactivated.

# Detection of cosmic rays using microwave radiation at the Pierre Auger Observatory

ROMAIN GAÏOR[1], FOR THE PIERRE AUGER COLLABORATION[2].

[1] *Laboratoire de Physique Nucléaire et de Hautes Energies, Universités Paris 6 et Paris 7, CNRS-IN2P3, Paris, France*
[2] *Full author list: http://www.auger.org/archive/authors_2013_05.html*

*auger_spokespersons@fnal.gov*

**Abstract:** The discovery of microwave radiation from the passage of charged particles has opened a new window for the detection of ultra high energy cosmic rays. The main potential advantages of this technique are the possibility to instrument a large area with a duty cycle of detection close to 100% and no atmospheric attenuation, all this using relatively cheap equipment. Cosmic ray detection in the GHz band is being pursued at the Pierre Auger Observatory with three different set-ups: MIDAS and AMBER are prototypes of an imaging parabolic dish detector, while EASIER instruments the surface detector units with a radio receiver of wide angular coverage. The status of microwave R&D activities at the Auger Observatory, including the first detections of cosmic ray air showers by EASIER, will be reported.

## 1 Introduction

The Pierre Auger Observatory [1] detects Ultra High Energy Cosmic Rays (UHECR) using a hybrid detector. The surface detector array (SD) is composed of 1660 water Cherenkov detectors that sample the air shower at the ground. The fluorescence detector (FD) consists of 27 telescopes installed at five sites and measures the shower development in the atmosphere by observing the fluorescence light. Recently the Auger Collaboration has undertaken the development of new detection techniques to enhance the current detection capability of the Observatory and serve as a test-bed for next generation experiments. Among these, radio detection techniques play a crucial role. The VHF band, between 30 and 80 MHz is extensively studied with the AERA [2] setup. Radio detection in microwave band is another alternative. It was triggered by the observation of a signal in the 1.5-6 GHz band upon the passage of an electron beam in a anechoic chamber [3]. The emission mechanism, interpreted as Molecular Bremsstrahlung Radiation (MBR), is expected to produce an unpolarized and isotropic signal. Moreover, the power emitted in microwaves was measured to scale quadratically with the beam energy. The expected emission from air showers together with the transparency of the atmosphere at these frequencies would allow the measurement of the shower longitudinal development with an almost 100% duty cycle. Three projects, AMBER, EASIER and MIDAS are being developed to measure this emission and prototypes are now operated at the Pierre Auger Observatory. We will describe the status of these developments and then report on the first detection of radio signals in microwave band in coincidence with air shower detected by the regular SD array and discuss their possible origin.

## 2 Microwave detection at the Pierre Auger Observatory

AMBER and MIDAS are imaging telescopes like an FD, instrumenting an array of feed horn antennas at the focus of a parabolic dish. EASIER is an alternative design to a radio telescope: it is embedded in the SD, observing the shower from the ground with a wide angle antenna pointing to the zenith. The locations of the three prototypes at the Pierre Auger Observatory are depicted in Fig. 1. All three take advantage of the available commercial equipment for TV satellite reception. They all use horn antennas as receivers, in C-band (3.4-4.2 GHz) and Ku-band (10.95-14.5 GHz) for AMBER, and only C-band for MIDAS and EASIER. A Low-Noise Block down-converter (LNB) is used to shift the central frequency below 2 GHz and amplify the signal. The RF signal is then transformed using a power detector whose output is a DC voltage proportional to the logarithm of the input power. The signal thus integrated can be acquired with sampling rates below 100 MHz. The three prototypes benefit from the commissioning at the Pierre Auger Observatory because of the radio quiet environment and the possible coincident detection with the SD or the FD. We present

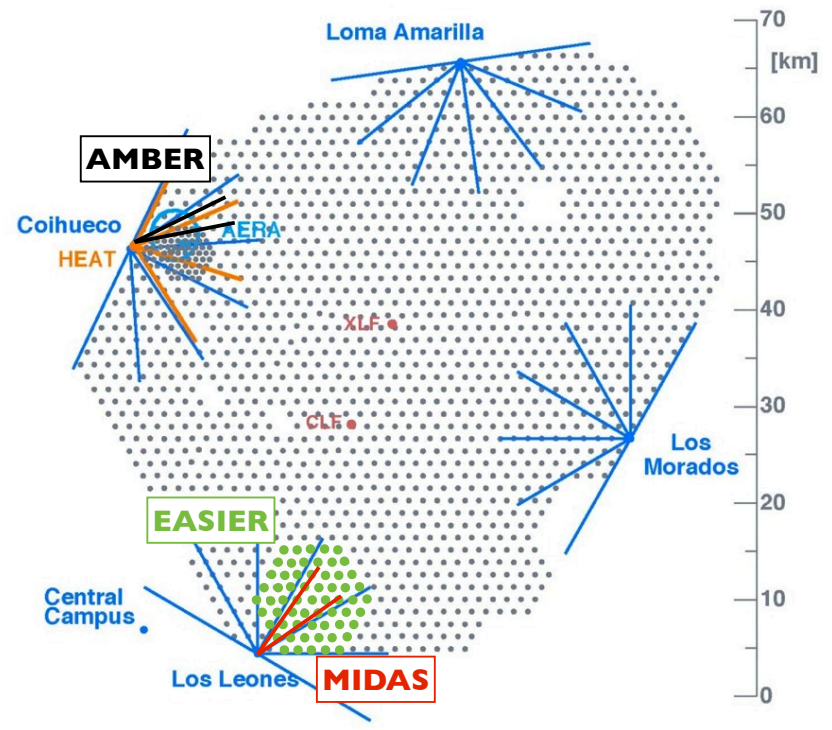

**Fig. 1**: Locations of the three microwave detection prototypes at the Pierre Auger Observatory. The field of view of AMBER and MIDAS are respectively delimited by the black and red lines. The EASIER array is represented by the green dots.

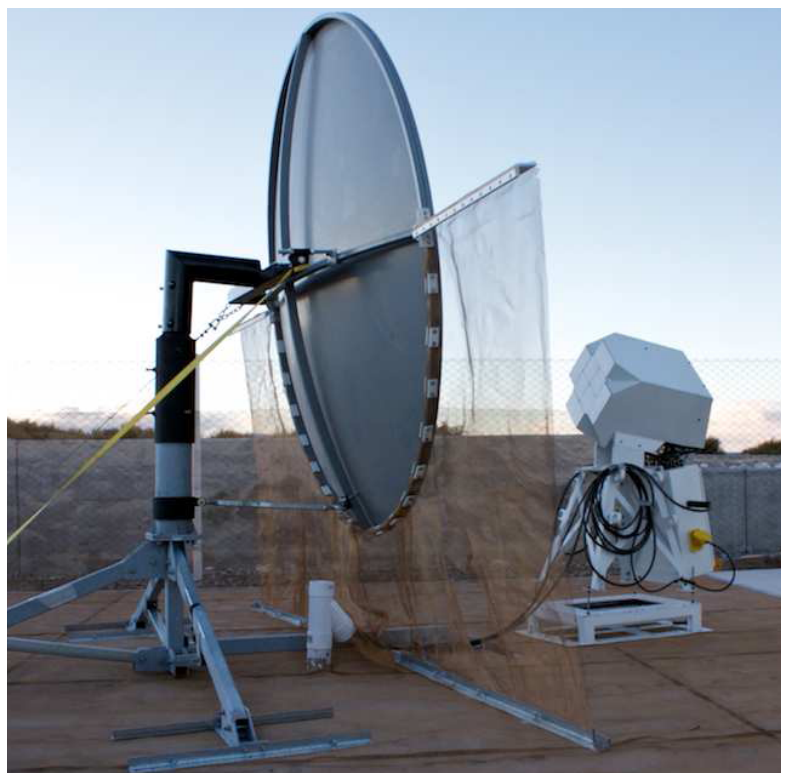

**Fig. 2**: AMBER detector at the Coihueco FD site.

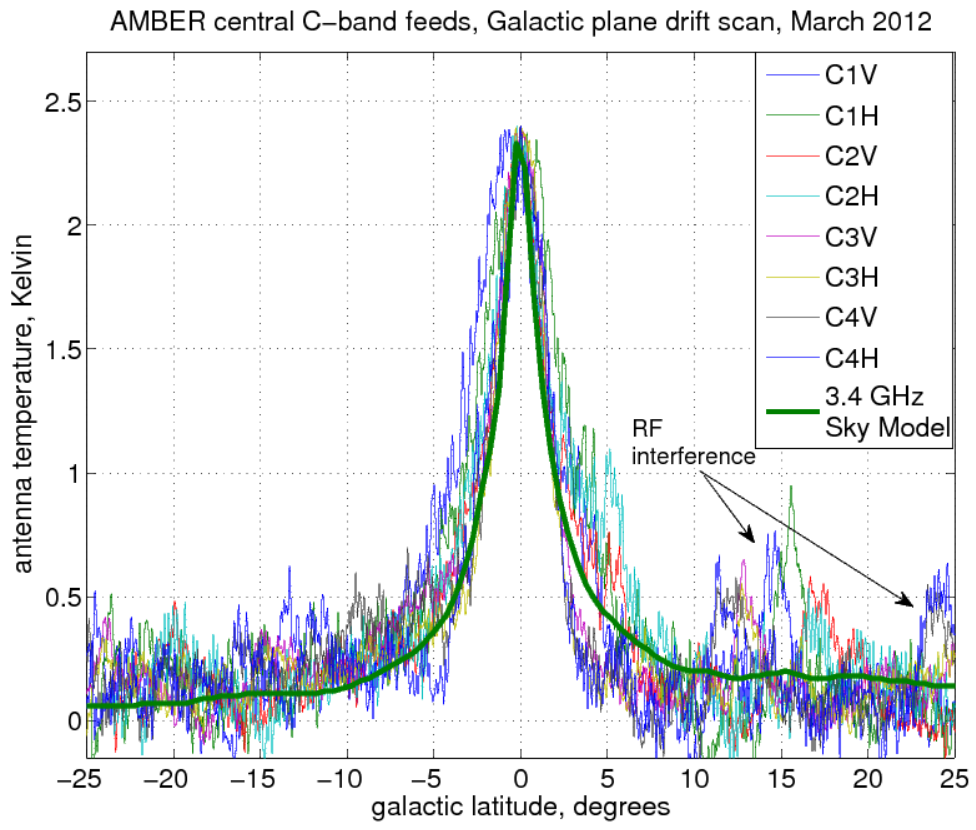

**Fig. 3**: Temperature elevation during the crossing of the Galactic Plane in the AMBER field of view.

here separately each prototype. More technical details can be found in a previous contribution [4].

## 2.1 AMBER

AMBER (Air shower Microwave Bremsstrahlung Experimental Radiometer) is a radio telescope instrumented with a 2.4 m off-axis parabolic dish imaging a section of 14 °x14 ° of the sky at 30 ° elevation angle with 16 pixels. The dish and the receivers are shown in Fig. 2. The four central pixels are dual polarized and dual band (C-band and Ku-band) and the 12 outer pixels are single polarized in C-band. The power detector output of each channel is sampled at 100 MHz with FADCs. The dish and the feeds were calibrated separately using the Y-factor method. The combined noise temperature was measured to be 45 to 65 K for the C-band pixels and around 100 K in the Ku-band.

AMBER was originally operated at the University of Hawaii with a self-triggered system. During this period, the validation of the optical performance of the telescope was performed by measuring the Sun transit. A search for signals induced by air shower was also performed, however the environment was found to be too noisy and no unambiguous event was found.

AMBER was shipped to Argentina and is now installed at the Coihueco FD site pointing in direction of the SD infill array [5] (cf. Fig. 1). AMBER uses a modified version of the SD trigger at the three-level trigger [6] that performs a fast geometrical reconstruction of the SD events and retrieves the time at which the shower crossed its field of view. Uncertainties in this reconstruction are compensated for by pulling an appropriately long trace (currently 150 us), from a large circular buffer of 5 s for each channel. This fast reconstruction is found to be valid within 10 ° in the shower direction and 500 m in the core position, uncertainties that are accounted for in the length of the trace being read out.

The triggering system requires a precise synchronization between the timing of Auger and AMBER detectors. It was tested twice on separate occasions by instrumenting first one, and then three surface detectors with C-band antennas and power detectors. A strong RF pulse was used to create a trigger in the antenna-equipped detectors and at the same time recorded in AMBER. An agreement in the order of 1 $\mu$s was found in the single tank test.

A calibration method based on the microwave signal emitted by the galactic plane is also in development to complement the Sun transit calibration. The baseline of the central C-band pixels averaged over 20 minutes is shown in Fig. 3 as a function of the galactic latitude. AMBER has acquired more than 18 months of data, and the data analysis is underway. An upgrade of the camera is under development to improve the sensitivity by 40% by lowering the noise temperature of the electronics and by increasing the efficiency of the focal surface. The field of view is also planned to be extended to 17 °.

## 2.2 EASIER

EASIER (Extensive Air Shower Identification with Electron Radiometer) is a radio detector array integrated with the Auger SD as illustrated in Fig. 4. Each detector is composed of a C-band horn antenna oriented towards the zenith covering a large field of view, 3 m above the ground. The output is sampled at 40 MHz by one of the six FADC channels initially used for the anode signal of one of the three PMTs. In this way, whenever an air shower triggers the SD, the radio trace is automatically recorded through the same stream as the SD data.

In April 2011, seven tanks were instrumented with an EASIER prototype and the first clear UHECR radio detection in this band was performed by one of those prototypes in June 2011. An extension of 54 units was carried out in April 2012. EASIER is now an array of 61 detectors with 33 antennas oriented with a North-South polarization and the other 28 ones with East-West polarization.

Calibration of the EASIER antennas is still underway. The simulation of the antenna pattern shows a half power beam width of around 100 ° and a maximum gain of 5 dBi.

The EASIER antennas collect the air shower signal from the ground. In such conditions the signal emitted along the air shower is compressed in time. The exact signal enhancement due to this compression depends on the arrival direction and distance to the shower axis. The smaller effective area with respect to a telescope is thus compensated by this compression effect.

EASIER has been taking data in a very stable way for two years for the first set up and one year for the second and up to now, it has recorded a total of three unambiguous radio signals in coincidence with an air shower detected by the SD array. The data selection, the radio signals and their possible origin are discussed in Section 3.

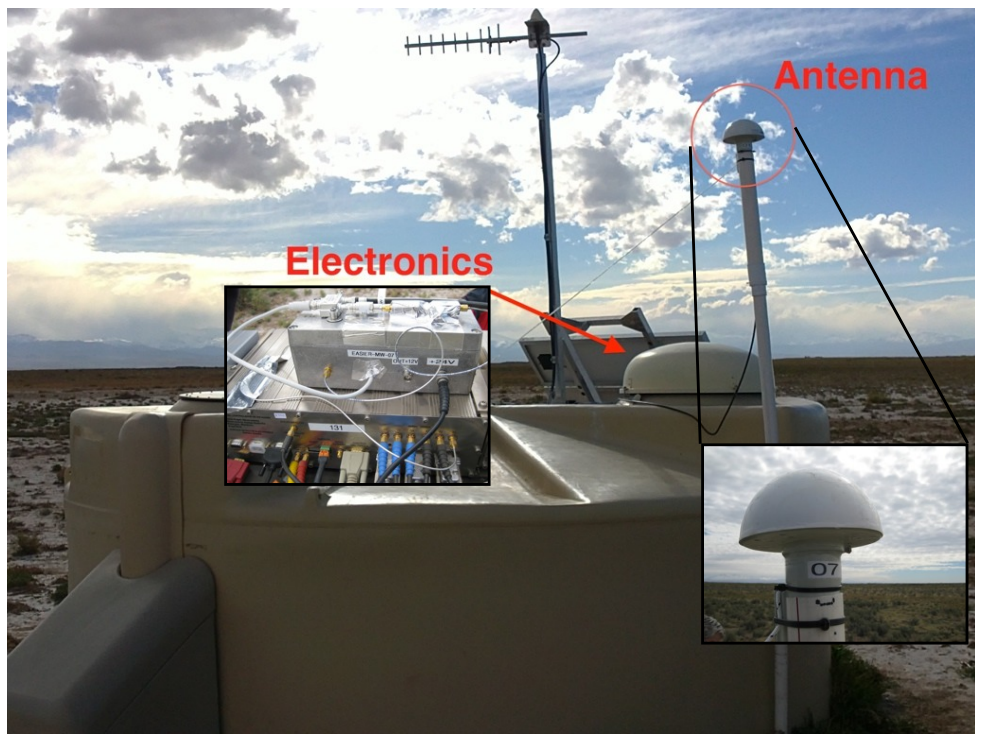

**Fig. 4**: EASIER detector installed at a surface detector.

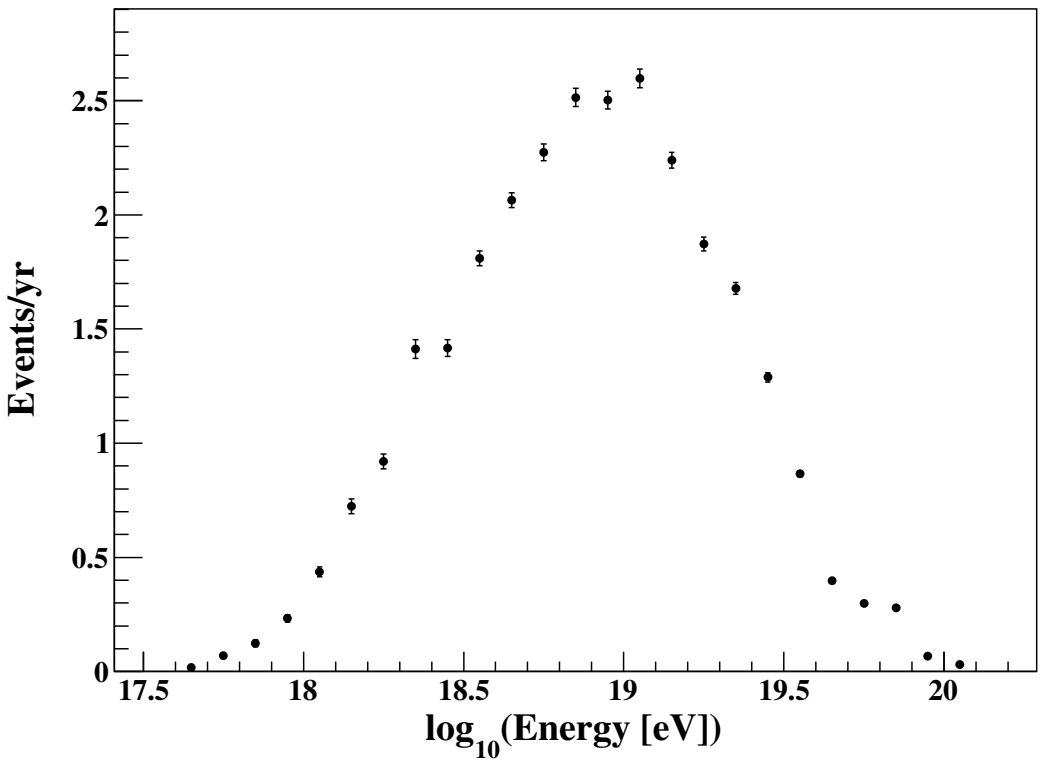

**Fig. 5**: Simulated event rate computed for the MIDAS detector [7] accounting for a linear scaling with energy of the observed signal in [3].

### 2.3 MIDAS

MIDAS (MIcrowave Detection of Air Showers) is a radio telescope instrumented with a 5 $m^2$ parabolic dish and a 53 pixels camera at its focal plane. Each pixel is a C-band LNBF covering approximately 1.3°x1.3° of the sky, for a total field of view of approximately 20°x10°. Each channel is digitized by a 14 bit FADC at 20 MHz sampling rate. MIDAS incorporates its own triggering logic. A First Level Trigger (FLT) at the pixel level is issued if the running sum of 20 data samples exceeds a predefined threshold. The FLT remains active for 10 $\mu s$ and the value of the threshold is adjusted to keep a FLT rate of 100 Hz. The Second Level Trigger (SLT) searches for four-fold patterns corresponding to the expected topology of a cosmic ray air shower in the overlapping FLT pixels. There are 767 expected patterns compatible with a cosmic ray air shower track giving an accidental SLT rate of $3 \times 10^{-4}$ Hz [7].

The telescope efficiency was calculated by performing a complete electromagnetic simulation of the MIDAS detector. The effective area at the central pixel was found to be 9.1 $m^2$ and falls to 20% of this value at the borders of the camera.

The MIDAS detector was originally installed at the University of Chicago. During this period of commissioning, the Sun was used as a calibrated source. Firstly the electromagnetic simulations were validated measuring the Sun transit over the camera. Secondly, as the flux of the Sun is monitored by several radio observatories, it was used to compute an absolute calibration. The system temperature of the central pixel was found to be $65 \pm 3$ K and similar values were obtained for the other pixels.

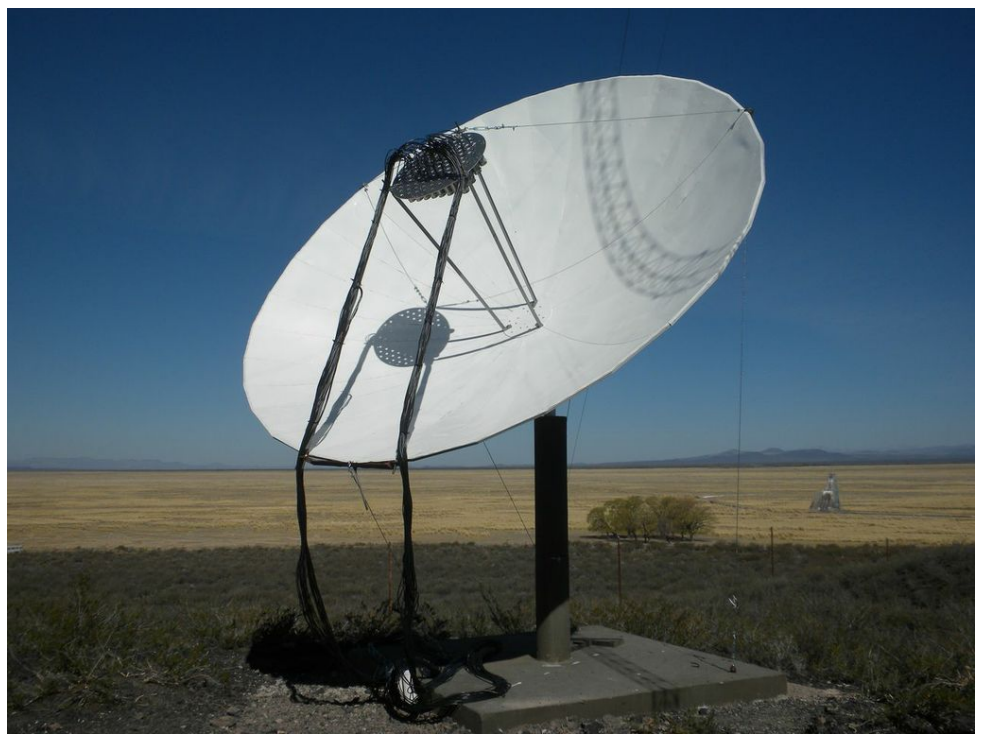

**Fig. 6**: MIDAS detector installed at the Pierre Auger Observatory next to the Los Leones FD building.

The data taking at the University of Chicago validated the principle of MIDAS and showed a stable behavior regardless of the weather conditions. No clear event candidate was found, thus excluding a quadratic scaling with the air shower energy [8] of the microwave signal measured in the beam experiment mentioned in the introduction. In the hypothesis of a linear scaling, a realistic simulation of the MIDAS detector yields a total of $\simeq 30$ events per year. The expected energy spectrum is shown in Fig. 5.

MIDAS is now installed at the Pierre Auger Observatory, next to the FD building Los Leones (cf. Fig. 6) and has been taking data since the beginning of 2013.

## 3 First detections of air showers in microwave band

The first detection of an air shower in microwave was performed in June 2011 by one of the EASIER detectors. It was in coincidence with an air shower registered by the SD that had an energy of 13.2 EeV and a zenith angle of 29.7°. The recorded GHz signal of this event is shown in Fig. 7 together with the PMT traces. The maximum of the signal was found to be more than 11 times larger than the noise fluctuations and occured just one time bin (25 ns) before the signal in the water Cherenkov detector.

A search for signals of air showers has been performed in the data of the extended EASIER array analyzing the maximum of the trace within a window of 200 ns around the station trigger. The normalized distribution of the maximum in this time window is shown in Fig. 8 in red and in $\sigma$ units (where $\sigma = (\text{maximum} - \text{mean})/(\text{standard deviation})$). On the same figure the distribution shown in blue represents the trace maximum found outside the selection window (and thus expected to be uncorrelated with the shower). We present in Table 1 the characteristics of the three events that lay above the noise distribution, i.e. above 8 $\sigma$.

All the air showers that gave rise to a radio pulse landed close to an EASIER antenna with an E-W polarization. The maximum distance from the antenna to shower axis is around 270 m for a rather inclined shower (55°). The detected radio pulses are not longer than 75 ns and their maximum occurs just before the start time of the PMT signal of the corresponding water Cherenkov detector. These characteristics make the interpretation of the signal difficult and are similar to the ones of the event candidates reported by another microwave experiment, CROME [9] at KASCADE site. On one side, at such close distance any emission from an air shower, even an isotropic one, is compressed in time and the signal is shortened and amplified, as seen in the EASIER data. On the other side, the viewing angle of the showers is close to the Cherenkov angle and the compression effect would also increase the observed frequency. One cannot discard an emission at lower frequencies shifted to the C-band. For instance, the radiation from the transverse current due to the geomagnetic deflections of the charged particles observed in the VHF [10] band could be the underlying emission process. An excess of detected events from the southern direction would point to a geomagnetic origin, but a larger data set is required to make relevant polarization comparisons.

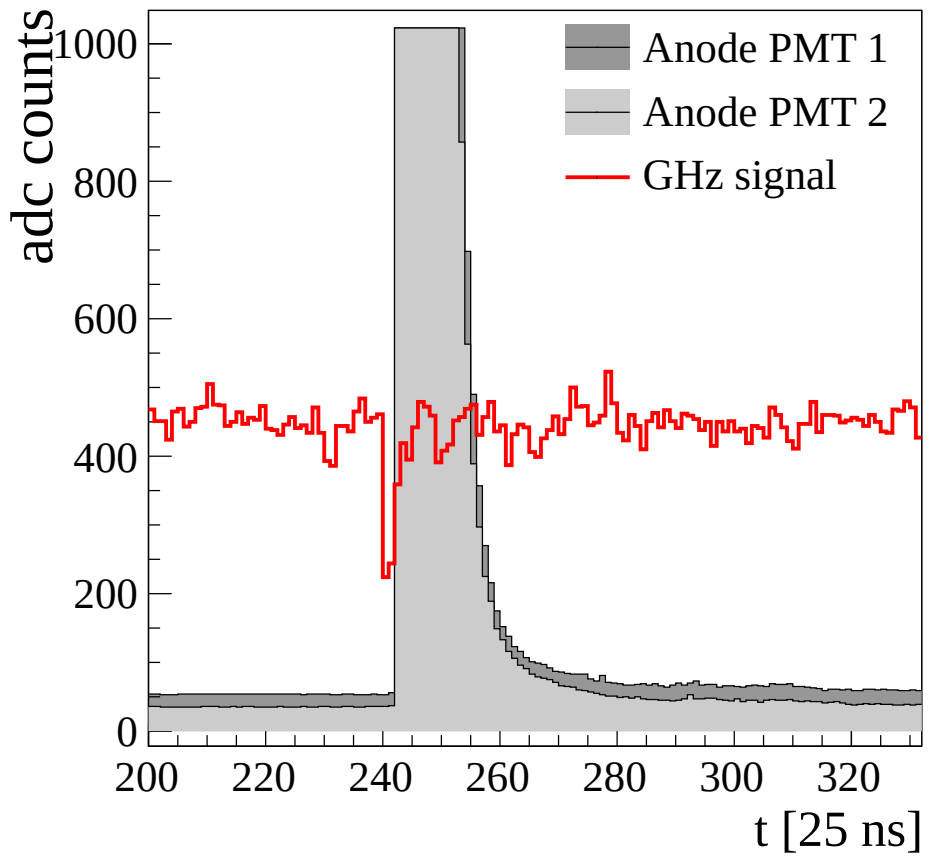

**Fig. 7**: Radio trace (in red) of the first event recorded by an EASIER detector with the signal of two low gain PMT channels (gray). The PMT signals are saturated as expected for a shower with the core at 136 m of the detector.

| Event ID | E [EeV] | $\theta$ [°] | $\phi$ [°] | d [m] | pol. |
|---|---|---|---|---|---|
| 12046376 | 13.2 | 29.7 | 344.6 | 136 | E-W |
| 20830870 | 17.1 | 55.3 | 33.8 | 269 | E-W |
| 21050180 | 2.6 | 47.4 | 290 | 193 | E-W |

**Table 1**: Main characteristics of the air showers detected in GHz range. Here E stands for energy; $\theta$ and $\phi$ for the zenith and azimuthal angle, d for the distance the shower axis ans pol. for the polarization of the antenna)

Further studies will be focused on the search for a fainter but longer signal and from more distant air showers. The current development of simulations of the MBR and other processes and detectors simulation as well as future results from the test beam experiments AMY [11] and MAYBE [12] will enable a better understanding of the observed emissions. Furthermore, the recent installation of MIDAS, the ongoing analysis of the AMBER data and its future upgrade will help in disentangling the origin of the emission process.

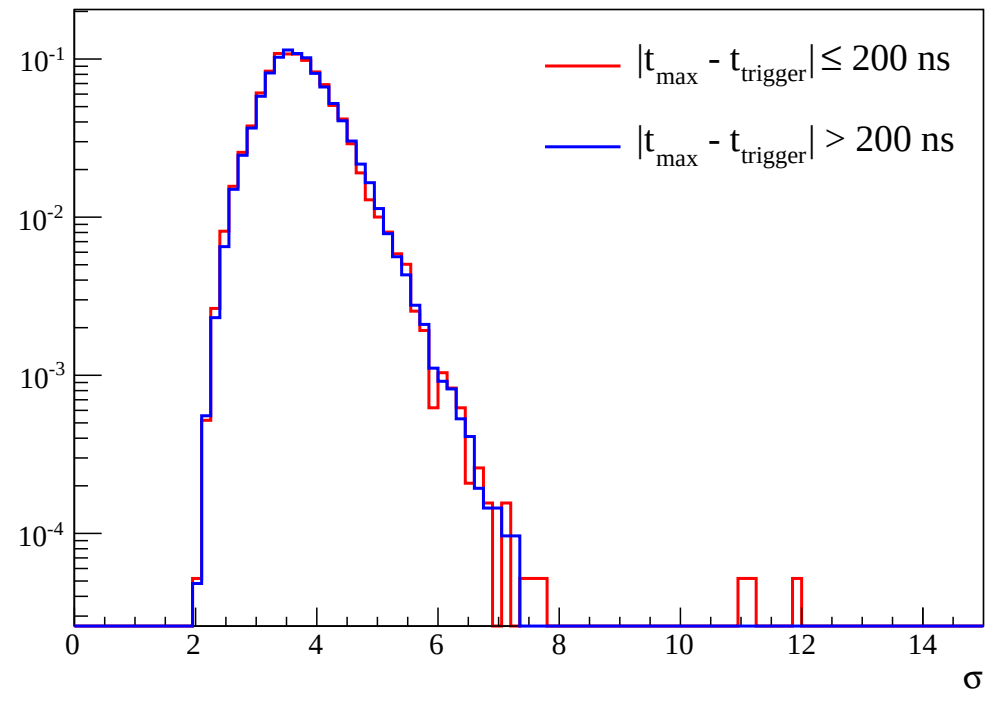

**Fig. 8**: Distribution of maxima found in a 200 ns time window around the trigger (in red) and outside this window (in blue).

## 4 Conclusions

The effort undertaken in microwave detection of air showers within the Pierre Auger Collaboration resulted in the installation of three prototypes at the Observatory site. All of them are now in the phase of stable data taking. The first three unequivocal radio signals detected in the GHz range by EASIER in coincidence with air showers detected by Auger SD were reported. However, because of their characteristics one cannot draw a conclusion on the emission mechanism. The viability of this technique remains an open question and the unique conditions offered by the Pierre Auger Observatory site should allow it to be addressed in the near future.

# Measuring Atmospheric Aerosol Attenuation at the Pierre Auger Observatory

Laura Valore[1] for the Pierre Auger Collaboration[2].

[1] *University of Naples Federico II and INFN Naples*
[2] *Full author list: http://www.auger.org/archive/authors_2013_05.html*

*auger_spokespersons@fnal.gov*

**Abstract:** The Fluorescence Detector (FD) of the Pierre Auger Observatory provides a nearly calorimetric measurement of the primary particle energy, since the fluorescence light produced is proportional to the energy dissipated by an Extensive Air Shower (EAS) in the atmosphere. For this reason, the FD is used to calibrate the absolute energy scale of the Surface Detector (SD) by means of hybrid events. Of the major correction terms applied to the FD, atmospheric transmission through aerosols has the largest time variation. The corresponding correction to an EAS energy can range from a few percent to more than 40%, depending on the aerosol attenuation conditions, the distance of the shower, and the energy. We report on 9 years of hourly aerosol optical depth profile measurements, including revised statistical and systematic error estimates, that are propagated through EAS reconstruction. To accumulate these hourly aerosol optical depth profiles, the Central Laser Facility (CLF) and the eXtreme Laser Facility (XLF) of the Auger Observatory generated more than 4 million laser tracks that were recorded by the FD telescopes. Finally we describe major upgrades in progress to the CLF and to the elastic LIDAR stations at the Pierre Auger Observatory. The main features of these complementary upgrades are discussed together with the expected results of their applications.

## 1 Atmospheric Aerosol Attenuation

Ultra High Energy particles entering the atmosphere produce a cascade of secondary particles, the Extensive Air Shower (EAS). During the development of an EAS, fluorescence light in the range 300–420 nm is emitted isotropically by excited air molecules. The Pierre Auger Observatory combines two well-established techniques to detect EAS, the detection of secondary particles at the ground (Surface Detector, SD) and the detection of fluorescence light emitted in the atmosphere (Fluorescence Detector, FD)[1]. The FD is composed of 24 telescopes positioned at 4 sites[1] overlooking the 1660 stations composing the SD covering an area of 3000 km$^2$. The FD provides a direct estimate of the energy of the primary particle without the need for simulations, therefore FD data are used to calibrate the absolute energy scale of the SD by means of hybrid events.

The direct measurement of the energy is possible since the amount of fluorescence light produced during the development of an EAS is proportional to the energy dissipated in the atmosphere by the EAS. The atmosphere is therefore comparable to a giant calorimeter, whose properties must be continuously monitored to ensure a reliable energy estimate. Atmospheric parameters influence both the production of fluorescence light and its attenuation to the FD telescopes. The molecular and aerosol scattering processes that contribute to the overall attenuation of light in the atmosphere can be treated separately. The molecular scattering is calculated once temperature, pressure and humidity are known from balloon and weather station measurements or model data[2]. The aerosol attenuation of light is the largest time dependent correction applied during air shower reconstruction, as aerosols are subject to significant variations on time scales as little as one hour. If the aerosol attenuation is not taken into account, the shower energy reconstruction is biased by 8 to 25% in the energy range measured by the Pierre Auger Observatory. On average, 20% of all showers have an energy correction larger than 20%, 7% of showers are corrected by more than 30% and 3% of showers are corrected by more than 40% [3]. Hourly vertical aerosol optical depth profiles are produced for each FD site for a correct reconstruction of FD events. 9 years of aerosol attenuation profiles, from January 2004 to December 2012, have been measured.

## 2 Laser Facilities

The Pierre Auger Observatory has a huge atmospheric monitoring system. Two laser facilities, the Central Laser Facility (CLF) and the eXtreme Laser Facility (XLF), both positioned nearly equidistant from three out of four of the FD sites (see figure 1), have been in operation for many years and provide vertical and inclined calibrated test beams. Sets of 50 vertical shots are produced every 15 minutes during FD data acquisition. The CLF [4], built in late 2003, operational since January 2004, is located at an altitude of 1416 m above sea level. The XLF was installed north of the CLF during 2008, closer to Loma Amarilla, at an altitude of 1397 m and has been producing stable laser shots since January 2010. Each facility uses a pulsed frequency tripled Nd:YAG laser (355 nm), whose wavelength is near the center of the spectrum of the fluorescence light, firing with an average energy of 6.5 mJ. A depolarizer is used to randomly polarize the laser light, to simulate the isotropic emission of the fluorescence light. CLF and XLF events are recorded by the FD telescopes and a specific GPS timing is used to distinguish laser from

1. in addition to the four FD sites, 3 high elevation telescopes (H.E.A.T.) are operating at a fifth site close to Coihueco

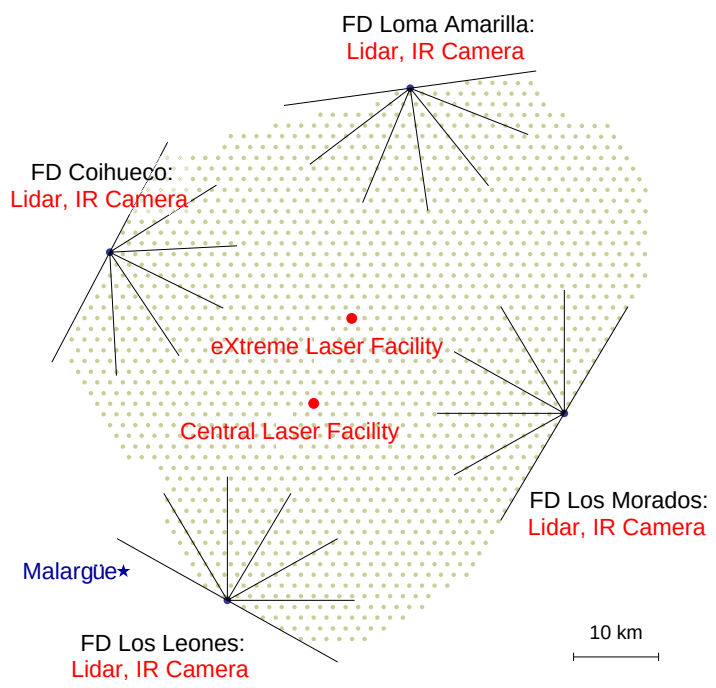

**Figure 1**: Map of the Pierre Auger Observatory. Some of the atmospheric monitoring devices are shown. The CLF and XLF are marked. The solid lines indicate the field of view of individual fluorescence telescopes.

EAS events. The amount of light scattered out of a 6.5 mJ laser beam by the atmosphere is roughly equivalent to the amount of fluorescence light produced by an EAS of $5 \times 10^{19}$ eV at a distance to the telescope of about 16 km, as shown in figure 2.

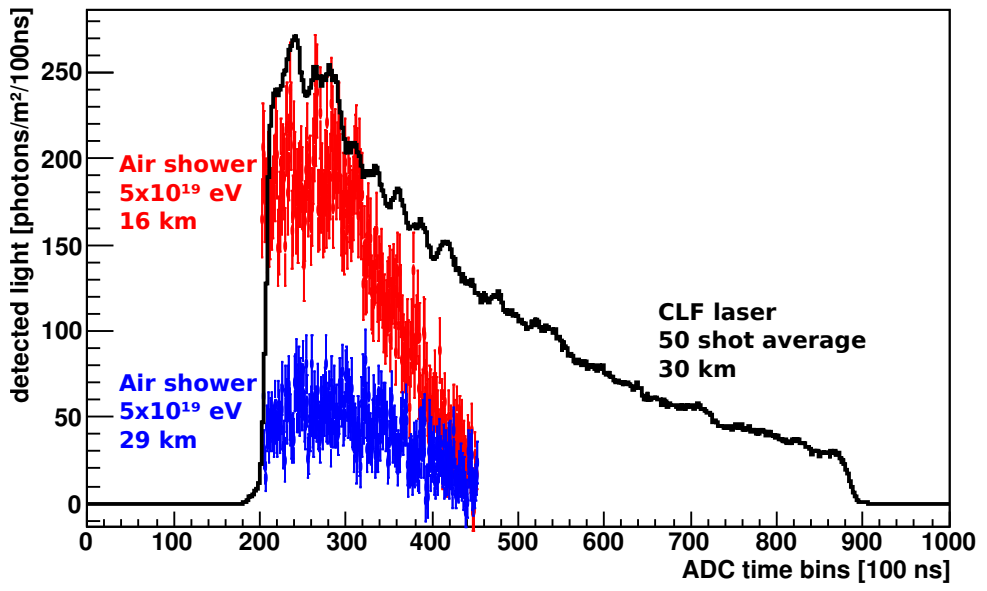

**Figure 2**: Comparison between a 50 shot average of vertical 6.5 mJ UV laser shots from the CLF and near-vertical cosmic ray showers measured with the FD. The cosmic ray profile has been flipped in time.

The laser energy of the CLF is monitored by a pyroelectric probe receiving a fraction of the laser beam for a relative calibration of each laser shot. Additionally, absolute calibrations are performed periodically, capturing the entire laser beam with an external radiometer before sending the laser light to the sky. The periodic absolute calibration permits to correct the sky energy for the effects related to dust accumulation on some of the optics of the laser bench. The XLF is equipped with a combined system of a pick-off probe for relative calibration, together with an automated calibration system which performs absolute calibrations on a nightly basis using a robotic arm moving a calibration probe in the beam path of the XLF laser.

## 3 Hourly aerosol optical depth profiles

Laser light is attenuated in the same way as fluorescence light as it propagates towards the FD. Therefore, analysis of the amount of laser light that reaches the FD as a function of time can be used to infer the attenuation due to aerosols between the position of the laser and each FD building. Two independent analyses have been developed to provide hourly aerosol characterization in the FD field of view using vertical laser shots : the Data Normalized Analysis and the Laser Simulation Analysis (more details can be found in [5]).

- The Data Normalized Analysis (DN) is based on the comparison of measured laser profiles with a reference clear night profile in which the light attenuation is dominated by molecular scattering.
- The Laser Simulation Analysis (LS) is based on the comparison of measured laser light profiles to simulations generated in various atmospheres in which the aerosol attenuation is described by a parametric model.

To minimize fluctuations, both analyses make use of average light profiles measured at the aperture of the FD buildings normalized to a fixed reference energy. Using measurements recorded on extremely clear nights where molecular Rayleigh scattering dominates, laser observations can be normalized without the need for absolute photometric calibrations of the FD or laser. These “reference clear nights” are identified using a procedure looking for profiles with maximum photon transmission and maximum compatibility with the shape of a profile simulated in conditions with negligible aerosol attenuation. One reference clear night per year is selected.

The Data Normalized Analysis is an iterative procedure that compares hourly average profiles to reference clear night profiles. The first step is to build the hourly profile, starting from the 4 sets of 50 shots. During this procedure, clouds positioned above the vertical laser beam are identified and the height of the lower layer of the cloud is set. Assuming that the atmosphere is horizontally uniform, the Vertical Aerosol Optical Depth $\tau_{\mathrm{aer}}^{\mathrm{DN}}(\mathrm{h})$ is measured as

$$\tau_{\mathrm{aer}}^{\mathrm{DN}}(\mathrm{h}) = \frac{\ln \mathrm{N_{mol}}(\mathrm{h}) - \ln \mathrm{N_{obs}}(\mathrm{h})}{1+\mathrm{cosec}(\theta)}$$

where $\mathrm{N_{mol}}(\mathrm{h})$ is the number of photons from the reference clear profile as a function of height, $\mathrm{N_{obs}}(\mathrm{h})$ is the number of photons from the observed hourly profile as a function of height and $\theta$ is the elevation angle of each laser track segment. This calculation does not take into account the scattering of the laser beam itself due to aerosols. To overcome this, $\tau_{\mathrm{aer}}^{\mathrm{DN}}(\mathrm{h})$ is differentiated to calculate the aerosol extinction coefficient $\alpha(\mathrm{h})$ over short intervals in which the aerosol scattering conditions change slowly. The final $\tau_{\mathrm{aer}}^{\mathrm{DN}}(\mathrm{h})$ is estimated by re-integrating $\alpha(\mathrm{h})$ (figure 3). The aerosol attenuation profile is calculated from the FD site altitude up to the cloud lower layer height or the highest point in the FD field of view.

The Laser Simulation Analysis is based on a comparison of light profiles from 50 shots every quarter-hour to simulations generated varying the aerosol attenuation conditions. The aerosol attenuation is described by two parameters, the aerosol horizontal attenuation length $\mathrm{L_{aer}}$ and the aerosol scale height $\mathrm{H_{aer}}$. The former describes the light attenuation due to aerosols at ground level, the latter accounts for its dependence on the height. With this parameterization, the expression of the vertical aerosol optical depth $\tau_{\mathrm{aer}}^{\mathrm{LS}}(\mathrm{h})$ between points at altitude $\mathrm{h_1}$ and $\mathrm{h_2}$ is :

$$\tau_{\mathrm{aer}}^{\mathrm{LS}}(\mathrm{h_2}-\mathrm{h_1}) = -\frac{\mathrm{H_{aer}}}{\mathrm{L_{aer}}}\left[\exp\left(-\frac{\mathrm{h_2}}{\mathrm{H_{aer}}}\right) - \exp\left(-\frac{\mathrm{h_1}}{\mathrm{H_{aer}}}\right)\right]$$

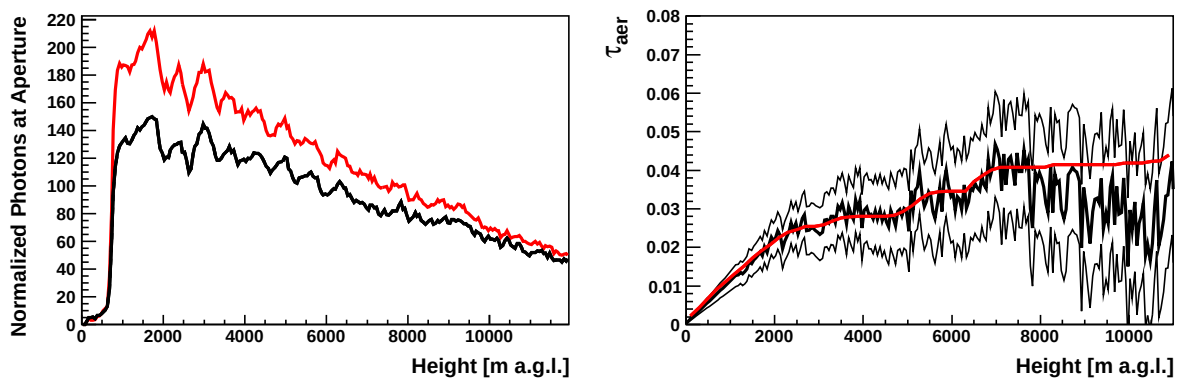

**Figure 3**: Light profiles (in red the reference clear profile, in black the measured one) and vertical aerosol optical depth measured using the Data Normalized Analysis with the FD at Los Morados during an average night.

A grid of 1540 profiles is simulated for each FD site, each month and at a reference energy, to normalize the measured profiles. Each measured profile is compared to the grid and the simulated profile closest to the measured event is identified and its associated parameters are used to calculate $\tau_{aer}^{LS}(h)$ (figure 4). During the procedure, clouds are identified and the aerosol attenuation profile is measured up to the cloud lower layer height.

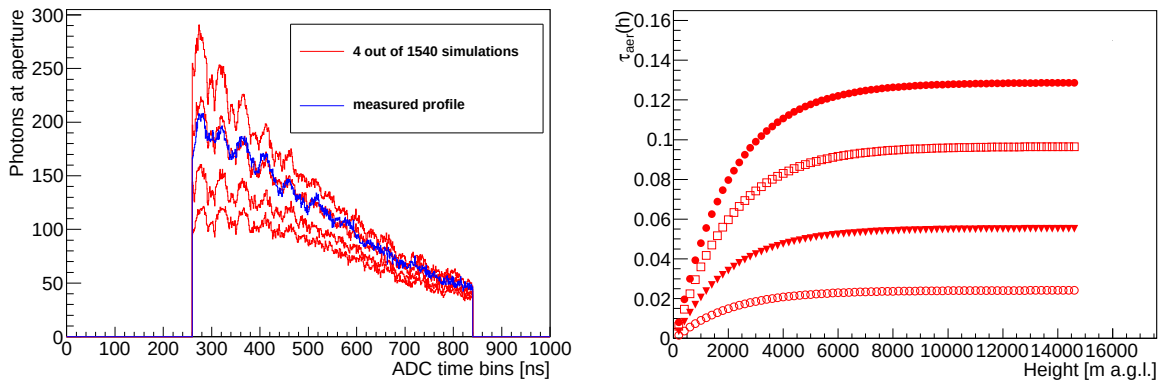

**Figure 4**: Left : four out of the 1540 simulated profiles of a monthly grid (red), superimposed on a measured profile (blue). Right : the four $\tau_{aer}^{LS}(h)$ profiles corresponding to the simulated CLF profiles.

## 4 Statistical and systematic error estimates

Various uncertainties were indentified in the methods for the determination of $\tau_{aer}(h)$ profiles. The uncertainties have been recently re-estimated and are now separated into systematic and statistical contributions. These assignments were based on whether the effect of the uncertainty would be correlated over the EAS data sample, or would be largely uncorrelated from one EAS to the next (see table 1). For more discussion see [6]. Since each method is based on the use of ratios of FD events, it is not sensitive to the absolute photometric calibration of either the laser or the FD. Consequently, the calibration correlated uncertainties in table 1 are those that describe how accurately drifts in the FD and laser energy calibrations were tracked over the period between reference nights. These nights are typically a year apart. For the CLF, the 1-2.5% value corresponds to different epochs over the 10 year life of the system and depends on how well the effect of dust accumulation on the optics downstream of the monitor probe was tracked. An estimate of the stability in the net depolarization of the laser beam is included in these numbers. The corresponding term for the XLF (1%) reflects the fact that this system has an automated calibration system that tracks beam energy and polarization. The uncorrelated error of the relative FD calibration was estimated to be 4%. It includes an estimate of the variability in FD calibration during the night. A 3% correlated uncertainty was estimated as due to the choice of the reference clear night. Finally the uncorrelated error due to the atmospheric fluctuations within the hour is estimated on a event-by-event basis and is about 3%. These errors are estimated for each of the two methods described. In the Laser Simulation Analysis a 2% uncorrelated uncertainty is added to take into account how well the parametric model used describes the real aerosol attenuation conditions. A study was performed on hybrid events to estimate the effect on reconstructed EAS energy and Xmax when moving $\tau_{aer}(h)$ up or down by its systematic uncertainty. It was found that the energy varies from +2.4% to -2.5%, and Xmax from 0.8 to -1.2 $g \cdot cm^{-2}$.

| | Correlated | Uncorrelated |
|---|---|---|
| Relative FD Calibration | 2% | 4% |
| Relative Laser Energy (CLF) | 1–2.5% | 2% |
| Relative Laser Energy (XLF) | 1% | 2% |
| Reference Clear Night | 3% | - |
| Atmospheric Fluctuations | - | $\sim 3\%$ |

**Table 1**: List of uncertainties in the determination of the $\tau_{aer}(h)$ profiles (see text).

## 5 2004–2012 Aerosol Attenuation Profiles

The hourly aerosol attenuation profiles over 9 years (from January 2004 to December 2012) have been measured using the two analyses described.

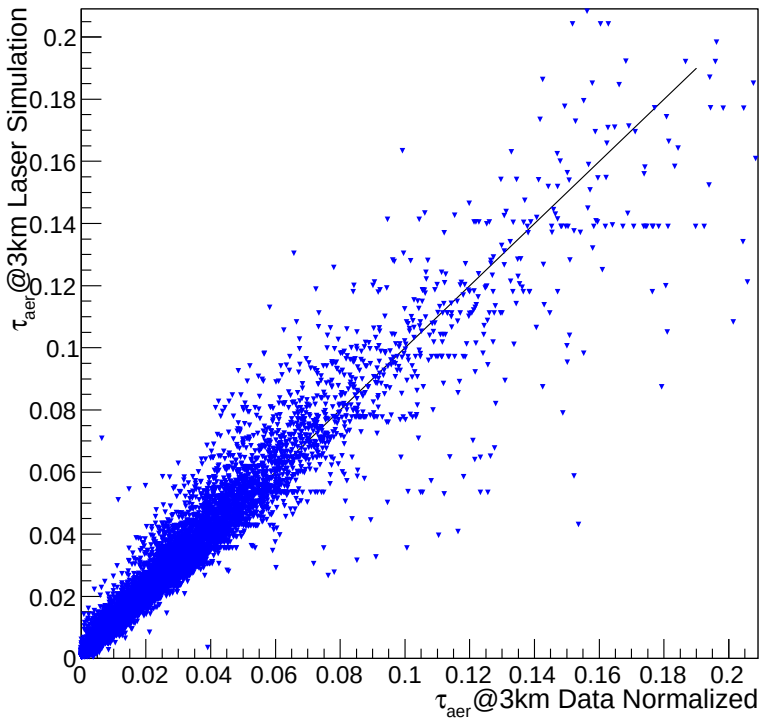

**Figure 5**: Comparison of Vertical Aerosol Optical Depth at 3 km above ground measured with the two analyses for the Coihueco site. 9 years of data are shown.

Due to the distance, XLF events were used to produce aerosol profiles for Loma Amarilla and CLF events were used for Los Leones, Los Morados and Coihueco. Results from the two analyses were compared and are fully compatible. In figure 5, the correlation of $\tau_{aer}^{DN}$ versus $\tau_{aer}^{LS}$ measured at 3 km above the ground level is shown for the Coihueco site for the period January 2004 to December 2012. Hourly profiles measured with the two analyses together with correlated and uncorrelated error bands in average aerosol attenuation conditions are shown in figure 6. The aerosol profiles measured are stored in the Pierre Auger Observatory Aerosol Database for the reconstruction of EAS data. The database is filled with results obtained with the Data Normalized analysis, while results from Laser Simulation analysis are used to fill gaps. A total of 10430 hours are stored in the aerosol database for the Los Leones site, 9302 for Los Morados, 2270 for Loma Amarilla and 10430 for Coihueco. In figure 7 $\tau_{aer}$ measured at 3 km above ground as a function of time is shown for each FD site.

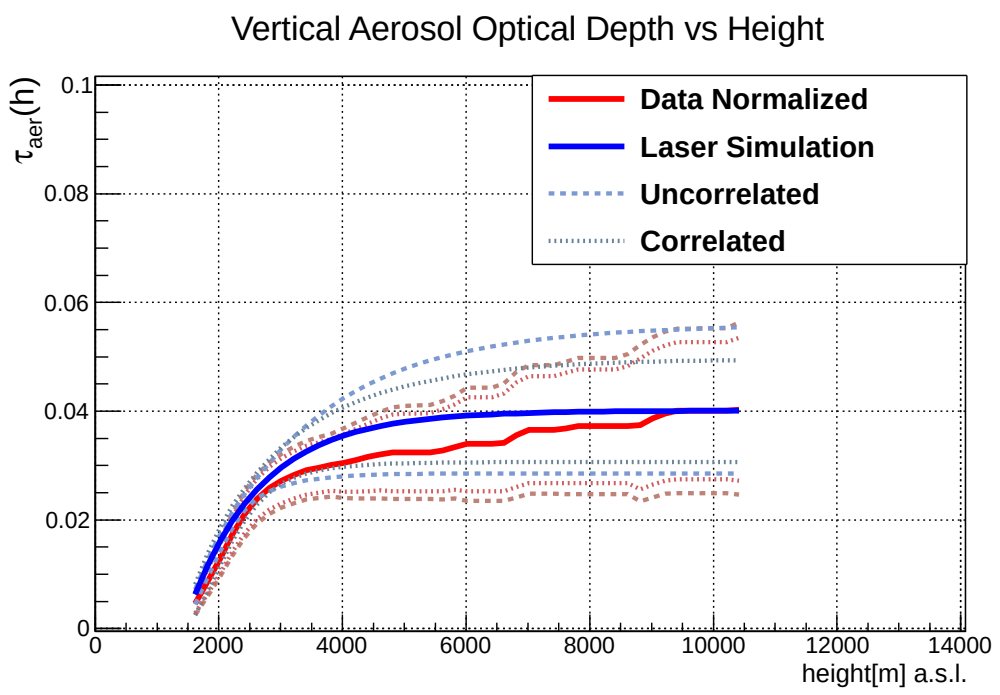

**Figure 6**: Hourly aerosol profiles measured with the Data Normalized (red) and Laser Simulation (blue) analyses in average conditions. Correlated and uncorrelated uncertainties are shown.

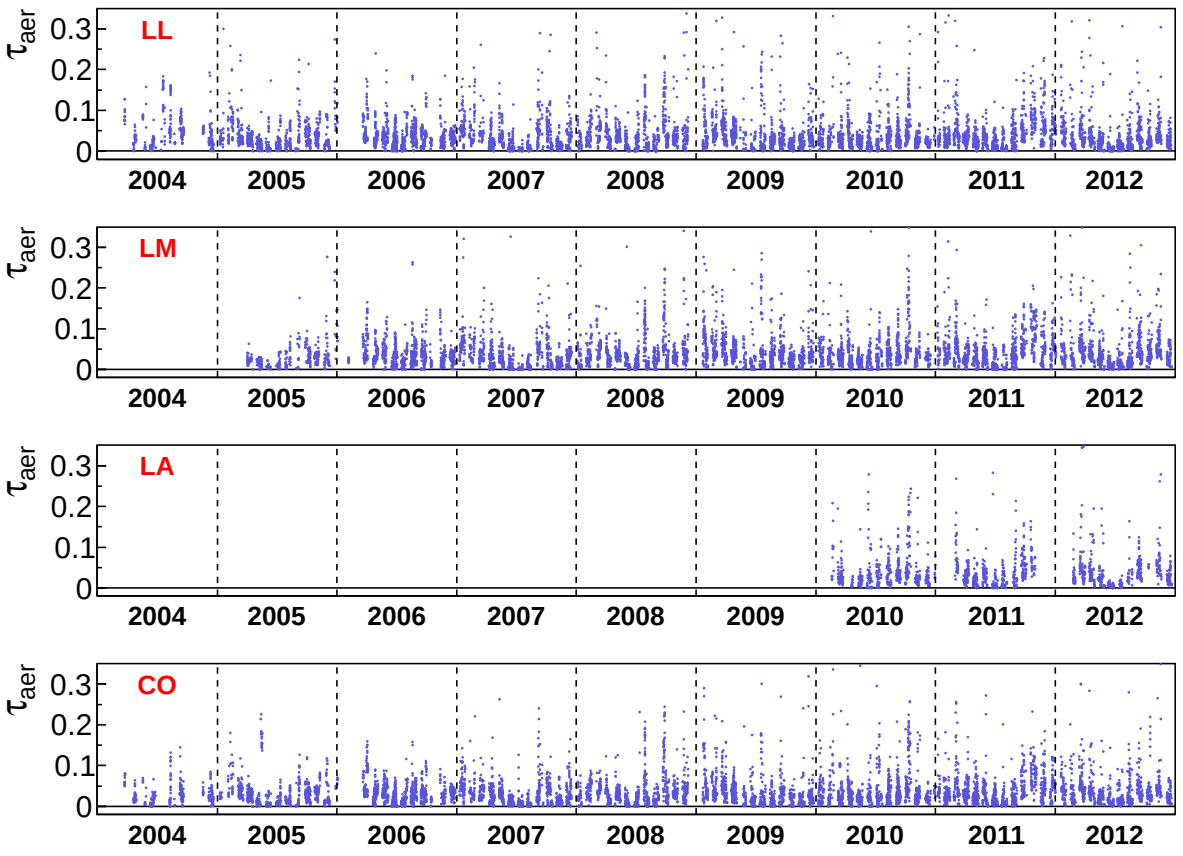

**Figure 7**: 9 years of $\tau_{aer}$ measured at 3km above ground.

## 6 Upgrades

A major update is in progress at the CLF site. Upgrades include the addition of a Raman LIDAR to the system to perform $\tau_{aer}(h)$ measurements independently of the methods described here, a solid state laser with better shot-to-shot stability, and an automated calibration system similar to the one presently in use at the XLF to improve the laser calibration reliability over long periods. Infrastructure upgrades include a 2000 liter thermal reservoir to reduce temperature fluctuations of the equipment, and a new shipping container shelter with better insulation and dust control. Completion is expected by July 2013.

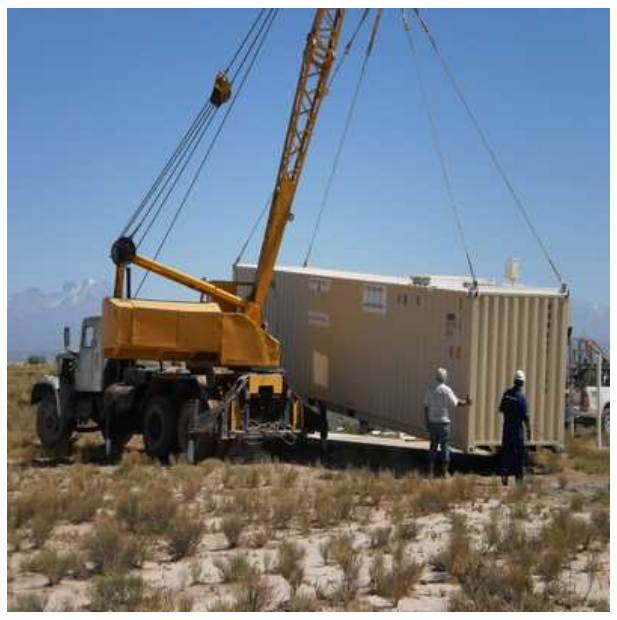

**Figure 8**: The new shelter for the upgraded CLF is placed in position at the site.

The atmospheric monitoring system of the Pierre Auger Observatory also includes 4 steerable elastic LIDAR stations[7], one for each FD site. LIDARs provide an independent estimation of the $\tau_{aer}(h)$, but only outside the FOV of the FD due to the high interference with data acquisition, therefore they are used to monitor the cloud cover. A new prototype with improved mechanics and alignment capabilities will be tested at the Loma Amarilla site. The new system has a one-meter-diameter f/1 composite mirror, and the capability of shooting the laser beam coaxially or with a parallax of 1.5 meters. This allows us to extend the sampled atmosphere down to 200 m, and the range up to 40 km. The new LIDAR is expected to provide very precise measurements of the aerosol optical depth. In figure 9, the schema of the full prototype and a picture of the box, the carousel and the mirror are visible. Installation will take place during 2013.

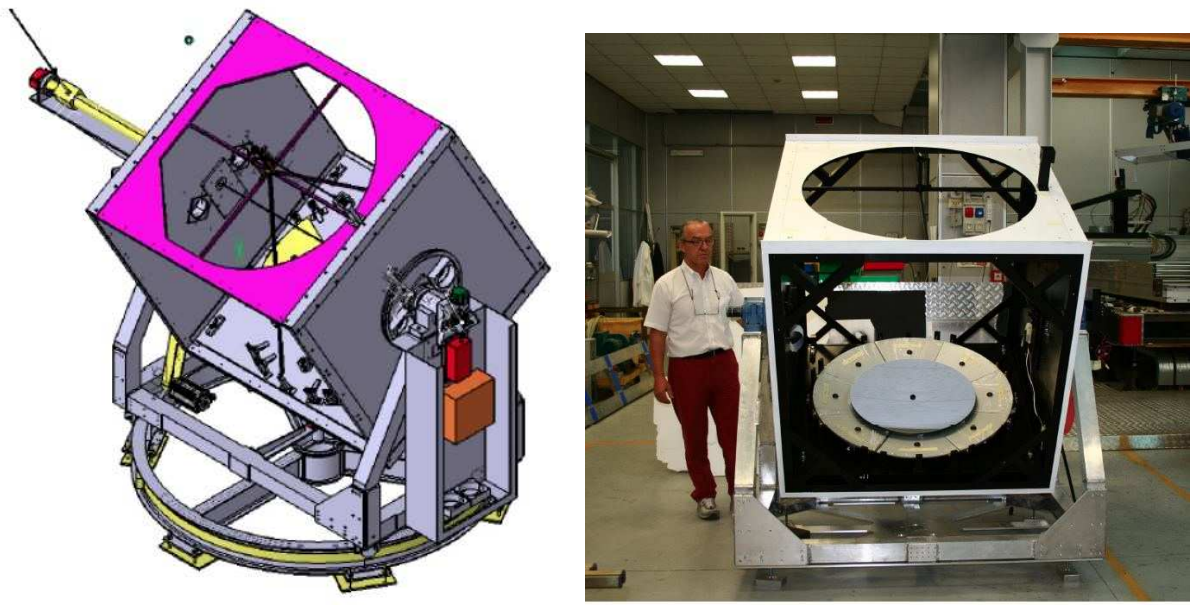

**Figure 9**: The new prototype of the LIDAR system.

**Acknowledgment:**The ICRC 2013 is funded by FAPERJ, CNPq, FAPESP, CAPES and IUPAP.

# Aerosol characterization at the Pierre Auger Observatory

M.I. MICHELETTI[1,2], FOR THE PIERRE AUGER COLLABORATION[3] AND J. DAVIDSON[4,5], M. DEBRAY[5,6], M. FREIRE[7], M. MAŠEK[8], R. PIACENTINI[1,7], M. ROSENBUSCH[4], H. SOMACAL[5,6]

[1] *Instituto de Fisica Rosario (IFIR)CONICET/UNR, Bv. 27 de Febrero 210 bis, Rosario, Argentina*
[2] *Facultad de Ciencias Bioquimicas y Farmaceuticas, Universidad Nacional de Rosario (UNR), Suipacha 531, Rosario, Argentina*
[3] *Full author list: http://www.auger.org/archive/authors_2013_05.html*
[4] *CONICET (Consejo Nacional de Investigaciones Cientificas y Tecnicas), Rivadavia 1917, C.A. de Buenos Aires, Argentina*
[5] *Gerencia de Investigacion y Aplicaciones, Comision Nacional de Energia Atomica, Av. Gral. Paz 1499, San Martin, Bs.As., Argentina*
[6] *Universidad Nacional de Gral. San Martin, M. de Irigoyen 3100, San Martin, Buenos Aires, Argentina*
[7] *Facultad de Ciencias Exactas, Ingenieria y Agrimensura, UNR, Av. Pellegrini 250, Rosario, Argentina*
[8] *Institute of Physics of the Academy of Sciences of the Czech Republic, Praha, Czech Republic*

*auger_spokespersons@fnal.gov*

**Abstract:** The Pierre Auger Observatory is a hybrid facility composed of surface and fluorescence detectors (denoted as SD and FD, respectively), where the FD system sets the energy scale in cosmic ray shower reconstruction. Atmospheric components attenuate the fluorescence light emitted by the de-excitation of atmospheric nitrogen, previously excited by the charged particles of the shower. Amongst these components, atmospheric aerosols are the ones with the largest fluctuations, being responsible for one of the major uncertainties in shower reconstruction with FD data. We present a detailed characterization of aerosols. They are collected at the Observatory and analyzed, in morphology and elemental composition, with experimental techniques used for the first time in a cosmic ray observatory: gravimetry, PIXE and SEM/EDX. An analysis of wind trajectories using the program HYSPLIT is used to understand the sources and the evolution of aerosols. The aerosols are further characterized by the FRAM, an optical telescope employed to perform CCD photometry of selected Landolt fields, in which we observe sets of precisely measured standard stars at various wavelengths. Using this photometric information we then compute the Ångström coefficients that also characterize the size of aerosols.

## 1 Introduction

The Pierre Auger Observatory, conceived to measure ultra-high energy cosmic rays, has a hybrid design, consisting of surface and fluorescence detectors. The FD system is composed of 27 telescopes distributed in four stations [1]. The FD detects fluorescence light (300-450 nm) produced by the interaction of the cosmic ray showers with atmospheric nitrogen. This light is attenuated on its path from the shower to the telescopes by atmospheric constituents, with aerosols (suspended particles) playing a role of major importance. Since the aerosols are highly variable in space and time, a continuous monitoring of them is necessary. A large network of aerosol monitors is installed at the Observatory. Some of them, like the lidars (4 elastic ones and a Raman one) or CLF and XLF (Central and eXtreme Laser Facilities), obtain the aerosol optical depth as a function of height [2]. On the other hand, the FRAM (F/Photometric Robotic Atmospheric Monitor) is a star monitor that measures the total aerosol optical depth from the top of the atmosphere to the ground. Elastic lidars and the FRAM can also measure the optical depth from the shower to the FD, as part of a rapid monitoring program [3]. To complement the setup, the HAM (Horizontal Attenuation Monitor) has been used to measure the horizontal attenuation length between the FD sites almost at ground level and the APF (Aerosol Phase Function Monitors) give the aerosol phase functions (describing the angular distribution of aerosol diffused light) [4]. For these devices, the aerosols work just as a medium interacting with the fluorescence light. But their measurements do not specify what the properties of the aerosols are (shape, size, composition), apart for data from the APF and FRAM that give information on their mean size. Their characteristics remain hidden and some assumptions are made to model their interaction with radiation. Therefore, a detailed aerosol characterization project, by direct measurement and analysis, is being performed at the Observatory to improve and complement the information supplied by the other monitors. Knowledge of aerosol characteristics permits one to infer their origin (sources, evolution). This can be correlated with a study of air mass trajectories which evaluates the behavior of aerosols in space and time.

## 2 Instruments and techniques for aerosol measurements

### 2.1 Direct aerosol sampling using an Andersen-Graseby 240 (A-G 240)

By using an A-G 240 dichotomous sampler, direct sampling of aerosols takes place at the Auger Observatory. This instrument is installed at Coihueco FD station (35°06'52.9" S, 69°36'02.7" W, 1712 m a.s.l.), at 6.3 m above ground level (AGL). It has a pump that drives air into it, sweeping along atmospheric particulate matter PM10 (with aerodynamic diameters d≤10$\mu$m). The air is divided in two fluxes, one carrying fine particles (d≤2.5$\mu$m) and the other carrying the coarse ones (2.5<d≤10$\mu$m), which are deposited in two filters of polycarbonate (Millipore®HTTP, diameter 37 mm, pore 0.4 $\mu$m). The sampling period is 24 h. The operational, or actual, flow rate $Q_a$ is 16.7 l/min. $Q_a$ was corrected to U.S. Environmental Protection Agency reference conditions (298 K and 760 mm Hg), to obtain the standard flow rate $Q_{std}$ used in calculations of mass and elemental concentrations. The aerosols captured in the filters are later analyzed by different experimental techniques: Gravimetry, PIXE, SEM/EDX (see Sec. 3.1, 3.2 and 3.3).

### 2.2 Concentration measurements using a Grimm 1.109

A portable laser aerosol spectrometer and dust monitor Grimm 1.109 was installed in the FD building of Coihueco in November 2010, at 1.715 m AGL, to perform local superficial aerosol concentration measurements. It operates with a dual technique: a) Continuous measurements of particle or mass concentration (particle/liter, $\mu$g/m$^3$), in fixed time intervals, for size channels from 0.22 to 32$\mu$m. The principle of operation is based on an internal laser (655 nm) and a model for the light dispersion produced by aerosols contained in the flux of air driven into the apparatus by a pump (flow rate 1.2 l/min) b) Collection of particles in a filter for later analysis.

### 2.3 Ångström coefficient measurements using FRAM

FRAM is a small optical telescope located about 20 m from the FD building at Los Leones (35°29’45.2” S, 69°26’58.9” W, 1430 m a.s.l.). The Schmidt-Cassegrain telescope with diameter of 300 mm is equipped with a type G2 CCD camera from Moravian Instruments and uses the Johnson & Bessel set of UBVRI filters. It is equipped with a wide-field camera (Moravian instruments G4) using a Nikkor 300 mm/f 2.8 camera lens with diameter of 12.5 cm. The telescope uses the equatorial Bisque Paramount ME mount and operates in robotic mode driven by the custom-made RTS2 software package [5]. It regularly observes selected fields of standard stars, so called Landolt fields, in various filters to derive extinction coefficients (or optical depths) at various wavelengths. The goal is to obtain the Ångström coefficient $\gamma$, used for parametrization of wavelength $\lambda$ dependence of the aerosol optical depth $\tau_A$: $\tau_A(\lambda) = \tau_{A0} \times (\lambda_0/\lambda)^\gamma$, where $\lambda_0$ is the reference wavelength and $\tau_{A0}$ is the aerosol optical depth measured for this wavelength. The Ångström coefficient is used in the cosmic ray shower reconstructions.

## 3 Instruments and techniques for aerosol analysis. Results.

### 3.1 Concentration analysis using Gravimetry

Filters are weighed with a Microbalance M3 (precision $\pm 1\mu$g) before and after collecting aerosols with the A-G 240 at Coihueco, to obtain the mass of particulate matter deposited during the samplings. Before weighing, filters are conditioned (humidity 50% and temperature 25°C during at least 24 h) and irradiated with an alpha source ($^{238}U$) to eliminate static charge on them during weighing. The PM2.5 and PM2.5-10 concentrations (in $\mu$g/m$^3$), are calculated as the ratio between the collected mass and the volume of air that passed through the sampler during the period of measurement. The PM10 concentration in the ambient air is computed as the sum of PM2.5 and PM2.5-10 concentrations. The total volume of air sampled is corrected to standard conditions ($V_{std}$) and it is determined from the standard total flow rate $Q_{std}$ and the sampling time (24 h). An analysis performed for a total of 36 days of measurements in the period June 2008-February 2009, gave a mean PM10 of 10.3$\mu$g/m$^3$ (standard deviation 6.5$\mu$g/m$^3$). PM2.5 and PM2.5-10 represent 31.1 % and 68.9 % of PM10. There is a trend towards increasing concentrations with warmer seasons. During the coldest days of the winter, very low concentration values are observed (of less than 2.5$\mu$g/m$^3$) because snow keeps aerosols captured at the soil surface. Concurrently, in winter, air masses arrive at the Observatory mostly from the Pacific Ocean, presenting a lower aerosol content (see Sec. 4).

### 3.2 Elemental analysis using PIXE and SEM/EDX

The PIXE technique [6] is performed on PM2.5 and PM2.5-10 samples of aerosols collected at the Auger Observatory with the A-G 240, to analyze their elemental composition (from S up), at the TANDAR Laboratory accelerator facility of the Comision Nacional de Energia Atomica, Argentina. The targets were irradiated with heavy ions $^{16}$O (7+ charge state) and the induced X-rays, characteristic of the elemental composition of the samples, were measured using an EG&G Ortec Si(Li) detector (sensitive area of 80 mm$^2$, 12.5$\mu$m Be window), with a resolution of 220 eV at 5.9 keV (Ka Mn line). More details about the experiment can be found elsewhere [7]. The X-ray spectra were analyzed with the WinQxas 1.40 computer code developed by IAEA. A PIXE analysis performed on 19 samples of each fraction corresponding to June-August 2008 showed that S, Cl, K, Ca, Ti, Mn and Fe, represent 25% and 13% of the PM2.5 and PM2.5-10 total mass, respectively. The rest of the mass is due to elements with low atomic number Z (not detected with our X-ray setup). S dominates in the fine fraction and Ca in the coarse fraction [7]. Elemental composition was also studied using a Scanning Electron Microscope (Philips SEM 515) with an Energy Dispersive X-ray system (EDAX Falcon PV 8200), provided with a Si(Li)-Be window detector. With this SEM/EDX arrangement, the detection of elements of Z higher than 11 (Na) is possible, complementing the PIXE results. Semiquantitative standardless analysis with ZAF factors for matrix correction was used for composition calculations. Si, Al, Ca, Mg and Fe, the typical mineral soil elements, are the major components, indicating that the aerosols present at the Auger Observatory consist mostly of suspended mineral dust from the soil of the Andean region. SEM observations indicate that the mass not detected by PIXE corresponds to

Si and Al (aluminosilicates)[7].

### 3.3 Morphological (shape and size) analysis using SEM images

SEM micrographs of the sampled aerosols, like the one shown in Fig. 1, are analyzed in morphology with software developed by using the ImageJ application (http://rsb.info.nih.gov/ij/). An analysis performed over 23 SEM images of June-August 2008 gave the result shown in Fig. 2 for the relative frequency of appearance of the different aerosols diameters, in the represented intervals (of 0.5$\mu$m, except the first one that ranges from 0.4 to 0.5$\mu$m, 0.4$\mu$m being the minimum aerosol size taken into account in the analysis, which corresponds to the pore size of the filters). Most of the analyzed aerosols (64.9%) were in the range 0.5-1$\mu$m. The shapes of the aerosols have been investigated through an approximation of circularity (defined as $4\pi\times$ Area / Perimeter$^2$) applied to the plane SEM images of the particles. It ranges between 0 and 1 (the latter for perfect circles). The analysis showed that 75% of the analyzed PM10 particles have circularity bigger than 0.5.

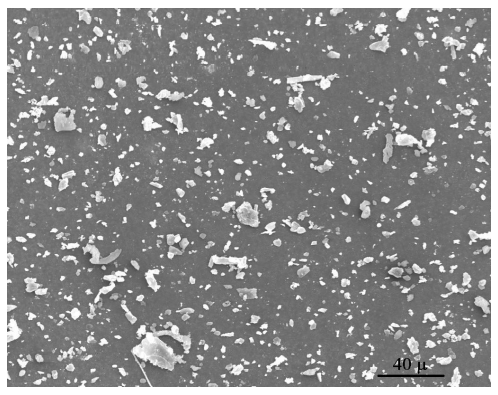

**Figure 1**: PM2.5-10 sample of 14 August 2008. Mass concentration: 13.6 $\mu$g/m$^3$.

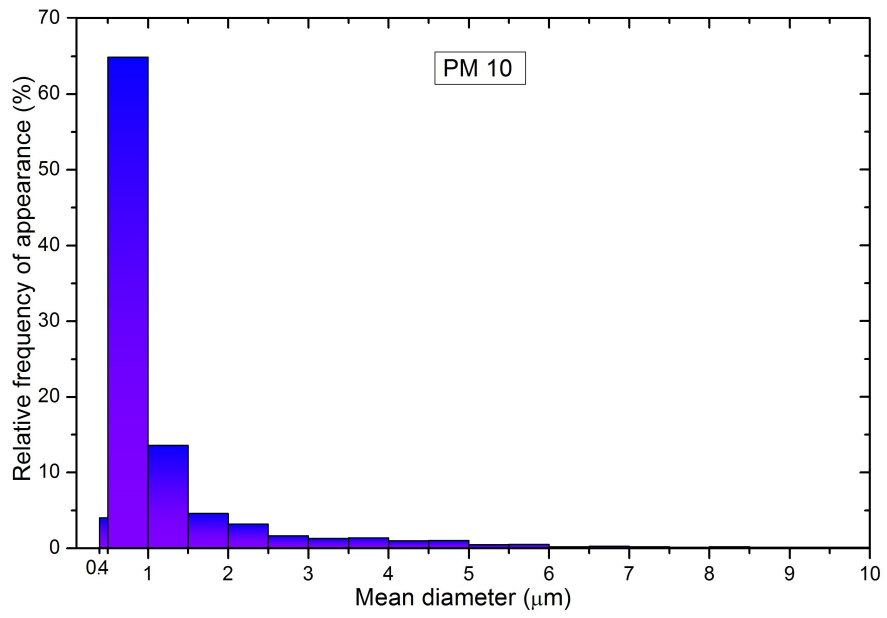

**Figure 2**: Relative frequency for different aerosol diameters. Period: June-August 2008.

### 3.4 Concentration and size distribution analysis using Grimm 1.109 data

Concentration measurements were obtained with Grimm 1.109 in Coihueco every 5 minutes, for different size channels, for June - August 2011. Mean mass concentrations for this period, for each size channel, normalized by the width of the size range, are shown in Fig. 3. The area under the histogram is the total mean concentration for June - August 2011. Table 1 gives the concentration values for some size ranges. PM2.5 and PM2.5-10 represent 21.2 % and 78.8 % of PM10. The particulate matter with aerodynamic diameter d>10$\mu$m contributes only 5.6% of the total concentration. The aerosol size range from 0.5 to 1$\mu$m represents only 2.2% of the total concentration (Table 1).

| | | |
|---|---|---|
| PMtotal | 9.0 | 100% |
| PM0.5 (d≤0.5$\mu$m) | 0.7 | 7.8% |
| PM>0.5 (d>0.5$\mu$m) | 8.3 | 92.2% |
| PM2.5 (d≤2.5$\mu$m) | 1.8 | 20.0% |
| PM10 (d≤10$\mu$m) | 8.5 | 94.4% |
| PM2.5-10 (2.5<d≤10$\mu$m) | 6.7 | 74.4% |
| PM>10 (d>10$\mu$m) | 0.5 | 5.6% |
| PM0.5-1.0 (0.5<d≤1$\mu$m) | 0.2 | 2.2% |

**Table 1**: Concentrations values (in $\mu$g/m$^3$) for different aerosol size ranges (in$\mu$m) and percent contribution of these size ranges in the total aerosol concentration.

By contrast, the size analysis performed on SEM images of filters collected with A-G 240 during June-August 2008 showed that the great majority of the analyzed aerosols, 64.9 %, have sizes in that range (Fig. 2). Comparing both results for the same season of the year -even if they are for different years- it is evident that while the aerosols of the 0.5-1$\mu$m range are the most abundant, their contribution to total mass concentration has little significance, due to their light mass. Instead, concentrations measured by Grimm show a peak in the 4-5$\mu$m range (of 1.7$\mu$g/m$^3$, representing 18.9% of the total mass concentration) while atmospheric particles are less abundant at this size range according to SEM image analysis.

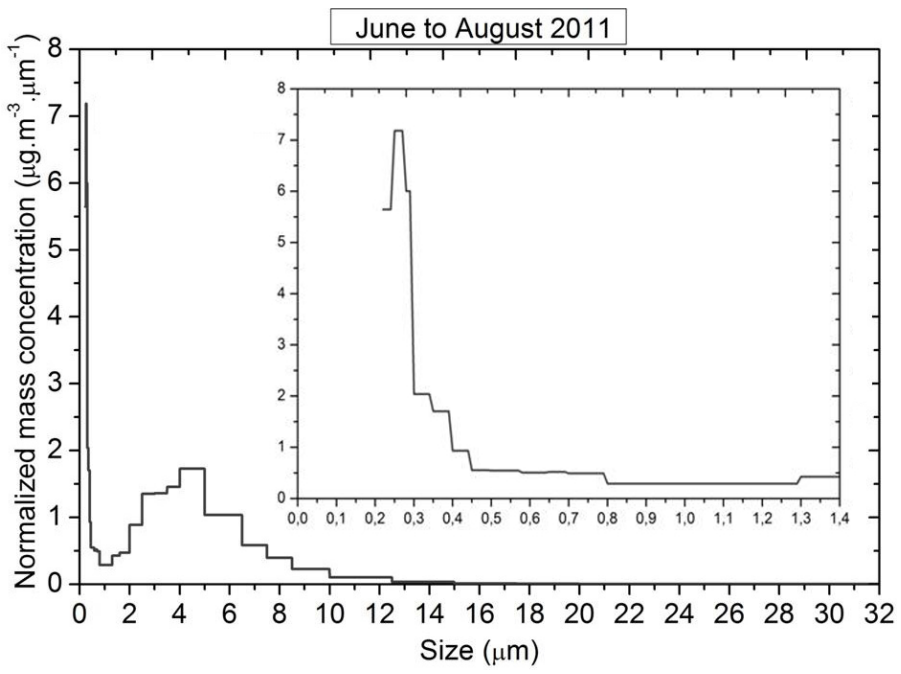

**Figure 3**: Left: Mean aerosol concentration for different size intervals, normalized by the width of the interval, obtained using the Grimm 1.109, June-August 2011. Right: idem with the smaller diameter range expanded.

### 3.5 Mean size analysis using FRAM data

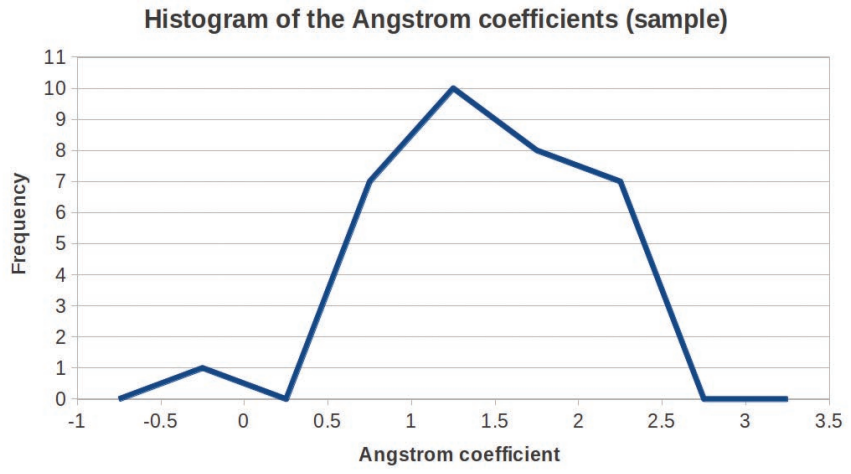

**Figure 4**: Histogram of FRAM measurements of the Ångström coefficient

FRAM measures the difference of the observed magnitude $m_{OBS}$ and catalogue magnitude $m_{CAT}$ of given stars within a selected standard field. Using the known airmass AM for the object, the total extinction coefficient $\kappa$ can be easily derived. The result is transformed into total optical depth $\tau$: $\kappa = (m_{OBS} - m_{CAT})/AM$ ; $\tau = (\sqrt[5]{100}/e)\kappa = 0.924\ \kappa$. The aerosol optical depth $\tau_A$ is obtained after subtraction of the computed molecular Rayleigh part. Since the absolute calibration of the FRAM telescope might be time dependent, the system is calibrated each night using observations of several different AM (within a few hours). Then, the observed dependence of $\kappa$ is fitted on the AM independently of the telescope calibration constants, obtaining a precise result on $\kappa$. The Ångström coefficient $\gamma$ is then obtained from the resulting aerosol optical depths for individual standard stars within one field, fitting the values in different wavelength filters. The Ångström coefficient varies from 0 to 4. Typical values: 0 (coarse particles, e.g. desert dust), 2 (fine particles, e.g. ash or automobile exhaust), 4 (molecules). This method of $\gamma$ determination using CCD photometry of Landolt fields has been used at FRAM only since December 2012. From the available measurements (December 2012 - March 2013) we have constructed the histogram for $\gamma$ (Fig. 4). The mean value of $\gamma$ is $1.1 \pm 0.7$, indicating that both coarse and fine particles are present at the Observatory, in agreement with conclusions in other parts of this work.

## 4 Sources and evolution study using trajectories of air masses, HYSPLIT

HYSPLIT is an air-modeling program to calculate air mass displacements from one region to another [8]. It was used to analyze forward and backward trajectories of air masses to infer the sources and evolution of aerosols present at the Auger Observatory. 48 hour backward trajectories were evaluated every hour, throughout the year, for 2008-2010.

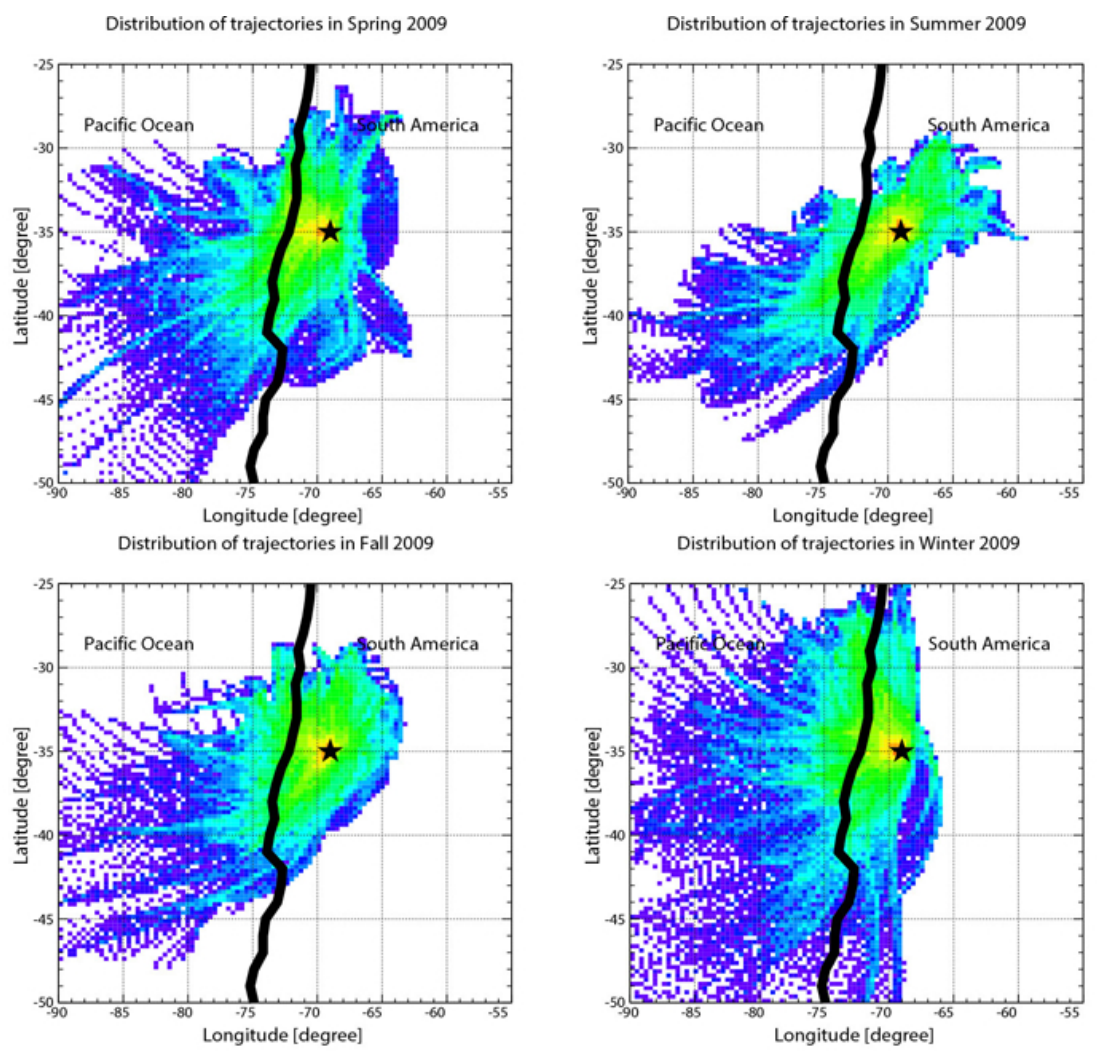

**Figure 5**: Seasonal analysis of 48 hours HYSPLIT backward trajectories evaluated every hour, arriving at the Auger Observatory at 500 m AGL.

A backward analysis performed at a start altitude of 500 m AGL showed that air masses originate mainly over the Pacific Ocean during the clear nights and travel principally through continental areas during the previous 48 h for hazy nights. Clear and hazy nights were identified using aerosol optical depth values at 3.5 km AGL, at Los Morados FD site, obtained from the CLF data during the mentioned period. Clear and hazy nights correspond to aerosol optical depths up to 0.01, for the former, and from 0.1 up, for the latter [9]. Elemental results previously described (Sec. 3.2), which indicated that most of the aerosols are soil suspended particles of the Andean region, explain the lower aerosol optical depth when the air masses have traveled mainly over the ocean during the previous 48 h. From the different monitors of the Observatory it is known that the presence of aerosols is lower in winter than in the rest of the year. A seasonal analysis performed with HYSPLIT during 2009 shows that in winter the backward trajectories of air masses spend more time over the ocean than in the other seasons (Fig.5).

## 5 Conclusions and Future Plans

The characterization of aerosols collected at the Auger Observatory is giving interesting information about their morphology and composition, thanks to the application of advanced analysis techniques used for the first time in a cosmic ray observatory. The results obtained from direct sampling and analysis complement information supplied by other aerosols monitors at the Observatory, which are evaluating the effect of these particles in fluorescence light attenuation. The results agree qualitatively with available FRAM data that estimate the mean size of the local aerosols, and can be combined with studies of air masses trajectories to infer the sources and evolution of these particles. This detailed aerosol characterization surpasses its application in cosmic rays showers reconstructions, being of major interest in other fields of study. A collaborative project, being designed with atmospheric scientists, is expected to give valuable information about the atmosphere at the southernmost latitudes of the globe.

**Acknowledgment:**The operation of FRAM is supported by grants of EU GLORIA (No. 283783 in FP7-Capacities program) and of the Czech Ministry of Education (MSMT-CR LG13007).

# Cloud Monitoring at the Pierre Auger Observatory

Johana Chirinos[1], for the Pierre Auger Collaboration[2].

[1] *Michigan Tech. University, Houghton-MI, USA*
[2] *Full author list: http://www.auger.org/archive/authors 2013 05.html*

*auger_spokespersons@fnal.gov*

**Abstract:** Several methods are used to detect night-time cloud cover over the 3000 km$^2$ Pierre Auger Observatory, including lidars and laser sources. Here, we describe two methods. Infrared cloud cameras, installed at each of the four fluorescence detector sites, detect the presence of cloud within the fields of view of each fluorescence telescope every 5 minutes. Operating since 2002, an upgrade to improved hardware is underway. Secondly, a method has been implemented to use GOES-12 and GOES-13 satellites to identify night-time clouds over the Observatory. It has been validated using the Observatory's Central Laser Facility, which determines cloud cover above this facility. We develop cloud probability maps for the 3000 km$^2$ of the Observatory twice per hour and with spatial resolution of 2.4 km by 5.5 km and a database with the cloud probabilities for further analysis.

## 1 Introduction

The Pierre Auger Observatory employs several systems capable of detecting night-time cloud over the 3000 km$^2$ viewed by the observatory's fluorescence detectors, including two laser facilities within the array and lidar systems at the fluorescence detector (FD) sites [1, 2, 3]. Clearly, cloud is capable of attenuating fluorescence light from parts of air showers, resulting in dips in the measured longitudinal development profiles. In contrast, if a shower passes through a cloud layer, its intense Cherenkov beam may be scattered by the cloud resulting in an *increase* in light received at a fluorescence detector. Thus cloud detection is vital for reliable measurements of longitudinal shower profiles, the depth of shower maximum and primary particle energies, but also in searches for any effects of exotic particle physics on air shower development.

In this paper we discuss two key cloud detection methods which are used by the Observatory. The strength of our cloud detection lies in our ability to combine measurements from instruments with different capabilities.

## 2 Infra-red Cloud Cameras

The Pierre Auger Observatory utilizes infra-red cameras, located on the roof of each of the four main FD buildings, to observe the night-time sky conditions over the array. Being co-located with the fluorescence detectors, it is possible to directly associate cloud camera pixel directions with FD pixel directions, thus flagging detected showers that may be affected by cloud. Conservatively, showers may be disregarded if cloud is detected within any FD pixel viewing an event, or alternatively cloud camera data (giving cloud *direction*) may be combined with lidar or Central Laser Facility data on cloud *height* to locate the cloud in three dimensional space.

Each camera was installed shortly after the corresponding FD building was completed, between 2004 and 2007. Beginning in early 2013, each of the four infra-red cameras are being replaced by new radiometric cameras, which will be discussed in Section 2.4. What follows is a discussion on the current analysis for the original cameras.

### 2.1 Original Infra-red Cameras

Initially installed at the Pierre Auger Observatory were four Raytheon ControlIR 2000B infra-red cameras. The cameras were designed to measure infra-red light in the $7-14$ $\mu$m wavelength band, suitable for distinguishing warm clouds from the cold clear sky. Images captured by the cameras consist of $320\times240$ pixels spanning a $48^\circ\times36^\circ$ field of view. Every five minutes during FD operation the cameras capture so-called *field-of-view* image sequences, which consist of five images covering the fields of view of the six FD telescopes. Each of the field-of-view sequences are used to evaluate the cloud conditions within the field of view of each FD pixel at the time of image capture. Additionally, a *full-sky* sequence of images is acquired every fifteen minutes which views the entire hemisphere above the FD site; the purpose being to aid shift-operators in determining real-time weather conditions near each FD site.

### 2.2 Image Artefacts

Over time, the quality of each camera deteriorates due to constant exposure to weather as well as simple wear and tear. Some image artefacts develop during operation which can be removed by periodic flat-fielding of the cameras. More difficult to handle are certain artefacts that have developed for some of the cameras.

Visible to some degree in most cameras is a consistent curved streaking artefact, possibly caused by a displacement of the internal *chopper* wheel during the panning of the camera (Fig. 1, left). Despite the artefacts retaining an unvarying shape from image to image, their baseline intensity appears to vary sporadically over long time scales as well being dependent on the local temperature at the camera, making it hard to apply a numerical correction for the artefacts. The development of these streaking artefacts has made previous algorithms, which utilise local weather conditions to estimate an expected camera signal for a clear sky, largely ineffective. For this reason we have developed a new method which can account for the most serious of artefacts yet still provide reliable cloud information.

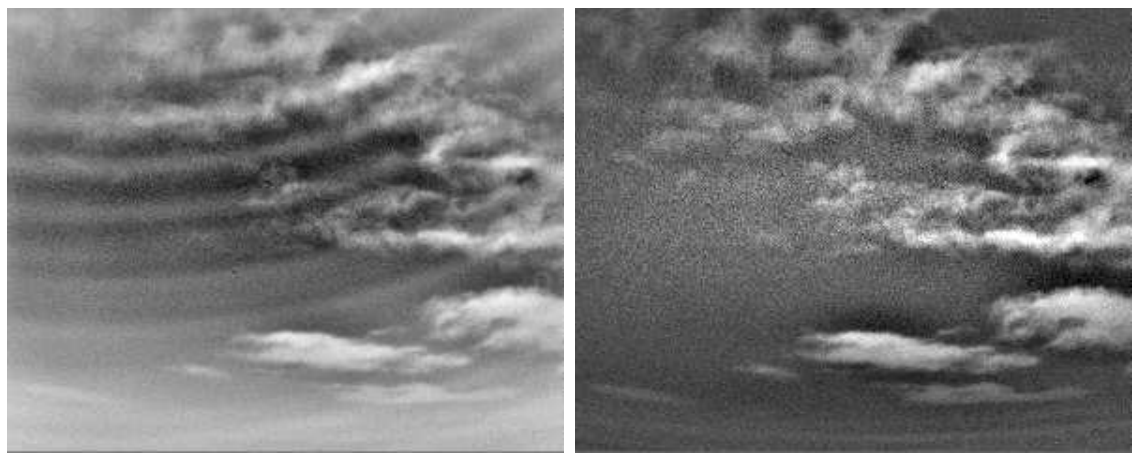

**Figure 1**: *Left*: Raw image from a *field-of-view* sequence taken from Coihueco on 2009/07/19 at 08:25 UTC. *Right*: The same image after having a clear sky template image removed from it. A signal threshold is then applied to create a binary cloud mask which is then mapped to the FD telescope pixels.

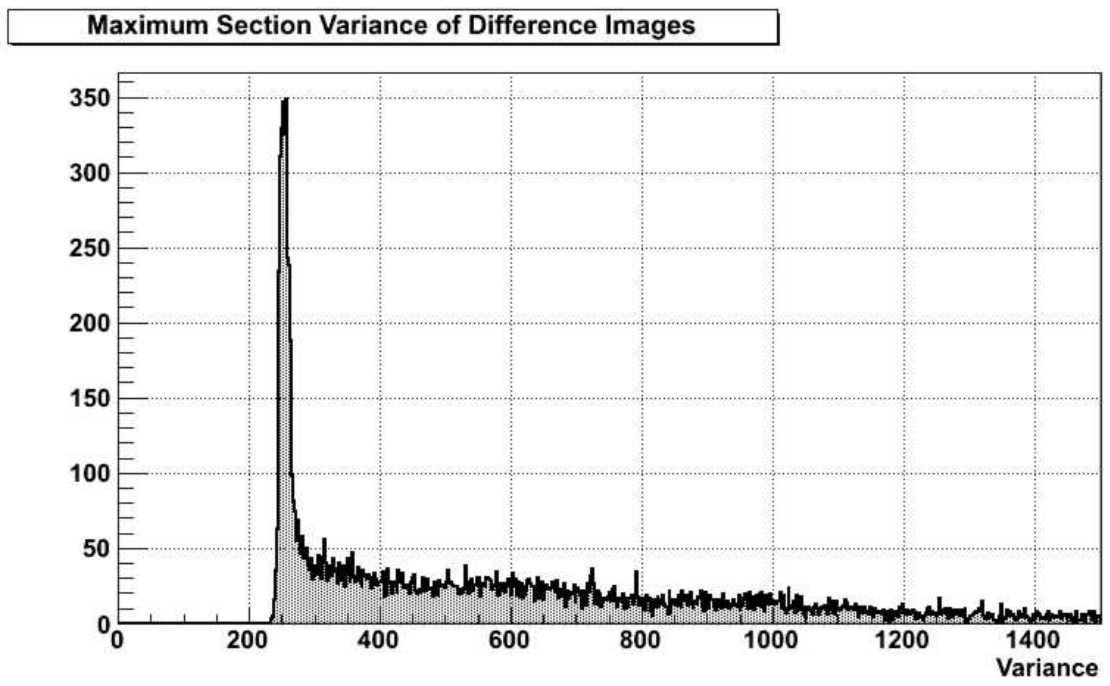

**Figure 2**: Distribution of the maximum section variance from the difference of consecutive images in time. A noticeable peak can be seen which is associated with clear sky images. Applying a cut corresponding to the tail of the peak can be used to distinguish clear and cloudy images.

### 2.3 Cloud Mask Generation

We have developed a new technique which can discriminate between pixels observing clear or cloudy sky within poor-quality images. The technique relies upon building a library of clear sky images which can be used as templates to be removed from each cloud covered image. The difference between the clear and cloudy images should result in enhancement of the cloud affected pixels.

To first determine which images are clear of cloud, each field-of-view sequence is compared to the next sequence of images in time. Any significant difference between the two sets of images may indicate that cloud is present. A small difference indicates the images may be clear. By splitting the difference images into $n \times m$ sections, small changes can be detected to provide higher precision in discriminating between clear sky and cloud.

Figure 2 shows the distribution of the maximum variance from all sections in many difference images. A clear peak can be seen associating with images with a low maximum section variance, indicating images that are most likely clear (no object movement). The tail of this peak can be used to differentiate between clear and cloudy images when combined with a mean signal cut to account for overcast images. A library of clear images is collected for a given (roughly two weeks) FD observation period.

For every other image, a weighting algorithm is applied to find the best match clear-sky template based on similar temperatures and time-of-day. The difference of the two images should result in a flat baseline image with areas of increased signals associated with cloud (Figure 1, right). Applying a simple threshold on the resultant image reveals the designated cloud and clear pixels. By averaging the information from each cloud camera pixel which shares the same direction as an FD pixel, we generate a *cloud index* for each FD pixel which represents the fraction of cloud in its field of view of that time. The result is a cloud camera database for use in shower analysis that contains cloud indices for every FD pixel at five-minute intervals.

### 2.4 New Infra-red Cameras

We are replacing the existing cloud cameras with Gobi-384 uncooled radiometric microbolometer array infra-red cameras. These cameras operate in the $8-14$ $\mu$m wavelength band with a field of view of $50° \times 37.5°$ and produce images consisting of $384 \times 288$ pixels (Figure 3). The design of these new cameras prevents the occurrence of the main artefacts associated with the existing cameras, while also enabling absolute infra-red brightness temperature measurements of the sky. The cameras will be controlled using new LabView based software, to perform the same image capture sequences as described in Section 2.1. Using LabView for the hardware control system will allow a greater level of automation to be implemented into the image capture and camera calibration processes.

### 2.5 Future Work

The radiometric nature of the new cameras will allow for new cloud detection techniques. Of particular interest is the comparison of observed infra-red brightness temperatures with simulations, using GDAS (Global Data Assimilation System) [4] atmospheric profile data and atmospheric radiation transfer software. This will introduce an atmospheric profile dependent cloud detection threshold, which will help resolve the issue of high levels of water vapour being falsely detected as cloud, given that the camera wavelength range includes a water vapour emission band.

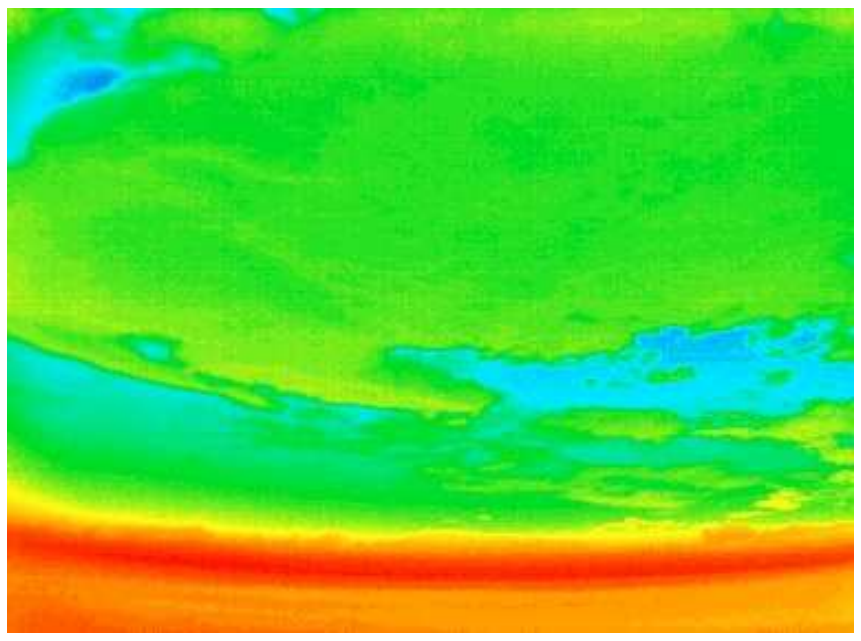

**Figure 3**: Example of a sky image taken with the radiometric Gobi-384 IR camera at the Los Leones FD site. Temperatures range from cooler (blue) to warmer (red).

## 3 Satellite Based Cloud Identification Method

### 3.1 Satellite

Besides the ground equipment at the Pierre Auger Observatory, we have developed a new method to identify night-time clouds [5] from satellite images provided by the Geostationary Operational Environmental Satellites - GOES [6] (GOES-12, which was replaced by GOES-13 in April 2010). The satellite is stationed at 75 degrees West longitude. Its Imager instrument captures images of the South American continent every 30 minutes. Satellite images are produced in one visible band, and four infrared bands centered at wavelengths 3.9, 6.5, 10.7, and 13.3 $\mu$m and labeled Band 2, Band 3, Band 4 and Band 6, respectively.

When the pixels from the infrared band are projected on the ground at the Pierre Auger Observatory, the distance between the center of each pixel is about 2.4 km longitudinally and 5.5 km latitudinally. The visible band resolution is higher.

The raw data are publicly available from the NOAA website [7]. We selected a rectangular region centered at the Pierre Auger Observatory (S 35.6°, W 69.6°). These files contain information for 538 pixels. The data for each pixel in each Band i contain the latitude and longitude of the pixel center and, after calibration and some calculations, we obtain the pixel brightness temperature Ti. The brightness temperatures are the basic quantities for cloud determinations.

### 3.2 Ground-truthing with CLF

The most suitable ground instrument of the Auger Observatory for comparison with the satellite method is the Central Laser Facility (CLF). The cloudiness of the pixel encompassing the CLF can be monitored by the CLF. Every 15 minutes, the CLF produces a series of 50 vertical laser shots which are observed by all four FD stations. From the observed profile, clouds or clear sky over the CLF can be identified.

Typically, each satellite image is bracketed in time by two CLF shots, one 9 minutes before and other 6 minutes after the timestamp of the satellite image. The CLF pixel is tagged as "clear pixel" ("cloudy pixel") if the two bracketing CLF profiles were both identified as "clear CLF" ("cloudy CLF") states. This is to eliminate short-term cloud cover changes.

We used one year of data. For these studies, we arbitrarily chose data from 2007.

### 3.3 Cloud Identification Method

For our method, we selected the difference between the unattenuated brightness temperatures (T2-T4) and the highly attenuated brightness temperature (T3). These satellite-based variables show a separation between "clear pixel" and "cloudy pixel" and only a mildly dependence on ground temperature, minimizing the dependence on daily, weekly or seasonal temperature variations of the method.

In Figure 4, we plot T3 vs. T2-T4 using data of the CLF pixel in 2007. The tagged "clear pixel" (open blue circles) congregate in the upper left quadrant. The tagged "cloudy pixel" (red stars) form an anti-correlated linear feature occupying the center.

This study can be extended for any of the other 538 pixels of the satellite image provided that we can get T2, T3 and T4 for that pixel and we consider that the geographical and meteorological conditions of the other 538 pixels are similar to the conditions of the CLF pixel.

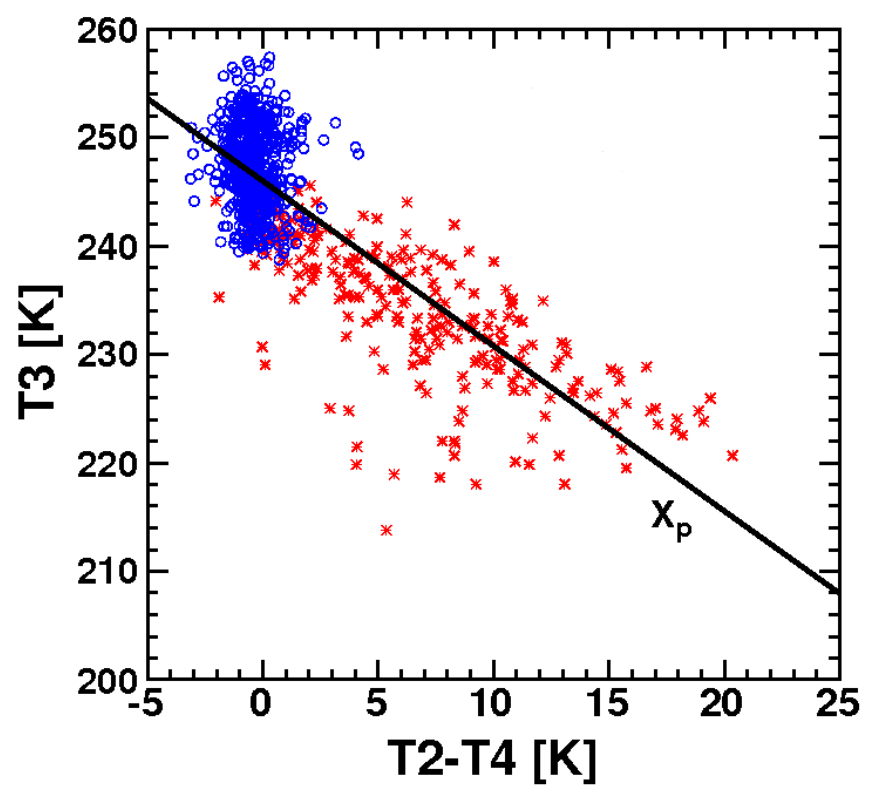

**Figure 4**: T3 vs. T2-T4 of the CLF pixel in 2007. Open blue circles (red stars) were tagged "clear pixels" ("cloudy pixels") from the CLF study. $X_p$ is the principal axis of the fitted line.

We project the data from Figure 4 onto the principal axis $X_p$ described by the fitted line to the overall distribution. In Figure 5, we show one-dimensional histograms of the clear (black thick line) and cloudy (red dashed line on the right) tagged data with respect to the position along the principal axis $X_p$. Also shown is a clear pixel "normalized" histogram (blue thin line on the left) scaled to have the same area as the cloudy pixel histogram. Suitably normalized, these histograms represent probability distribution functions, yielding the probability of identifying a cloudy (clear) pixel for a given value of the principal axis coordinate. Using information from both the reduced clear histogram and the cloudy histogram, we assign a cloud probability for each bin along the principal axis $X_p$ by dividing the number of cloudy entries by the sum of the cloudy and clear entries.

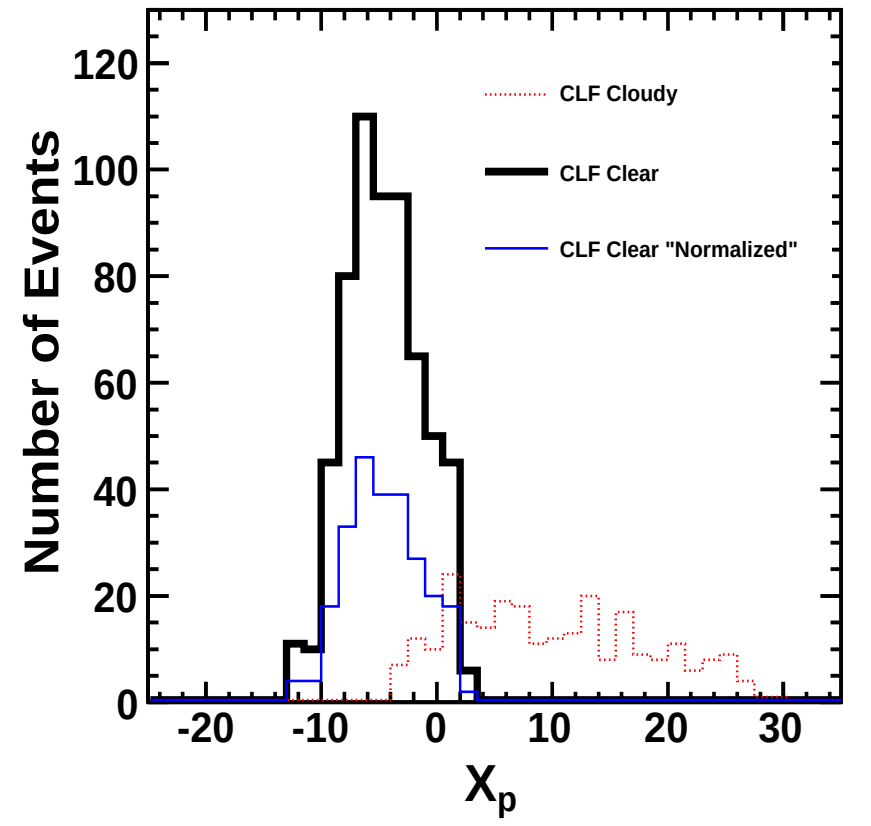

**Figure 5**: Clear (black thick line), clear "normalized" (blue thin line on the left), and cloudy (red dashed line on the right) tagged distributions on principal axis $X_p$.

### 3.4 Applications for the Auger Observatory

We have generated cloud probability maps (see Figure 6) for each satellite image available from all the FD running nights since 2004. In addition, using these maps, nightly animated maps were created. These maps (especially the animated versions) are useful in visualizing the cloud cover during particular cosmic ray events. This helps us to distinguish shower profiles distorted by clouds from unusual shaped shower profiles that could correspond to exotic or rare events.

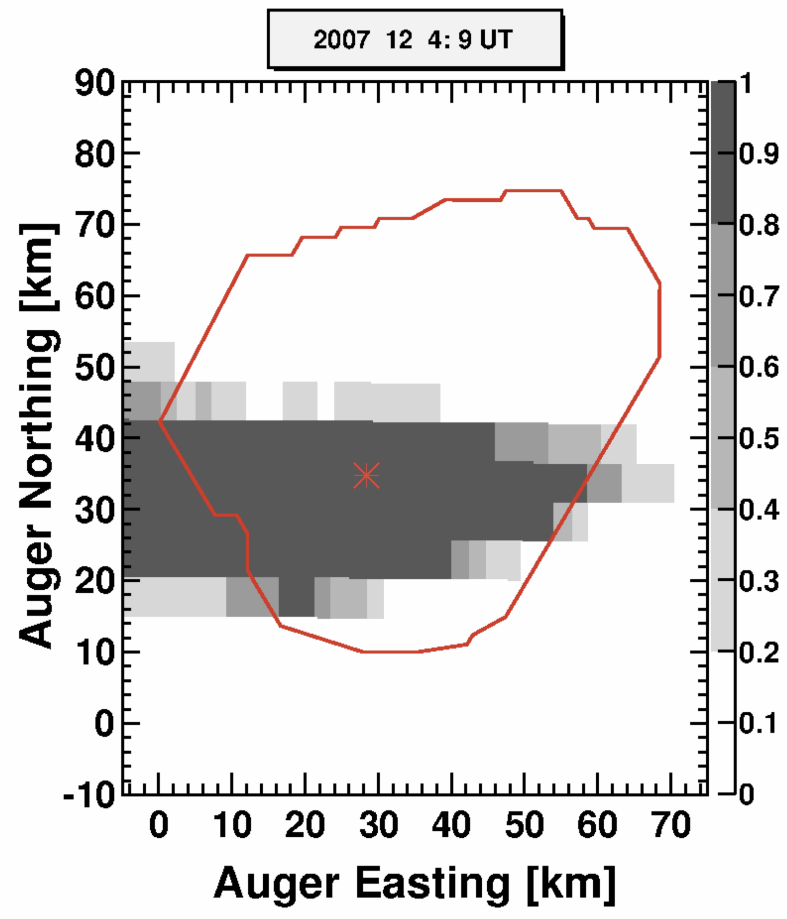

**Figure 6**: Example of a cloud probability map of the Pierre Auger Observatory. Pixels are colored in accordance with the gray scale to the right of the maps. Shown are the borders of the SD (red) and the CLF (red star).

The cloud probabilities for every pixel of the satellite images since 2004 are provided in one of the Auger atmospheric databases for further reconstruction analysis of the cosmic ray data. In the database, the cloud probability is digitized with cloud probability indexes ranging from 0 to 4. For all the cases when the cloud probability of the corresponding bin is between 0 and 20 %, we consider the pixel as clear and we assign 0 to the cloud probability index (CP = 0). For all the cases when the cloud probability of the corresponding bin is between 80 and 100 %, we consider the pixel as cloudy and we assign 4 to the cloud probability index (CP = 4). We assign cloud probability indexes between 1 and 3 for the intermediate cases.

Using this database, the plan is to increase the efficiency of the data analysis cuts for the cloudy nights for the cosmic ray analysis. Also, candidates of exotic events will be vetoed, when these events developed within cloudy pixels.

### 3.5 Reliability of the method

As we can see in Figure 5, there is a small overlap in the distributions. One contribution to the overlap may come from the fact that the CLF data and the satellite image are not precisely simultaneous. Another contribution comes from the fact that the CLF laser beam illuminates an area less than 100 m across as compared to the size of square kilometers of the satellite pixel. The rare clouds higher than the maximum field of view of the FD (14 km) contribute to this overlap since they can be identified by the satellite but not by the CLF. The spatial uncertainty in the satellite pixel location could contribute also to the overlap since the raw data at NOAA do not include the spatial correction in the coordinates of the satellite pixel.

Another reason for the overlap could be that the satellite may be less sensitive than the CLF to certain types of clouds such as the optically thin clouds. Optically thin clouds are not important for distortions due to absorption, but could indeed act as side-scatterers. The CLF is a perfect device to simulate such side scattering. It is possible by averaging over 50 CLF tracks to detect clouds that would cause a less than significant effect on a single cosmic ray event but would still cause a statistically significant change to the averaged CLF profile. However, the goal of this study is only to identify night-time clouds and not to discriminate between cloud type or altitude.

For the atmospheric database, we defined CP = 0 for clear pixels and CP = 4 for cloudy pixels. With these cuts, we are conservative for not incorporating false positives. CP = 0, means cloud probabilities for each corresponding bin of less than 20 %. Using all the bins below this cut, we get a total cloud probability of 4 %. CP = 4 means cloud probabilities for each corresponding bin of more than 80 %. Using all the bins above this cut, we get a total cloud probability of 99 %. However, our method could be applied with different cuts, depending on the needed goal.

## 4 Conclusions

We have described two of the four cloud detection methods employed at the Pierre Auger Observatory. Infra-red cloud cameras situated at each of the four fluorescence sites can directly map cloud directions onto the FD pixel directions which can then be used to veto cloud-affected air showers. The new GOES-satellite based method provides straightforward cloud detection over the entire array that can also be used to veto events. The Observatory relies on the combination of *all* of its cloud detection instruments, including the central laser facilities and lidars at the FD sites, for cloud information. This is especially the case for studies very sensitive to cloud effects, such as the search for exotic physics in the development of air shower cascades.

**Acknowledgment:**This work was in part supported by the Department of Energy (Contract No. DE-FG02-99ER41107) and other agencies involved in funding of the construction and operation of the Pierre Auger Observatory.

# Observation of Elves at the Pierre Auger Observatory

AURELIO TONACHINI[1] FOR THE PIERRE AUGER COLLABORATION[2].

[1] *Università degli Studi di Torino and INFN Sezione di Torino, Via Pietro Giuria 1, 10125 Torino, Italy*
[2] *Full author list: http://www.auger.org/archive/authors_2013_05.html*

*auger_spokespersons@fnal.gov*

**Abstract:** The 'elves' are transient luminous events generated by the sudden excitation of the lower ionosphere, caused by lightning. Exploiting a time resolution of 100 ns and a space resolution of about 1°, the Fluorescence Detectors of the Pierre Auger Observatory in Argentina can provide 2D imaging of elves, originating at distances of several hundred kilometers, with unprecedented accuracy. Using 60 elves event candidates from prescaled data taken in the period 2008-2011 by the Fluorescence Detectors of the Pierre Auger Observatory, we have redesigned the third level trigger of the experiment, in order to acquire elve data with much higher efficiency in the coming years of operation. Preliminary results from the first months of data taking with the upgraded trigger will be given.

## 1 Introduction

Transient luminous events (TLEs) like elves, sprites, halos and jets are luminous emissions detectable well above thunderstorms. Lightning discharges generate electromagnetic pulses (EMPs) which accelerate free electrons. TLEs are different kinds of optical flashes due to the interaction of these electrons with atmospheric species. Elves appear as rapidly expanding rings of light, and are a consequence of the heating and ionization of the lower boundary of the ionosphere. The light emission attains diameters of several hundred kilometers [1], while the typical duration is less than 1 ms.

The wavelength range of light emission in elve events extends from UV to near-infrared. Prompt emissions are due to the electron impact with nitrogen and oxygen molecules. These emissions are followed by chemical reactions which produce a dim chemiluminescence [2].

The first clear detection of elves, made using a high speed photometer pointed at altitudes in coincidence with the observation of sprites [3]. The following observations from ground were made with linear arrays of photomultiplier tubes (PMTs) [4, 5]. In particular, the PIPER instrument [6] adopts two or more orthogonal arrays of PMTs with a time resolution of 40 $\mu$s. The arrays can be combined in order to reconstruct 2D images.

The ISUAL mission, which ran from 2004 to 2007 onboard the FORMOSAT-2 satellite, studied systematically TLEs from space. The data collected allowed one to study the global rate and occurrence conditions [7]. The global elve occurrence rate estimated by ISUAL is 35 events per minute. Elves are thus identified as the dominant kind of TLEs. There is also a clear relation between their occurrence and the temperature of the sea surface, which favors the warmest zones of the Earth. The elve occurrence rate, in fact, increases dramatically when the sea surface temperature exceeds 26° Celsius. Globally, it has been shown that there are ten times more elves above the Ocean than on land.
Further progress in understanding and modeling elves may be achieved using the data recorded by the air fluorescence detector (FD) of the Pierre Auger Observatory [8]. The observatory is located near the city of Malargüe, Argentina (69° W, 35° S, 1400 m a.s.l.). The FD comprises four observation sites overlooking a 3000 km$^2$ water Cherenkov surface array. Each site is in turn made of six independent telescopes, each one with a field of view (FOV) of 30° × 30° in azimuth and elevation (see Fig. 1). In each site, thus, the combination of the FOV of six telescopes covers 180° in azimuth. Since elves detectable by the Pierre Auger Observatory are located far away from the observatory, there is a good probability to see them simultaneously with two or more FD sites (stereo mode).

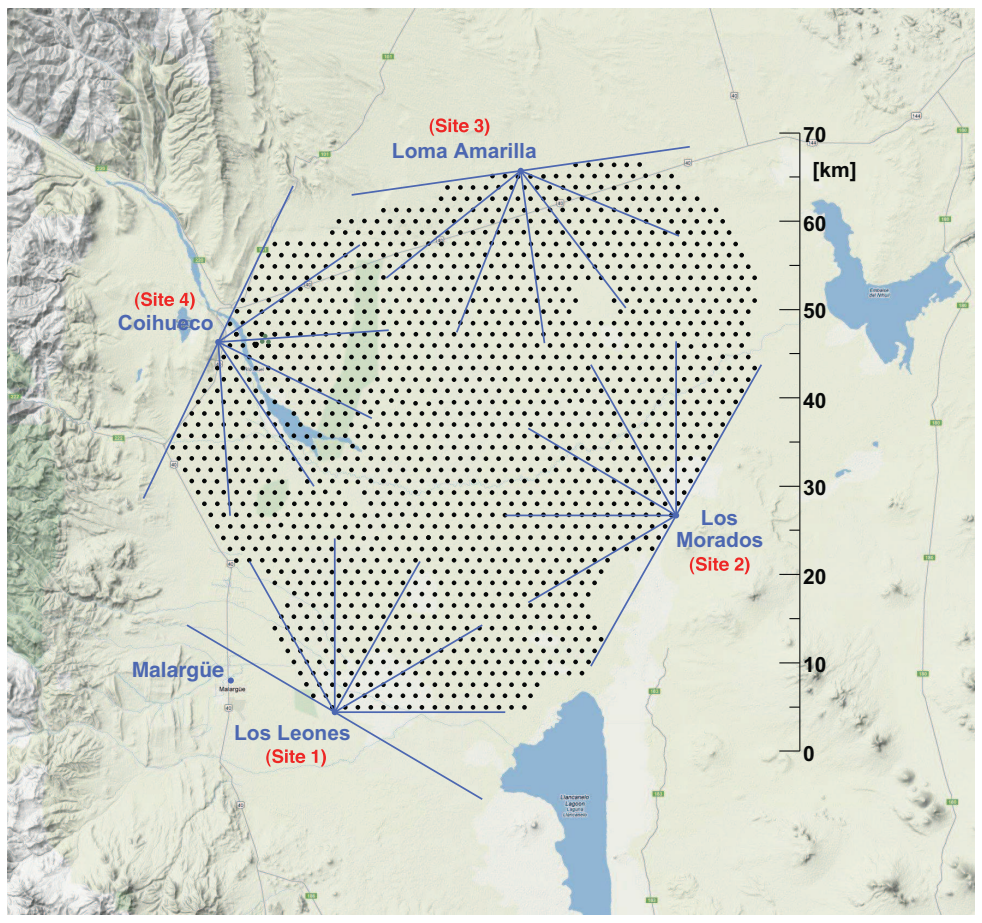

**Fig. 1**: Map of the Pierre Auger Observatory. The surface detector (small dots) is overlooked by the fluorescence detector. The blue lines show the field of view of each FD telescope.

In each telescope, incoming light passes through a large UV filter window before being focused by a 10 square meter mirror on a grid of 440 photomultiplier tubes (PMTs). The range of wavelengths passing the filter goes from ~290 to ~410 nm. Signals in each PMT are digitized at 10 MHz.

The FD geometry, resolution, and its 100 ns sampling rate make Auger telescopes suitable for studying such fast developing TLEs as elves. Furthermore, the location of the Pierre Auger Observatory allows the detection of elves both on land and above the Pacific Ocean, allowing one to study in detail the difference among the two classes in terms of occurrence rate and signal development.

## 2 Design of the elve trigger

The event selection of the FDs is based on a three-stage trigger. The first level trigger operates at pixel level, and keeps the PMT trigger rate at 100 Hz by adjusting the threshold. The second level trigger checks the $22 \times 20$ pixel camera for five-pixel track segments. This basic selection is passed by both cosmic ray showers and elve-like events. The third level trigger (TLT) is software based. It is designed to efficiently filter out lightning events, and is based on the number of triggered pixels of one camera at the same time (called multiplicity). To achieve a very high efficiency a dedicated study with a sample of true showers and lightning events was performed. The multiplicity-based TLT was installed in the end of 2007 as a replacement of a previous and less efficient version.

Before that time, three elves passed by chance the lightning cuts, and were tagged as cosmic ray showers. After their serendipitous discovery, these events were studied in detail from the point of view of the evolution of the signal in time and space [9].

Based on these events a deep search in a prescaled sample (1 in a 100) of minimum bias events has been done. The result after many iterations was a set of conditions that selects elves very efficiently. Each event that is preliminarily classified as lightning noise is thus analyzed in order to recognize if the light front expands radially. Firstly, a fast pulse analysis is done on the triggered pixels. Once the first triggered pixel is identified, pulse start times of the triggered pixels falling on the same row and the one of the pixels falling on the same column are checked.

For the sample of pixels of the same column, the algorithm requests that at least three pixels before *and* three pixels after the central one have a pulse. Moreover, 80% of them must show an increasing pulse arrival time.

For the sample of pixels of the same row, instead, the algorithm just requests three pixels to the left of the central one *or* three pixels to the right of it. Any of the two arms must show an increasing pulse arrival time in 80% of the pixels.

Considering that elves release a large amount of light compared to cosmic rays, an additional cut on the pulse amplitude has been introduced in order to remove unwanted noise: among the triggered pixels, at least one must have a pulse amplitude greater than 50 ADC channels.

Running this selection procedure over the prescaled data recorded from 2008 to 2011, 58 elves have been found. 39 of them have the centre contained inside the camera, and thus are well reconstructable (see Table 1). In the remaining 19 candidates, the centre occurs in an adjacent bay (which has been randomly discarded) or outside the field of view. An example of an elve detected with the FD is shown in Fig. 2.

As expected, the elve rate is not uniform over time, but shows a substantial increase in the warmest months, when usually violent thunderstorms take place (Fig. 3). For the events with the centre well visible within the FD camera, the direction of the electromagnetic pulse which triggered the elve is easily reconstructable. Fig. 4 shows that most of the events detected so far come from the warmest regions of Argentina, while none of these events took place above the sea.

| FD site | Elves | Centre contained |
|---|---|---|
| 1. Los Leones | 21 | 17 |
| 2. Los Morados | 6 | 3 |
| 3. Loma Amarilla | 12 | 9 |
| 4. Coihueco | 19 | 10 |
| **Total** | **58** | **39** |

**Table 1**: Number of elves found in the FD prescaled data, grouped by site. the site number refers to Fig. 1.

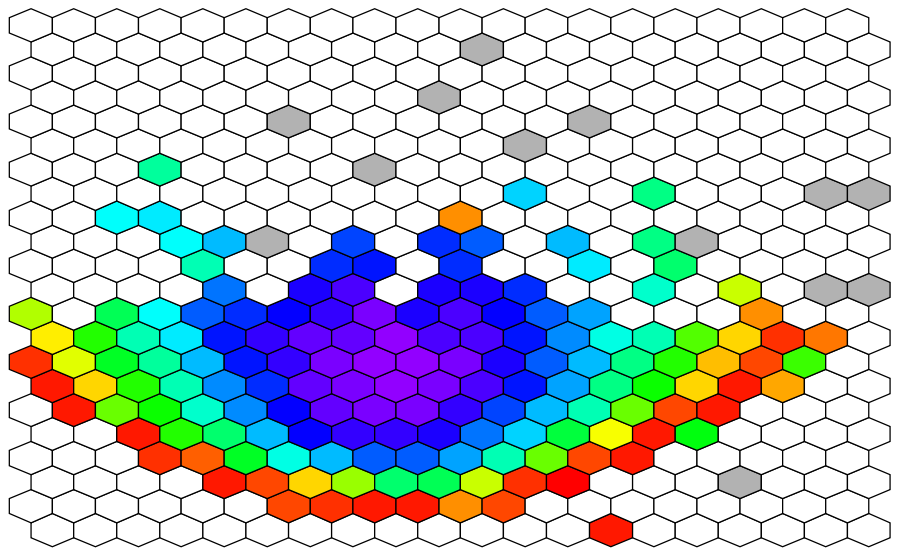

**Fig. 2**: Pixels of an FD telescope triggered by an elve. The event expands radially from its centre, as shown by the colour scale (pulse start time increases from violet to red).

The elve trigger, as described in the previous section, has been fully integrated in the TLT as of March 2013. During the data taking periods of March and April 133 events were tagged as elve candidates, and only six of them (4.5%) are false positives.

## 3 Elve reconstruction

The light observed with the FDs is emitted by the D region of the ionosphere, which has been excited by elec-trons accelerated by a lightning-launched electromagnetic pulse. The signal travels from the source to the ionosphere at nearly the speed of light; fluorescence light is emitted at altitudes around 90 km and travels towards the detector. The light observed by different PMTs of a fluorescence detector in the same time frame corresponds to paths traveled in the same amount of time. One can describe this as the intersection of an expanding ellipsoid, with the EMP source and the FD at the foci, and a sphere concentric to the Earth with a radius given by the D layer altitude plus the Earth radius (see Fig. 5). In this model, the first light is observed at $t_0$, when the ellipsoid is tangent to the sphere. If the altitude of the D layer were well known, one would have a strong geometrical constrain on the position of the EMP source. In reality, this altitude fluctuates by several kilometers. Fig. 5 illustrates how the position of the EMP found in such a way is sensitive to this fluctuation. For $t > t_0$ one observes a closed curve, whose lateral expansion is symmetric. The front moving towards the FD appears to move faster than that moving in the opposite direction, since at higher elevation angles the portion of D region observed with the same field of view is less.

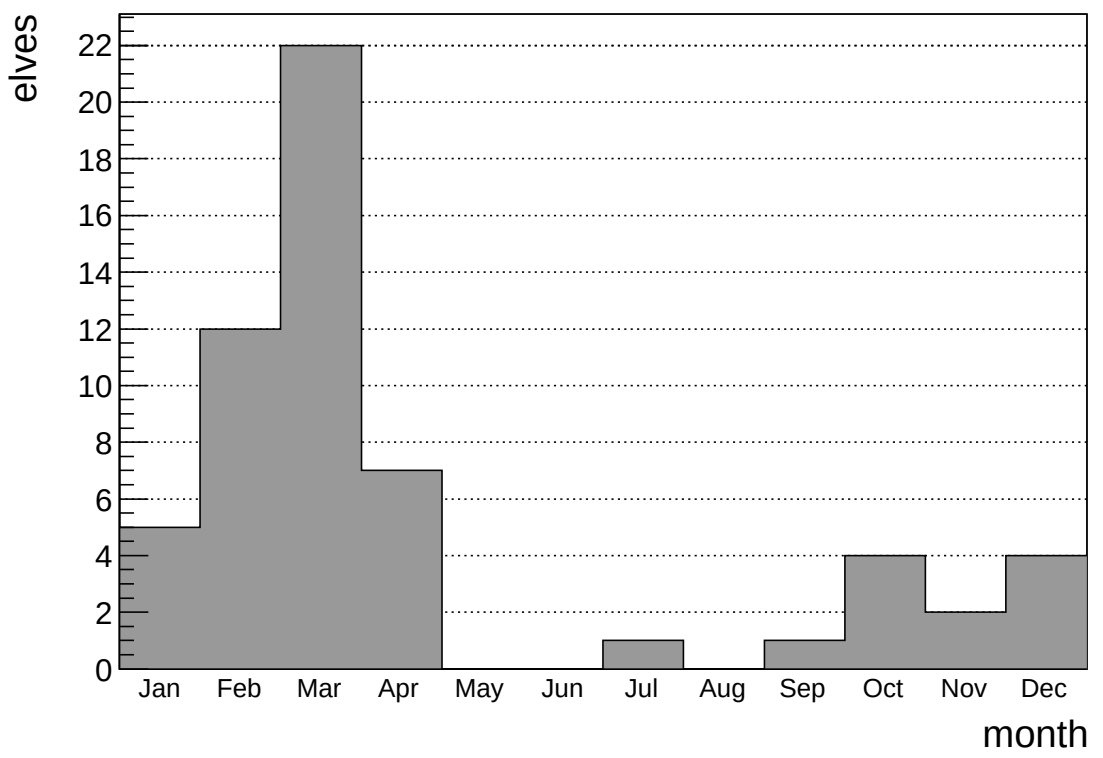

**Fig. 3**: Elve rate per month from the 58 events found in the prescaled data. Most of the events were detected during austral summer.

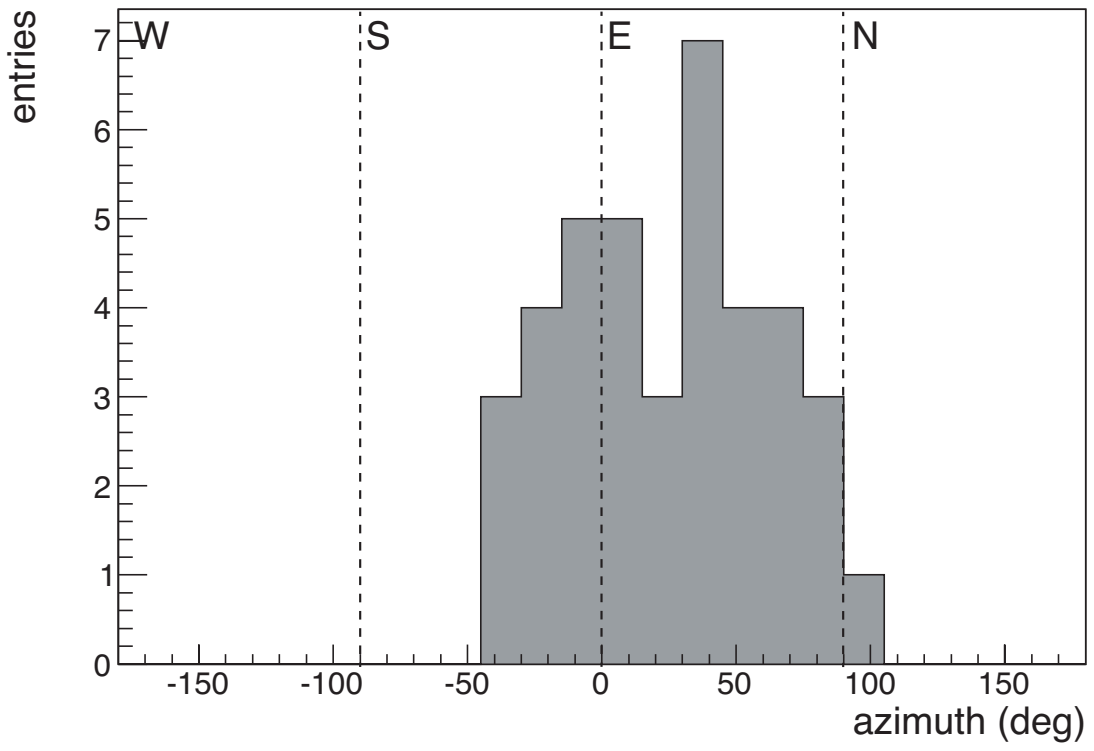

**Fig. 4**: Azimuth direction of the EMP location. The direction is in degrees anticlockwise with respect to the East. Most of the events came from the North-East. None of them came from the direction of the Pacific Ocean .

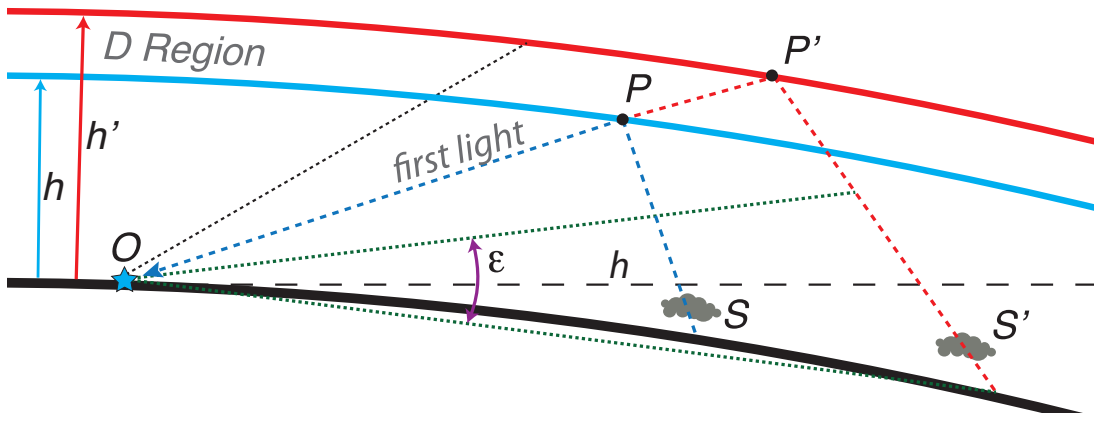

**Fig. 5**: Schematic view of an EMP generated by a thunderstorm at $S$, which interacts with the D region of the ionosphere. The light emitted by the ionosphere (in blue) is detected by the fluorescence detector at $O$. The observed signal time t is the sum of the time needed by the pulse to move from $S$ to the interaction point $P$ and the time needed by the emitted light to travel from $P$ to $O$. If the D region is higher, the first light is emitted by $P'$ instead of $P$, and the source $S'$ is much farther. The D region altitude can be retrieved from the overall development of the elve.

### 3.1 Signal processing

For those PMTs that pass the first level trigger, a trace of 1000 bins of 100 ns each is recorded. A pulse search is run on each trace, and the pulse start and end are found by maximizing the signal to noise ratio. For reconstructing the development of elve events it is particularly important to have a precise determination of the starting time of the signal. For this reason, small pulses whose maxima are below three standard deviations with respect to the background or very short pulses ($\Delta t < 3\mu$s) are not considered.

For the other ones, a 2.1 $\mu$s running average is applied in order to smooth the traces. The pulse start time previously determined is then moved back until the smoothed signal is less than $5\sigma$ above its pedestal. The uncertainty associated with this point is determined by searching for the time at which the signal falls below $3\sigma$, and taking the time difference with respect to the start point.

Once the first triggered pixel is found, the duration of the pulse recorded is measured. This pulse width can be related either to the duration of the EMP, or to the thickness of the light emitting layer.

### 3.2 Elve location

In order to have a precise measurement of the azimuthal direction of the EMP source, a parabolic fit of the lateral expansion is done considering all triggered pixels with elevation within 0.75° (half the PMT field of view) with respect to the first triggered PMT elevation.

Once the azimuth and elevation of the first light are determined, the model still depends on two parameters: the D layer height $h$, and the elevation $\varepsilon$ of the EMP source with respect to the horizon. In order to determine the exact location of the source and the D layer altitude at the same time, $h$ is variated between 50 to 110 km, and $\varepsilon$ between -8.0° and +5.0°. For each step in $h$ and $\varepsilon$ the model is used to calculate the expected times of the light pulses observed by the PMTs, which are then compared to the values recorded by the triggered PMTs.

## 4 Results

The reconstruction algorithm has been tested on the 39 elves contained in the prescaled data. The distance of the EMP source to the fluorescence detector varies from about

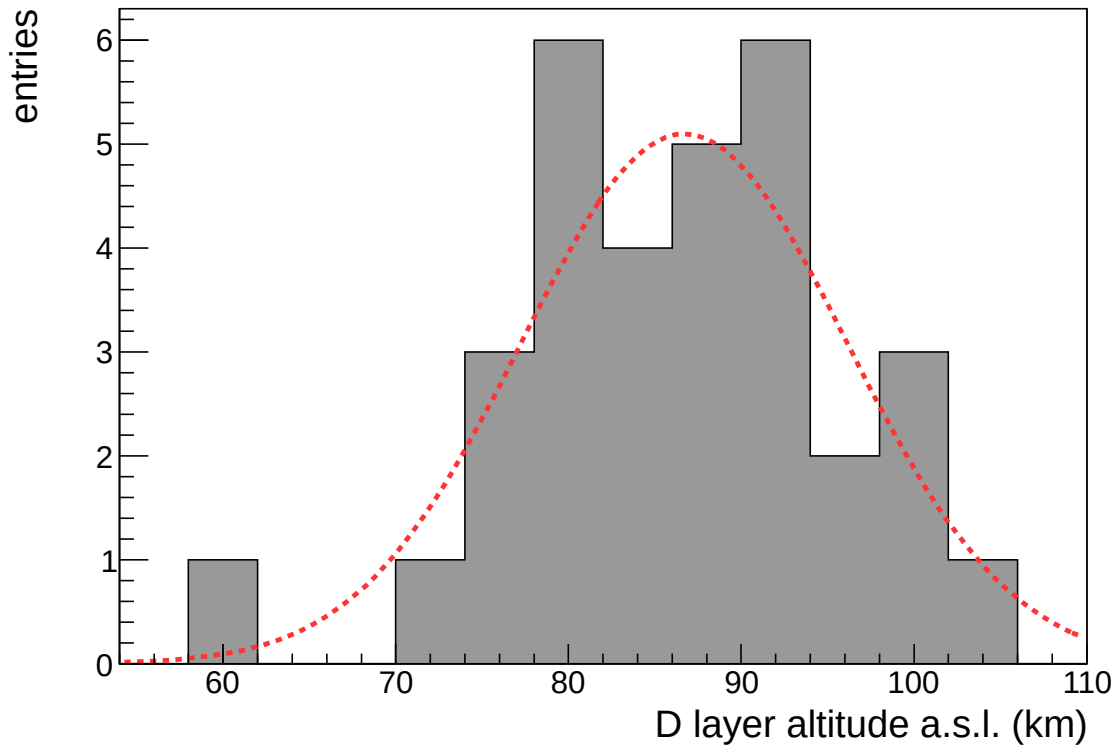

**Fig. 6**: The distribution of the D layer altitudes has mean value at ~86 km and RMS of 9 km.

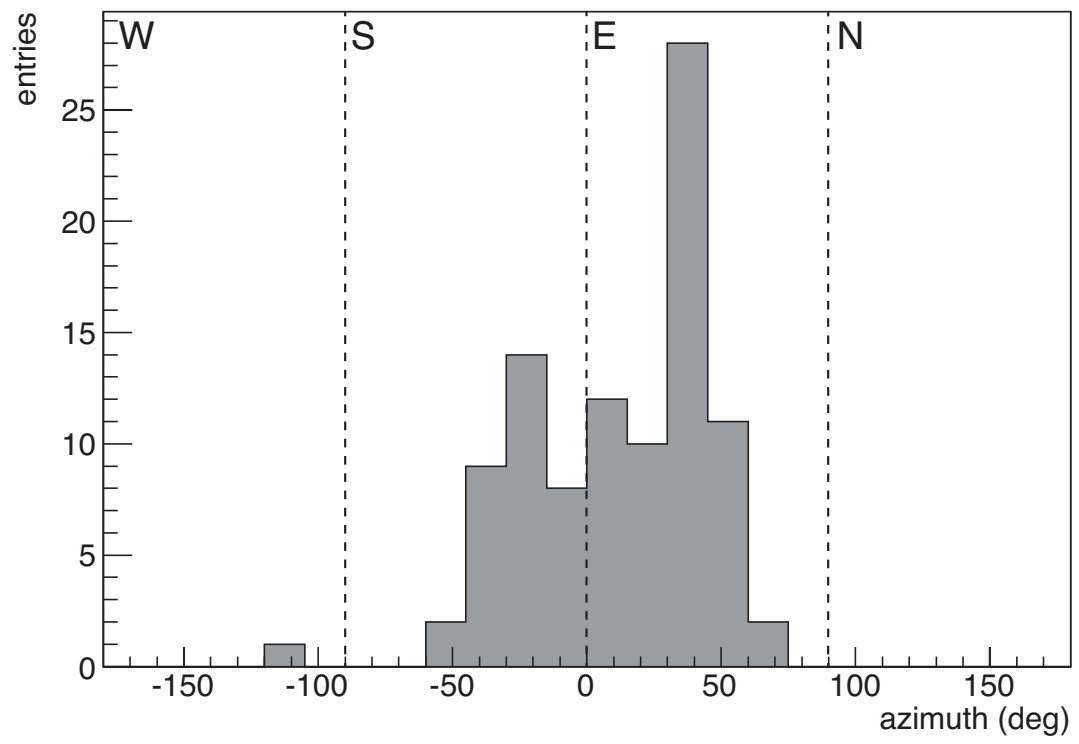

**Fig. 7**: Azimuth direction of the EMP location for the events recorded in March and April 2013. The direction is in degrees anticlockwise with respect to the East. Most of the events came from the East with a slightly different distribution with respect to Fig. 4.

| FD site | Elves | Reconstructed | Stereo |
|---|---|---|---|
| 1. Los Leones | 41 | 29 | 26 |
| 2. Los Morados | 8 | 2 | 7 |
| 3. Loma Amarilla | 32 | 26 | 13 |
| 4. Coihueco | 52 | 40 | 22 |

**Table 2**: Number of elves found in the FD measured data, grouped by site. The site number refers to Fig. 1. Three of these elves were seen by three FDs simultaneously.

200 km to 900 km. The elevation angle of the first light observed varies respectively from $\sim 23^\circ$ to $\sim 10^\circ$. The D layer altitude, measured for each elve, is distributed as shown in Fig. 6, with mean at $\sim$86 km.

The same algorithm was used to process the signals of elves recorded with the new third level trigger of the fluorescence detector. Table 2 reports the number of events detected by each FD site. Many events have been observed simultaneously by more than one detector (stereo mode). Events have been seen at elevations as low as 6°, corresponding to distances as far as 1000 km from the Pierre Auger Observatory. Most of the events occurred in East direction, while no events have been detected above the Pacific Ocean so far (see Fig. 7).

Fig. 8 shows the distribution of the pulse durations measured from the signal recorded by the first triggered PMT of the elves.

## 5 Conclusions

A dedicated trigger for recording elves has been recently implemented as part of the third level trigger of the fluorescence detector of the Pierre Auger Observatory. 133 events have been already collected, with a low number of false positives. For the first time a detector is recording 2D images of elves with a time resolution 50 times better than the previous observations. Moreover, many events are recorded simultaneously by two or more detectors placed at several tens of kilometers one from the other, thus providing a stereo view of elves.

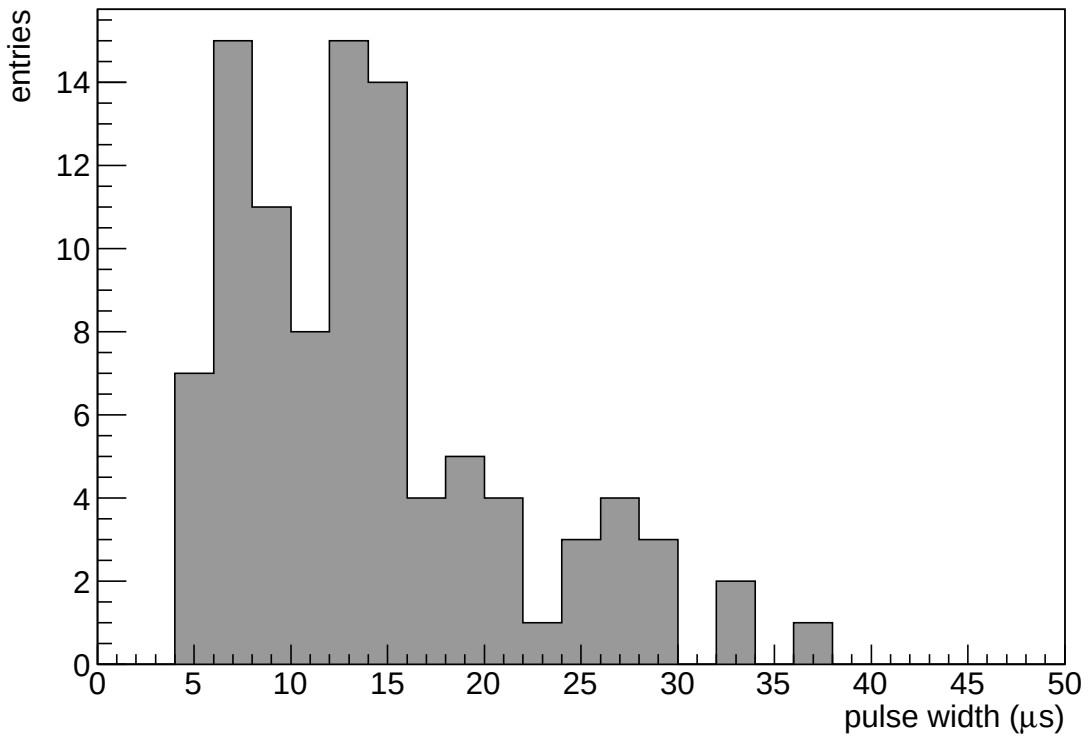

**Fig. 8**: Pulse width distribution for the events recorded in March and April 2013. This observable can be related to the EMP duration or the thickness of the excited layer in the ionosphere.

# Education and Outreach Activities of the Pierre Auger Observatory

GREGORY R. SNOW[1] FOR THE PIERRE AUGER COLLABORATION[2].

[1] *Department of Physics and Astronomy, University of Nebraska, USA*
[2] *Full author list: http://www.auger.org/archive/authors_2013_05.html*

*auger_spokespersons@fnal.gov*

**Abstract:** The scale and scope of the physics studied at the Pierre Auger Observatory continue to offer significant opportunities for original outreach work. Education, outreach and public relations of the Auger Collaboration are coordinated in a separate task whose goals are to encourage and support a wide range of education and outreach efforts that link schools and the public with the Auger scientists and the science of cosmic rays, particle physics, and associated technologies. The presentation will focus on the impact of the collaboration in Mendoza Province, Argentina. The Auger Visitor Center in Malargüe has hosted over 80,000 visitors since 2001, and a fourth collaboration-sponsored science fair was held on the Observatory campus in November 2012. The Rural Schools Program, which is run by Observatory staff and which brings cosmic-ray science and infrastructure improvements to remote schools, continues to broaden its reach. Numerous online resources, video documentaries, and animations of extensive air showers have been created for wide public release. Increasingly, collaborators draw on these resources to develop Auger related displays and outreach events at their institutions and in public settings to disseminate the science and successes of the Observatory worldwide.

## 1 Introduction

Education and public outreach (EPO) have been an integral part of the Pierre Auger Observatory since its inception. The collaboration's EPO activities are organized in a separate Education and Outreach Task that was established in 1997. With the Observatory headquarters located in the remote city of Malargüe, population 23,000, early outreach activities, which included public talks, visits to schools, and courses for science teachers and students, were aimed at familiarizing the local population with the science of the Observatory and the presence of the large collaboration of international scientists in the isolated communities and countryside of Mendoza Province. As an example of the Observatory's integration into the local culture, the collaboration has a tradition of participating in the annual Malargüe Day parade since 2001 with collaborators marching behind a large Auger banner (see Fig. 1). Close contact with the community fosters a sense of ownership and being a part of our scientific mission. The Observatory's EPO efforts have been documented in previous ICRC contributions [1]. We report here highlights of recent activities.

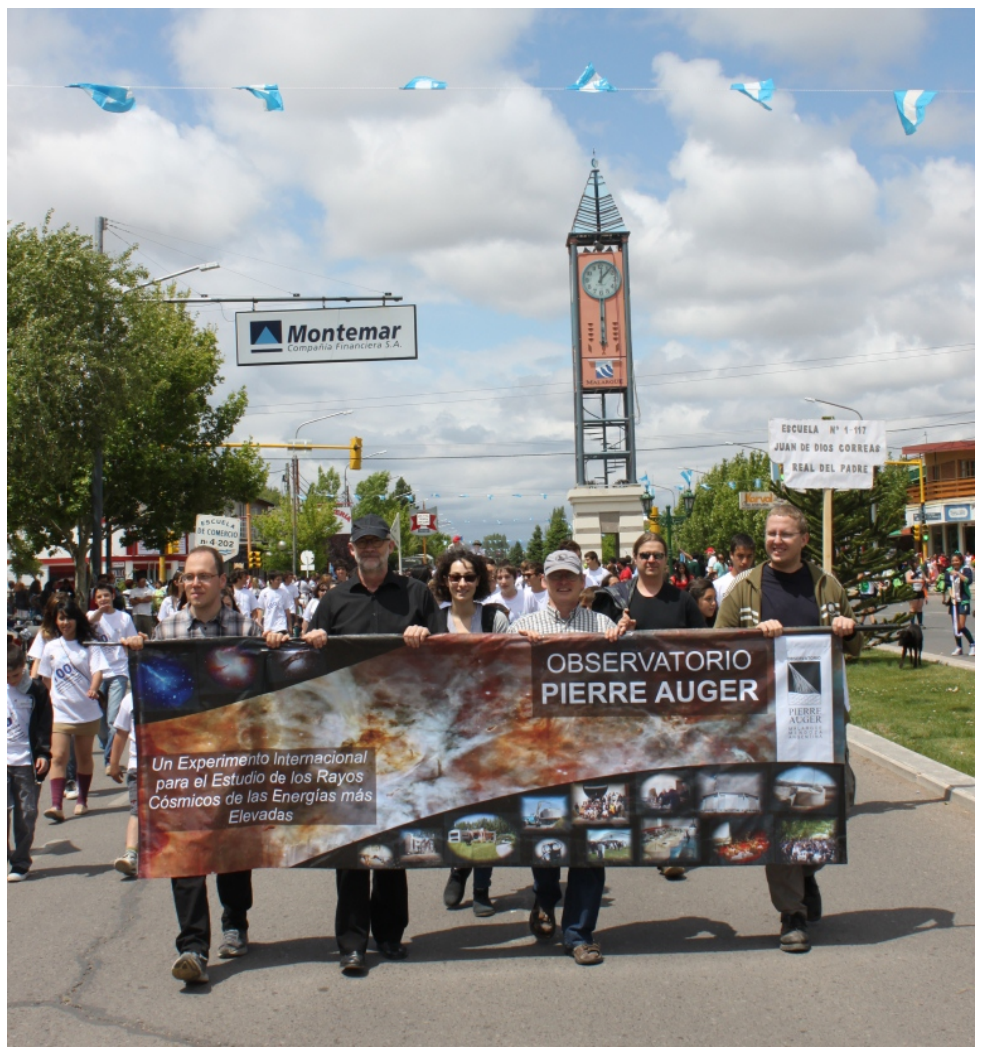

**Figure 1**: Auger collaborators and Science Fair participants in the November 2012 Malargüe Day parade.

## 2 The Auger Visitor Center in Malargüe

The Auger Visitor Center (VC), located in the central office complex in Malargüe, continues to be a popular attraction. Through Feb. 22, 2013, the VC has hosted 79,924 visitors. Fig. 2 shows the number of visitors logged per year from Nov. 2001 through Feb. 2013. The noticeable increase of visitors since 2008 occurred after the opening of a nearby planetarium [2] in August of that year. The VC is managed by a small staff led by Observatory employee Analía Cáceres; she and other collaborators share the task of giving presentations and tours to visitors and school groups. During the Feb.-March 2013 collaboration meeting, a small group arriving at the VC included the 80,000$^{th}$ visitor, see Fig. 3. Auger sourvenirs were presented to the group, senior collaborators welcomed them, and local television and radio covered the visit to highlight this attendance milestone. A recent addition to the visitor experience is an AERA [3] radio detector station outside the office complex.

## 3 The Rural Schools Program and Education Fund

The Rural Schools Program, initiated by the Observatory staff who volunteer their time, continues to bring information about the Observatory and needed infrastructure im-provements to remote schools that have difficulty bringing their students to the Observatory. Fig. 4 shows three technicians at the Peregrina Cantos school in Bardas Blancas, 60 km from Malargüe, working on the school's internet connectivity using hardware purchased from the Education Fund to which Auger collaborators and institutions contribute. In November 2012, several Auger collaborators met with a group of headmasters from 10 remote schools to discuss increased communication with the Observatory as well as visits to the schools by the Rural School team.

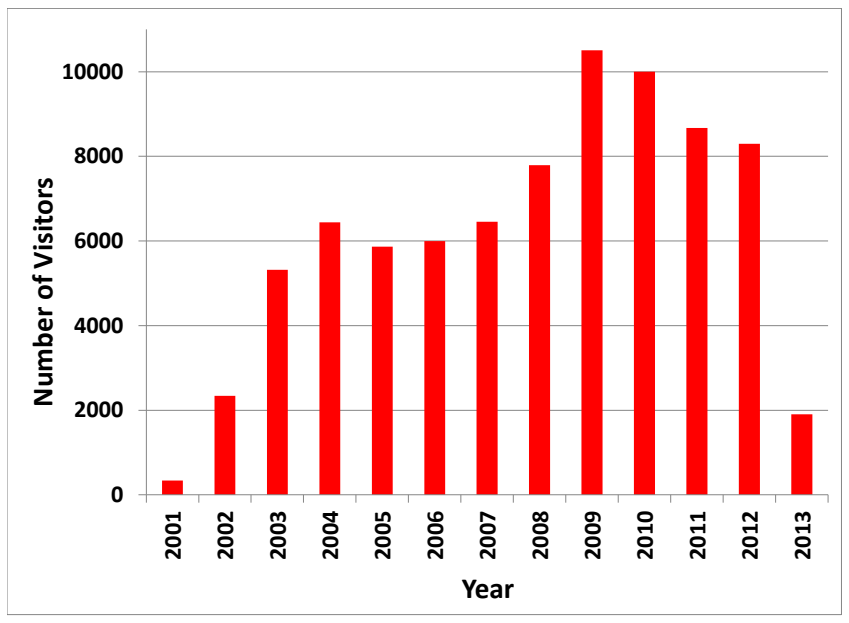

**Figure 2**: Number of visitors logged by year at the Auger Visitor Center from November 2001 through February 2013.

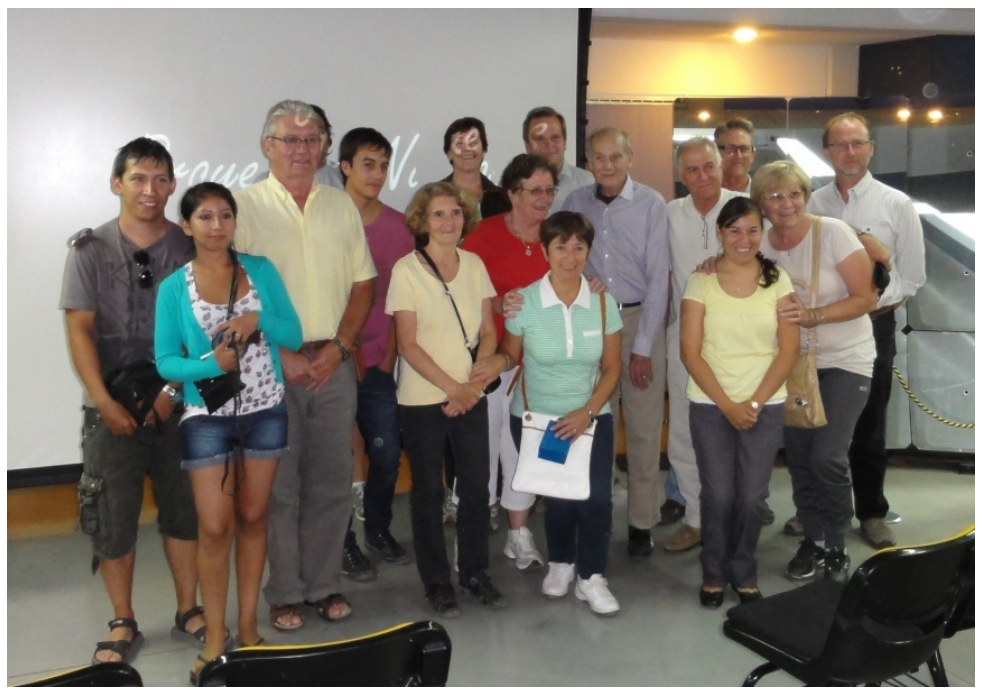

**Figure 3**: Auger collaborators with a visitor group which passed the 80,000 visitor milestone.

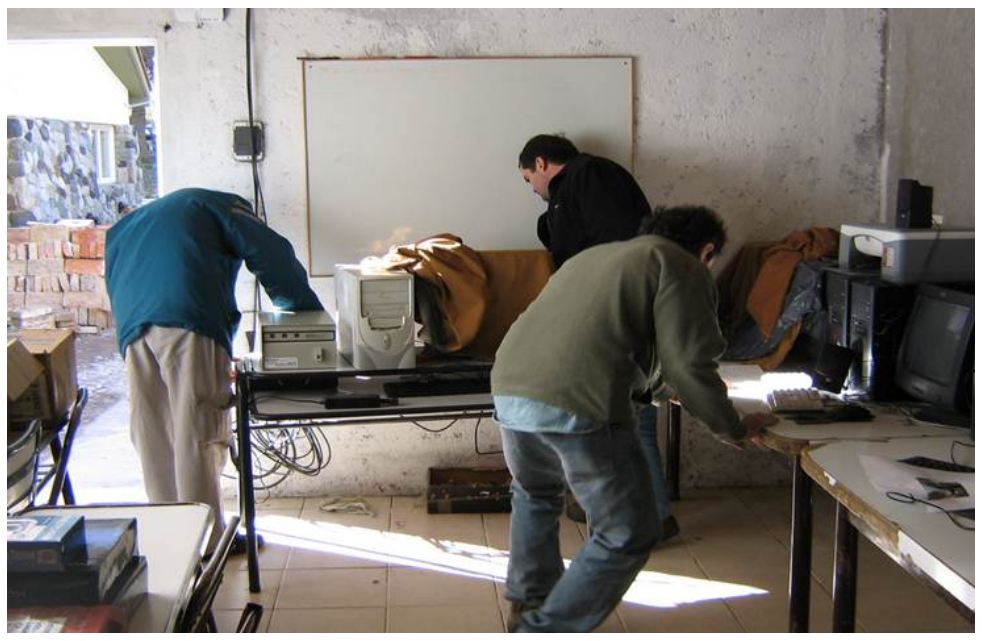

**Figure 4**: Auger technicians working on internet hardware at a school in Bardas Blancas.

## 4 The 2012 Auger Science Fair

The Observatory hosted its fourth biannual Science Fair in the Assembly Building November 15-17, 2012, this Fair dedicated to the 100$^{th}$ anniversary of Victor Hess's balloon flights which are often referred to as the discovery of cosmic rays. See Figs. 5 and 6. Thirty-six student teams from all over Mendoza Province, with ages ranging from primary school through high school, presented research projects in the areas of natural science, exact science, and technology. Auger collaborators and a few invitees served as judges for the student projects, and prizes were awarded to the top teams in several categories in the closing ceremony on November 17. On the 16$^{th}$, the Science Fair participants had the opportunity to walk in the Malargüe Day parade along with Auger collaborators, everyone attended a presentation about the Observatory in the Visitor Center, and a pizza party and *asado* were held in the evening. The November 2012 Science Fair owes its success to the Observatory staff, the collaborators who served as judges, the Municipality of Malargüe, the participating teachers and students, and special mention goes to the lead local organizers: Miguel Herrera, Fabian Amaya, and Alicia Piastrellini.

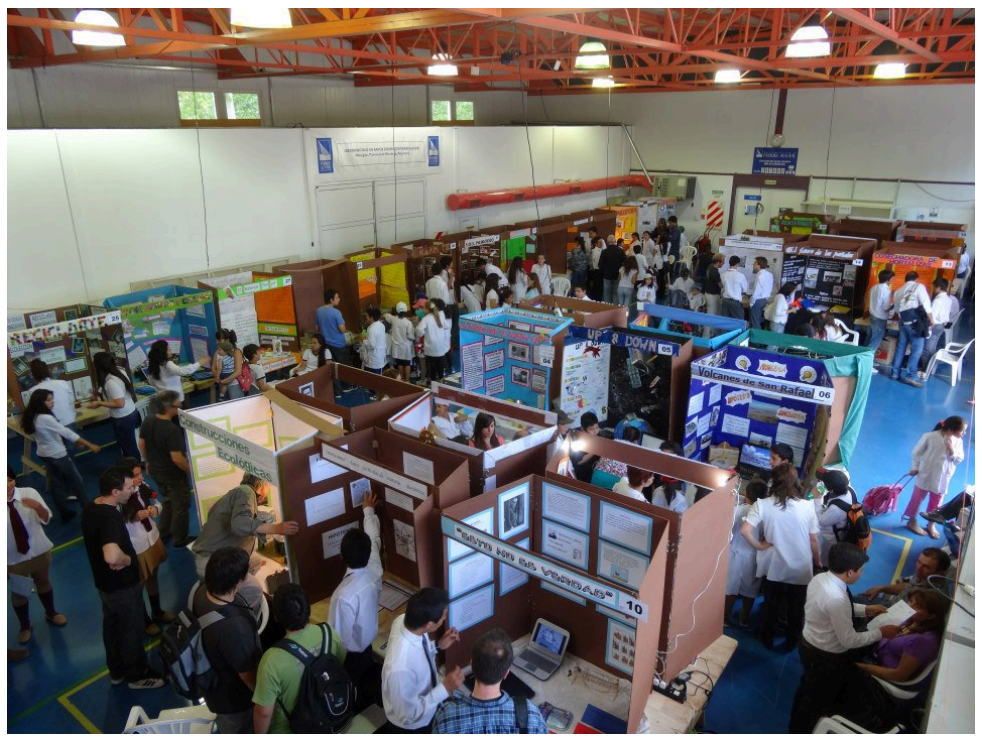

**Figure 5**: November 2012 Science Fair in progress in the Assembly Building.

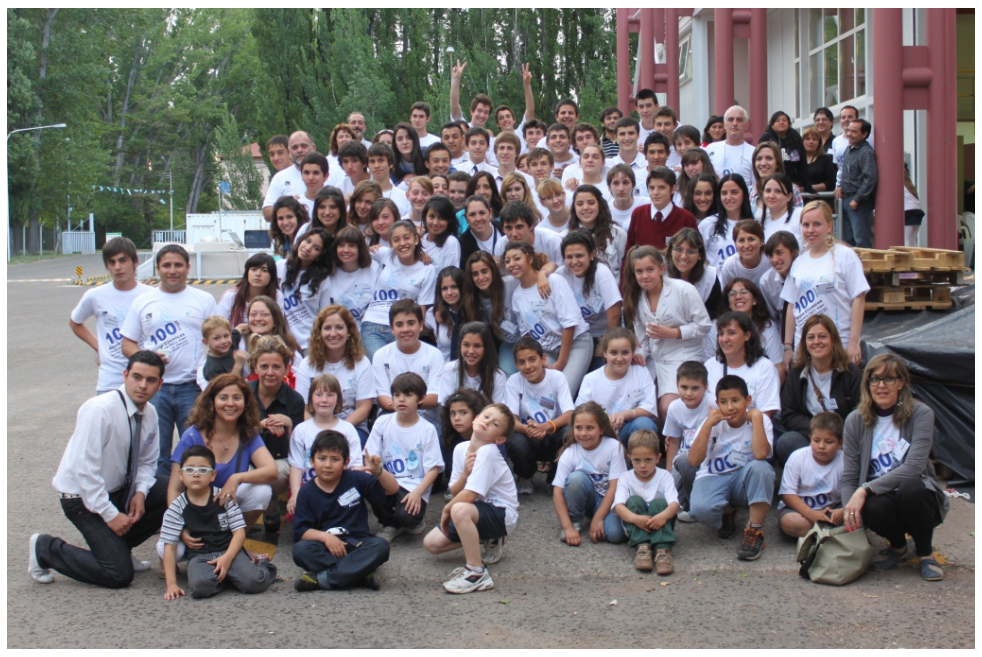

**Figure 6**: Science Fair organizers and participants.

## 5 Selected Outreach Activities in Member Countries

**La Brújula in Mendoza**

The Observatory is currently being featured at the first exhibition of science and technology, *La Brújula* ("The Compass") [4], which highlights scientific research projects based in Mendoza Province, Argentina, and runs from May 15 to June 16, 2013, in the Cultural Center Julio Le Parc in the provincial capital. Figs. 7 and 8 show a Fluorescence Detector prototype mirror and a Surface Detector station at the exhibition, which are accompanied by signage explaining cosmic ray research and the hybrid detection techniques employed by the Auger Observatory. During its first four days, over 200,000 people attended *La Brújula*. The exhibition is sponsored by Mendoza Province, the federal Ministry of Science and Technology (MINCyT), and the National Atomic Energy Commision (CNEA). The Auger exhibit is similar to the successful presence of the Observatory at the Technopolis exhibition in Buenos Aires last year.

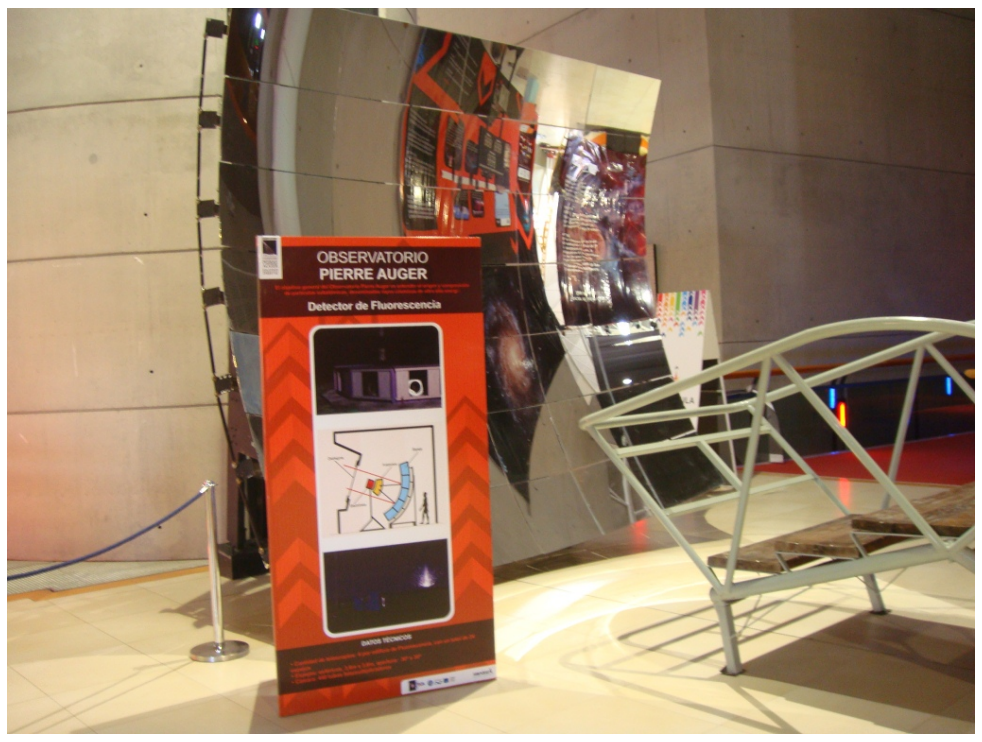

**Figure 7**: Prototype Fluorescence Detector mirror installed at *La Brújula* in Mendoza.

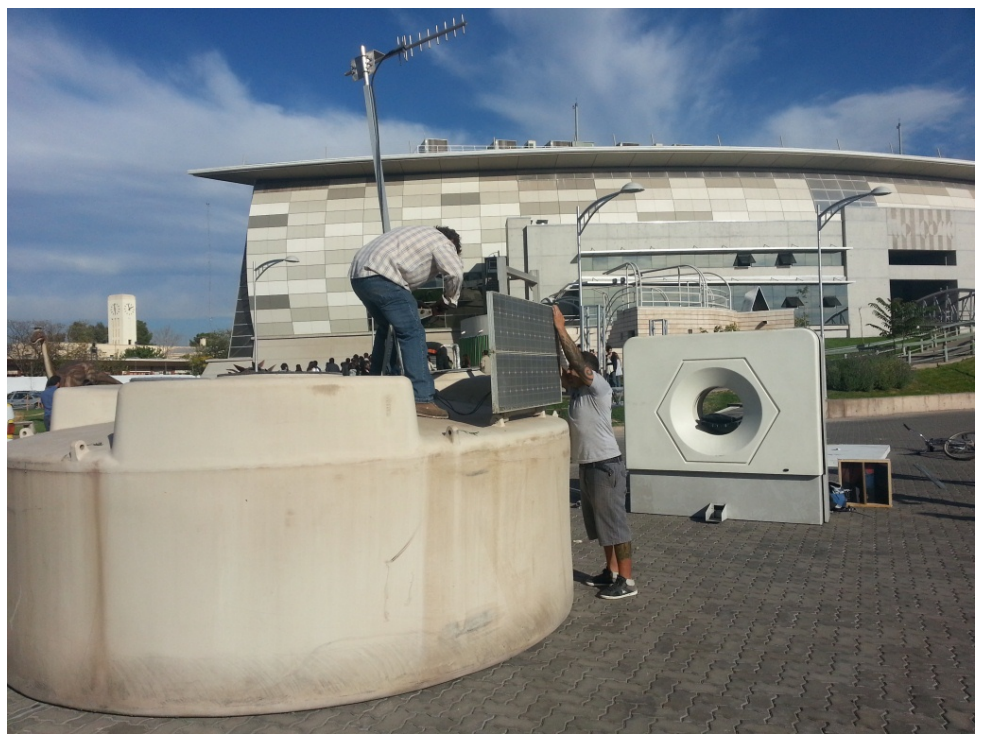

**Figure 8**: Surface Detector station being prepared for CNEA's Auger exhibit at *La Brújula*.

**Auger Observatory public data set**

The online release of extensive air shower data [5] continues to draw attention from around the world. Betwen May 1, 2012, and April 30, 2013, the web site which currently offers Surface Detector (SD) information from 30,000 air showers had 3776 visits among which 70% were new to the site. Auger collaborators in several countries have employed the public data in exercises for students and teachers in educational settings and workshops. As an example, Auger collaborators and physics education reseachers at the University of São Paulo developed a hands-on exercise to construct a 3-D model of a single air shower using simple, inexpensive materials and SD data (station positions and recorded signal strengths and times) provided in the data set [6]. Fig. 9 shows two high-school physics teachers working on their model during a School of Modern Physics in 2011, and Fig. 10 shows the finished model.

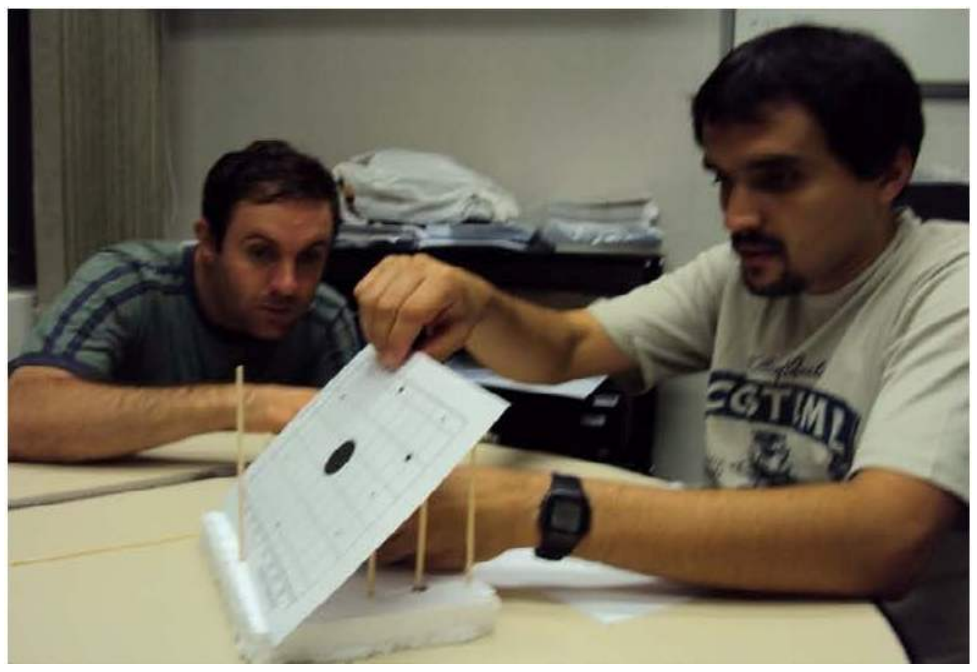

**Figure 9**: High school teachers constructing a 3-D model of an extensive air shower.

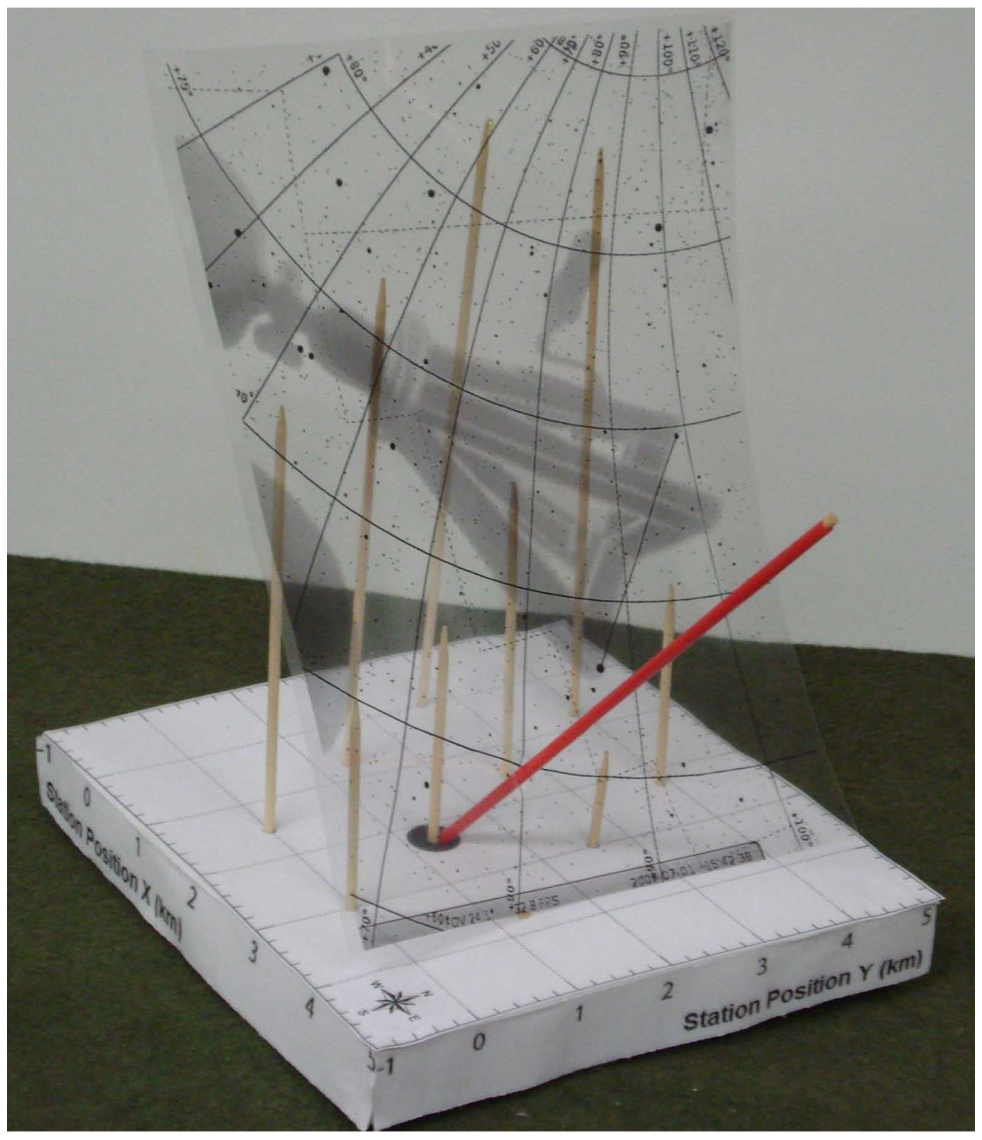

**Figure 10**: Completed 3-D model of an air shower front (plastic sheet), with wooden stick lengths representing the arrival times for SD stations included in the event. The red stick indicates the shower arrival direction pointing to the shower core.

Auger collaborators continue to develop tools and lesson plans for the use of the public data set. Collaborators at RWTH Aachen University have developed a user-friendly analysis package named Visual Physics Analysis (VISPA) [7] which has been field tested by large numbers of undergraduate students using the public data set as input. As an example, students have used VISPA to learn about the supergalactic coordinate system by plotting cosmic ray arrival directions from the public data set, as shown in Fig. 11. Collaborators at the Laboratory of Instrumentation and Experimental Particle Physics (LIP) in Portugal have also developed a detailed users guide for the public data set.

**High altitude balloons**

During the halftime of a University of Nebraska football game against Arkansas State University on September 15,

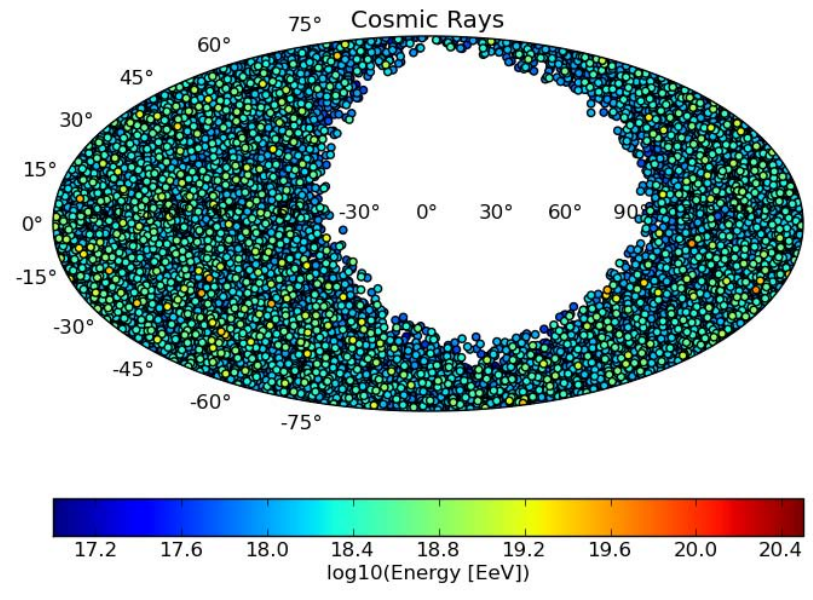

**Figure 11**: Cosmic ray arrival directions from the Auger public data set plotted in supergalactic coordinates using VISPA.

2012, 85,000 spectators viewed the launch of three high-altitude balloons from the football stadium. Suspended below the balloons were a number of "pods" containing experiments designed by high school students from Lincoln and Omaha and undergraduate students from the University of Nebraska-Lincoln (UNL). UNL physics students were responsible for measuring radiation vs. altitude using a small-area portable Geiger counter in one of the pods, in commemoration of the 100$^{th}$ anniversary of the manned balloon flights taken by Victor Hess. The balloons were equipped with cameras, GPS receivers, and radio transmitters to relay data to a ground station in real time (see Fig. 12). The balloons all reached about 95,000 feet (30,000 meters) altitude before bursting, and parachutes carried the payloads to the ground, to be retrieved by chase vehicles. The balloon round-trip time was about 90 minutes. NASA astronaut Clayton Anderson, a Nebraska native who has flown on the International Space Station, attended the event, signed autographs, spoke on local radio about the importance of science education, and helped launch the balloons from the football field. A two-minute video clip of the event can be found on YouTube [8].

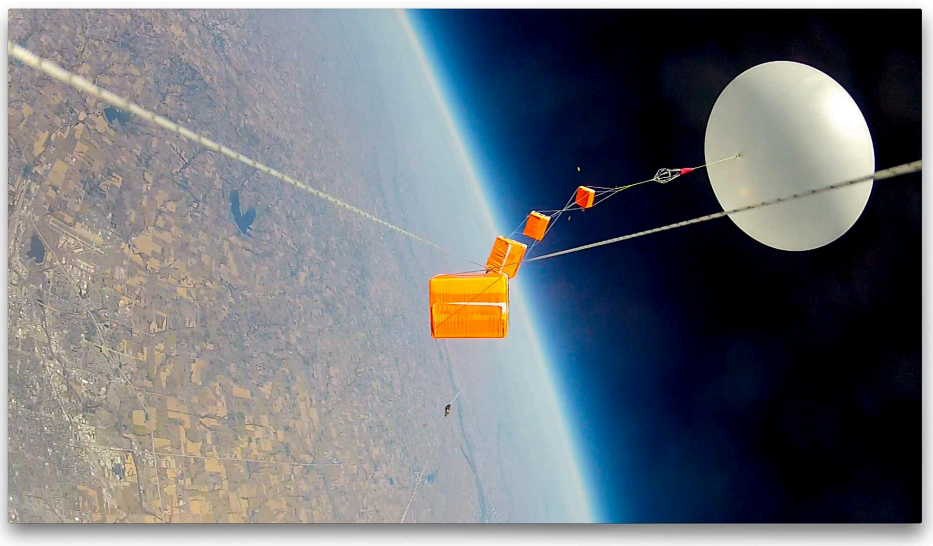

**Figure 12**: View looking up at one of the ballons and the suspended experimental pods. The prominent lines that extend to the edge of the photo are the strings attaching the camera pod to the orange pods above it.

Fig. 13 shows the Geiger counter counts per minute vs. altitude, showing the increase in measured radiation with altitude, as seen by Victor Hess at lower altitudes. Data points from both the ascent and descent are plotted. For altitudes above 60,000 feet, one observes more scatter in the data points and a slight decrease in the average radiation counts. This is attributed to the transition from detecting extensive air shower particles at lower altitudes to detecting primary cosmic-ray particles at higher altitudes.

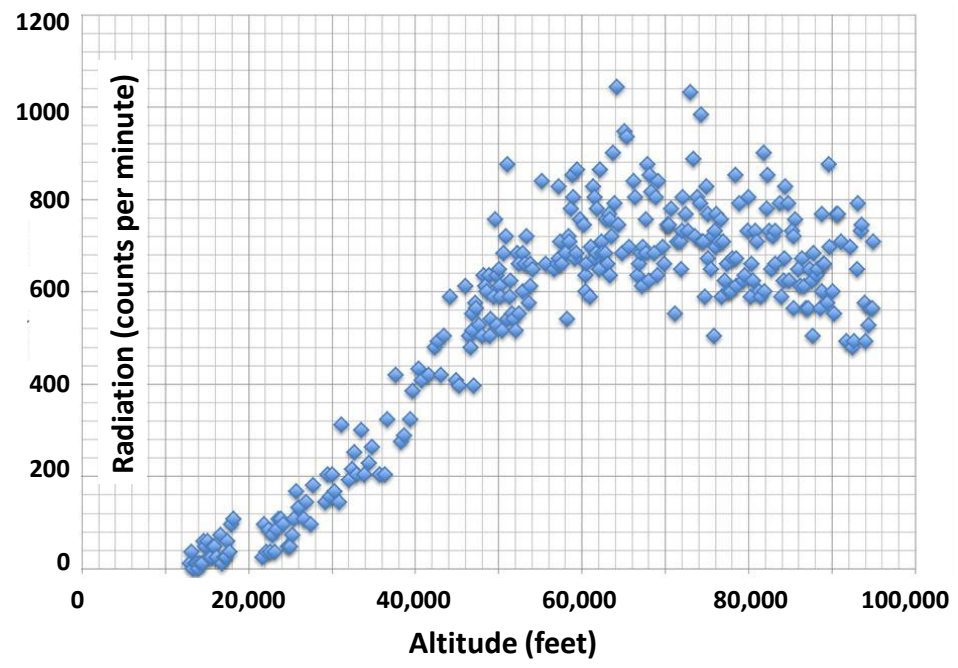

**Figure 13**: Measured Geiger counter counts per minute vs. altitude.

**Partnership with the Helmholtz Alliance**

The Auger Observatory has formed a partnership with the outreach arm of Germany's Helmholtz Alliance for Astroparticle Physics (HAP) [9] led by Astrid Chantelauze who visited the Observatory and covered the Nov. 2012 Science Fair. The Observatory profits from HAP's online resources and its extensive use of social media to communicate news and events related to several astroparticle experiments and theory.

## 6 Conclusions

The Auger Observatory continues to provide unique education and outreach opportunities which expose people of all ages to the excitement of astroparticle physics. Its Visitor Center, Rural Schools Program, and Science Fairs have great local impact near Malargüe, while collaborators around the world ensure that the Observatory's science and successes have international reach.

# 8 Acknowledgements

The successful installation, commissioning, and operation of the Pierre Auger Observatory would not have been possible without the strong commitment and effort from the technical and administrative staff in Malargüe.

We are very grateful to the following agencies and organizations for financial support: Comisión Nacional de Energía Atómica, Fundación Antorchas, Gobierno De La Provincia de Mendoza, Municipalidad de Malargüe, NDM Holdings and Valle Las Leñas, in gratitude for their continuing cooperation over land access, Argentina; the Australian Research Council; Conselho Nacional de Desenvolvimento Científico e Tecnológico (CNPq), Financiadora de Estudos e Projetos (FINEP), Fundação de Amparo à Pesquisa do Estado de Rio de Janeiro (FAPERJ), Fundação de Amparo à Pesquisa do Estado de São Paulo (FAPESP), Ministério de Ciência e Tecnologia (MCT), Brazil; AVCR, MSMT-CR LG13007, 7AMB12AR013, MSM0021620859, and TACR TA01010517 , Czech Republic; Centre de Calcul IN2P3/CNRS, Centre National de la Recherche Scientifique (CNRS), Conseil Régional Ile-de-France, Département Physique Nucléaire et Corpusculaire (PNC-IN2P3/CNRS), Département Sciences de l'Univers (SDU-INSU/CNRS), France; Bundesministerium für Bildung und Forschung (BMBF), Deutsche Forschungsgemeinschaft (DFG), Finanzministerium Baden-Württemberg, Helmholtz-Gemeinschaft Deutscher Forschungszentren (HGF), Ministerium für Wissenschaft und Forschung, Nordrhein-Westfalen, Ministerium für Wissenschaft, Forschung und Kunst, Baden-Württemberg, Germany; Istituto Nazionale di Fisica Nucleare (INFN), Ministero dell'Istruzione, dell'Università e della Ricerca (MIUR), Gran Sasso Center for Astroparticle Physics (CFA), CETEMPS Center of Excellence, Italy; Consejo Nacional de Ciencia y Tecnología (CONACYT), Mexico; Ministerie van Onderwijs, Cultuur en Wetenschap, Nederlandse Organisatie voor Wetenschappelijk Onderzoek (NWO), Stichting voor Fundamenteel Onderzoek der Materie (FOM), Netherlands; Ministry of Science and Higher Education, Grant Nos. N N202 200239 and N N202 207238, The National Centre for Research and Development Grant No ERA-NET-ASPERA/02/11, Poland; Portuguese national funds and FEDER funds within COMPETE - Programa Operacional Factores de Competitividade through Fundação para a Ciência e a Tecnologia, Portugal; Romanian Authority for Scientific Research ANCS, CNDI-UEFISCDI partnership projects nr.20/2012 and nr.194/2012, project nr.1/ASPERA2/2012 ERA-NET, PN-II-RU-PD-2011-3-0145-17, and PN-II-RU-PD-2011-3-0062, Romania; Ministry for Higher Education, Science, and Technology, Slovenian Research Agency, Slovenia; Comunidad de Madrid, FEDER funds, Ministerio de Ciencia e Innovación and Consolider-Ingenio 2010 (CPAN), Xunta de Galicia, Spain; The Leverhulme Foundation, Science and Technology Facilities Council, United Kingdom; Department of Energy, Contract Nos. DE-AC02-07CH11359, DE-FR02-04ER41300, DE-FG02-99ER41107, National Science Foundation, Grant No. 0450696, The Grainger Foundation USA; NAFOSTED, Vietnam; Marie Curie-IRSES/EPLANET, European Particle Physics Latin American Network, European Union 7th Framework Program, Grant No. PIRSES-2009-GA-246806; and UNESCO.